\documentclass[11pt,a4paper,chapterprefix,bibtotoc]{scrartcl}
\pdfoutput=1
\addtokomafont{sectioning}{\rmfamily\boldmath}
\usepackage[german,english]{babel}
\usepackage{fullpage}
\usepackage{afterpage}
\usepackage{graphicx}
\usepackage[tight]{subfigure}
\usepackage{atlasphysics}
\usepackage{amsmath}
\usepackage{amssymb}
\usepackage{mathrsfs}
\usepackage{multirow}
\usepackage{lscape}
\usepackage{rotating}
\usepackage{slashed}
\usepackage{url}
\usepackage{cite}
\usepackage{xspace}
\usepackage{booktabs}
\usepackage{placeins}
\usepackage{flafter}
\usepackage{notes2bib}
\usepackage[hang,flushmargin]{footmisc}
\usepackage{hyperref}

\hypersetup{
  colorlinks=true,
  linkcolor=blue,
  citecolor=blue,
  urlcolor=blue
}

\newcommand{\herwigpp}  {\textsc{HERWIG}++\xspace}

\newcommand{\geant}     {\textsc{Geant4}\xspace}
\newcommand{\jimmy}     {\textsc{JIMMY}\xspace}

\newcommand{\Pythia}    {\textsc{PYTHIA}\xspace}

\newcommand{\W}         {\Wboson\xspace}
\newcommand{\Z}         {\Zboson\xspace}

\newcommand{\antikt}{anti-$k_{\rm T}$}
\newcommand{\Antikt}{Anti-$k_{\rm T}$}

\newcommand{\invpb}     {~\ensuremath{{\rm pb}^{-1}}\xspace}

\newcommand{\invfb}     {~\ensuremath{{\rm fb}^{-1}}\xspace}

\newcommand{\rtrk}             {\ensuremath{r_{\rm trk}}\xspace}
\newcommand{\rtrksubjet}       {\ensuremath{r^{\rm subjet}_{\rm trk}}\xspace}
\newcommand{\Rtrk}             {\ensuremath{R_{\rm trk}}\xspace}

\newcommand{\DeltaR}     {\ensuremath{\Delta R}}

\newcommand{\ptjet}    {\ensuremath{\pt^\mathrm{jet}}\xspace}

\newcommand{\pttrk}   {\ensuremath{\pt^{\mathrm{track}}}}

\newcommand{\pTtrkjet}{\ensuremath{p^{\mathrm{track \; jet}}_{\rm T}}}
\newcommand{\mtrkjet}{\ensuremath{m^{\mathrm{track \; jet}}}}

\newcommand{\Npv}     {\ensuremath{N_{\rm PV}}\xspace}

\newcommand{\bjets}{{\ensuremath{b}\mbox{\rm-jets}}}

\newcommand{\Nsjn}{$N$-subjettiness\xspace}

\newcommand{\vcut}{\ensuremath{v_{\rm cut}}\xspace}

\input{tools/boostedsymbols.tex}
\newcommand{\comment}[1]{}

\newcommand{\figref}[1]       {Figure~\ref{fig:#1}}

\renewcommand{\eqref}[1]         {(\ref{eq:#1})}
\newcommand{\eqsref}[2]         {(\ref{eq:#1}) and (\ref{eq:#2})}
\newcommand{\eqsrange}[2]         {(\ref{eq:#1})--(\ref{eq:#2})}

\newcommand{\tabref}[1]       {Table~\ref{tab:#1}}

\newcommand{\secref}[1]       {Section~\ref{sec:#1}}

\newcommand{\secsref}[2]      {Sections~\ref{sec:#1} and \ref{sec:#2}}

\setcapindent{0pt}   % figure and table captions
\setcapwidth[c]{0.9\textwidth}
\graphicspath{{figures/}}

\begin{document}
%\linenumbers

\pagenumbering{roman}
%\pagenumbering{gobble}

%%%%%%%%%%%%%%%%%%%%%%%%%%%%%%%%%%%%%%%%%%%%%%%%%%%%%%%%%%%%%%%%%%%%%%%%%%%%%%
% Deckblatt
%%%%%%%%%%%%%%%%%%%%%%%%%%%%%%%%%%%%%%%%%%%%%%%%%%%%%%%%%%%%%%%%%%%%%%%%%%%%%%
\thispagestyle{empty}

\begin{center}
{\Large \bf Habilitationsschrift \\
zur \\
Erlangung der Venia legendi \\
f\"ur das Fach Physik \\
der \\
Ruprecht-Karls-Universit\"at \\
Heidelberg \\
}
\end{center}
\vfill
\begin{center}
\Large \bf
vorgelegt von \\
Sebastian Sch\"atzel \\
aus Kiel \\
\vspace{0.5em}
2013
\end{center}

\clearpage
\thispagestyle{empty}
\vglue20em
\clearpage

%%%%%%%%%%%%%%%%%%%%%%%%%%%%%%%%%%%%%%%%%%%%%%%%%%%%%%%%%%%%%%%%%%%%%%%%%%%%%%
% Titel
%%%%%%%%%%%%%%%%%%%%%%%%%%%%%%%%%%%%%%%%%%%%%%%%%%%%%%%%%%%%%%%%%%%%%%%%%%%%%%
\thispagestyle{empty}

\vglue15em
\begin{center}
{\bf \Large Boosted Top Quarks and Jet Structure}
\end{center}

\clearpage
%\thispagestyle{empty}
%\vglue20em
%\clearpage

%%%%%%%%%%%%%%%%%%%%%%%%%%%%%%%%%%%%%%%%%%%%%%%%%%%%%%%%%%%%%%%%%%%%%%%%%%%%%%
% Abstract
%%%%%%%%%%%%%%%%%%%%%%%%%%%%%%%%%%%%%%%%%%%%%%%%%%%%%%%%%%%%%%%%%%%%%%%%%%%%%%
\thispagestyle{empty}

\vglue20em
{\bf \Large Abstract}
\begin{center}
\begin{abstract}
The Large Hadron Collider (LHC) is the first machine that provides high enough energy to produce large numbers of boosted
top quarks. The decay products of these top quarks are confined to a cone in the top quark flight
direction and can be clustered to a single jet. Top quark reconstruction then amounts to analysing the
structure of the jet and looking for subjets that are kinematically compatible with top quark decay.
Many techniques have been developed recently to best use these topologies to identify top quarks in a
large background of non-top jets.
This article reviews the results obtained using LHC data recorded in the years 2010--2012
by the experiments ATLAS and CMS. Studies of Standard Model top quark production and searches for new
massive particles that decay to top quarks are presented.
\end{abstract}
\end{center}

\clearpage

%%%%%%%%%%%%%%%%%%%%%%%%%%%%%%%%%%%%%%%%%%%%%%%%%%%%%%%%%%%%%%%%%%%%%%%%%%%%%%
% Inhaltsverzeichnis
%%%%%%%%%%%%%%%%%%%%%%%%%%%%%%%%%%%%%%%%%%%%%%%%%%%%%%%%%%%%%%%%%%%%%%%%%%%%%%
%\thispagestyle{empty}
%\setcounter{page}{1}
\tableofcontents

\clearpage
\pagenumbering{arabic}

%%%%%%%%%%%%%%%%%%%%%%%%%%%%%%%%%%%%%%%%%%%%%%%%%%%%%%%%%%%%%%%%%%%%%%%%%%%%%%
% Text
%%%%%%%%%%%%%%%%%%%%%%%%%%%%%%%%%%%%%%%%%%%%%%%%%%%%%%%%%%%%%%%%%%%%%%%%%%%%%%

\section{Introduction}
The Large Hadron Collider (LHC) at the European particle physics research
centre CERN in Geneva, Switzerland, is a discovery machine at the energy
frontier. A primary goal, the observation of the Higgs boson, has already been
achieved~\cite{Aad:2012tfa,Chatrchyan:2012ufa}. Other important research topics are searches for deviations from
predictions of the Standard Model of particle physics (SM).
This review is concerned with processes that involve the heaviest known
elementary particle, the top quark. Due to its mass it is expected to play
an important role in electroweak symmetry breaking.

The top quark is reconstructed via its decay products which are collimated
if the top quark Lorentz factor $\gamma = E/m$ is large.\footnote{Throughout
this text, natural units are used with $c = 1 = \hbar$.}
The top quark is then boosted and the decay products are confined to a cone with an opening
angle that depends inversely on $\gamma$. All particles inside the cone can be
clustered into a jet, the structure of which reflects the top quark decay
pattern. The first paper on this topic was published in 1994 by Seymour~\cite{Seymour:1993mx}.

The LHC is the first machine that provides high enough energy to produce
large numbers of boosted top quarks. The study of boosted topologies has seen
an explosion of interest after it had been shown in 2008 by Butterworth et
al.~\cite{Butterworth:2008iy} that the boosted signature makes possible the use
of hadronic decay channels in searches at the LHC. These channels have often
the highest branching ratio but had been deemed infeasible
before because of the large background at a hadron-hadron machine.
In the years following, many aspects of jet structure have been investigated
in the light of the identification of boosted top quarks and boosted \W, \Z, and Higgs bosons.
The crucial point
has always been how well the jets that include the decay products of a heavy particle can be
distinguished from background jets that originate from hard light quarks or gluons,
so-called {\em QCD jets} (events in which the jet activity is given solely by QCD jets are referred to as {\em multijet} events).

After the start of the LHC, the two multipurpose experiments ATLAS and CMS
began studying the behaviour of jet structure techniques in the real world.
Before these techniques could be used in analyses, a number of basic and technical works
had to be carried out, some of which are listed in the following.
The jets in top quark reconstruction are much larger than the ones used to reconstruct the
kinematics of single partons. These large jets, so-called {\em fat jets}, first need to be
calibrated.
Then the precision with which simulations can model the jet structure
observed in the detector had to be quantified so that comparisons with predictions
became meaningful.
 The quality of predictions of the parton shower and hadronisation
needed to be assessed which is especially important in the context of
large jets that contain several hard partons, some of which may be connected by
colour strings.
In addition, the situation is complicated
by the presence of overlay signals that result from slow detector read-out
and additional particles due to multiple inelastic proton-proton ($pp$)
interactions. The size of the large jets makes them especially susceptible to this
so-called {\em pile-up} energy.

The techniques have been studied using SM processes in $pp$ collisions at centre-of-mass
energies $\sqrt{s}=7$ and $8\TeV$. Background samples that are dominated by jets which
do not contain top quark decay products are easily obtained and the jet structure
was studied extensively.
Samples of events with a top quark and an anti-top quark (\ttbar pair)
were obtained through a conventional selection,
i.e., without applying jet structure techniques.
These events were used to test the performance of boosted top quark reconstruction
methods and to evaluate systematic uncertainties. This made
first applications of jet structure techniques in physics analyses possible.
These were searches for new \TeV-scale particles that decay to highly
energetic top quarks. To this day, these types of searches feature prominently
in the analysis of ATLAS and CMS data.

After the repair of the LHC, in collisions at $\sqrt{s}=13\TeV$ in 2015,
the cross sections of other important
processes will be high enough to allow the application of jet structure techniques.
One example is the associated production of a Higgs boson with a \ttbar pair~\cite{Plehn:2009rk}.
The measurement of the production cross section of this process will allow the extraction of the coupling strength
between the Higgs boson and the top quark. This is the largest Higgs coupling in the SM and deviations
from the SM prediction would indicate New Physics.

Jet structure and its application to identify bosons and top quarks is a new and extremely rich field,
both on the phenomenological and on the experimental side. Many questions
need to be addressed and new developments are emerging from the collaboration of
theorists and experimentalists. The most important annual meeting of the community
is the BOOST workshop, of which reports are published in~\cite{Abdesselam:2010pt,Altheimer:2012mn}.
A theoretical review of jet structure methods is given in~\cite{Plehn:2011tg}.
For a review of mostly conventional LHC top quark analyses see~\cite{Schilling:2012dx}.

This article reviews the current state of boosted top quark reconstruction
using jet structure techniques and its application in physics analyses.
For future searches, a new method is presented that overcomes current experimental
limitations in the regime of very high top quark energies. The review closes with an
outlook on the future of the field.

\section{Motivation}
This section introduces the basic ideas behind jet structure methods
and how they can be used to find top quarks.
With the currently available data, the strongest impact of these
techniques is in searches for physics beyond the SM. Two New Physics models are described that
have been used in the analyses discussed in \secref{searches}.

\subsection{Top quark production and decay}
\label{sec:topprod}
\begin{figure}[hbt]
\centering
\includegraphics[width=0.8\textwidth,angle=0]{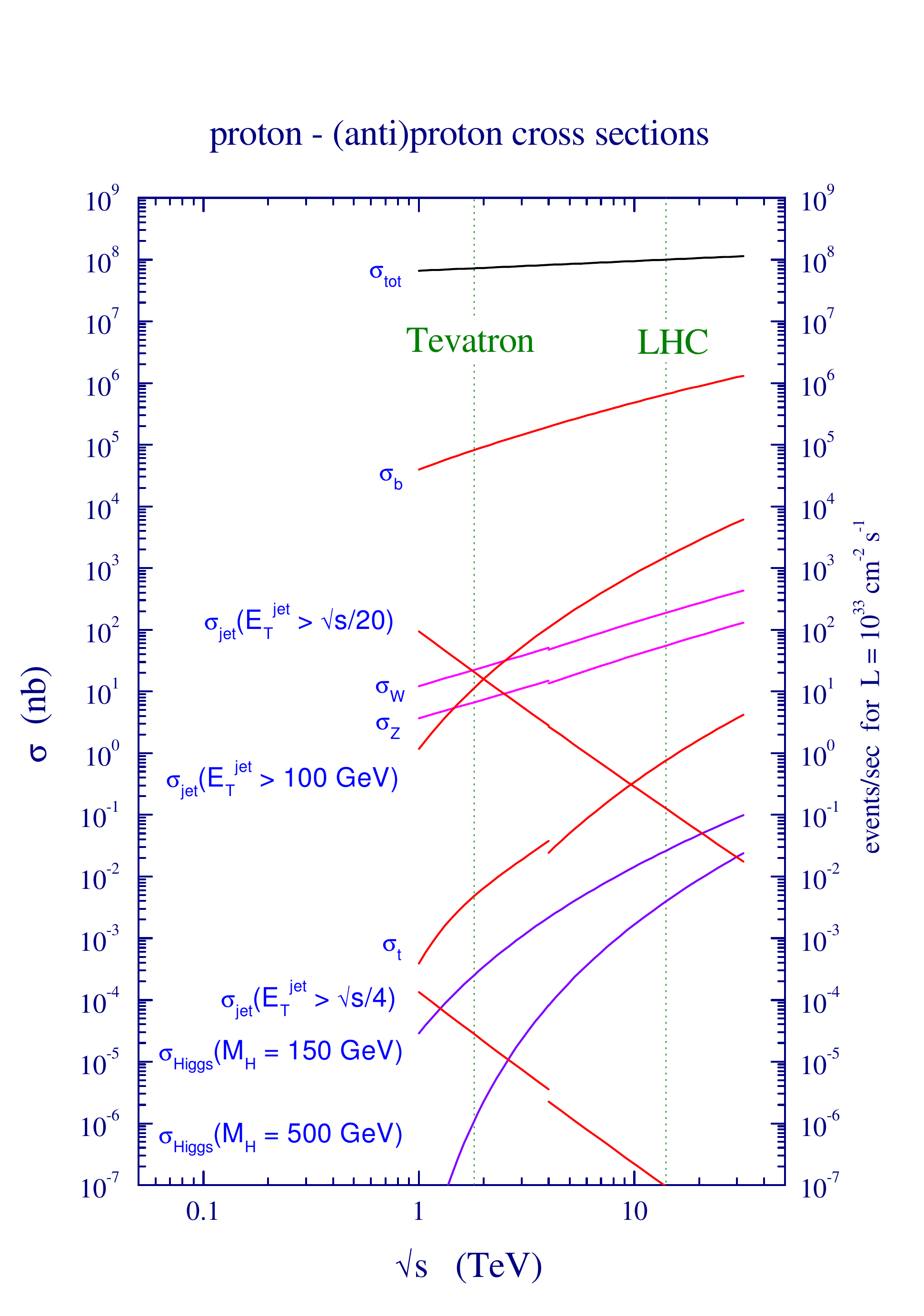}
\caption{Predictions for proton-(anti-)proton SM production cross sections as a function of the centre-of-mass energy $\sqrt{s}$.
The Tevatron energy ($1.96\TeV$) and the design energy of the LHC ($14\TeV$) are indicated.
From~\cite{Campbell:2006wx}.}
\label{fig:Huston_plot}
\end{figure}

\begin{figure}[hbt]
\centering
\includegraphics[width=0.8\textwidth,angle=0]{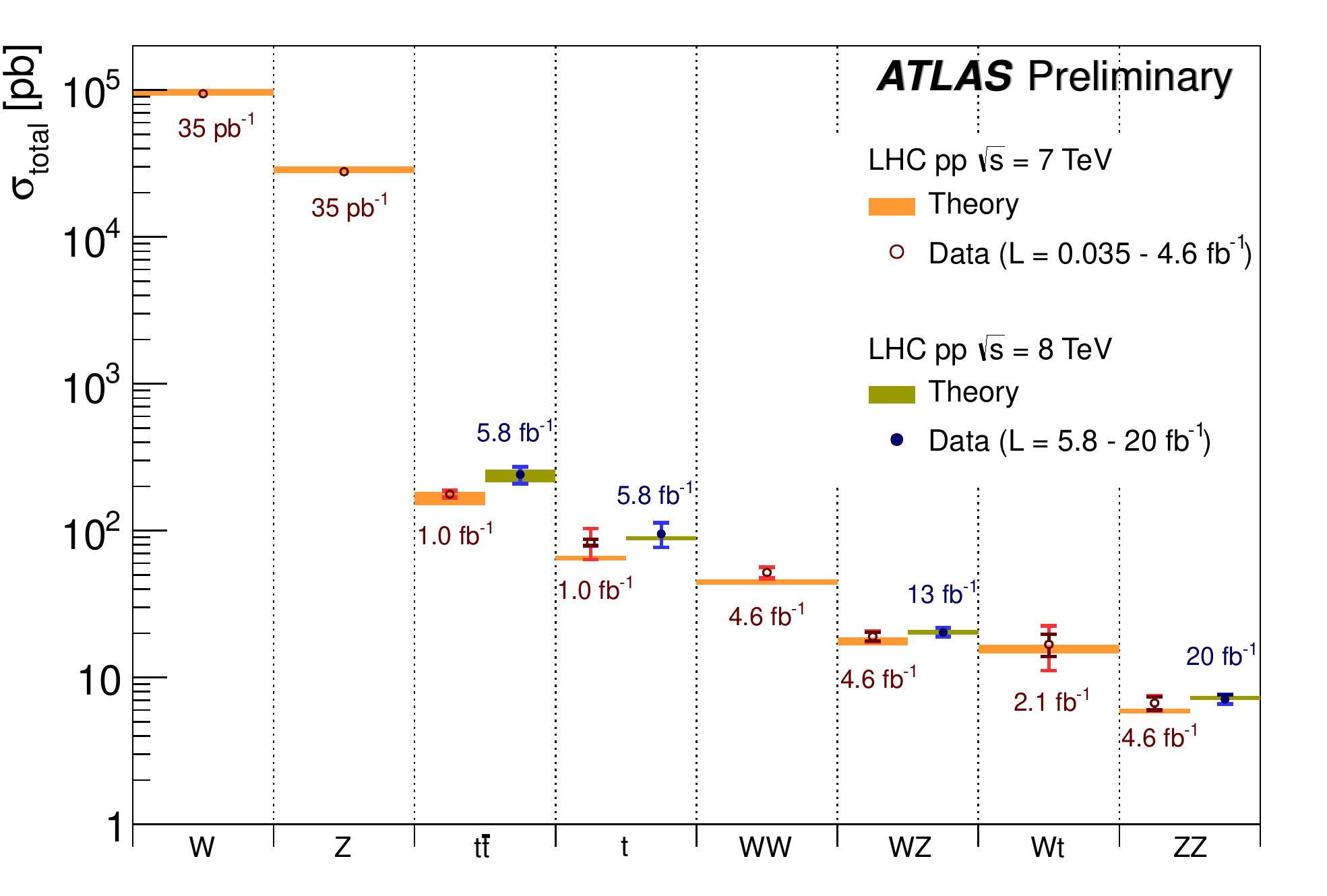}
\caption{Measurements of several SM production cross sections. Also shown are theory predictions.
From~\cite{SM_xs_summary}.}
\label{fig:SM_xs_summary}
\end{figure}

The top quark is the heaviest particle in the SM.
The current Particle Data Group top quark mass is $173.07\pm0.52 ({\rm stat.}) \pm0.72 ({\rm syst.})\GeV$~\cite{Beringer:1900zz,PDG2013web},
obtained from measurements at the Tevatron.
The probability to produce top quarks in particle collisions is small because
of the large mass.
Predictions for (total) SM production cross sections in $pp$ and $p\bar{p}$ collisions
are shown in \figref{Huston_plot}
and ATLAS measurements of SM production processes are summarised in \figref{SM_xs_summary}.
The cross sections are dominated by soft collisions in which little energy is exchanged
and the outgoing particles do not acquire large momenta transverse to the beam line.
The total inelastic $pp$ cross section at $\sqrt{s} = 7\TeV$ was measured to
be $70(7)$~mb\footnote{The notation $70(7)$ is short for $70\pm 7$. In general,
the value in parentheses denotes the uncertainty in the last digits.}~\cite{Aad:2011eu}.
At this energy, hard parton scattering, leading to jets with transverse momentum (\pt) larger than $100\GeV$, has a cross section
of $\approx 4\times 10^5$~pb, approximately five orders of magnitude smaller.
The \W boson production cross section is $\approx 10^5$~pb.
The pair-production cross section for top quarks at approximate next-to-next-to-leading
order (NNLO) in QCD is $167$~pb when using the MSTW2008 NNLO proton parton
densities~\cite{Martin:2009iq} and the \hathor Monte Carlo program~\cite{Aliev:2010zk}.
This is approximately a factor $1/2000$ smaller than the jet cross section.
This illustrates that top quark physics at the LHC has to fight large
backgrounds from jets and/or \W boson production.

The top quark decay width predicted by the SM at next-to-leading order (NLO) is
$\approx\!1.3\GeV$, corresponding to a lifetime of $\approx\!0.5\times10^{-24}\,$s~\cite{QuadtLiss2010}.
The CKM matrix element $V_{tb}$ is estimated to be larger than 0.999, indicating
that almost all top quarks decay according to $t\rightarrow Wb$.
The \W boson decays in $67\%$ of the cases to two quarks (hadronic decay), the branching
ratio to a neutrino and a lepton is $11\%$ for each lepton flavour.
The tau lepton decays in $34\%$ of the cases to an electron or a muon and these
cases look experimentally like direct \W boson decays to electron or muon.
The electron and muon channels (including the corresponding $\tau$ decays)
are collectively referred to as leptonic decay.

For decays of pairs of top quarks\footnote{Throughout the text, the word `quark'
is used to denote also the antiquark. In addition, in decays like $t\to b q q$
it is understood that the quarks from \W boson decay are of different flavour.}
(\ttbar) the decays are to $45\%$ hadronic (both top quarks decay hadronically) and to
$35\%$ semileptonic ($e/\mu$+jets, one top quark decays leptonically).
The rest of the decays are dileptonic decays and hadronic tau decays.

\subsection{Boosted particle decays}

\begin{figure}[hbt]
\centering
\includegraphics[width=0.8\textwidth,angle=0]{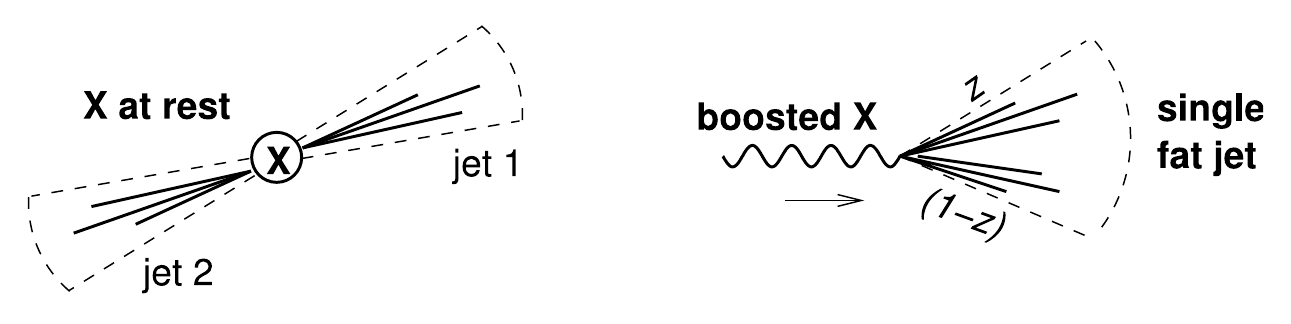}
\caption{Schematic diagrams of the decay of a particle $X$ to two jets
when the particle is at rest (left) and boosted (right). In the boosted case,
the two jets merge into a single fat jet.
The fraction of the particle momentum taken by one of the jets is denoted by $z$.
From~\cite{Salam_boost}.}
\label{fig:Gavin_boost}
\end{figure}

The LHC can produce particles with kinetic energies much larger than the electroweak
scale. In the laboratory frame, the decay products of such a particle are collimated in the particle
flight direction. This poses new experimental challenges compared to
decays at rest. The difference between decay at rest and boosted decay is illustrated in
\figref{Gavin_boost}. A particle $X$ decays to two jets. If $X$ is at rest,
the two jets are well separated and will be detected as two distinct jets.
If $X$ is boosted, the two jets are collimated in the forward direction.
If the boost of $X$ is large enough, the two jets merge into a large single jet
({\em fat jet}). The structure of this fat jet contains information about
the decay.

For a two-body decay, the distance of the decay products in
rapidity-azimuth space\footnote{The rapidity of a particle is
defined as $y = 0.5 \times \ln [ ( E + p_z )/( E - p_z ) ]$, in which $E$
denotes the particle energy and $p_z$ is the component of the momentum along
the beam direction. The azimuthal angle $\phi$ is measured in the plane
transverse to the beam direction and the polar angle $\theta$ is measured with
respect to the beam direction. The pseudorapidity is defined as
$\eta = 0.5 \times \ln [ ( p + p_z )/( p - p_z ) ] = -\ln \tan(\theta/2)$.
Transverse momentum and energy are defined as
$\pt = p \times \sin \theta$ and $\ET = E \times \sin \theta$, respectively.
The distance between two objects in rapidity-azimuth space $(y,\phi)$ is given
by $\DR = \sqrt{(\Delta y)^2 + (\Delta \phi)^2}$ and the distance in
pseudorapidity-azimuth space $(\eta,\phi)$ is denoted by $\DeltaReta$.
The two distances are identical for massless objects.} $(y,\phi)$ is given by
\begin{equation}
\DeltaR \approx 2m/\pt \label{eq:size}
\end{equation}
in which $m$ and \pt are the mass and transverse momentum of the
decaying particle, respectively.
For a \W boson with $\pt = 200\GeV$, the distance is $\DR = 0.8$ and
$\DR=0.5$ for $\pt = 320\GeV$. The conventional jets used in the LHC experiments
cover distances $\DR=0.4$--$0.6$. With these jets, the two decay products of
a highly energetic \W boson cannot be resolved and conventional reconstruction
techniques fail. The same is true for the decays of other boosted particles,
like \Z bosons, Higgs bosons, and top quarks.

\begin{figure}[hbt]
\centering
\includegraphics[width=0.45\textwidth,angle=0]{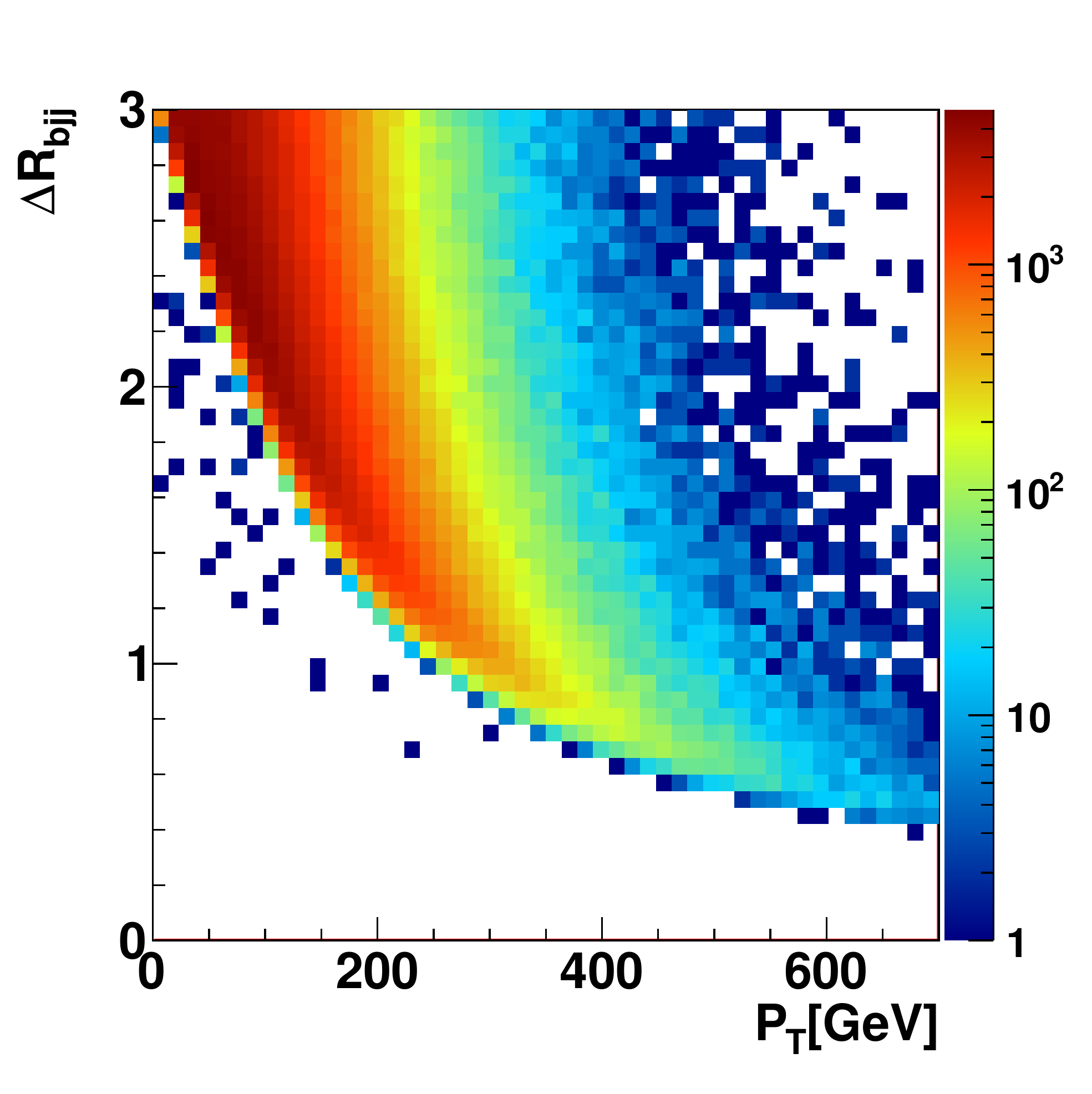}
\caption{The minimal size of a \ca jet that contains the decay quarks
 in the decay $t \rightarrow bqq$ as a function of the top quark \pt. From~\cite{Plehn:2010st}.}
\label{fig:topsize}
\end{figure}

The minimal size of a \ca jet that contains the decay quarks in hadronic top quark decay, $t\to bqq$,
is shown in \figref{topsize} as a function of the top quark \pt.\footnote{The
size $\DeltaR_{bjj}$ is defined as follows: the two closest quarks
$k$ and $l$, separated by the distance $\DeltaR_{kl}$, are combined by adding their four-momenta to obtain a vector $m$.
The distance between $m$ and the third quark $n$ is calculated and
$\DeltaR_{bjj}$ is defined to be the maximum of this distance and $\DeltaR_{kl}$.}
This size corresponds to the minimal radius parameter $R$ that would have to be used
in a jet to capture the three quarks (cf. \secref{jetalgorithms}).
Even jets with $R=1.5$ catch only a small fraction of top quark decays with $\pt=200\GeV$.

\label{sec:top_decay}

Conventional techniques that
rely on the detection of isolated decay products fail when those products are
collimated and merged into single reconstructed objects, such as jets.
It is the analysis of the internal structure of these objects that offers
a way to identify and measure boosted particles.

\begin{figure}[hbt]
\centering
\includegraphics[width=0.5\textwidth,angle=0]{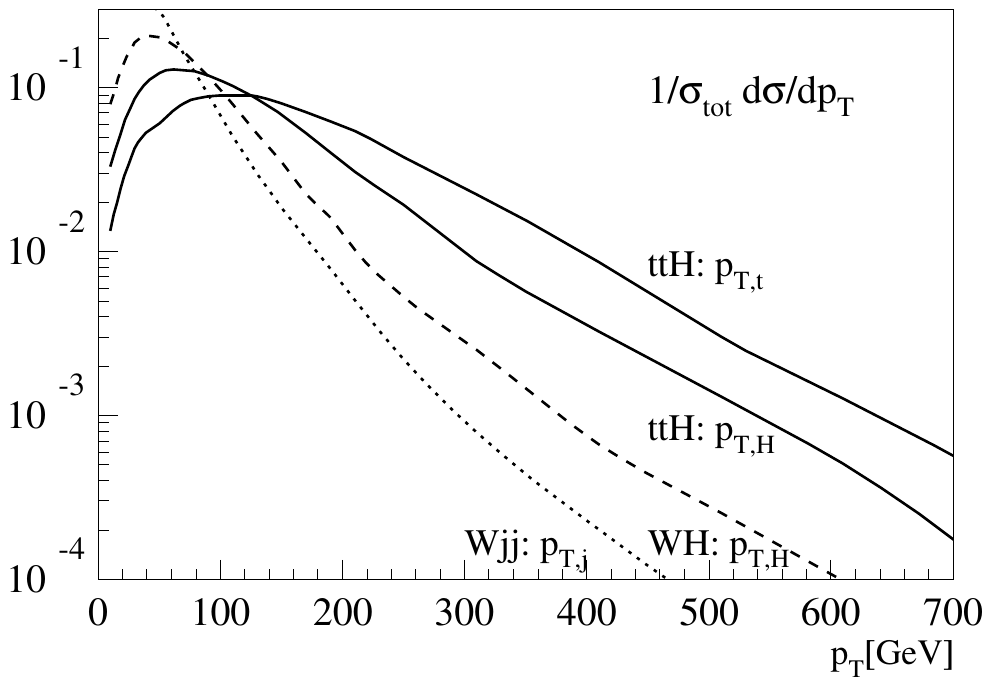}
\caption{Normalised transverse momentum distributions in simulated LHC events
at $\sqrt{s} = 14\TeV$  for associated \ttbar pair plus Higgs boson production (ttH) and background
processes (\Wpjets production, Wjj, and W+H production). Shown are the
\pt distributions of the top quarks ($p_{{\rm T,t}}$), the Higgs boson ($p_{{\rm T,H}}$)
and the leading \pt jet ($p_{{\rm T,j}}$) as predicted by \madgraphherwigpp.
From~\cite{Plehn:2009rk}.}
\label{fig:fj_lH_pt}
\end{figure}

Boosted techniques are also useful if the background falls more rapidly
with \pt than the signal.
An example of this kind is the analysis of
associated Higgs boson production with a \ttbar pair~\cite{Plehn:2009rk}.
Shown in \figref{fj_lH_pt} are the \pt spectra of the involved particles:
the distributions for the Higgs boson and the top quarks are harder
than those for the background. An analysis in the boosted regime can therefore
have the advantage of an enhanced signal-to-background ratio (S/B).

\subsection{Higgs mass fine-tuning}

Boosted particles are also frequently encountered in extensions of the SM. In these
theories and models, new heavy particles are proposed with masses at or above the
\TeV{} scale. The SM particles to which these new states decay are highly boosted,
making jet structure techniques ideal discovery tools. The new theories
are introduced to overcome shortcomings of the SM, such as the hierarchy problem
and the Higgs mass fine-tuning.

In the SM, electroweak symmetry breaking (EWSB) occurs due to the introduction
of a scalar weak isospin Higgs doublet. One component of the doublet has a non-vanishing
vacuum expectation value (VEV). Upon expanding the complex doublet Higgs field in
four real fields about the VEV, three of the fields are massless (the Nambu-Goldstone bosons, NGB)
and one field gains mass (the Higgs boson).
In the unitary gauge, the three NBGs are {\em eaten} by the vector bosons $W^+, W^-, Z$ and give them mass.

One of the puzzles of the SM is the fine-tuning of the radiative corrections
to the Higgs mass. These large corrections appear because the Higgs boson
is a scalar. By contrast, fermion masses are protected by a {\em custodial} symmetry as follows~\cite{Hill:2002ap}.
The fermion spinors can be decomposed into left- and right-handed
components (using the projection operators $P_{R,L} = (1\pm \gamma^5)/2$).
In the massless limit, the free Lagrangian decomposes into two terms, one for each chiral component:
\begin{equation}
\Lagr = \overline{\psi} i \dslash \psi = \overline{\psi_L} i \dslash \psi_L + \overline{\psi_R} i \dslash \psi_R \,. \label{eq:chiralSym}
\end{equation}
 This Lagrangian is
invariant under two independent global symmetry transformations.
For example, for the massless electron in QED, the symmetries are
$U(1)_L$ and $U(1)_R$, which act only on the left- and right-handed components, respectively.
The theory is {\em chiral} because it distinguishes between left and right handedness
and $U(1)_L \times U(1)_R$ is called {\em chiral symmetry}.
An explicit mass term $-m \overline{\psi}\psi$ would couple both chiralities and break $U(1)_L \times U(1)_R$.
It turns out that the chiral symmetry also forbids a finite electron mass to
be generated by radiative corrections: all corrections to the mass are multiplicative
and therefore only relevant if the mass is non-zero.
The chiral symmetry is said to be the custodial symmetry that protects the electron mass.

Typically, scalar particles do not have
a custodial symmetry and perturbative corrections can produce a large mass.
Important exceptions are~\cite{Hill:2002ap}:
(i) Nambu-Goldstone bosons which are protected by the spontaneously broken
global symmetry; (ii) composite scalars which form at a strong scale
could receive only additive corrections to their mass of order this scale;
(iii) scalars which have fermion partners are protected by the chiral
symmetry of their partner, like in Supersymmetry (SUSY)~\cite{Fayet:1976et,Fayet:1977yc,Farrar:1978xj,Fayet:1979sa,Dimopoulos:1981zb}.

\begin{figure}[hbt]
\centering
\includegraphics[width=0.8\textwidth,angle=0]{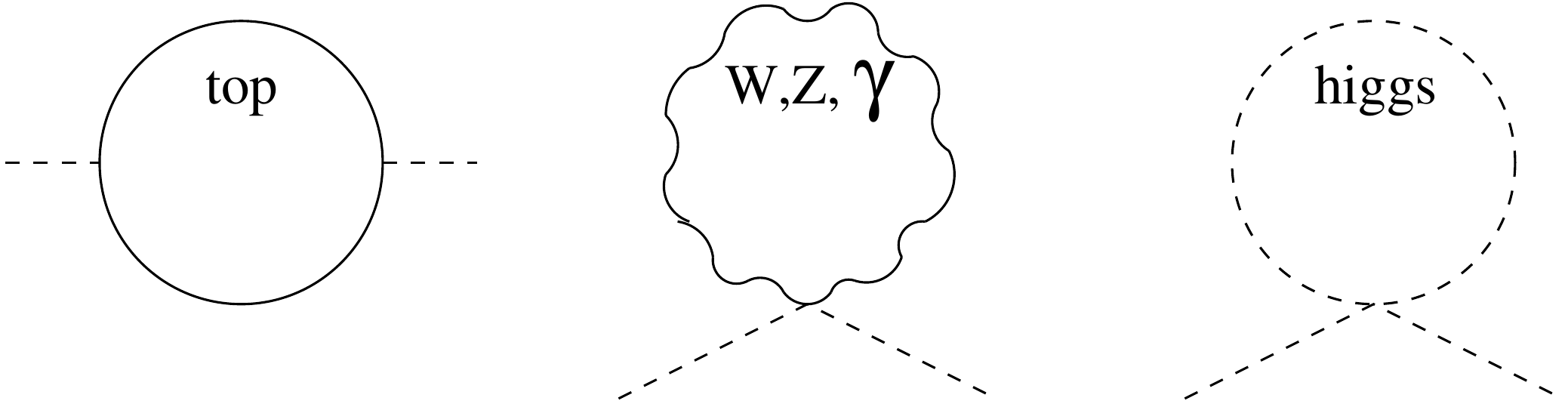}
\caption{The largest radiative corrections to the Higgs boson mass in the SM.
From~\cite{Schmaltz:2005ky}.}
\label{fig:mH_corrections}
\end{figure}

The largest radiative corrections to the SM Higgs boson mass are shown in
\figref{mH_corrections}. They result from loops of top quarks,
\W and \Z bosons, and the Higgs self coupling.
The momentum integration in the loops is cut off at some scale $\Lambda$.
Numerically, the corrections to $m_H^2$ are~\cite{Schmaltz:2005ky}
\begin{eqnarray}
& & {\rm at\ } \Lambda = 10\TeV: \nonumber \\
{\rm top\ quark \ loop} &\quad
-\frac{3}{8 \pi^2} \lambda_t^2 \Lambda^2\ &\approx\ -(2\TeV)^2
\label{eq:loop_top} \\
SU(2) {\rm\ gauge\ boson\ loops} &\quad
{9 \over 64 \pi^2} g^2 \Lambda^2\ &\approx \ (700\GeV)^2
\label{eq:loop_ew} \\
{\rm Higgs\ boson\ loop} & \quad
{1 \over 16 \pi^2} \lambda^2 \Lambda^2\ &\approx \ (500\GeV)^2
\label{eq:loop_higgs}
\end{eqnarray}

with $\lambda_t$ the top quark Yukawa coupling to the Higgs boson,
$g = e/\sin\theta_W$ and $\lambda$ the Higgs self coupling.
At $\Lambda = 10\TeV$, which is approximately the centre-of-mass energy of the LHC,
the observed Higgs boson mass of $125\GeV$ is obtained from
a bare mass $m_0$ and corrections:
\begin{eqnarray}
m_{\rm obs}^2 & = & m_0^2 + \Delta m_H^2 \nonumber \\
              & = & m_0^2 - (2\TeV)^2 + (700\GeV)^2 + (500\GeV)^2 \nonumber \\
              & = & m_0^2 - (256-31-16) \, (125\GeV)^2 \nonumber \\
              & = & m_0^2 - 209\, m_{\rm obs}^2 \nonumber \\
        m_0^2 & = & 210\, m_{\rm obs}^2
\end{eqnarray}
The ratio of the bare mass to the mass correction is
\begin{equation}
m_0/|\Delta m_H| = \sqrt{210}/\sqrt{209} = 1.002 \,.
\end{equation}
This means that the bare mass has to be fine-tuned to the mass correction at the level of $0.2\%$.
The subtraction of two finely tuned large variables is unnatural and
an impetus to develop new theories. The fact that the level of fine-tuning
is already problematic at $10\TeV$ prompts hopes of finding New
Physics at the LHC.
Of course, if the cut-off scale is taken to be the Planck scale of $10^{19}\GeV$
then the problem is all the worse.
The top quark is at the heart of this problem because it contributes the largest mass correction.
The corrections due to the other fermions are much smaller because their Yukawa
couplings are $\ll 1$.

Different extensions of the SM exist that tackle the fine-tuning problem.
Supersymmetry introduces partner particles which differ by $1/2$ in the spin quantum number
such that their loop contributions cancel those of the SM particles.
Little Higgs models~\cite{ArkaniHamed:2001nc} (and references in~\cite{Schmaltz:2005ky})
generate the Higgs boson as a (pseudo-)Nambu-Goldstone boson
of a new approximate global symmetry that is collectively broken. The Higgs boson mass
is then protected by this symmetry, to the extend that the divergence at the 1-loop level is only logarithmic
and not quadratic as in \eqsrange{loop_top}{loop_higgs}.

Other models that will be tested with the data presented in this review are
technicolor and warped extra dimensions.

\subsection{Technicolor}
\label{sec:technicolor}

In a simplified model of QCD with only $u$ and $d$ quarks,
a mechanism was observed~\cite{Weinberg:1975gm,Susskind:1978ms}
that dynamically creates a scalar
as a composite particle. The mass of this scalar is protected because the scalar is
composed of two fermions.
The description below follows~\cite{Hill:2002ap}.

In the massless limit, the
two quarks are arranged in a doublet
and the Lagrangian is invariant under transformations of the
chiral symmetry $SU(2)_L \times SU(2)_R$.
At low energies, the strong coupling is large and binds quarks and antiquarks
into a composite scalar state (quark condensate).
This state can be taken to have a non-vanishing VEV and spontaneously
break the chiral symmetry. The quark condensate plays a role analogous to that of the
Higgs doublet in SM EWSB. Expanding the quark condensate about the VEV, three
massless quark-antiquark states occur that can be identified with the
three pions. The pions are the NGBs of the spontaneous breaking of the chiral symmetry.
This finding is the idea behind technicolor~\cite{Weinberg:1975gm,Susskind:1978ms}: the Higgs field is
not fundamental but a composite, a condensate of fermions.

Technicolor is a new force that is modelled after QCD
and exists at scales larger than the electroweak scale.
At the electroweak scale, the techniquarks condensate
to a scalar field. This field breaks the technicolor symmetry
and the technipions are eaten by the \W and \Z bosons.

As in the SM, the fermion masses $m_f$ are given by Yukawa couplings $\lambda_f$:
$m_f=\lambda_f v/\sqrt{2}$, in which $v/\sqrt{2}$ is the Higgs VEV.
The parameter $v$ is related to the Fermi constant $G\approx 1.166\times 10^{-5}\GeV^{-2}$
and is given by $v = 1/\sqrt{G \sqrt{2}} \approx 250\GeV$~\cite{MandlShaw}.

The top quark is special because its mass corresponds approximately
to the VEV so that $\lambda_t \approx 1$.
This has inspired EWSB models in which
the top quark plays a special role, such as topcolor and topcolor-assisted technicolor~\cite{Hill:1991at,Hill:1994hp}.
The following summary is based on the introduction in~\cite{Harris:1999ya}.
Topcolor is a new force, given by $SU(3)_1 \times SU(3)_2$,
in which group 1 couples the first two generations and group 2 the third generation
and the coupling in group 2 is much stronger.
The breaking of global $SU(3)_1 \times SU(3)_2$ to the SM $SU(3)_C$ produces
eight NGBs, the {\em topgluons}, which couple mainly to \bbbar and \ttbar.
To remove the degeneracy between top and bottom quarks,
a new neutral gauge boson, the topcolor \Zprime, is introduced. It provides an
attractive interaction between \ttbar and a repulsive interaction
between \bbbar. This is achieved by introducing a new $U(1)_1 \times U(1)_2$
symmetry which is broken to the SM $U(1)_Y$. The \Zprime is the gauge boson
of the $U(1)_i$. Different \Zprime models can be obtained by changing the
assignment of the generations to the two groups~\cite{Harris:1999ya}.
The width of the topcolor \Zprime boson is typically $\Gamma = 1.2\% \times m_{\Zprime}$.

\subsection{Warped Extra Dimensions}
\label{sec:wed}
\begin{figure}[hbt]
\centering
\includegraphics[width=0.5\textwidth,angle=0]{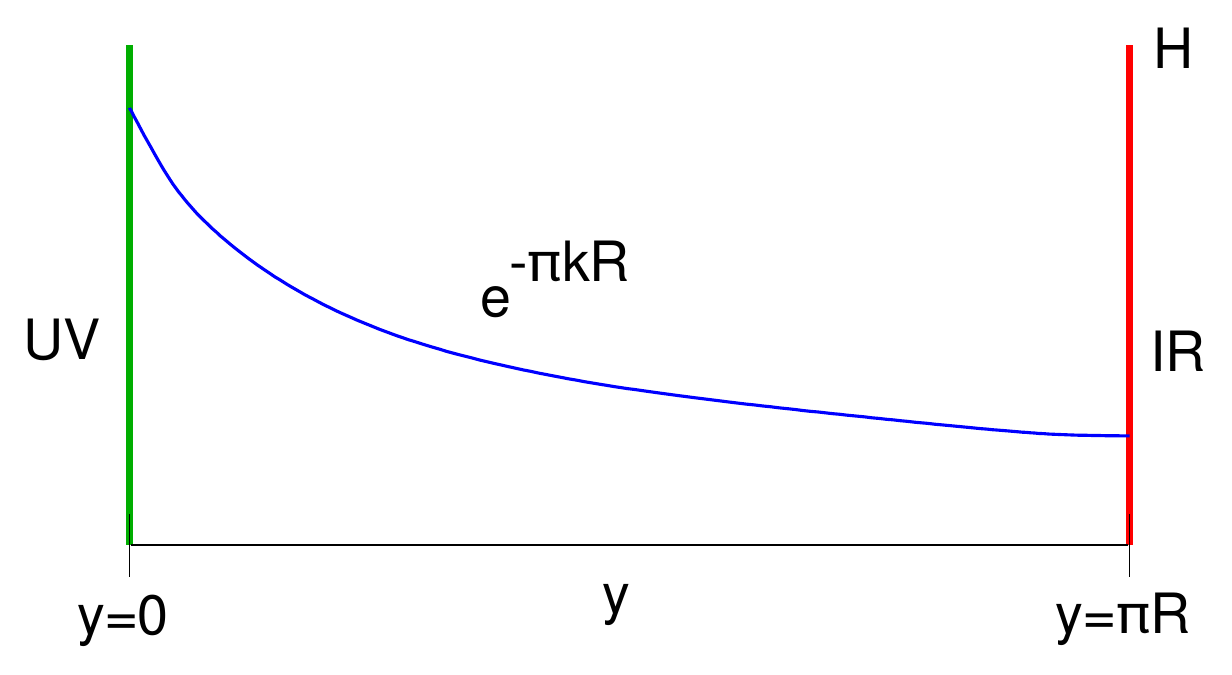}
\caption{Schematic view of Randall-Sundrum spacetime. Two four-dimensional branes
are separated in a fifth dimension which is denoted by the coordinate $y$.
Masses on the {\em infrared brane} (IR) at $y=\pi R$ are reduced by a {\em warp factor} $\exp(-\pi k R)$ with respect to the same mass
on the {\em ultraviolet brane} (UV).
The quantity $R$ is the compactification radius of the fifth dimension and $k$ the spacetime curvature.
The Higgs boson (H) is located on the IR brane.
After~\cite{Morrissey:2009tf}.}
\label{fig:branes}
\end{figure}

Another example for a theory beyond the SM is that of warped extra dimensions~\cite{Randall:1999ee,Randall:1999vf}.
Introductory overviews
of the theory are for example given in~\cite{Morrissey:2009tf,GiudiceWells2010}.
A fifth dimension, denoted by the coordinate $y$, separates two four-dimensional
{\em branes}:
the {\em ultraviolet (UV) brane} at $y=0$ and the {\em infrared (IR) brane} at $y=\pi R$.
Here $R$ is the compactification radius. The space between the branes is called {\em bulk}.
The four-dimensional metric depends on $y$ through a factor $\exp(-2 k |y|)$ in which
$k$ is the spacetime curvature.
The SM particles live on the IR brane (the Higgs boson) or near it (other particles).
A mass on the IR brane is smaller compared to the same mass on
the UV brane by a factor $\exp(-\pi k R)$ (called the {\em warp factor}). A
schematic view is shown in \figref{branes}.

The Planck mass \Mpl on the UV brane is $\approx\!10^{19}\GeV$.
On the IR brane it is only $1\TeV$ if $k R = 11$.
Effectively, fine-tuning of the radiative corrections to the Higgs mass
is avoided by lowering the cut-off scale to $1\TeV$.

The fifth dimension is assumed to be periodic ($y \in ( - \pi R, \pi R ]$ with
$y=-\pi R = \pi R$).
All fields in the five dimensions (specified by five coordinates, $x^\mu$ and $y$)
can be Fourier-expanded in a series of fields that depend only on $x^\mu$~\cite{Sellerholm}:
\begin{equation}
F(x^\mu,y) = \sum_{n=-\infty}^{\infty} \ F_n(x^\mu) e^{i n y/R}\,.
\end{equation}
The $F_n$ are the Kaluza-Klein (KK)~\cite{Kaluza:1921tu,Klein:1926tv} excitations of $F$.
Their masses $m_n$ are given by $m^2_n = n^2/R^2 + m_0^2$ in which
$m_0$ is the mass of the {\em zero mode}, the SM particle.
For new KK particles to exist with masses of the order $1\TeV$,
the radius $R$ has to be of the order $10^{-19}$~m.

At the LHC, the KK particle with the largest production cross section
is the first excitation of the gluon.
The KK gluon (\gKK) is the most strongly coupled KK particle and is produced resonantly
in the $s$-channel from two quarks. It is localised near the IR brane.

\begin{figure}[hbt]
\centering
\subfigure[]{
   \includegraphics[width=0.33\textwidth,angle=-90]{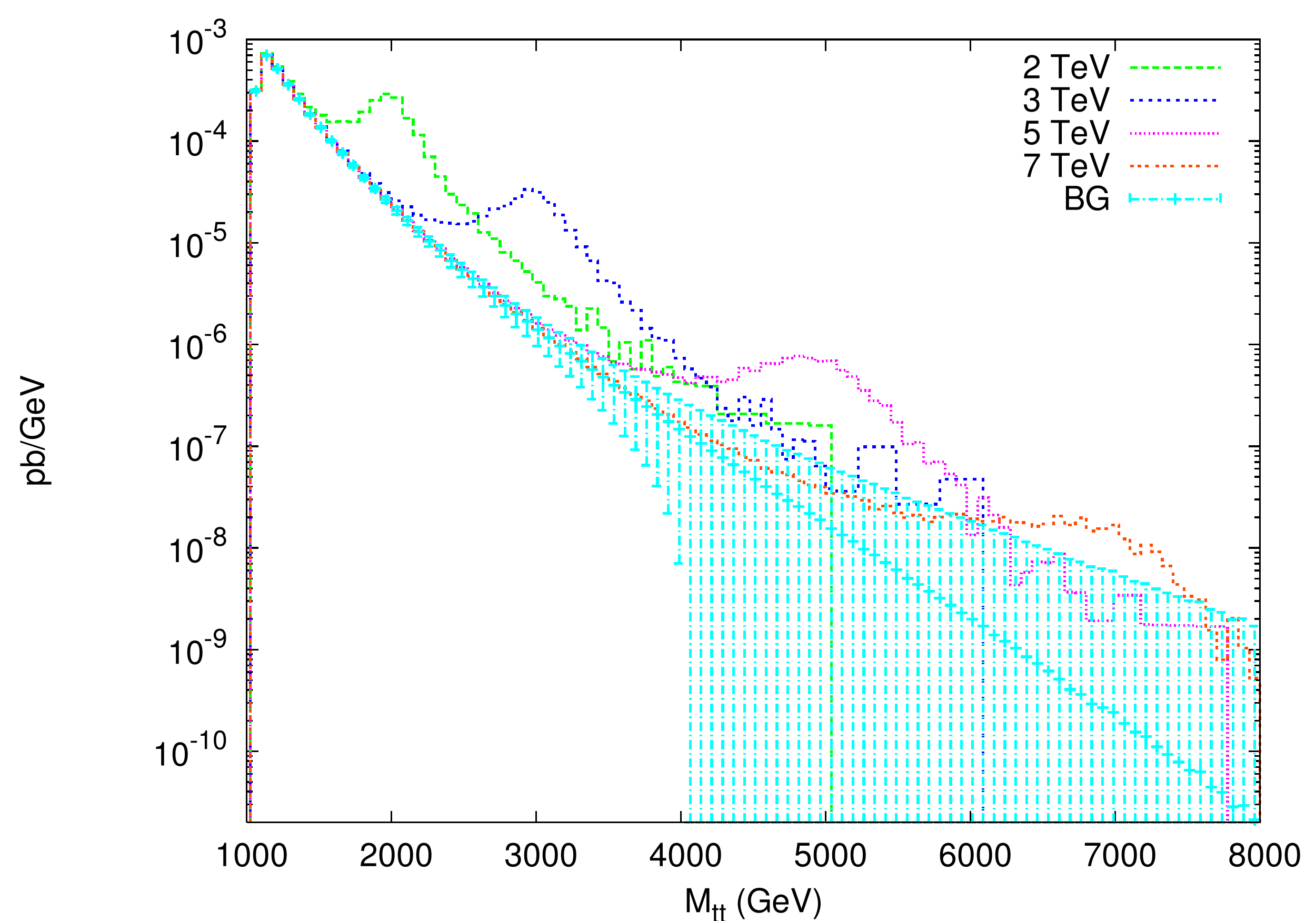}
}
\subfigure[]{
   \includegraphics[width=0.33\textwidth,angle=-90]{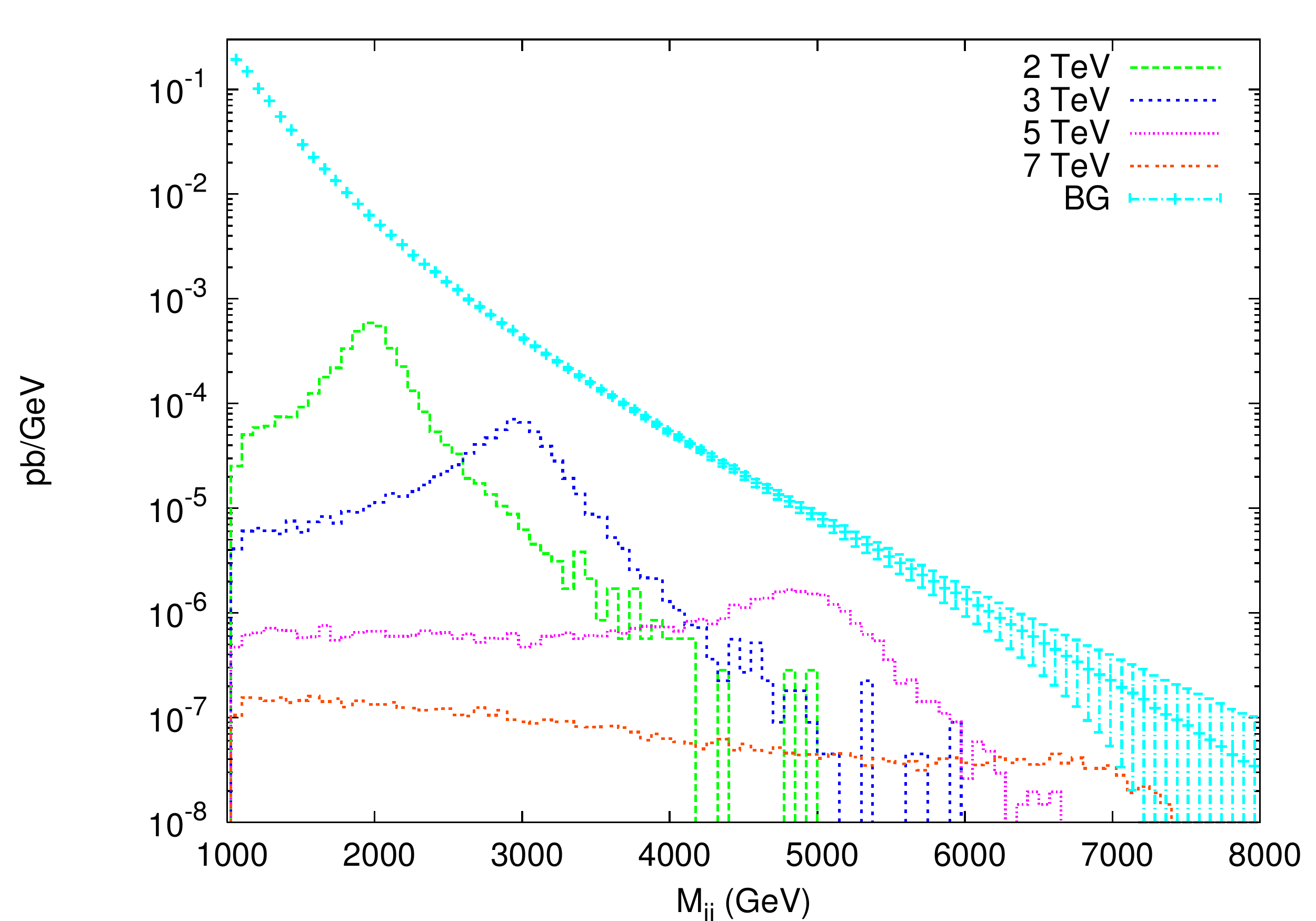}
}
\caption{Invariant mass of \ttbar pairs from decays of Kaluza-Klein gluons with different
masses (generated with \madgraph). The distributions are compared to those from SM background (\pythia):
a) \ttbar production and b) multijet production for which
the invariant mass of the leading \pt dijets is plotted ($|\eta^{\rm jet}|<0.5$
and leading $\pt > 500\GeV$).
The events are produced in LHC collisions at $\sqrt{s}=14\TeV$ and
the background uncertainties correspond to $100\invfb$.
From~\cite{Lillie:2007yh}.}
\label{fig:RS_resonances}
\end{figure}

The SM particles live at different distances to the IR brane.
The distances are free para\-meters of the theory and are adjusted manually
to obtain the observed masses.
The masses are determined by the overlap of the particle wave function with
that of the Higgs boson which lives on the IR brane.
The top quark is the SM fermion closest to the IR brane because it has the largest mass.
Because the \gKK also lives near the IR brane, the consequence is that
it prefers to decay to \ttbar pairs.
Shown in \figref{RS_resonances} are distributions of the invariant mass
of the \ttbar pair in LHC collisions for different \gKK masses $m_{\gKK}$ from $2$
to $7\TeV$.
The width of the KK gluon is $\Gamma_{\gKK}/m_{\gKK} = 17\%$.
For these high resonance masses, the top quark \pt is $\approx\!m_{\gKK}/2$ and
the decay products are strongly collimated. Each top quark is therefore
reconstructed as a single jet.
Also shown is the background from SM \ttbar production above which
the signal clearly stands out (left panel).
The dominating background is that from QCD dijet production (also
referred to as multijet production)
in which
two partons scatter and produce two high \pt outgoing partons. After the
QCD shower (gluon radiation and splitting) and fragmentation into hadrons,
the final state consists of two or more jets.
This background is shown in the right
panel. It exceeds the signal by approximately one order of magnitude.
To discover the signal, the multijet background has to be suppressed
by analysing the internal structure of the jets.

\section{Jet Structure}
This section first briefly summarises the jet algorithms that are used in the results presented
in this review. It then goes on to explain how jet structure differs for signal and background
and introduces methods that exploit these differences.
The susceptibility of jets to corrections from hadronisation and underlying event (UE)
is discussed before methods are introduced that remove contributions from UE and pile-up
from a jet ({\em grooming}).
The analysis of the internal structure of jets is also referred to as {\em substructure} analysis,
and the words {\em structure} and {\em substructure} are used synonymously in this context.

%%%%%%%%%%%%%%%%%%%%%%%%%%%%%%%%%%%%%%%%%%%%%%%%%%%%%%%%%%%%%%%%%%%%%%%%%%%%%%%%

\subsection{Jet algorithms}
\label{sec:jetalgorithms}
Jets are collimated sprays of particles and algorithms
are used to define the geometrical size and the kinematics of the combined object.
Conventionally, jets are used to get an estimate of the kinematics of partons
that underwent a hard scattering process or which originate from the
decay of a heavy particle. These high-energy partons surround themselves with
a parton cloud by radiating gluons which can split into gluons or quark pairs.
After hadronisation, the original parton momentum is distributed among many
particles. A jet algorithm tries to find the original parton momentum by
iteratively combining the momenta of nearby partons and in that sense reverse the parton splitting.

The most natural definition of a jet is based on a cone within which most of these particles are contained
and the first jet algorithms used this concept~\cite{Sterman:1977wj,Arnison:1983gw}.
Another class of algorithms is based on the iterative recombination of neighbouring particles.
These algorithms are easier made {\em infrared-safe} such that they arrive at the same
hard jets when an additional soft gluon is added to the event.
A discussion of jet algorithms can be found in~\cite{Salam:2009jx}.
All results discussed in this text use recombination algorithms.

All jet algorithms operate on a list of four-momenta, which can correspond to
particles or detector quantities like
tracks or calorimeter clusters, which will generically be
referred to as {\em constituents} in the following.
The combination algorithms merge two neighbouring constituents into one
by combining their momenta. For the results discussed in this text,
the {\em $E$-scheme} is used, in which the four-momenta are added, leading to massive jets.
The objects that result from the merging are called {\em protojets} if they
are not the final jets.

A distinction is made between {\em inclusive} and {\em exclusive} clustering:
in the case of the former, a distance parameter $R$ is specified and
the constituents or protojets $i$ and $j$ that are nearest in terms of a chosen distance scale are combined
as long as $\DRij < R$. All resulting jets are separated by $\DR \ge R$.
Exclusive clustering, on the other hand, ends when a specified number of jets has been obtained.
Each jet is represented by a four-momentum vector, the $\theta$ and $\phi$ components
of which define a jet axis.

The merging order is determined by the definition of the distance scale which specifies
which neighbours $i, j$ are closest and hence will be merged next.
Three common choices are
\begin{itemize}
\item the separation \DRij. In this case, the neighbours nearest in $(y,\phi)$ space are clustered first
and the procedure is referred to as \Ca (\ca) algorithm~\cite{Dokshitzer:1997in, Wobisch:1998wt, WobischPhD}.
\item $\min(p_{{\rm T},i}, p_{{\rm T},j}) \times \DRij$ (\kt algorithm~\cite{Catani:1991hj, Ellis1993, Catani1993}).
Compared to \ca, this clusters low \pt constituents earlier.
\item $\min(1/p_{{\rm T},i}, 1/p_{{\rm T},j}) \times \DRij$ (\akt algorithm~\cite{Cacciari:2008gp}).
This clusters high \pt constituents earlier.
\end{itemize}

The \kt algorithm aims at reversing the angular ordered parton shower implemented
in the \herwig generator\cite{Marchesini:1991ch}.
Jets reconstructed with the \akt algorithm have a cone-like shape with the covered
area given by $\approx \pi R^2$ whereas \ca and \kt jets tend to have
more irregular shapes~\cite{Cacciari:2008gp} as discussed in \secref{perf_mass}.
Regardless of this fact, the distance parameter $R$ is commonly referred to
as {\em jet radius} for all jet algorithms.

%%%%%%%%%%%%%%%%%%%%%%%%%%%%%%%%%%%%%%%%%%%%%%%%%%%%%%%%%%%%%%%%%%%%%%%%%%%%%%%%

\subsection{Jet structure in signal and background}

\begin{figure}[hbt]
\centering
\includegraphics[width=0.7\textwidth,angle=0]{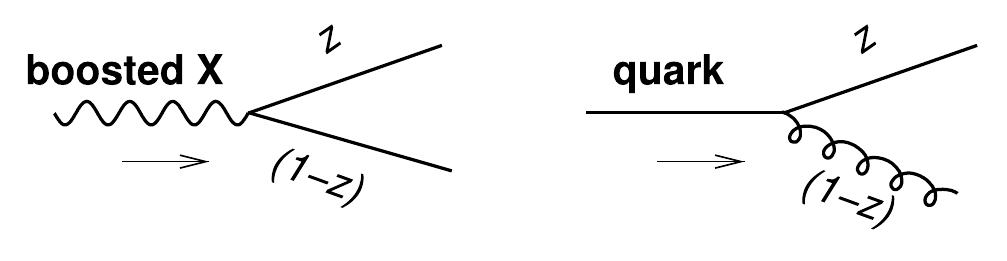}
\caption{Schematic diagrams of (left) the boosted decay of a particle $X$ to two
quarks and (right) gluon radiation off a quark. In the right diagram, the fraction of the initial quark
momentum retained by the quark is denoted $z$. The momentum fraction of the
radiated gluon is $1-z$. Similarly, in the left diagram, the two quarks from the decay of $X$ carry
momentum fractions $z$ and $1-z$. From~\cite{Salam_boost}.}
\label{fig:Gavin_splitting}
\end{figure}

To analyse differences in the (fat) jet structure between signal and background, it is instructive to
compare the kinematics of the signal decay with QCD parton splitting processes.
Schematic diagrams of these processes are shown in \figref{Gavin_splitting}.

It is of interest how the parent particle energy is distributed among the two outgoing
particles. For gluon radiation in the collinear approximation,
the probability that the quark retains a fraction
$z$ of its momentum is given in leading order by the Altarelli-Parisi splitting function~\cite{Altarelli:1977zs}
\begin{equation}
P_{gq} = \frac{4}{3} \frac{1+z^2}{1-z}\,.  \label{eq:splitting}
\end{equation}
Most of the gluons are therefore soft ($z \rightarrow 1$). For the signal, the
decay is not as asymmetric. For example, the decay amplitude of the Higgs boson
for $H\rightarrow b\bar{b}$
is flat in $z$~\cite{Salam_boost}.
An efficient way to suppress background is therefore to reject
configurations with large $z$. This is the idea behind the mass drop
technique~\cite{Butterworth:2008iy}.

\subsubsection{Mass drop}

\label{sec:md}
The mass drop (MD) criterion was developed to identify
the decay $H\rightarrow b\bar{b}$ against a large multijet background~\cite{Butterworth:2008iy}.
The idea is to use boosted Higgs bosons for which the \bbbar pair is collimated
and contained inside a \ca fat jet.
To find the subjets that correspond to the \bjets from the Higgs decay,
the MD algorithm searches for a merging $i+j \rightarrow p$
for which the combined mass $m_p$ is significantly larger than either one of $m_i$ and $m_j$.
Fat jets that originate from hard light quarks or gluons are unlikely to
display this pattern because the splitting function (\ref{eq:splitting}) prefers
soft radiation.

An iterative procedure is used because \ca clustering is by smallest angular separation
and the last two protojets are not necessarily the wanted subjets.
The algorithm starts with a fat jet $p$ and proceeds as follows:
\begin{enumerate}
\item the last clustering of $p$ is undone to obtain two protojets $i$ and $j$,
labelled such that $m_i>m_j$.
If $p$ cannot be split because it is a constituent then the fat jet is discarded.
When applying the MD algorithm to calorimeter clusters, the detector
resolution becomes relevant. In~\cite{ATLAS:2012am}, the jet $p$ was also
discarded if $\DeltaRij < 0.3$.
\item If $m_i/m_p < \mu$ and $\sqrt{v} \equiv \DeltaRij \times \min(p_{{\rm T},i}, p_{{\rm T},j})/m_p > \sqrt{\vcut}$
then $i$ and $j$ are identified as the wanted subjets and the procedure ends. Otherwise the procedure continues with step 1
but now using the leading mass subjet as input ($p=i$). With $\DeltaRij \approx 2 m_p/p_{{\rm T},p}$,
the second requirement reads $\min(p_{{\rm T},i}, p_{{\rm T},j})/p_{{\rm T},p} \gtrsim \sqrt{\vcut}/2$ and
implies a minimum \pt for the softer protojet.
\end{enumerate}

If two subjets can be found in this way then the original (fat) jet
satisfies the MD criterion.
In~\cite{Butterworth:2008iy}, the parameters are $\mu = 0.67$, implying
a mass drop of at least $33\%$, and $\vcut = 0.09$, i.e.,
the softer protojet has to have at least $\approx\!15\%$ of the combined \pt.

By changing the parameters, the procedure can be adapted to the decay of other massive particles,
like \W or \Z bosons or the top quark.
One can also continue the mass drop procedure
to identify two successive decays of massive particles, like in $t\rightarrow Wb \rightarrow qqb$.

\subsubsection{\kt splitting scales}

The splitting function (\ref{eq:splitting}) is also the motivation for
cuts on {\em \kt splitting scales}~\cite{Butterworth:2002tt} as explained
in~\cite{Salam:2009jx}: for quasi-collinear splitting to two partons $i$ and $j$,
the squared invariant mass of the two partons is given by~\cite{Ellis:2007ib,Salam:2009jx}
\begin{equation}
m^2 \approx \pTi \, \pTj \, \DR_{ij}^2\,. \label{eq:mass}
\end{equation}
Transverse momenta are used in this expression because jets are detected centrally in the LHC detectors
where \pt is a good approximation of the full momentum. With $j$ denoting the softer parton and $\pt = \pTi + \pTj$,
\eqref{mass} can be rewritten as
\begin{equation}
m^2 \approx x\,(1-x) \, \pT^2 \, \DR_{ij}^2\,,
\end{equation}
with the fraction $x = \pTj/\pT < 0.5$. For a signal-like flat
$x$ distribution with an average value of $0.25$, the mass is
$m \approx 0.43 \, \pT\,\DR_{ij}$, which corresponds to \eqref{size}.
In the case of gluon radiation,
$x$ is the \pt fraction carried by the gluon and corresponds to $1-z$ in \eqref{splitting}.

The \kt splitting scale corresponds to the distance scale of the \kt algorithm and is given by
\begin{equation}
\Dij \equiv \min(p_{{\rm T},i}, p_{{\rm T},j}) \, \DRij \, = x \, \pt \, \DR_{ij} \approx \sqrt{\frac{x}{1-x}} \, m\,.
\end{equation}

For gluon radiation, $x\rightarrow 0$ and the scale is small. For heavy particle
decays with a more uniform $x$ distribution, this is not the case and a cut can
be used to suppress the background. For a flat $x$ distribution,
$\sqrt{d_{ij}} \approx 0.58\, m$. For top quark and vector boson decay,
values of approximately half the parent particle mass are observed for $\sqrt{d_{ij}}$.

The \kt algorithm clusters high \pT objects late, so that the scale of the last merging, \DOneTwo,
and the one of the second-to-last merging, \DTwoThr,
are sensitive to the hard structure of the jet. For top quark decay, \DOneTwo
is peaked near half the top quark mass and \DTwoThr is peaked near half the \W boson mass.

%%%%%%%%%%%%%%%%%%%%%%%%%%%%%%%%%%%%%%%%%%%%%%%%%%%%%%%%%%%%%%%%%%%%%%%%%%%%%%%%

\subsection{Jet energy corrections and contamination}

\begin{figure}[hbt]
\centering
\includegraphics[width=0.45\textwidth,angle=-90]{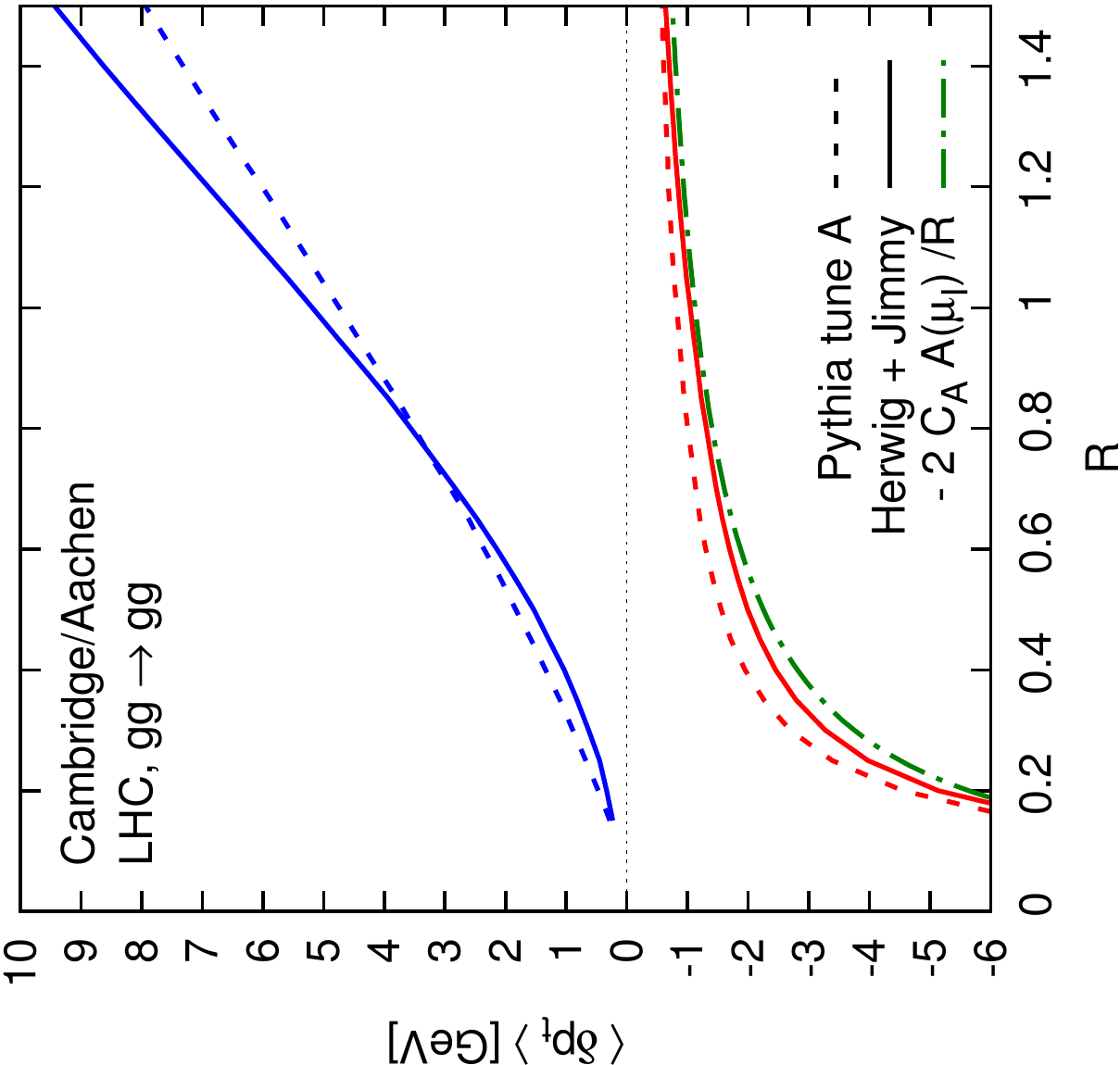}
\caption{The correction from underlying event ($\delta p_t>0$) and
hadronisation ($\delta p_t<0$) to the \pt of \ca jets as a function of
the radius parameter $R$ for $gg \rightarrow gg$ scattering in $pp$
collisions at $\sqrt{s}=14\TeV$.
The parton-level jet (after the parton shower) is required to have
$55< \pt < 70 \GeV$.
Shown are predictions from \pythia and from \herwig with the \jimmy underlying event model
and an analytical result for the hadronisation correction.
From~\cite{Dasgupta:2007wa}.}
\label{fig:UEvsR}
\end{figure}

The size of a jet is determined by the radius parameter $R$.
The larger $R$, the larger the area in $(\eta, \phi)$ that is covered by the jet
and the more underlying event (UE) energy will be picked up.
The underlying event are the particles scattered in interactions that are not
related to the hard scatter. These additional interactions are predominantly
soft and the energies are small compared to those involved in the hard scatter.
Nevertheless, these energies lead to a shift in the reconstructed jet energy compared
to the energy of the hard scatter parton. The shift in \pt due to the UE is shown in
\figref{UEvsR} (upper curves) as a function of the radius parameter $R$ for \ca jets
with $55< \pt < 70 \GeV$ at the parton level in
$gg \rightarrow gg$ scattering at the LHC with $\sqrt{s}=14\TeV$. The shift
is evaluated using two different UE models (\pythia and \herwig). It is approximately
proportional to $R^2$ and for $R=1$ amounts to $\approx\!5\GeV$ which is almost
$10\%$ of the jet \pt. The correction depends on the collision energy. For
Tevatron $p\bar{p}$ collisions at $1.96\TeV$, the correction at $R=1$ is
$1$--$2\GeV$, depending on the model.

Another effect that depends on $R$ is the energy lost outside the jet
by hadronisation. Hadronisation denotes the transition from coloured partons to
colour-neutral hadrons. In this process,
new partons emerge between colour-connected partons and the energy is
in part re-assigned. If one compares a jet built from the partons
with a jet of the same size built from the hadrons, the hadron jet has smaller
energy (or \pt) because some of the parton energy is lost in hadrons that
are not captured in the jet. This loss is larger when the jet $R$ is smaller
as shown in \figref{UEvsR} (lower curves). The \pt shift can be calculated
analytically~\cite{Dasgupta:2007wa} and is approximately proportional to $1/R$.
For $R=1$ the shift is $\approx -1\GeV$ for \ca jets. In magnitude this
shift is only $20\%$ of the UE shift which works in the opposite direction.
The parameter $R$ that minimises the quadratic sum of the
UE and hadronisation corrections
is $R=0.41$ for quark jets and $R=0.54$ for gluon jets~\cite{Dasgupta:2007wa}.
The standard jet sizes in the ATLAS and CMS experiments ($R=0.4$ and $R=0.5$, respectively)
are driven by such optimisations. For fat jets ($R\ge0.8$), the
UE corrections are more important than for these standard jets.

Experimentally, jets are contaminated by {\em pile-up} which denotes the case
that several hard interactions appear in the same event. This happens for two
reasons: first, when the luminosity of the collider is sufficiently high (thereby giving a
high probability for two hard interactions to occur in the same bunch crossing)
and second, when the detector readout is slow such that events see remnants of
signals from earlier events. These two contributions are sometimes referred to as
in-time and out-of-time pile-up, respectively.
This pile-up energy is larger than the UE contribution and scales with the
area of the jet.

Jet substructure analysis tries to identify jets from top quark decay (or decay of other particles)
inside a large (fat) jet. By exploiting kinematic relations between the decay partons,
background can be suppressed. However, these relations no longer hold if the jet
kinematics are changed by UE and pile-up contributions. In other words,
these contaminations have to be removed to clearly ``see'' the jet substructure.
Different techniques have been devised in this regard and are referred to
as {\em jet grooming}.

%%%%%%%%%%%%%%%%%%%%%%%%%%%%%%%%%%%%%%%%%%%%%%%%%%%%%%%%%%%%%%%%%%%%%%%%%%%%%%%%

\subsection{Jet grooming}
The process of {\em jet grooming} is the removal of unwanted constituents from
a fat jet ($R\ge0.8$). Different procedures have been developed and the ones relevant
for this text are described in the following.

\subsubsection*{Trimming}
\label{sec:trimming}
Contributions from underlying event and pile-up are usually soft, i.e., have small energy,
compared to those from the high-\pt hard scatter. The {\em trimming} procedure~\cite{Krohn:2009th}
creates \kt subjets with a radius parameter $\Rsub<R$ (typically $\Rsub=0.3$) using all
constituents of the fat jet. Then the constituents of the subjets which carry
less than a fraction \fcut (typically $5\%$) of the fat jet \pt are removed {\em (trimmed)} from
the fat jet. This method does not correct for unwanted contributions that overlap
with the hard subjets. It is therefore most useful for substructure
variables that are very sensitive to soft contributions. An example is the fat jet mass,
to which even low \pt constituents contribute significantly if they lie at large
angles with respect to the hard constituents.

\subsubsection*{Pruning}
\label{sec:pruning}
The jet pruning procedure~\cite{Ellis:2009su,Ellis:2009me} removes
soft protojets at large angles in every jet clustering step.
At every merging step of two protojets, $i+j \rightarrow p$, one calculates
\begin{equation}
z \equiv \min(p_{{\rm T},i}, p_{{\rm T},j})/p_{{\rm T},p}
\end{equation}
and discards the softer protojet if
\begin{equation}
z < \zcut   \quad {\rm and}  \quad \Delta R_{ij} > \Dcut\,.
\end{equation}
Otherwise the merging is applied.
A protojet is therefore discarded if the other protojet carries much more \pt and
the distance between the two protojets is large.
The jet obtained using this conditional clustering is called a
{\em pruned jet}. In~\cite{Ellis:2009su} the cut values are $\zcut = 0.1$ for the \ca algorithm and
$0.15$ for the \kt algorithm, and $\Dcut = m_J/p_{{\rm T},J}$ is the ratio of the
mass of the unpruned jet to its \pt.

\subsubsection*{Filtering}
\label{sec:filtering}
\begin{figure}[hbt]
\centering
\includegraphics[width=0.9\textwidth,angle=0]{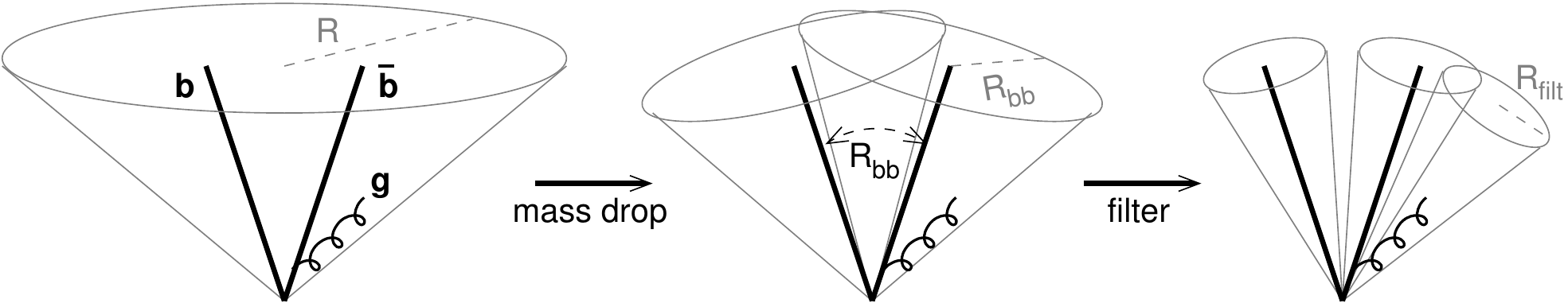}
\caption{Illustration of the mass drop filtering technique.
A fat jet containing the two bottom quarks from the decay $H\to b\bar{b}$
is broken down into two hard subjets (mass drop). The constituents
of the subjets are reclustered with a radius parameter \Rfilt that is small compared
to the subjet distance $R_{bb}$ and only the three highest \pt small-$R$ jets are kept (filtering).
The third jet captures gluon radiation.
From~\cite{Butterworth:2008iy}.}
\label{fig:massdrop}
\end{figure}

For filtering, the constituents of a jet are inclusively clustered using
a {\em filter radius} that is small compared to the size of the jet.
Only $N$ filter jets with the largest \pt are kept.
The combination of a mass drop criterion with filtering
was first used in~\cite{Butterworth:2008iy} and has become known as {\em mass drop filtering}.
It is illustrated in \figref{massdrop} for $H\to b\bar{b}$.
It is used in a number of substructure algorithms that have been suggested since,
such as the \htt~\cite{Plehn:2009rk,Plehn:2010st} which is described in \secref{htt_algo}.

\section{Experimental Setup}
This section describes the devices used for the experimental
results in this review.

\subsection{LHC}

The Large Hadron Collider (LHC) started colliding protons in December 2009.
The results shown in this review are
obtained using $pp$ collisions at
centre-of-mass energies of $\sqrt{s}=7\TeV$ (2011) and $8\TeV$ (2012).
The LHC was designed for $\sqrt{s}=14\TeV$ but operation at higher energies
was not possible because cable connections between dipole magnets
that were soldered at room temperature
can develop high resistivity due to mechanical stress when cooled down to superconducting temperature.
This can lead to electric arcs when the current is large.
This happened on 19 September 2008 when a magnet quenched and an electric arc developed and punctured
the enclosure that held the liquid helium.
The helium expanded to the gaseous state and was released into the vacuum that
thermally insulates the beam pipe. Upon expansion, the helium volume increased by a factor 1000
and the resulting
pressure destroyed several magnets. The LHC had to be shut down
for a year for repairs. After the restart, the magnet currents were kept below a safety threshold,
thereby limiting the bending power and consequently the beam energy.
Safely going to higher collision energies requires the replacement of all soldered
connections with clamped splices. This work is ongoing since spring 2013 in the so-called
Long Shutdown 1. After the replacement, the LHC is expected to collide protons with $\sqrt{s} = 13$--$14\TeV$
starting in spring 2015.

In 2011 and 2012, the number of protons per bunch was $1.5$--$1.7\times 10^{11}$ with
1380 bunches in the machine~\cite{LHCstatus}.
The bunch separation was $50\,$ns.
The instantaneous luminosity \instL reached values of $3.5\times 10^{33}\,$Hz/cm$^2$
in 2011 and $7.7\times 10^{33}\,$Hz/cm$^2$ in 2012.

\begin{figure}[hbt]
\centering
\subfigure[]{
   \includegraphics[width=0.48\textwidth,angle=0]{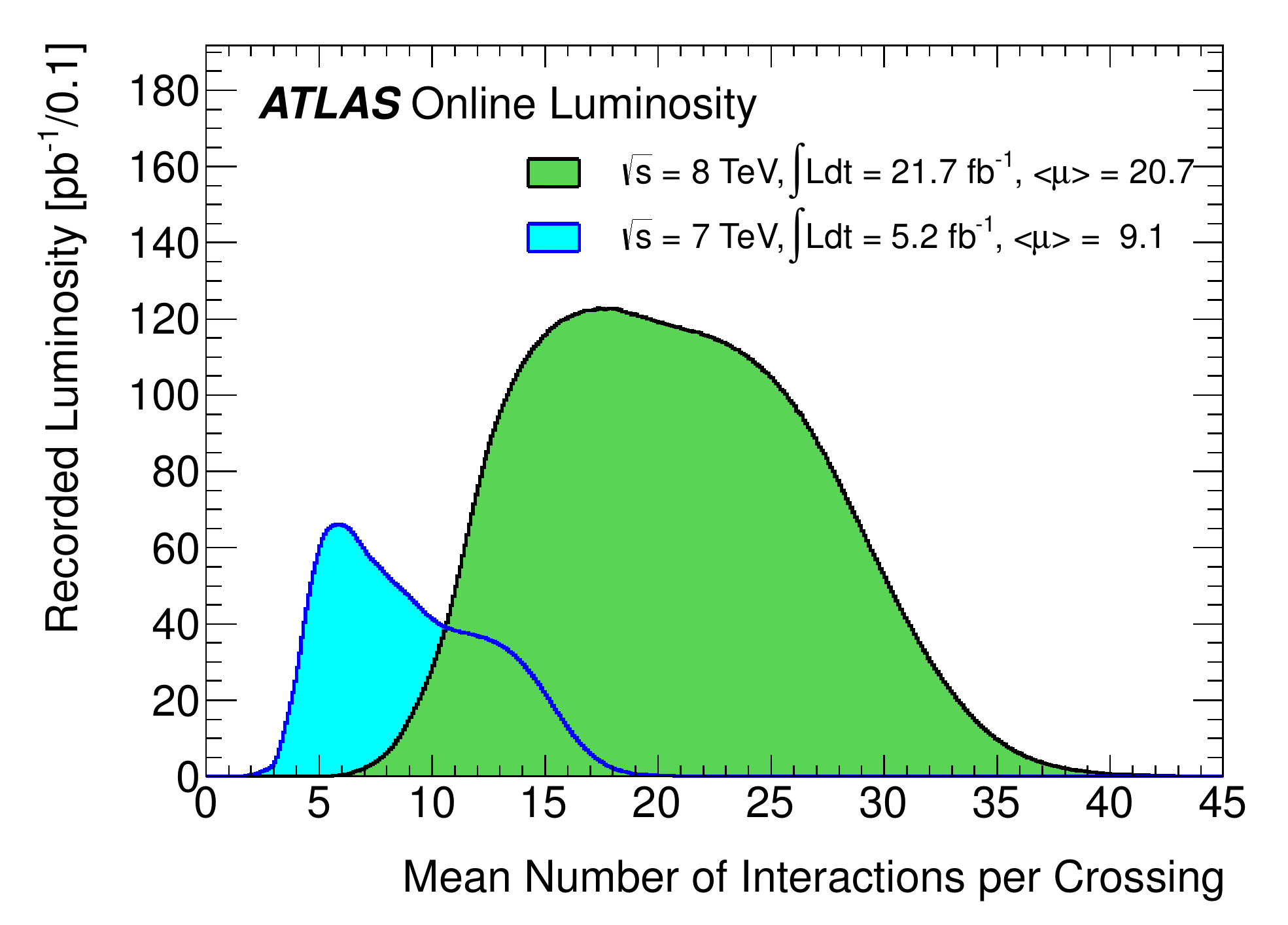}
} \\
\noindent
\subfigure[]{
   \includegraphics[width=0.80\textwidth,angle=0]{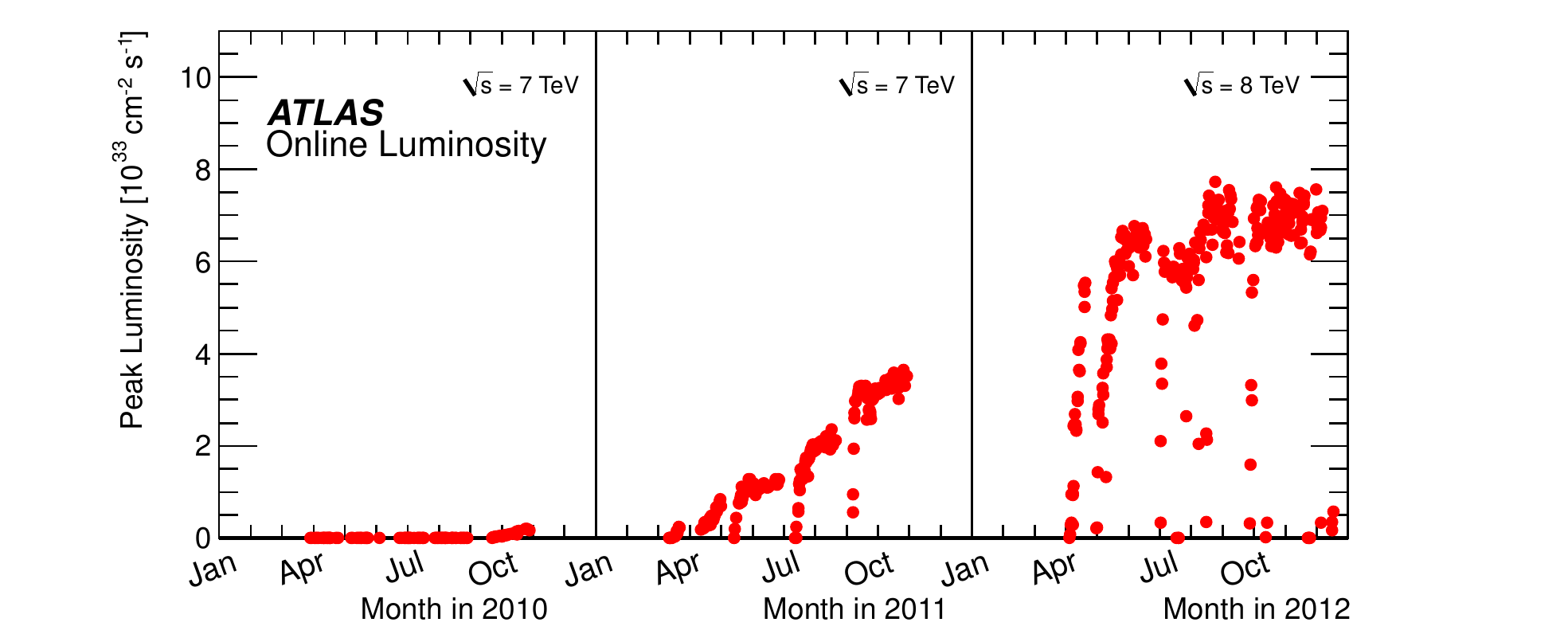}
}
\caption{a) The luminosity-weighted distribution of the average number of
inelastic $pp$ interactions per bunch crossing \avmu for 2011 and 2012 ATLAS data.
b) The peak instantaneous luminosity as a function of time. From~\cite{mu_lumi}.}
\label{fig:mu_lumi}
\end{figure}

\label{sec:pileup}
The variable \avmu denotes the average number of inelastic $pp$ interactions
per bunch crossing. It is calculated from the inelastic cross section $\sigma_{\rm inel}$, \instL, and
the average frequency $f_{\rm bunch}$ of bunch crossings in the LHC:
\begin{equation}
\avmu =  \sigma_{\rm inel} \ \frac{ \instL }{ f_{\rm bunch} }  \, .
\end{equation}
The value used by ATLAS for the inelastic cross section is $71.5\,$mb at $\sqrt{s} = 7\TeV$
and $73.0\,$mb at $8\TeV$.
\figref{mu_lumi} shows the \avmu distribution and the maximum instantaneous
luminosity as a function of time. The average \avmu was 9.1 in 2011 and 20.7 in 2012.

\subsection{ATLAS}
\label{sec:atlas_detector}
\label{clusters}

\begin{figure}[hbt]
\centering
\includegraphics[width=0.8\textwidth,angle=0]{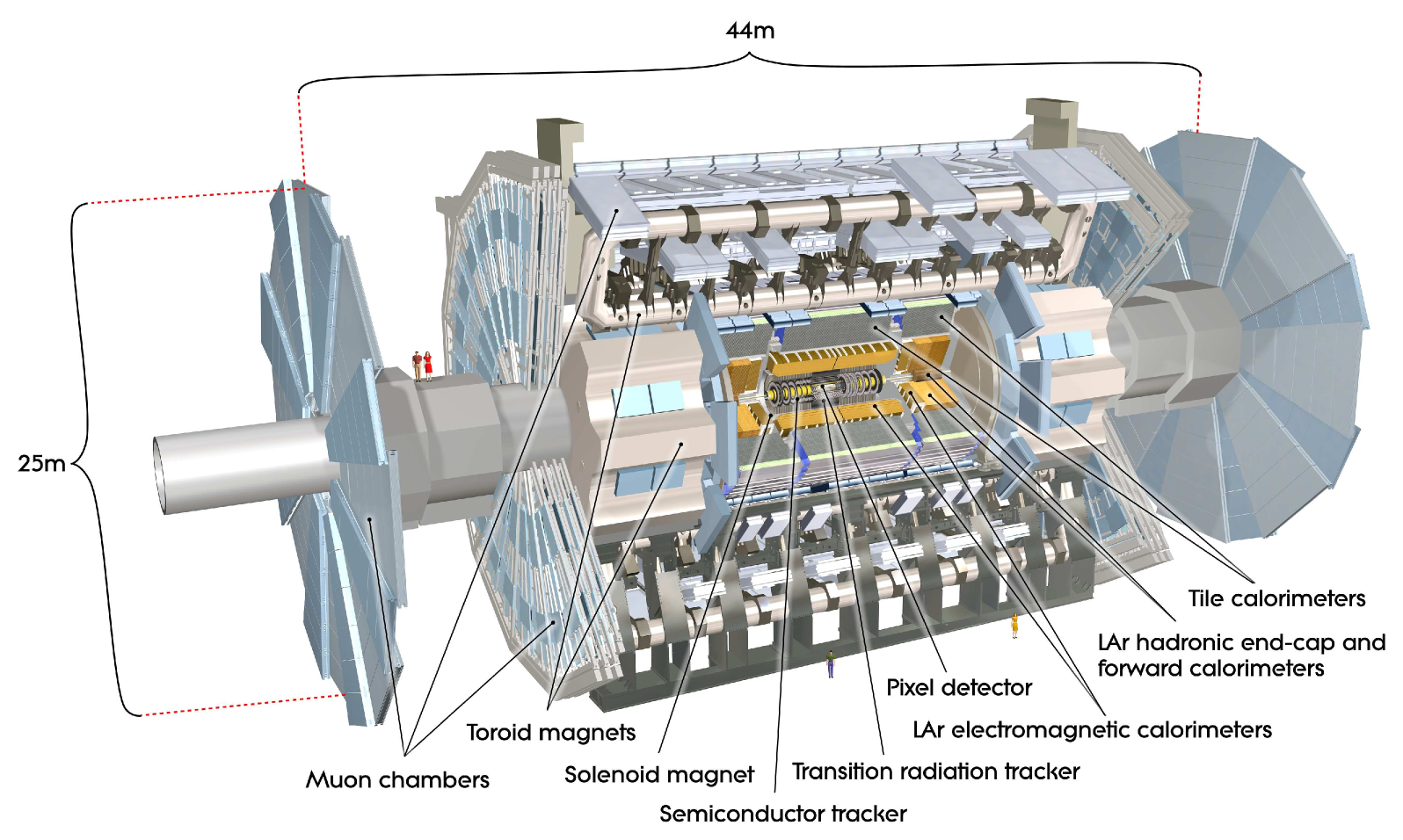}
\caption{Schematic view of the ATLAS detector. From~\cite{Aad:2008zzm}.}
\label{fig:atlas}
\end{figure}

A schematic view of the ATLAS detector is shown in \figref{atlas}.
A full description of it can be found in~\cite{Aad:2008zzm,Collaboration:2010knc}.
The parts relevant to the discussion of the results presented in this text are summarised below.

Closest to the interaction point is the inner tracking detector (ID) which consists
of a silicon part (pixel and strips) and a transition radiation detector (TRT).
The ID spans the full azimuthal range and $|\eta|<2.5$, and is immersed
in a magnetic field of 2~T that is provided by a coil outside of the ID volume.

Hits in the ID are used to construct tracks of charged particles.
The angular resolution of the ATLAS inner tracking detector for
charged particles with $\pt = 10\GeV$ and $\eta = 0.25$
is $\approx 10^{-3}$ in $\eta$ and $\approx 0.3$~mrad in $\phi$~\cite{Aad:2008zzm} with a track construction efficiency
larger than $78\%$ for charged particles with $\pt>500\MeV$~\cite{Aad:2010ac}.
The momentum resolution for charged pions is $4\%$ for momenta $p<10\GeV$,
rising to $18\%$ at $p=100\GeV$~\cite{Aad:2008zzm}.

Surrounding the magnet coil is the electromagnetic calorimeter (ECAL) which
consists of a barrel part ($|\eta|<1.475$)
and two endcap parts ($1.375<|\eta|<3.2$).
It is a sandwich calorimeter with lead absorber plates and kapton electrodes
immersed in liquid argon (LAr). The electrode cell size in $(\eta,\phi)$ varies from
$0.025\times 0.025$ to $0.1\times 0.1$, depending on the layer and $\eta$.
The hadronic calorimeter (HCAL) in the barrel ($|\eta|<1.7$)
uses scintillating tiles while in the endcaps ($1.5<|\eta|<3.2$)
the ECAL technology is used. HCAL cell sizes vary from $0.1\times 0.1$ to $0.2\times 0.2$.

Topological cell clusters are formed around
seed cells with an energy $|E_{\rm cell}|>4\sigma_{\rm noise}$ by adding
the neighbouring cells with $|E_{\rm cell}|>2\sigma_{\rm noise}$, and then
all surrounding cells~\cite{Aad:2012vm}.
The minimal transverse size for a cluster of hadronic calorimeter cells is therefore
$0.3\times0.3$ and is reached if all significant activity is
concentrated in a single cell. Two particle jets leave
distinguishable clusters if each jet hits only a single cell
and the jet axes are separated by at least $\DeltaReta=0.2$,
so that there is one empty cell between the two seed cells.\footnote{A splitting algorithm
has to be used in this case to divide this big cluster into two.}
The finest angular resolution of the hadronic calorimeter is therefore $\DeltaReta=0.2$
which is much coarser than the resolution of the tracking detector given above.

\begin{figure}[hbt]
\centering
\includegraphics[width=0.45\textwidth,angle=0]{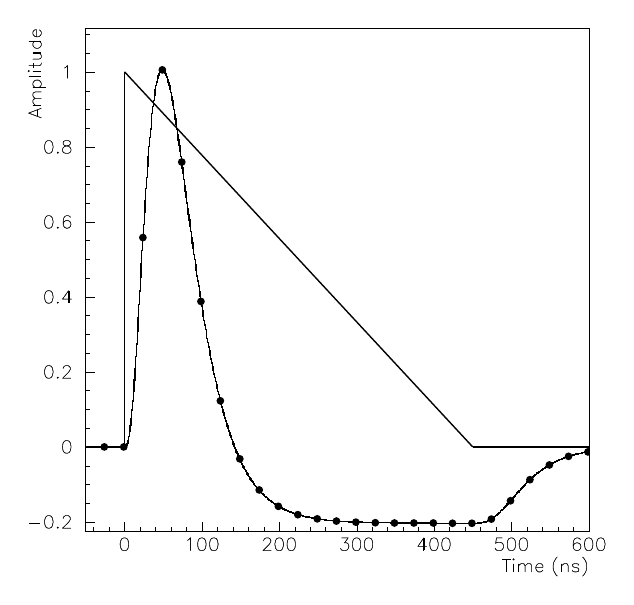}
\caption{Amplitude vs. time for a triangular current pulse in one of the ATLAS LAr calorimeter cells.
From~\cite{Aad:2008zzm}.}
\label{fig:LArsignal}
\end{figure}

The LAr system is slow and signals from several inelastic $pp$ interactions can
overlap. The signal from one of the cells in the barrel is shown in \figref{LArsignal}.
A long tail of several hundred ns is visible. With a bunch spacing
of $50\,$ns and many interactions per bunch crossing, it is likely that the
same cell again detects activity while signals from previous events are still being
processed. The bipolar shape was designed such that the negative signal from
earlier events cancels pile-up signals from current events. This cancellation
holds for $\instL = 10^{34}\,$Hz/cm$^2$ at $\sqrt{s} = 14\TeV$. For other
luminosities and collision energies the system is susceptible to (out-of-time)
pile-up.

% Muon system
Muons are detected in a spectrometer that covers $|\eta|<2.7$ with a toroidal
magnetic field that is perpendicular to the momentum of central muons.

\subsection{CMS}

\begin{figure}[hbt]
\centering
\includegraphics[width=0.8\textwidth,angle=0]{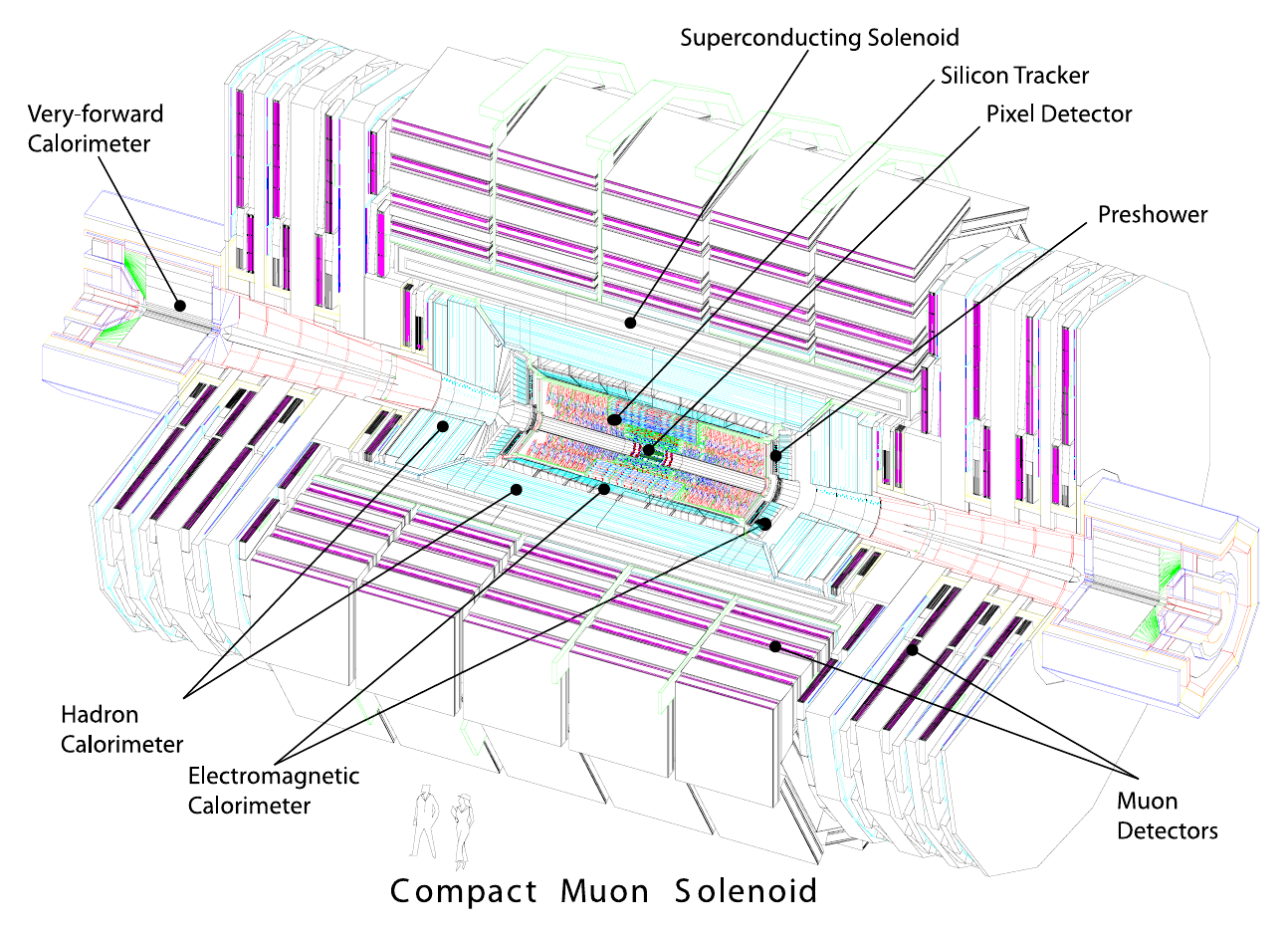}
\caption{Schematic view of the CMS detector. From~\cite{Chatrchyan:2008aa}.}
\label{fig:cms}
\end{figure}

The CMS detector is shown schematically in \figref{cms}. A detailed description
is given in~\cite{Chatrchyan:2008aa}.
Tracking is provided by silicon pixel and strip detectors inside a 3.8~T
magnetic field. The magnet coil has a diameter of six metres and surrounds
the barrel and endcap calorimeters ($|\eta|<3$).
The ECAL consists of scintillating lead tungstate crystals.
The HCAL is of a sandwich type with alternating layers of brass and scintillator
tiles. Outside the magnet coil are gaseous detectors that are used to measure muons.
The use of scintillator technology for the calorimeters makes the CMS data less
susceptible to pile-up than ATLAS data.

\label{sec:pflow}
To reconstruct particles, CMS uses the {\em particle flow} approach, which
correlates information from the inner tracking detector, the calorimeters, and the muon detector.
Clusters are reconstructed separately in the preshower detector, the ECAL, and the HCAL.
Tracks and clusters are linked if they can be geometrically matched.
The tracks are extrapolated to the calorimeter and if the end point is within the cluster or within a margin of one cell around the cluster,
a track-cluster link is established.
Clusters in different calorimeters are linked if the cluster in the finer calorimeter
is within the cluster of the coarser calorimeter.

The particle flow algorithm~\cite{CMS-PAS-PFT-09-001} first reconstructs muons, then electrons and Bremsstrahlung photons,
and finally charged hadrons. The remaining entries are then taken to be photons and neutral hadrons.
After every step, the detector entries associated with a particular particle type are removed before continuing with the next type.
Muons are identified through a global fit of hits in the inner detector and the muon detector.
The energy left by muons is $0.5(5)\GeV$ in the ECAL and $3(3)\GeV$ in the HCAL.
Electrons are found in a fit that includes emission of Bremsstrahlung photons at
layers of the tracking detector and cuts on calorimeter variables.
Charged hadrons are identified by comparing the \pt of a track
with the calibrated energy of the associated cluster (track-cluster link).
If the \pt and the energy are compatible,
a charged hadron is reconstructed using the track \pt and the mass of a charged pion.
A complication arises when several ECAL clusters are associated to the track.
These clusters can correspond to the electromagnetic part of a hadronic shower, in which case
they should be included in the charged hadron energy, or to photons.
The clusters are ordered in distance from the extrapolated track position and
the energy of the closest clusters is added to the track-associated cluster
until the \pt of the track is reached.
If the cluster energy exceeds the track \pt then a photon is reconstructed
with the energy measured in the ECAL and a neutral hadron with the HCAL energy.
The preference is given to photons because they constitute on average $25\%$ of
the jet energy while only $10\%$ is carried by neutral hadrons and not all
of their energy is deposited in the ECAL.
The cluster calibration is obtained from simulation and is validated using
test-beam data and collision data with isolated charged hadrons~\cite{CMS-PAS-PFT-10-002}.
The photon energy is validated using $\pi^0$ decays~\cite{CMS-PAS-PFT-10-002}.

\section{Monte Carlo Generation and Detector Simulation}
This section briefly describes the simulation tools that are used to obtain
predictions for jet substructure.
The degree to which jet substructure observables can be predicted using
generated events and a simulation of the detector response greatly
determines their usefulness in physics analyses. The precision of
these predictions enters as a systematic uncertainty in analyses.

\subsection{Monte Carlo generators}
Different Monte Carlo (MC) generators are used to obtain predictions at the particle level.
The multipurpose generator \pythia is used in versions 6~\cite{pythia} and 8~\cite{pythia8}.
\pythia calculates hard $2\rightarrow2$ parton scattering at leading order (LO) in perturbative QCD.
Higher orders are emulated using a parton shower~\cite{pythiapartonshower}.
In version 6, the evolution variable in the parton shower can be chosen to be
either virtuality (mass) or transverse momentum. In version 8, the evolution is in
transverse momentum and dipole showering is possible for the final
state.\footnote{Different parton shower evolutions are discussed in~\cite{Sjostrand:2006su}.}
The partons are hadronised
using Lund string fragmentation~\cite{Andersson:1983ia}. \pythia has a multiple interactions
model for the underlying event (UE). The hadronisation and UE model parameters are tuned to minimum bias data.
ATLAS tunes are described in~\cite{AUET2B,AMBT1}.

Another LO generator is \herwig~\cite{Marchesini:1987cf,Marchesini:1991ch,Corcella:2002jc}.
It uses an angular-ordered parton shower to emulate higher order effects.
The final state partons are hadronised using cluster fragmentation~\cite{Webber:1983if}.
\herwig is commonly combined with the \jimmy generator~\cite{jimmy} for multiple interactions.
The hadronisation and UE parameters are tuned to data, for example in~\cite{AUET1}.
\herwigpp~\cite{Herwigpp} is a replacement for \herwig and is written in C++.
It includes an underlying event model~\cite{Bahr:2008dy}.

\mcatnlo~\cite{mcatnlo} was the first program to combine next-to-leading order (NLO) QCD calculations
with a parton shower without double-counting. It uses the \herwig parton shower.
\powheg~\cite{Frixione:2007vw} also combines an NLO matrix element with a parton shower
but any parton shower program can be interfaced.

In the NLO MC programs, the matrix element calculations are limited
to a small number of outgoing particles. {\em Multileg} generators were created
that calculate final states with more particles at LO. An example is \amegicpp~\cite{Krauss:2001iv}
which was integrated in the \sherpa framework~\cite{Gleisberg:2003xi,Gleisberg:2008ta}
to supplement it with a parton shower and evolve the events to the hadron level.
\sherpa uses virtuality-ordered parton showering, cluster fragmentation,
and an underlying event model similar to that of \pythia.
Other multileg generators are \alpgen~\cite{Mangano:2002ea} and
\madgraph~\cite{Stelzer:1994ta,Alwall:2007st}.
The program \acermc~\cite{Kersevan:2004yg} uses matrix elements from \madgraph
and has optimised phase space sampling for a selection of SM processes.

\subsection{Detector simulation}

Detectors for particle physics experiments are very complicated setups.
To correct experimental effects introduced through the measurement
apparatus, the generated particles at the stable hadron level are passed through
a simulation of a detector.

The best detector simulation is obtained when every interaction of each
particle with the detector material is calculated separately, usually with the
program \geant~\cite{Agostinelli:2002hh}.
 This type of
simulation is referred to as {\em full simulation}
and requires detailed knowledge about
the detector geometry and material.
Pile-up is simulated by overlaying the hard scattering event with
minimum bias events that are produced using \pythia.
All predictions in this review use full simulation, except where indicated.

Full simulation is part of the intellectual property of the detector collaboration
and usually not available to non-collaborators.
Without access to the full simulation, phenomenologists use {\em generic}
detector simulations to estimate the impact of detector effects. The generic simulations
are based on simplified virtual versions of
existing detectors such as ATLAS or CMS. Key figures like geometry, acceptance, granularity,
tracking and calorimeter resolutions are taken from information published
by the collaborations. In this way, a decent simulation
is possible that achieves predictions within approximately
$20\%$ of the real response. Examples of generic simulation frameworks
are \acerdet~\cite{RichterWas:2002ch}, \Delphes~\cite{Ovyn:2009tx,deFavereau:2013fsa}, and \PGS~\cite{PGS}.

\section{Jet Reconstruction}
This section summarises jet reconstruction and calibration in ATLAS and CMS.
Both collaborations use the \fastjet program~\cite{Cacciari:2005hq,Cacciari:2011ma} to cluster input objects into
jets.

\subsection{Jets in the ATLAS detector}

Different stages in the construction and use of ATLAS jets are discussed below.
First the inputs to jets are described, followed by summaries of the
calibration of calorimeter jets and the evaluation of systematic uncertainties
associated with the modelling of the jet response in simulation.
Finally, the procedure used to identify jets originating from hard
bottom quarks ({\em $b$-jets}) in ATLAS is discussed.

\subsubsection{Inputs to jet construction}

For the analysis of ATLAS data, jets are formed using the four-momenta of
different input quantities.
At the particle level, jets are obtained from all particles with a lifetime
of at least 10~ps. These particle jets are used to calibrate the calorimeter jet response.

The standard detector level jets in ATLAS are built from topological calorimeter clusters
with positive energy. Depending on the cluster energy density, likelihoods are calculated that the
cluster results from electromagnetic or hadronic interactions and a correction is applied to
the cluster energy based on simulations of single pion interactions with the
calorimeter (a process called {\em local cluster weighting}, LCW).
The clusters are taken to be massless.

Jets are also constructed using tracks. The resulting track jets are used to validate
the calibrations obtained through calorimeter simulation.
The tracks have to fulfil quality requirements such as a minimal number of
hits in the silicon detector and small longitudinal and transverse impact parameters
with respect to the hard scattering vertex which is chosen to be the one with
largest $\sum p_{\rm T,track}^2$. For jet reconstruction, tracks must have $\pt > 500\MeV$
and the mass is set to the charged pion mass to obtain a four-momentum for jet clustering.

\subsubsection{Jet calibration}
\label{sec:jetcalib}
The standard ATLAS jets that are used in conventional analyses are constructed
with the \akt algorithm with $R=0.4$ or $R=0.6$.
A long chain of sophisticated methods is applied to calibrate the jets
and to derive the systematic uncertainties associated with the simulation of
the jets~\cite{Aad:2011he,ATLAS-CONF-2013-004}. The methods are refined for
every data-taking year because higher statistical measurement precision allows
the application more data-driven approaches.
Only parts of the full chain are used for the substructure jets, i.e.,
fat jets and their subjets, because of manpower constraints. The uncertainties
are therefore larger for the substructure jets.
Only the substructure jet calibrations and uncertainties are discussed in the following.

The jets are calibrated using a simulation of the calorimeter jet response
by comparing the energy and pseudorapidity of a particle jet to that of a
matching calorimeter cluster jet~\cite{Aad:2011he}.
The mass of fat jets is calibrated in an analogous way.
The procedure matches particle jets to calorimeter jets geometrically and
determines the distribution of the ratio of the reconstructed quantity (energy, pseudorapidity, mass)
and the particle level quantity. A Gaussian fit is performed to the core
of the distribution to obtain a correction factor.
Simulations of multijet events are used for the correction.
Fat jets are required to be isolated (typically $\DeltaReta>1.0$) and the particle/detector level jet matching
uses $\DeltaReta < 0.3$. For subjets, the isolation criterion is removed
and a tighter matching is used ($\DeltaReta < 0.1$).

Different approaches are used for 2011 and 2012 ATLAS data to suppress pile-up
contributions.
For the 2011 substructure jets, an implicit average pile-up correction is applied
by overlaying minimum bias events with the hard multijet events
that are used to calculate the detector-to-particle-level correction factor.
In the correction to the particle level, pile-up due to earlier collisions (and
the slow calorimeter) is therefore removed because it is not part of the particle level jet.
The 2012 procedure is described in~\cite{ATLAS-CONF-2013-084}:
before calibrating the subjets, energy depositions that originate from pile-up are removed
to a large extent by applying an area correction~\cite{Cacciari:2007fd} to each
jet~\cite{ATLAS-CONF-2013-083}.  In this correction, the product $\rho \times A_{\rm T}$ is subtracted
from the jet \pt, in which $\rho$ is the median \pt density of the event and $A_{\rm T}$ is the transverse
component of the jet area which is evaluated using ghost association~\cite{Cacciari:2008gn}.
The median \pt density is defined as
\begin{equation}
\rho = {\rm median} \left\{ p_{{\rm T},i}^{\rm jet}/A^{\rm jet}_i \right\}
\end{equation}
in which the index $i$ enumerates the jets found when clustering the event
with the \kt algorithm with $R=0.4$ and requiring positive jet energy but no
minimal jet \pt.

\begin{figure}[hbt]
\centering
\subfigure[]{
   \includegraphics[width=0.45\textwidth,angle=0]{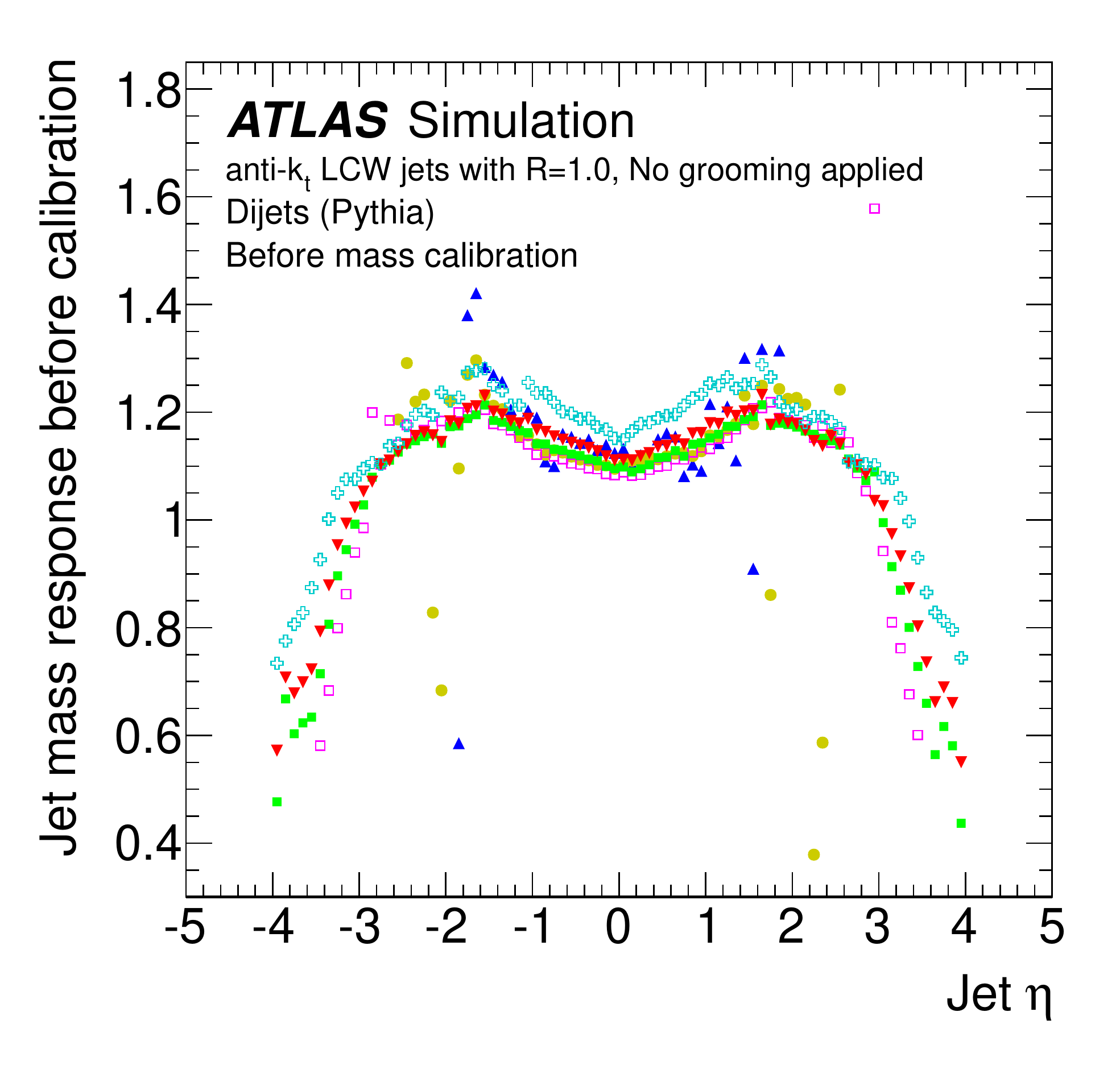}
}
\subfigure[]{
   \includegraphics[width=0.45\textwidth,angle=0]{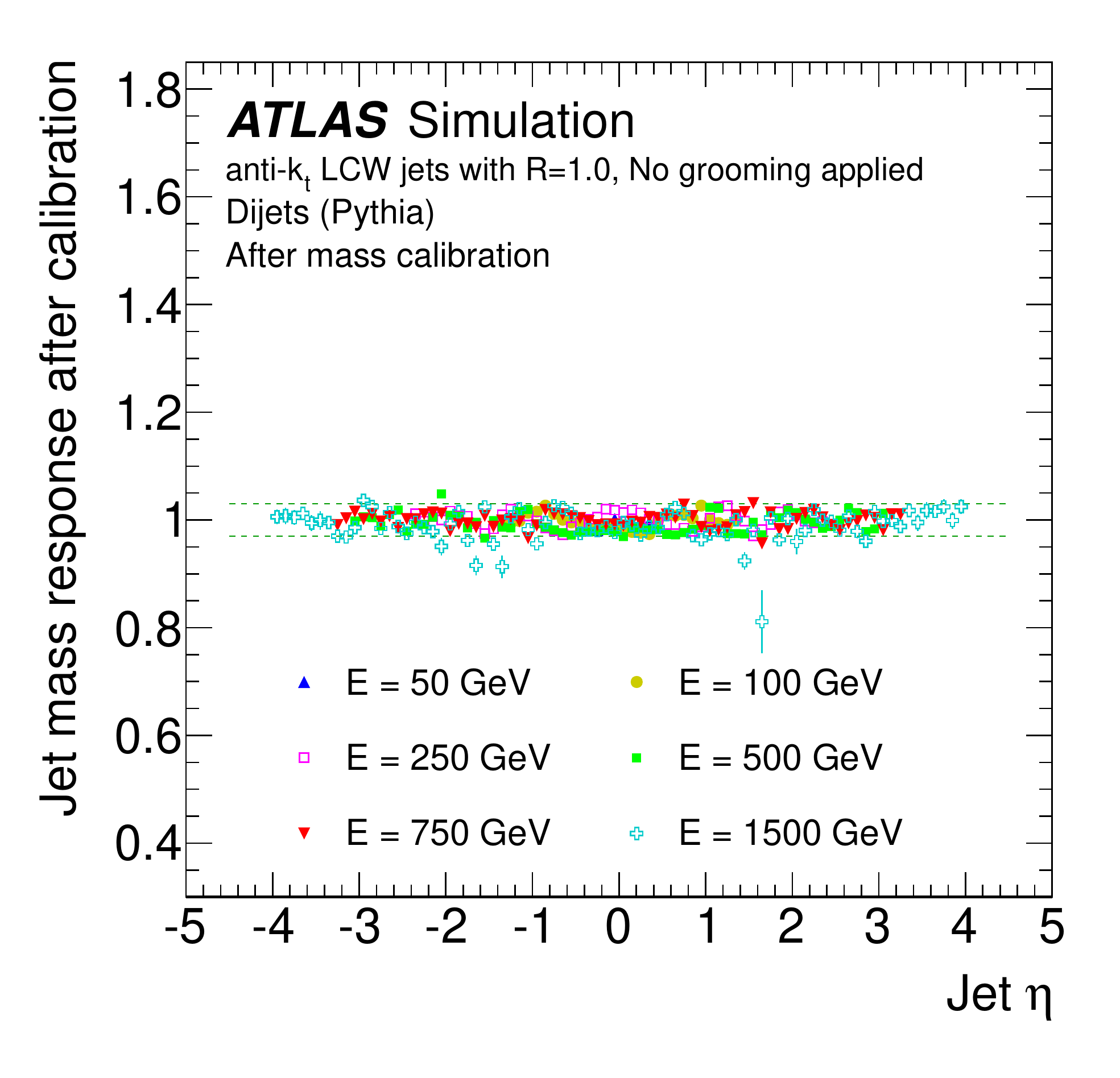}
}
\caption{Mass response of the ATLAS calorimeter for \akt $R=1.0$ jets (a) before and (b) after jet mass calibration.
The response is defined as the mean of Gaussian fit to the core of the ratio of the
reconstructed jet mass to the mass of a geometrically matched particle jet.
From~\cite{Aad:2013gja}.}
\label{fig:mass_calib}
\end{figure}

The 2011 ATLAS mass response for \akt $R=1.0$ fat jets before and after
calibration is shown as a function of $\eta$
in \figref{mass_calib}.
Before calibration, the mass in the central region ($|\eta|<2$) is too large
by $10$--$20\%$ because of pile-up contributions and noise. After correction,
the particle jet mass is reconstructed within $3\%$ for all energies.

The \ca subjets used in the \htt are calibrated as follows. To be able to provide
calibrations for the filtering step with its dynamic distance parameter $\Rfilt$,
the calibration constants are derived for jets with $R=0.2, 0.25, \ldots, 0.5$.
When applying the calibration in the \htt,
the constants are used that correspond to $\Rfilt$ or the next largest $R$ if
no constants exist for the dynamically calculated \Rfilt.

\subsubsection{Validation of jet calibration using tracks}

Uncertainties in the jet calibration are determined from the quality of the
modelling of the calorimeter jet \pt and mass. The direct ratio $\ptjet({\rm
MC})/\ptjet({\rm data})$ is sensitive to mis-modelling of jets at the particle
level (the same is true for the mass). To reduce this effect, the calorimeter jet \pt is normalised to the
\pt of tracks associated with the jet (or to the \pt of a geometrically matched track jet)
because the uncertainty on the track \pt is small compared to the uncertainty on the calorimeter energy.

To evaluate the calorimeter-associated uncertainties of fat jets, jets built from
tracks are geometrically matched to calorimeter jets ($\DeltaReta < 0.3$).
The ratios \rtrkjetpt and \rtrkjetm are defined as the calorimeter-to-track ratios
\begin{equation}
\rtrkjetpt = \frac{ \ptjet }{ \pTtrkjet }\,, \qquad  \rtrkjetm = \frac{ \mjet }{ \mtrkjet }\,.
\end{equation}

\begin{figure}[hbt]
\centering
\subfigure[]{
   \includegraphics[width=0.45\textwidth,angle=0]{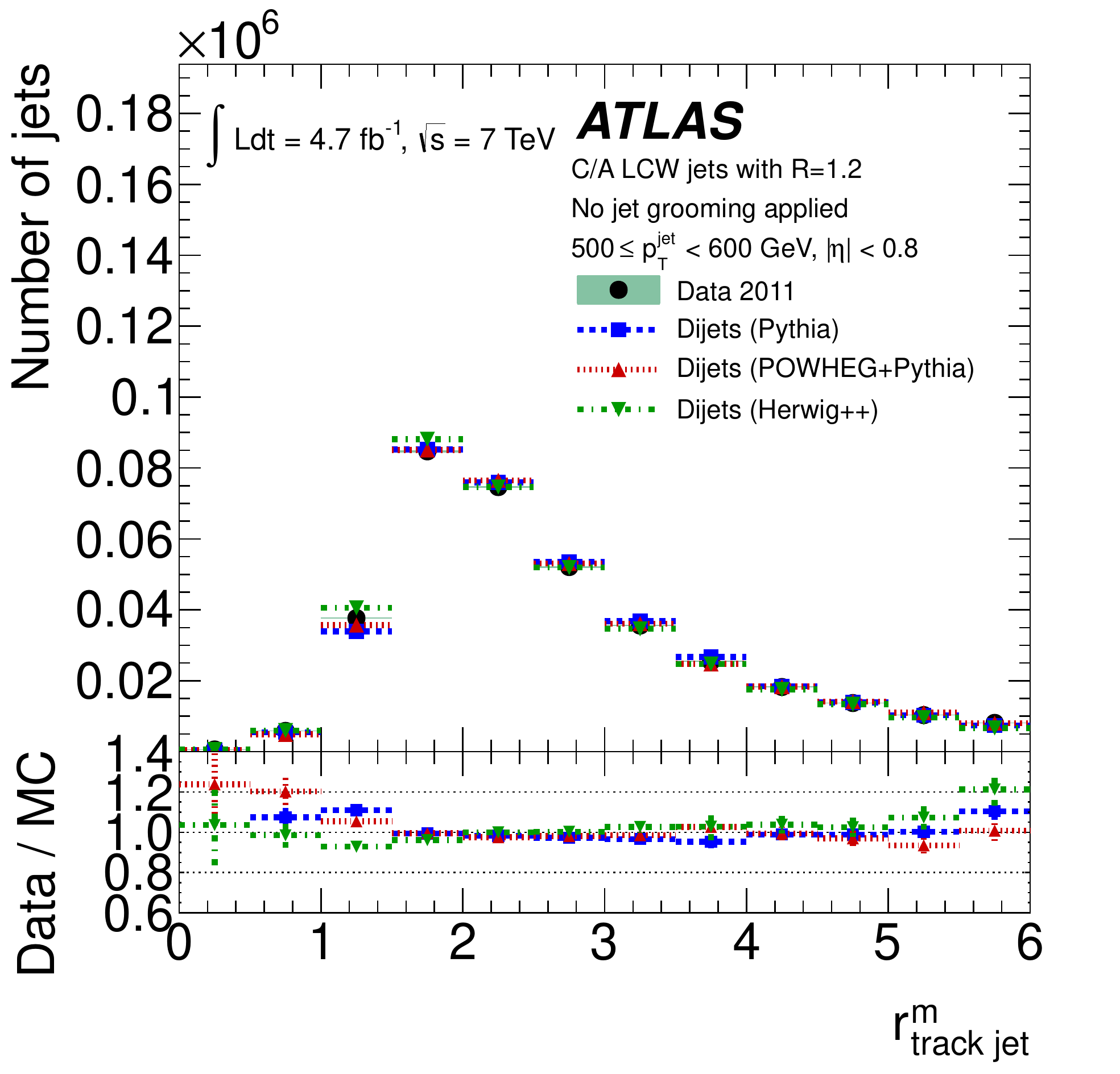}
}
\subfigure[]{
   \includegraphics[width=0.45\textwidth,angle=0]{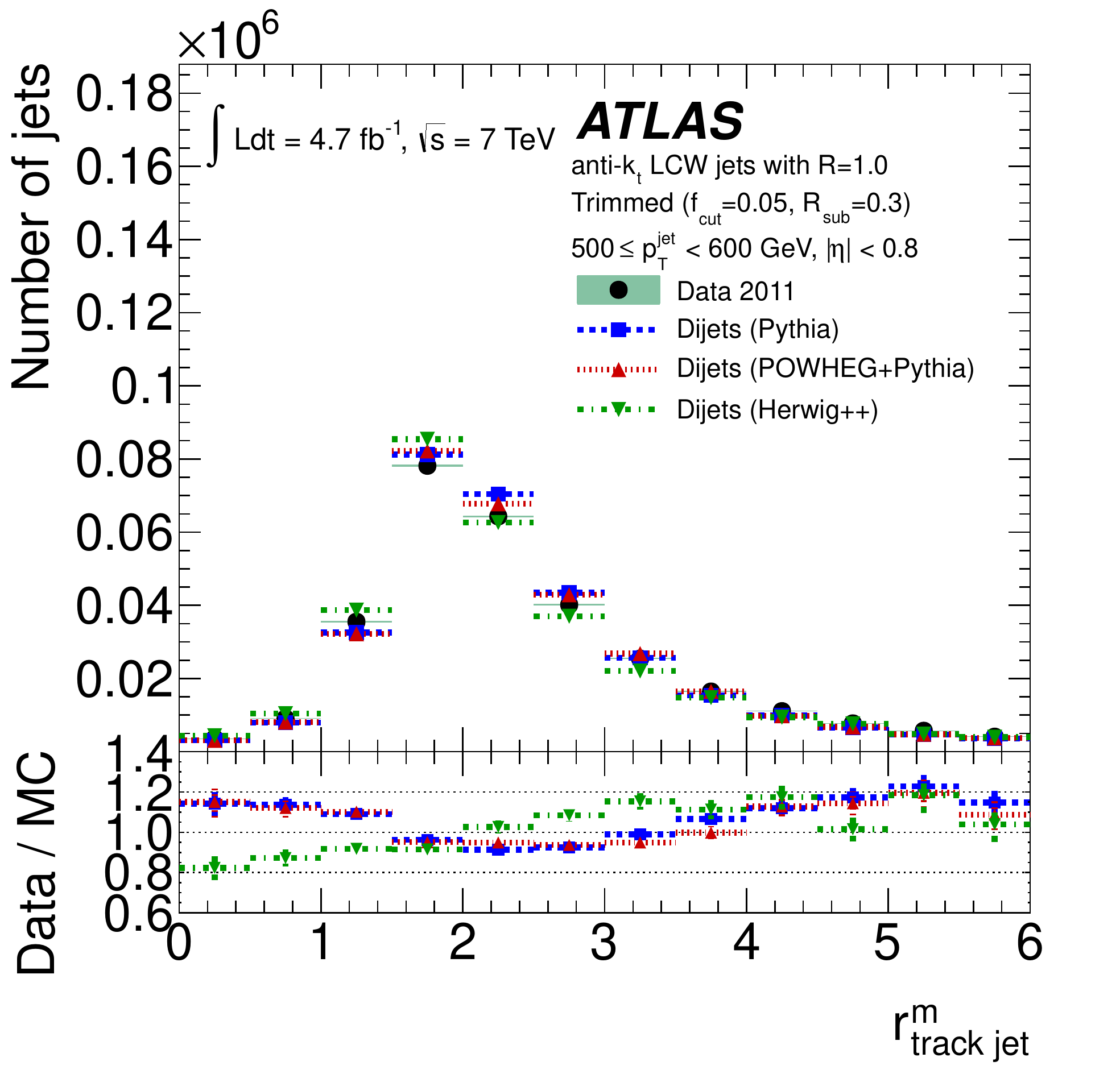}
}
\caption{Ratio of the calorimeter jet mass to the mass of geometrically matched
track jets for (a) \ca $R=1.2$ jets without grooming and (b) \akt $R=1.0$ jets, trimmed with
$\fcut=5\%$ and $\Rsub = 0.3$.
From~\cite{Aad:2013gja}.}
\label{fig:mass_rtrk}
\end{figure}

\figref{mass_rtrk} shows \rtrkjetm for ungroomed \ca fat jets and trimmed \akt fat jets ($\fcut = 5\%$, $\Rsub=0.3$).
The ratios are larger than unity because only charged particles contribute to
the track jets. The distribution for the ungroomed \ca jets is well described by the simulations.
The ratio for the trimmed jets is described within $20\%$.
Data-to-simulation double ratios are defined using the average values of \rtrkjetpt and \rtrkjetm:
\begin{equation}
\Rtrkjetpt = \frac{\langle \rtrkjetpt({\rm data}) \rangle }{ \langle \rtrkjetpt({\rm MC}) \rangle} \, ,
\qquad \Rtrkjetm = \frac{\langle \rtrkjetm({\rm data}) \rangle }{ \langle \rtrkjetm({\rm MC}) \rangle}\, .
\end{equation}
The double ratio is calculated in bins of \pt and \eta{} and for different Monte Carlo generators.
The deviation from unity is
used as an estimate of the systematic simulation uncertainty.
The uncertainty is in the range of $4$--$8\%$, depending on \pt, \eta, the jet
algorithm, and on whether the jet is trimmed or not. This uncertainty includes
uncertainties related to the tracking efficiency which arise from the imperfect
knowledge of the material distribution in the tracking detector.

To evaluate the subjet energy scale uncertainty, tracks are matched
to calorimeter jets using {\em ghost association}~\cite{Cacciari:2007fd, Cacciari:2008gn}
as follows. For every track, a ghost is created by setting the \pt to a small value ($10$~eV) and using
the track $\eta$ and $\phi$ at the calorimeter surface. The energy of the ghost
is set to $1.001$ times its momentum to ensure a positive ghost mass. The ghost tracks are added to the
calorimeter jet clustering but do not change the jet because their energy
is negligible. If the ghost track ends up in the jet
then the original track is taken to be associated with the jet.
The jets are
required to lie within $|\eta| < 2.1$ to ensure coverage of the
associated tracks by the tracking detector.
The impact of pile-up is reduced because only tracks coming from the hard scattering vertex
are used.
The ratio \rtrk is defined as the ratio of the sum of the \pt of the tracks
associated with the jet to the \pt of the jet:
\begin{equation}
\rtrksubjet \equiv \frac{ \sum \pttrk }{ \ptsubjet }
\end{equation}

\begin{figure}[hbt]
\centering
\subfigure[]{
   \includegraphics[width=0.45\textwidth,angle=0]{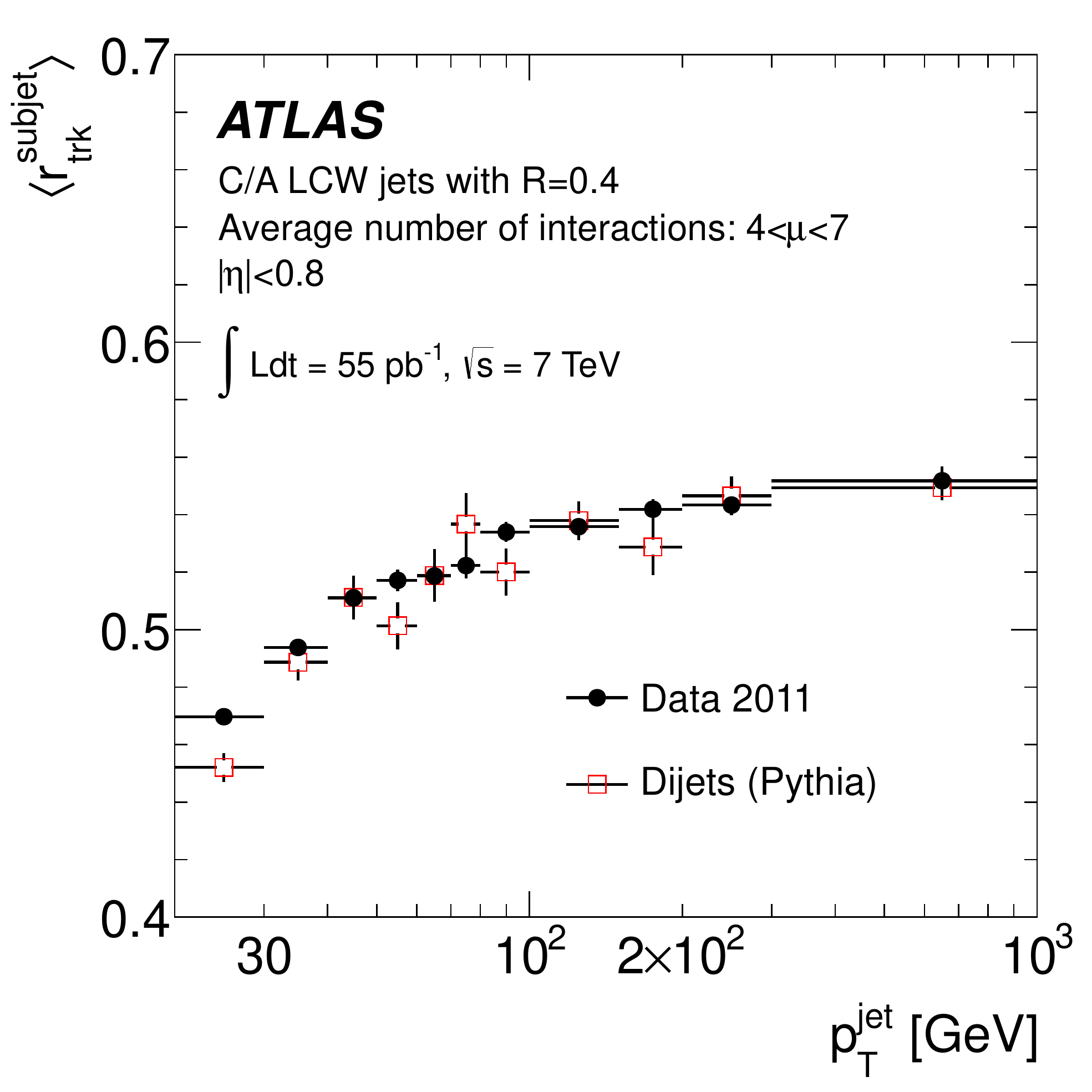}
}
\subfigure[]{
   \includegraphics[width=0.45\textwidth,angle=0]{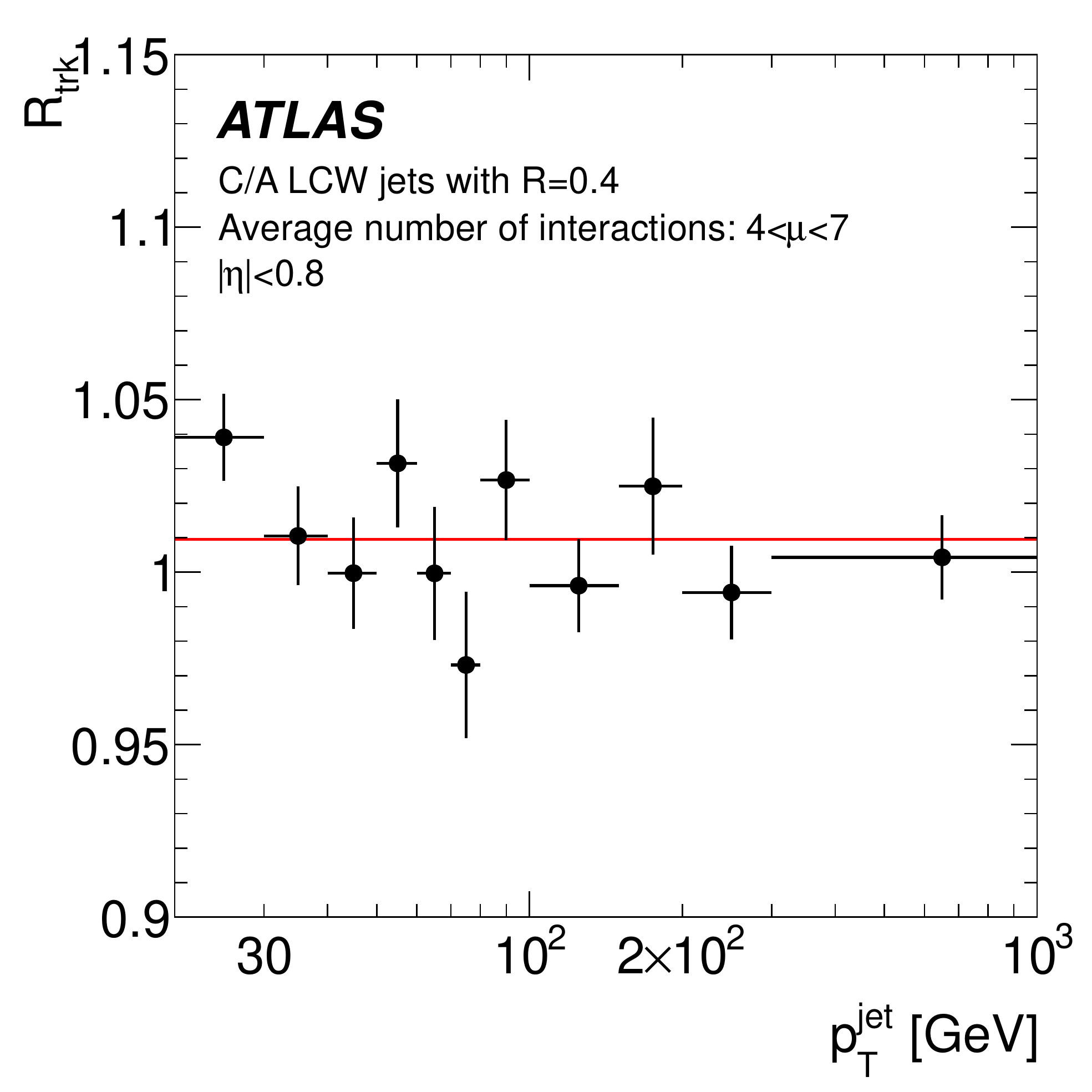}
}
\caption{Comparison of the calibrated Cambridge/Aachen $R=0.4$ calorimeter
jet \pt with the \pt of tracks matched to the jet for ATLAS data recorded in
2011 and for multijet simulations from \pythia. The average number $\avmu$ of
inelastic $pp$ interactions per bunch crossing ranges from $4$ to $7$.
(a) The average ratio \rtrk as a function of the calibrated jet \pt.
(b) The double ratio $\Rtrk = {\langle \rtrksubjet \rangle}_{\rm data} /
{\langle \rtrksubjet \rangle }_{\rm MC}$. The horizontal line indicates the
uncertainty-weighted average.
From~\cite{Aad:2013gja}.}
\label{fig:cartrk}
\end{figure}

\figref{cartrk}a shows the average \rtrk as a function of jet \pt
for \ca $R=0.4$ jets with calibrated \pt between $30$ and $40\GeV$ for $4 < \avmu < 7$.
The track-to-calorimeter \pt ratio \rtrksubjet would be equal to 2/3 if all produced
particle were pions because the tracking detector responds only to charged pions.
However, this ratio is changed by the production of other mesons and baryons.
The fraction of charged particles is well simulated by \pythia as evident
from the description in \figref{cartrk}a.

The double ratio $\displaystyle \Rtrk \equiv {\langle {\rtrksubjet}  \rangle}_{\rm data} / {\langle {\rtrksubjet}  \rangle }_{\rm MC}$
is shown in \figref{cartrk}b.
The largest deviation from unity is
$4\%$ at low \pt with a statistical uncertainty of $1\%$. No rise of the
uncertainty with jet \pt is observed.
Similar results are obtained when varying
the jet radius parameter between $0.2$ and $0.5$ and for high pile-up conditions
($13<\mu<15$). Including systematic uncertainties on the tracking performance,
the subjet energy uncertainty varies between $2.3\%$ and $6.8\%$, depending on
\pt, \eta, and the jet radius.

A sample of semileptonically decaying \ttbar pairs is used to study the
energy scale in events with heavy flavour jets and a boosted topology in which
subjets are close-by. Events are selected from the 2011 dataset as described
in \secref{httperformance} with a \ca $R=1.5$ jet with $\pt > 200\GeV$.
According to simulation, $\approx\!50\%$ of the events have
semileptonically \ttbar pairs in which the hadronically
decaying top quark has $\pt > 200\GeV$ and the remaining events are dominated by
\Wpjets production. The subjet energy uncertainty determined from this
sample varies between $2.4\%$ and $5.7\%$.
ATLAS members can find a detailed description of
the calibration of the \htt subjets and the evaluation of the energy scale and
\pt resolution uncertainties in~\cite{htt_jes, htt_jer}.

\subsubsection{$b$-jets}

The $b$-tagging algorithm MV1 used for the results presented in this article
is based on a neural net that combines information from
significances of impact parameters and decay length,
the total invariant mass of the tracks at the vertex, the
fraction of the total jet energy carried by tracks that is associated
with tracks from the vertex, the multiplicity of two track vertices, and
the direction of the $b$-hadron determined from the subsequent decay vertex of the charmed hadron.
The algorithm tags \akt $R=0.4$ jets.

\begin{figure}[hbt]
\centering
\subfigure[]{
   \includegraphics[width=0.45\textwidth,angle=0]{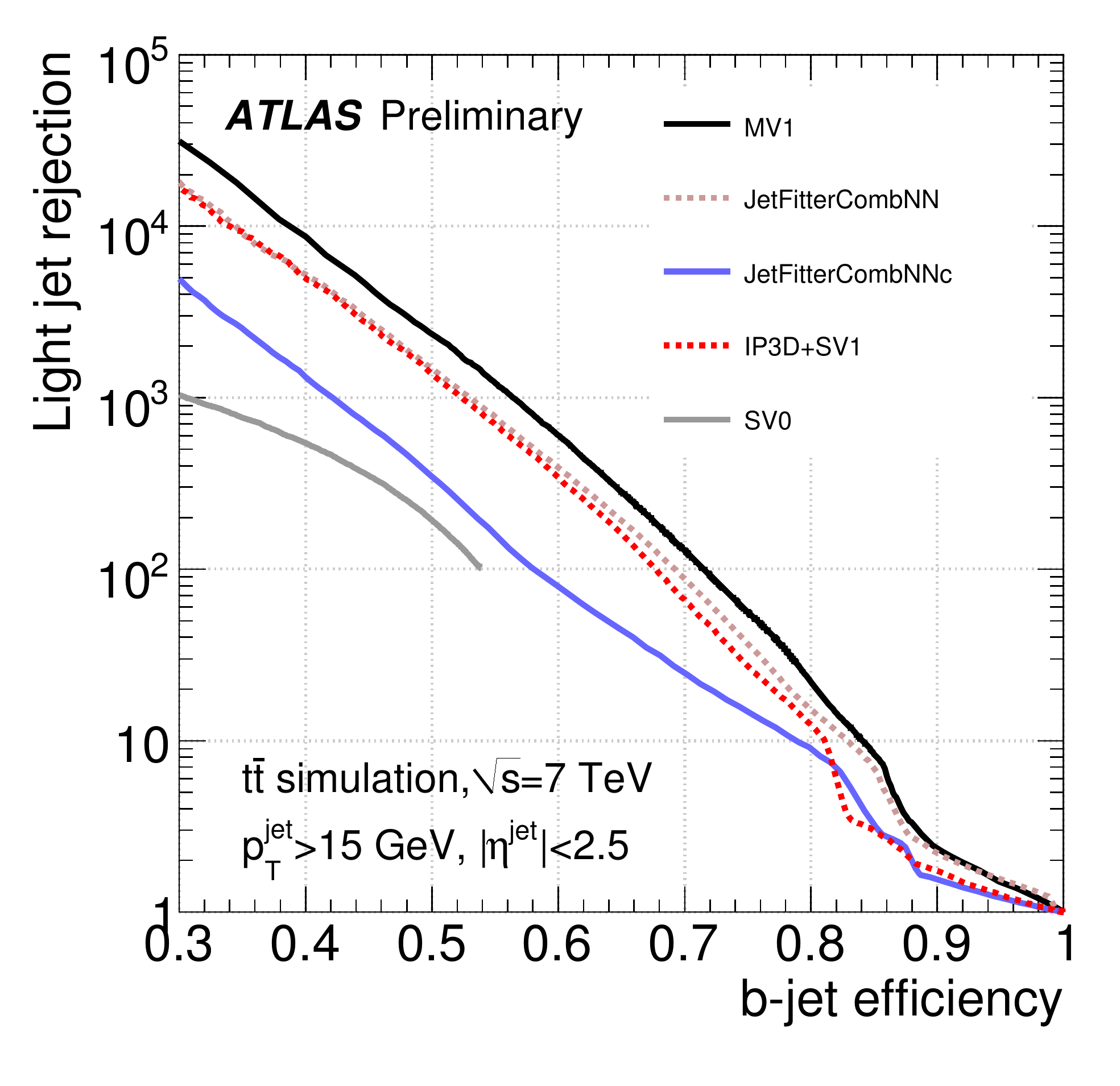}
}
\subfigure[]{
   \includegraphics[width=0.45\textwidth,angle=0]{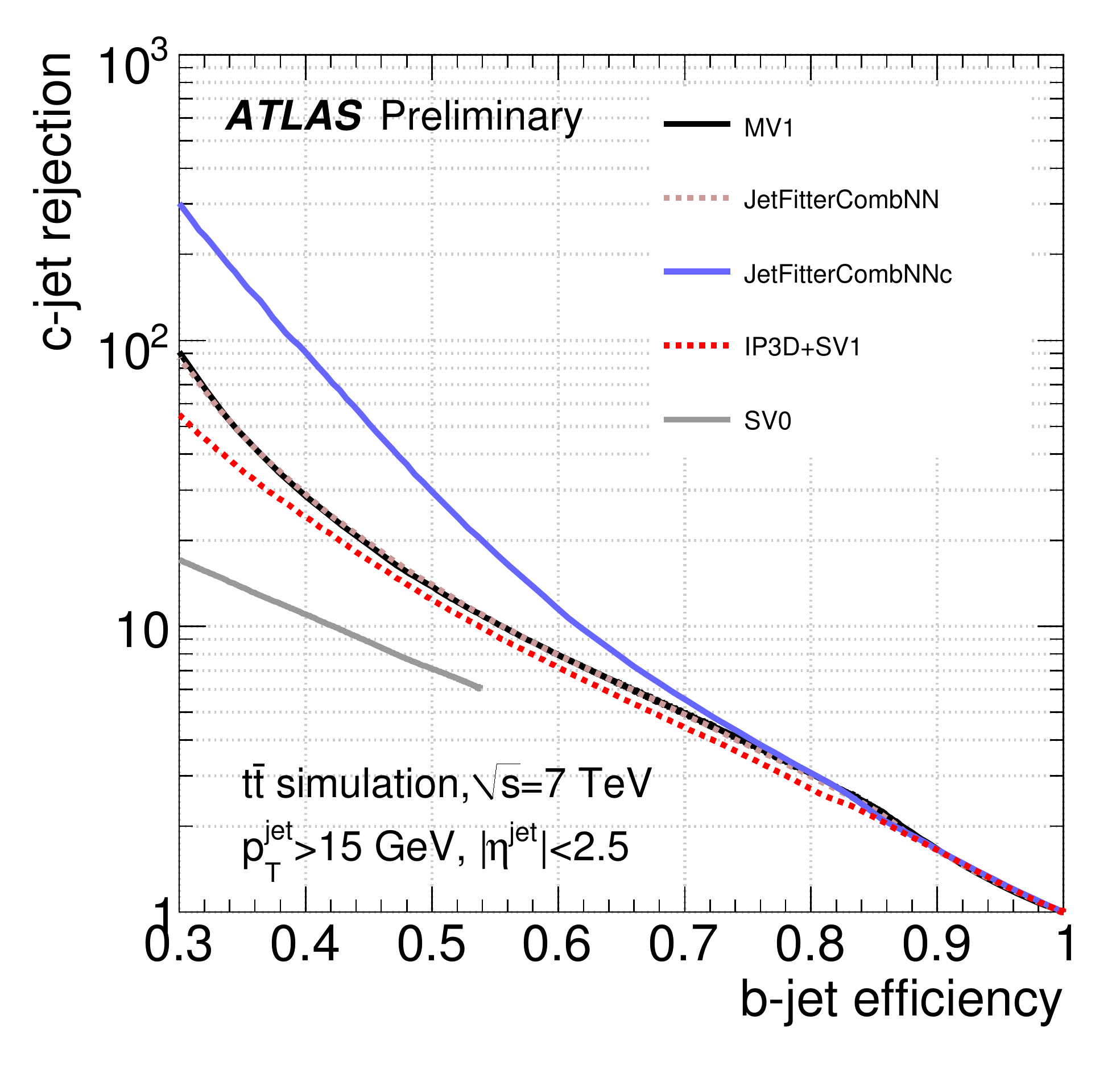}
}
\caption{The performance of $b$-tagging algorithms in ATLAS in simulated \ttbar events.
(a) Rejection of light quark jets vs. efficiency and (b) Rejection of charmed jets vs. efficiency.
From~\cite{ATLAS-CONF-2012-043}.}
\label{fig:btageff}
\end{figure}

The MV1 efficiency is shown in \figref{btageff} as a function of the rejection of
light quark jets and charmed jets for simulated \ttbar events. Rejection is defined as the inverse of the mis-tagging rate (fake rate).
At an efficiency of $70\%$, the fake rate is $\approx\!0.8\%$ for light quark jets
and $\approx\!20\%$ for $c$-jets.

Systematic simulation uncertainties on the $b$-tagging efficiency and on the
mis-tagging rate are obtained from data using different methods.
The most precise method determines the uncertainty for jets with $\pt < 300\GeV$ from
\ttbar events~\cite{ATLAS-CONF-2012-097} and fits the observed $b$-tag multiplicity.
The study is carried out using a semileptonic \ttbar selection and a leptonic selection.
The flavour composition before the tag is obtained from simulation.
The expected number of $b$-tags is given by the product of the number of signal and background jets
of a particular flavour with the efficiency to tag this jet.
The mis-tag rates are taken from simulation but with data-driven scale factors applied~\cite{ATLAS-CONF-2012-097}.
The efficiency is obtained through a fit to the measured $b$-tag multiplicity.
If one would consider only one data sample then the equation could be solved
for the efficiency instead of fitting. However, several channels are considered:
$e$+jets, $\mu$+jets, $ee$, $\mu\mu$, and $e\mu$. The \ttbar production
cross section is known to $10\%$ and the normalisation is left floating in the fit within this uncertainty.
Similarly, the background normalisation is determined in the fit.

\begin{figure}[hbt]
\centering
\subfigure[]{
   \includegraphics[width=0.45\textwidth,angle=0]{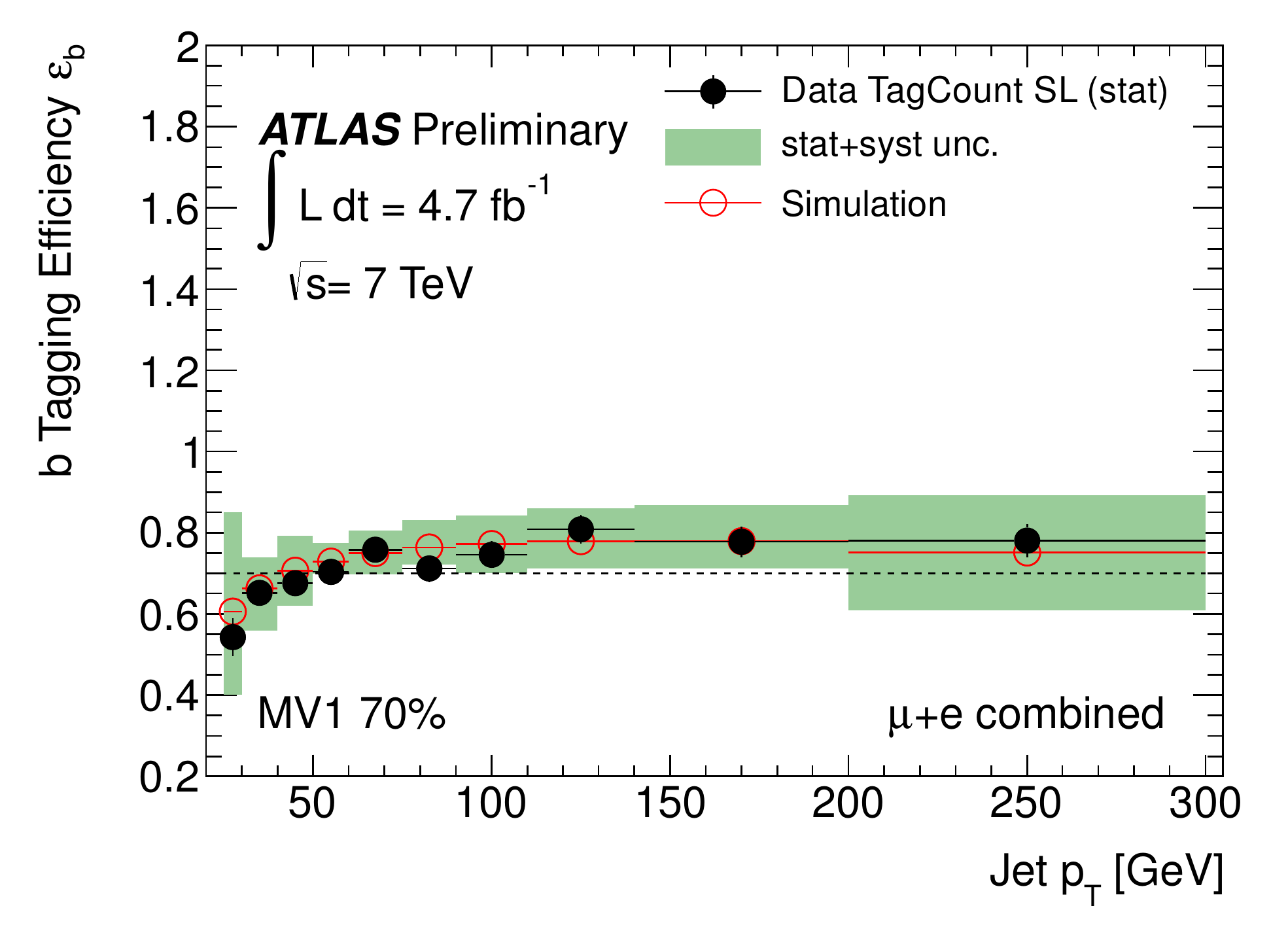}
}
\subfigure[]{
   \includegraphics[width=0.48\textwidth,angle=0]{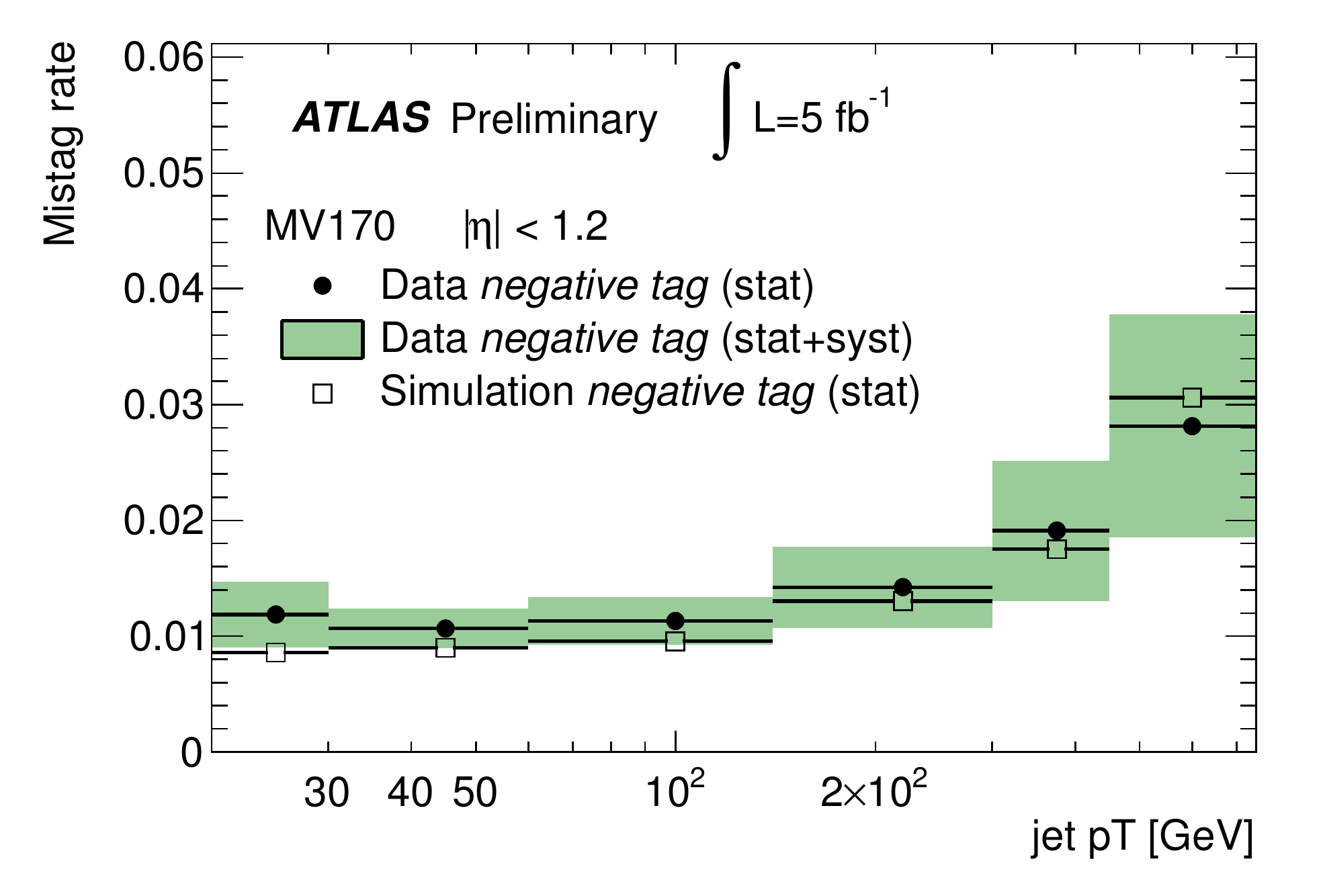}
}
\caption{Efficiency and mis-tag rate of the MV1 algorithm operating at the $70\%$ working point
for \akt $R=0.4$ jets as a function of the jet \pt.
(a) The $b$-tag efficiency obtained from a fit of the $b$-tag multiplicity in a selection of \ttbar events.
The simulation is from \mcatnlo. From~\cite{ATLAS-CONF-2012-097}.
(b) The mis-tag rate obtained from negative impact parameters and decay lengths.
The simulation is from \pythia. From~\cite{ATLAS-CONF-2012-040}.}
\label{fig:mv1_eff_pt}
\end{figure}

The obtained MV1 $b$-tag efficiency is shown in \figref{mv1_eff_pt}a as a function of the \akt $R=0.4$ jet \pt
for the $70\%$ efficiency working point settings.
The efficiency is $55\%$ at $\pt = 25\GeV$ and reaches $75\%$ for $\pt = 100\GeV$.
The relative uncertainty in the \pt range $200$--$300\GeV$ is $25\%$. A different method that uses
a kinematic fit to the \W boson and top quark masses to determine the $b$-jet yields an uncertainty of $10\%$ in this \pt range but
has slightly larger errors at low \pt.
The uncertainty for $200$--$300\GeV$ is assumed to be valid also for larger \pt but
additional uncertainties are added in quadrature as discussed below.
The efficiency decreases with \pt for $\pt>200\GeV$.

The rate at which light quark jets are tagged by the MV1 algorithm has been studied in~\cite{ATLAS-CONF-2012-040}.
Mis-tags occur from fake secondary vertices which
result from tracks being reconstructed at displaced locations due to the finite tracking resolution,
from material interactions, or from long-lived particles (e.g., $K^0, \Lambda$).
For the former of these sources of mis-tags, the signed impact parameter distribution and the signed
decay length distribution should be symmetric around zero and a
mis-tag rate is calculated from the negative tail of these distributions
in multijet events. This rate is then corrected for contributions from material interactions
and long-lived particles
which occur predominantly at positive impact parameters and decay lengths.
The measured mis-tag rate is shown in \figref{mv1_eff_pt}b. It is $1\%$ for jet $\pt<100\GeV$
and rises to $3\%$ for $750\GeV$. The uncertainty is $\approx\!30\%$ for $300$--$450\GeV$.

Additional uncertainties on the $b$-tagging efficiency and fake rate are estimated
using simulation and are added in quadrature to the uncertainties determined at lower \pt.
The largest contribution comes for the loss of tracking efficiency
in the core of jets where adjacent hits created by two charged particles in the pixel detector
are merged such that only one track is reconstructed.%~\cite{ATL-COM-INDET-2011-028}. % internal note
This effect is relevant for jet $\pt > 500\GeV$ and is propagated to the
$b$-tagging algorithm using simulation. The resulting uncertainty on the $b$-tagging
can be as large as $50\%$ for $\pt>800\GeV$.

\subsection{Jets in the CMS detector}

The CMS analysis discussed in this article uses jets reconstructed
from charged hadrons reconstructed with the particle flow approach (cf. \secref{pflow}).
The hadrons have to be consistent with originating from the hard scattering vertex
(the primary vertex with the largest $\sum p_{\rm T,track}^2$) to reduce pile-up contributions.

The jets are calibrated in different stages~\cite{Chatrchyan:2011ds}.
First, the jet area correction is applied to reduce pile-up contributions. The method
is based on the multiplication of the jet area with the average \pt density in the event
as discussed for ATLAS jets in \secref{jetcalib}.
Then the jet energy is corrected using simulation by comparing the jet to a geometrically matched particle jet ($\DeltaReta < 0.25$).
The correction for \akt $R=0.5$ jets is $+10\%$ at $\pt = 20\GeV$ and smaller at higher \pt.
The smallness of the correction is due to the use of the track \pt in the particle flow algorithm.
For comparison, the correction of ATLAS \ca $R=0.4$ jets which are calculated from clusters calibrated
to the response of single pions (LCW) is $+33\%$ when a similar jet area correction
has been applied before~\cite{ATLAS-CONF-2013-084}.
Next, the jets are intercalibrated in $\eta$ using the \pt balance of jets in the central
region ($|\eta|<1.3$) with more forward jets.   % important only for calo jets because |eta|<1.3 is the barrel
The measured \pt balance in $\gamma$+jets events and $\Z\!$+jets events with \ETmiss is
used to correct for small differences in simulation and data ({\em missing transverse energy projection fraction method}, MPF~\cite{Abbott:1998xw}).
This last correction factor is calculated for the central region.

\begin{figure}[hbt]
\centering
\subfigure[]{
   \includegraphics[width=0.45\textwidth,angle=0]{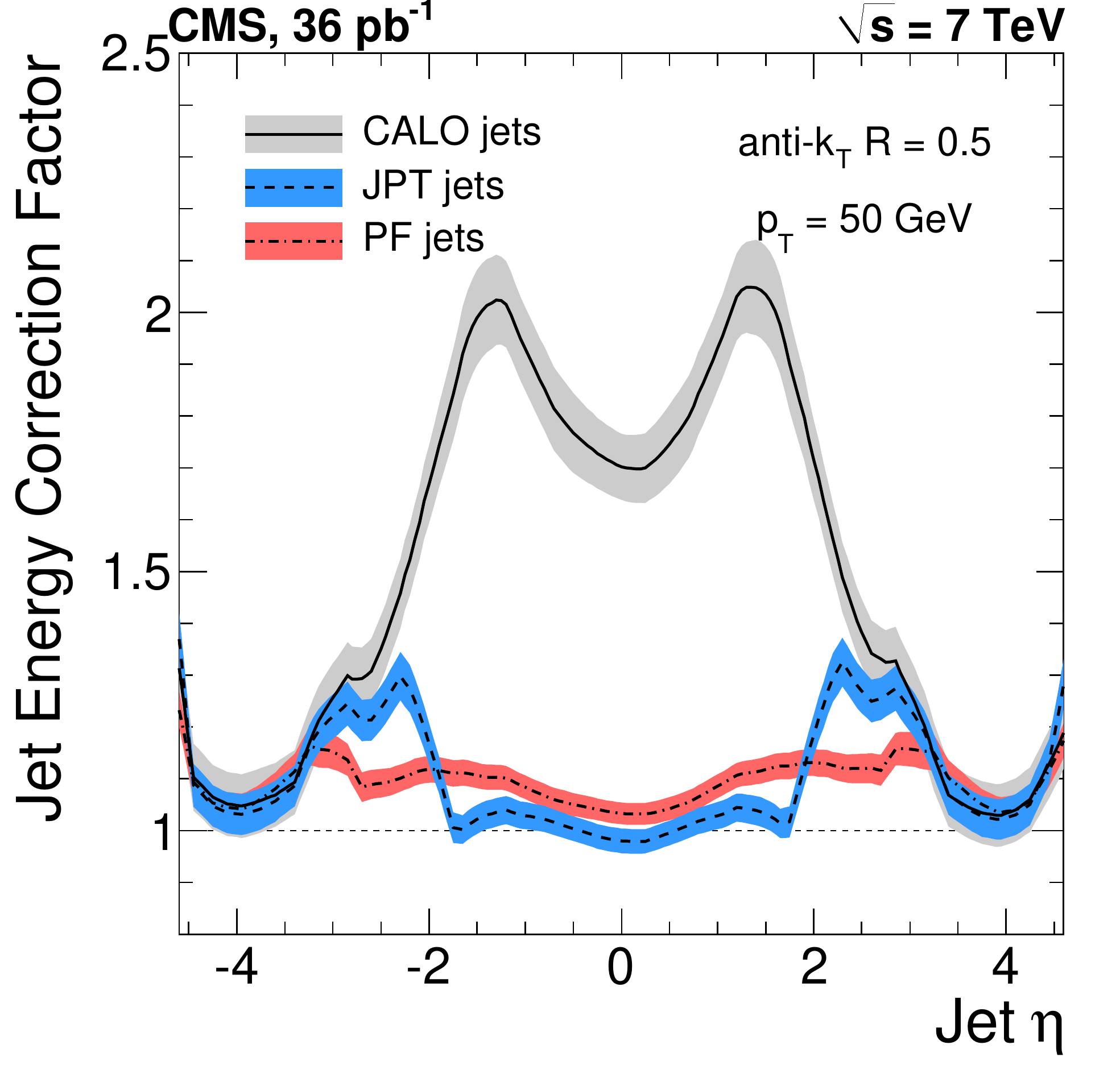}
}
\subfigure[]{
   \includegraphics[width=0.45\textwidth,angle=0]{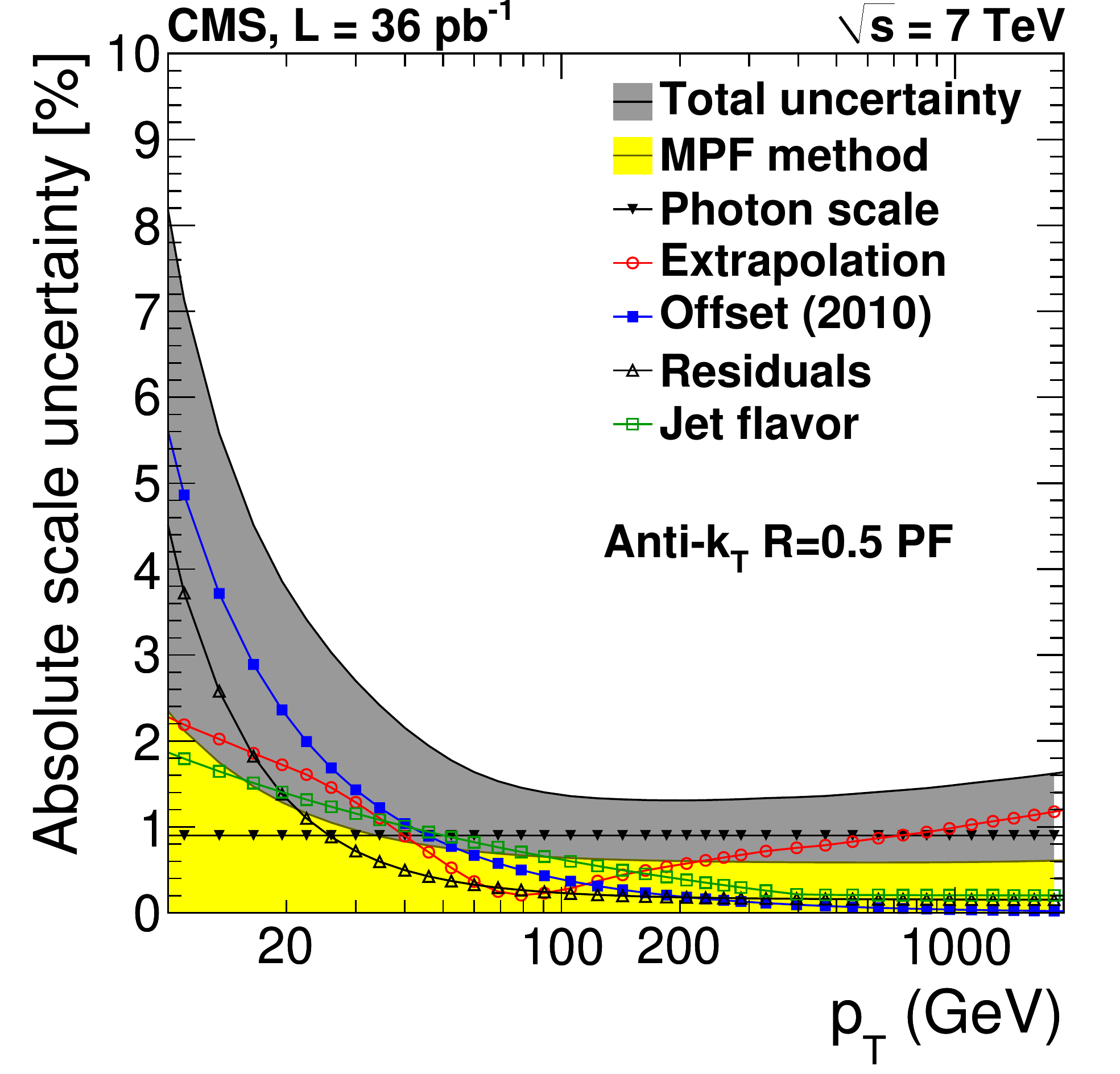}
}
\caption{
Calibration of the energy of CMS \akt $R=0.5$ jets.
(a) The energy calibration factor at jet $\pt = 50\GeV$ as a function of $\eta$
for jets constructed using particle flow (PF), calorimeter towers (CALO), and
the Jet-Plus-Track algorithm (JPT).
(b) The energy scale uncertainty for particle flow jets as a function of \pt.
From~\cite{Chatrchyan:2011ds}.}
\label{fig:cms_jetcalib}
\end{figure}

The full jet energy correction factor is shown in \figref{cms_jetcalib}a for \akt $R=0.5$ jets with $\pt=50\GeV$.
It is less than $1.13$ for $|\eta|<2.7$.
Also shown are the correction factors for jets constructed from calorimeter towers
and using the Jet-Plus-Track (JPT) algorithm~\cite{CMS-PAS-JME-09-002}, which corrects the
energy of calorimeter jets using track jets.
The corrections for the JPT algorithm are smaller in the central region with uncertainties
similar to the particle flow approach. The correction rises for $|\eta|>2$ because
the calorimeter jets extend beyond the acceptance of the tracking detector.
The correction factor for the calorimeter tower jets is larger than $1.5$ for $|\eta|<2$ and
reaches $\approx\!2$ in the barrel-to-endcap transition at $|\eta|=1.3$.

Uncertainties on the jet energy originate from a variety of sources.
The largest contribution at low \pt results from the uncertainty on the \pt density in the pile-up offset correction.
The photon energy scale is known to $\approx\!1\%$.
A $1\%$ uncertainty at $\pt = 20\GeV$ results for the MPF method from leakage of
forward particles ($|\eta|>5$), the uncertainty of which is taken from simulation
to be $50\%$ of the difference between the true response and the MPF response.
The jet energy scale uncertainty is shown in \figref{cms_jetcalib}b. It is $5\%$ at $\pt = 20\GeV$
and smaller than $3\%$ for $\pt > 50\GeV$.

\section{Boosted Top Quark Finders}
\label{sec:topfinders}
Boosted top quark identification algorithms operate on fat jets
which, for signal, are supposed to contain all decay products of the top quark.
The non-top contributions inside the fat jet are removed using jet grooming
procedures. If the remaining constituents fulfil certain kinematic requirements
(such as the jet mass being compatible with the top quark mass) they form a
top quark candidate.
The algorithms are referred to as {\em top taggers}, because of their ability to
tag labels {\em top} or {\em non-top} to the fat jets.

The taggers can be classified into two categories. The first
class, commonly called {\em energy flow taggers}, uses the spatial
distribution of the fat jet constituents and their energies to
calculate cut variables. The second approach is to
explicitly reconstruct the 3-prong top quark decay structure in the form of
subjets inside the fat jet.
The algorithms and variables used so far in ATLAS and CMS are introduced in the following.
Their grouping into the two categories is as follows:
\begin{center}
\renewcommand{\arraystretch}{1.5}
\begin{tabular}{ll}
    energy flow: & jet mass, \kt splitting scales, \Nsjn, \ttt \\
    3-prong:     & \jht, CMS Top Tagger, \htt \\
\end{tabular}
\end{center}
A detailed review of top taggers is given in~\cite{Plehn:2011tg}.
The top taggers in the present article aim at reconstructing hadronically decaying top quarks.

%%%%%%%%%%%%%%%%%%%%%%%%%%%%%%%%%%%%%%%%%%%%%%%%%%%%%%%%%%%%%%%%%%%%%%%%%%%%%%%%

\subsection{Jet mass}
\label{sec:perf_mass}

The mass of a jet is an inclusive substructure variable.
The structure information is encoded in a single number that results from
a convolution of energy deposits and angles.
The mass of a jet that contains
all top decay products can have a mass greater than the top quark mass \mt because of contributions
from other particles. The application of grooming techniques is particularly useful here.

To illustrate the distributions and the effect of grooming,
simulated signal events are used with pairs of top quarks,
and simulated background events from multijet production.
The top quarks result from the decay of a hypothetical new particle,
the topcolor \Zprime introduced in \secref{technicolor},
of mass $1.6\TeV$ into \ttbar. The top quarks are back-to-back in $\phi$
and their \pt is large and peaked at $800\GeV$ (Jacobian peak).
Two different algorithms are used to construct fat jets: the \akt algorithm
with $R=1.0$ and the \ca algorithm with $R=1.2$.
The figures are taken from~\cite{Aad:2013gja}, an extensive study of jet substructure methods with ATLAS.

\begin{figure}[hbt]
\centering
\subfigure[\akt $R=1.0$]{
   \includegraphics[width=0.45\textwidth,angle=0]{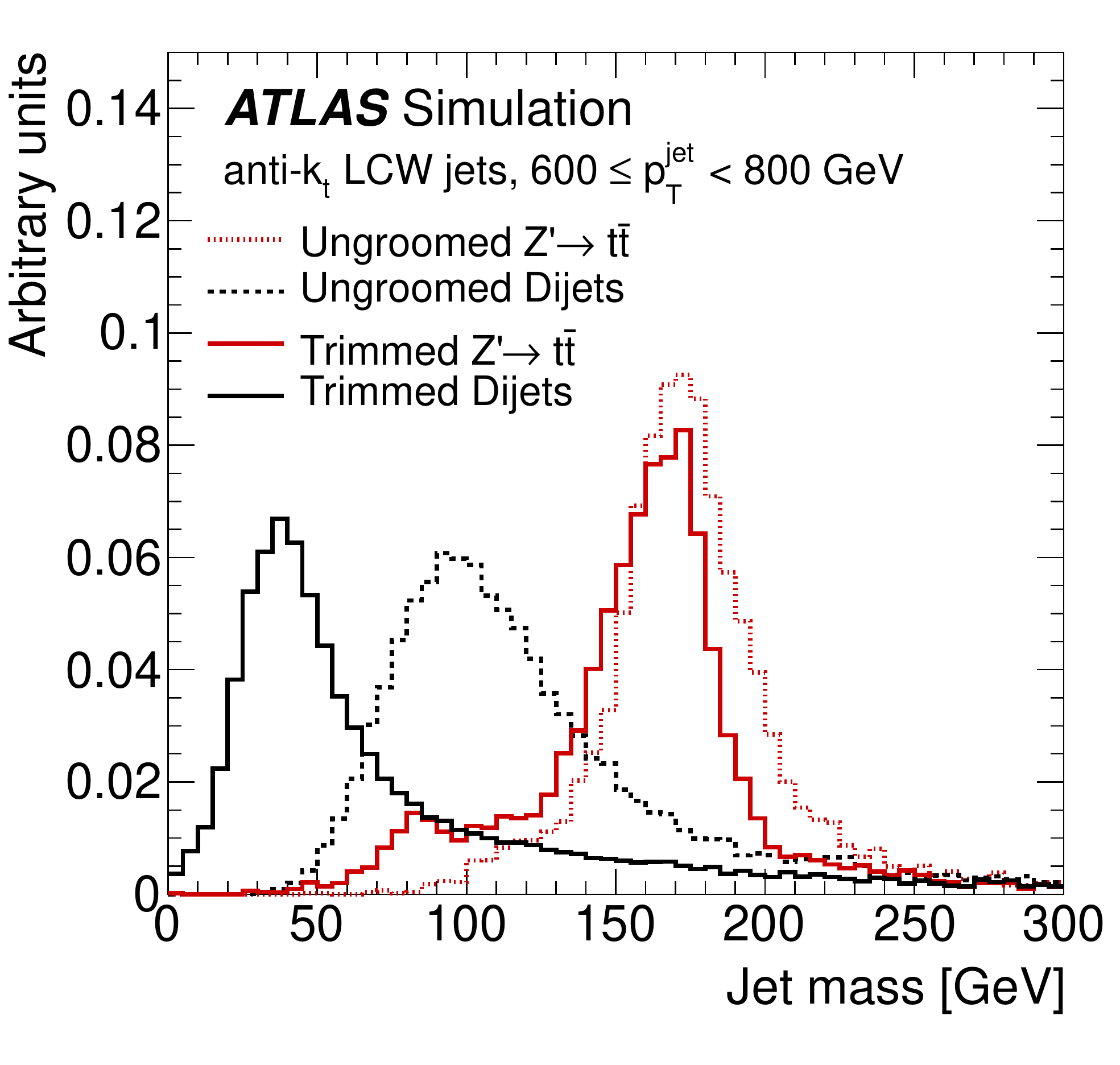}
}
\subfigure[\ca $R=1.2$]{
   \includegraphics[width=0.45\textwidth,angle=0]{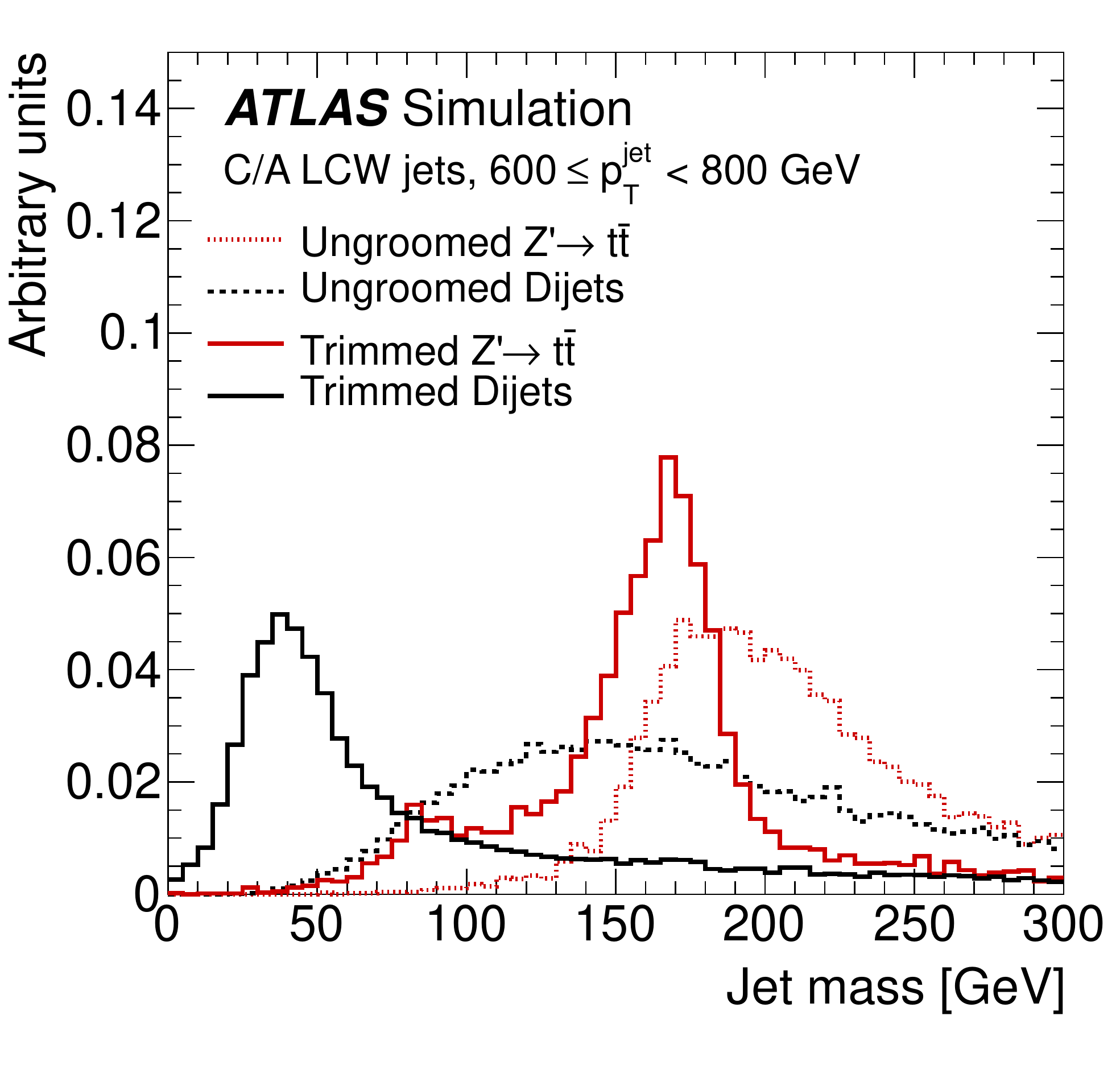}
}
\caption{The mass of the leading \pt fat jet in simulated ATLAS events before
and after applying the trimming procedure with $\fcut = 5\%$ and $\Rsub = 0.3$.
The fat jet \pt ranges from $600$ to $800\GeV$ and the jets have been reconstructed
using (a) the \akt algorithm with $R=1.0$ and (b) the \ca algorithm with $R=1.2$.
Shown are distributions for $\Zprime \rightarrow \ttbar$ events with $m_{\Zprime} = 1.6\TeV$,
simulated using \pythia and for multijet events from \powhegpythia. From~\cite{Aad:2013gja}.}
\label{fig:perf_mass}
\end{figure}

The fat jet mass for the chosen signal and background events is shown in \figref{perf_mass}.
The ungroomed mass of \akt jets in signal events is peaked near the top quark mass.
Trimming with parameters $\fcut = 5\%$ and $\Rsub = 0.3$ removes part of the constituents which shifts the distribution to lower
masses. The peak position shifts by $5$--$10\GeV$.
In some cases, the clusters from the $b$-jet are removed, giving rise to the shoulder at \mw.

The jet mass is smaller for background because the partons themselves
have negligible mass and the jet mass is generated geometrically in
quasi-collinear splitting (cf.~\eqref{mass}).
The trimming effect is larger for the background jets because
for them more soft particles contribute to the mass while the mass of signal
jets is dominated by three hard particle jets.
The original background distribution is peaked near $100\GeV$
and trimming shifts it down by $60\GeV$.
The separation of signal and background is therefore improved by the application
of trimming. To select events with top quarks, the trimmed fat jet mass
is required to lie in a mass window around the true top quark mass.
Other grooming techniques can be used instead of trimming.

\begin{figure}[hbt]
\centering
\includegraphics[width=0.52\textwidth,angle=0]{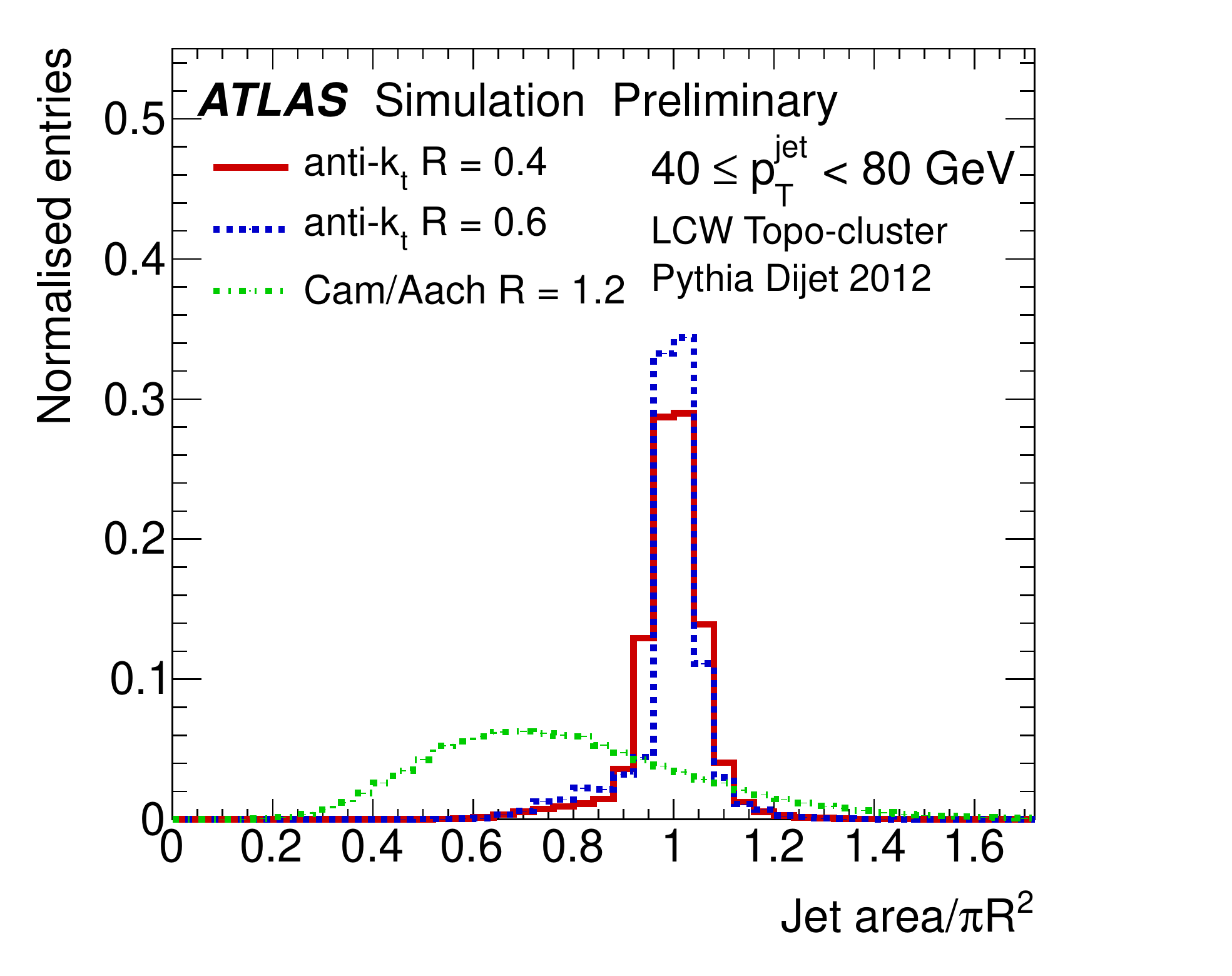}
\caption{Catchment areas of jets reconstructed with different jet algorithms from the same
simulated ATLAS events. The jet \pt at the local cluster weighting scale (LCW),
i.e., corrected to the level of single pions, is between $40$ and $80\GeV$.
From~\cite{ATLAS-CONF-2013-083}.}
\label{fig:jetareas}
\end{figure}

For \ca $R=1.2$ jets, the ungroomed mass distributions are broader
and the masses larger on average when compared with the \akt $R=1.0$ jets.
Trimming yields mass distributions similar to the ones of the trimmed \akt jets.
The ungroomed \ca jets therefore contain more soft constituents that lie at large angles to the
jet axis. This is what one naively expects from the larger radius parameter
but \ca jets are not cone-like.
\figref{jetareas} shows the jet catchment area~\cite{Cacciari:2008gn} for \akt and \ca jets
for $40<\pt<80\GeV$.
This area is calculated by including in the jet clustering a large number
of {\em ghosts} which are distributed according to a fine grid in $(y,\phi)$.
The ghosts carry no significant energy such that they do not affect the
kinematics of the jet. By looking at the ghosts which end up in the jet,
the catchment area of the jet can be determined.
The area of \akt jets is close to $\pi R^2$, with the peak
for $R=0.6$ being narrower than the one for $R=0.4$, which is why it is reasonable
to assume that the area will scale similarly when going to $R=1.0$.
For \ca $R=1.2$ jets the distribution is much broader and
the most probable area is $0.7 \times \pi \times 1.2^2 \approx \pi$, which corresponds to the area of
\akt $R=1.0$ jets. For approximately equal areas,
the irregular shape of a \ca jet implies that it
contains more constituents at large angles to the jet axis which leads to a larger
mass.
The average area for \ca jets is also slightly larger because the distribution
is asymmetric with a more pronounced large area tail.

\subsection{\kt splitting scales}

\begin{figure}[hbt]
\centering
\subfigure[]{
   \includegraphics[width=0.45\textwidth,angle=0]{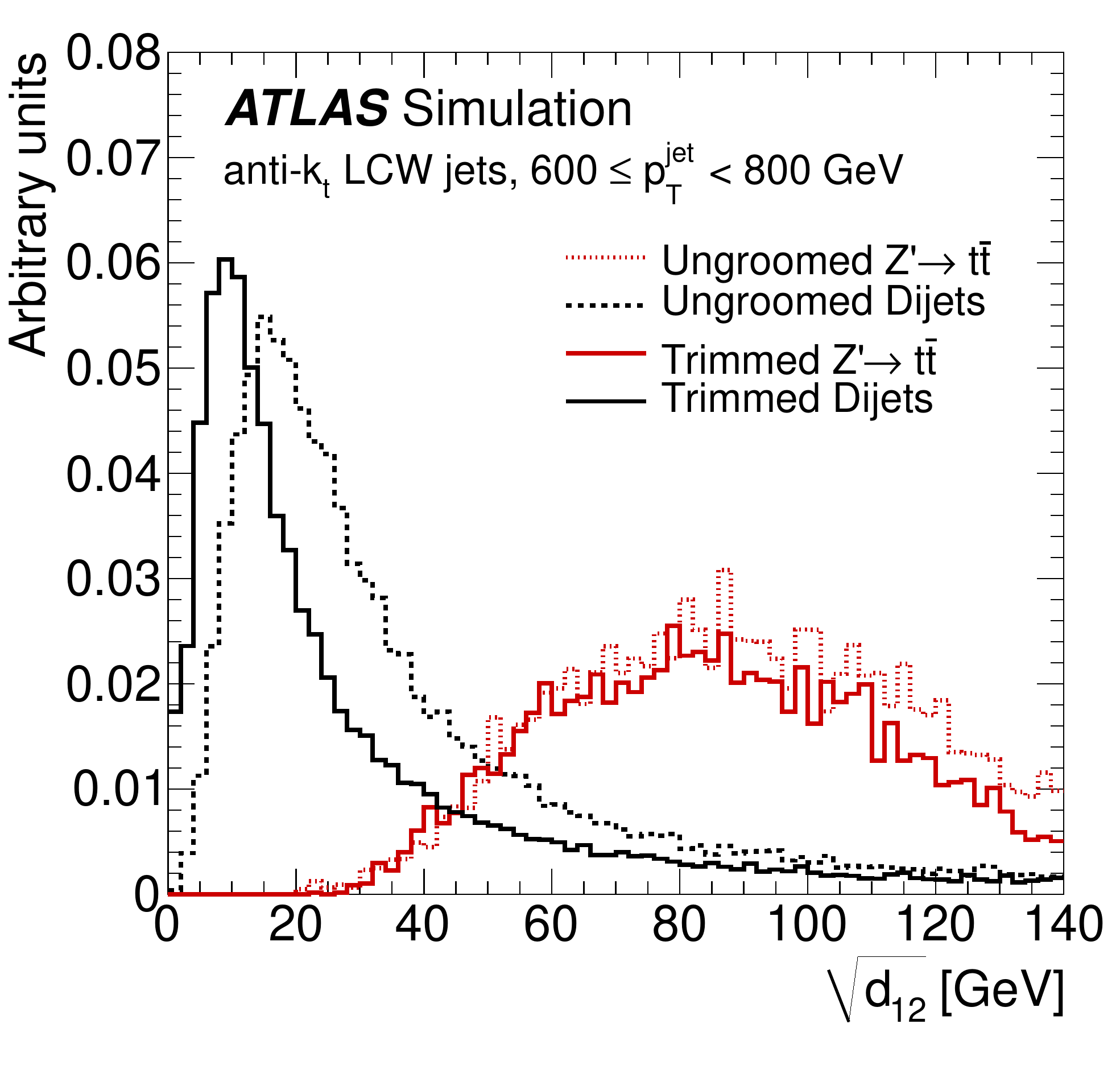}
}
\subfigure[]{
   \includegraphics[width=0.45\textwidth,angle=0]{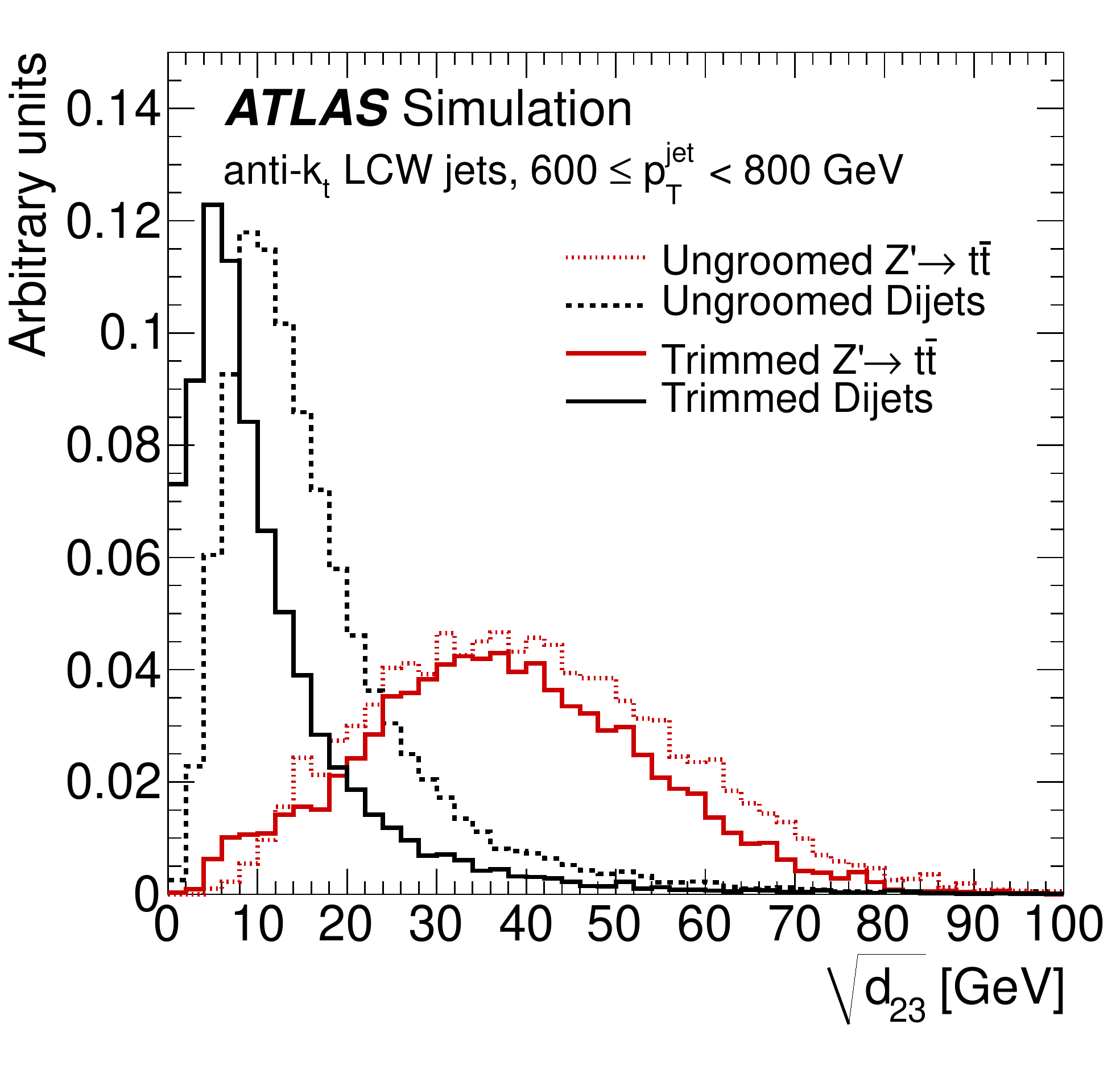}
}
\caption{The (a) first and (b) second \kt splitting scale of the leading \pt \akt $R=1.0$ fat jet
in simulated ATLAS events before and after applying the trimming procedure with $\fcut = 5\%$ and $\Rsub = 0.3$.
The fat jet \pt at the local cluster weighting scale (LCW) ranges from $600$ to $800\GeV$.
Shown are distributions for $\Zprime \rightarrow \ttbar$ events with $m_{\Zprime} = 1.6\TeV$,
simulated using \pythia and for multijet events from \powhegpythia. From~\cite{Aad:2013gja}.}
\label{fig:perf_scales}
\end{figure}

\figref{perf_scales} shows the first and second \kt splitting scales for
\akt fat jets whose constituents have been reclustered using the \kt algorithm.
For the top jets, the first scale \DOneTwo shows a broad peak near $\mt/2$ and
\DTwoThr peaks at $\approx\!m_W/2$. For background jets the values are much lower
because the transverse momenta in their mergings are more asymmetric.
Trimming affects signal jets only slightly but has a significant impact on the
background jets.
The distributions for \ca fat jets are similar.
Top quark jets can be selected by requiring, e.g., $\DOneTwo > 40\GeV$ and
$\DTwoThr>20\GeV$ for the trimmed jets.

\subsection{\Nsjn}
\label{sec:nsjn}
\Nsjn is a variable that quantifies to what extend the constituents of a jet
align along $N$ subjet axes. The constituents of a jet with radius $R$
are clustered exclusively into $N$ subjets using the \kt algorithm.
The \Nsjn $\tau_{N}$ is defined as the normalised \pt-weighted sum of the
constituent distances to the nearest subjet axis:
\begin{equation}
\tau_N \equiv \frac{1}{d_0} \, \sum_{k=1}^M \left( p_{{\rm T},k} \times \DeltaR^{\rm min}_k \right) \, . \label{eq:nsubjettiness}
\end{equation}
The sum is over all $M$ constituents $k$, and $\DeltaR^{\rm min}_k$ is the
distance of $k$ to the nearest subjet axis.
The normalisation is given by the variable $d_0$ which is the sum of
the \pt of all constituents multiplied by $R$.

For $N = M$ each constituent represents a subjet and $\tau_M = 0$.
For $N = 1$ all constituents are clustered into a single \kt jet and in almost all cases
$\tau_1<1$.

If the internal jet structure follows an $N$-prong pattern, then $\tau_N$ is
significantly smaller than $\tau_{N-1}$ because the constituents align more
closely with the axes of the $N$ subjets. A useful variable to identify hadronic top quark
decay is therefore $\tau_{32} \equiv \tau_3/\tau_2$ while $\tau_{21} \equiv \tau_2/\tau_1$
is used for 2-prong decays, like that of \Z or \W bosons.

\begin{figure}[hbt]
\centering
\subfigure[]{
   \includegraphics[width=0.45\textwidth,angle=0]{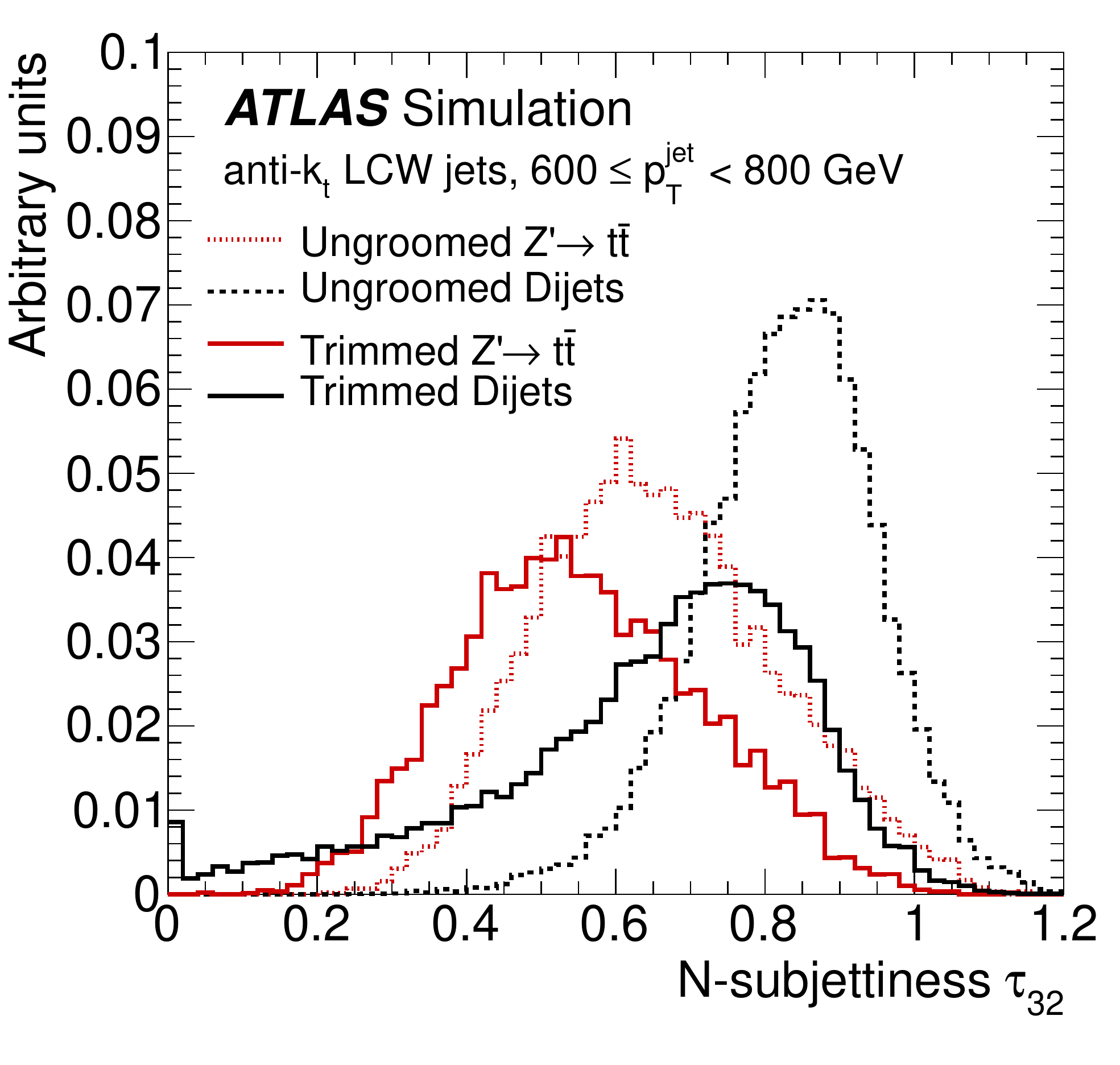}
}
\subfigure[]{
   \includegraphics[width=0.45\textwidth,angle=0]{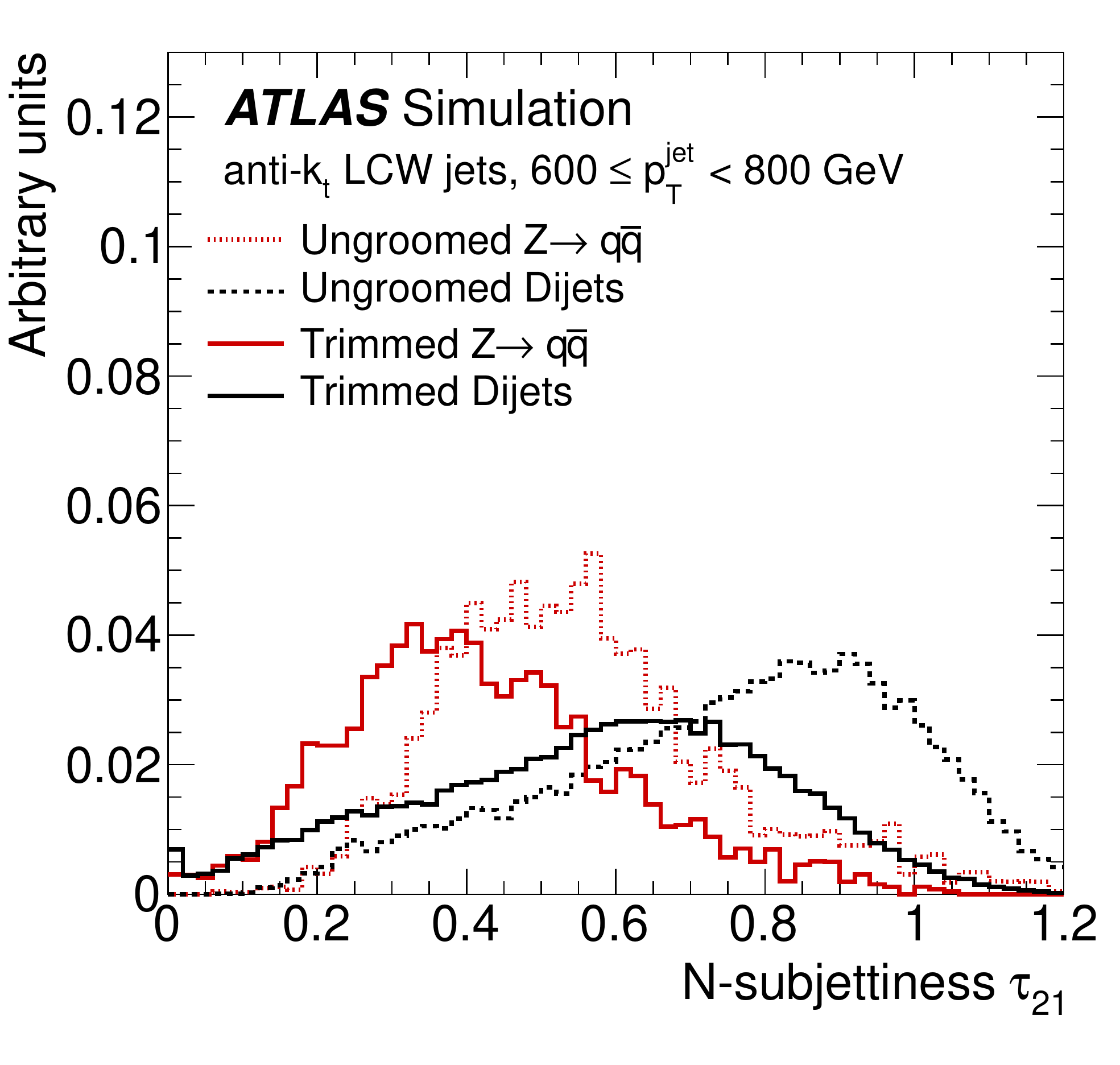}
}
\caption{\Nsjn ratios $\tau_{32}$ (a) and $\tau_{21}$ (b) for the leading \pt \akt $R=1.0$ fat jet
in simulated ATLAS events before and after applying the trimming procedure with $\fcut = 5\%$ and $\Rsub = 0.3$.
The fat jet \pt at the local cluster weighting scale (LCW) ranges from $600$ to $800\GeV$.
Shown are distributions for $\Zprime \rightarrow \ttbar$ events with $m_{\Zprime} = 1.6\TeV$,
simulated using \pythia and for multijet events from \powhegpythia. From~\cite{Aad:2013gja}.}
\label{fig:perf_nsubjettiness}
\end{figure}

Distributions of $\tau_{32}$ for signal and background fat jets are shown in
\figref{perf_nsubjettiness}a. Jets with hadronic top quark decay products are better described
by three than by two subjets and hence have small average $\tau_{32}$. For
background jets the additional subjet does not reduce $\tau_3$ much compared to $\tau_2$.
Trimming affects signal and background jets similarly.
A typical cut to enrich top quark decays is $\tau_{32}<0.65$ for trimmed fat jets.
Trimmed jets are used for the cut because the \Nsjn of trimmed jets is better
described by simulation than that of ungroomed jets, as will be shown in \secref{measurements_nsjn}.

\figref{perf_nsubjettiness}b shows $\tau_{21}$ but here the signal is given by
hadronic SM \Z boson decays. These jets are better described by two subjets than
the background jets, resulting in a smaller value of $\tau_{21}$. Again, trimming
does not help to separate signal and background.
The distributions for \ca fat jets look similar~\cite{Aad:2013gja} (not shown).

%%%%%%%%%%%%%%%%%%%%%%%%%%%%%%%%%%%%%%%%%%%%%%%%%%%%%%%%%%%%%%%%%%%%%%%%%%%%%%%%

\subsection{\ttt}

The \ttt method~\cite{Almeida:2010pa,Almeida:2011aa} compares the energy distribution inside a jet
with a template at the parton level. The agreement is quantified in terms
of the energy difference between the energies of the three final state partons
in $t\rightarrow b q q$ and the energies
of the calorimeter clusters in the vicinity of the partons.
The method represents a typical energy flow tagger.

The technique has first been applied to measured jets in an ATLAS analysis
that searched for \ttbar resonances~\cite{Aad:2012raa}, which is discussed
in \secref{ttt_ana}. In this analysis, the {\em overlap} of a parton
configuration with the measured calorimeter cluster energies is given by
\begin{equation}
{\rm overlap} = \exp \Biggl[ - \sum_{\underset{i=1}{\rm parton}}^3 \frac{1}{2\sigma_i^2}
\Bigl(E_i - \!\!\underset{<0.2}{\sum_{\DeltaReta(\mathrm{clus},i)}}\!\!\!
E_{\mathrm{clus}} \Bigr)^{2} \Biggr] \, , \label{eq:overlap}
\end{equation}
in which $\sigma_i \equiv E_i/3$ and the minimum cluster \pt is $2\GeV$.
Clusters are matched to partons if the distance in $(\eta, \phi)$ space
$\DeltaReta \equiv \sqrt{(\Delta \eta)^2 + (\Delta \phi)^2}$ is less than $0.2$,
which corresponds to the approximate angular calorimeter resolution.
The overall performance of the method has been found to be insensitive
to the particular choices of the matching cone size, the normalisation variable
$\sigma_i$, and the minimum cluster \pt.

A parton configuration is a specific arrangement of the four-momenta of the
quarks. Millions of configurations have to be calculated and stored as templates
to cover the full three-body decay phase space to sufficient granularity to
allow separation of background.
The overlap is calculated for each template according to \eqref{overlap} and
the maximum overlap of all templates is called \ovthree.

\begin{figure}[hbt]
\centering
\includegraphics[width=0.5\textwidth,angle=0]{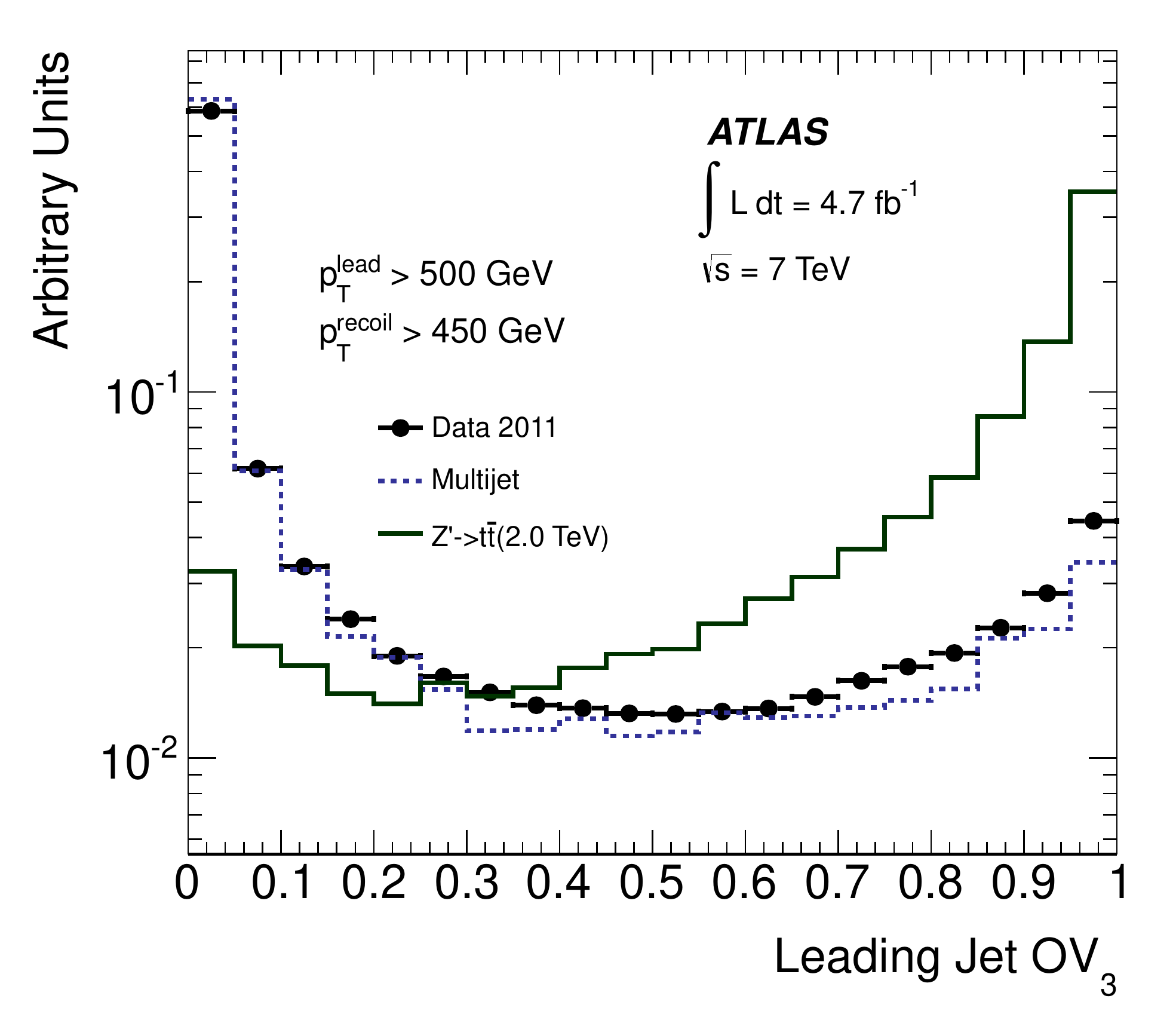}
\caption{Normalised distribution of the \ttt overlap variable \ovthree
for the leading \pt \akt $R=1.0$
fat jet with $\pt>500\GeV$ at the local cluster weighting scale (LCW)
in an ATLAS data sample that is dominated by
multijet events. Also shown are distributions for simulated events from \pythia with multijets and
with a hypothetical \Zprime boson of mass $2\TeV$ that decays to \ttbar. From~\cite{Aad:2012raa}.}
\label{fig:ttt_ov3}
\end{figure}

The distribution of \ovthree is shown in \figref{ttt_ov3} for the leading \pt
\akt $R=1.0$ fat jet with $\pt>500\GeV$ in ATLAS multijet events and in simulated
events with a heavy hypothetical \Zprime boson that decays to \ttbar.
The overlap is on average larger for the \Zprime events and a cut $\ovthree>0.7$ is
used to enhance signal over background in the resonance search.
The background distribution rises towards large values of \ovthree because some background
jets look top-like by maximising the overlap accidentally.
The description of the data is not perfect, especially at high values, and that
is the reason why the background in the resonance analysis is not taken from
simulation but from data in side-bands.

A second variable used in the tagger to enhance the signal contribution
is the fat jet mass which is required to be within $\pm50\GeV$ of the
top quark mass of $172.3\GeV$.
The mass is sensitive to pile-up contributions as discussed in \secref{perf_mass}.
The average mass shift is measured in data as a function of the energy flow
away from the jet and a correction is applied~\cite{Alon:2011xb,Aaltonen:2011pg}.

The combination of cuts on \ovthree and the fat jet mass results in an efficiency of $\approx\!75\%$ for
tagging top quarks and a fake rate of $\approx\!10\%$ for tagging fat jets that
originate from hard light quarks or gluons. Both cuts are found to be
equally important in the background suppression.

%%%%%%%%%%%%%%%%%%%%%%%%%%%%%%%%%%%%%%%%%%%%%%%%%%%%%%%%%%%%%%%%%%%%%%%%%%%%%%%%

\subsection{Johns Hopkins Top Tagger}
\label{sec:jhtagger}
The Johns Hopkins Top Tagger~\cite{Kaplan:2008ie} was the first public
prong-based top quark tagger. It was adopted
with small changes by CMS (cf. \secref{cms_tagger}).
 A \ca fat jet is iteratively declustered with
the goal of identifying three or four
hard subjets. This is done by searching for the last two mergings of hard
protojets.

The last clustering of the fat jet is undone, yielding two protojets.
If the \pt of each protojet exceeds a fraction $\delta_p$ (typically $5\%$ or $10\%$)
of the fat jet \pt then both protojets are kept and represent the last
hard merging.
It frequently happens, because \ca ordering is by angular separation,
that a soft protojet (or even a constituent)
is combined with a hard protojet in the last step. The hard structure of the fat jet is then
represented by the structure of the hard protojet.
Therefore, mergings where one of the protojets has $\pt<\delta_p \times p_{\rm T}^{\rm fat\,jet}$
are skipped and the search for the last hard merging continues by decomposing the
hard protojet. This procedure continues until the last hard merging has been found
or the two protojets are both soft, too close to each other (specified by
a parameter $\delta_r$ of the algorithm), or the
protojet is a constituent. In the last three cases, the fat jet is considered to be irreducible.
The two protojets of the last hard merging are then decomposed following
the same procedure. If for one (both) of them a last hard merging is found, the
resulting three (four) protojets are taken as the hard subjets of the fat jet.

Kinematic cuts are applied to reject fat jets that do not contain the
decay products of a top quark. The invariant mass obtained when combining the subjets should
be near the top quark mass, two subjets should give the \W boson mass and
the reconstructed \W boson helicity angle should be consistent with top quark decay.
The efficiency for tagging top quarks when using $R=0.8$ fat jets with
$1.0<\pt<1.1\TeV$ is $\approx\!40\%$ and the fake rate for quark and gluon jets
is below $2\%$. The inefficiency results from losses due to the subjet finding procedure
and the kinematic cuts.
A detailed description of the algorithm and of its performance
is given in~\cite{Kaplan:2008ie}.

%%%%%%%%%%%%%%%%%%%%%%%%%%%%%%%%%%%%%%%%%%%%%%%%%%%%%%%%%%%%%%%%%%%%%%%%%%%%%%%%

\subsection{CMS Top Tagger}
\label{sec:cms_tagger}

The CMS Top Tagger\cite{CMS-PAS-JME-09-001} is based on the Johns Hopkins Top Tagger.
Fat jets are built using the \ca algorithm with $R=0.8$ and three or four subjets
are required to be identified using the procedure described in \secref{jhtagger} with $\delta_p=5\%$.

The invariant mass is calculated for all pairs of the
three leading \pt subjets and all of these pairwise masses are required to exceed $50\GeV$.
This requirement is motivated by the fact that in more than $60\%$ of all hadronic top quark decays the smallest of these
invariant masses is formed by the \W boson decay jets.

\begin{figure}[hbt]
\centering
\includegraphics[width=0.58\textwidth,angle=0]{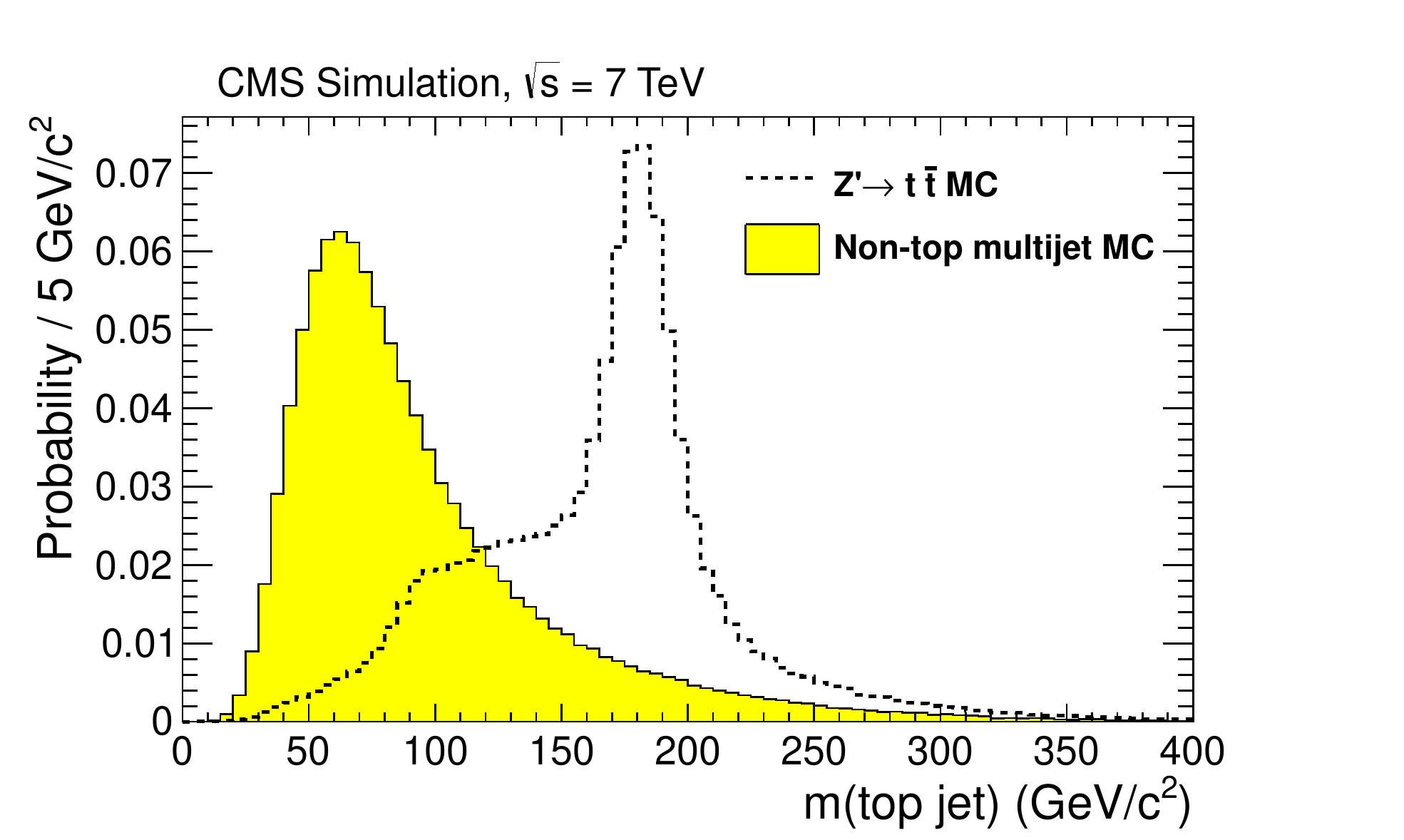}
\caption{Normalised distributions of the CMS Top Tagger top quark candidate mass
in simulated $\Zprime \rightarrow \ttbar$ events (\madgraphpythia) and multijet events (\pythia).
From~\cite{Chatrchyan:2012ku}.}
\label{fig:cms_tagger_mass_MC}
\end{figure}

The tagger has first been applied in a search for \ttbar resonances~\cite{Chatrchyan:2012ku}
which is discussed in \secref{CMS_hadronic} and in which the fat jet mass is
required to be in the range from $140$ to $250\GeV$.
The top quark candidate mass is shown in \figref{cms_tagger_mass_MC} for simulated
$\Zprime \rightarrow \ttbar$ events and for multijet events.
A clear signal peak near the top quark mass is obtained while the background
distribution is smoothly falling for masses larger than $60\GeV$.

%%%%%%%%%%%%%%%%%%%%%%%%%%%%%%%%%%%%%%%%%%%%%%%%%%%%%%%%%%%%%%%%%%%%%%%%%%%%%%%%

\subsection{\htt}
\label{sec:htt_algo}
The \htt algorithm~\cite{Plehn:2009rk,Plehn:2010st,Plehn:2011sj} identifies the hard
substructure of a \ca $R=1.5$ fat jet and tests
it for compatibility with the 3-prong pattern of hadronic top quark decay.
The tagger has been developed to find mildly boosted top quarks with $\pt>200\,\GeV$ in
events with an associated production of \ttbar pairs with a Higgs boson.
The algorithm uses internal parameters that can be changed to optimise the
performance. A detailed explanation of the algorithm is given
in the appendix of~\cite{Plehn:2010st} and only a brief summary is given here.
The tagger uses three main steps.

First, the hard substructure of the fat jet is identified and soft contributions
are removed. This step starts by undoing the last fat jet
clustering. If one of the two resulting subjets carries at least $80\%$ of the
fat jet mass then the second subjet is discarded because it likely represents
a collinearly radiated gluon. This approach is similar to applying a mass drop criterion
with $\mu=0.8$ (cf. \secref{md}).
The remaining subjets are then iteratively
decomposed in the same manner until all subjets have
a mass below $\mcut$ (default value in ATLAS: $50\GeV$).
These subjets constitute the hard fat jet substructure and are referred to
as {\em substructure objects} to distinguish them from the final subjets which
are obtained in the third step.
Because of the mass cut-off, the substructure objects typically
correspond to several particles (or calorimeter cells).

In the second phase, all combinations of three substructure objects ({\em triplets})
are tested for compatibility with hadronic top quark decay.
Energy contributions from underlying event and pile-up are removed using a filtering
procedure that adapts to the distance of the substructure objects:
small radius parameter $\ca$ jets are built from the constituents of the substructure objects
using a filter radius parameter $\Rfilt$ that is given by half of the smallest
pair-wise distance in the triplet, but at most $\Rfilt^{\rm max}$.
Only the $\Nfilt$ largest \pt filter jets are kept for further analysis.
If there are fewer than $\Nfilt$ filter jets then all are kept.
All other constituents in the triplet under consideration are discarded.

In the third \htt step, the constituents of the kept filter jets are clustered
into three top quark subjets using the exclusive \ca algorithm.
Under the hypothesis that these subjets correspond to the three products
from $t\to b q q$ decay, kinematic cuts are applied on these subjets to
reject non-top background.
The kinematic constraints are the \W boson mass and the Lorentz structure of the
$t\to bW$ decay (helicity angle).

The helicity angle $\theta^*$ is determined in the rest frame of the \W boson as the
angle between the top quark momentum and the momentum of one of the \W boson decay products (which are back-to-back).
The distribution of the cosine of this angle is asymmetric, see for example~\cite{Aad:2012ky}.
For background the distribution is more uniform and a typical cut to enhance
the top quark signal is $\cos \theta^* < 0.6$.
Experimentally, the boost to the \W boson rest frame is associated with large uncertainties
from jet reconstruction which reduce the effectiveness of a direct cut on the reconstructed angle.
Instead, the kinematic constraints are formulated in terms of ratios of invariant
subjet masses.

With $p_i$ ($i=1,2,3$) denoting the four-momenta of the decay quarks in $t \to b qq$,
the square of the top quark mass is given by
\begin{eqnarray}
m_t^2 = m_{123}^2 = (p_1 + p_2 + p_3)^2 & = & (p_1 + p_2)^2 + 2 (p_1+p_2) \cdot p_3 + m_3^2 \\
                                & = & m_1^2 + m_2^2 + m_3^2 + 2 p_1\cdot p_2 + 2 p_1 \cdot p_3 + 2 p_2\cdot p_3\,.
\end{eqnarray}
In the limit of negligible subjet masses ($m_i^2 = 0$), the equation simplifies to
\begin{eqnarray}
m_{123}^2 & = & (p_1 + p_2)^2 + (p_1 + p_3)^2 + (p_2 + p_3)^2 \\
          & = & m_{12}^2 + m_{13}^2 + m_{23}^2\,. \label{eq:sphere}
\end{eqnarray}
This is the equation of a sphere with radius $m_{123} = m_t$.
In the limit $m_i^2 = 0$, the kinematics of the decay $t\to b qq$ is fully specified
for each point on the surface of this sphere.
The kinematics is therefore described by the two angles of spherical coordinates,
\begin{eqnarray}
\cos \theta & = & m_{23}/m_{123}\,,    \label{eq:theta} \\
\phi & = & \arctan(m_{13}/m_{12})\,. \label{eq:phi}
\end{eqnarray}

In the \htt, the final three exclusively clustered subjets are identified
with the top quark decay products, such that $p_1$ denotes the four-momentum
of the leading \pt subjet, $p_2$ that of the subleading \pt subjet, and $p_3$ that of the
sub-subleading \pt subjet.

\begin{figure}[hbt]
\centering
\subfigure[\ttbar]{
   \includegraphics[width=0.31\textwidth,angle=0]{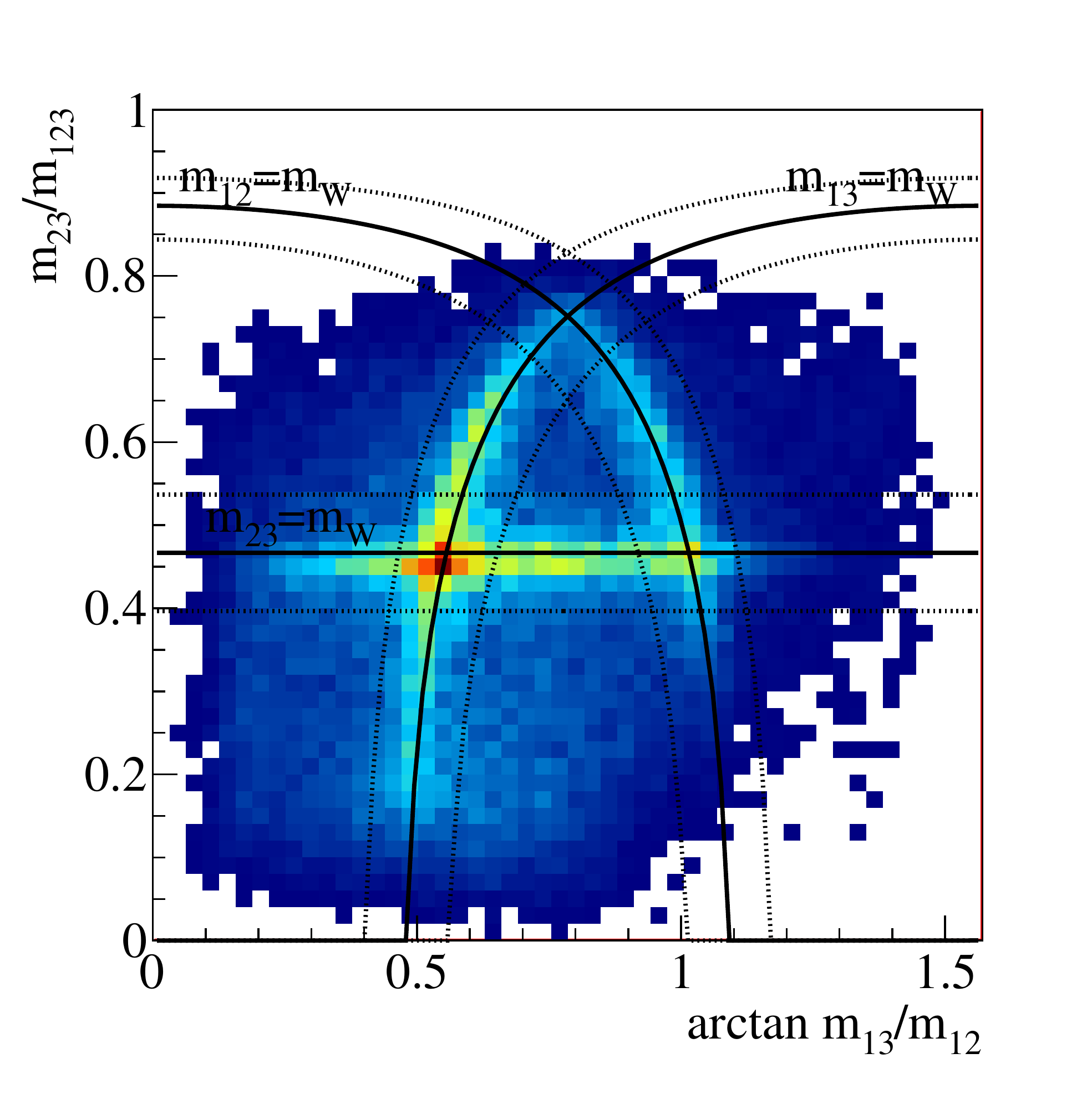}
}
\subfigure[\Wpjets]{
   \includegraphics[width=0.31\textwidth,angle=0]{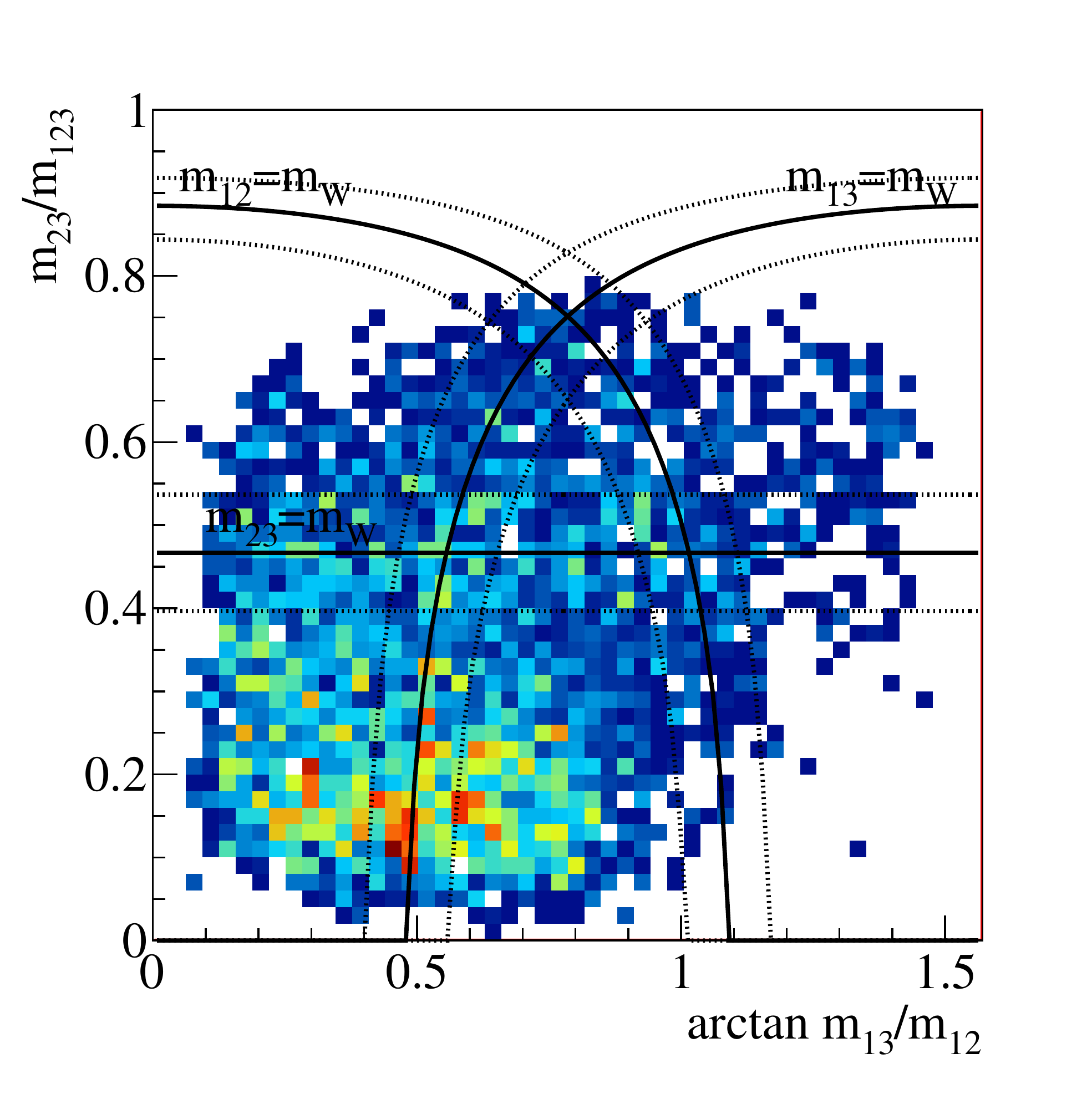}
}
\subfigure[Multijets (QCD)]{
   \includegraphics[width=0.31\textwidth,angle=0]{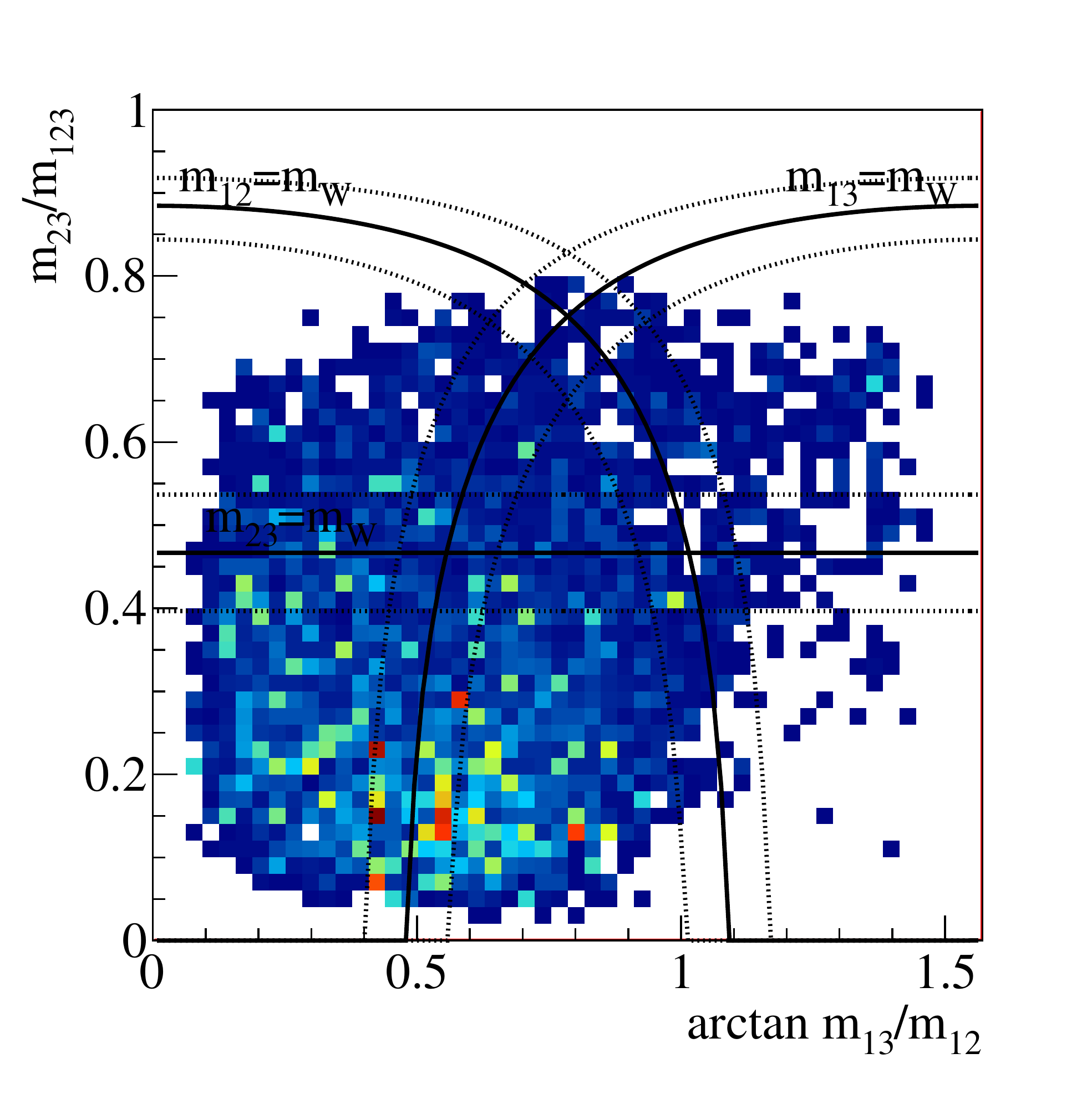}
}
\caption{Distributions of kinematics of the final three subjets used in the \htt
for (a) SM \ttbar production, (b) \Wpjets events, and (c) multijet events.
Shown is the surface of the sphere defined by
$m_{123}^2 = m_{12}^2 + m_{13}^2 + m_{23}^2.$
The quantity $m_{123}$ is the invariant mass of all three subjets.
The quantities $m_{ij}$ are invariant masses of two of the subjets which are
ordered in \pt such that $p_{{\rm T},1} > p_{{\rm T},2} > p_{{\rm T},3}$.
From~\cite{Plehn:2010st}.}
\label{fig:htt_bg_map}
\end{figure}

\figref{htt_bg_map} shows how the subjet kinematics is distributed on the surface of the
sphere given by \eqref{sphere} in terms of the coordinates $\cos \theta$ and
$\phi$ as defined in \eqsref{theta}{phi} for fat jets in different event samples.
Shown are the distributions for signal events (SM \ttbar production) and background
events (\Wpjets and multijets (QCD)).
The signal distribution resembles the shape of the letter A.
As written on the plots in the figure, the arms of the letter A can be identified
with the relations
\begin{eqnarray}
m_{23} & \approx & \mw \quad {\rm (horizontal~arm)}, \label{eq:horizarm} \\
m_{12} & \approx & \mw \quad {\rm (right~arm)}, \label{eq:rightarm} \\
m_{13} & \approx & \mw \quad {\rm (left~arm)}, \label{eq:leftarm}
\end{eqnarray}
in which $\mw = 80.4\GeV$ denotes the \W boson mass.
Most of the background is located below $m_{23}/m_{123} = 0.35$.
Kinematic cuts are now defined that reject most of the background while retaining
the signal.

The relations in \eqsrange{horizarm}{leftarm} are formulated in terms of a
\W boson mass window which is specified by $R_\mp = (1 \mp \fwidth)\, \mw/\mt$
in which the width is given by $\fwidth$ which has a default value of $15\%$ and
$m_t = 172.3\GeV$ (this value stems from a previous best global value of the measured top quark mass):
\begin{eqnarray}
m_{23} \approx \mw: & \quad & R_- < \frac{m_{23}}{m_{123}} < R_+ \label{eq:horizarmR} \\
m_{12} \approx \mw: & \quad & R_-^2 \left( 1 + \left(\frac{m_{13}}{m_{12}}\right)^2 \right) < 1 - \left(\frac{m_{23}}{m_{123}}\right)^2 < R_+^2 \left( 1+\left( \frac{m_{13}}{m_{12}}\right)^2 \right) \label{eq:rightarmR} \\
m_{13} \approx \mw: & \quad & R_-^2 \left( 1 + \left(\frac{m_{12}}{m_{13}}\right)^2 \right) < 1 - \left(\frac{m_{23}}{m_{123}}\right)^2 < R_+^2 \left( 1+\left( \frac{m_{12}}{m_{13}}\right)^2 \right) \label{eq:leftarmR}
\end{eqnarray}

With the relations $m_{ij} \approx \mw$ used as short-hand notations for the mass windows specified in \eqsrange{horizarmR}{leftarmR},
the kinematic selection cuts for the \htt subjets are given by
\begin{eqnarray}
&  & \left( m_{23} \approx \mw \quad {\rm and} \quad 0.2 < \arctan(m_{13}/m_{12}) < 1.3 \right) \\
{\rm or} & & \left( m_{12} \approx \mw \quad {\rm and} \quad m_{23}/m_{123} > 0.35 \right) \\
{\rm or} & & \left( m_{13} \approx \mw \quad {\rm and} \quad m_{23}/m_{123} > 0.35 \right) \,.
\end{eqnarray}
The \htt therefore requires one of the three possible subjet pairs to reconstruct
the \W boson mass but it does not explicitly reconstruct the \W boson or identify the $b$-quark jet.

The top quark candidate four-momentum vector is given by the sum of the vectors
of the $\Nfilt$ largest \pt filter jets.\footnote{At the
particle level, the sum of the
four-momentum vectors of the $\Nfilt$ largest \pt filter jets is identical to
the sum of the momenta of the three final subjets. At the detector level,
however, the jets have to be calibrated. The calibration constants are
determined as a function of the radius parameter (cf. \secref{jetcalib}).
For the filter jets this parameter is known because they are clustered inclusively.
For the exclusive jets it is not immediately clear which constants should be chosen.
One possibility to assign a radius parameter to an exclusively clustered jet
is to use the minimal radius that would yield the same jet in inclusive clustering.
To remove the uncertainty associated with choosing a radius parameter for the
exclusive jets, the momenta of the (calibrated) filter jets are used.}
The top quark candidate mass has to lie in a window around the measured top quark
mass, usually $140$--$200\GeV$.
If all criteria are met, the fat jet is considered to be tagged.
If several top quark candidates are found, the one with mass closest to $172.3\GeV$ is used.

The parameter settings used in this text are summarised in \tabref{HTTsettings}.
The parameters listed in the `default' column gave the best expected significance
in a search for \ttbar resonances in the hadronic decay channel~\cite{Aad:2012raa} which
is discussed in \secref{htt}.

\begin{table}[hbt]
  \begin{center}
    \begin{tabular}{|c|p{1.7cm}|p{1.7cm}|c|c|}
      \hline
      parameter & \hfill default\hglue0.3cm \hfill & default-30 & tight & loose \\
      \hline
      \hline
      \mcut (\GeV) & \multicolumn{1}{|c|}{$50$} & \multicolumn{1}{|c|}{$30$} & $30$ & $70$ \\
      \hline
      $\Rfilt^{\rm max}$ & \multicolumn{2}{|c|}{$0.3$\hspace{0.5cm}} & $0.2$ & $0.5$ \\
      \hline
      \Nfilt & \multicolumn{2}{|c|}{$5$} & $4$ & $7$ \\
      \hline
      $\fwidth$ & \multicolumn{2}{|c|}{$15\%$}  & $10\%$ & $20\%$ \\
      \hline
    \end{tabular}
    \caption{\htt parameter settings.}
  \label{tab:HTTsettings}
  \end{center}
\end{table}

\section{Measurements of Jet Structure}
This section presents measurements of jet substructure variable
distributions at the detector level and distributions corrected to the level of stable
particles (lifetime longer than 10~ps). The impact of pile-up energy is evaluated for ungroomed
and groomed jets.

%%%%%%%%%%%%%%%%%%%%%%%%%%%%%%%%%%%%%%%%%%%%%%%%%%%%%%%%%%%%%%%%%%%%%%%%%%%%%%%%

\subsection{Data samples}
\label{sec:measurements_datasamples}
The studies presented in this section have been performed using data collected with the ATLAS detector
in $pp$ collisions in 2010~\cite{ATLAS:2012am}, 2011~\cite{Aad:2013gja}, and 2012~\cite{ATLAS-CONF-2013-084}.
The increased luminosity in each year allowed the study of different event topologies.

In 2010, the integrated luminosity was $35(1)\invpb$. This dataset is used to study
the structure of non-top fat jets that occur in dijet or multijet production because
no significant number of boosted top quarks were produced.
An advantage of this dataset is that pile-up was very limited: approximately
$22\%$ of the events had only one reconstructed primary vertex (with at least five tracks).
This allows to study fat jets in the absence of pile-up and how the structure changes
in the presence of a few additional vertices.
Fat jets are reconstructed using the \ca algorithm with $R=1.2$ and the \akt algorithm with $R=1.0$.
To show the effect of grooming, mass drop filtering (cf. \secsref{md}{filtering}) is applied to the \ca jets.
The mass drop criterion with $\mu = 0.67$ and $\vcut = 0.09$ is used to identify
two subjets which must be separated by $\DeltaROneTwo \ge 0.3$.
Compared to the original jet, a jet that satisfies this MD criterion
exhibits a 2-prong structure like that expected from hadronic heavy particle decay (\W, \Z, or Higgs bosons):
two hard subjets (each with $\pt > 15\%$ of the sum of the two subjet transverse momenta)
and a significant mass increase when the two subjets are combined.
The fraction of jets in this data sample that contain decay products of
heavy particles is negligible.
A jet that survives the splitting/filtering criteria is a background jet that
displays the 2-prong pattern by accident.

Filtering is applied
to the jets that pass the MD criterion to reject pile-up and underlying event activity:
the separation \DeltaROneTwo of the two subjets identified by the MD procedure
is used to determine the filter radius
$\Rfilt = \min(0.3, \DeltaROneTwo/2)$ and the three leading \pt filter jets
(or all if fewer than three are found) when clustering all constituents of the
original fat jet are combined to obtain a {\em split/filtered} fat jet.
The data are corrected for detector effects using matrix-based
unfolding with simulated events from \pythia. \sherpa events are used to estimate systematic uncertainties.

\begin{figure}[hbt]
\centering
\includegraphics[width=0.45\textwidth,angle=0]{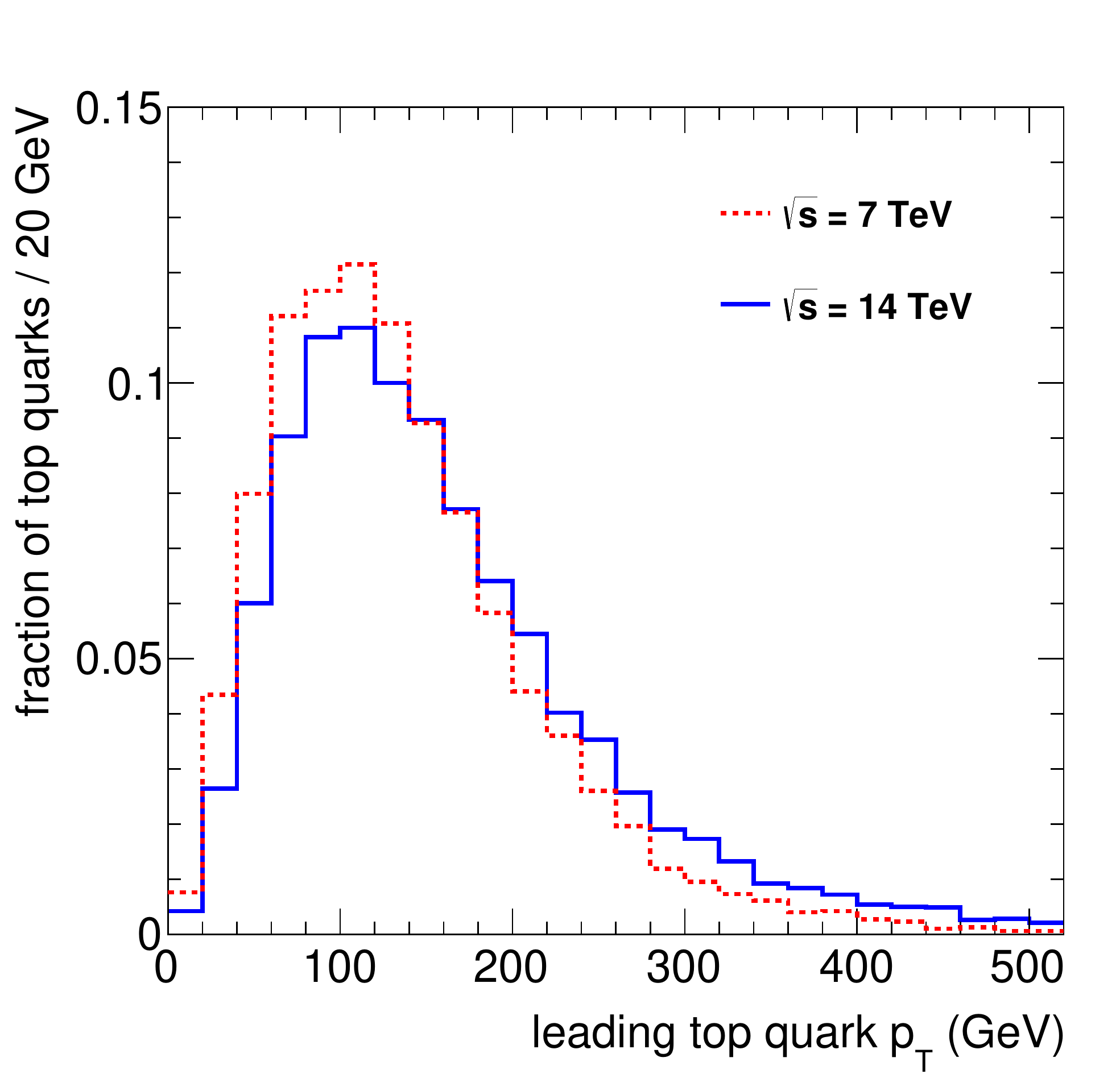}
\caption{Normalised distributions of the leading top quark \pt in \ttbar events simulated
with \pythia8 for two LHC centre-of-mass energies.}
\label{fig:toppt}
\end{figure}

The 2011 data correspond to $4.7(1)\invfb$.
In this dataset, a significant number of top quarks with $\pt > 200\GeV$ are present.
The \pt spectrum of the leading \pt top quark in SM \ttbar events generated with \pythia8 is shown in
\figref{toppt}. The fraction of leading \pt top quarks with $\pt > 200\GeV$
is $18\%$ for $\sqrt{s}=7\TeV$ and $27\%$ for $\sqrt{s} = 14\TeV$.
Using this fraction in combination with the approximate NNLO cross section for \ttbar production of
167~pb (cf. \secref{topprod}),
the number of \ttbar events with top quark $\pt > 200\GeV$ in the 2011 data sample
is predicted to be approximately $142,\!000$.
Events with \ttbar pairs from 2011 will be used in the \htt performance
studies in \secref{httperformance}. For the substructure variables discussed
in the present section, multijet events from 2011 and \ttbar events from 2012 are used.
The multijet events differ from those in the 2010 data by the much larger
pile-up energy (average $\avmu = 9.1$).
The main goal is therefore the study of fat jets (\akt $R=1.0$) with $\pt>350\GeV$,
both ungroomed (i.e., including the pile-up energy) and trimmed (using the
parameters $\fcut=5\%$ and $\Rsub=0.3$ introduced in \secref{trimming}).

From the 2012 data sample, corresponding to $20.3(6)\invfb$ at $\sqrt{s}=8\TeV$,
SM \ttbar events, in
which one \W boson decays to a neutrino and a muon, are selected as follows.
Events must contain exactly one muon with $\pt>25\GeV$ and $|\eta|<2.5$
and have missing transverse momentum $\ETmiss>20\GeV$.
The scalar sum of \ETmiss and the transverse mass $\WmassT$ of the
leptonic \W boson candidate must be larger than $60\GeV$, where
$\WmassT=\sqrt{2\, p_{{\rm T},\mu}\,\ETmiss\,(1-\cos\Delta\phi)}$ in which $\Delta\phi$ is
the azimuthal angle between the muon-\pt and \ETmiss vectors.
To reduce contamination from \Wpjets events, each event must contain at least
one $b$-tagged \akt $R=0.4$ jet with $\pt>20\GeV$ and $|\eta|<2.5$
within $\DeltaReta = 1.5$ of the muon.
Substructure variable distributions will be shown in the following for \akt $R=1.0$ jets
which have been trimmed like the 2011 jets.
The \ttbar simulation is divided into two categories, {\em contained} and {\em non-contained}.
A hadronically decaying top quark is said to be contained
if all decay quarks are separated by less than $\DeltaReta=1.0$ from the top quark flight direction.

%%%%%%%%%%%%%%%%%%%%%%%%%%%%%%%%%%%%%%%%%%%%%%%%%%%%%%%%%%%%%%%%%%%%%%%%%%%%%%%%

\subsection{Jet mass}

\begin{figure}[hbt]
\centering
\subfigure[]{
   \includegraphics[width=0.45\textwidth,angle=0]{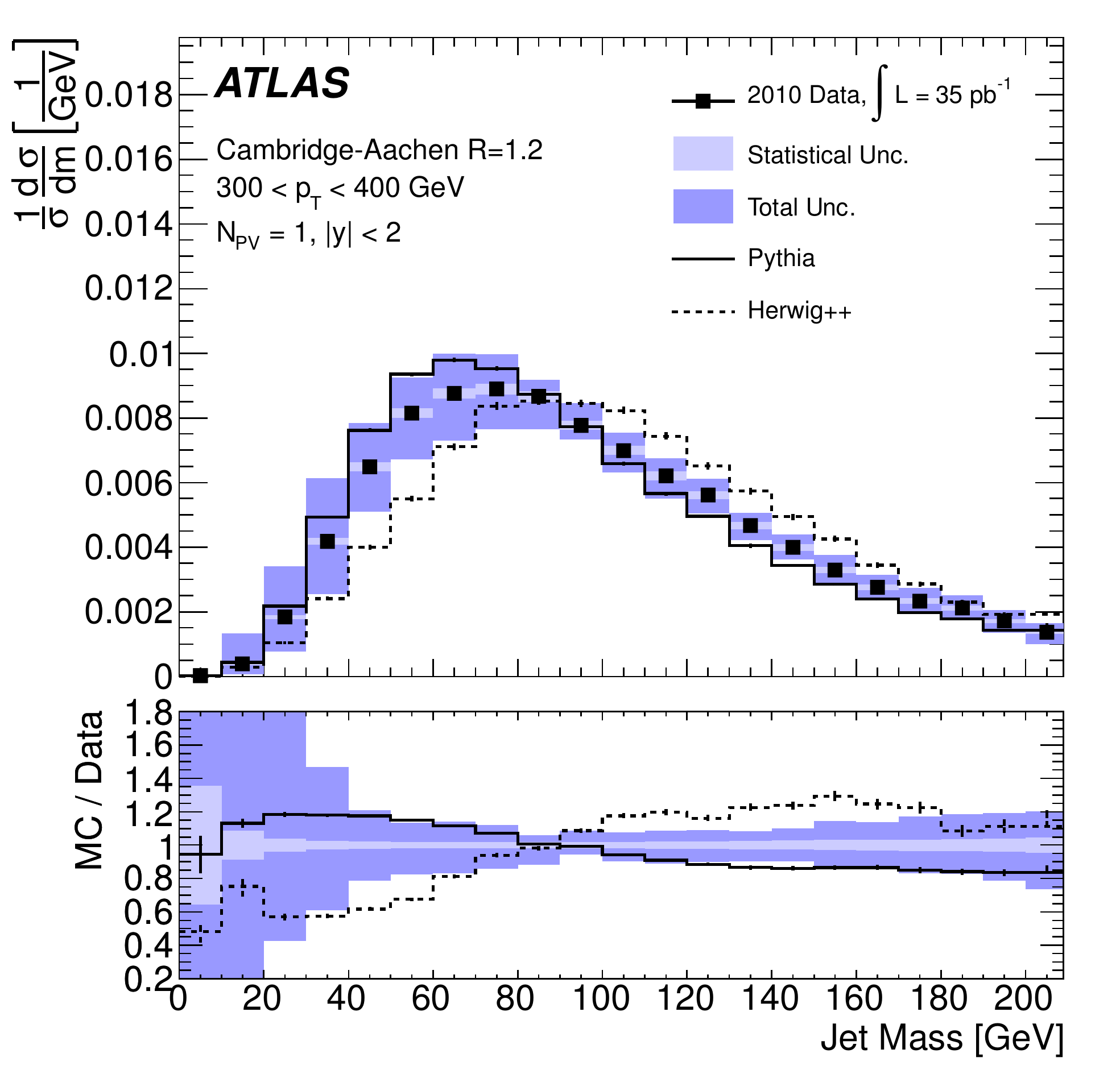}
}
\subfigure[]{
   \includegraphics[width=0.45\textwidth,angle=0]{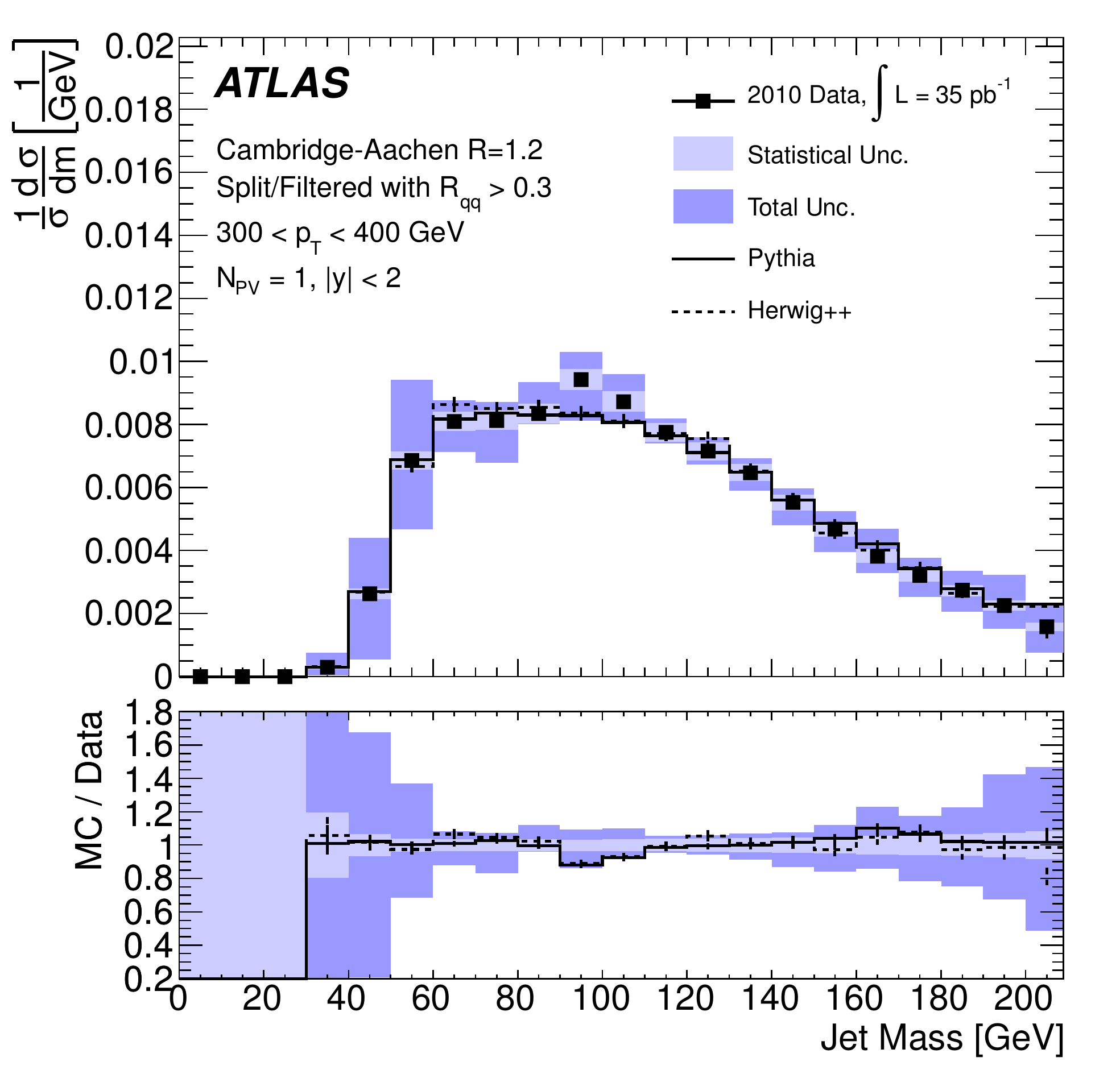}
}
\caption{Unfolded normalised distribution of the mass of \ca $R=1.2$ jets with $300<\pt<400\GeV$ a) before and b)
after splitting and filtering in an inclusive jet sample without pile-up (number of primary vertices $\Npv=1$).
From~\cite{ATLAS:2012am}.}
\label{fig:unfolded_mass}
\end{figure}

\figref{unfolded_mass} shows the normalised particle level measurement of the
mass of \ca $R=1.2$ jets before (left panel) and after splitting and filtering (right panel)
for $300<\pt<400\GeV$ in events with only
one primary vertex, i.e., without pile-up.
The mass is peaked between $70$ and $80\GeV$ for the original jets
and between $90$ and $100\GeV$ for the split/filtered jets. The higher mass is
explained by the requirement $\DeltaROneTwo \ge 0.3$ used in the splitting.
The \pythia prediction for the original jet mass spectrum (the one before splitting and filtering)
tends to be too soft, with the ratio to the measured
distribution varying between $1.2$ and $0.8$. Although this is within the
measurement uncertainties, a clear trend is visible. The \herwigpp prediction is too
hard, within $+30\%$ and $-40\%$ of the data. The level of description by the
simulation is similar for jet \pt between $200$ and $600\GeV$ and for \akt $R=1.0$ jets (not shown).
The shape of the mass distribution of split/filtered jets in \figref{unfolded_mass}b
is well described, indicating that the problem lies in the simulation of soft hadrons.
Potential candidates for improvement are the
underlying event models and their tunes, and the fragmentation models.

\begin{figure}[hbt]
\centering
\subfigure[]{
   \includegraphics[width=0.45\textwidth,angle=0]{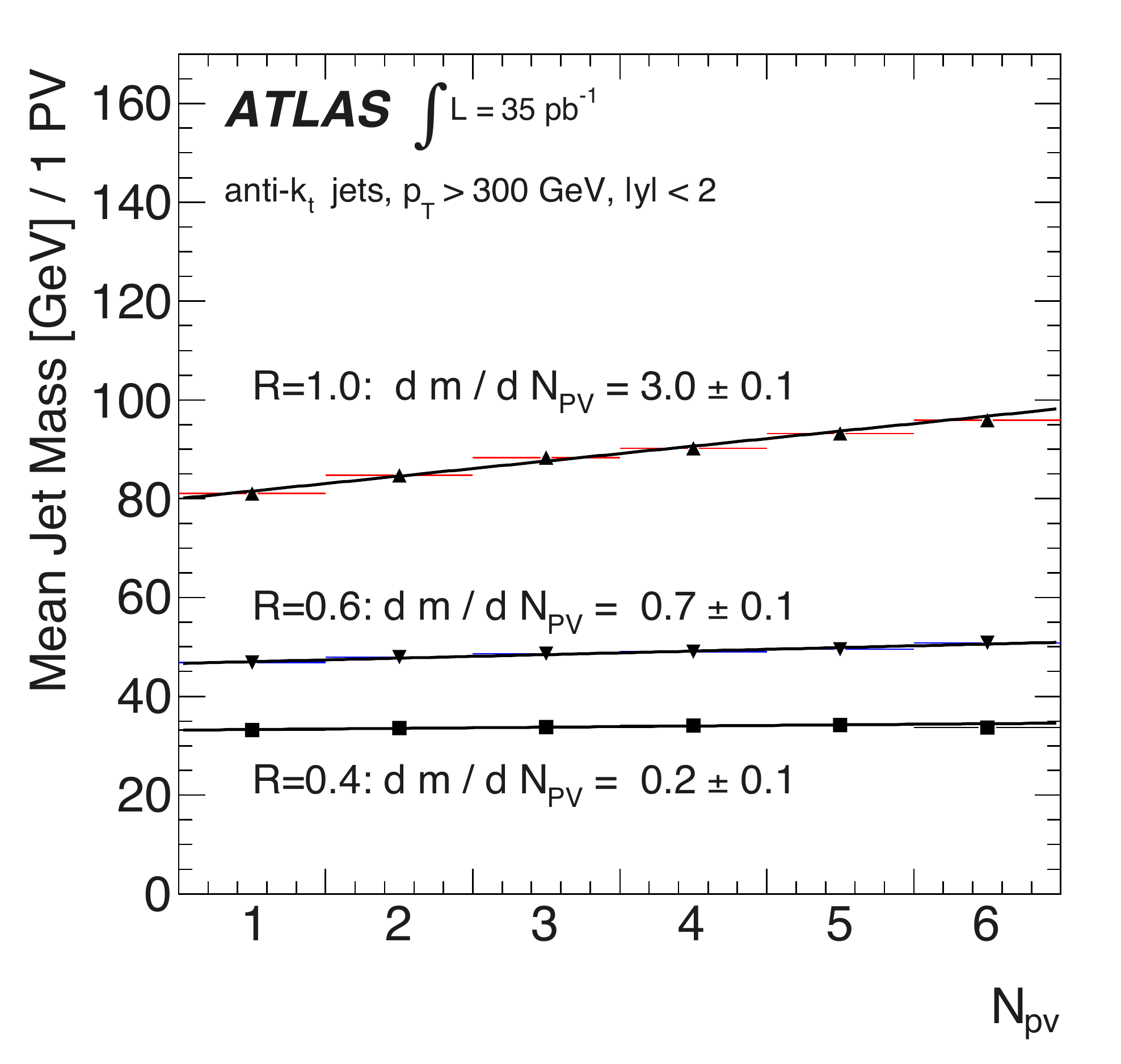}
}
\subfigure[]{
   \includegraphics[width=0.45\textwidth,angle=0]{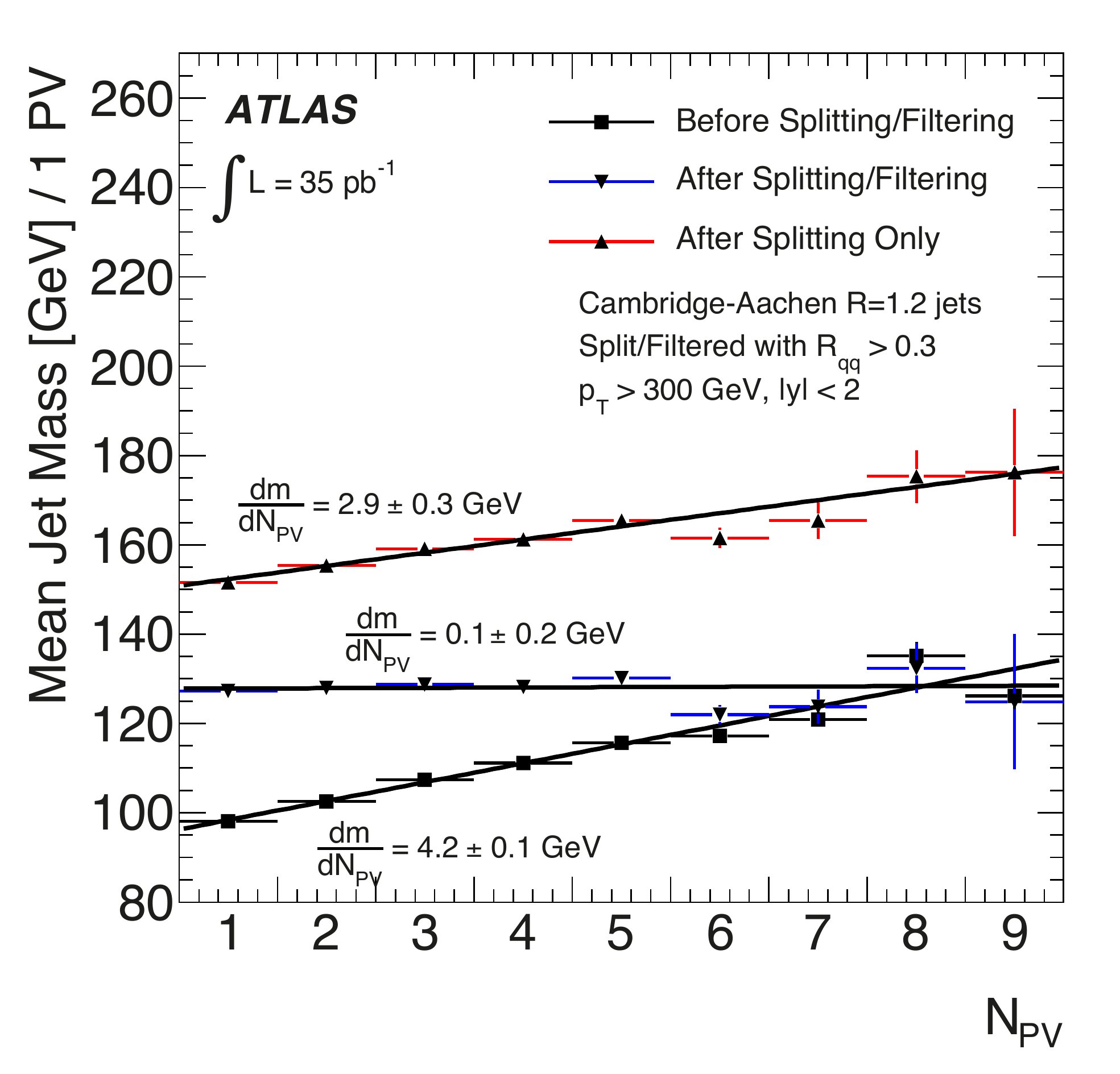}
}
\caption{The average detector level mass of jets with $\pt > 300\GeV$
as a function of the number of reconstructed primary vertices in the event.
Shown are a) \akt jets with different radius parameters $R$ and b) C/A $R=1.2$
jets before grooming, after the mass drop criterion ({\em splitting}), and
after mass drop filtering ({\em splitting/filtering}). From~\cite{ATLAS:2012am}.}
\label{fig:mass_npv}
\end{figure}

The average detector level jet mass is shown in \figref{mass_npv}a for
\akt jets with different radius parameters as a function of the number \Npv of
reconstructed primary vertices in the event. The number \Npv is a measure of
the number of inelastic $pp$ interactions in the event.
For $R=0.4$, the jet mass is $\approx\!35\GeV$
and does not depend on \Npv. For $R=0.6$ ($1.0$), the mass is $55\GeV$ ($80\GeV$)
without pile-up and increases by $0.7(1)\GeV$ ($3.0(1)\GeV$) per primary vertex.
As described in~\cite{ATLAS:2012am}, the jet mass pile-up dependence
$d\langle m \rangle/d\Npv$ is approximately proportional to $R^3$: the ratios
of the fitted slopes $s_R$ are
\begin{align}
s_{1.0}/s_{0.6}  &=  4.3 \pm 0.5 & ((1.0/0.6)^3 &= 4.6), \nonumber \\
s_{1.0}/s_{0.4}  &=  13 \pm 3 & ((1.0/0.4)^3 &= 15.6), \nonumber \\
s_{0.6}/s_{0.4}  &=  3.0 \pm 0.8 & ((0.6/0.4)^3 &= 3.4)\nonumber \,,
\end{align}
in good agreement with $R^3$ scaling. This scaling is explained by two factors.
First, the jet area of \akt jets is $\approx\!\pi R^2$ (cf.~\secref{perf_mass} and \figref{jetareas}).
The amount of pile-up energy in a jet is proportional to the jet area because
the pile-up energy is approximately equally distributed in $(y,\phi)$.
Another power of $R$ results from the separation of the constituents (cf.~\eqref{mass}).

The pile-up dependence of the \ca $R=1.2$ jet mass is shown in \figref{mass_npv}b.
The mass is $100\GeV$ and rises with $4.2(1)\GeV$ per vertex.
The jet mass is also shown for the jets that satisfy the
mass drop criterion before filtering (labelled in the figure as {\em after splitting only}).
These jets have a larger mass
($150\GeV$ at $\Npv=1$) because the $\DeltaROneTwo \ge 0.3$ cut enhances configurations
with a geometrically induced higher mass (cf.~\eqref{mass}).
The dependence on the number of vertices is reduced for these jets, but still
significant at $2.9(3)\GeV$ per vertex.
Efficient pile-up removal is achieved in the filtering step: after applying
filtering to the split jets (label {\em after splitting/filtering}), the
slope is consistent with zero. The mass of the split/filtered jets
is reduced ($\approx\!130\GeV$) because
only the three leading \pt filter jets are used.

\begin{figure}[hbt]
\centering
\subfigure[]{
   \includegraphics[width=0.45\textwidth,angle=0]{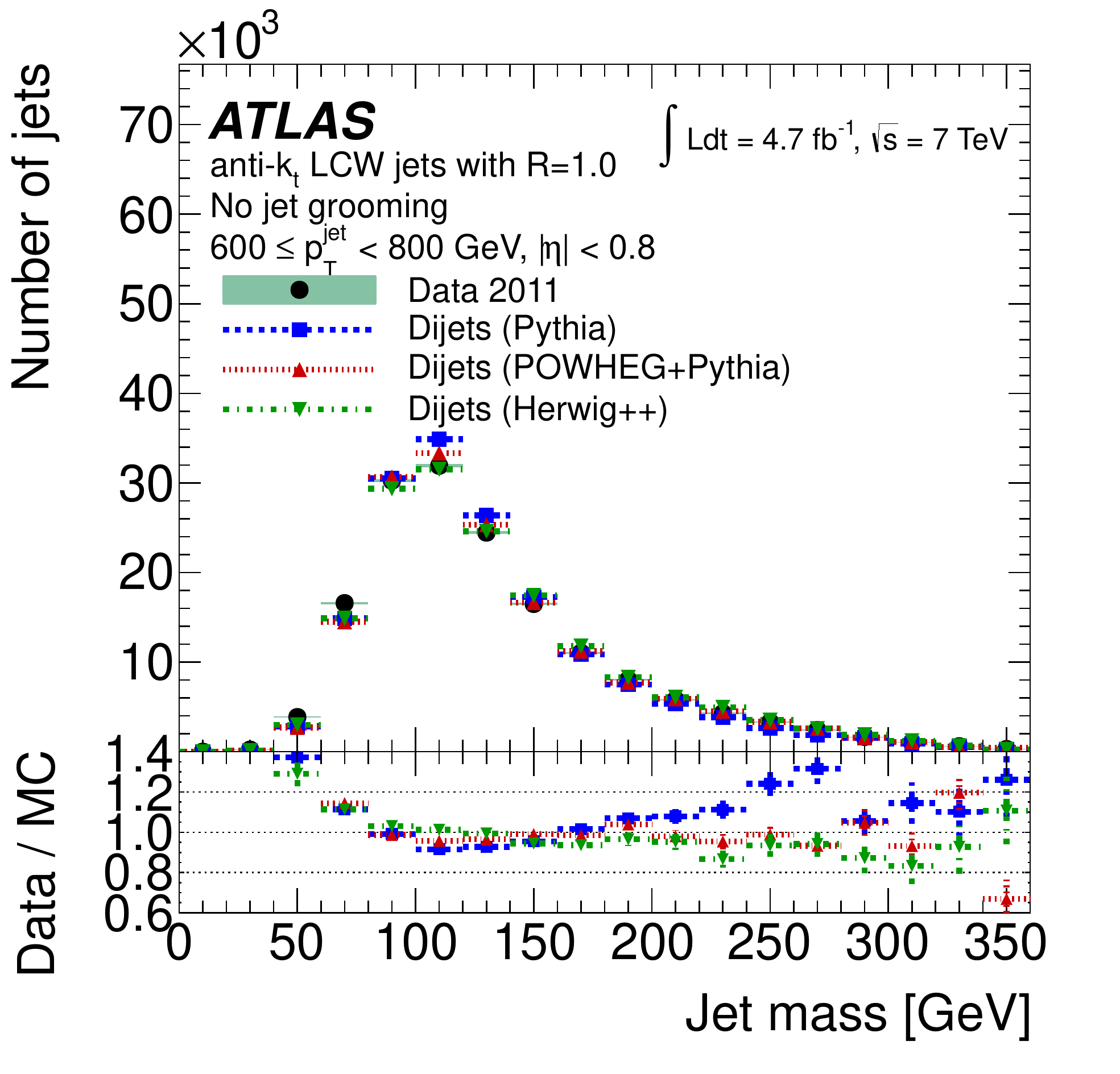}
}
\subfigure[]{
   \includegraphics[width=0.45\textwidth,angle=0]{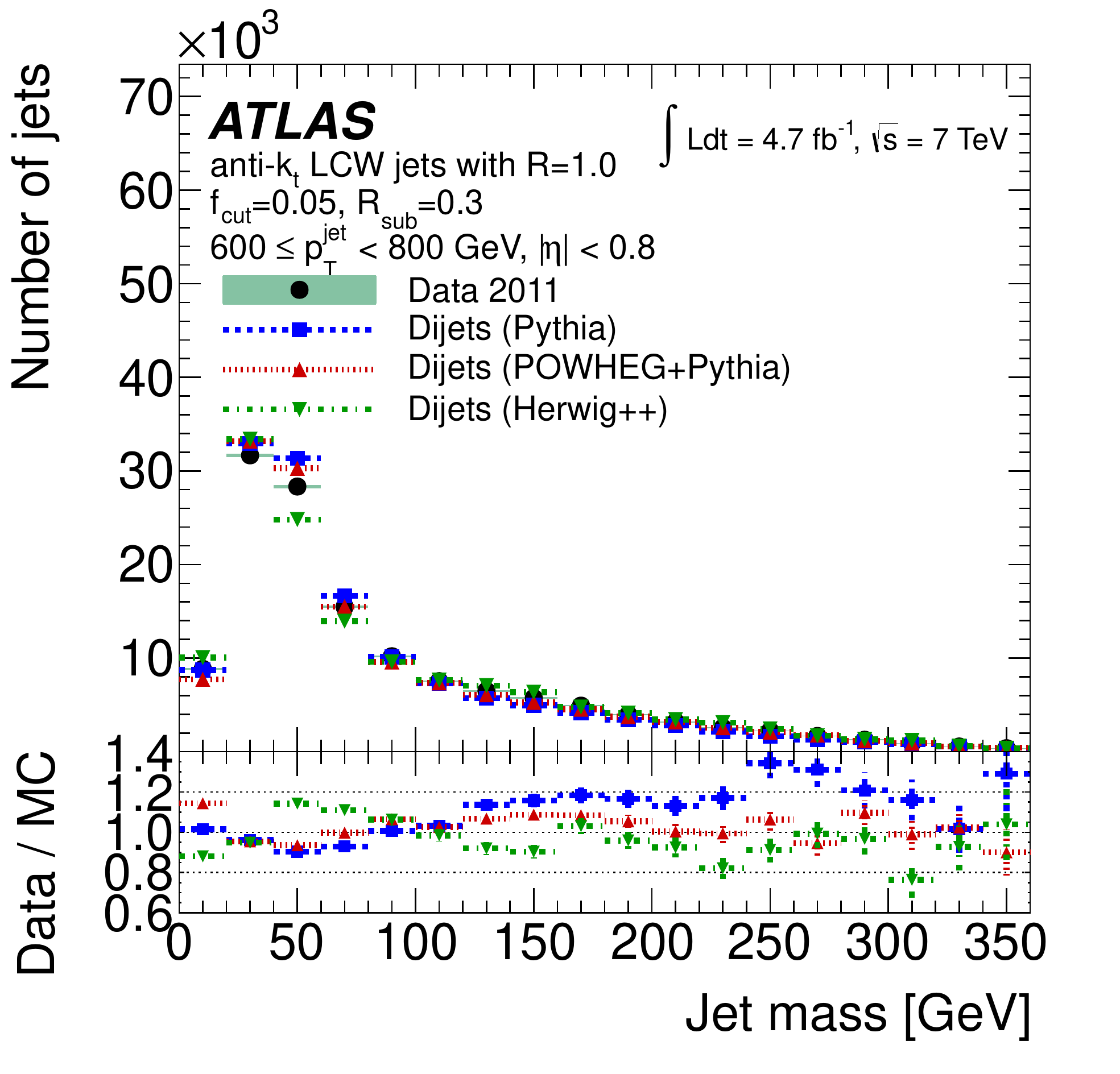}
}
\caption{Detector level distributions of the mass of
\akt $R=1.0$ jets with $600<\pt<800\GeV$ in multijet events with an average
pile-up \avmu of $9.1$.
a) ungroomed jets and b) trimmed jets using $\fcut = 5\%$ and $\Rsub = 0.3$. From~\cite{Aad:2013gja}.}
\label{fig:perf_mass2011}
\end{figure}

Detector level distributions of the jet mass are shown \figref{perf_mass2011}
for jets with $600<\pt<800\GeV$ in multijet events with an average pile-up \avmu of 9.1 (2011 data).
Before grooming, the jet mass is peaked at $100$--$120\GeV$ and the
description by the Monte Carlo generators is within $\pm10\%$ for the bulk of the
distribution.
The shift of the peak position with respect to \figref{unfolded_mass} is in part due to the
larger jet \pt. In \figref{perf_mass2011}, \herwigpp uses a refined UE model (colour reconnection model)~\cite{Gieseke:2012ft}
with parameters tuned to LHC data (tune UE7-2). The description of the jet mass
by \herwigpp is improved by this model compared to \figref{unfolded_mass}
where the default UE model was used. Also the description by \pythia is improved,
through the use of the parameter set AUET2B~\cite{AMBT1,AUET2B}
which is tuned to LHC data and results in a better prediction than the AMBT1 set~\cite{Aad:2010ac} used in
\figref{unfolded_mass} which was tuned to $e^+e^-$ and Tevatron data.
After trimming, the
peak is at $20$--$40\GeV$, with the Monte Carlo distributions within $\pm20\%$
of the data points up to $250\GeV$.
The fact that the ungroomed and the trimmed jet masses are well
described by the simulation, where trimming reduces the mass by $\approx\!80\GeV$,
demonstrates that
pile-up contributions to
jets as well their removal is well modelled in the simulation.

\begin{figure}[hbt]
\centering
\subfigure[]{
   \includegraphics[width=0.45\textwidth,angle=0]{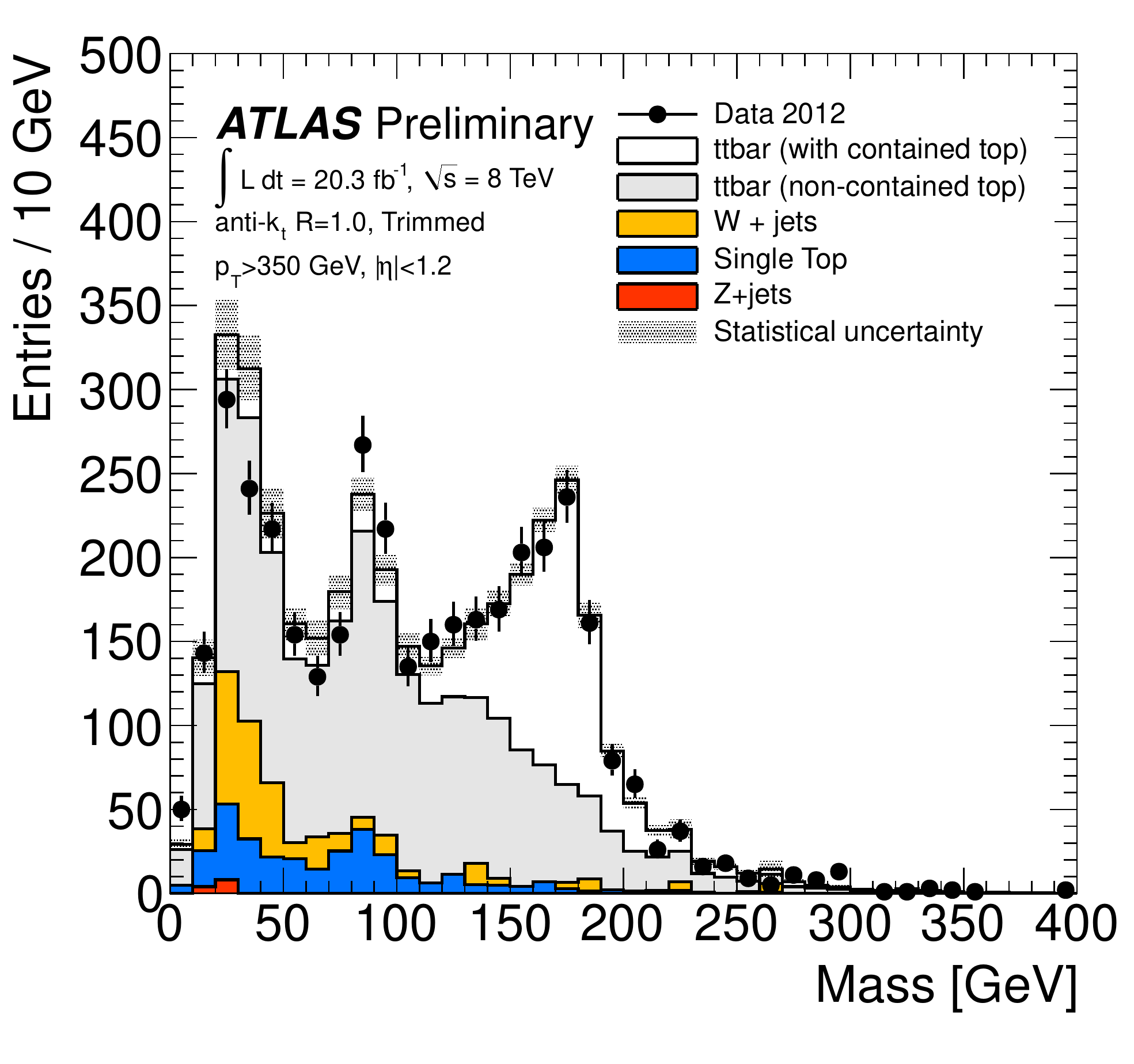}
}
\subfigure[]{
   \includegraphics[width=0.45\textwidth,angle=0]{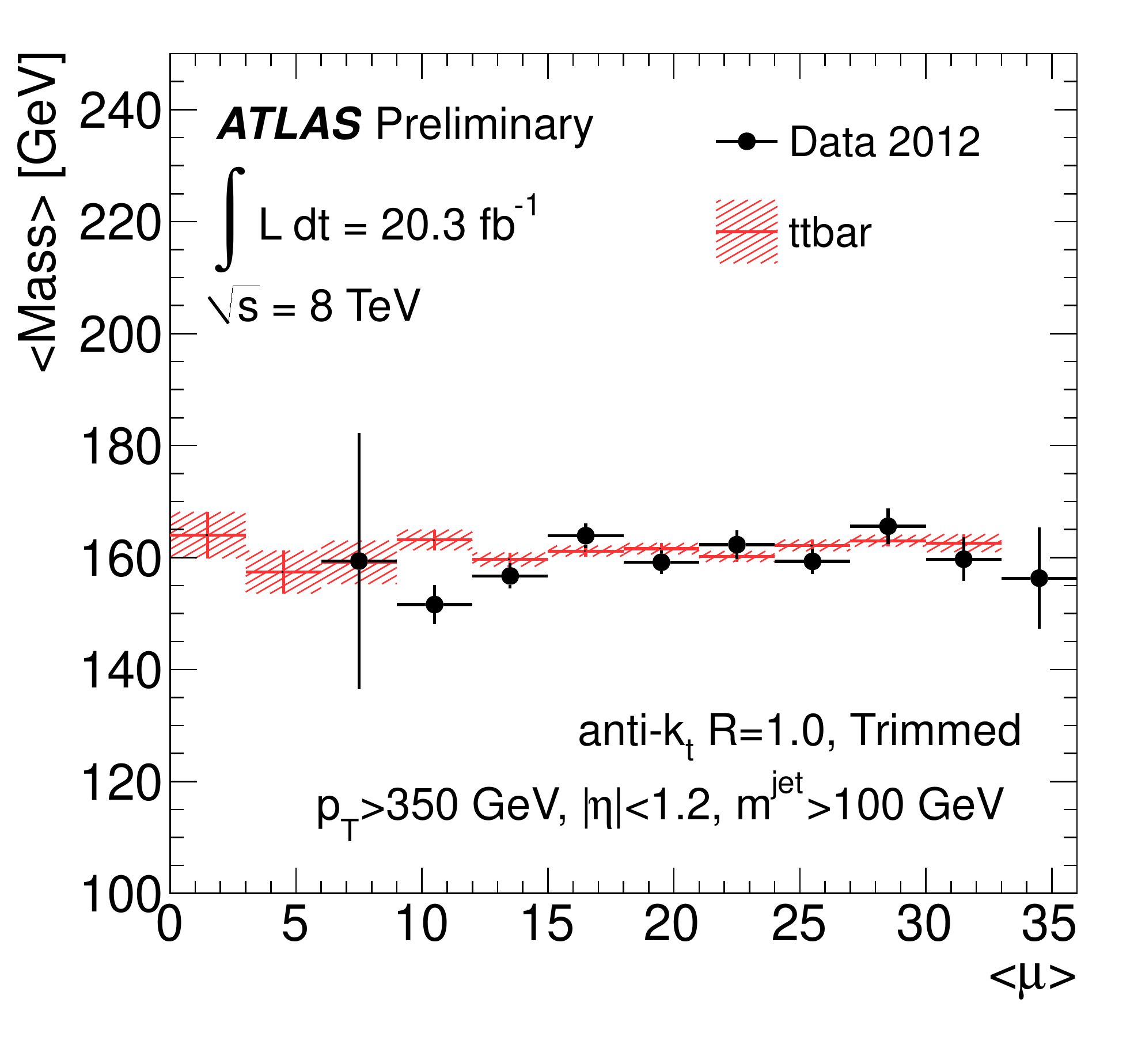}
}
\caption{a) Detector level distributions of the mass of
trimmed \akt $R=1.0$ jets ($\fcut=5\%$, $\Rsub=0.3$) with $\pt>350\GeV$ in a
selection of semileptonic \ttbar decays.
{\em Contained} refers to hadronically-decaying top quarks for which
all three decay quarks are separated by less than $\DeltaReta = 1.0$ from
the top quark flight direction. The shaded band represents the statistical simulation uncertainty.
b) The average of the distribution
of masses larger than $100\GeV$ as a function of the
average number of inelastic $pp$ collisions per bunch crossing.
The \ttbar simulation is obtained from \powheg with \pythia for the parton shower
and hadronisation.
From~\cite{ATLAS-CONF-2013-084}.}
\label{fig:tt_trimmed_mass}
\end{figure}

\begin{figure}[hbt]
\centering
\subfigure[$N_{\rm subjets} = 1$]{
   \includegraphics[width=0.45\textwidth,angle=0]{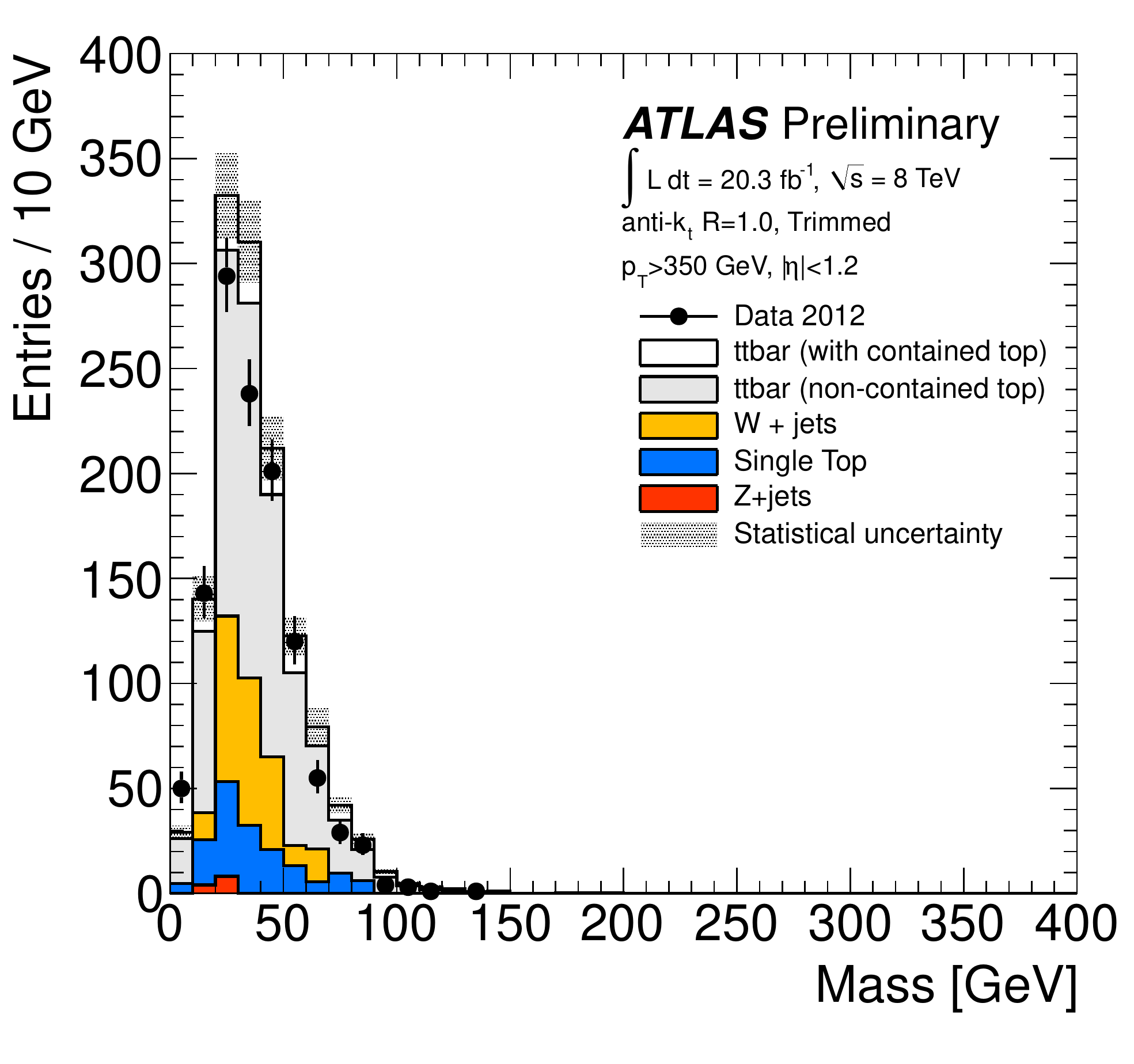}
}
\subfigure[$N_{\rm subjets} = 2$]{
   \includegraphics[width=0.45\textwidth,angle=0]{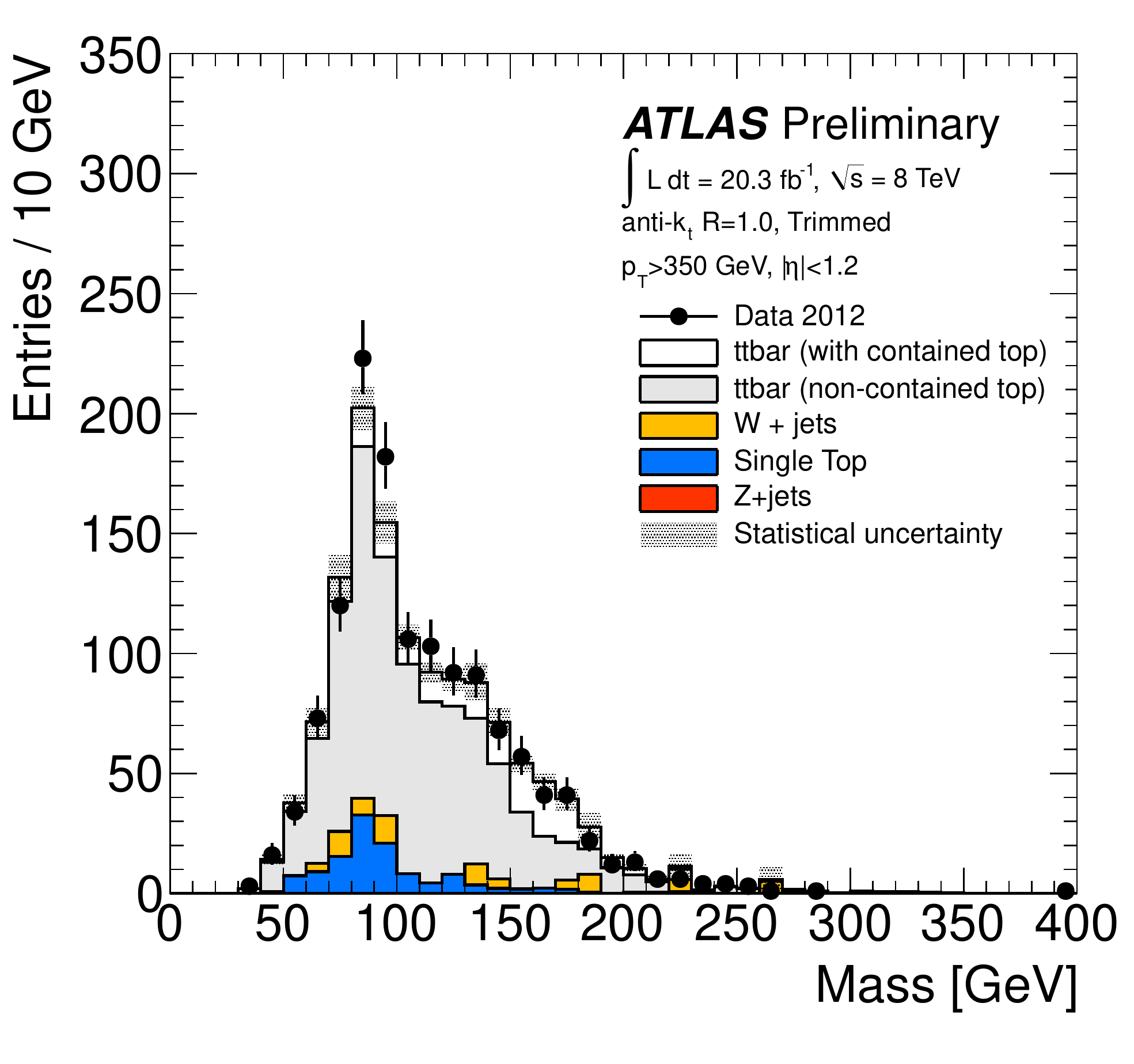}
} \\
\noindent
\subfigure[$N_{\rm subjets} = 3$]{
   \includegraphics[width=0.45\textwidth,angle=0]{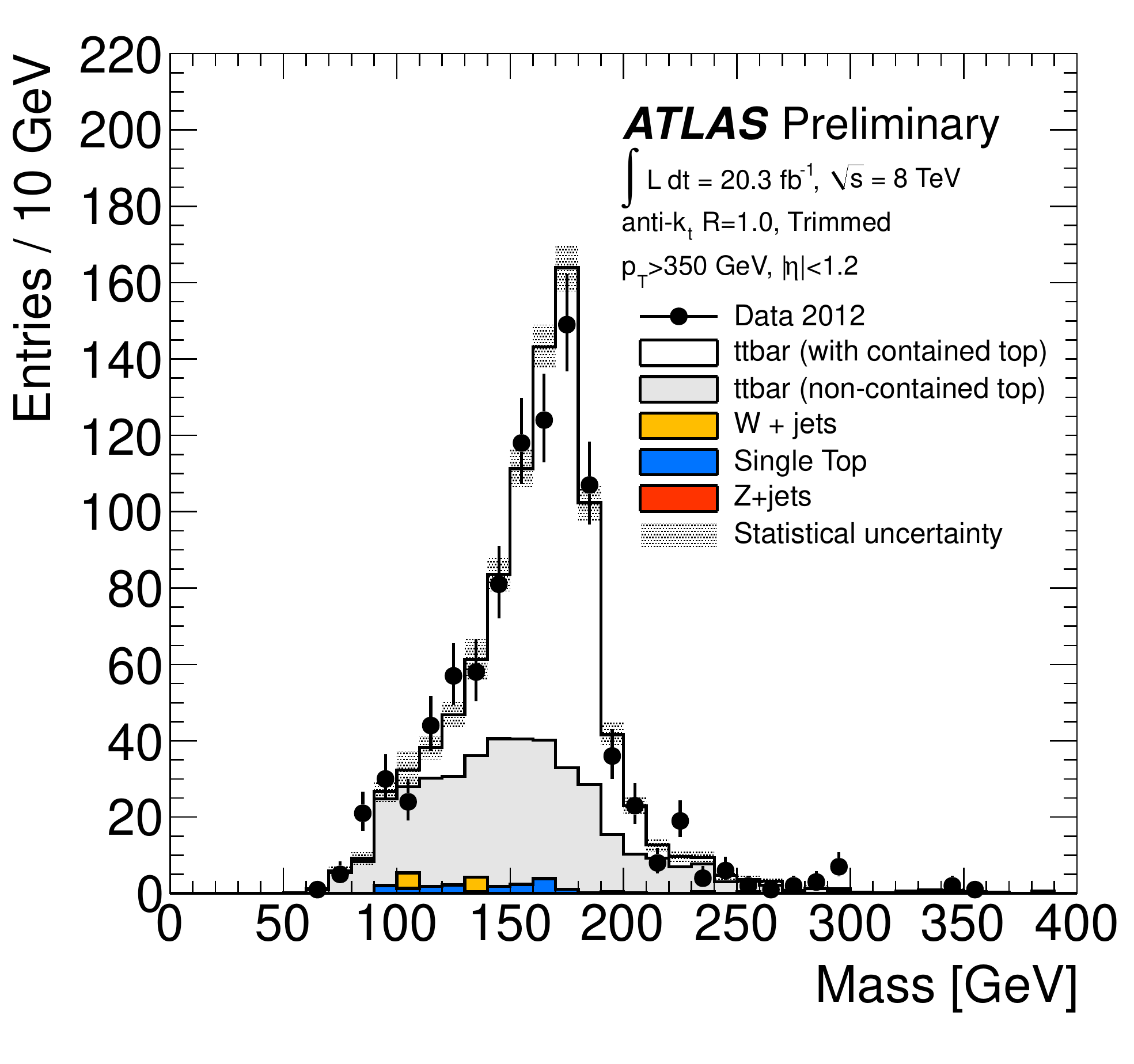}
}
\subfigure[$N_{\rm subjets} \ge 4$]{
   \includegraphics[width=0.45\textwidth,angle=0]{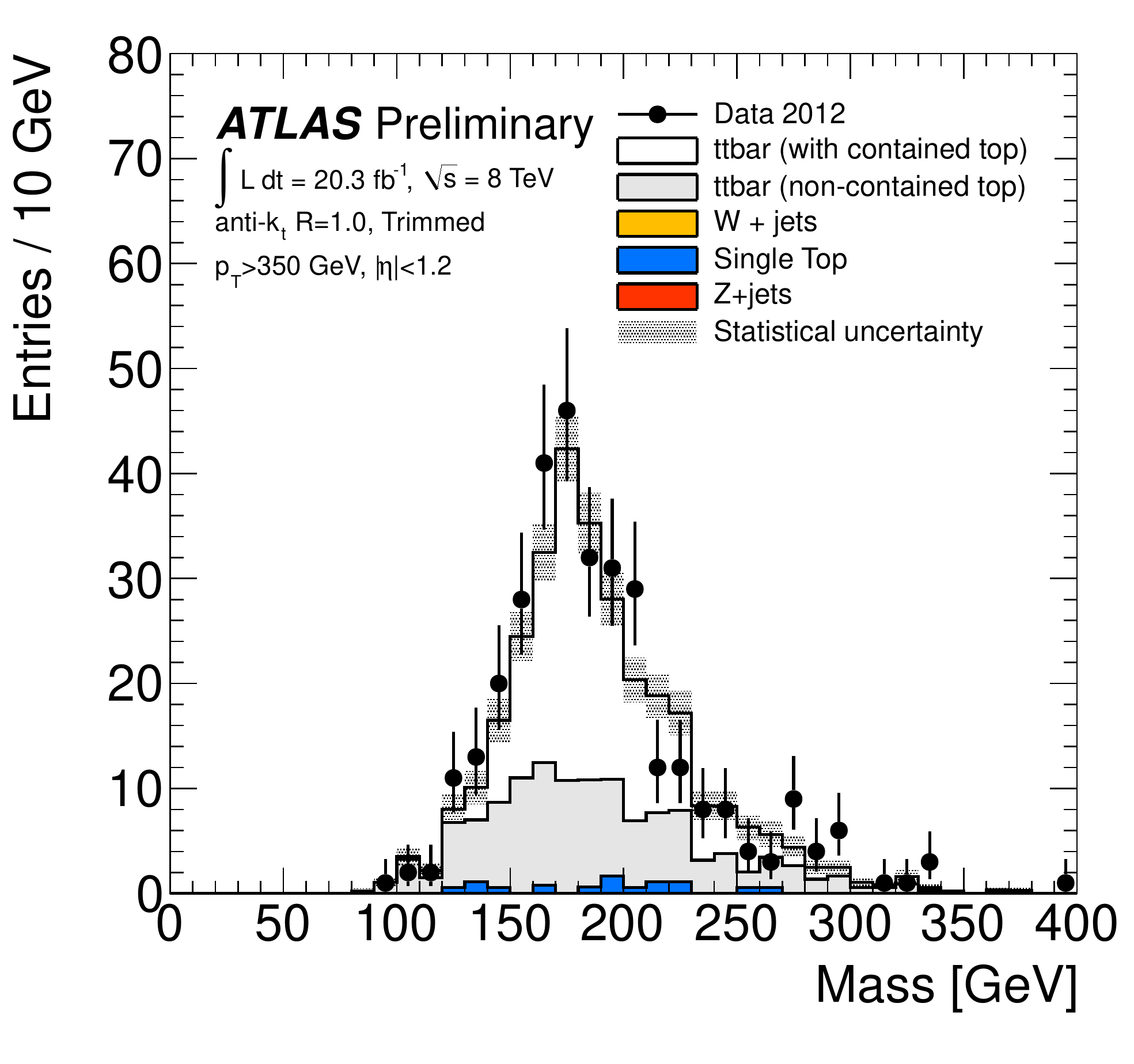}
}
\caption{Detector level distributions of the mass of
trimmed \akt $R=1.0$ jets ($\fcut=5\%$, $\Rsub=0.3$) with $\pt>350\GeV$ in a
selection of semileptonic \ttbar decays for different \kt $R=0.3$ subjet multiplicities.
{\em Contained} refers to hadronically-decaying top quarks for which
all three decay quarks are separated by less than $\DeltaReta = 1.0$ from
the top quark flight direction. The shaded band represents the statistical simulation uncertainty.
The \ttbar simulation is obtained from \powhegpythia.
From~\cite{ATLAS-CONF-2013-084}.}
\label{fig:tt_trimmed_mass_nsubjet}
\end{figure}

\figref{tt_trimmed_mass}a shows the mass distribution for trimmed \akt $R=1.0$ fat jets
in the event sample obtained with the \ttbar selection.
The peak at the top quark mass results from hadronically decaying top quarks
for which all decay products are captured in the fat jet.
The peak at the \W boson mass represents the cases in which the $b$-jet is not part of the fat jet.
The average mass above $100\GeV$ is shown in \figref{tt_trimmed_mass}b as
a function of \avmu.
Within the statistical uncertainty of $\approx\!5\GeV$, no pile-up dependence is present.
\figref{tt_trimmed_mass_nsubjet} shows the jet mass distribution for different
subjet multiplicities. The subjets are reconstructed with the \kt
algorithm using $R=0.3$ and correspond to the subjets that remain after trimming
and have $\pt > 17.5\GeV = 5\% \times 350\GeV$.
All distributions are well described by the simulation as is the
integrated distribution in \figref{tt_trimmed_mass}a.
The prediction for \ttbar production is obtained from \powheg with \pythia for the parton shower
and hadronisation.
For exactly one subjet, the fat jet mass peaks at $20$--$30\GeV$ and
the fat jet results from hadronic top quark decays with more than one decay jet
not captured, background from \Wpjets events, or production of single top quarks.
With two subjets, the distribution peaks at the \W boson mass,
indicating that the subjets correspond to the two \W boson decay jets. A shoulder
towards higher masses is also visible, resulting from $t\to bqq$ decays in which
the subjets correspond to one \W boson decay jet and the $b$-jet, which has on
average larger \pt. A \W boson peak is also visible for events with single top quarks,
which therefore seem to contain a hadronically decaying \W boson while still
satisfying the requirements for an isolated muon and significant \ETmiss.
Fat jets with three subjets have a mass that peaks near the top quark mass
and result mostly from SM \ttbar events in which all top quark decay jets
are contained in the fat jet.
For four or more subjets, the high mass tail
is more pronounced but is still described by \ttbar production, indicating that
energy from underlying event and/or pile-up is picked up.
These studies suggest that the mass of trimmed \akt $R=1.0$ jets with exactly three subjets
is well suited for the identification of hadronically decaying top quarks.

%%%%%%%%%%%%%%%%%%%%%%%%%%%%%%%%%%%%%%%%%%%%%%%%%%%%%%%%%%%%%%%%%%%%%%%%%%%%%%%%

\subsection{\kt splitting scales}

\begin{figure}[hbt]
\centering
\subfigure[\DOneTwo]{
   \includegraphics[width=0.45\textwidth,angle=0]{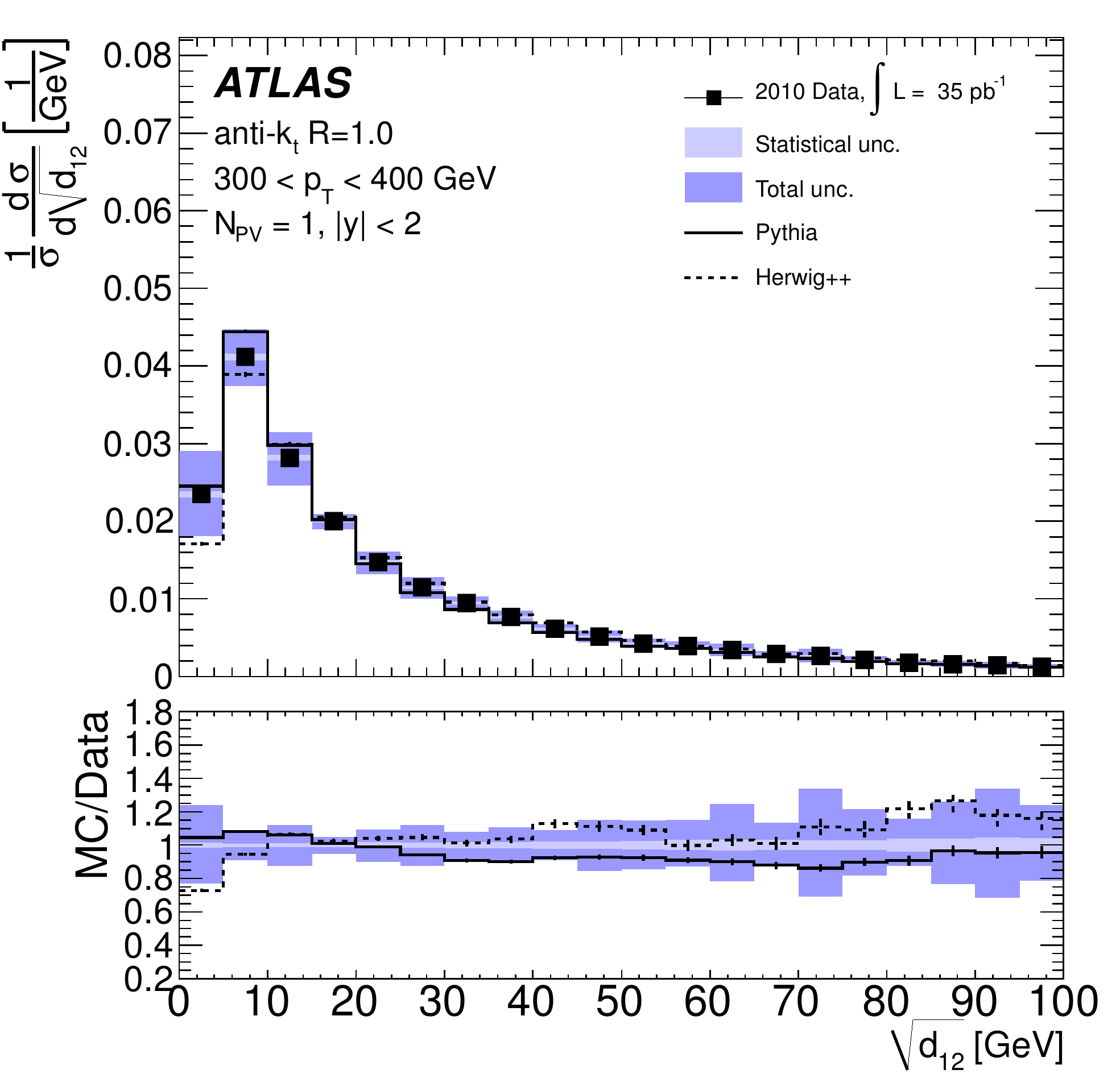}
}
\subfigure[\DTwoThr]{
   \includegraphics[width=0.45\textwidth,angle=0]{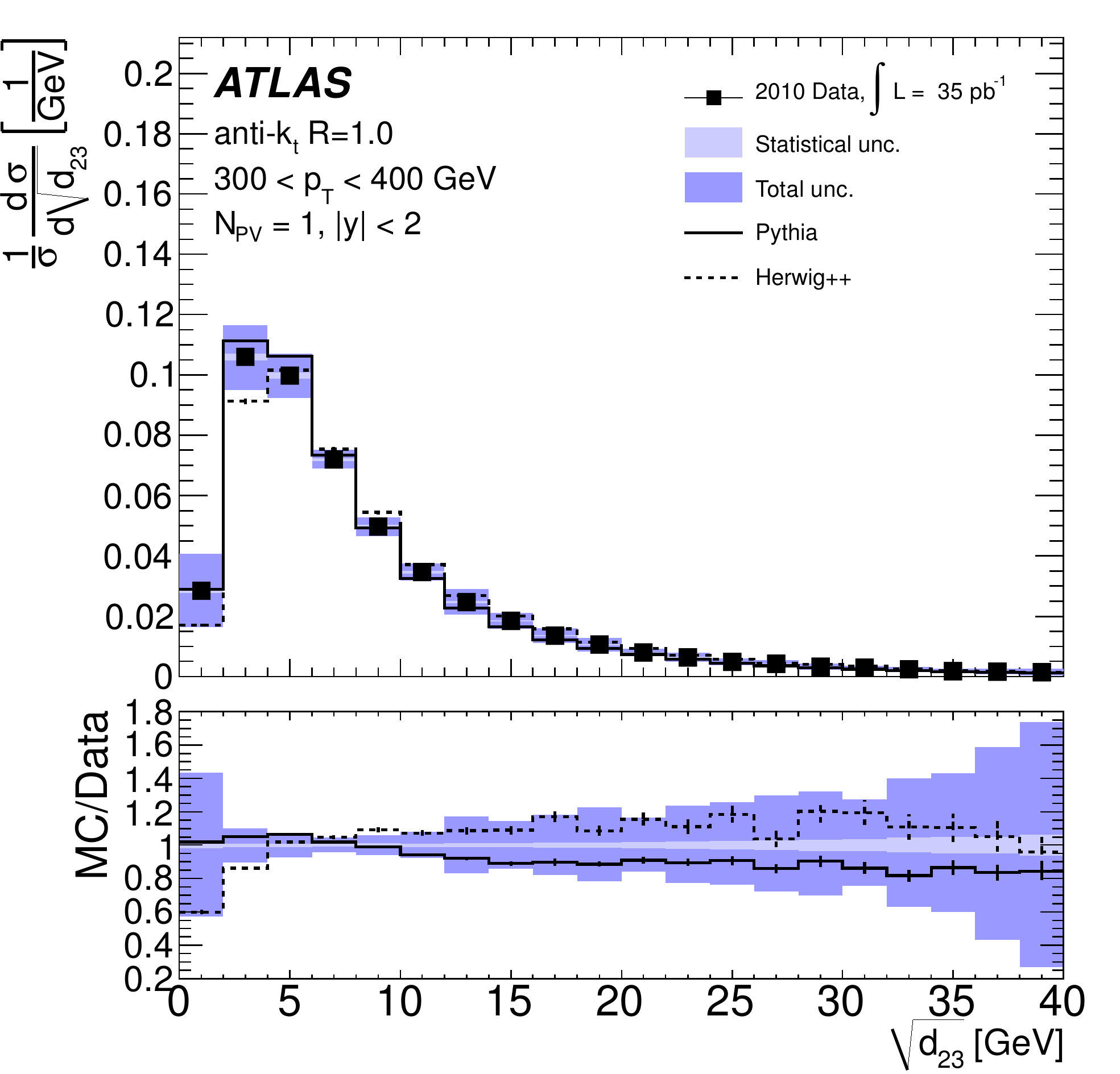}
}
\caption{Unfolded normalised distributions of \kt splitting scales of
\akt $R=1.0$ jets with $300<\pt<400\GeV$ in an inclusive jet sample without pile-up.
From~\cite{ATLAS:2012am}.}
\label{fig:unfolded_splitscales}
\end{figure}

Normalised distributions of the \kt splitting scales \DOneTwo and \DTwoThr are
shown in \figref{unfolded_splitscales} for jets in multijet events without pile-up overlay energy.
The distributions peak in the ranges $5$--$10\GeV$ and $2.5$--$7.5\GeV$, respectively.
The uncertainties from unfolding (related mainly to the uncertainty with which
the detector jet energy scale and resolution can be simulated)
are $\pm20\%$ in the ranges $5$--$70\GeV$ and $2.5$--$25\GeV$,
respectively, which contain the bulk of the data. The predicted distributions describe the data within these
uncertainties.
Trends are however visible, with the \herwigpp spectrum being too hard and the one by
\pythia too soft.

\begin{figure}[hbt]
\centering
\subfigure[\DOneTwo]{
   \includegraphics[width=0.45\textwidth,angle=0]{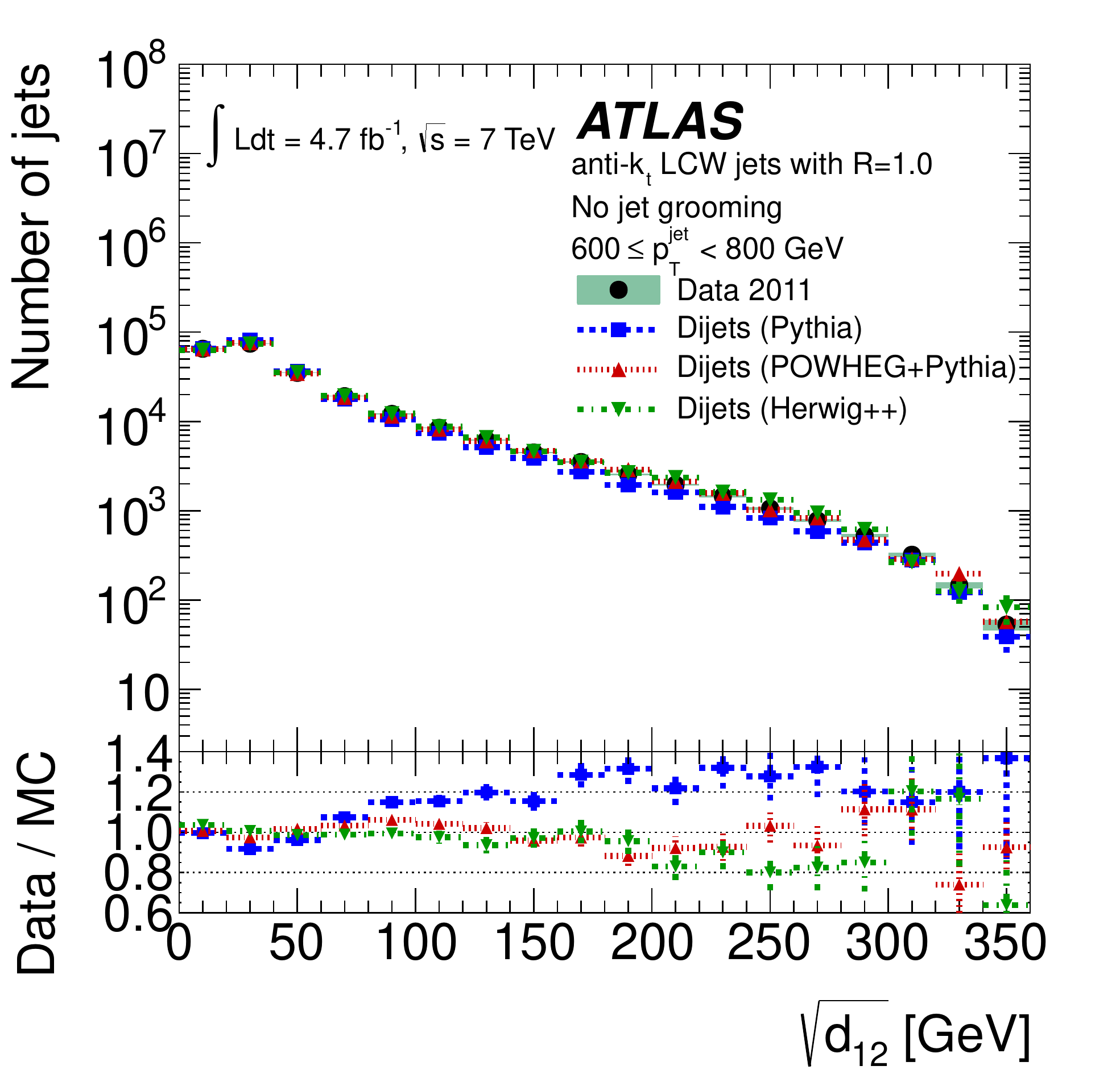}
}
\subfigure[\DTwoThr]{
   \includegraphics[width=0.45\textwidth,angle=0]{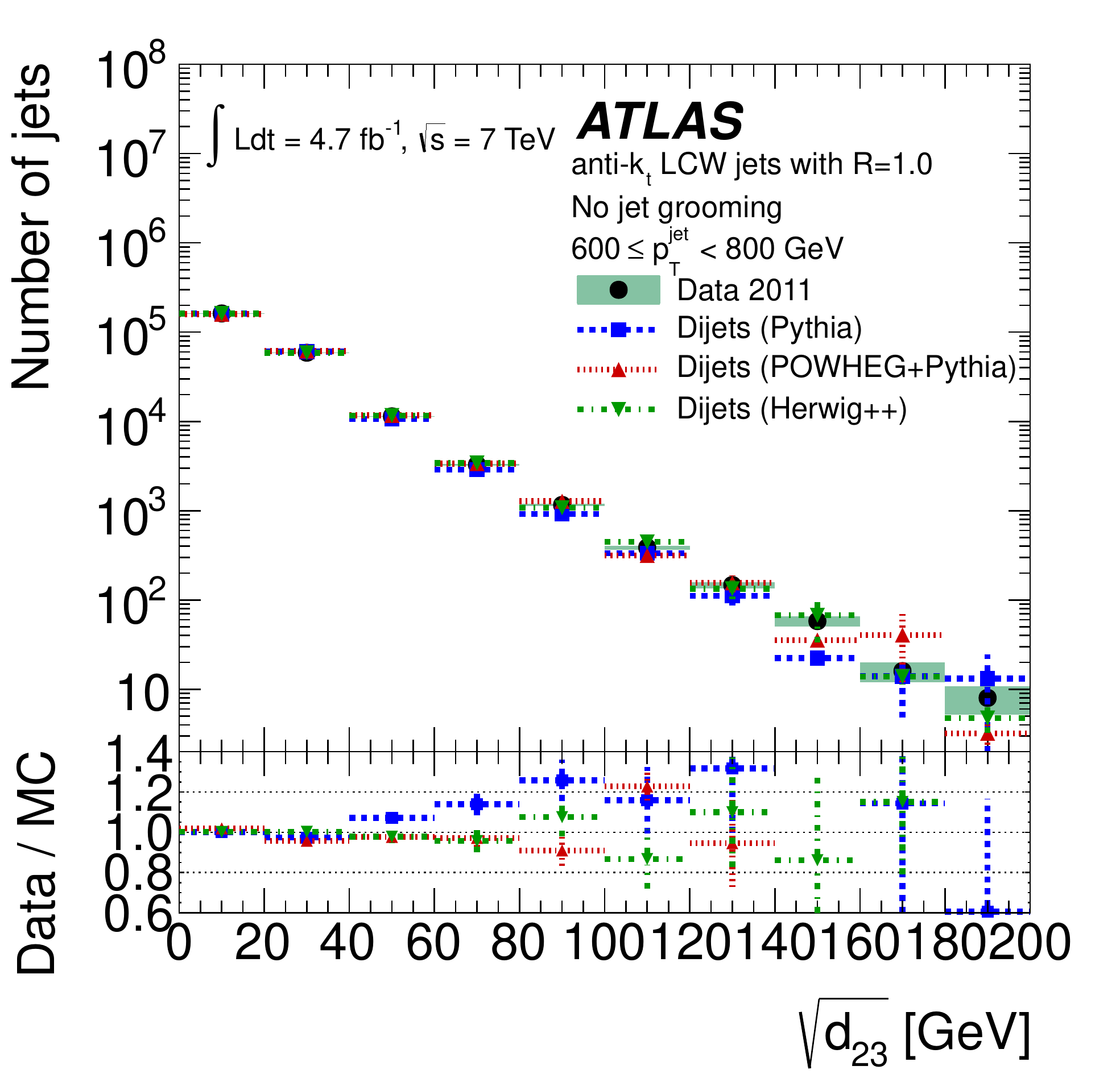}
}
\caption{Detector level distributions of \kt splitting scales of
\akt $R=1.0$ jets with $600<\pt<800\GeV$ in multijet events with an average
pile-up \avmu of $9.1$. From~\cite{Aad:2013gja}.}
\label{fig:perf_splitscales}
\end{figure}

Detector level distributions of the splitting scales are shown in \figref{perf_splitscales}
for jets in multijet events with an average pile-up \avmu of 9.1.
Compared to the no-pile-up case in \figref{unfolded_splitscales},
the horizontal axis now extends to much larger values of the scales
and a logarithmic vertical axis is chosen to highlight the tails of the
distributions.
For $\DOneTwo < 100\GeV$ and $\DTwoThr < 40\GeV$, the quality of the descriptions
is similar to the no-pile-up case and describes the data within $10$--$20\%$.
A typical cut for top quark identification is $\DOneTwo > 40\GeV$ and
the prediction at this value is within $10\%$ of the data. In the tails, \powheg and \herwigpp
give a better prediction than \pythia.
The \pythia spectra are too soft with the relative difference to the data points
being $\approx\!30\%$ for $\DOneTwo > 160\GeV$ and $\DTwoThr>80\GeV$.
The \herwigpp spectrum tends to be
too hard for $\DOneTwo>200\GeV$. Grooming has little impact on the splitting scales and the quality of the
simulation is similar for trimmed jets~\cite{Aad:2013gja} (not shown).

\begin{figure}[hbt]
\centering
\subfigure[\DOneTwo]{
   \includegraphics[width=0.45\textwidth,angle=0]{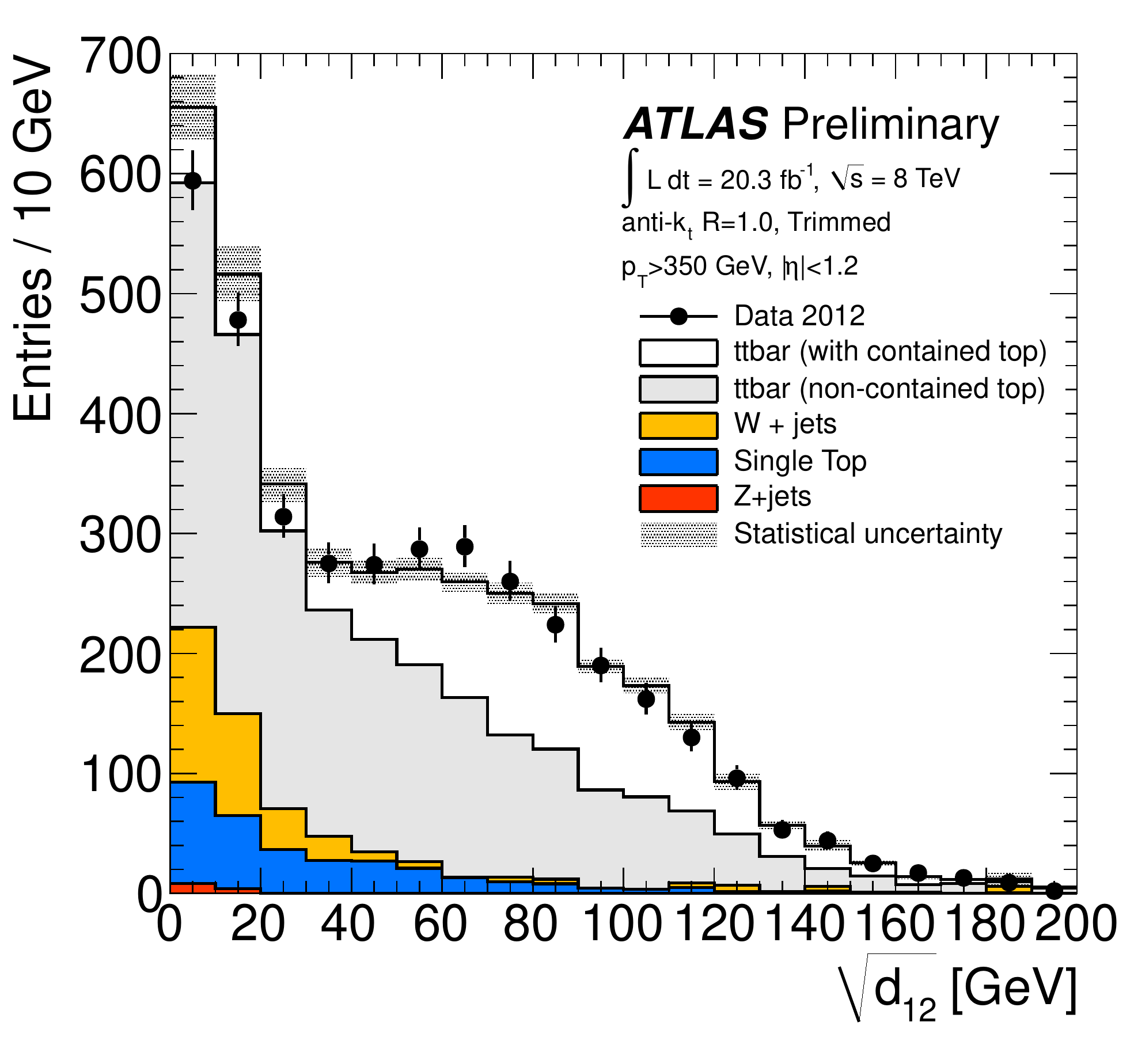}
}
\subfigure[\DTwoThr]{
   \includegraphics[width=0.45\textwidth,angle=0]{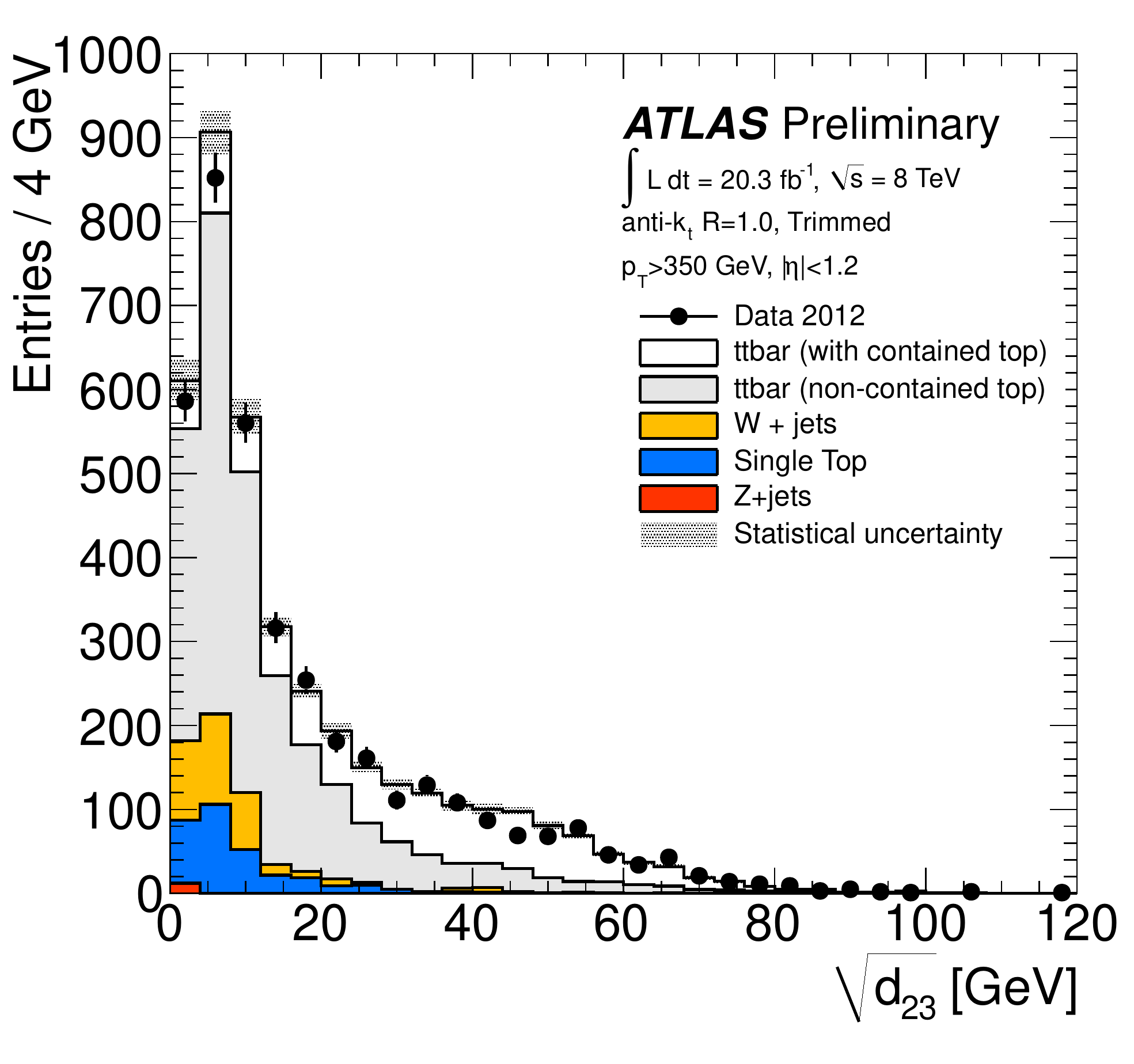}
}
\caption{Detector level distributions of the \kt splitting scales for
trimmed \akt $R=1.0$ jets ($\fcut=5\%$, $\Rsub=0.3$) with $\pt>350\GeV$ in a
selection of semileptonic \ttbar decays.
{\em Contained} refers to hadronically-decaying top quarks for which
all three decay quarks are separated by less than $\DeltaReta = 1.0$ from
the top quark flight direction.
The shaded band represents the statistical simulation uncertainty.
The \ttbar simulation is obtained from \powhegpythia.
From~\cite{ATLAS-CONF-2013-084}.}
\label{fig:tt_splitscales}
\end{figure}

The distributions of the splitting scales for jets in
the \ttbar event selection are shown in \figref{tt_splitscales}.
They are well described by the simulation. For {\em contained} top quarks,
the \DOneTwo (\DTwoThr) distribution peaks at approximately $80$--$90\GeV$ ($35$--$40\GeV$)
as expected.
The majority of top quarks is however not contained because the radius parameter
of $R=1.0$ is not large enough to capture all decay products of top quarks
with $\pt>350\GeV$ (cf. \figref{topsize}). The splitting scales for non-contained
top quarks are naturally smaller. Also these cases are well described.

%%%%%%%%%%%%%%%%%%%%%%%%%%%%%%%%%%%%%%%%%%%%%%%%%%%%%%%%%%%%%%%%%%%%%%%%%%%%%%%%

\subsection{\Nsjn}
\label{sec:measurements_nsjn}

\begin{figure}[hbt]
\centering
\subfigure[$\tau_{32}$]{
   \includegraphics[width=0.45\textwidth,angle=0]{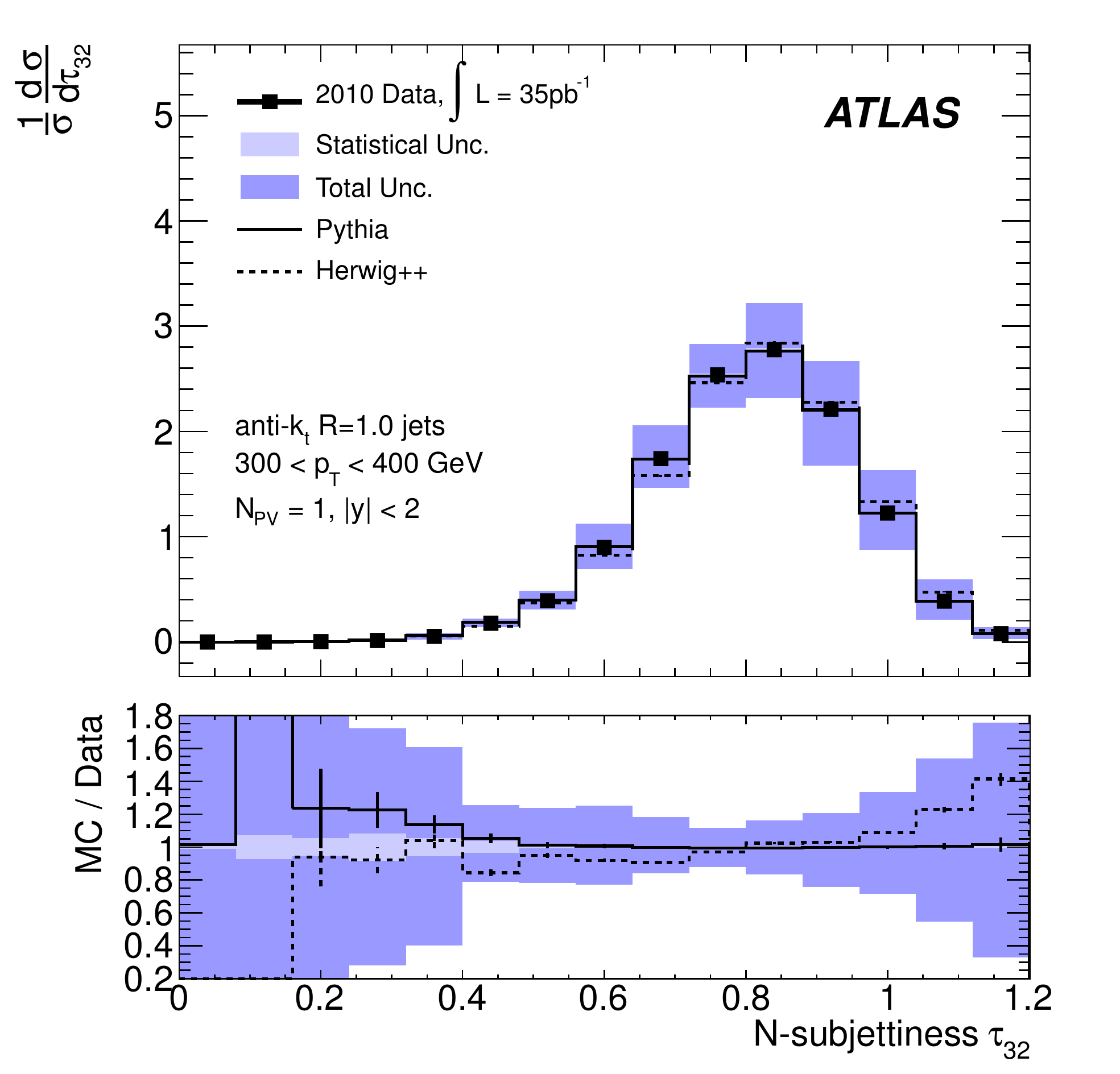}
}
\subfigure[$\tau_{21}$]{
   \includegraphics[width=0.45\textwidth,angle=0]{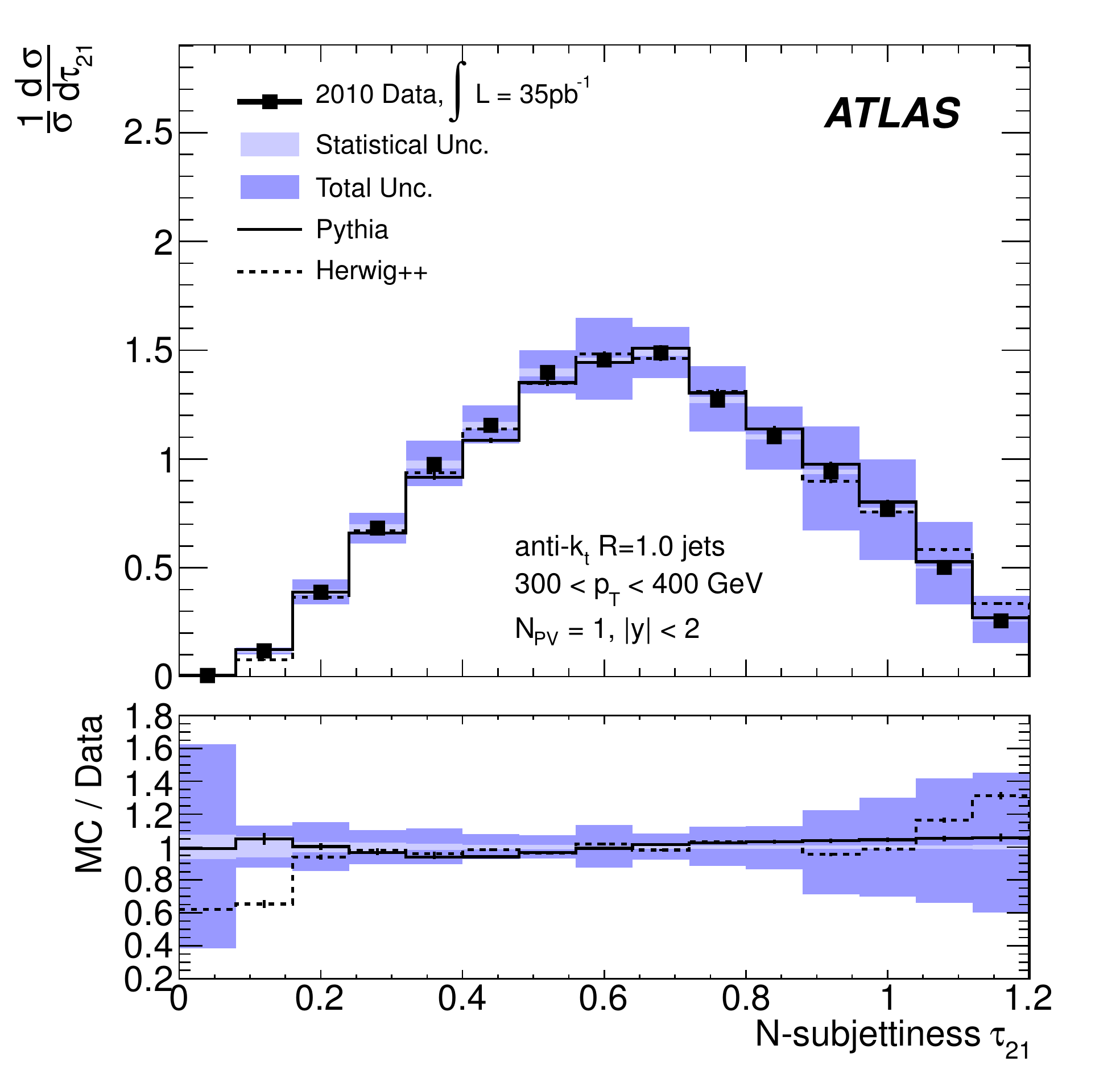}
}
\caption{Unfolded normalised distributions of \Nsjn ratios for
\akt $R=1.0$ jets with $300<\pt<400\GeV$ in an inclusive jet sample without pile-up.
From~\cite{ATLAS:2012am}.}
\label{fig:unfolded_nsjn}
\end{figure}

Normalised distributions of the measured \Nsjn ratios $\tau_{32}$ and $\tau_{21}$ at the particle level
are
shown in \figref{unfolded_nsjn} for jets in multijet events without pile-up overlay energy.
The most probable values are $\approx\!0.85$ for $\tau_{32}$ and $\approx\!0.7$ for $\tau_{21}$.
The bulk of each distribution is described by simulations within the measurement uncertainties,
which amount to $\pm20\%$ and $\pm10\%$, respectively.

\begin{figure}[hbt]
\centering
\subfigure[]{
   \includegraphics[width=0.45\textwidth,angle=0]{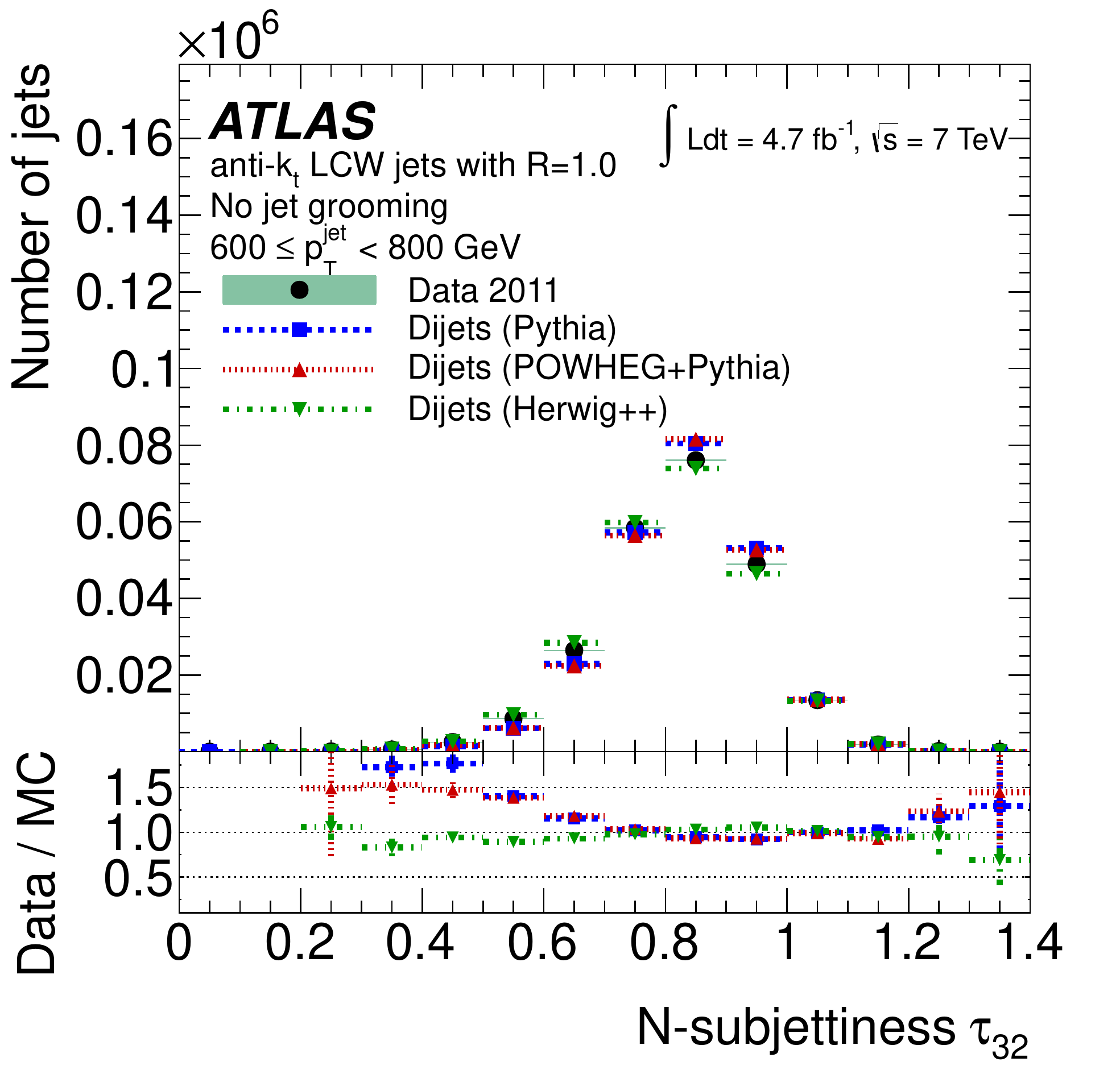}
}
\subfigure[]{
   \includegraphics[width=0.45\textwidth,angle=0]{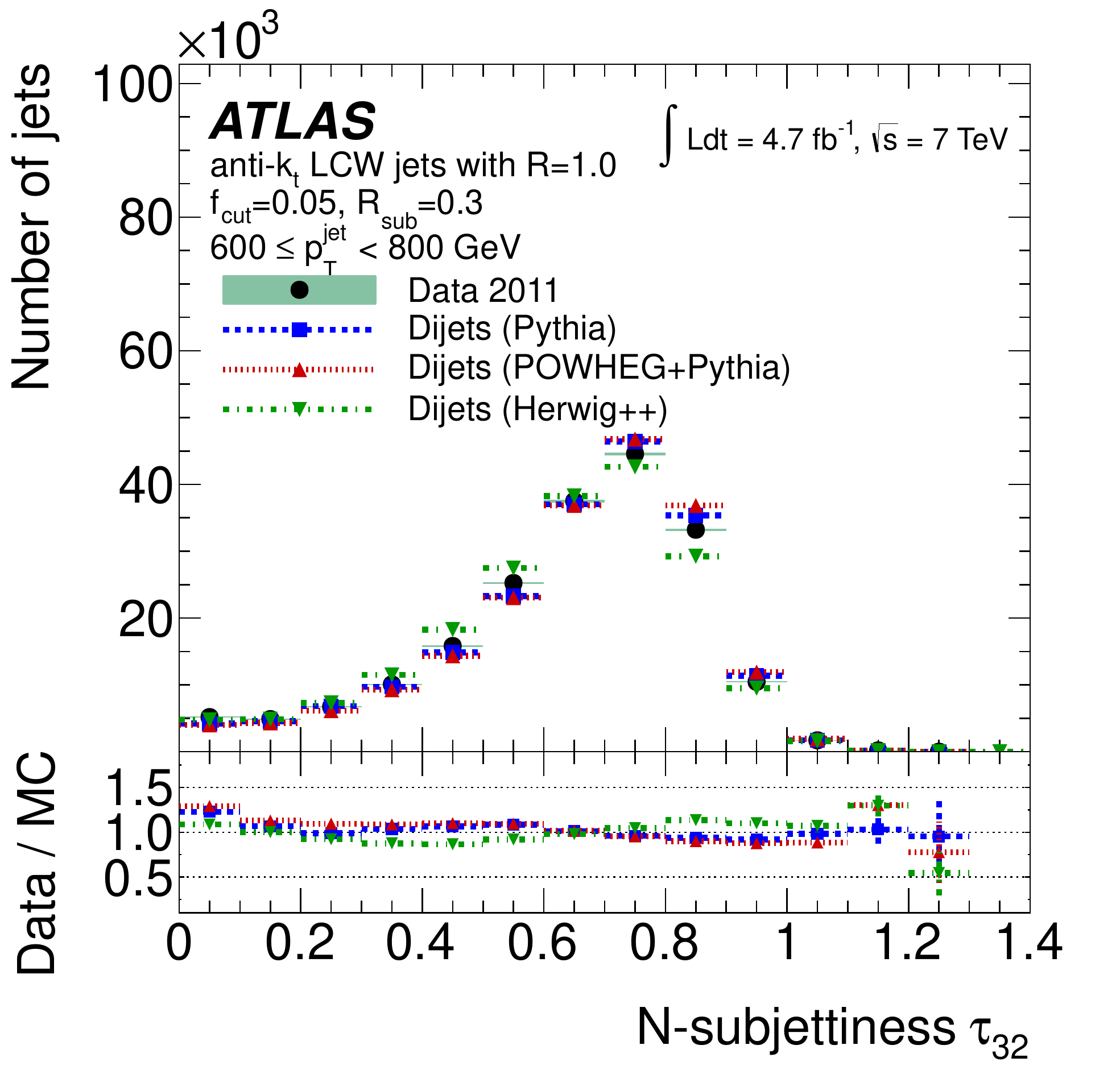}
}

\caption{Detector level distributions of the \Nsjn ratio $\tau_{32}$ for
\akt $R=1.0$ jets with $600<\pt<800\GeV$ in multijet events with an average
pile-up \avmu of $9.1$
a) before trimming and b) after trimming with $\fcut = 5\%$ and $\Rsub = 0.3$. From~\cite{Aad:2013gja}.}
\label{fig:perf_tauThreeTwo}
\end{figure}

Detector level distributions of the \Nsjn ratio $\tau_{32}$ are shown in \figref{perf_tauThreeTwo}
for jets in multijet events with an average pile-up \avmu of 9.1.
The distribution for ungroomed jets peaks at $0.8$--$0.9$ (panel a) like in the
no-pile-up case.
\herwigpp (with the UE model parameters tuned to LHC data) gives the best description
while \pythia and \powhegpythia underestimate the tail for $\tau_{32}<0.7$.
Trimming shifts the distributions by $0.1$ to lower values and increases the tail (panel b).
Trimming therefore makes the non-top jets more top-like. This has already
been observed in simulation in \secref{nsjn}.
A benefit of trimming is, however, that it improves the description of the distribution by
simulation: all generators yield a description within $10\%$ of the data.

\begin{figure}[hbt]
\centering
\subfigure[]{
   \includegraphics[width=0.45\textwidth,angle=0]{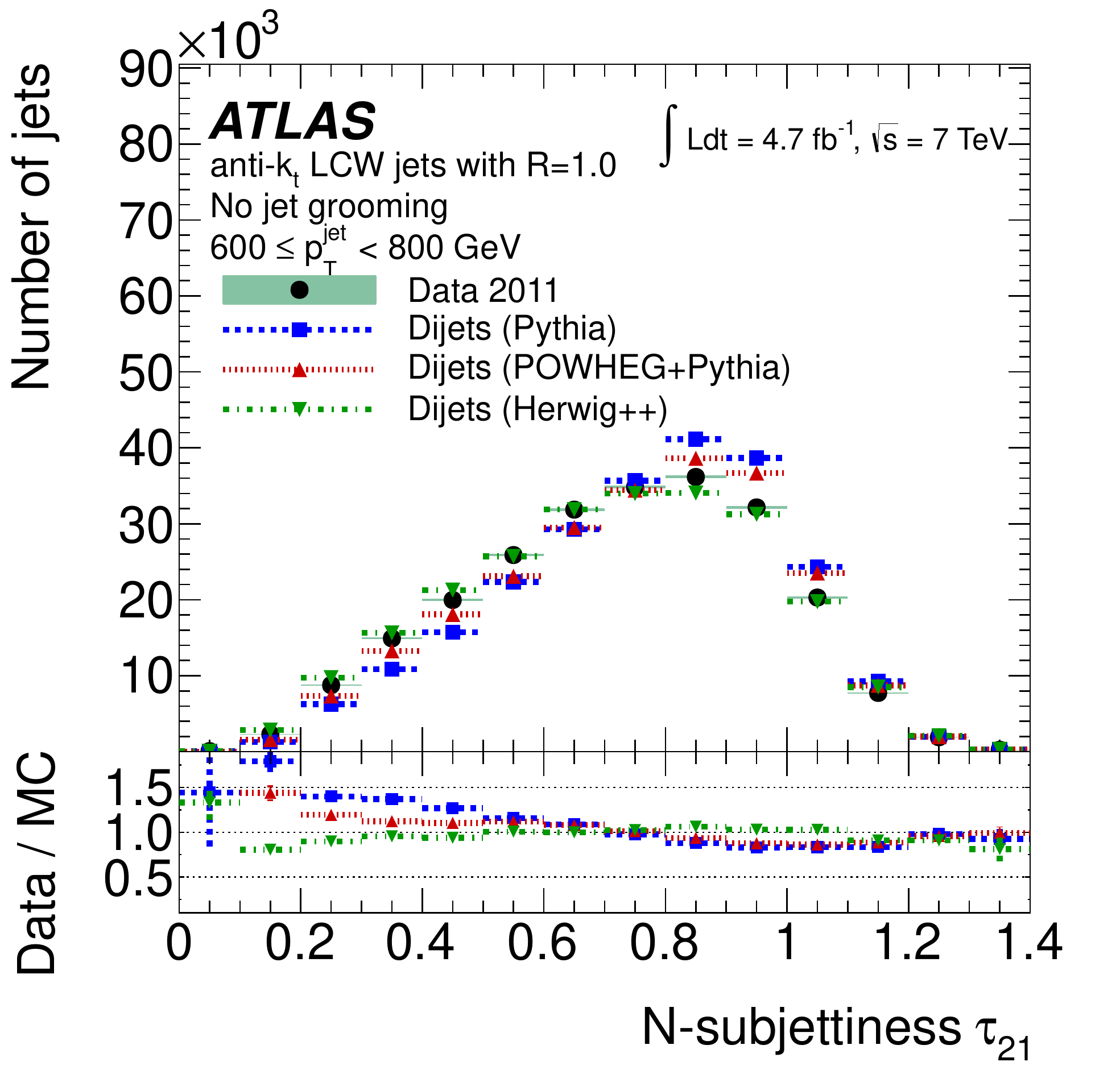}
}
\subfigure[]{
   \includegraphics[width=0.45\textwidth,angle=0]{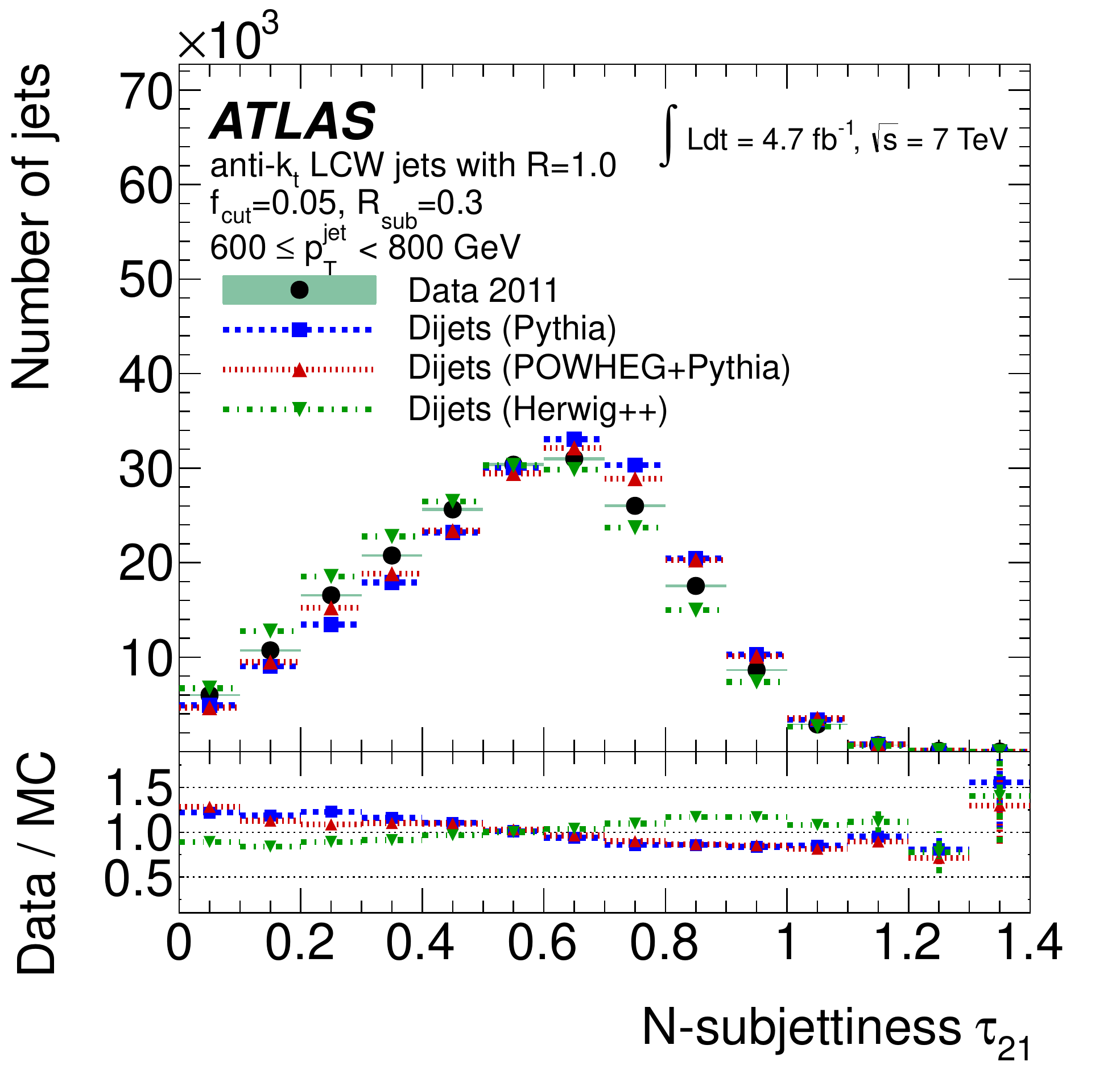}
}
\caption{Detector level distributions of the \Nsjn ratio $\tau_{21}$ for
\akt $R=1.0$ jets with $600<\pt<800\GeV$ in multijet events with an average
pile-up \avmu of $9.1$
a) before trimming and b) after trimming with $\fcut = 5\%$ and $\Rsub = 0.3$. From~\cite{Aad:2013gja}.}
\label{fig:perf_tauTwoOne}
\end{figure}

A similar observation is made for $\tau_{21}$ which is shown in \figref{perf_tauTwoOne}.
Also here, \herwigpp gives the best description before trimming.
After trimming, the \herwigpp distribution is shifted to lower values, although
still within $\pm20\%$ of the data points.
The \pythia prediction underestimates the data in the tail ($\tau_{21}<0.6$)
by $30\%$ or more and is shifted towards high values.
Trimming reduces the discrepancy in the tail to $20\%$ but the distribution
is still shifted. This shift is also seen for \powheg with trimming having little effect.

\begin{figure}[hbt]
\centering
\subfigure[$\tau_{32}$]{
   \includegraphics[width=0.45\textwidth,angle=0]{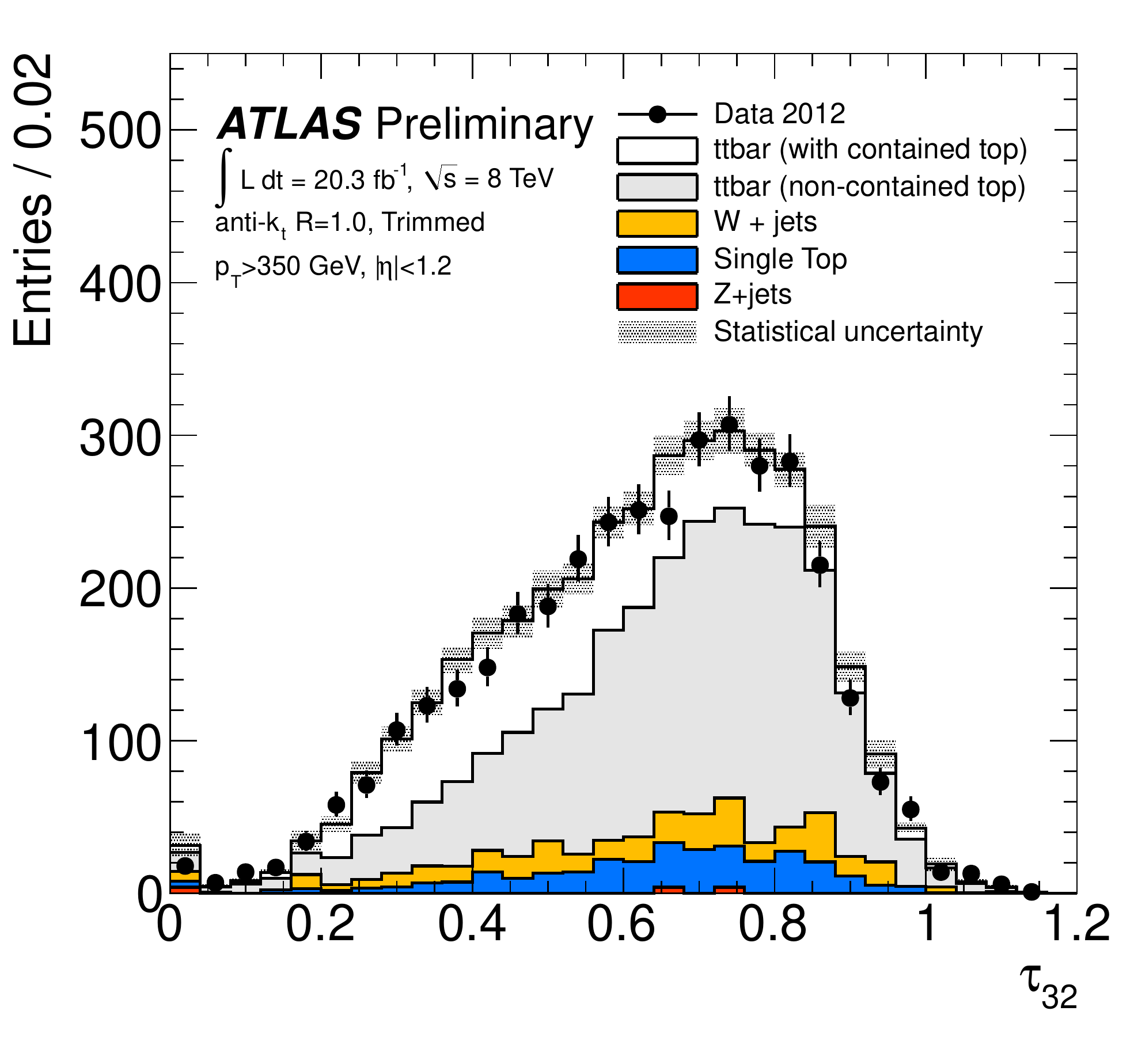}
}
\subfigure[$\tau_{21}$]{
   \includegraphics[width=0.45\textwidth,angle=0]{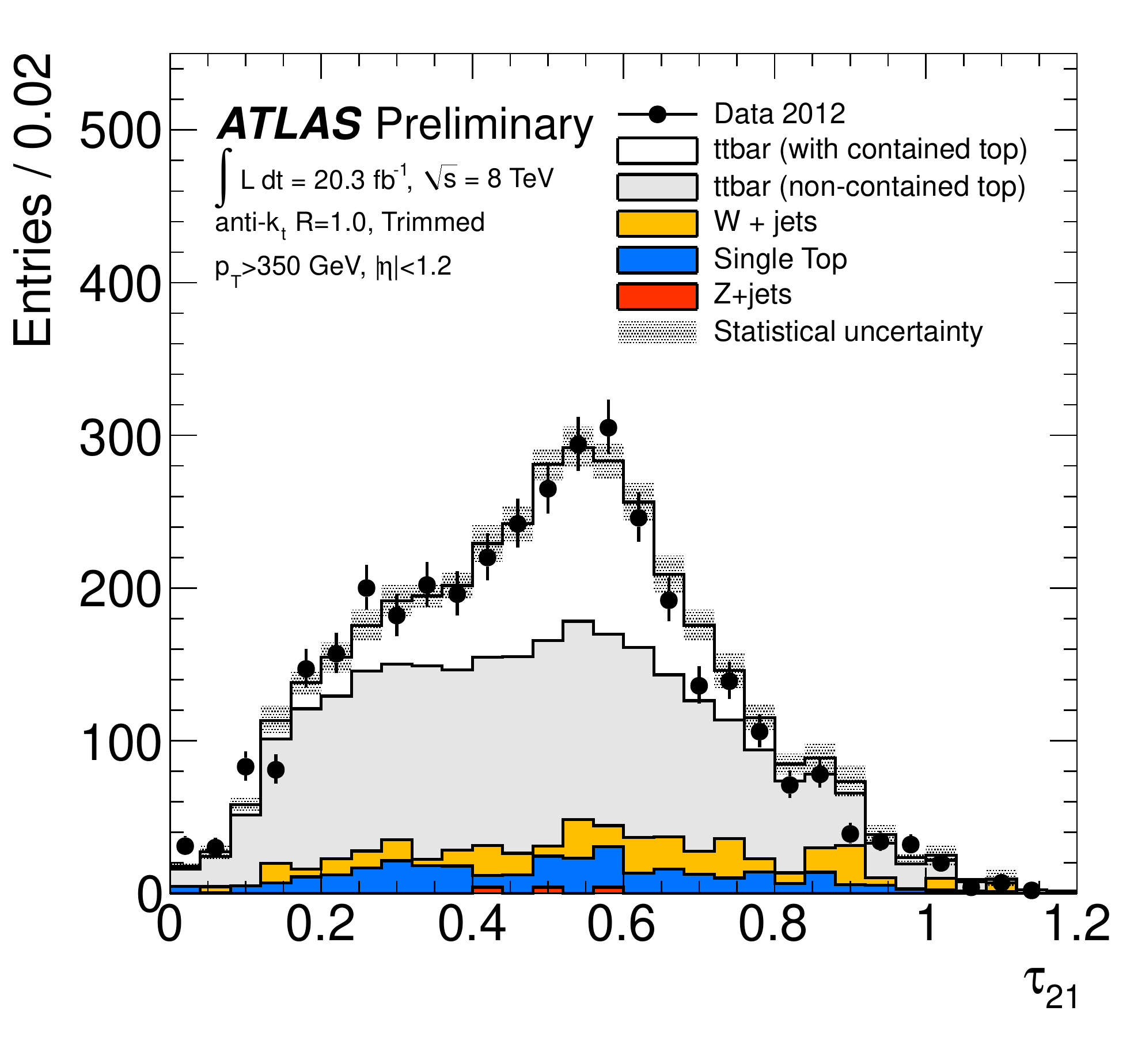}
}
\caption{Detector level distributions of \Nsjn ratios for
trimmed \akt $R=1.0$ jets ($\fcut=5\%$, $\Rsub=0.3$) with $\pt>350\GeV$ in a
selection of semileptonic \ttbar decays.
{\em Contained} refers to hadronically-decaying top quarks for which
all three decay quarks are separated by less than $\DeltaReta = 1.0$ from
the top quark flight direction. The shaded band represents the statistical simulation uncertainty.
The \ttbar simulation is obtained from \powhegpythia.
From~\cite{ATLAS-CONF-2013-084}.}
\label{fig:tt_nsjn}
\end{figure}

The \Nsjn ratios $\tau_{32}$ and $\tau_{21}$ are shown in \figref{tt_nsjn} for
fat jets in the \ttbar event selection.
The distributions are well described by the simulation. Fat jets containing
all top quark decay jets have the smallest $\tau_{32}$.
This is the expected
behaviour because most of the containing fat jets have three or four subjets while
the other fat jets have one or two (cf.~\figref{tt_trimmed_mass_nsubjet}).
The 3-prong or 4-prong structure of the fully-containing fat jets
leads to reduced $\tau_3$ with respect to $\tau_2$, and therefore a small
value of $\tau_{32} = \tau_3/\tau_2$.
Similarly, fat jets with only two subjets are expected to have the smallest $\tau_{21}$.
\figref{tt_nsjn}b is compatible with this expectation: the non-containing fat
jet contribution at low values of $\tau_{21}$ presumably originates from
fat jets with two subjets while that at larger values stems from fat jets with
only one subjet.

\subsection{Summary of measurements of substructure variables}

Jet grooming is a powerful tool to remove the
pile-up dependence of substructure variables as was shown for the jet mass.
Similar pile-up stability is obtained for the other variables~\cite{Aad:2013gja} (not shown).
The leading order plus parton shower generators \herwigpp and \pythia
give a good description of the hard substructure of inclusive fat jets.
The distributions of the mass and \Nsjn ratios of trimmed jets, as well
as the \kt splitting scales, are described within $20\%$ or better.
The soft substructure is less well modelled, as shown for the mass and \Nsjn
of ungroomed jets, where the difference to the measured data points can be
as big as $40\%$.
The quality of the modelling by the NLO generator \powheg is similar,
suggesting that the NLO effects in jet substructure are well modelled by the
parton shower approach. The structure of hadronic top quark decay is well
described by \powheg.

\section{Performance of Boosted Top Quark Reconstruction}
In this section, the signal efficiency and background rejection of different
top tagging approaches are examined. First, the \htt performance is
studied in data and simulation using \ttbar and multijet events.
Then the \htt performance is compared to that of
cuts on substructure variables (trimmed jet mass, \kt splitting scales,
and \Nsjn) in a simulated scenario of high \pt top quarks originating
from a massive \ttbar resonance.
The section closes with a performance comparison of the \htt and substructure
variable cuts in events with a high multiplicity of jets and intermediate top quark
\pt, as it is common in Supersymmetry.

\subsection{\htt performance}
\label{sec:httperformance}
The \htt performance has been studied with
event samples enriched in SM \ttbar production. These events are selected
from 2011 and 2012 ATLAS data by requiring a leptonically decaying top quark as
described for 2012 in \secref{measurements_datasamples}.
The selection for 2011 is similar, with the exceptions that no $b$-jet was required
and that events had to have four small-$R$ jets with $\pt>20\GeV$.

\begin{figure}[hbt]
\centering
\subfigure[]{
   \includegraphics[width=0.45\textwidth,angle=0]{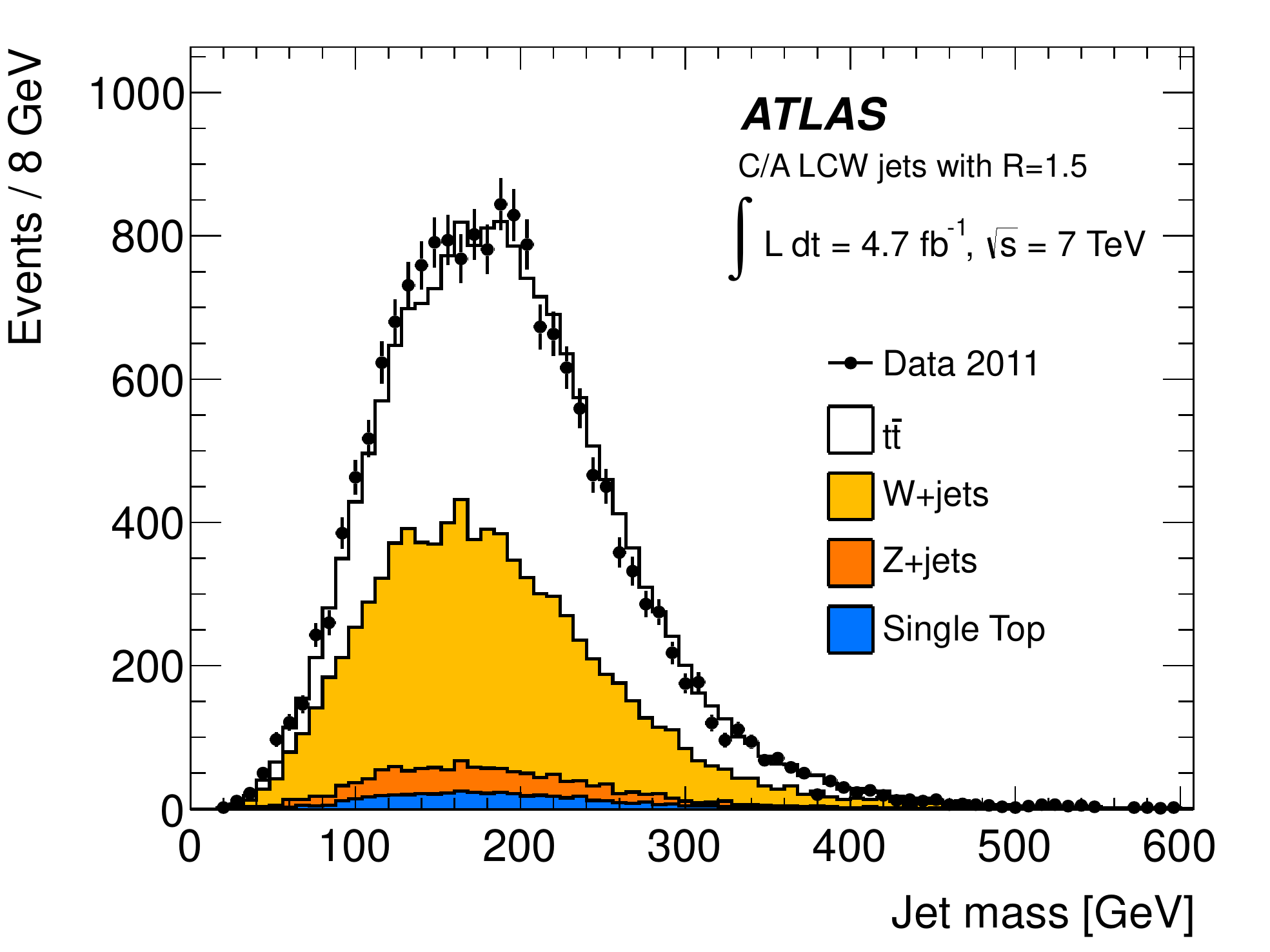}
}
\subfigure[]{
   \includegraphics[width=0.45\textwidth,angle=0]{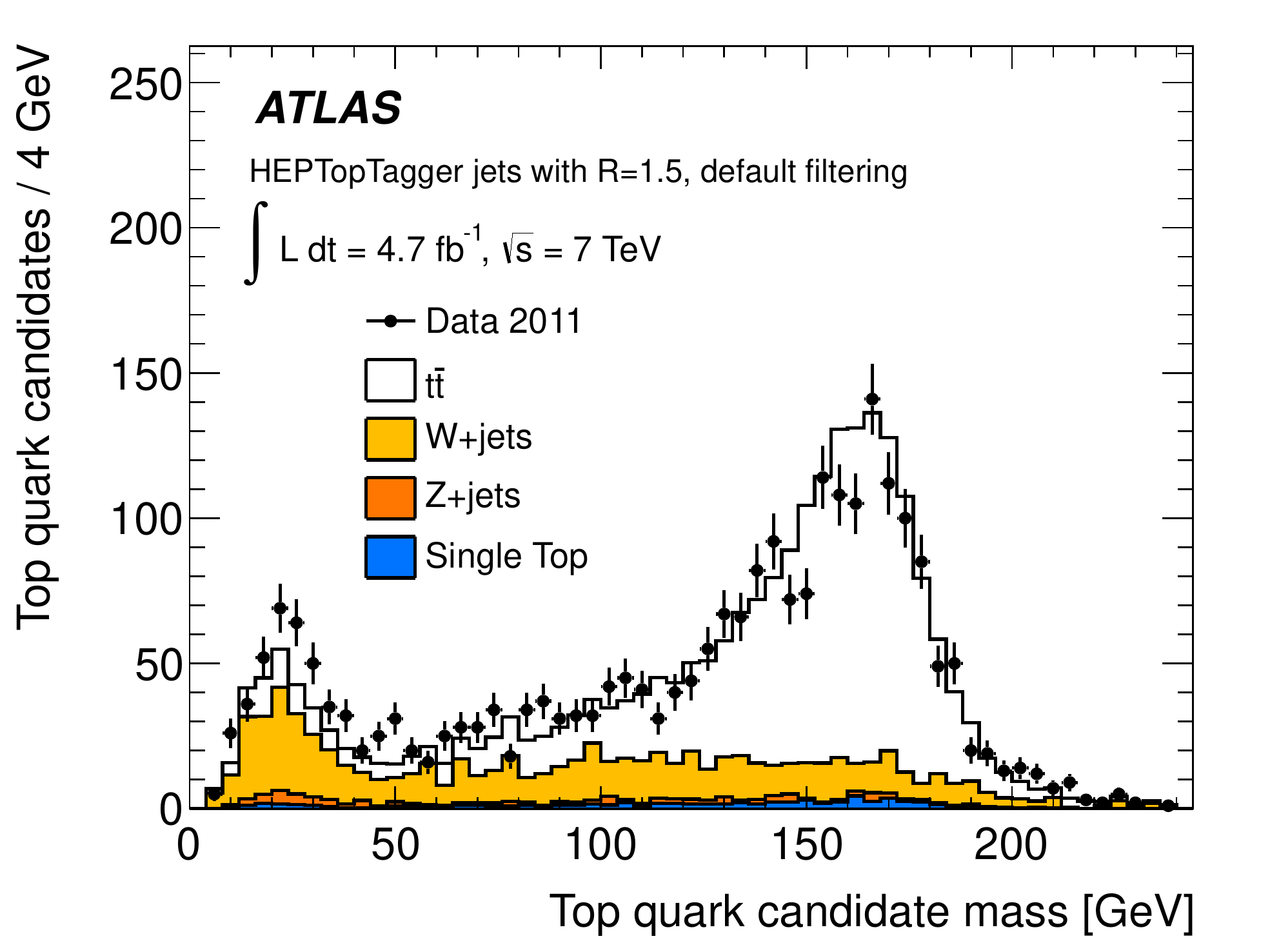}
} \\
\noindent
\subfigure[]{
   \includegraphics[width=0.45\textwidth,angle=0]{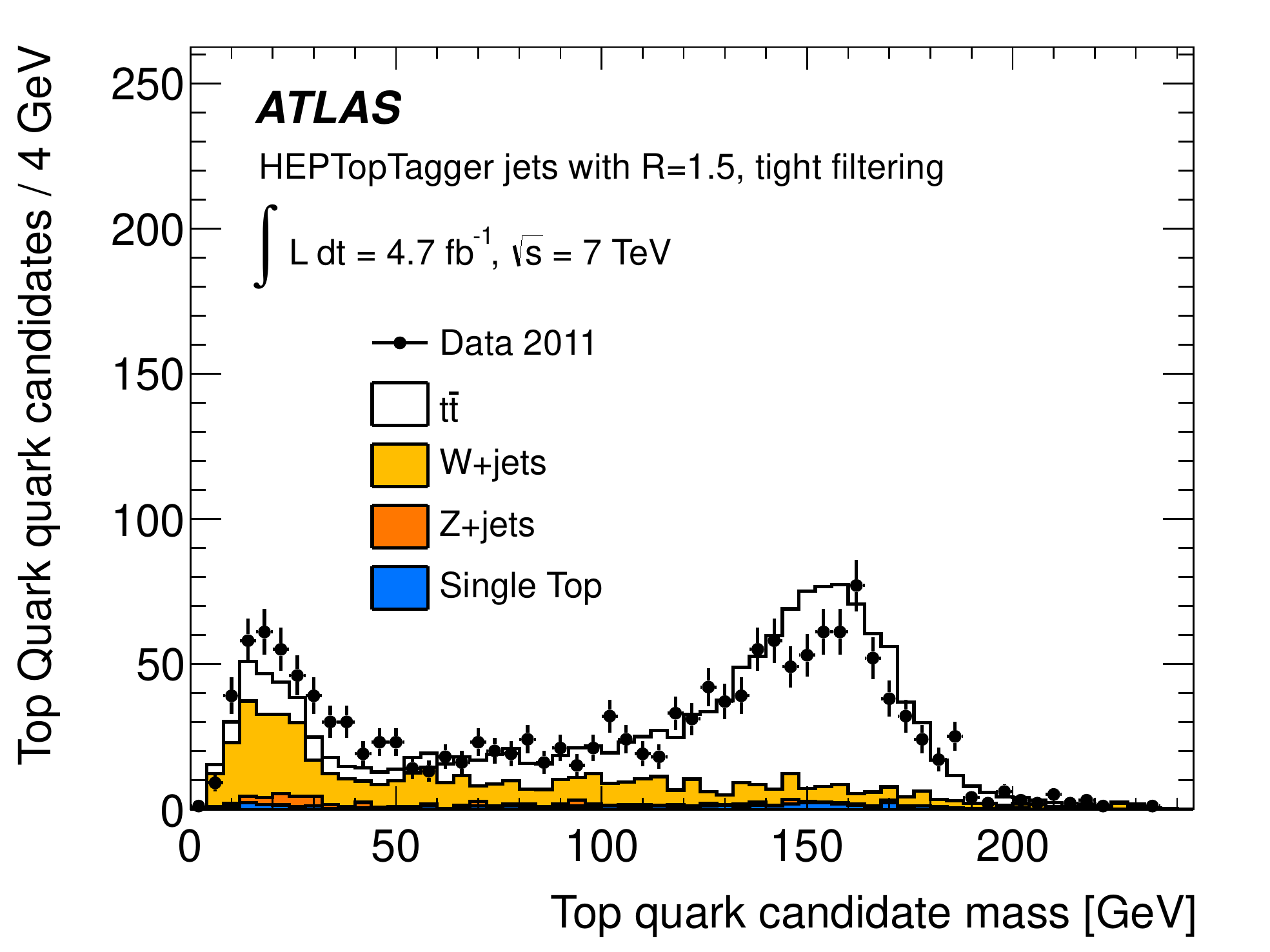}
}
\subfigure[]{
   \includegraphics[width=0.45\textwidth,angle=0]{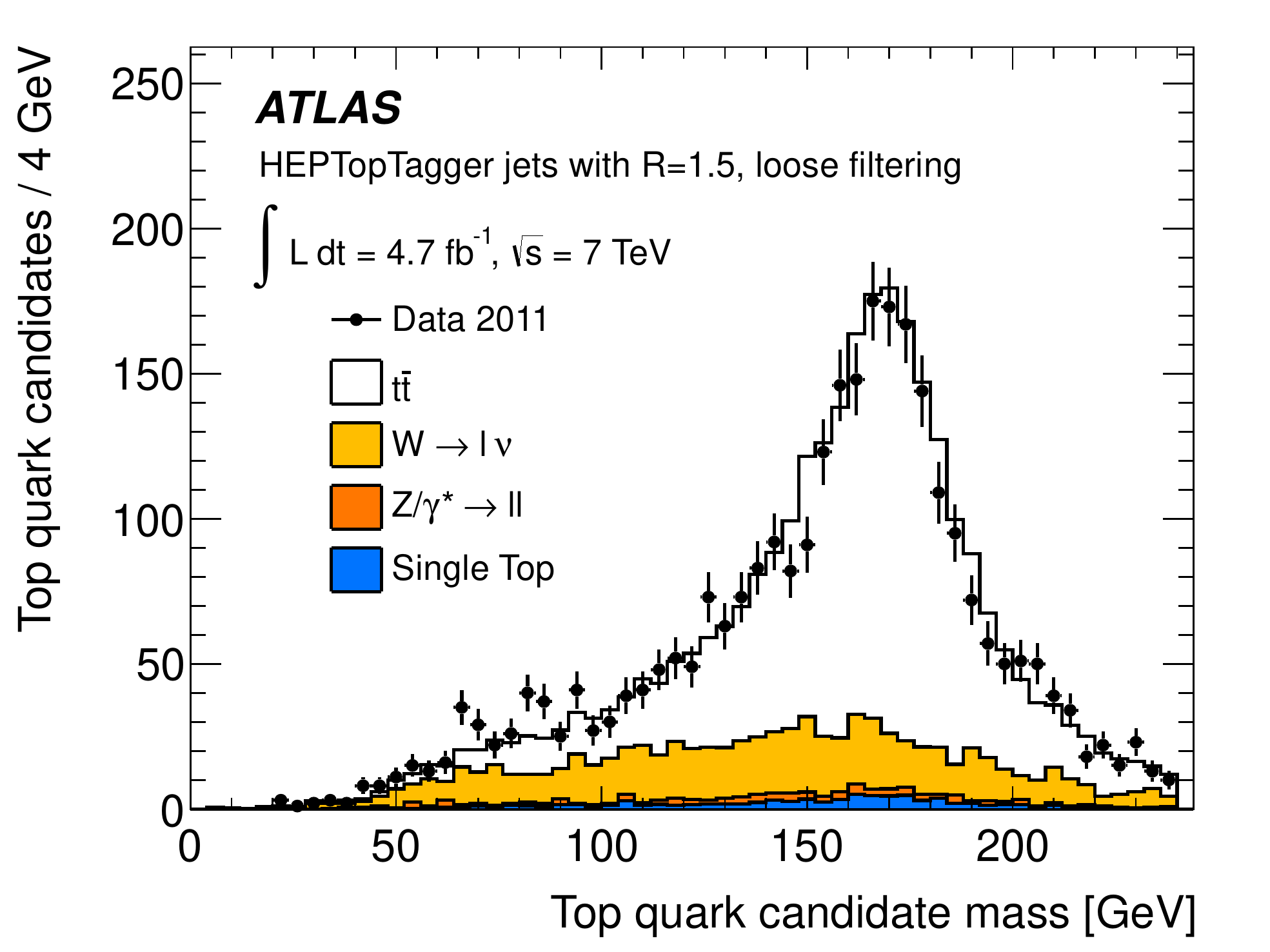}
}
\caption{Detector level distributions in a
selection of semileptonic \ttbar decays of (a) the mass of \ca $R=1.5$ jets with $\pt>200\GeV$ and
of the \htt candidate mass using (b) the default-30, (c) tight, and (d) loose \htt settings
of \tabref{HTTsettings}. The \ttbar simulation is from \mcatnlo and the \Wpjets
simulation from \alpgenherwig. From~\cite{Aad:2013gja}.}
\label{fig:htt_perf_m}
\end{figure}

The mass of the \htt input jets is shown in \figref{htt_perf_m}a.
These \ca $R=1.5$ fat jets have $\pt > 200\GeV$ and the distribution is peaked near the top quark mass
with a large width of $\approx\!160\GeV$.
The simulated distribution is shifted by $\approx\!8\GeV$ to higher masses.
This difference to the observed mass is within the typical jet mass simulation uncertainty of $5$--$6\%$
found for fat jets~\cite{Aad:2013gja}.
Approximately $50\%$ of the fat jets sample are SM \ttbar events and the
other half is dominated by \Wpjets events.
The \htt top quark candidate masses are
shown in \figref{htt_perf_m}b--d for different \htt parameter settings.
The {\em default-30} settings are used in \figref{htt_perf_m}b which
shows a peak at the top quark mass that falls rapidly towards larger masses.
The tail towards lower masses results from cases where not all energy
associated with the top quark decay is recovered. These losses occur in the
filtering and when not all decay products are contained in the fat jet.
The effect on the mass distribution of tightening and loosening the filtering
is shown in \figref{htt_perf_m}c and d: tight settings enhance the tail with respect to the peak
while loose settings reduce it. Which settings are optimal depends on the analysis
and on the amount of background that needs to be rejected.
The top candidate mass $m_t$ above $50\GeV$ is well described by the simulation.
The \ttbar events are generated using \mcatnlo and \alpgen is used for \Wpjets events.
The \htt increases the purity from $\approx\!50\%$ \ttbar processes before
tagging to $86\%$ in the mass window $140<m_t<200\GeV$.

The simulation underestimates the data at low $m_t$. This phase space represents the
{\em deep substructure}, where three subjets with $\pt>20\GeV$ combine to a
small invariant mass of $10$--$40\GeV$. The subjet energy scale simulation uncertainty is expected
to play an important role here and this uncertainty is not shown in the figure.
The peak at low mass is removed when using the loose settings (\figref{htt_perf_m}d).
This is due to an increase
of the $\mcut$ parameter from $30$ to $70\GeV$.
As described in \secref{htt_algo}, this parameter controls the
maximal mass of the substructure objects which enter the filtering procedure.
Lowering this mass implies a larger number of substructure objects.
A triplet of
substructure objects is combined to a top quark candidate. If $\mcut$ is too small
then a fat jet that contains a hadronically decaying top quark is split into
four or more substructure objects which share contributions from the three decay jets.
In that case, a triplet of these objects cannot reconstruct the top quark.
The value that was used in the \htt analysis described in \secref{htt}
is $\mcut=50\GeV$ which removes the peak in the top candidate mass distribution
as shown below.

\begin{figure}[hbt]
\centering
\subfigure[]{
   \includegraphics[width=0.45\textwidth,angle=0]{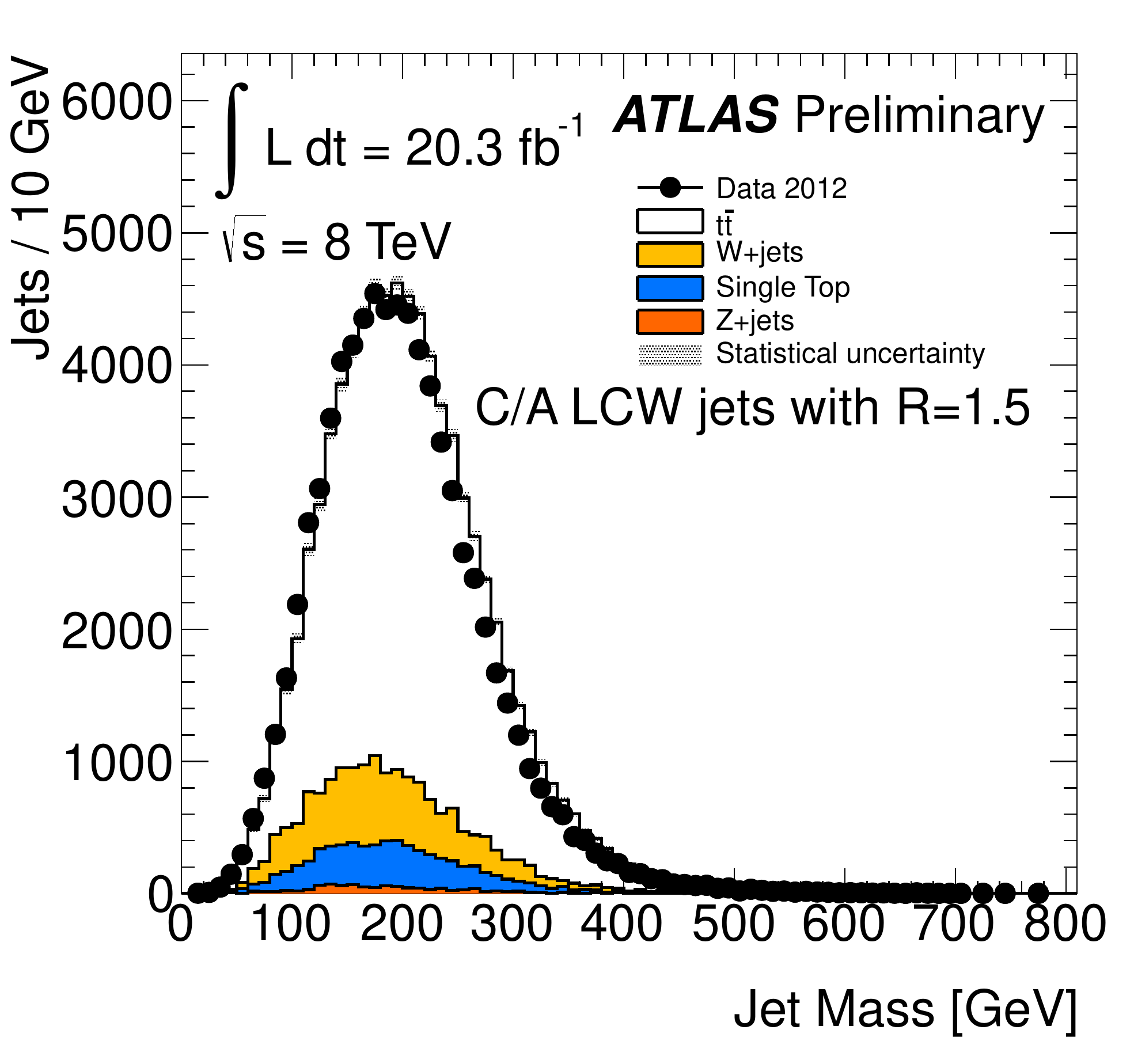}
}
\subfigure[]{
   \includegraphics[width=0.45\textwidth,angle=0]{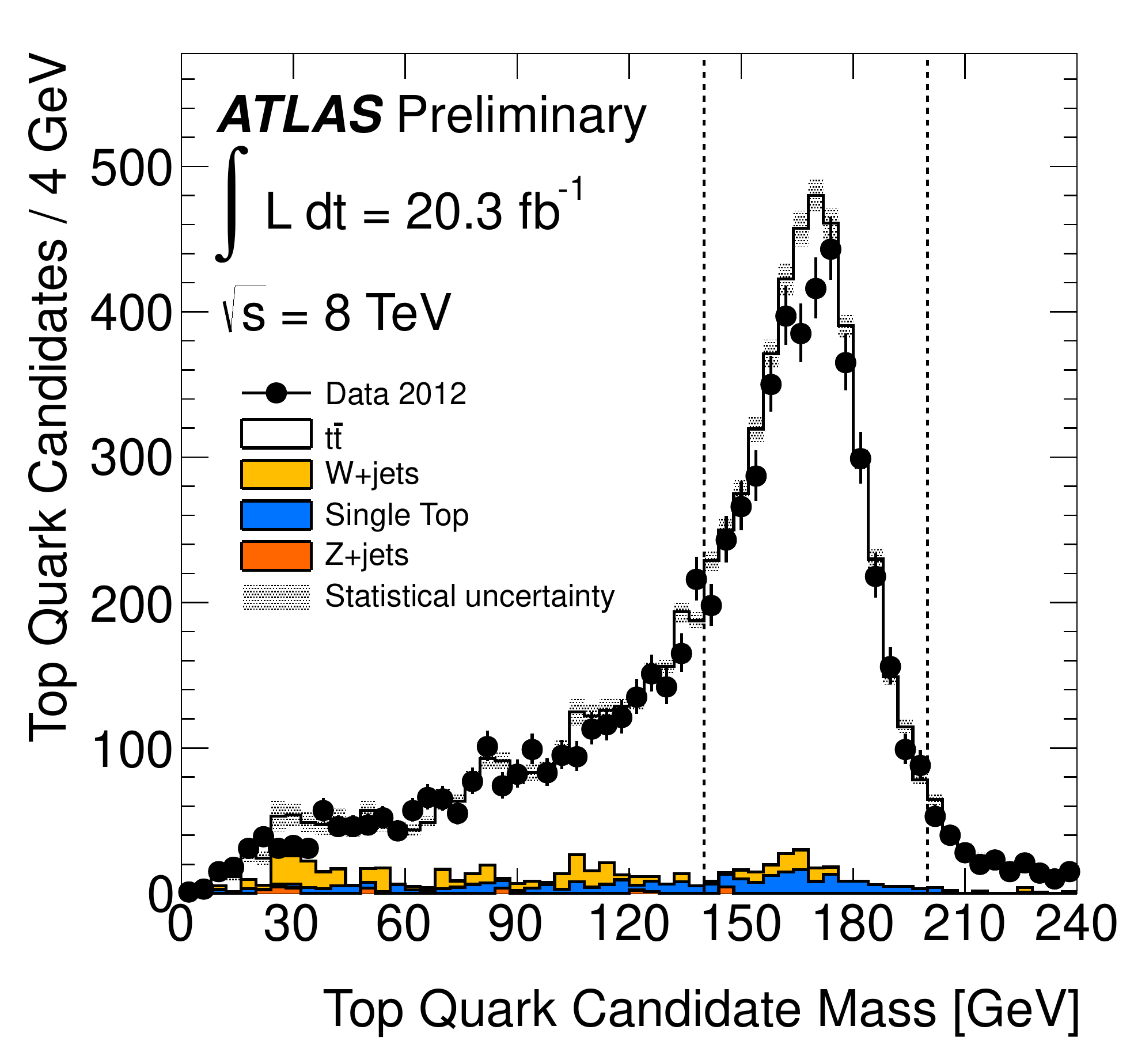}
}
\caption{Detector level distributions in a
selection of semileptonic \ttbar decays with a $b$-tag of (a) the mass of \ca $R=1.5$ jets with $\pt>200\GeV$ and
(b) the \htt candidate mass using the default \htt settings of \tabref{HTTsettings}.
The \ttbar simulation is from \powhegpythia, the \Wpjets events from \alpgenpythia,
and single-top production in the $s$-channel with a leptonically decaying \W boson is from \mcatnlo.
From~\cite{ATLAS-CONF-2013-084}.}
\label{fig:htt_perf_m_2012}
\end{figure}

\figref{htt_perf_m_2012}a shows the fat jet mass distribution for
ATLAS events from 2012 that pass the
semileptonic \ttbar event selection, including a $b$-tag to reduce the \Wpjets background.
The top quark purity of this sample of fat jets is $86\%$.
The fat jet mass exhibits a broad peak with a maximum at $180\GeV$.
The distribution predicted by the simulation is shifted towards higher masses.
Between $200$ and $350\GeV$ the shift is approximately $10\GeV$, which
corresponds to a relative change of $5\%$ at a mass of $200\GeV$.
As for 2011, this difference is covered by the typical fat jet simulation uncertainty~\cite{Aad:2013gja}.
The \htt candidate mass, reconstructed using the default settings listed in \tabref{HTTsettings},
is shown in \figref{htt_perf_m_2012}b. No peak is visible at low mass due to the use
of $\mcut = 50\GeV$. The top candidate mass is well described by the simulation.
The \htt increases the top quark purity to $98\%$ in the candidate mass window $140<m_t<200\GeV$.
The number of top quark candidates reconstructed in this mass window
in the data is $4210$. As proposed in~\cite{Schaetzel:2013vka}, the top quark
mass peak can be used to calibrate the \htt subjets and to evaluate the subjet
energy simulation uncertainty.

\begin{figure}[hbt]
\centering
\subfigure[]{
   \includegraphics[width=0.45\textwidth,angle=0]{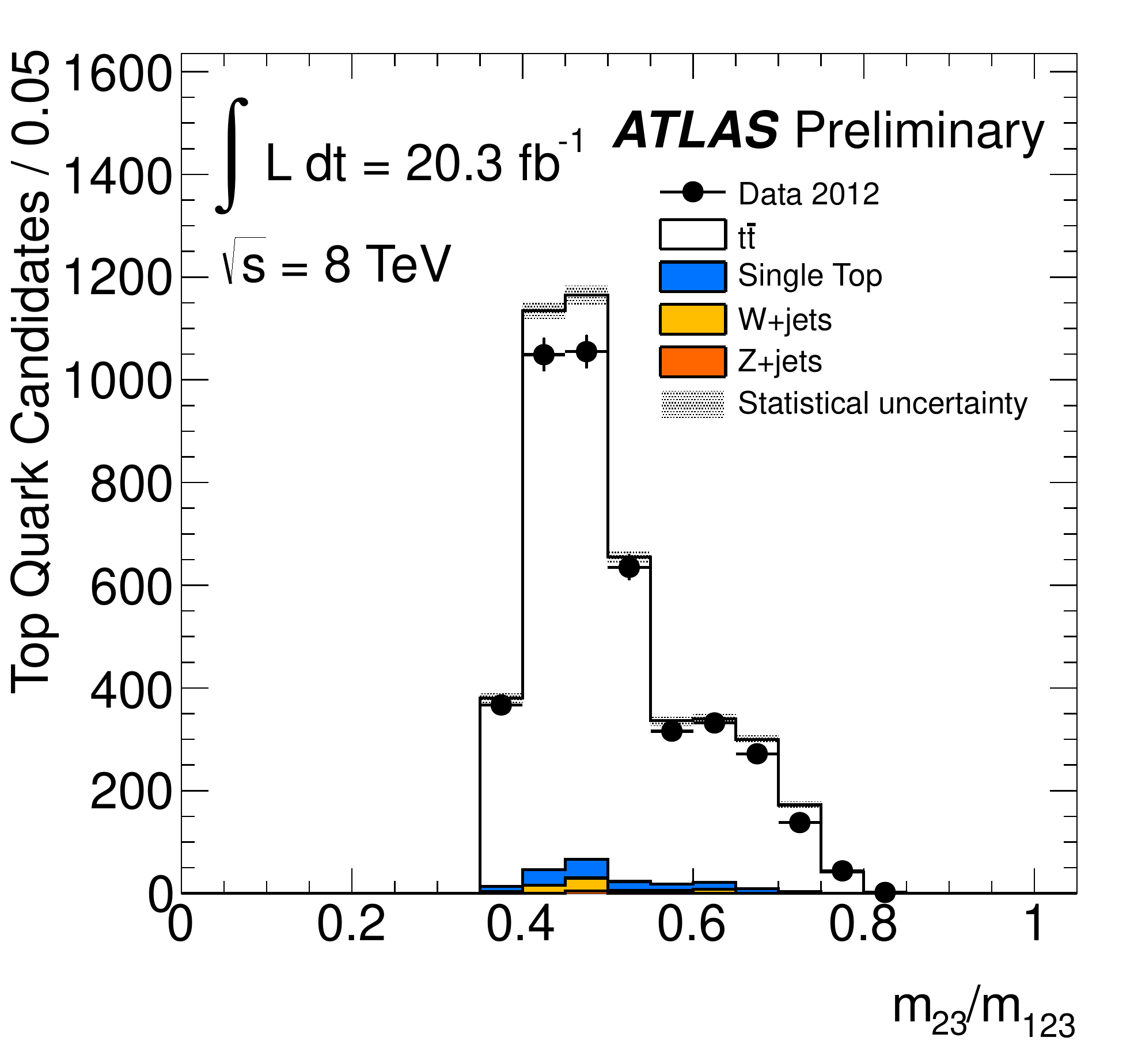}
}
\subfigure[]{
   \includegraphics[width=0.45\textwidth,angle=0]{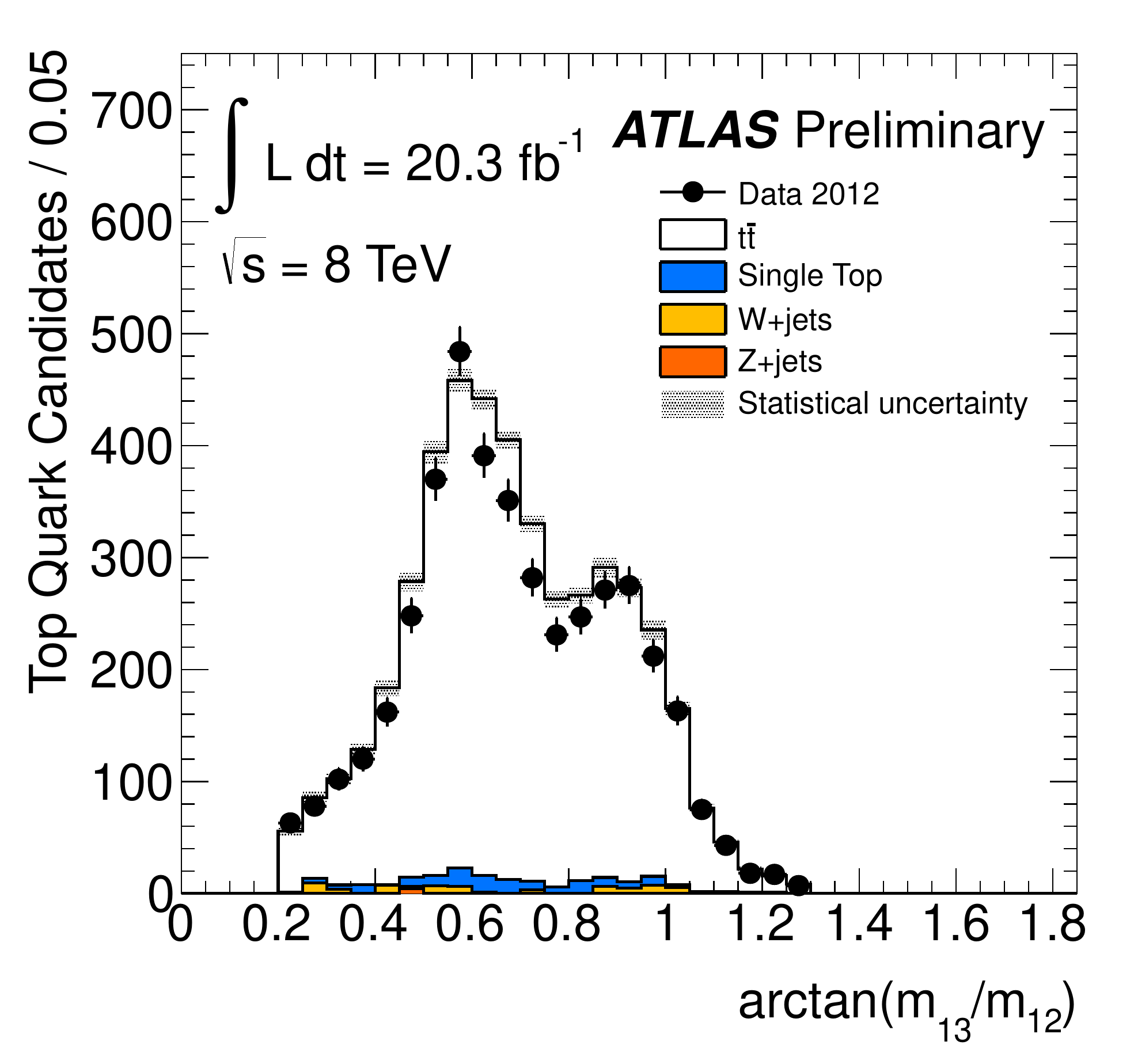}
}
\caption{Detector level distributions of the \htt substructure variables for top quark candidates
that pass the \htt procedure and have a mass between $140$ and $200\GeV$
in a selection of semileptonic \ttbar decays with a $b$-tag. (a) The subjet invariant mass
ratio $m_{23}/m_{123}$. (b) The quantity $\arctan(m_{13}/m_{12})$.
The quantities $m_{ij}$ are invariant masses of two of the three final subjets
in the \htt procedure which are
ordered in \pt such that $p_{{\rm T},1} > p_{{\rm T},2} > p_{{\rm T},3}$.
The \ttbar simulation is from \powhegpythia, the \Wpjets events from \alpgenpythia,
and single-top production in the $s$-channel with a leptonically decaying \W boson is from \mcatnlo.
From~\cite{ATLAS-CONF-2013-084}.}
\label{fig:HTTsubst}
\end{figure}

Distributions of substructure variables used by the \htt to apply kinematic cuts
on the three identified subjet are shown in
\figref{HTTsubst} for the top quark candidates with $140<m_t < 200\GeV$.
These variables are invariant mass ratios of combinations
of the three (exclusive) subjets identified in the last step of the \htt procedure (cf. \secref{htt_algo}).
For example, the variable $m_{23}$ is the invariant mass of the
subleading \pt and the sub-subleading \pt subjet, and the variable $m_{123}$ is the mass
of all three subjets combined. The ratio $m_{23}/m_{123}$ displayed in panel (a)
shows a peak at $m_W/m_t$. This indicates that in most of the cases the leading \pt subjet corresponds
to the $b$-quark (but this information is not used in the \htt).
The distribution is well described by
the simulation as is the quantity $\arctan(m_{13}/m_{12})$ in panel (b).

\begin{figure}[hbt]
\centering
\subfigure[]{
   \includegraphics[width=0.45\textwidth,angle=0]{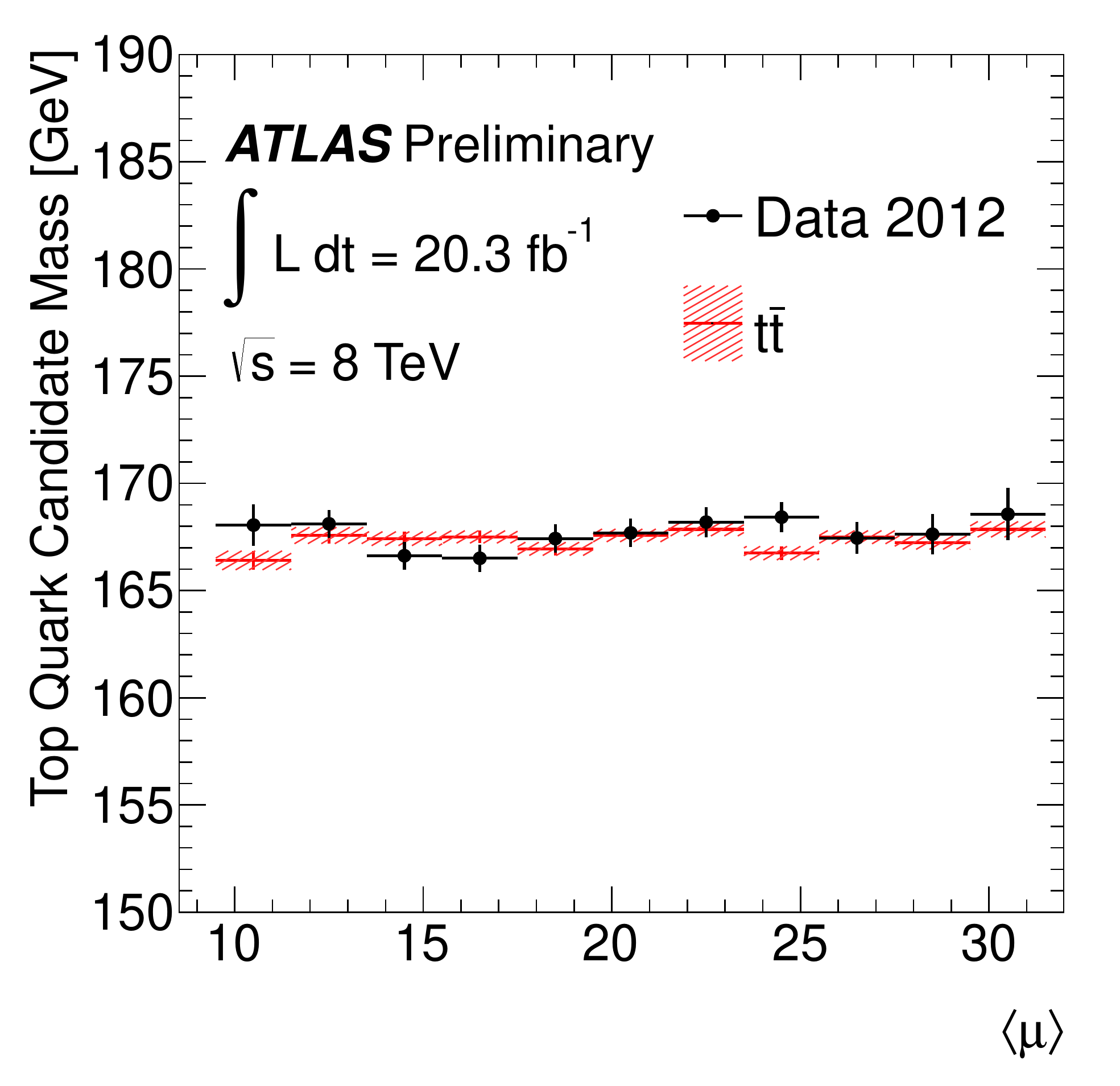}
}
\subfigure[]{
   \includegraphics[width=0.45\textwidth,angle=0]{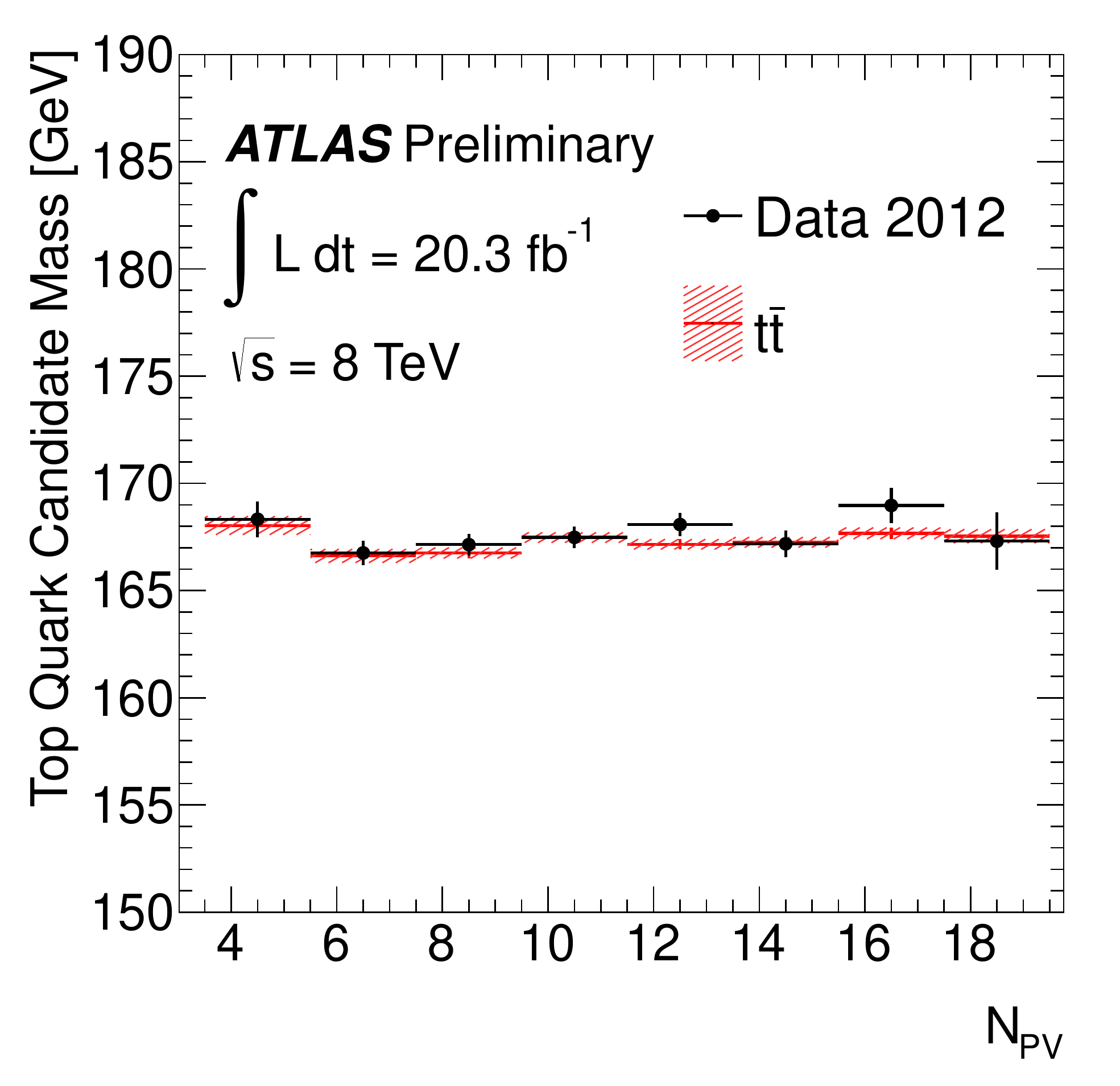}
}
\caption{\htt top candidate mass as a function of (a) the average number of interactions
per bunch-crossing $\avmu$ and (b) the number of reconstructed primary vertices in the
event. The mass is defined as the average mass in the window $140<m_t<200\GeV$
and the error bars correspond to the statistical uncertainty of the average mass.
The \ttbar simulation is from \powhegpythia.
From~\cite{ATLAS-CONF-2013-084}.}
\label{fig:HTTpileup}
\end{figure}

\figref{HTTpileup} shows the reconstructed top candidate mass for different
pile-up conditions. The average candidate mass in the
window $140<m_t<200\GeV$ is plotted as a function of the average number of interactions
per bunch-crossing $\avmu$ in panel (a) and as a function of the number of reconstructed
primary vertices \Npv in (b). The $\avmu$ and \Npv intervals are chosen such that
each average mass is calculated from at least 100 entries.
Within the statistical uncertainty
of the average mass of approximately $\pm1\GeV$, the reconstructed top
quark mass is not affected by
pile-up energy in the full accessible range up to $\mu=31$ and $\Npv=19$.
This is also seen for the generated semileptonic \ttbar events that have been
passed through a full ATLAS detector simulation under the same pile-up conditions.
The numerical value of the reconstructed top quark mass depends on the choice of the
mass window and is well predicted by the simulation.

\begin{figure}[hbt]
\centering
\subfigure[]{
   \includegraphics[width=0.45\textwidth,angle=0]{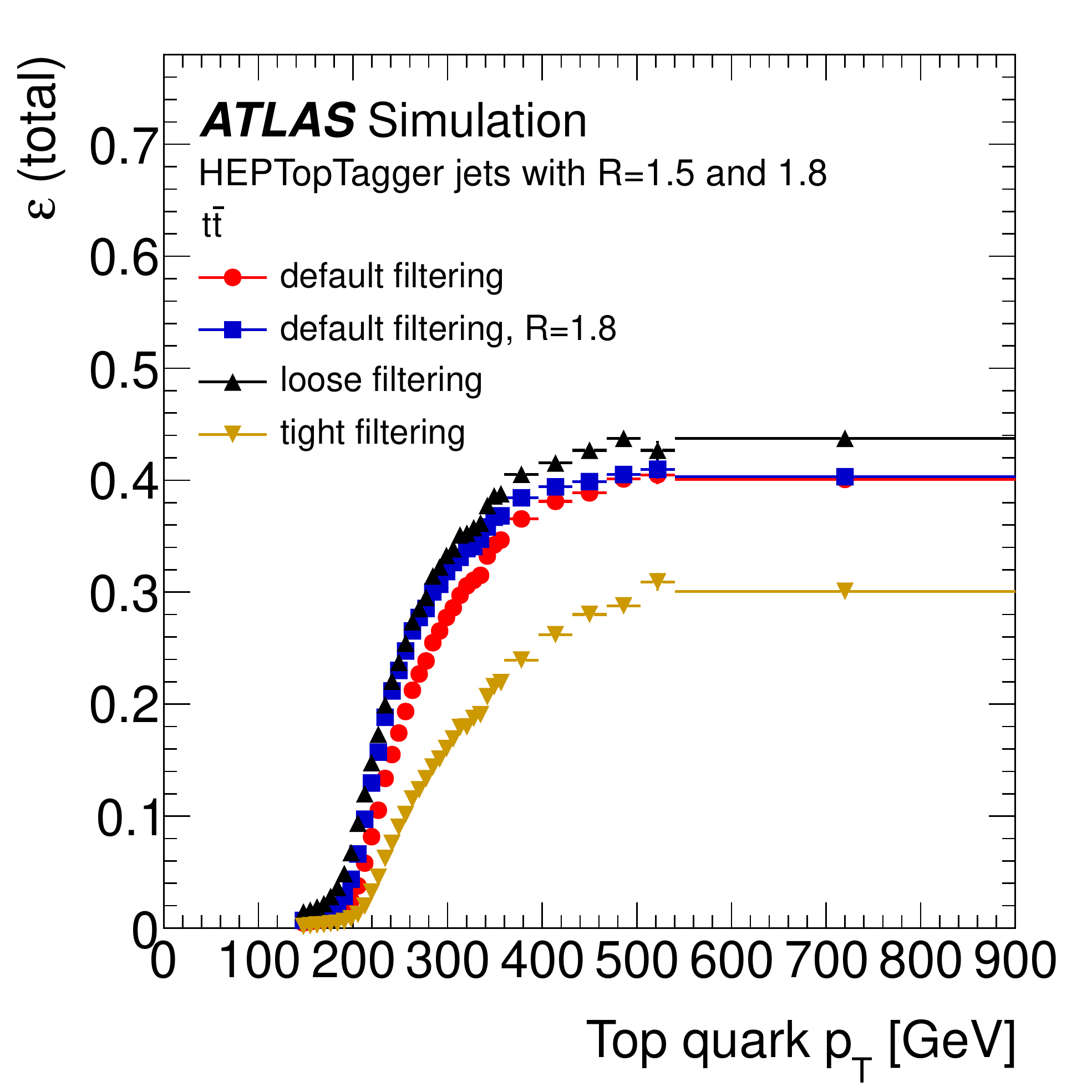}
} \\
\noindent
\subfigure[]{
   \includegraphics[width=0.45\textwidth,angle=0]{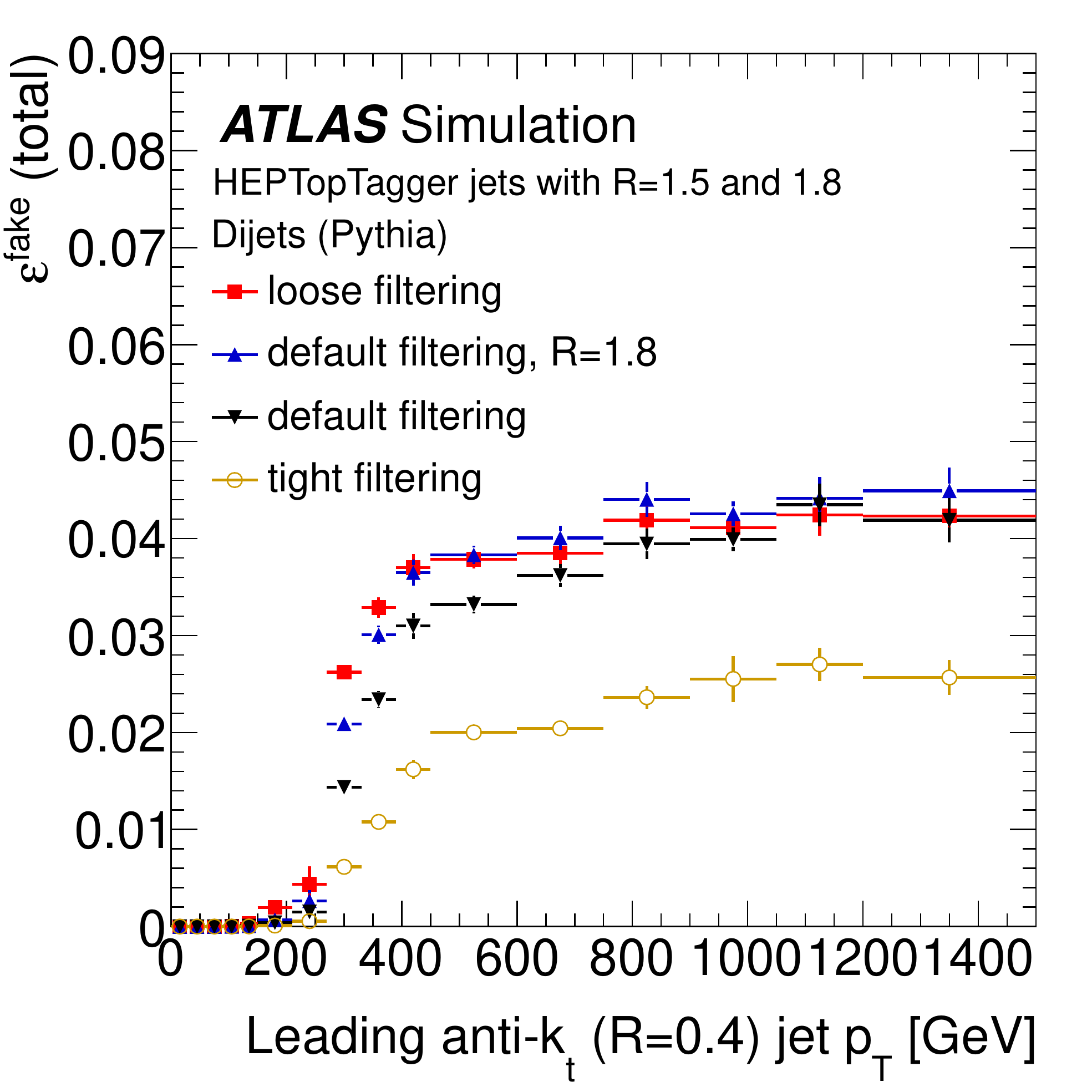}
}
\subfigure[]{
   \includegraphics[width=0.45\textwidth,angle=0]{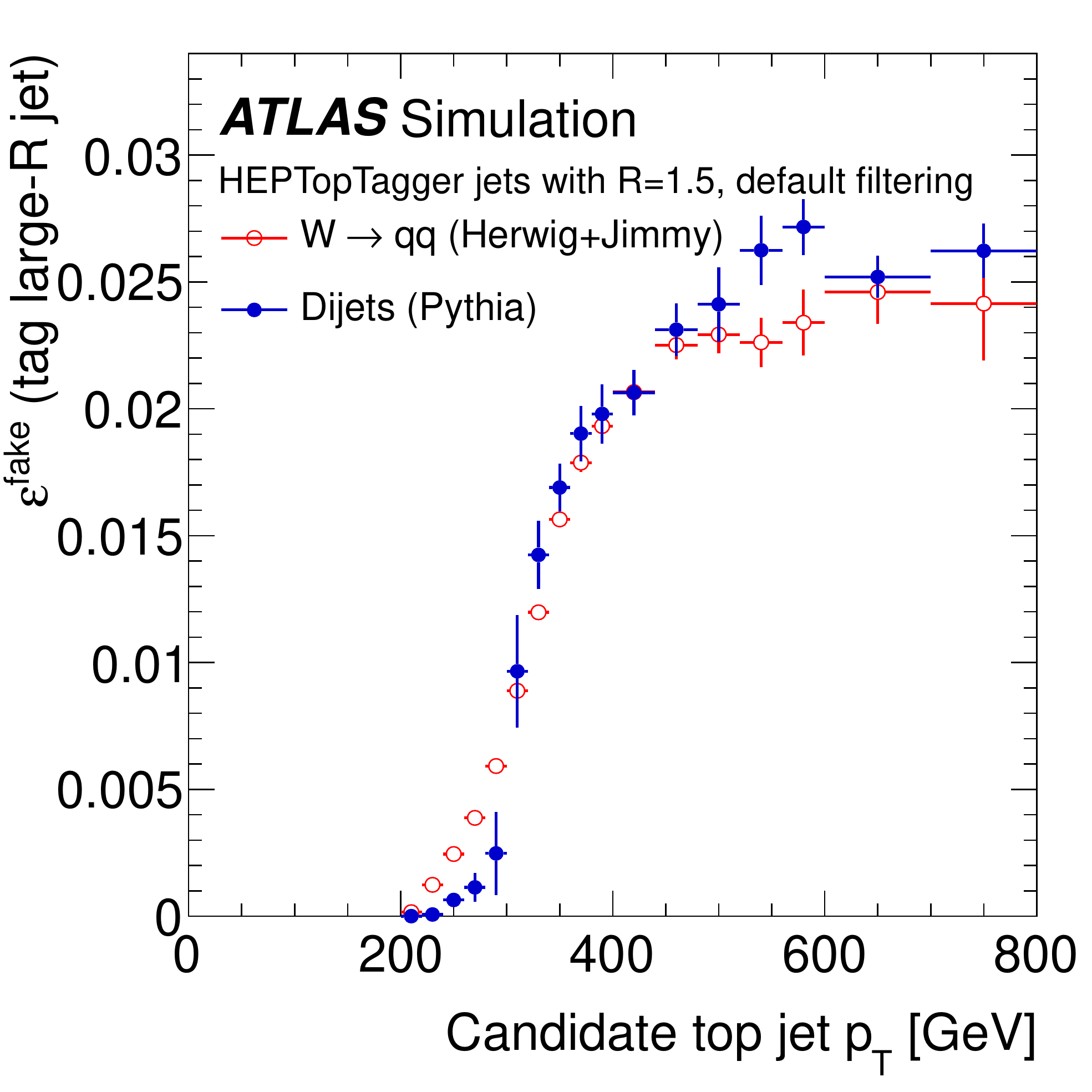}
}
\caption{Simulated \htt efficiency for different parameter settings in (a) semileptonic \ttbar decays (\mcatnlo),
(b) multijet events, and (c) multijet and \Wpjets events.
The efficiency in (a) is per top quark and shown as a function of the top quark \pt.
The fake rate in (b) is an event-level rate and is shown as a function of the leading \akt $R=0.4$ jet \pt in the event.
The fake rate in (c) is per fat jet and shown as a function of the fat jet \pt.
The {\em default filtering} efficiencies correspond to the default-30 settings in
\tabref{HTTsettings}. From~\cite{Aad:2013gja}.}
\label{fig:htt_eff}
\end{figure}

As shown above, the simulation describes the data well before and after applying the
\htt. Simulated \htt efficiencies are presented in \figref{htt_eff}.
The top quark tagging efficiency per top quark is shown in panel (a) as a function of the
top quark transverse momentum \ptt. The efficiency is a product of the efficiency to find a fat jet
in the event and the efficiency to tag the fat jet. The former efficiency is
close to $100\%$ for $\ptt>300\GeV$.
The efficiency shows a turn-on at $\ptt\approx200\GeV$ and rises to a plateau
of $\approx\!40\%$ for $\ptt > 500\GeV$. The turn-on is a geometric effect,
with more top quark decay products contained in the fat jet at larger \pt.
This is evident from a comparison with the efficiency obtained when using \ca $R=1.8$ fat jet which
is larger in the turn-on but equal to the $R=1.5$ efficiency in the plateau.
The plateau is given by the internal \htt inefficiencies associated with
the procedure that identifies the hard substructure (mass drop), the filtering, and
the kinematic selection. Different choices of the \htt parameters allow to
{\em tighten} or {\em loosen} the top quark selection to achieve an optimal
signal-to-background ratio (S/B). Typically tight settings are needed when the background
is large and the fake rate needs to be kept as small as possible.
Loose settings are possible if the background is small or has been reduced to a low level
through other selection cuts. In this case a high efficiency is preferred.
The tight and loose parameters are listed in \tabref{HTTsettings}.
The tight settings include more aggressive filtering, through the use of a smaller filtering radius
and a smaller number of kept filtered jets, and a smaller \W boson mass window.
For the loose settings, the filtering is relaxed (more jets are accepted) and the
\W boson mass window is widened. The tight \htt settings reduce the efficiency
to $30\%$ while the loose settings increase it to $44\%$.

The \htt fake rate is shown in \figref{htt_eff}b for multijet events. This is a
per-event rate with no event selection cuts other than the \htt applied. The rate is shown as a function
of the leading \akt $R=0.4$ jet \pt in the event to facilitate comparisons between
taggers that use different definitions of fat jets. The fake rate shows a turn-on
at $\approx\!200\GeV$ and rises to $\approx\!4\%$ at $800\GeV$ and continues
to rise slowly with \pt. This turn-on is an effect of increased hard QCD radiation
at large parton \pt which increases the chance to fake the three prong structure
of hadronic top quark decay.
With tight settings, the fake rate is reduced from $4\%$ to $2.5\%$ at high \pt.
The fake rate with loose settings differs from that obtained with the default
settings only in the turn-on.

\figref{htt_eff}c shows the fat jet tagging efficiency as a function
of the fat jet \pt for multijet events and \Wpjets events in which the \W boson
decays hadronically. For both types of events,
the fake rate is $\approx\!2.5\%$ for fat jet $\pt > 500\GeV$, while the turn-on
begins $\approx\!50\GeV$ earlier for \Wpjets events. This is expected because
$W\rightarrow qq$ decay provides a hard 2-prong structure
that is not present in most QCD jets.

%%%%%%%%%%%%%%%%%%%%%%%%%%%%%%%%%%%%%%%%%%%%%%%%%%%%%%%%%%%%%%%%%%%%%%%%%%%%%%%%

\subsection{Performance comparison of top tagging approaches}

{\em Parts of this section are taken from section~6 of~\cite{ATLAS-CONF-2013-084}, having been written by the author.}
\vglue0.8em
\noindent
This section compares the performance of the \htt with that of cuts on substructure variables
as presented in~\cite{ATLAS-CONF-2013-084}.
The comparison is made in the context of searches for resonances in
the \ttbar invariant mass spectrum using the benchmark model
of a \Zprime boson with a mass of $1.75\TeV$. This mass is just above the current
published exclusion limit~\cite{Aad:2013nca} discussed in \secref{ljets}.

The boosted top quark tagging efficiency and the rejection of jets arising from light quarks or gluons
are determined from simulation. High \pt top quarks are obtained from
\pythia~8 \Zprime events with $m_{\rm Z'} = 1.75\TeV$, in which the \Zprime boson decays
exclusively to \ttbar.
To estimate the background rejection, a simulated inclusive jet sample from \pythia~8 is used.

The direct comparison at the fat jet level is difficult because of different
fat jet definitions: the \htt uses \ca fat jets while the variables are calculated for trimmed \akt jets.
The cleanest comparison would be at the event level
but the approach taken in~\cite{ATLAS-CONF-2013-084} is to geometrically match the fat jets.
Samples of truth level particle jets are defined for signal and background
by requiring at least one \ca $R=1.2$ jet
and at least one trimmed \akt $R=1.0$ jet, both with $\pt > 150\GeV$ and $|\eta| < 1.2$.
The trimming parameters are $\fcut = 5\%$ and $\Rsub = 0.3$.
For the signal, the truth level requirements include a hadronically decaying top quark
with $\pt > 150 \GeV$ and $|\eta| < 1.2$ within $\DeltaReta < 0.75\times R$ of
the leading \pt fat jet of either type.
At the reconstruction level, events are required to contain at least one \ca $R=1.2$ jet
and at least one trimmed \akt $R=1.0$ jet, both with $\pt > 550\GeV$ and $|\eta| < 1.2$.
Both reconstructed fat jets are required to be matched within $\DeltaReta < 0.75\times R$
to the corresponding truth level particle jet of the same algorithm and $R$ parameter.
The reconstructed \ca and \akt fat jets are also required to be within $\DeltaReta <0.75$
of one another to ensure that the performance of the different taggers is measured
using input objects from the same top quark decay.

\begin{figure}[hbt]
\centering
\includegraphics[width=0.45\textwidth,angle=0]{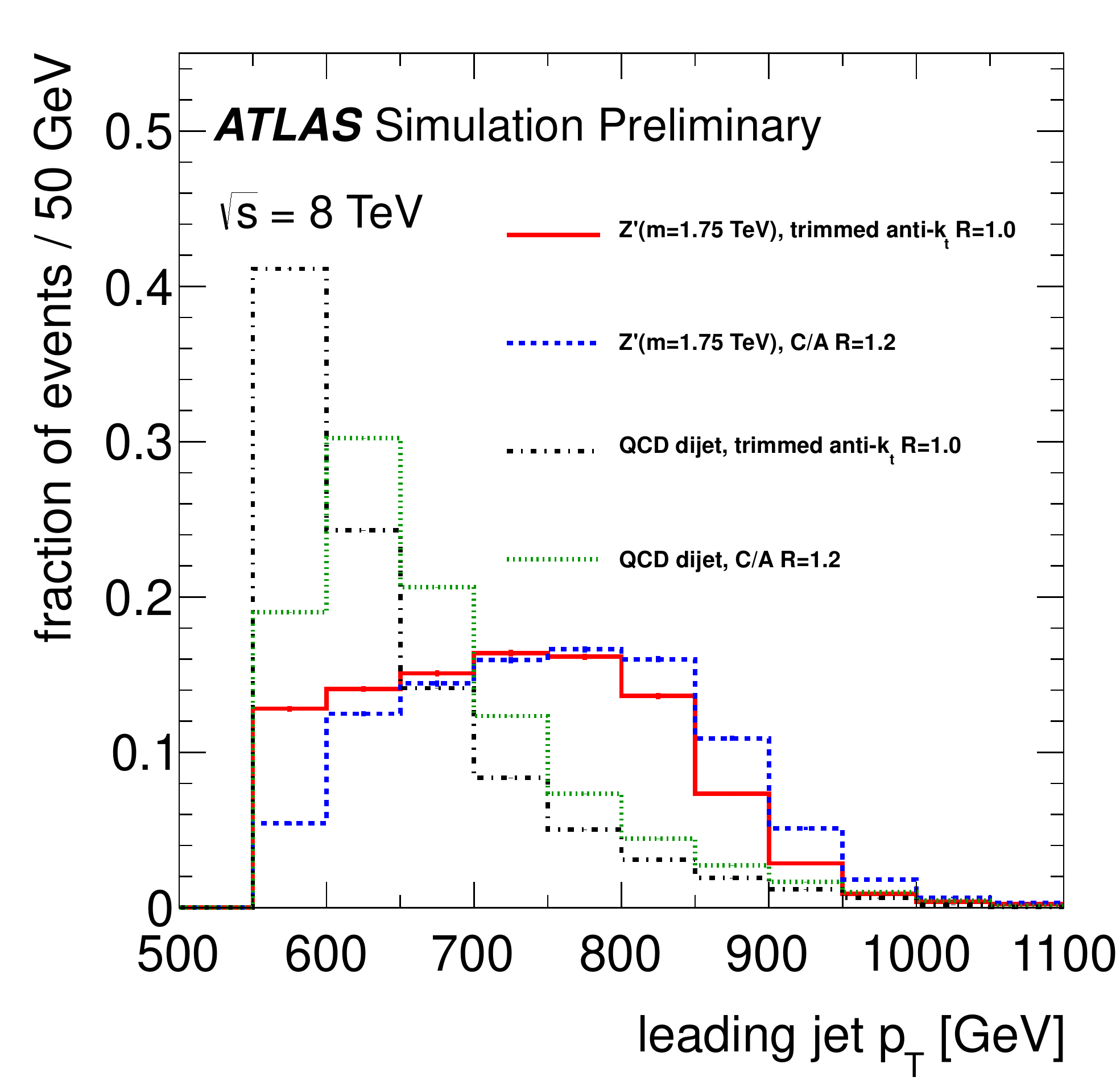}
\caption{The fat jet \pt spectra for signal ($\Zprime \rightarrow \ttbar$) and background
(multijets) that are used for the tagging comparison. The events are generated with \pythia~8.
From~\cite{ATLAS-CONF-2013-084}.}
\label{fig:fatjetinput}
\end{figure}

The shape of the detector level fat jet \pt distribution is shown
in \figref{fatjetinput} for signal and background. For the signal, the spectrum
reflects the Jacobian peak structure of the top quark \pt distribution and peaks
near $m_{\rm Z'}/2$. The \pt of the \ca jets is greater on average than the
\pt of geometrically matched trimmed \akt jets.

\begin{figure}[hbt]
\centering
\includegraphics[width=0.8\textwidth,angle=0]{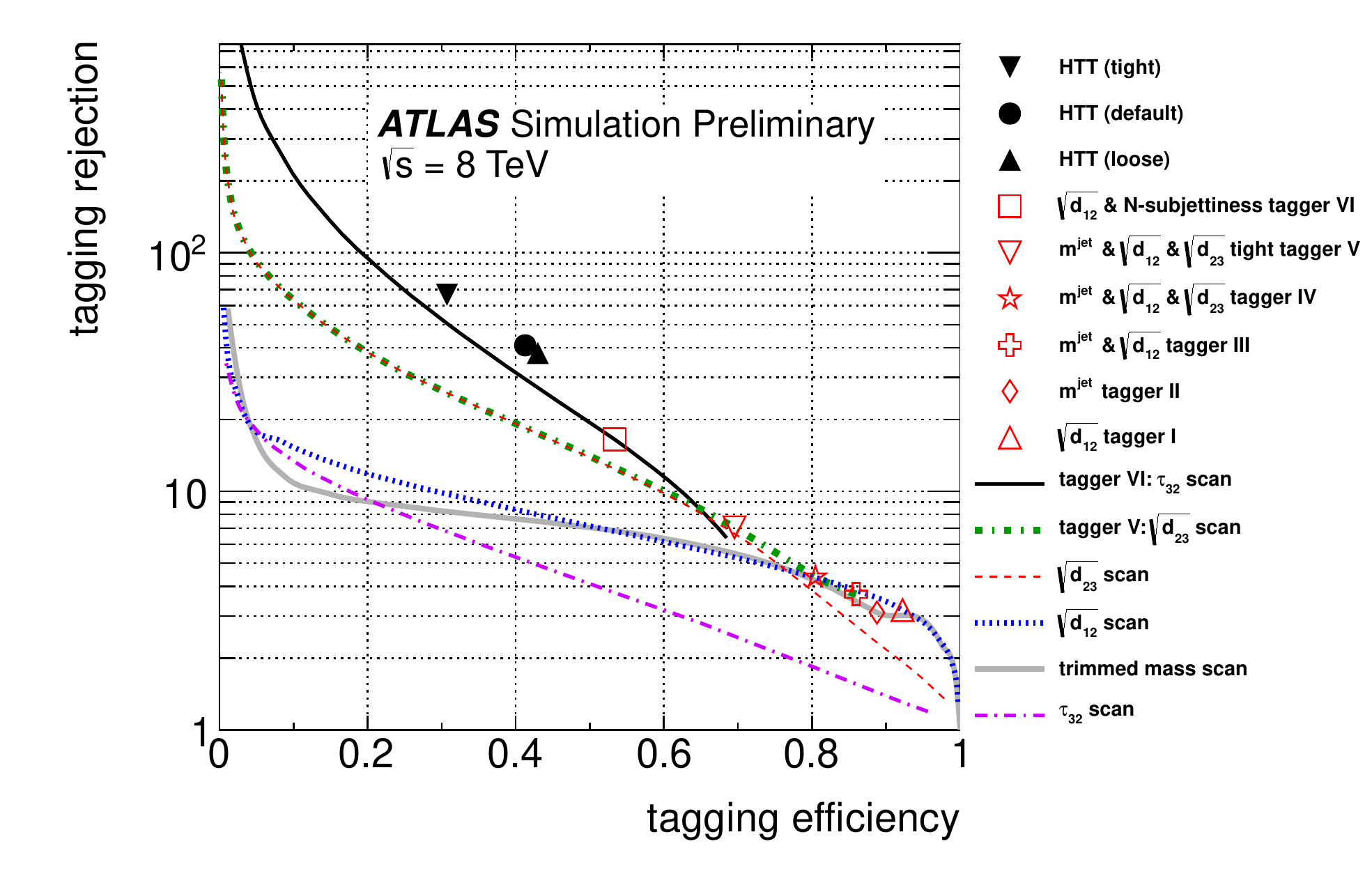}
\caption{Comparison of the simulated fat top jet tagging efficiency and fat light quark/gluon
jet rejection (defined as the inverse of the fake rate) of different taggers.
The top jets originate from the decay of a \Zprime{} boson of mass $1.75\TeV$ and
the background is given by multijet events. All events are generated with \pythia~8.
The fat jet \pt is required to be greater than $550\GeV$.
All substructure taggers and
scans use trimmed \akt $R=1.0$ jets. The \htt (points labelled `HTT') uses \ca $R=1.2$ jets that are geometrically
matched to the \akt jets. From~\cite{ATLAS-CONF-2013-084}.}
\label{fig:roc}
\end{figure}

Six different substructure variable based taggers
are compared, all based on trimmed \akt $R=1.0$ jets:
\begin{itemize}
\item substructure tagger I: $\sqrt{d_{12}} > 40 \GeV$ (labelled `$\sqrt{d_{12}}$ tagger I' in \figref{roc})
\item substructure tagger II: trimmed \akt $R=1.0$ mass $\mjet > 100\GeV$  (labelled `$\mjet$ tagger II')
\item substructure tagger III: $\mjet > 100\GeV$, $\DOneTwo > 40\GeV$ (labelled `$\mjet$ \& $\sqrt{d_{12}}$ tagger III')
\item substructure tagger IV: $\mjet > 100\GeV$, $\DOneTwo > 40\GeV$, $\DTwoThr > 10 \GeV$ \\(labelled `$\mjet$ \& \DOneTwo \& \DTwoThr tagger IV')
\item substructure tagger V: $\mjet > 100\GeV$, $\DOneTwo > 40\GeV$, $\DTwoThr > 20 \GeV$ \\(labelled `$\mjet$ \& \DOneTwo \& \DTwoThr tight tagger V')
\item substructure tagger VI: $\DOneTwo > 40\GeV$, $0.4 < \tau_{21} < 0.9$, $\tau_{32} < 0.65$ \\(labelled `\DOneTwo \& N-subjettiness tagger VI')
\end{itemize}

Tagger III has been used in searches for \ttbar resonances in the semileptonic
channel~\cite{Aad:2013nca,ATLAS-CONF-2013-052}. The others are compared
to study the performance of different combinations of substructure variables.
The \htt performance is evaluated using the default, tight, and loose parameters specified in \tabref{HTTsettings}.
The \ca fat jet radius parameter $R$ was lowered from the nominal $1.5$, for which the \htt was designed and optimised,
to $1.2$. One reason is that the large \pt of the top quarks under study leads to a
collimation of the decay products to a smaller cone. The second motivation is to
make the radius parameter more similar to the one used for the trimmed fat
jets.

\figref{roc} shows a comparison of the efficiency $\varepsilon$ for tagging fat jets that
contain a top quark and the rejection of fat jets originating from
hard light quarks or gluons. The rejection is defined as the
inverse of the fake rate.
The best rejection of all studied taggers is obtained with the \htt when using
the tight settings, which, however, also results in the smallest efficiency.
Loosening the \htt parameters improves the efficiency at a cost in rejection.
The performances obtained from the default and loose parameters
differ only slightly.

The substructure tagger III has a high efficiency and a small rejection. Such
a working point is typically used in analyses in which the background has already
been reduced by measures not related to jet substructure, such as the requirement of
a final state lepton.

Tagger I is tagger III without the requirement on the trimmed fat jet mass
and yields a higher efficiency but worse rejection. The cut on $\DOneTwo$
has been varied from the nominal $40\GeV$ for tagger I, resulting in the
continuous curve labelled `$\DOneTwo$ scan', which crosses the points for taggers III and IV.
This indicates that,
for the signal and background fat jets under study here,
the mass cut can be dropped from tagger III without compromising its performance.
Adding a cut $\DTwoThr>10\GeV$ (as in tagger IV) also does not help in separating
signal from background. A larger cut value of $20\GeV$ in combination with
the mass cut, however, increases the rejection as can be seen from the position
of the tagger V point in relation to the curve.
The variation of the \mjet cut in tagger II (`trimmed mass scan') gives a
performance very similar to the one from the $\DOneTwo$ cut, indicating that
the two variables are strongly correlated.

Also investigated was
the impact of a standalone cut on $\DTwoThr$. The performance that results
from varying the minimal required $\DTwoThr$ (`$\DTwoThr$ scan') offers
better rejection than a cut on $\DOneTwo$ for $\varepsilon \lesssim 75\%$,
almost reaching that of
tagger V which uses a combination of cuts on $\DTwoThr$, $\DOneTwo$, and the mass,
suggesting that the latter two cuts can be dropped at minimal expense.
The $\DTwoThr$ cut in tagger V has been varied while keeping the other cuts fixed (`tagger V: $\DTwoThr$ scan'),
resulting in a performance which differs from that of $\DTwoThr$
only for $\varepsilon \gtrsim 70\%$ where it gives better rejection.
Variations of the other cuts in tagger V
(one at a time, fixing the other cut-offs to their nominal values) do not improve
the performance compared to the $\DTwoThr$ variation.

A standalone cut on the $N$-subjettiness variable $\tau_{32}$ (`$\tau_{32}$ scan')
yields a significantly worse performance than cuts on splitting scales.
Combining $N$-subjettiness and splitting scale information (tagger VI), however,
gives the best performance of all substructure variable based taggers for
$\varepsilon \lesssim 60\%$. By varying the $\tau_{32}$ cut in tagger VI
(`tagger VI: $\tau_{32}$ scan'), rejections close to those of the \htt can be achieved:
for $\varepsilon \approx 40\%$ ($30\%$) the rejection is
$\approx 0.75$ ($0.7$) times that of the \htt.

For the present study, the performance of taggers based on substructure variables
is driven by cuts on splitting scales. For $\varepsilon \gtrsim 75\%$,
a standalone cut on the first splitting scale $\DOneTwo$ gives the best performance.
At lower efficiencies, the second splitting scale $\DTwoThr$ is the variable
that in a standalone cut best separates signal from background. Adding
$N$-subjettiness information increases the performance, whereas an additional
cut on the trimmed fat jet mass does not help significantly.

The studied substructure variable taggers achieve higher efficiencies than
the \htt. Compared at the same efficiency, however, they yield a smaller rejection,
making the \htt the algorithm of choice for analyses in which the multijet background is
large. Finding the best performance of the substructure taggers that are based on
several variables requires a multi-dimensional variation of the cut values.
The top quark taggers presented here provide a broad spectrum of
working points, and the best choice for any particular analysis depends on the
needed efficiency and rejection.

%%%%%%%%%%%%%%%%%%%%%%%%%%%%%%%%%%%%%%%%%%%%%%%%%%%%%%%%%%%%%%%%%%%%%%%%%%%%%%%%

\subsection{Top tagging in high jet multiplicity environments}
\label{sec:highmultstudy}
In this section the performance of top tagging approaches
is studied in LHC events with a large number of jets.
Such an environment is typically encountered in scenarios of New Physics models
where new massive particles decay with large hadronic branching fractions.
For the present study, $\sqrt{s} = 14\TeV$ $pp$ collisions are used and small-$R$
jets are constructed using the \akt algorithm with $R=0.4$.
Fat jets are constructed using the \ca algorithm with $R=1.5$,
and the efficiency and rejection of the \htt is compared with those of
cuts on the fat jet mass, \pt, and
the multiplicity of hard substructure objects inside the fat jet.
In a real analysis, these substructure techniques would be combined
with other requirements to increase the signal sensitivity. Many such requirements,
e.g., cuts on \ETmiss and $b$-tagging, factorise because the
\htt does not use the corresponding variables and the fat jet variables are not
or only weakly correlated with those variables. The approach taken in this study is therefore
to not consider other cuts and to look at the significance improvement rather
than absolute values of $S/\sqrt{B}$.

\begin{figure}[hbt]
\centering
\subfigure[]{
   \includegraphics[width=0.40\textwidth,angle=0]{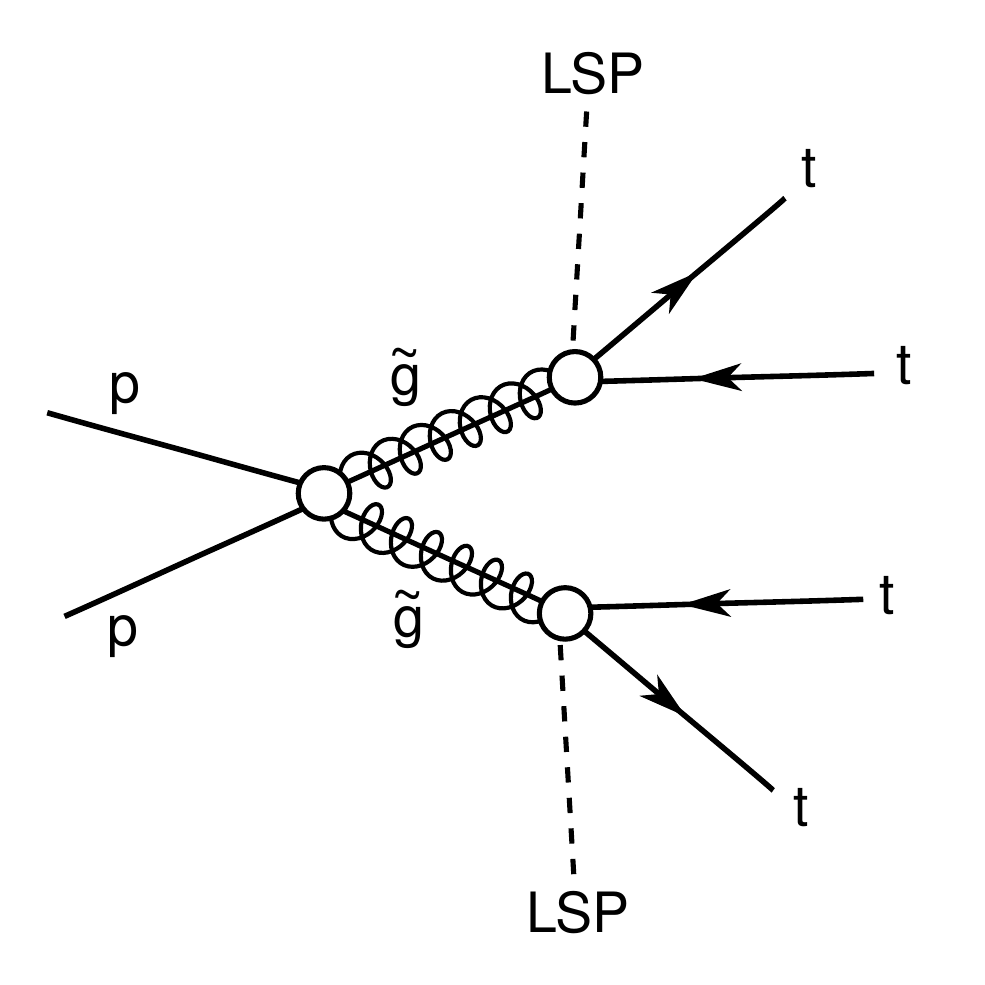}
}
\subfigure[]{
   \includegraphics[width=0.45\textwidth,angle=0]{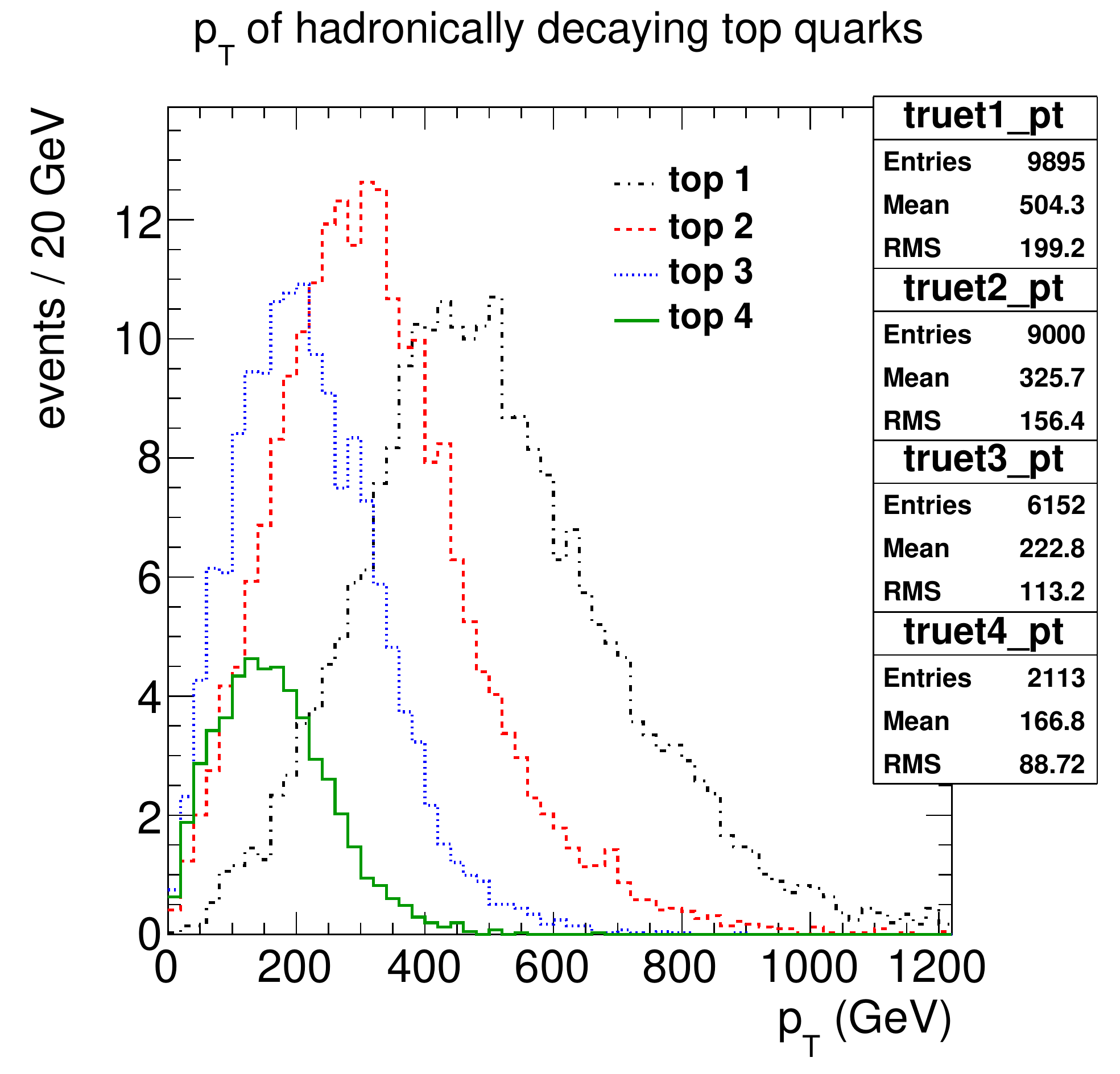}
}
\caption{(a) Feynman diagram for the process $pp \rightarrow \tilde{g}\tilde{g} \rightarrow 4t \, 2\lsp$.
(b) Top quark \pt for $10\invfb$ of $\sqrt{s}=14\TeV$ $pp$ collisions with $m_{\tilde{g}} = 1.3\TeV$, $m_{\tilde{t}} = 2.5\TeV$, $m_{\lsp} = 100\GeV$.
The events are generated with \madgraphherwigpp.}
\label{fig:gluinogluino}
\end{figure}

A supersymmetric extension of the SM ($R$-parity conserving SUSY) is used as
the benchmark model for the comparison. Gluinos are pair-produced
and decay to four top quarks and two neutralinos, as
shown in \figref{gluinogluino}a.
Each gluino decays according to
$\tilde{g} \rightarrow  \bar{t} \, \tilde{t} \rightarrow \bar{t} t \lsp$,
in which \lsp is the lightest supersymmetric particle (LSP).
The masses chosen for the study are
$m_{\tilde{g}} = 1.3\TeV$, $m_{\tilde{t}} = 2.5\TeV$, and $m_{\lsp} = 100\GeV$
with the top squark $\tilde{t}$ from gluino decay being off-shell. These signal events
are generated using \madgraphherwigpp.

\begin{figure}[hbt]
\centering
\subfigure[signal]{
   \includegraphics[width=0.45\textwidth,angle=0]{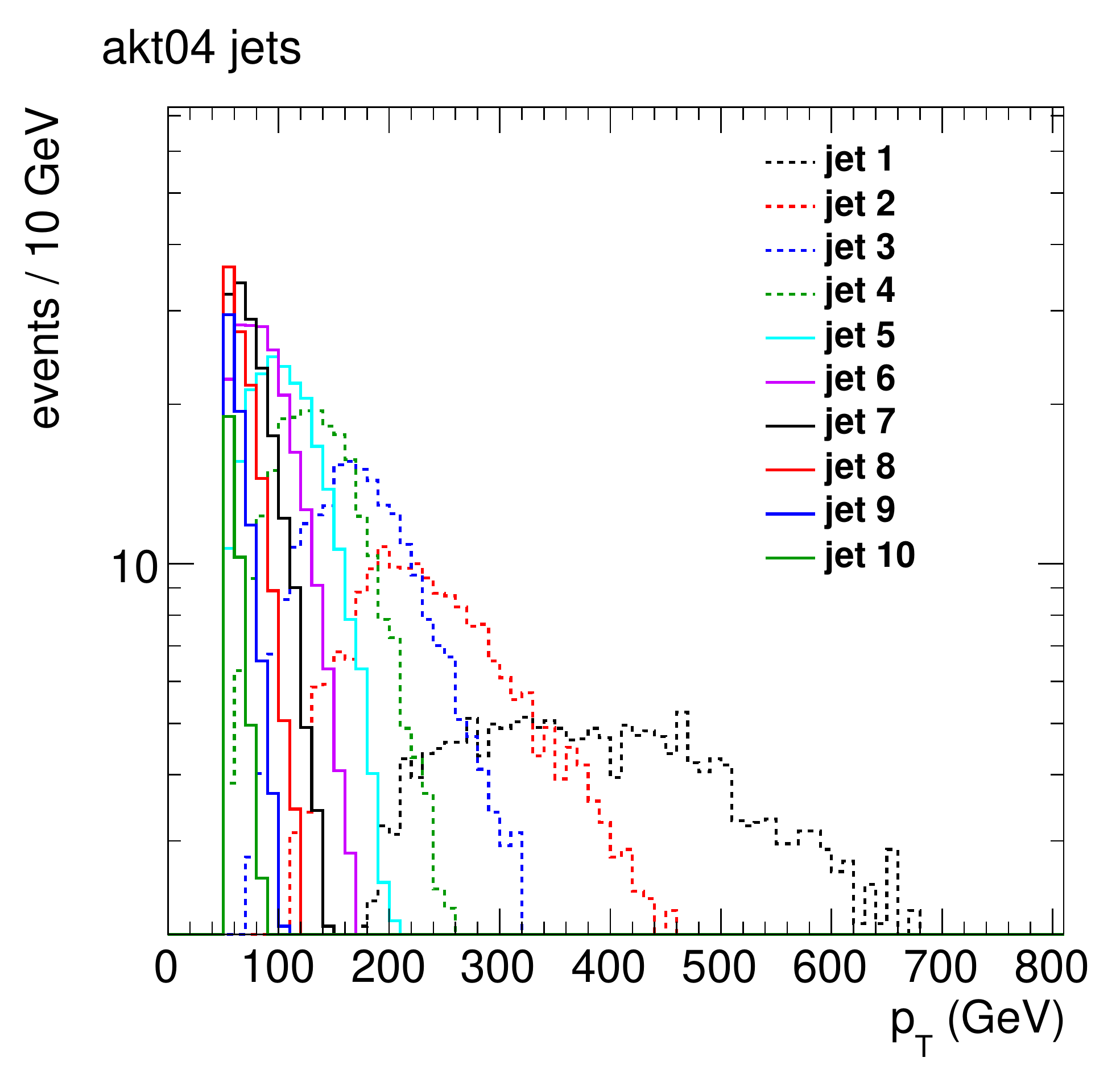}
}
\subfigure[background]{
   \includegraphics[width=0.45\textwidth,angle=0]{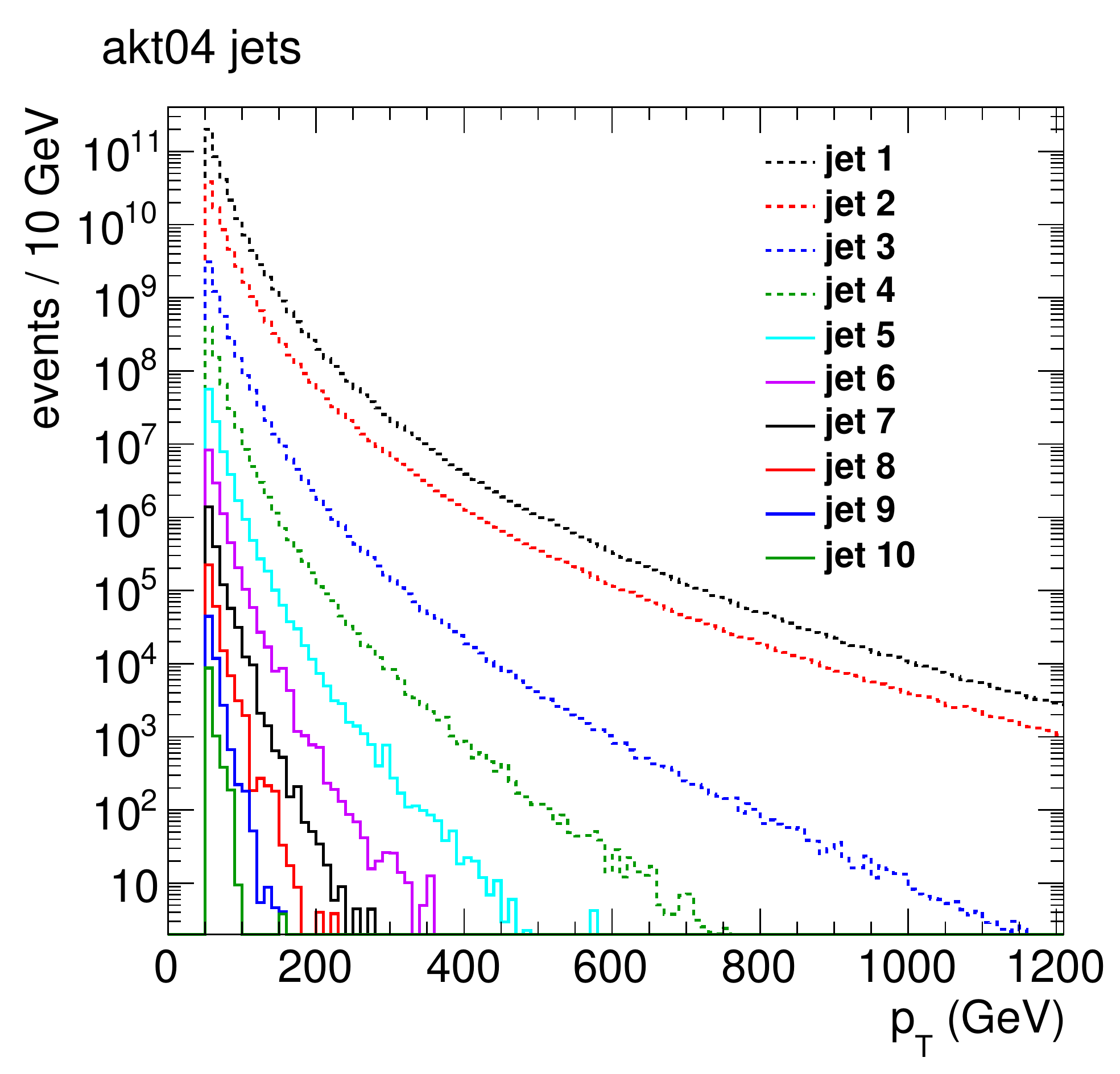}
}
\caption{The \pt distributions of the ten leading \pt \akt $R=0.4$ jets in events of (a) the SUSY benchmark model (from \madgraphherwigpp)
and (b) the multijet background (from \pythia~8). The events
correspond to $10\invfb$ of $\sqrt{s}=14\TeV$ $pp$ collisions
and
have been passed through the \delphes simulation of the ATLAS detector.}
\label{fig:susyjets}
\end{figure}

The \pt spectrum of the top quarks is shown in \figref{gluinogluino}b were the
event counts correspond to $10\invfb$.
The quarks are moderately boosted, with the average leading \pt being $504\GeV$
and the average subleading \pt being $326\GeV$.
The events are processed with the \delphes simulation of the ATLAS detector.
The \pt distribution of \akt $R=0.4$ jets reconstructed in the events is shown in \figref{susyjets}a.

The background for the study are multijet events generated with \pythia~8.175
and the \pt spectra are shown in \figref{susyjets}b. A leading-order $2\!\!\rightarrow\!\!2$ parton generator,
supplemented with a parton shower to approximate higher orders particle emission,
may not be the optimal tool to simulate the multijet background.
Multileg generators like \sherpa (up to 6 partons in the matrix element)
and \madgraph (up to 4 partons) might
be better choices although both also rely on the parton shower for higher multiplicity.
The best approach is to take the multijet distributions from data as done
in the high jet multiplicity ATLAS analysis discussed in \secref{searchsusy}.
The simulations used in this study do not include pile-up. No grooming techniques
are therefore used to clean the fat jets.

\begin{figure}[hbt]
\centering
\subfigure[signal]{
   \includegraphics[width=0.48\textwidth,angle=0]{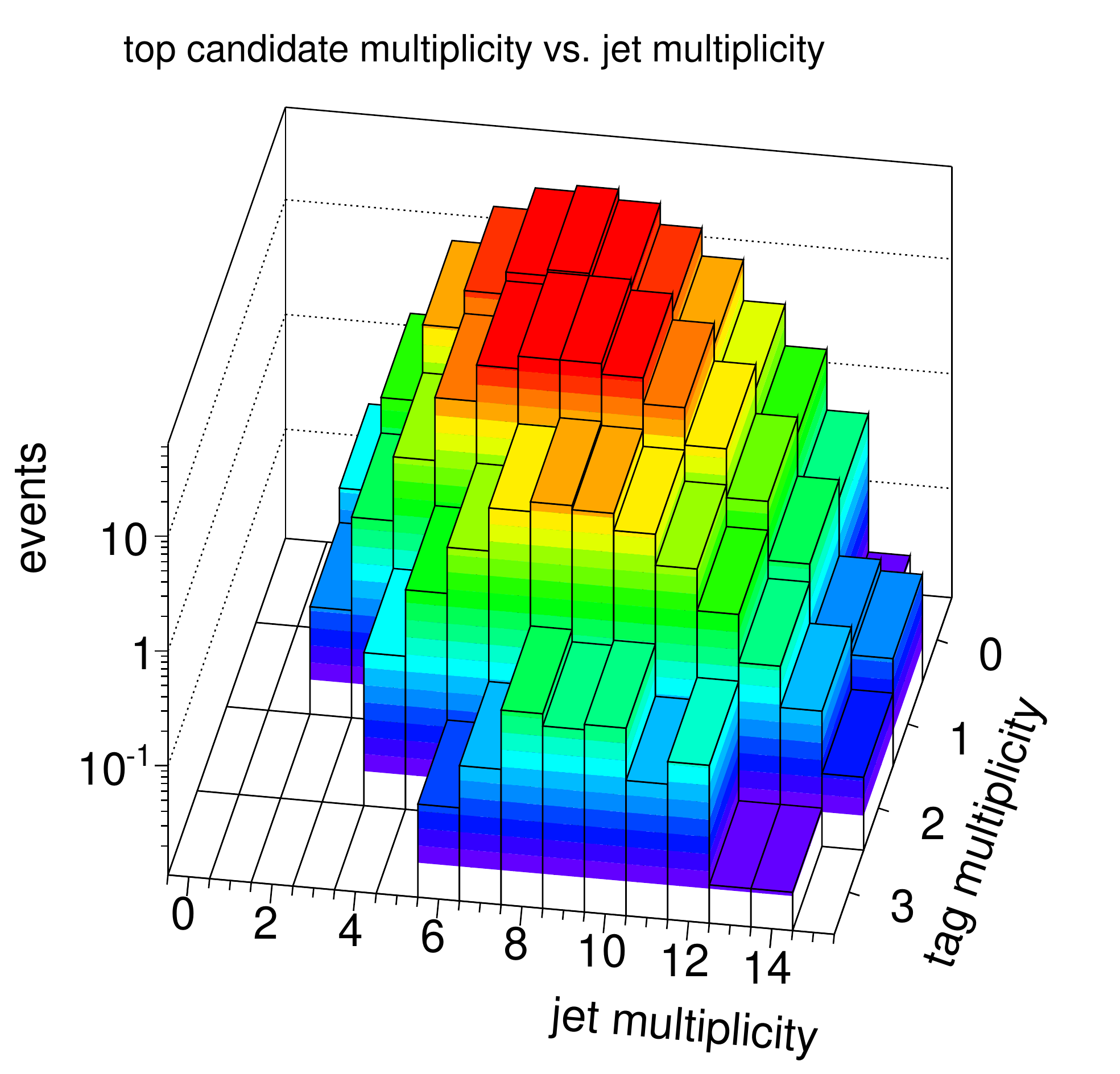}
}
\subfigure[background]{
   \includegraphics[width=0.48\textwidth,angle=0]{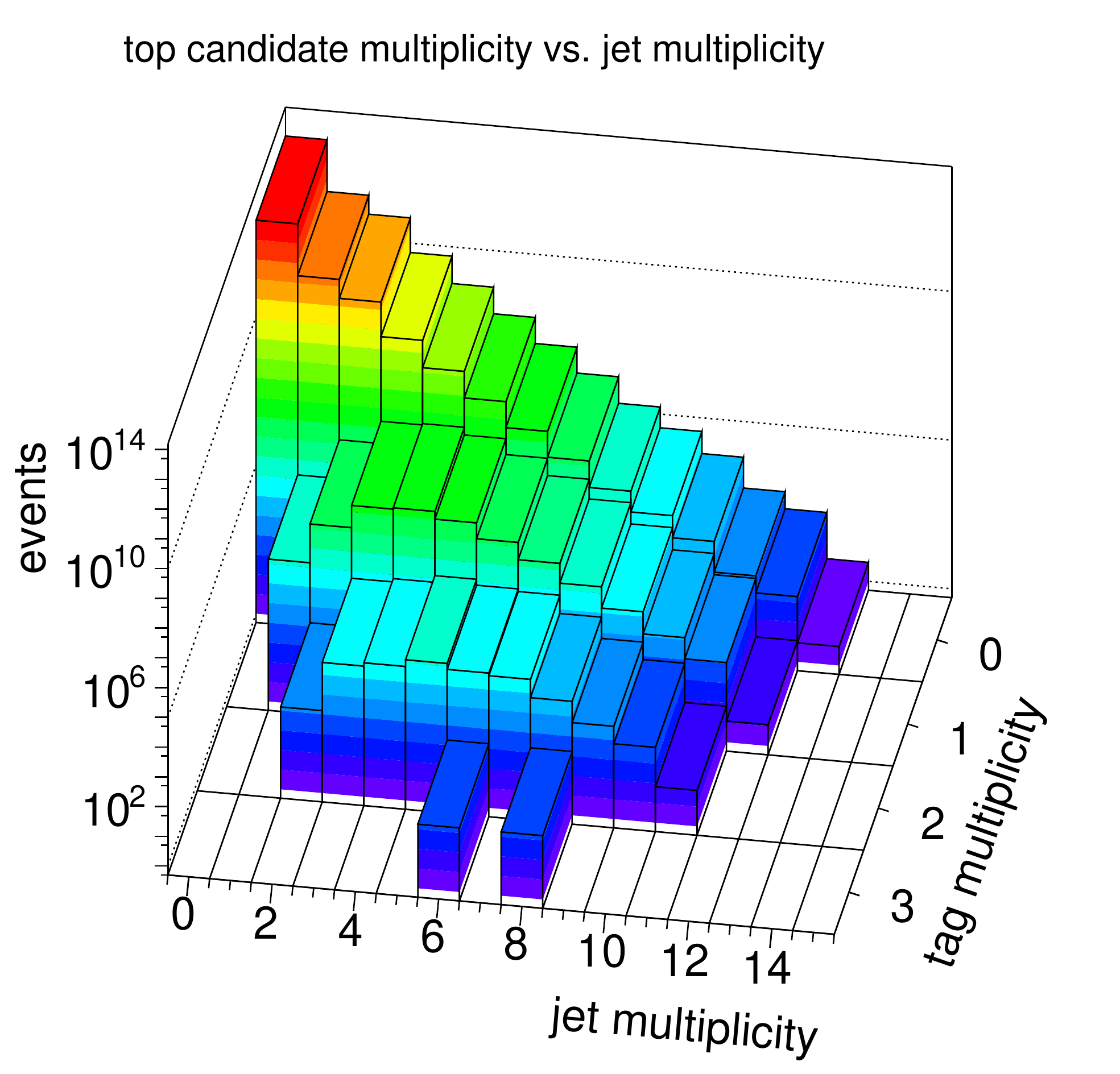}
}
\caption{The \htt top quark candidate multiplicity as a function of the \akt
$R=0.4$ jet multiplicity in $10\invfb$ of $\sqrt{s}=14\TeV$ $pp$ collision events
 that have been passed through the \delphes simulation of the ATLAS detector for
(a) the SUSY benchmark model (from \madgraphherwigpp) and (b) multijet background events (from \pythia~8).
}
\label{fig:master}
\end{figure}

The \htt top quark candidate multiplicity is shown in \figref{master} as a function of the multiplicity \njet
of \akt $R=0.4$ jets with $\pt > 50\GeV$ for signal and background events when using the default
\htt settings. The average \njet in signal events is $7.6$ while the background \njet spectrum
is steeply falling. For signal and background,
the \njet distribution is shifted towards higher values the more top tags
are found. In other words, the probability for a top tag
rises with the number of jets in the event.
This behaviour can be understood as follows. The \htt looks for hard substructure
objects inside the fat jet. It then tries all permutations of three of the substructure
objects and looks if this triplet combines to a top quark. E.g., the (filtered) triplet has to
fulfil the top quark mass constraint and two subjets have to fulfil the \W boson mass constraint.
The larger the small-$R$ jet multiplicity, the higher the number of hard substructure
objects and the probability increases that a triplet is found
which satisfies the kinematic cuts ({\em hits the mass windows}).

The observed feature, that the tagging probability rises with the
jet multiplicity is not limited to the \htt. It is a general feature that
affects all top taggers. The reason is that the more jets are in the event,
the larger the probability that a combination of jets looks like hadronic top quark decay.
The chance that the structure of a fat jet resembles hadronic top quark decay
is higher if the fat jet has more subjets.

Experimentally, a minimal \njet cut is more natural than a requirement of exactly
\njet small-$R$ jets. Measuring an exact number of jets is harder and the systematic
uncertainties are larger. Therefore, in the rest of this section, distributions
like those in \figref{master} are
integrated over \njet, starting from a minimal value.

\begin{figure}[hbt]
\centering
\subfigure[]{
   \includegraphics[width=0.48\textwidth,angle=0]{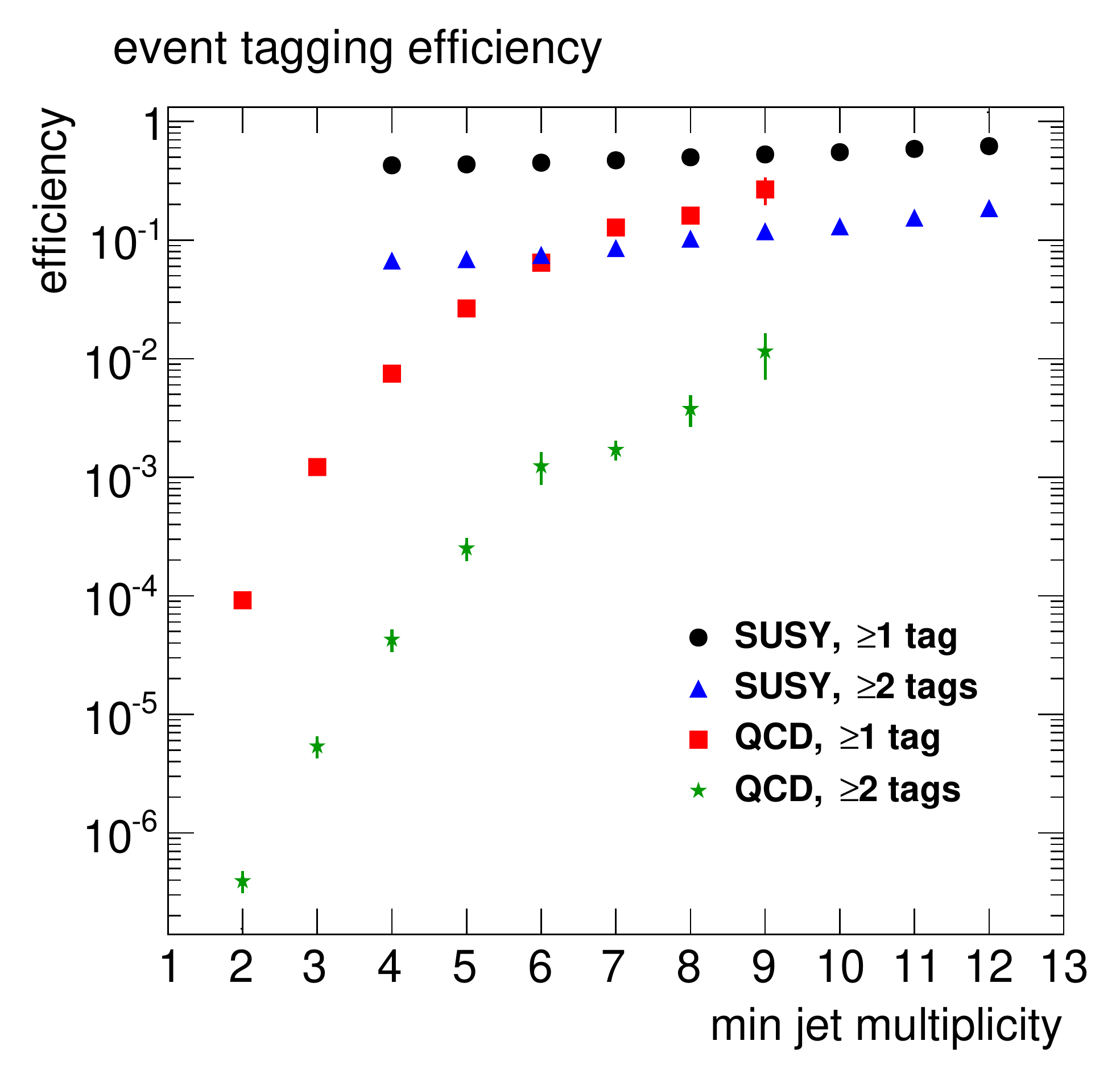}
}
\subfigure[]{
   \includegraphics[width=0.48\textwidth,angle=0]{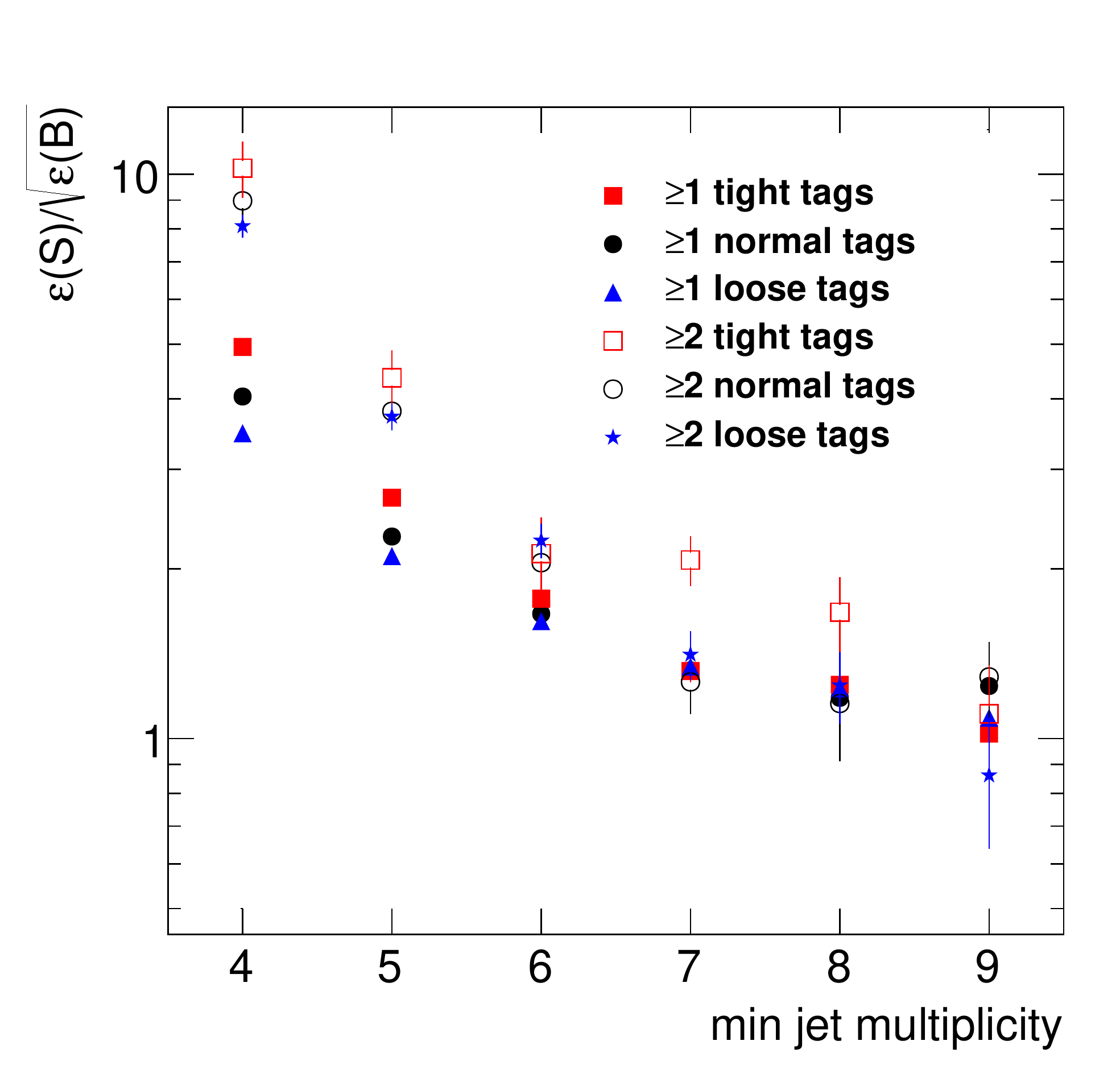}
}
\caption{(a) The \htt efficiency to tag signal events (SUSY, from \madgraphherwigpp) and background events (QCD jets, from \pythia~8)
with at least one or at least two tags as a function of the minimal required number
of \akt $R=0.4$ jets with $\pt>50\GeV$. The default \htt settings from \tabref{HTTsettings} are used.
(b) The ratio of the signal efficiency to the square root of the background efficiency
for different \htt settings. The events have been passed through the \delphes simulation of the ATLAS detector.}
\label{fig:highmult_eff}
\end{figure}

\figref{highmult_eff}a shows the \htt efficiency for at least one and at least
two top tags in signal and background events as a function of the minimal number of required jets.
The efficiency to find at least one top tag in SUSY events rises
from $44\%$ at $\njet \ge 4$ to $60\%$ for $\njet \ge 12$.
The efficiency to have at least a double-tag rises from $6\%$ to $18\%$.
The rise is due to mass window hits which are more likely at higher
jet multiplicity. This effect is more dramatic for the fake rate which
rises from $10^{-4}$ at $\njet \ge 2$ to $16\%$ at $\njet \ge 8$ for single-tags.

The rise of the signal efficiency is much smaller than that of the fake rate.
The explanation is that for signal, the fat jets contain
top quark decays and the efficiency is already high at low \njet.
The plateau efficiency of the \htt is $\approx\!40\%$ when using full detector
simulation (\secref{httperformance}). The \htt inefficiency of $\approx\!60\%$ is due to
mass drop filtering and mass cuts. High jet multiplicity in
signal events helps reduce this inefficiency.
E.g., a signal event that would have escaped detection because the top quark candidate
has a mass larger than the maximum allowed value can be tagged if
the excess energy is removed in the form of a radiated hard gluon. This gluon then gives
rise to another jet in the event.
The efficiency increase from such processes is relatively small because the
mass windows are already designed for high signal efficiency.

For background events, however, QCD radiation makes the difference between
almost no fat jet structure, where most of the energy
is carried by one subjet, and structure that resembles top quark decay.
The fake rate therefore depends strongly on the number of small-$R$ jets in the event.

The ratio of the signal efficiency to the square root of the background efficiency
is shown in \figref{highmult_eff}b. This ratio corresponds to the improvement in the
significance $S/\sqrt{B}$ from applying the \htt.
The \htt is most useful at low jet multiplicity. For $\njet \ge 4$, the improvement is
by a factor of $\approx\!10$ for double-tags. Single-tags with tight \htt settings
give a factor of $5$ and default and loose settings give $4$ and $3.5$, respectively.
The best performance is obtained with tight settings which suppress the background
most. However, the background jet multiplicity spectrum is steeply falling
and requiring more jets helps more than \htt tags as long as the signal is preserved.
Most of the signal events contain at least seven $R=0.4$ jets and for $\njet \ge 7$ the improvement from a tight double-tag
is a factor of $\approx\!2$.

\begin{figure}[hbt]
\centering
\subfigure[]{
   \includegraphics[width=0.48\textwidth,angle=0]{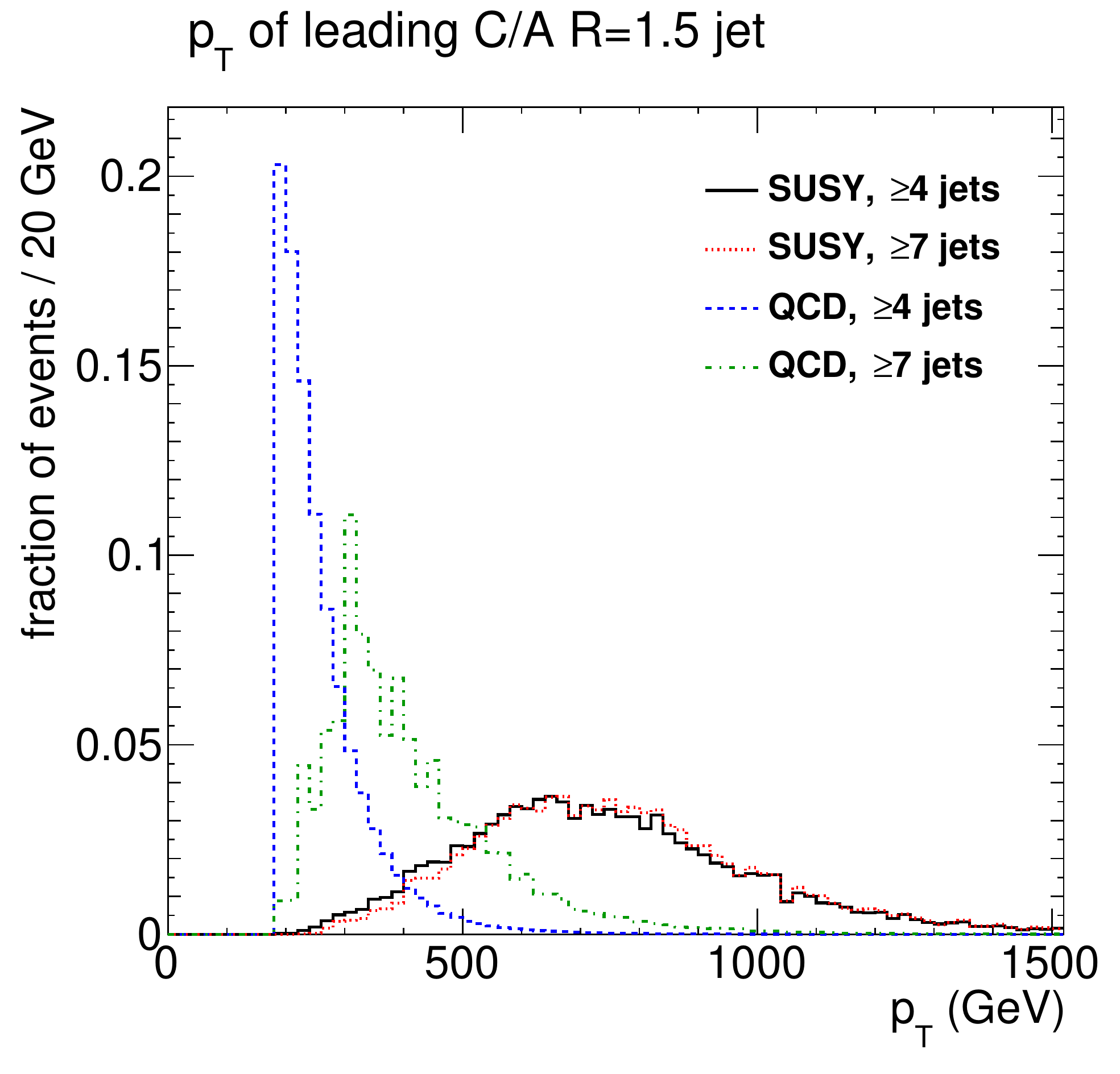}
}
\subfigure[]{
   \includegraphics[width=0.48\textwidth,angle=0]{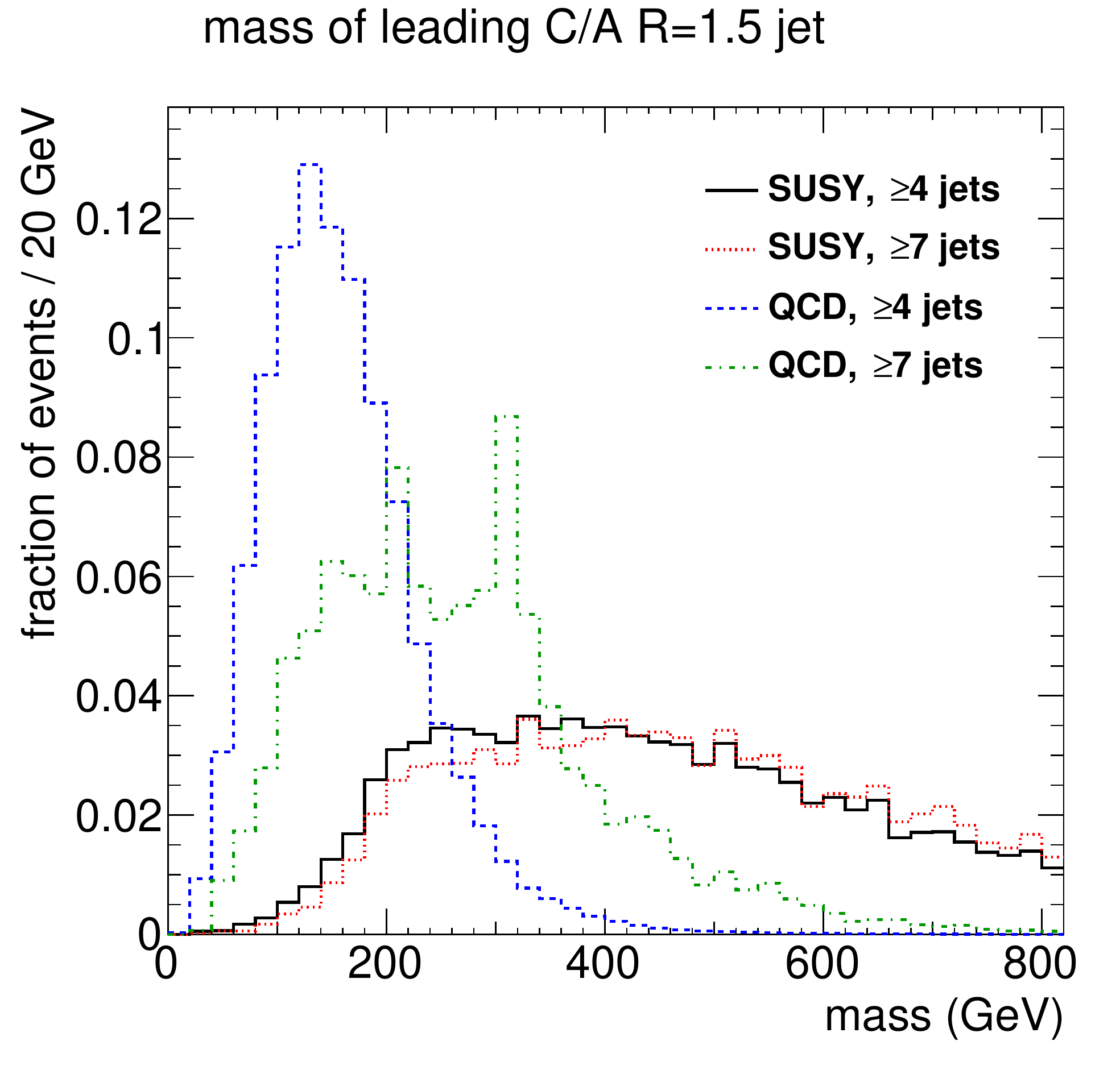}
} \\
\noindent
\subfigure[]{
   \includegraphics[width=0.48\textwidth,angle=0]{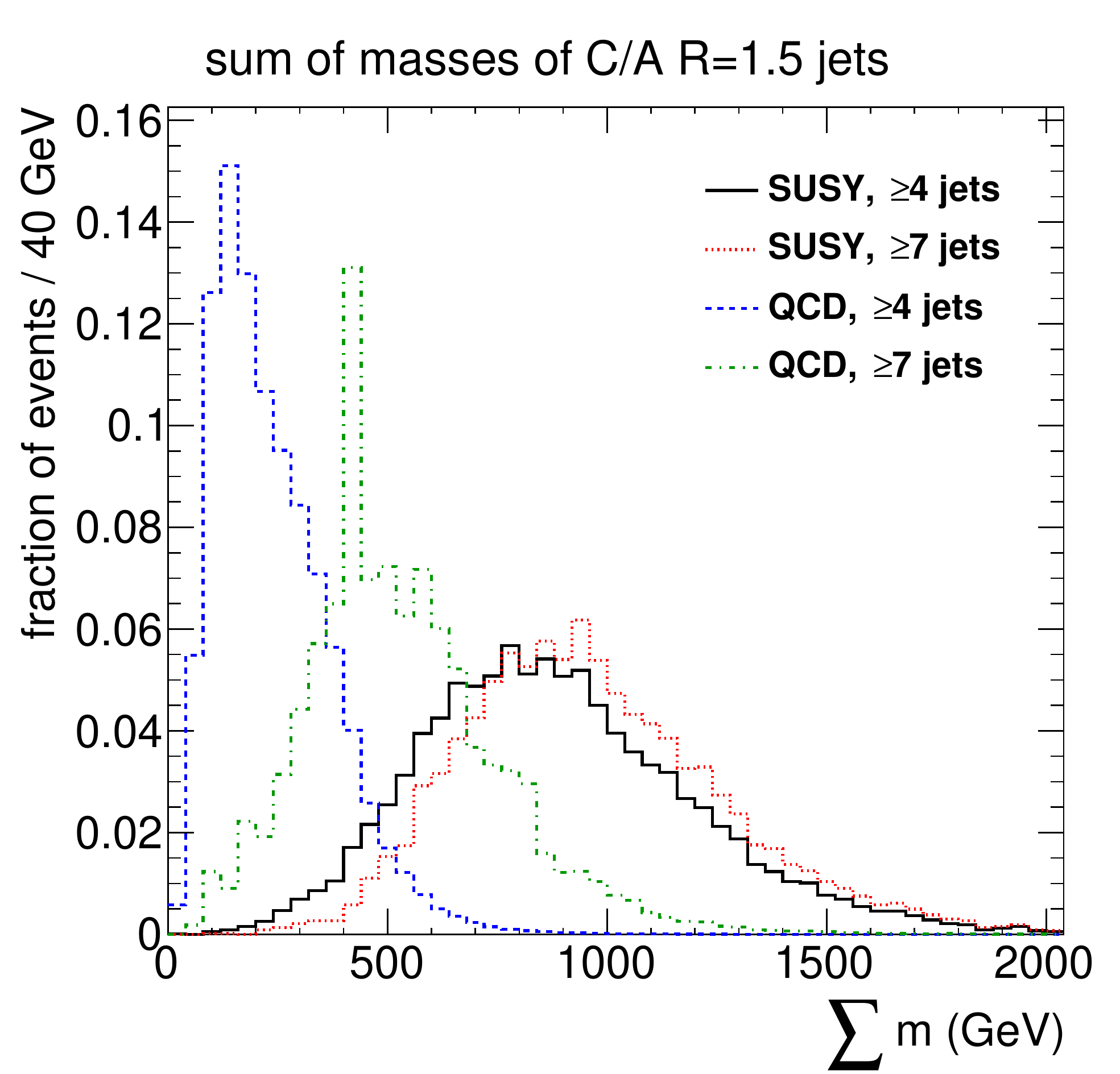}
}
\subfigure[]{
   \includegraphics[width=0.48\textwidth,angle=0]{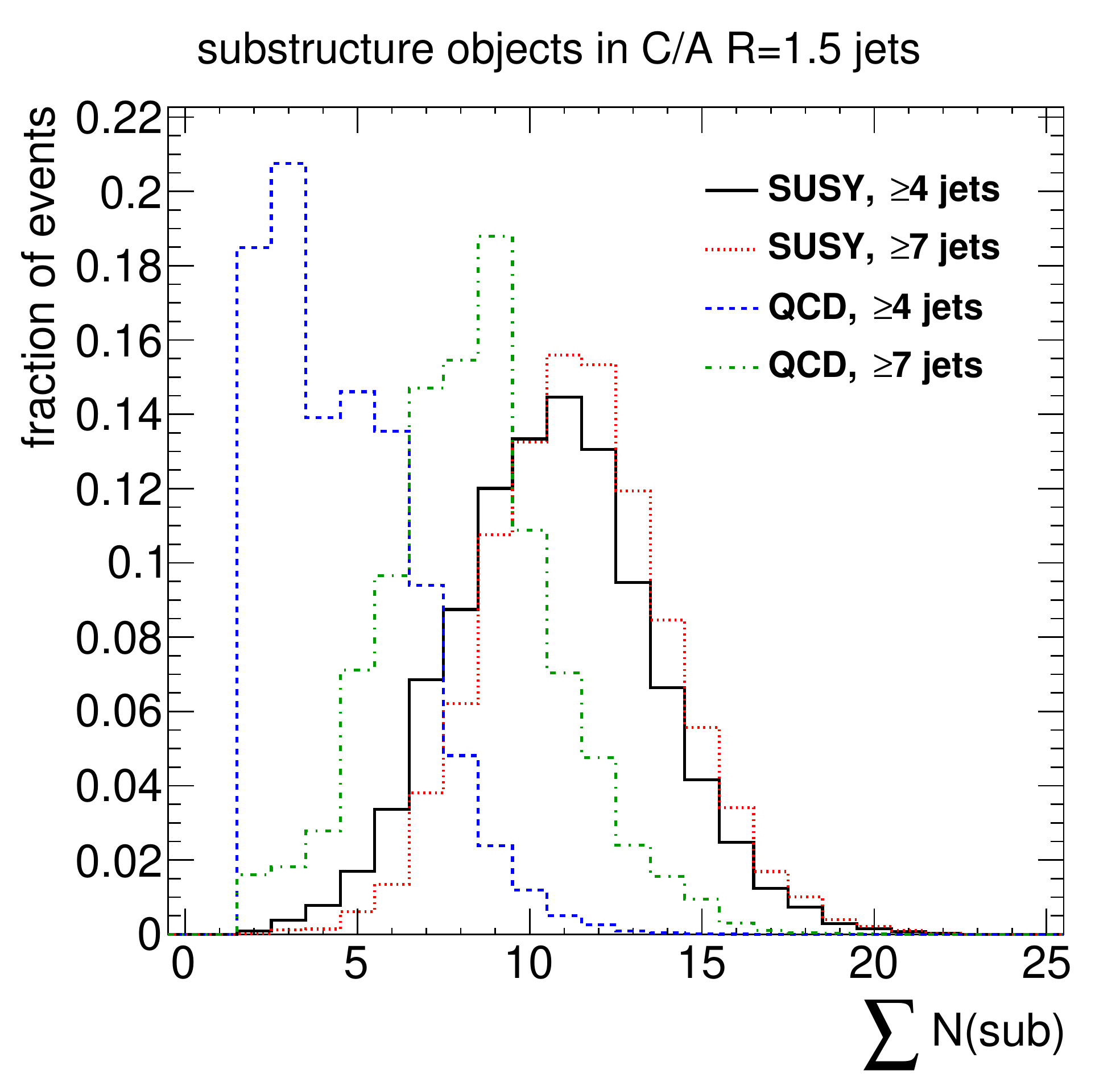}
}
\caption{Normalised distributions in signal (SUSY, from \madgraphherwigpp) and background (QCD, from \pythia~8) events of
\ca $R=1.5$ fat jet quantities for two \akt $R=0.4$ jet multiplicities:
(a) leading fat jet \pt, (b) mass of the leading \pt fat jet, (c) sum of masses of all fat jets, (d) sum of the number of substructure
objects in all fat jets. The substructure objects are obtained by the \htt after the mass drop requirement with
$\mu=80\%$ and $\mcut=50\GeV$). The spikes in the background distribution
for $\njet \ge 7$ are statistical fluctuations. The events have been passed through the \delphes simulation of the ATLAS detector.}
\label{fig:highmult_fj}
\end{figure}

For comparison with the performance of the \htt, cuts on fat jet quantities,
detailed below,
are used to separate signal and background events.
\figref{highmult_fj} shows the distributions of the \ca $R=1.5$ fat jet quantities.
The distributions are shown for $\njet \ge 4$ and $\njet \ge 7$.
Most of the signal events have $\njet \ge 7$ (\figref{master}a).
This implies that the signal fat jets for $\njet \ge 4$ and
those for $\njet \ge 7$ have almost the same number of subjets.
The subjets determine the kinematics of the fat jet.
This is the reason why the distributions of the kinematic variables
and the number of substructure objects do not change much
for the signal fat jets
when \njet is increased.
The situation is quite different for fat jets in background events.
These events have a falling jet multiplicity spectrum (\figref{master}b)
and the fat jets in events with $\njet \ge 4$ have fewer subjets
than those in events with $\njet \ge 7$. Consequently, the kinematics
of background fat jets becomes more signal-like at high \njet.

\figref{highmult_fj}a shows the leading fat jet \pt. As expected, the
background distribution is less well separated from the signal at high jet multiplicities.
A cut $\pt>500\GeV$ is used to enhance signal events.
This cut and the cuts on the other fat jet quantities are
chosen to ensure a high signal efficiency of at least $50\%$.
The leading \pt fat jet mass is shown in panel (b) and the sum of the masses
of all fat jets in the event in panel (c). Cuts $m>350\GeV$ and $\sum m > 800\GeV$
are used, respectively. The fat jet mass sum has been proposed in~\cite{Hook:2012fd}
to select top quark events in high multiplicity environments.
The sum $\sum \nsub$ of the number of substructure objects obtained by the \htt after the mass drop requirement
($\mu=80\%, \mcut=50\GeV$) in all fat jets in the event is shown in (d).
A cut $\sum \nsub > 10$ is used.
This variable is
inspired by \cite{Hedri:2013pvl} in which the counting of subjets has been
proposed as an effective discriminant.
Systematic uncertainties on the fat jet mass and \pt, and on the \htt subjet calibrations
have only a small impact on the efficiency: the relative uncertainty on the efficiency
is a few percent for all investigated methods.

\begin{figure}[hbt]
\centering
\includegraphics[width=0.48\textwidth,angle=0]{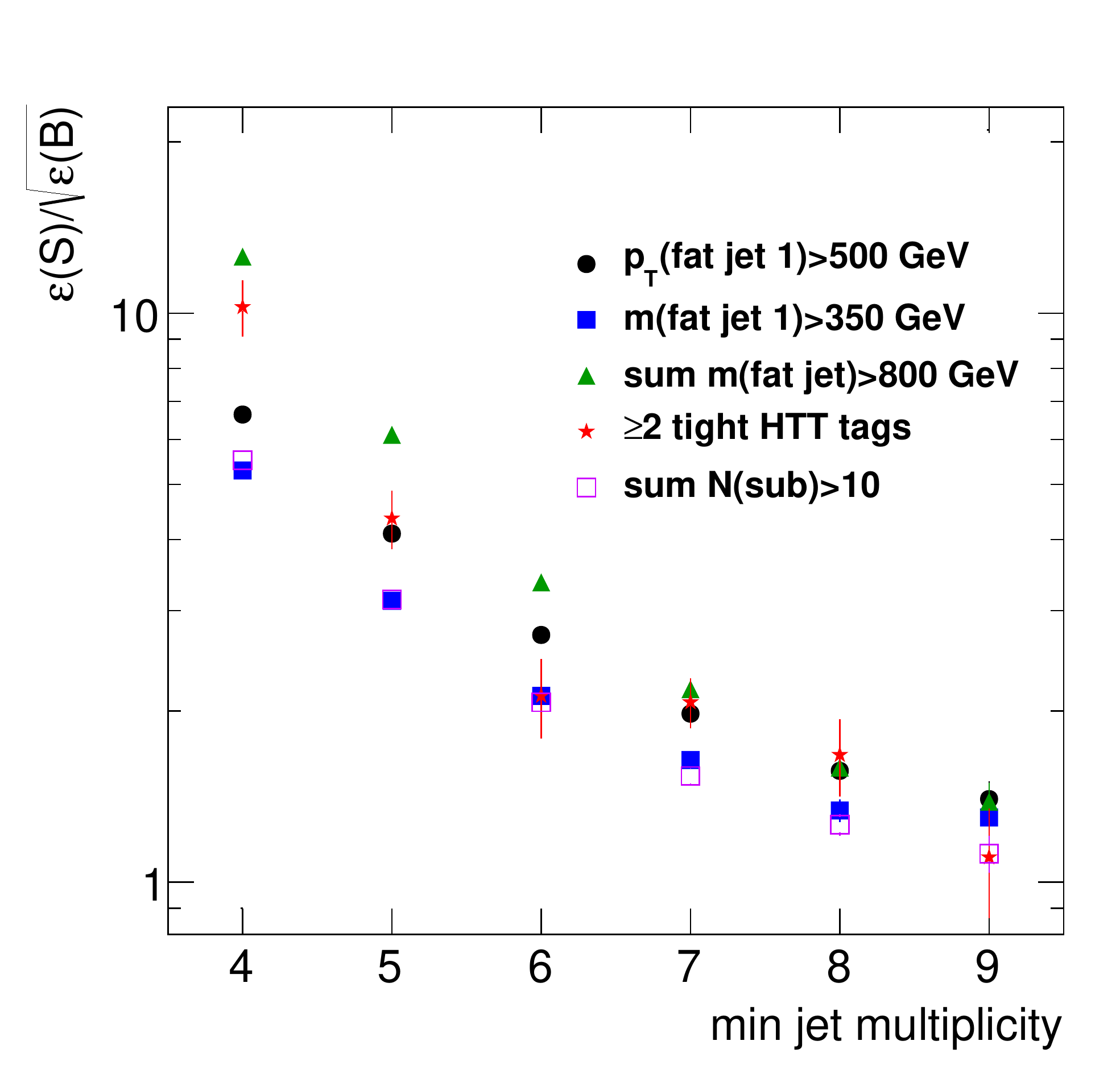}
\caption{The ratio of the efficiency for signal and the square root of the background efficiency
as a function of the minimal \akt $R=0.4$ jet multiplicity in the event for different
top tagging methods.
The SUSY signal events are generated with \madgraphherwigpp and the background multijet events with \pythia~8.
The events have been passed through the \delphes simulation of the ATLAS detector.
}
\label{fig:highmult_improv}
\end{figure}

The significance improvement with these cuts is shown in \figref{highmult_improv}.
The leading \pt fat jet mass cut and the cut on the number of substructure objects
yield similar improvements which are the smallest of all tested approaches.
The best improvement is obtained with the cut on the sum of the fat jet masses.
At small \njet, this method is superior to all other methods, reaching an improvement
by a factor of more than ten for $\njet \ge 4$.
For $\njet \ge 7$ and at higher multiplicities, the cuts on the
leading fat jet \pt and the tight \htt double-tag yield similar results.
The significance is improved by the fat jet mass sum method by factors
2.2, 1.6, 1.4 for $\njet \ge 7$, $8$, $9$, respectively.
This method is used in the high multiplicity ATLAS analysis discussed in \secref{searchsusy}.

\section{Searches for New Physics using Boosted Top Quarks}
\label{sec:searches}

This section discusses analyses that have been carried out using jet
structure techniques to identify boosted top quarks. The first analyses
of this type at the LHC looked for \ttbar resonances,
 both in the semileptonic decay channel, in which
one of the top quarks decays into a \W boson that decays into a lepton and a neutrino
and the other \W boson decays hadronically, and in the (fully) hadronic channel.

The analyses use different methods to identify boosted top quarks
and are compared using the upper limits they give for the production of
new particles in two benchmark models of New Physics:
topcolor (cf. \secref{technicolor}) and warped extra dimensions (\secref{wed}).
Both models predict new particles (the \Zprime boson and the Kaluza-Klein gluon, respectively)
that decay to \ttbar.
The first analysis discussed in \secref{ljets} compares a boosted top quark selection
with a conventional top quark selection and details the advantages of either method.

Analyses that searched for a topcolor \Zprime boson were also carried out at the
Tevatron~\cite{Abazov:2011gv,Aaltonen:2011ts,Aaltonen:2011vi,Aaltonen:2012af}.
They used conventional top quark reconstruction because Tevatron top quarks
are not boosted enough to benefit from substructure techniques.
A narrow leptophobic topcolor \Zprime boson is excluded by the Tevatron data
for masses smaller than $\approx\!900\GeV$ at 95\% confidence level.

The section also discusses a search for supersymmetry which employs
boosted top quark identification.

%%%%%%%%%%%%%%%%%%%%%%%%%%%%%%%%%%%%%%%%%%%%%%%%%%%%%%%%%%%%%%%%%%%%%%%%%%%%%%%%

\subsection{\ttbar resonances in the semileptonic decay channel}
\label{sec:ljets}
The best published limits on \ttbar resonances decaying in the semileptonic
channel are reported in~\cite{Aad:2013nca} using $4.7$~fb$^{-1}$ of ATLAS data taken
at $\sqrt{s}=7\,\TeV$.
The analysis aims at reconstructing the decay chain
$\ttbar \rightarrow b l \nu \, b q q$
with one top quark decaying
leptonically and the other hadronically.
A fat jet selection (called {\em boosted} selection)
is compared with a conventional top quark selection that is based on small-$R$ jets ({\em resolved} selection).

The boosted selection requires events to have at least one
\akt $R=1.0$ fat jet with $\pt>350\GeV$, $|\eta|<2$, and $m>100\GeV$.
The constituents of the leading \pt fat jet are reclustered using the \kt jet algorithm
and the \kt splitting scale $\DOneTwo$ is required to be larger than
$40\GeV$. The events are triggered with an \akt $R=1.0$ jet trigger with a threshold
\pt of $300\GeV$ which is $99\%$ efficient for the fat jets used in the analysis.

The resolved selection uses \akt $R=0.4$ jets.
Events are required to contain at least three jets with
$\pt>25\GeV$ and $|\eta|<2.5$. One of these jets must have a mass larger than $60\GeV$,
the rationale being that it contains the two \W boson decay jets.
If no jet passes this mass requirement, then an additional small-$R$ jet is
required. The events are triggered by a single lepton trigger.

For both selections, every event has to have at least one \akt $R=0.4$ jet
with $\pt>25\GeV$ and $|\eta|<2.5$ that is tagged as a $b$-jet.
Decays of \W bosons to electrons or muons are selected
by applying common lepton and missing transverse energy (\ETmiss) requirements.

For the boosted selection, the four-momentum of the leading \pt fat jet
is taken to correspond to the four-momentum of the hadronically decaying top quark candidate.
The four-momentum of the semileptonically decaying top quark candidate is reconstructed
using the lepton, the small-$R$ jet closest to the lepton, the \ETmiss vector,
and a \W boson mass constraint to calculate the longitudinal component
of the missing momentum. The invariant mass \mtt is then calculated from the
four-momenta of the two reconstructed top quarks.

For the resolved selection, all possible associations of \akt $R=0.4$ jets with
$\pt>20\GeV$ to two top quark candidates are tried and the one that gives the smallest $\chi^2$ when
applying top quark and \W boson mass constraints is chosen to calculate \mtt.

\begin{figure}[htb]
\centering
\subfigure[]{
   \includegraphics[width=0.48\textwidth,angle=0]{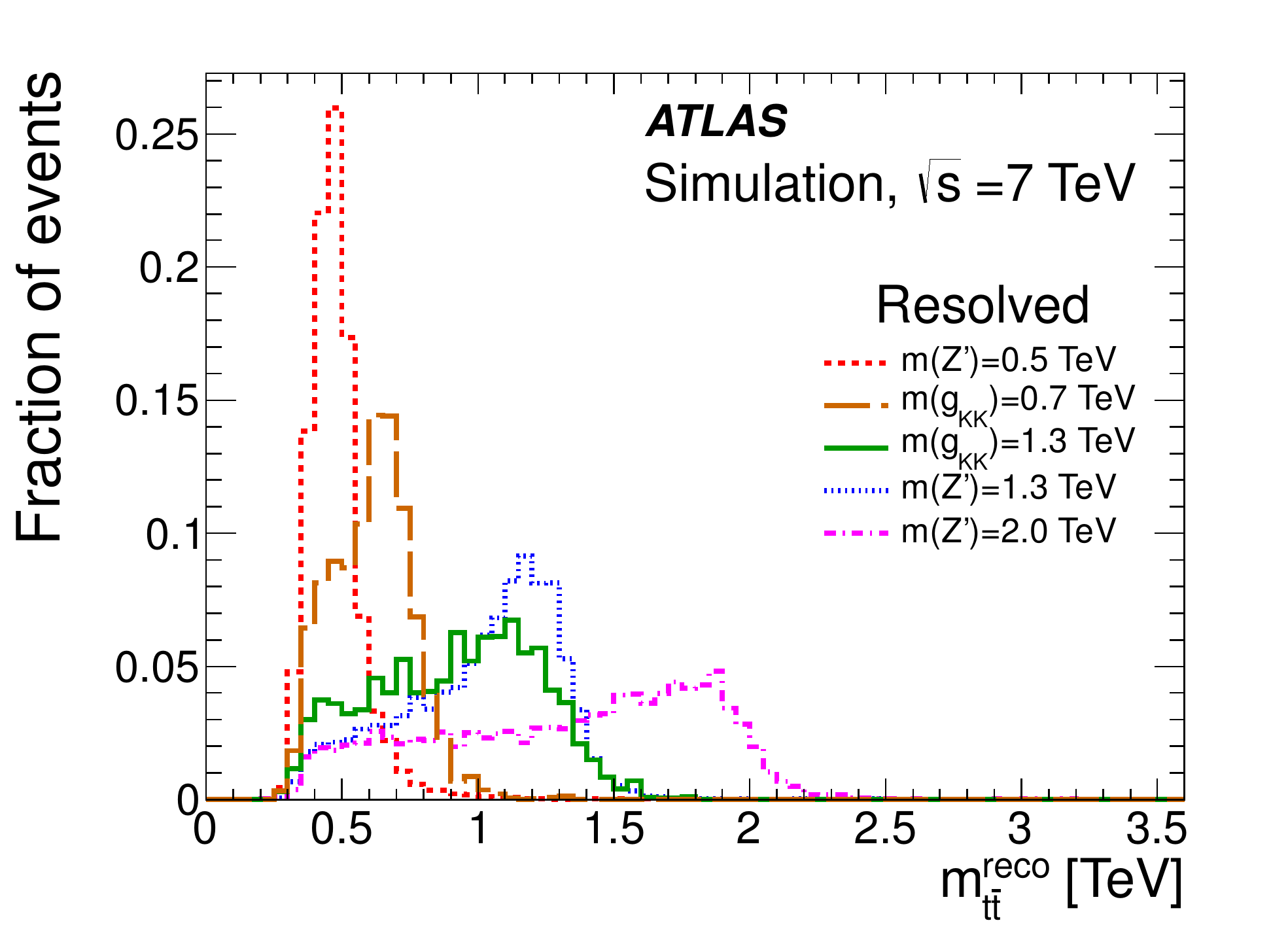}
}
\subfigure[]{
   \includegraphics[width=0.48\textwidth,angle=0]{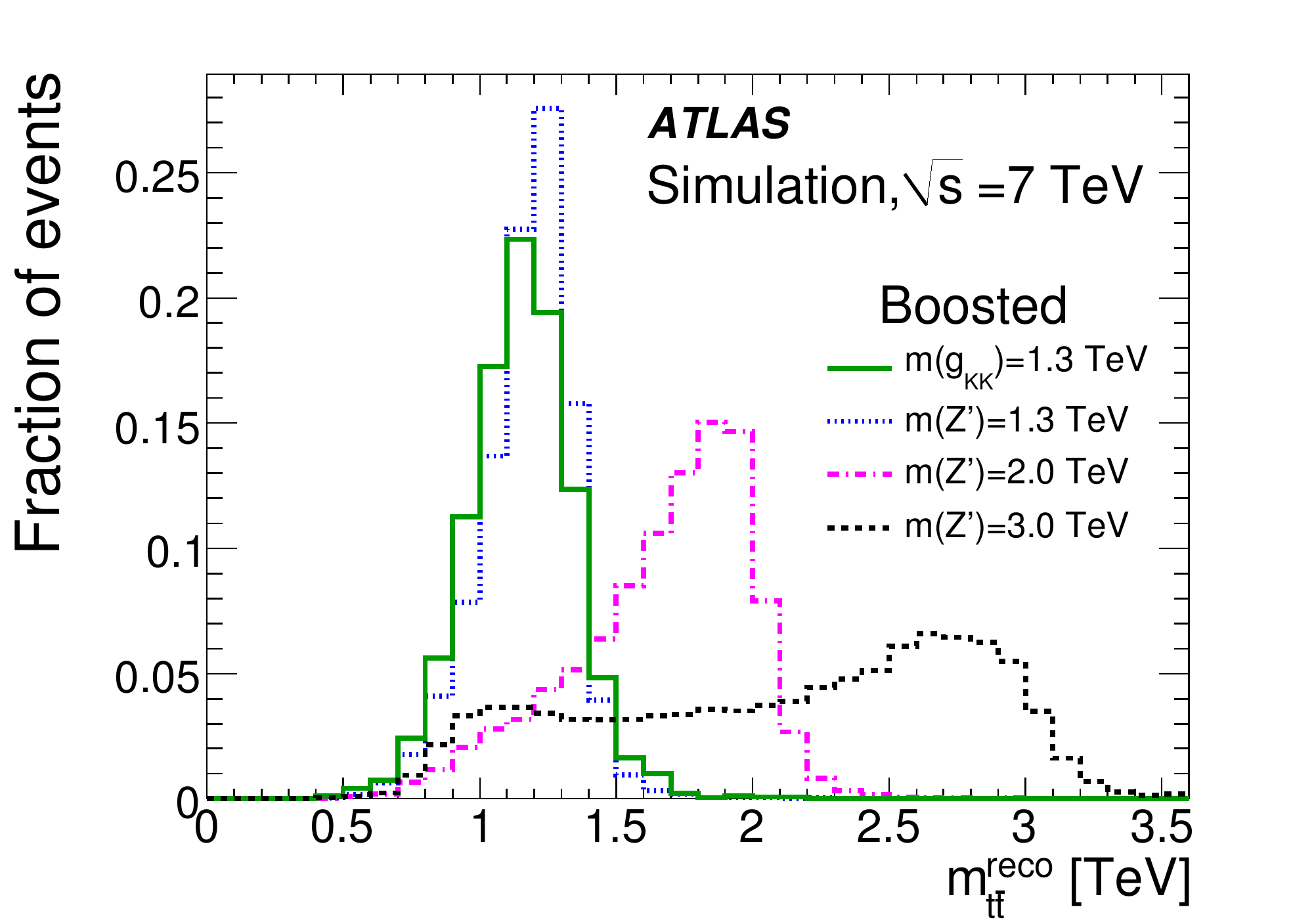}
}
\caption{The reconstructed invariant mass \mtt using simulated ATLAS events with
decays $\Zprime (\gKK) \rightarrow \ttbar \rightarrow b l \nu \, b q q$ for reconstructions
based on (a) small-$R$ jets and (b) fat jets.
The \Zprime events are generated with \pythia and the \gKK events with \madgraphpythia.
From~\cite{Aad:2013nca}.}
\label{fig:ljets_masssimulation}
\end{figure}

\figref{ljets_masssimulation} shows the reconstructed \mtt spectrum for simulated events
with decays of \Zprime bosons and Kaluza-Klein gluons of different masses.
The mass resolution of the fat jet method is much better for masses
larger than $0.8\TeV$.
The efficiency of the boosted selection for finding \ttbar pairs
is comparable to that of the resolved selection for
$\mtt^{\rm true} \approx 1\TeV$ and performs better for larger masses.

In~\cite{Aad:2013nca}, the following strategy is adopted to maximise the sensitivity: if an event
passes the boosted selection then \mtt is calculated using the fat jet;
if the event fails the boosted selection but passes the resolved selection
then \mtt is reconstructed from the small-$R$ jets. The analysis is also split
into an electron and a muon channel. The selection for muons is twice as efficient
because the probability to fake a muon signal is lower and therefore a less stringent
cut could be applied on \ETmiss, and because electron calorimeter clusters have
to be separated from hadronic clusters based on cuts which leads to an inherent
smaller reconstruction efficiency. The efficiency to select events with
\Zprime boson decays with muons is larger than $8\%$ for $m_{\Zprime} > 1.5\TeV$
when using fat jets.

\begin{figure}[htb]
\centering
\includegraphics[width=0.48\textwidth,angle=0]{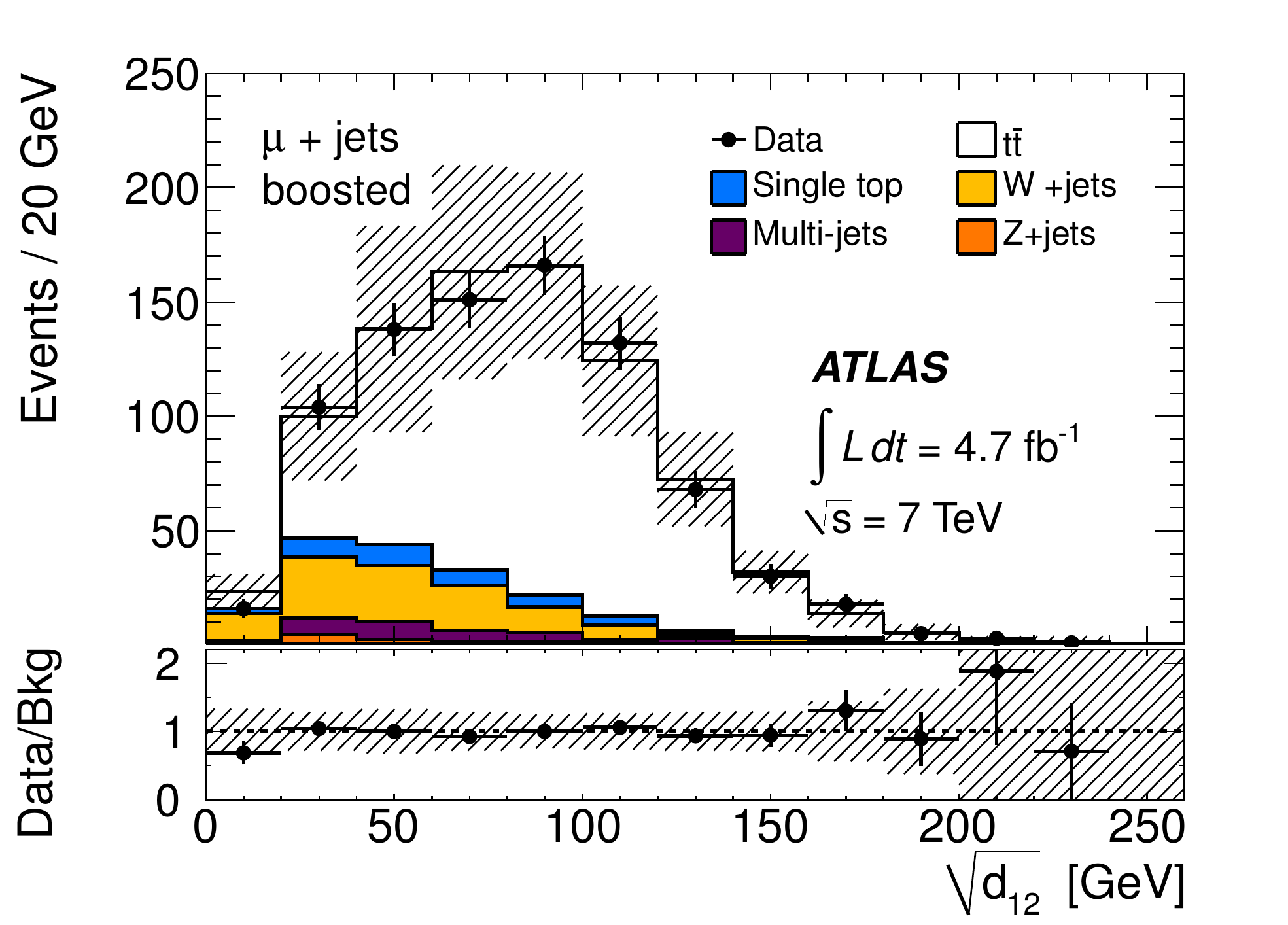}
\caption{The \kt splitting scale \DOneTwo, calculated by ATLAS for the constituents of
\akt $R=1.0$ jets in \ttbar events with a muon. The simulated \ttbar events are from
\mcatnlo. The \Wpjets events are generated with \alpgenherwig.
The multijet contribution is determined from data. From~\cite{Aad:2013nca}.}
\label{fig:ljets_d12}
\end{figure}

\figref{ljets_d12} shows the \kt splitting scale \DOneTwo for the leading \pt fat jet in muon events which
pass the boosted selection (the cut on \DOneTwo has been omitted to display
the full spectrum). The background from
\Wpjets events and multijet events has a smaller fat jet splitting scale than \ttbar events
and is reduced by the cut $\DOneTwo>40\GeV$.
The SM \ttbar background is taken from simulation (\mcatnlo) as
is the \mtt shape of the \Wpjets background (\alpgenherwig).
The normalisation of the \Wpjets background is obtained by scaling to the observed
charge asymmetry in data. Processes other than \Wpjets give equal numbers of
positively and negatively charged leptons.
The MC is scaled by a factor $(N_{W^+} - N_{W^-})_{\rm Data} / (N_{W^+} - N_{W^-})_{\rm MC}$.
The uncertainty of this factor is $10$--$20\%$, depending on the selection,
and is compatible with unity. The multijet background is obtained from
data by deriving lepton efficiencies and mis-tag rates in control samples
that are signal-enriched (dileptons in \Z boson mass window) or
dominated by multijets, respectively.
The largest contribution
to the relative systematic uncertainty of $25\%$ on the
yield of SM events is $17\%$ from the uncertainties on the energy and mass scales
of the fat jets.

\begin{figure}[htb]
\centering
\subfigure[]{
   \includegraphics[width=0.48\textwidth,angle=0]{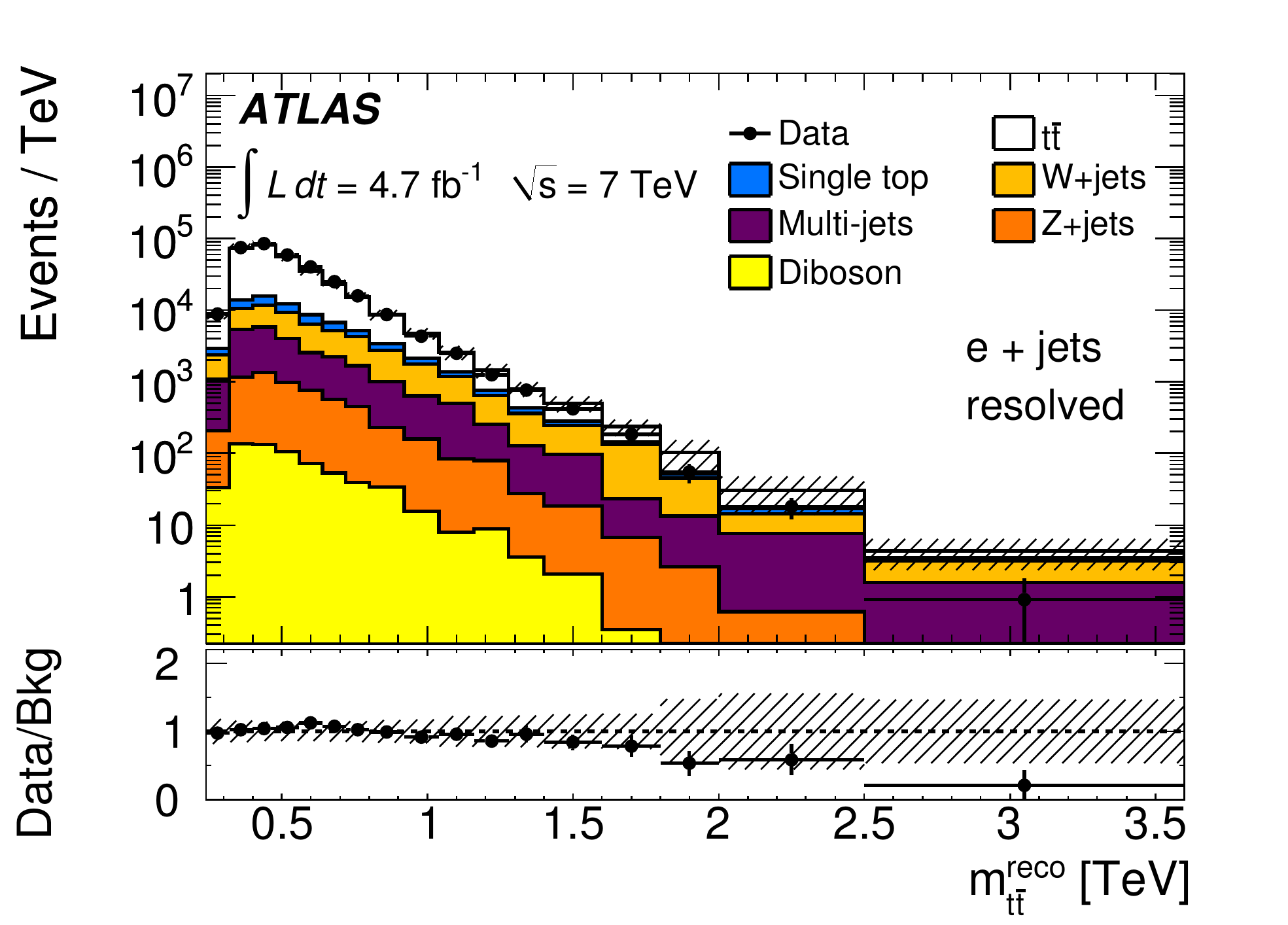}
}
\subfigure[]{
   \includegraphics[width=0.48\textwidth,angle=0]{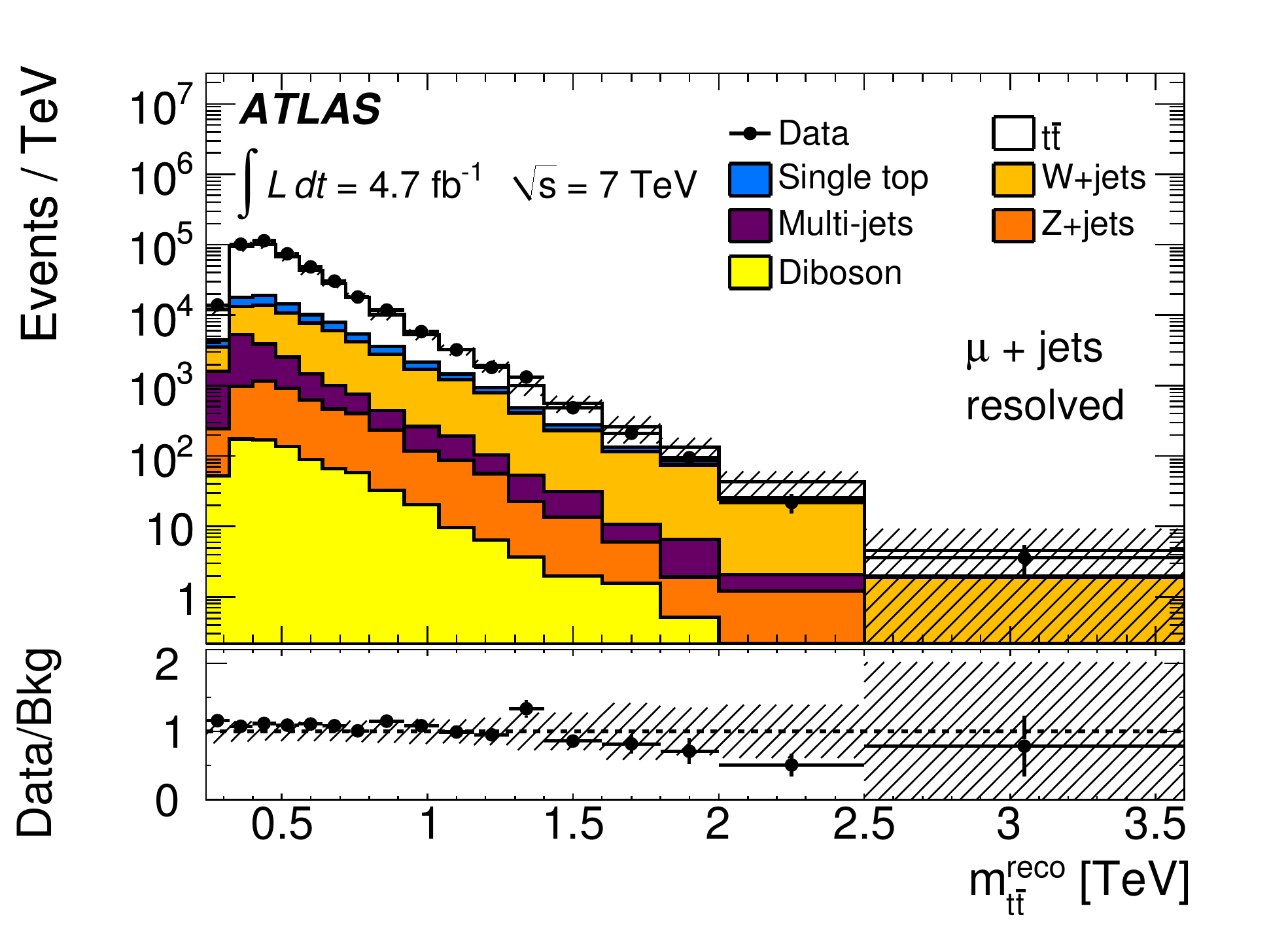}
} \newline
\subfigure[]{
   \includegraphics[width=0.48\textwidth,angle=0]{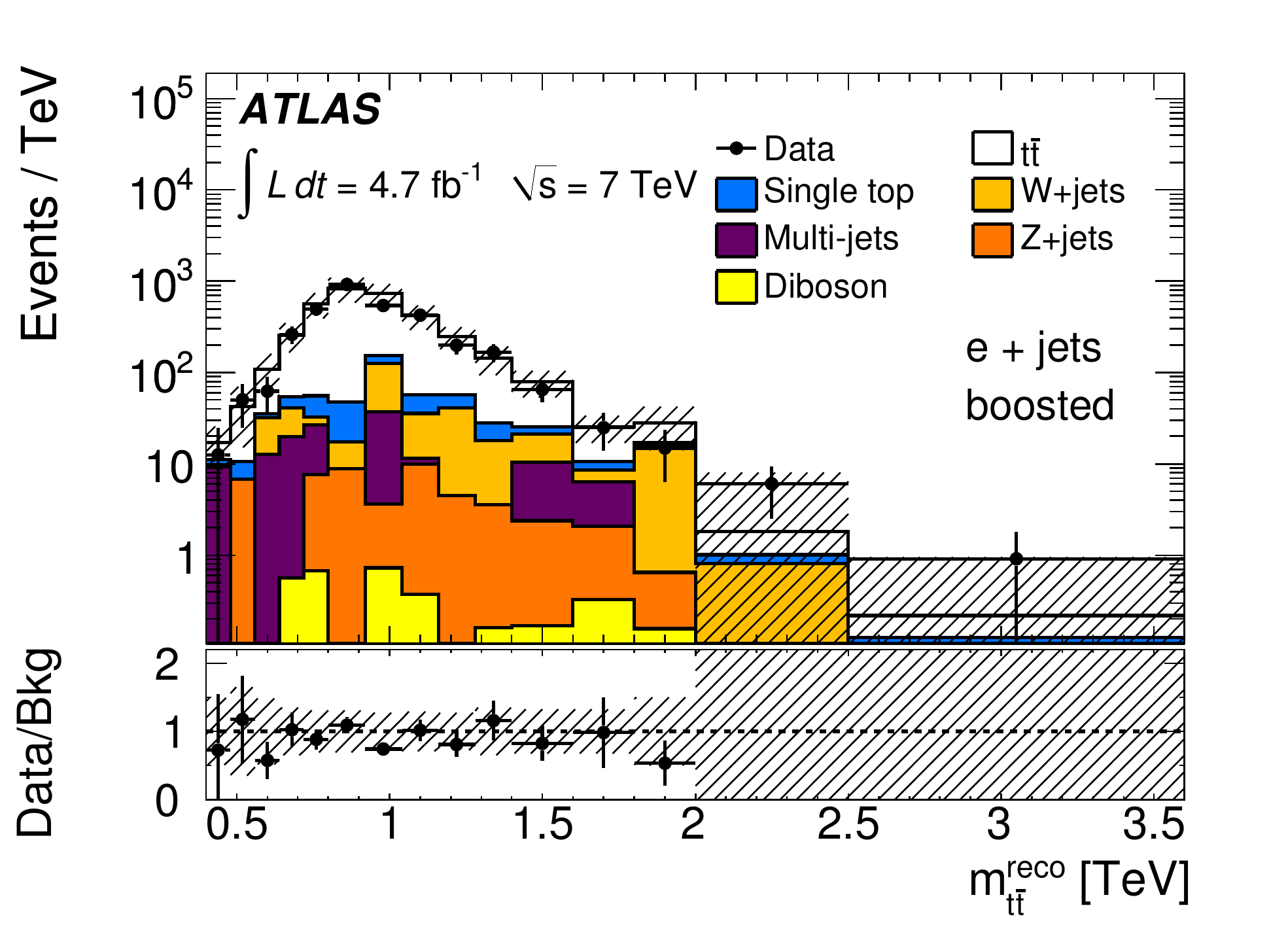}
}
\subfigure[]{
   \includegraphics[width=0.48\textwidth,angle=0]{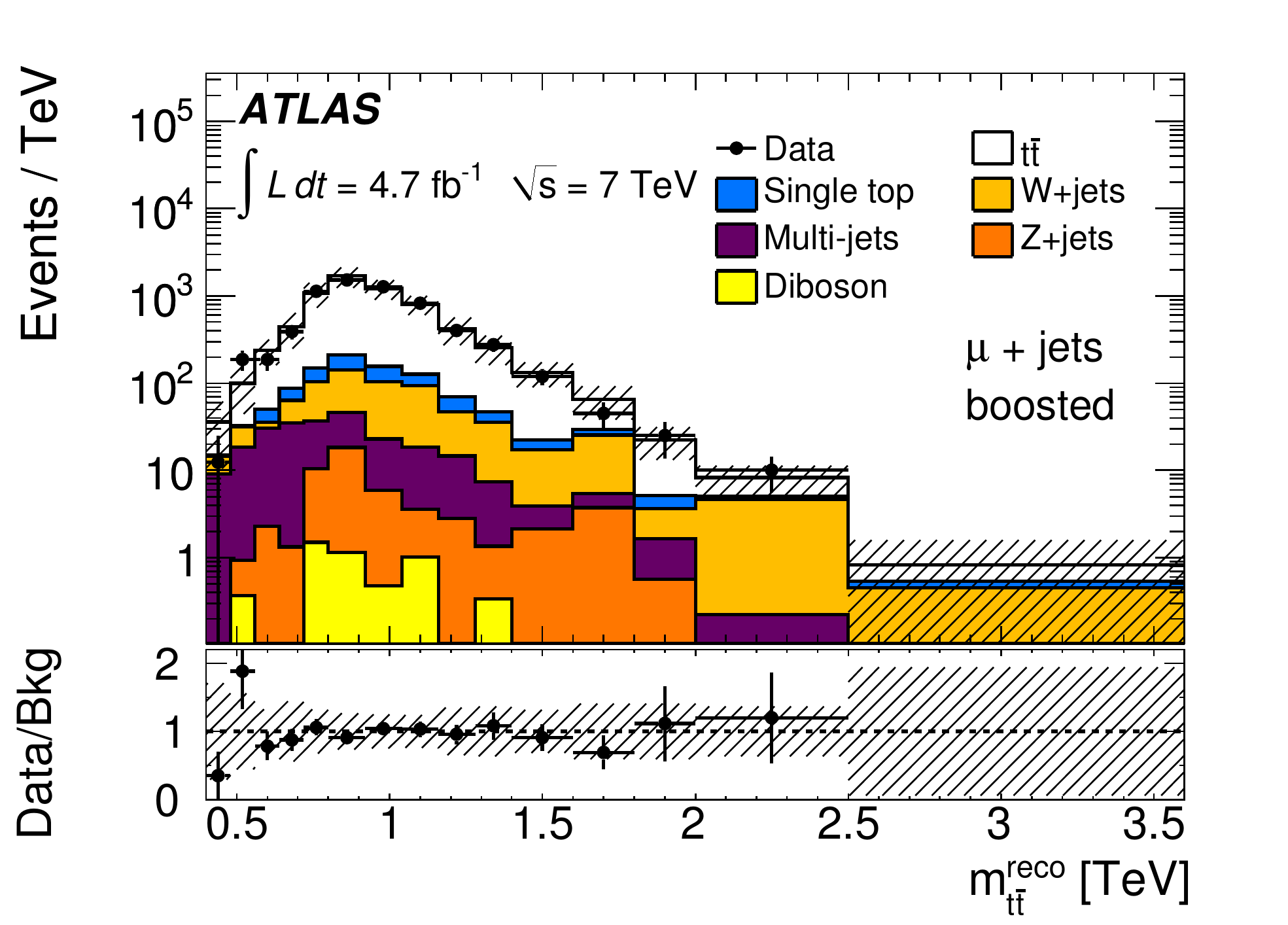}
}
\caption{The reconstructed invariant mass \mtt measured by ATLAS
in \ttbar events (a) with an electron and using a selection based on $R=0.4$ jets ({\em resolved}),
(b) with a muon and the resolved selection, (c) with an electron and using $R=1.0$ fat jets ({\em boosted} selection),
(d) with a muon and using the boosted selection.
The simulated \ttbar events are from
\mcatnlo. The \Wpjets events are generated with \alpgenherwig.
The multijet contribution is determined from data.
From~\cite{Aad:2013nca}.}
\label{fig:ljets_spectrum}
\end{figure}

The \mtt spectrum for electron and muon events is shown in \figref{ljets_spectrum} for the
resolved and the boosted selection.
Note that events which pass the boosted selection are not considered
for the resolved selection. However, even at high reconstructed invariant masses,
the resolved selection has more events.
This is due to the bad mass resolution of the resolved
selection which smears events from low true invariant masses to high reconstructed \mtt.

No deviation from the SM expectation is observed and limits are set on the
production cross section times branching ratio into \ttbar for two benchmark
models of New Physics.
The first is a leptophobic \Zprime boson which is predicted by technicolor
models~\cite{Hill:1994hp,Harris:1999ya,Harris:2011ez}.
Model IV of \cite{Harris:1999ya} is used, with $f_1 = 1$, $f_2 = 0$,
and the corrections to the Lagrangian discussed in~\cite{Harris:2011ez} are included.
The couplings of this new hypothetical particle are such that the width
is small in comparison to the mass, $\Gamma_{\Zprime}/m_{\Zprime} = 1.2\%$.
This width is also smaller than the detector \mtt resolution of $\approx\!10\%$.
The second benchmark is a Kaluza-Klein gluon (\gKK) that is predicted
by a Randall-Sundrum model of warped extra-dimensions~\cite{Lillie:2007yh,Lillie:2007ve,Agashe:2006hk,Djouadi:2007eg,Agashe:2007zd}.
The width of this particle is $\Gamma_{\gKK}/m_{\gKK} = 15.3\%$.

\begin{figure}[htb]
\centering
\subfigure[]{
   \includegraphics[width=0.48\textwidth,angle=0]{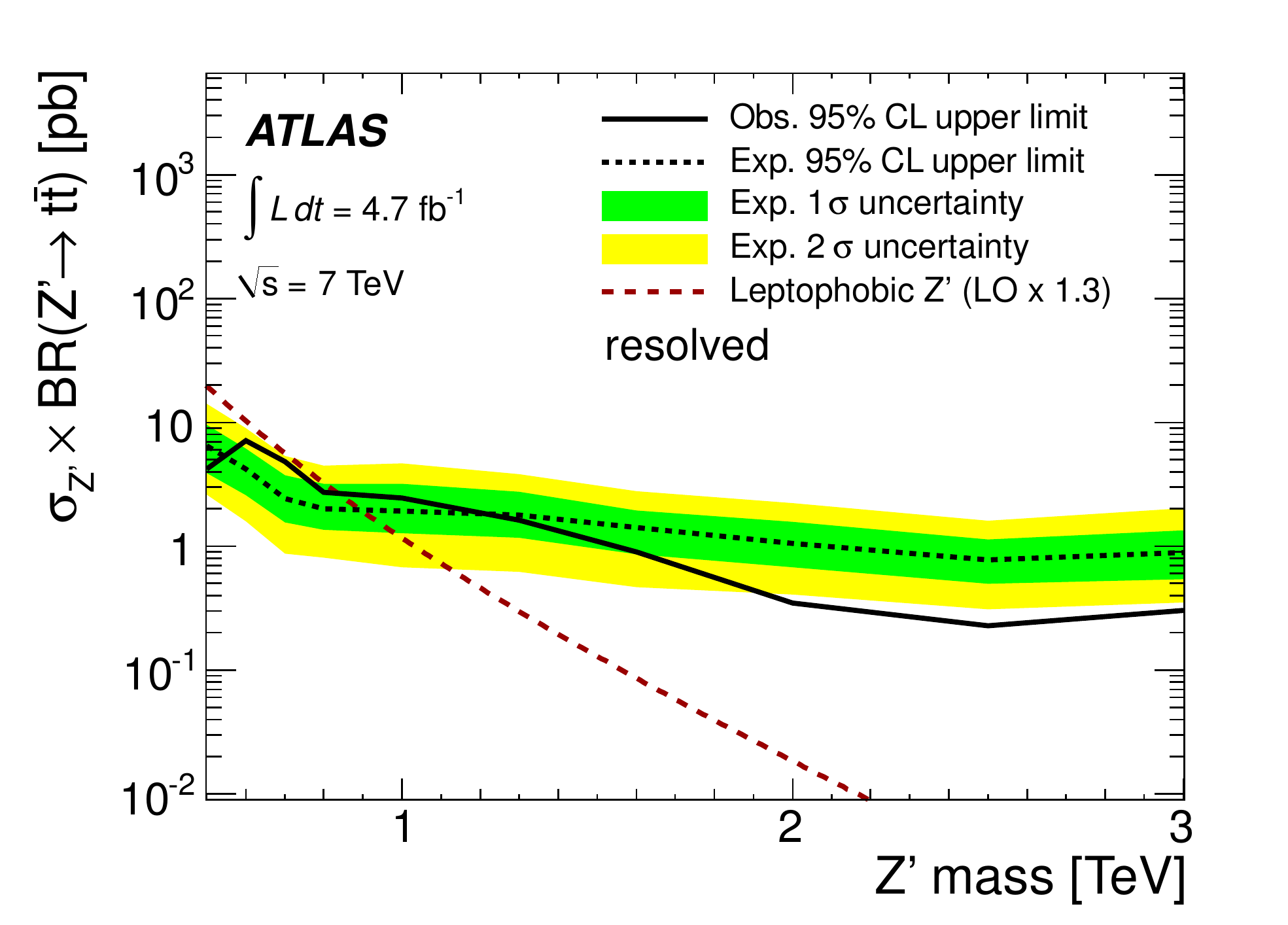}
}
\subfigure[]{
   \includegraphics[width=0.48\textwidth,angle=0]{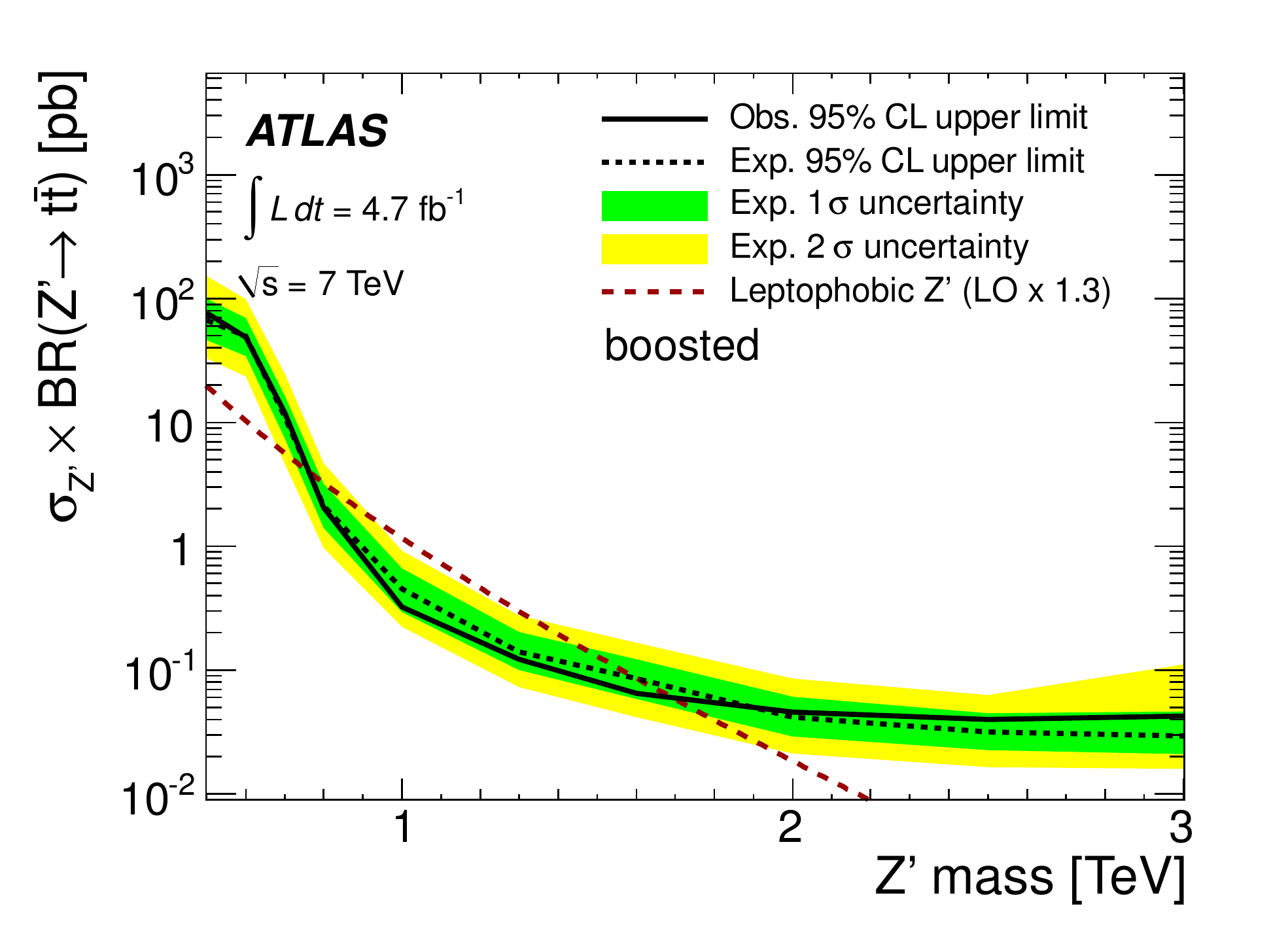}
} \newline
\subfigure[]{
   \includegraphics[width=0.48\textwidth,angle=0]{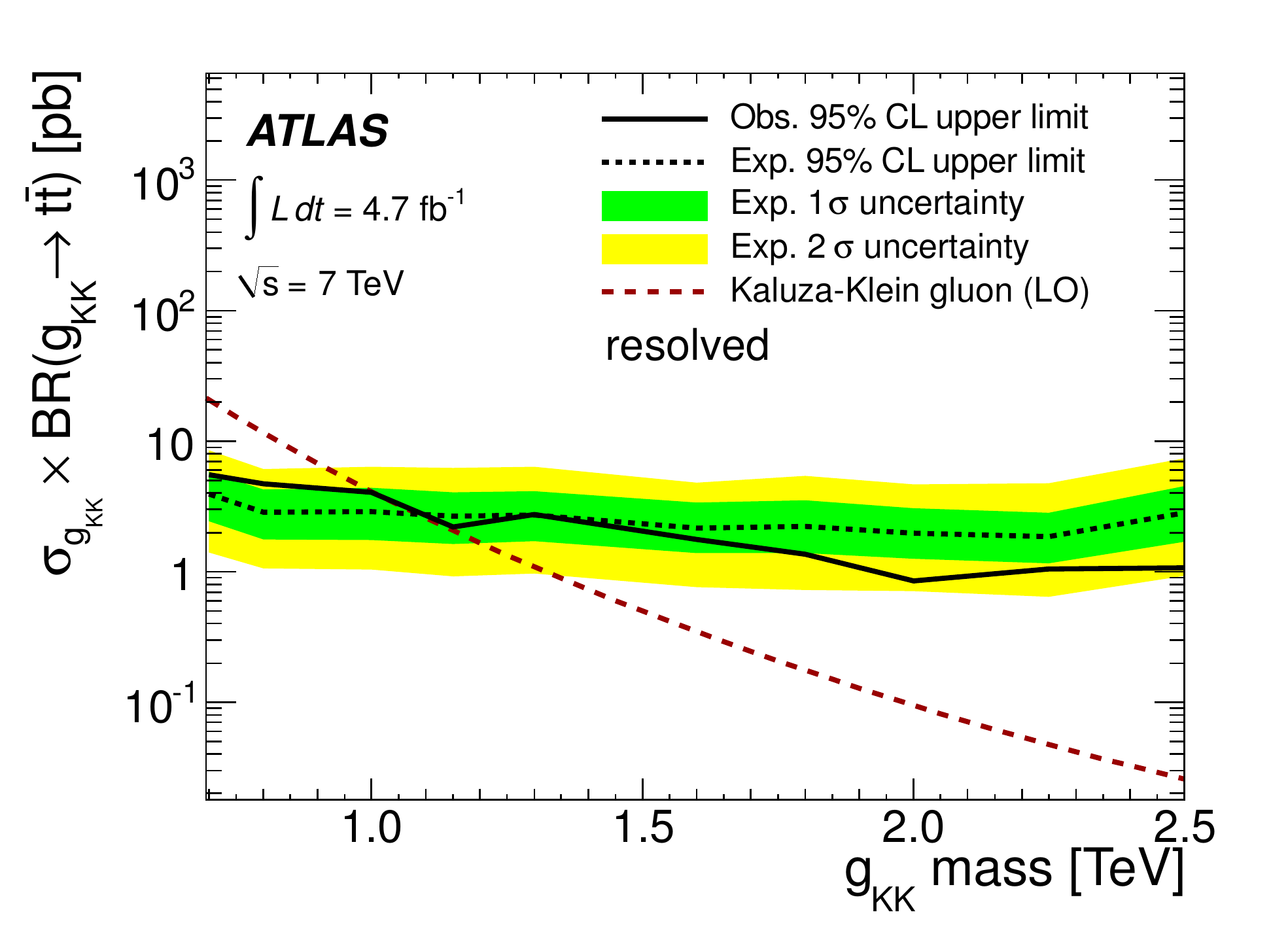}
}
\subfigure[]{
   \includegraphics[width=0.48\textwidth,angle=0]{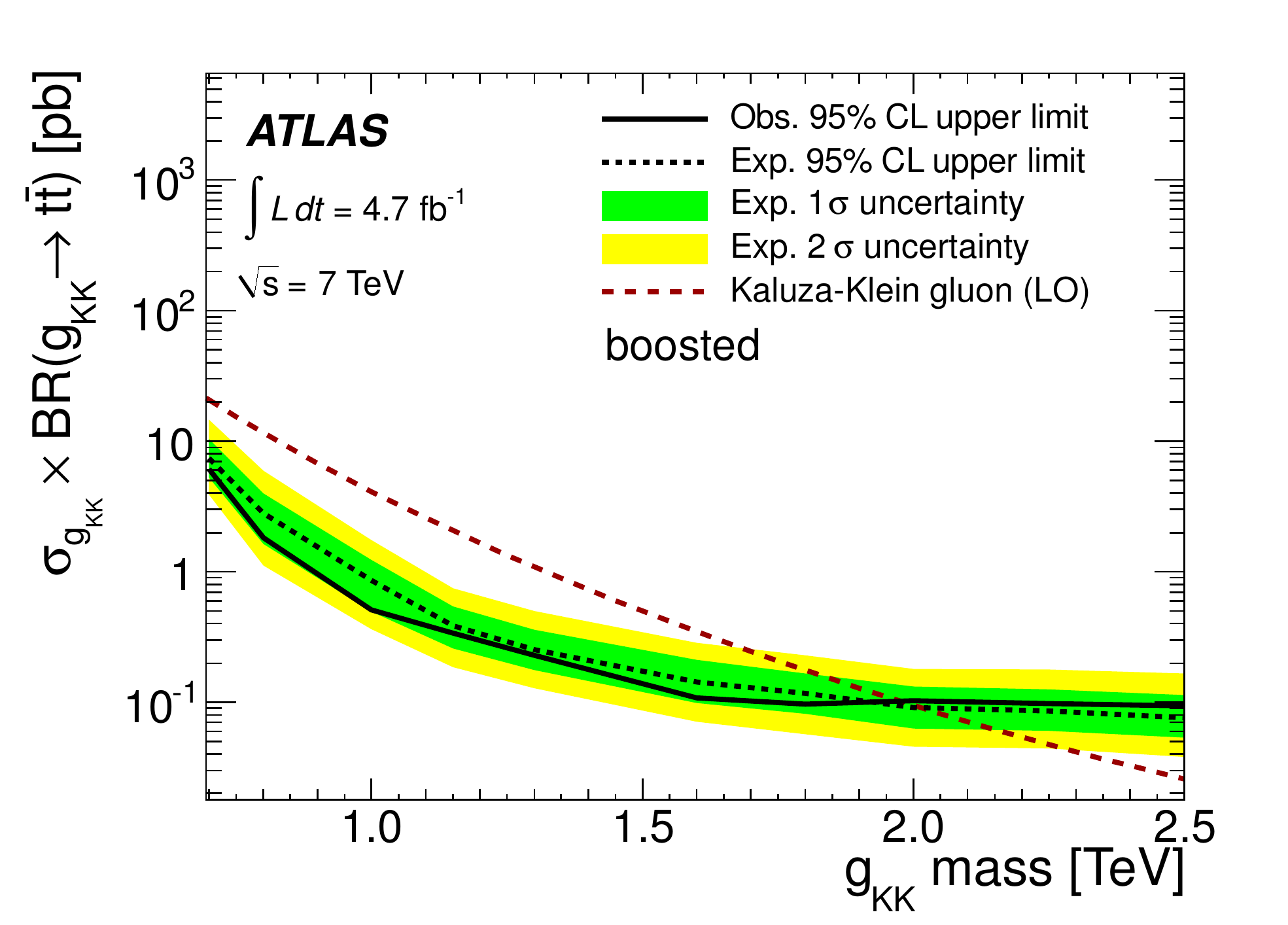}
}
\caption{The $95\%$ credibility level upper limit on the production cross section
times branching ratio into \ttbar for a leptophobic \Zprime boson with
width $\Gamma_{\Zprime}/m_{\Zprime} = 1.2\%$ using (a) the resolved selection
and (b) the boosted selection and for a Kaluza-Klein gluon with
width $\Gamma_{\gKK}/m_{\gKK} = 15.3\%$ using (c) the resolved selection and
(d) the boosted selection.
Events that pass the boosted selection are not considered for the resolved
selection.
The limits are obtained from an analysis
of semileptonic \ttbar decays. The expected limit is obtained by
using background-only pseudo-experiments.
From~\cite{Aad:2013nca}.}
\label{fig:ljets_boosted_resolved_limits}
\end{figure}

To illustrate the impact of the substructure technique,
the limits obtained using either only the resolved selection or
only the boosted selection
are shown in \figref{ljets_boosted_resolved_limits} where the
electron and muon channels have been combined.
The expected limit from the boosted selection is better than that from
the resolved selection for masses larger than $0.8\TeV$.
The better limit is a result of the better mass resolution of the boosted
selection.

\begin{figure}[htb]
\centering
\subfigure[]{
   \includegraphics[width=0.48\textwidth,angle=0]{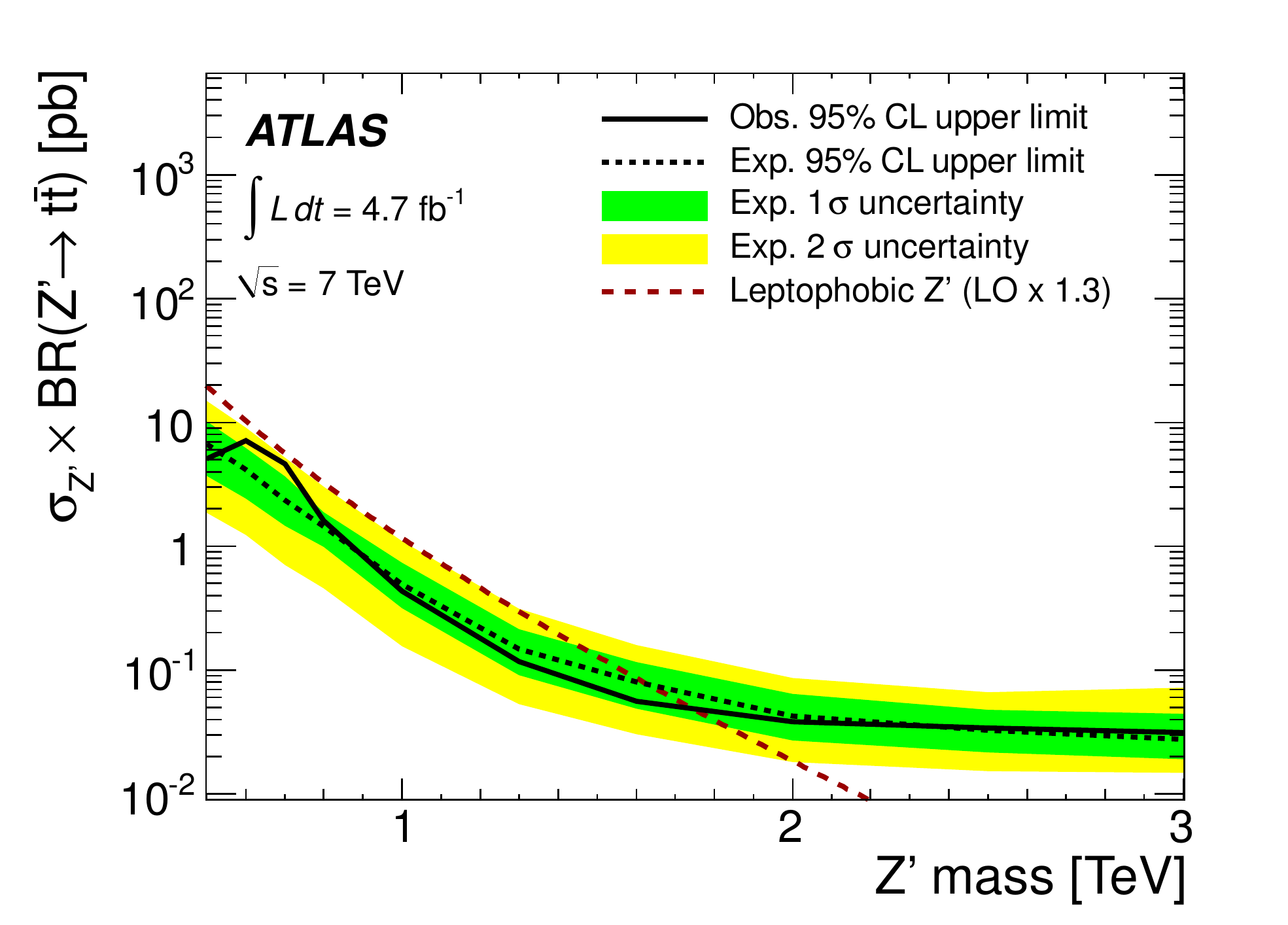}
}
\subfigure[]{
   \includegraphics[width=0.48\textwidth,angle=0]{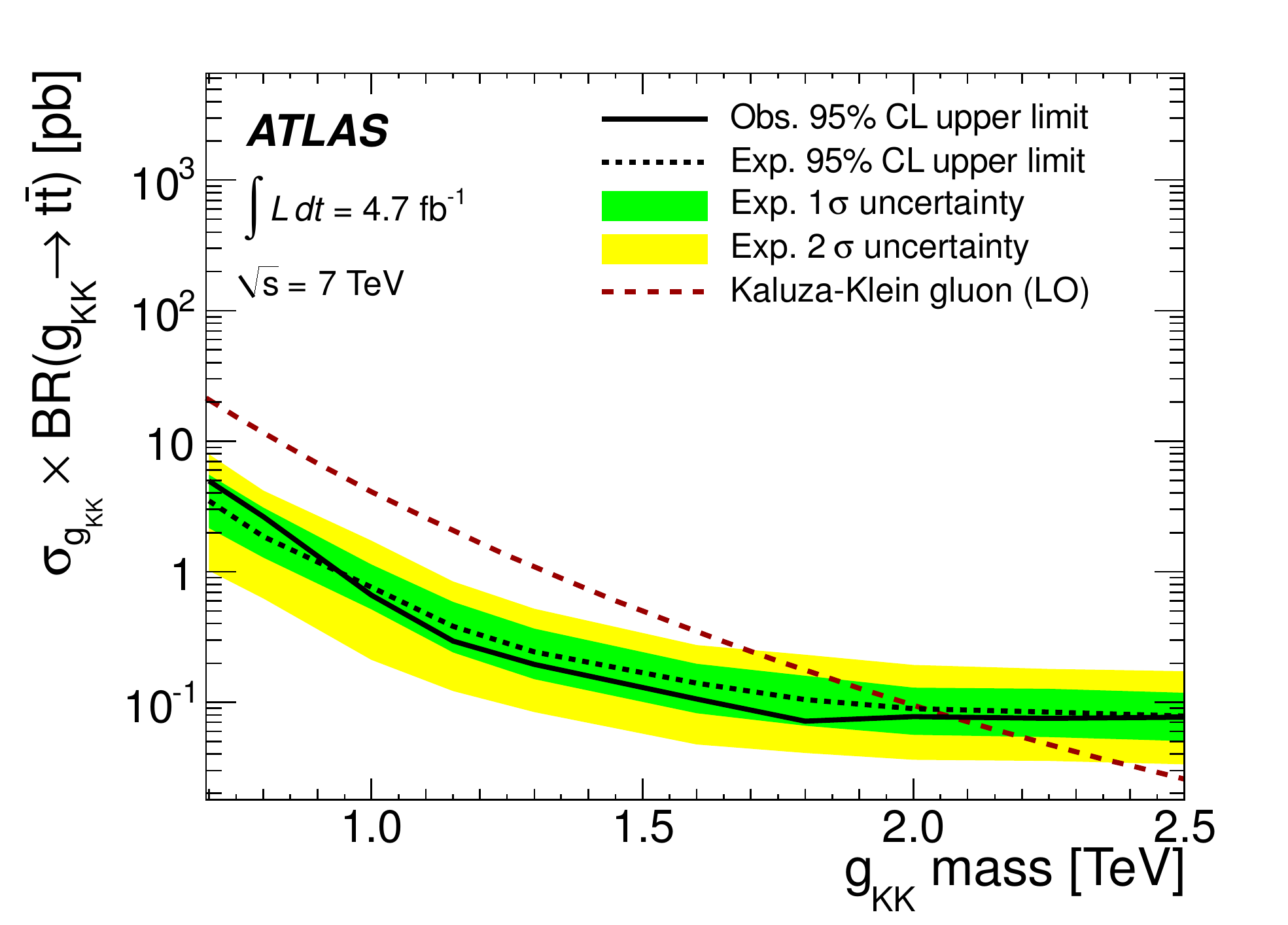}
}
\caption{The $95\%$ credibility level upper limit on the production cross section
times branching ratio into \ttbar for (a) a leptophobic \Zprime boson with
width $\Gamma_{\Zprime}/m_{\Zprime} = 1.2\%$ and (b) a Kaluza-Klein gluon with
width $\Gamma_{\gKK}/m_{\gKK} = 15.3\%$. The limits are obtained from an analysis
of semileptonic \ttbar decays. From~\cite{Aad:2013nca}.}
\label{fig:ljets_limits}
\end{figure}

The limit obtained when combining the electron and muon events and both the resolved
and the boosted selection is shown in \figref{ljets_limits}a for the \Zprime boson.
Such a particle is excluded to have a mass between $0.5$ and $1.74\TeV$ at $95\%$ credibility level (C.L.).
The limit on the Kaluza-Klein gluon is shown in \figref{ljets_limits}b and excludes masses
$0.7 < m_{\gKK} < 2.07\TeV$. No double counting occurs: the boosted and resolved selection are orthogonal
because only events which fail the boosted selection are considered for the resolved
selection. Also the electron/muon channels are orthogonal.
The final limit setting procedure uses all four \mtt spectra
of \figref{ljets_spectrum} separately, i.e., the histograms are not added for the combined
limit. Adding the histograms would spoil the high mass limit
because the smearing from low true to high reconstructed \mtt in the resolved selection
buries the high mass signal.

Using the data from 2011, the CMS finds limits that are less stringent than those from ATLAS:
the \Zprime boson is excluded by CMS for masses between $0.5$ and $1.49\TeV$ and
the KK gluon in the mass range $1.00$--$1.82\TeV$~\cite{Chatrchyan:2012cx}.\footnote{In the final phase
of the preparation of this text, an analysis of 2012 CMS data~\cite{Chatrchyan:2013lca} was submitted to the preprint archive.
These results are not included in this review.}

%%%%%%%%%%%%%%%%%%%%%%%%%%%%%%%%%%%%%%%%%%%%%%%%%%%%%%%%%%%%%%%%%%%%%%%%%%%%%%%%

\subsection{\ttbar resonances in the hadronic decay channel}

Both ATLAS and CMS have published analyses that use the hadronic decay channel
to search for \ttbar resonances. The analyses use different approaches to
top tagging and background estimation.

%%%%%%%%%%%%%%%%%%%%%%%%%%%%%%%%%%%%%%%%%%%%%%%%%%%%%%%%%%%%%%%%%%%%%%%%%%%%%%%%

\subsubsection{ATLAS \htt analysis}

\label{sec:htt}
ATLAS has analysed $4.7$~fb$^{-1}$ of $\sqrt{s}=7\TeV$ data~\cite{Aad:2012raa}
using both the \ttt and the \htt.
The ATLAS \htt analysis is described in detail in~\cite{Kasieczka2013}.
Events are triggered by requiring a large sum of jet \et
in the event or several jets with low \et. At the analysis level, each event
has to contain at least two \ca $R=1.5$ fat jets with $\pt>200\GeV$ and $|\eta|<2.5$.
Each fat jet serves as input to the \htt and top quark candidates with $\pt>200\GeV$
are kept. The two leading \pt candidates are used to calculate the invariant
mass \mtt.

\begin{figure}[htb]
\centering
\subfigure[]{
   \includegraphics[width=0.45\textwidth,angle=0]{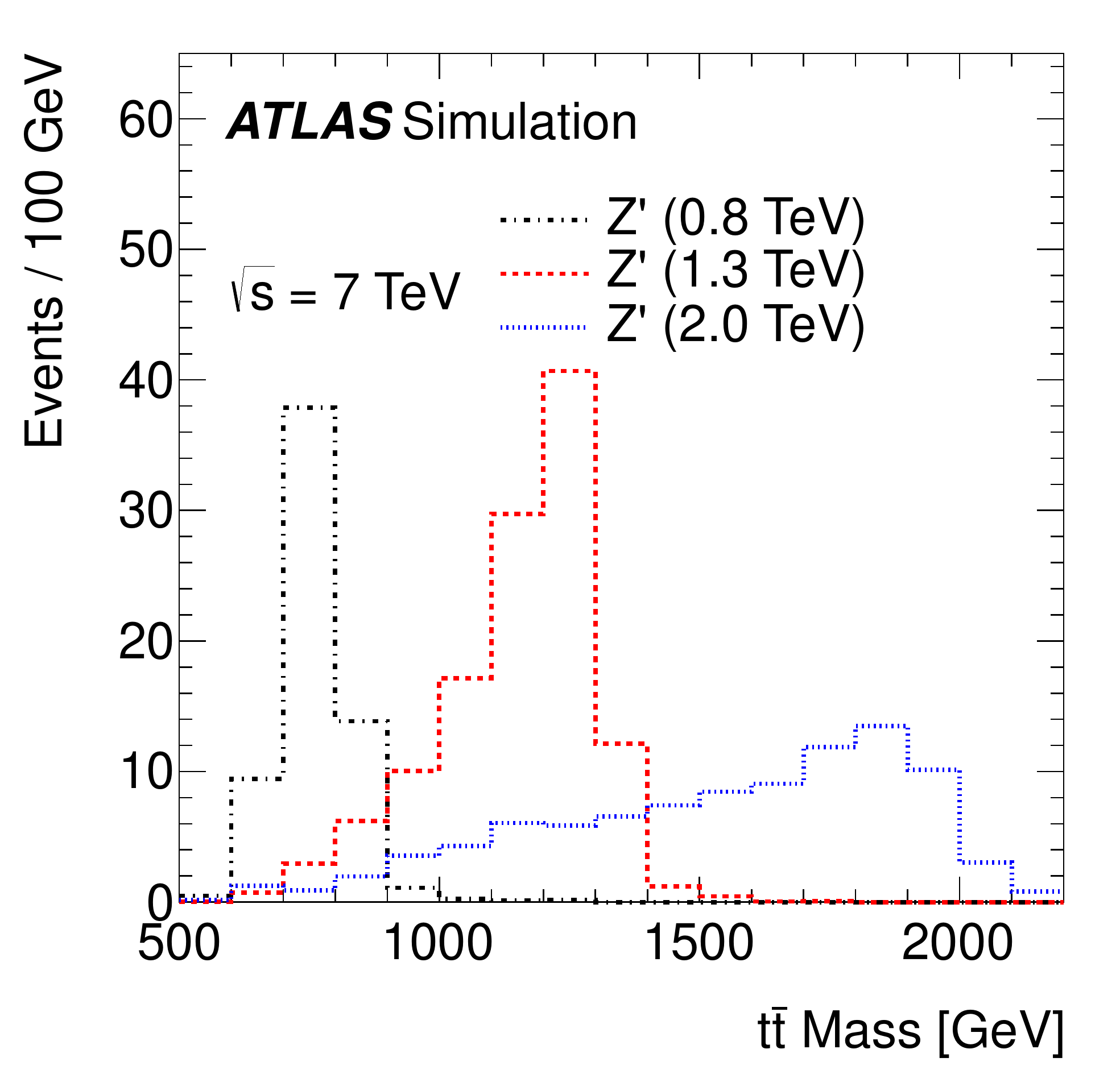}
}
\subfigure[]{
   \includegraphics[width=0.45\textwidth,angle=0]{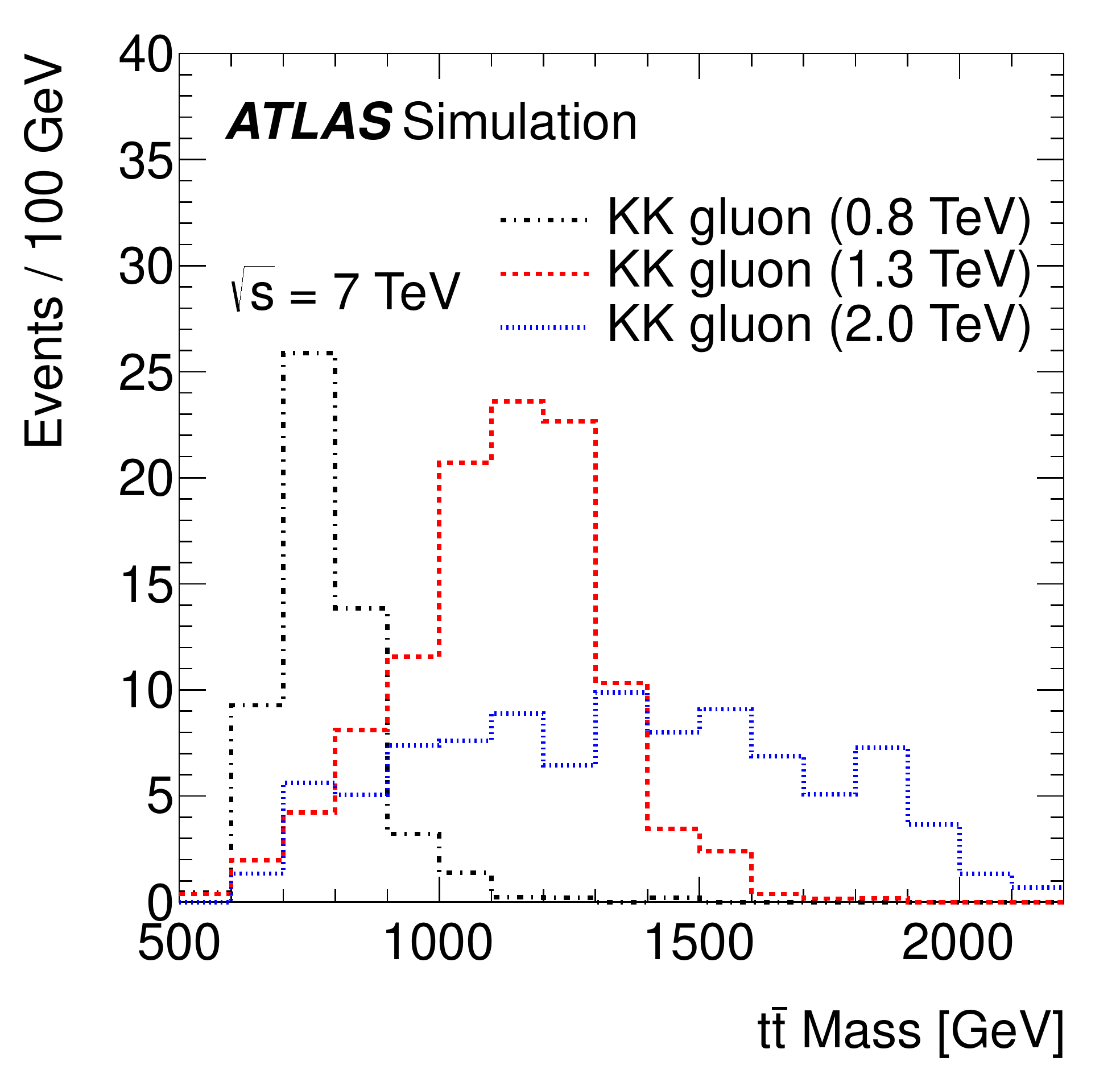}
}
\caption{The \htt reconstructed invariant mass \mtt using simulated ATLAS events with
decays $X \rightarrow \ttbar \rightarrow b q q \, b q q $ for (a) $X=\Zprime$
(from \pythia)
and (b) $X=\gKK$ (from \madgraphpythia). The production cross sections times branching ratios into \ttbar
are set to $1$~pb and the luminosity corresponds to $4.7\invfb$.
From~\cite{Aad:2012raa}.}
\label{fig:htt_masssimulation}
\end{figure}

\figref{htt_masssimulation} shows the \mtt spectrum for simulated events
with decays of \Zprime bosons and Kaluza-Klein gluons of different masses.
The \mtt resolution is better than in the \lpjets analysis (\figref{ljets_masssimulation}).
For a \gKK of mass $1.3\TeV$, the boosted \lpjets distribution has a full width at half maximum
of $500\GeV$ whereas with the \htt the width is $400\GeV$.
For a \Zprime boson of the same mass the widths are $400\GeV$ and $200\GeV$, respectively.
For a \Zprime of mass $2\TeV$ the two widths are equal ($600\GeV$).
These widths are dominated by detector resolution; the natural widths are $1.2\%$ and $15.3\%$ for
the \Zprime boson and the KK gluon, respectively.

Background from multijet events
is reduced by requiring a $b$-tag in the vicinity of each fat jet. A tagged \akt $R=0.4$ $b$-jet
must lie within $\DeltaReta = 1.4$ of each fat jet axis.

The efficiency to select events from \Zprime boson or \gKK decay depends on the
\pt of the top and bottom quarks and is less than $6\%$ for masses of the new
particles between $0.5$ and $2\TeV$.
Compared to the boosted selection in muon+jets events (cf. \secref{ljets}),
the efficiency is similar for masses below $1.3\TeV$ and smaller for larger masses.

\begin{table}[htb]
\centering
\begin{tabular}{|c|c|c|}
\hline
             & $1$ top tag & $\geq 2$ top tags \\
 \hline
 no $b$-tag    &   U (0.3\%)          &    V (2.4\%)    \\
 \hline
 1 $b$-tag     &   W (3.2\%)        &      X (24.3\%)     \\
 \hline
 $\geq 2$\ $b$-tags  &  Y (22.5\%) &    Z (80.9\%)    \\
 \hline
\end{tabular}
\caption{Regions used to validate the simulation and estimate the background
in the \htt analysis. The values in parentheses are SM \ttbar purities
determined from simulation. From~\cite{Aad:2012raa}.}
\label{tab:HTTregions}
\end{table}

\begin{figure}[htb]
\centering
\includegraphics[width=0.45\textwidth,angle=0]{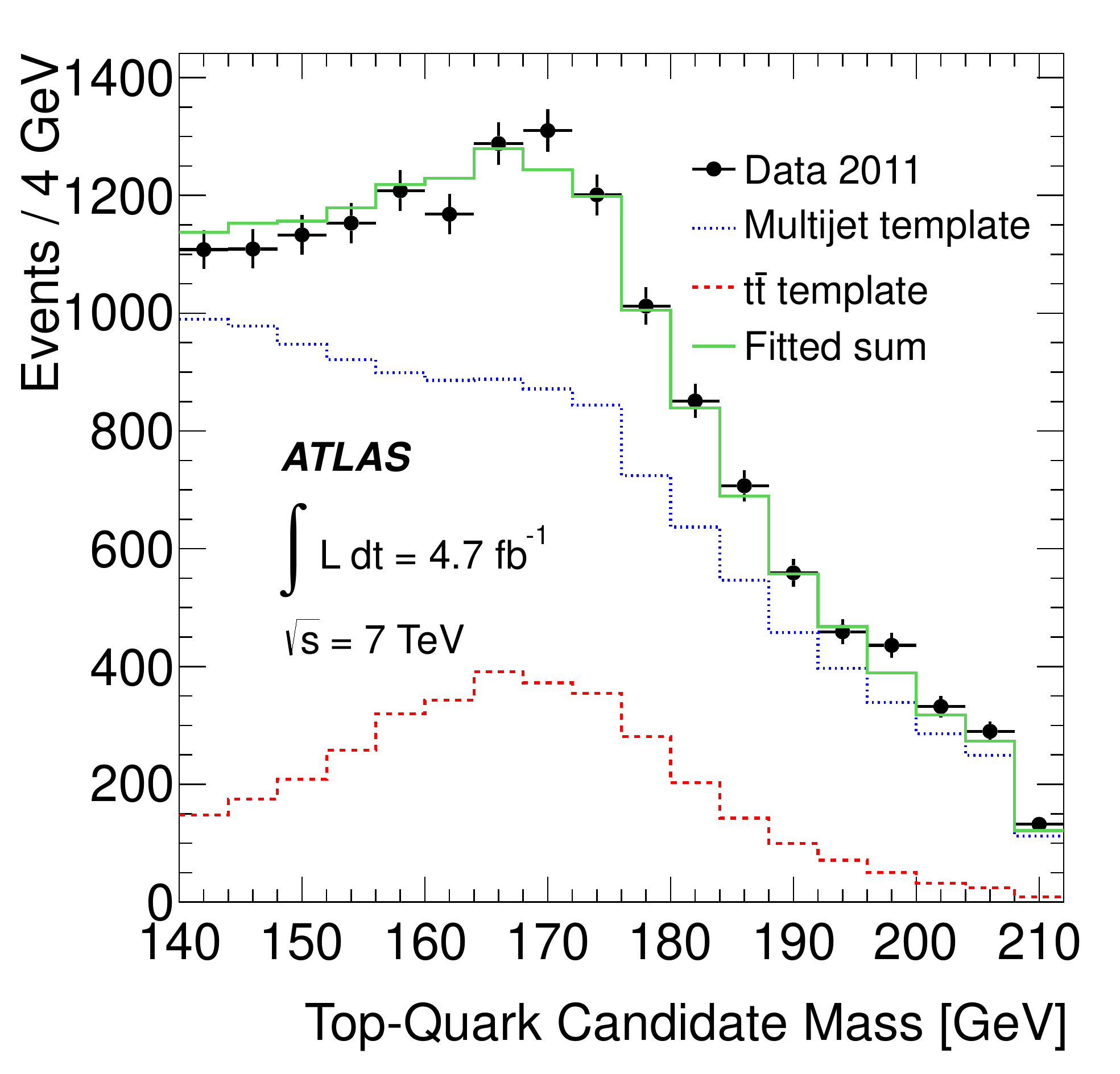}
\caption{The distribution of the \htt top quark candidate mass as measured in
control region Y. Also shown are the contributions from \ttbar simulation (\mcatnlo) and
the template from region W, the normalisations of which have been fitted such
that the sum describes the data. From~\cite{Aad:2012raa}.}
\label{fig:ttbarnorm}
\end{figure}

Control regions are defined by loosening the requirements on the number of
tagged top and bottom quarks as specified in \tabref{HTTregions}. These regions
are used to validate the simulation and estimate the background.
The dominating SM process ($80.9\%$) is \ttbar production and its contribution
to the \mtt spectrum is determined from simulation. The simulation is validated
in region Y, which contains exactly one top tag, by comparing the
measured top quark candidate mass distribution with that predicted by the
\ttbar simulation and a background template taken as the data distribution
in region W (after subtracting a small \ttbar contribution). The background
is dominated by multijet events and the shape of the reconstructed fake top mass
is significantly different from that of reconstructed top quarks.
A fit to the data is performed to determine the normalisations
of the background and \ttbar distributions as shown in
\figref{ttbarnorm}. The fitted sum describes the data distribution very well.
The normalisation obtained in this way for the \ttbar simulation corresponds
to the predicted normalisation within a statistical fit uncertainty of $9\%$.
This result is used to constrain the \ttbar normalisation in the analysis.

The rest of the SM contribution (mostly multijet events) is determined
from data by extrapolating from sideband regions.
For multijet events, the number of $b$-tags and the number of top tags are
uncorrelated to first order because the \htt does not use $b$-tagging information internally.
A small correlation is however present because the bottom and top tagging fake rates
both increase with the \pt of the jet under study (\akt $R=0.4$ and \ca $R=1.5$, respectively).
For the $b$-jets, this was shown in \figref{mv1_eff_pt}b. The \htt fake rate dependence
on the fat jet \pt is displayed in \figref{htt_eff}c.
Multijet events with a $b$-tag have, on average, larger $b$-jet \pt and, because of the
vicinity matching between the $b$-jet and the fat jet, this implies larger
fat jet \pt. The top quark fake rate is therefore larger in $b$-tagged multijet events.
The analysis could be made insensitive to this correlation by applying the
background estimation in bins of fat jet \pt (or in bins of \mtt because
high invariant masses are reached in multijet events only if the
fake top candidates are back-to-back). This has not been done in~\cite{Aad:2012raa}.

In~\cite{Aad:2012raa}, the background distribution as a function of a variable $x$ (e.g., fat jet \pt,
top quark candidate \pt, \mtt) in the signal region Z  is determined
from the average of the distributions in regions V and X, normalised to the
ratio of the event count in region Y to that in U and W, respectively:
\begin{equation}
  \frac{\text{d}n_Z}{\text{d}x} = \frac{1}{2} \left(\frac{n_Y}{n_U} \, \frac{\text{d}n_V}{\text{d}x}
   +\frac{n_Y}{n_W}\, \frac{\text{d}n_X}{\text{d}x}\right)\, ,  \label{eq:HTT_BG_estimate}
\end{equation}
with half the difference between the two averaged estimates serving as a relative
systematic uncertainty of $14\%$~\cite{Kasieczka2013}.
The simulated \ttbar
contribution has been subtracted from all regions in \eqref{HTT_BG_estimate}.
The shape of the spectrum of $x$ is taken from regions V and X which both contain two top quark candidates
so that the procedure is applicable to \mtt without kinematic corrections which
are necessary for the CMS analysis described below (\secref{CMS_hadronic}).
The correlation between $b$-tags and top tags implies that the distributions of
the fat jet \pt, the top quark candidate \pt, and \mtt are different in region Z
than in X and V because more $b$-tags imply larger \pt.

\begin{figure}[htb]
\centering
\subfigure[]{
   \includegraphics[width=0.45\textwidth,angle=0]{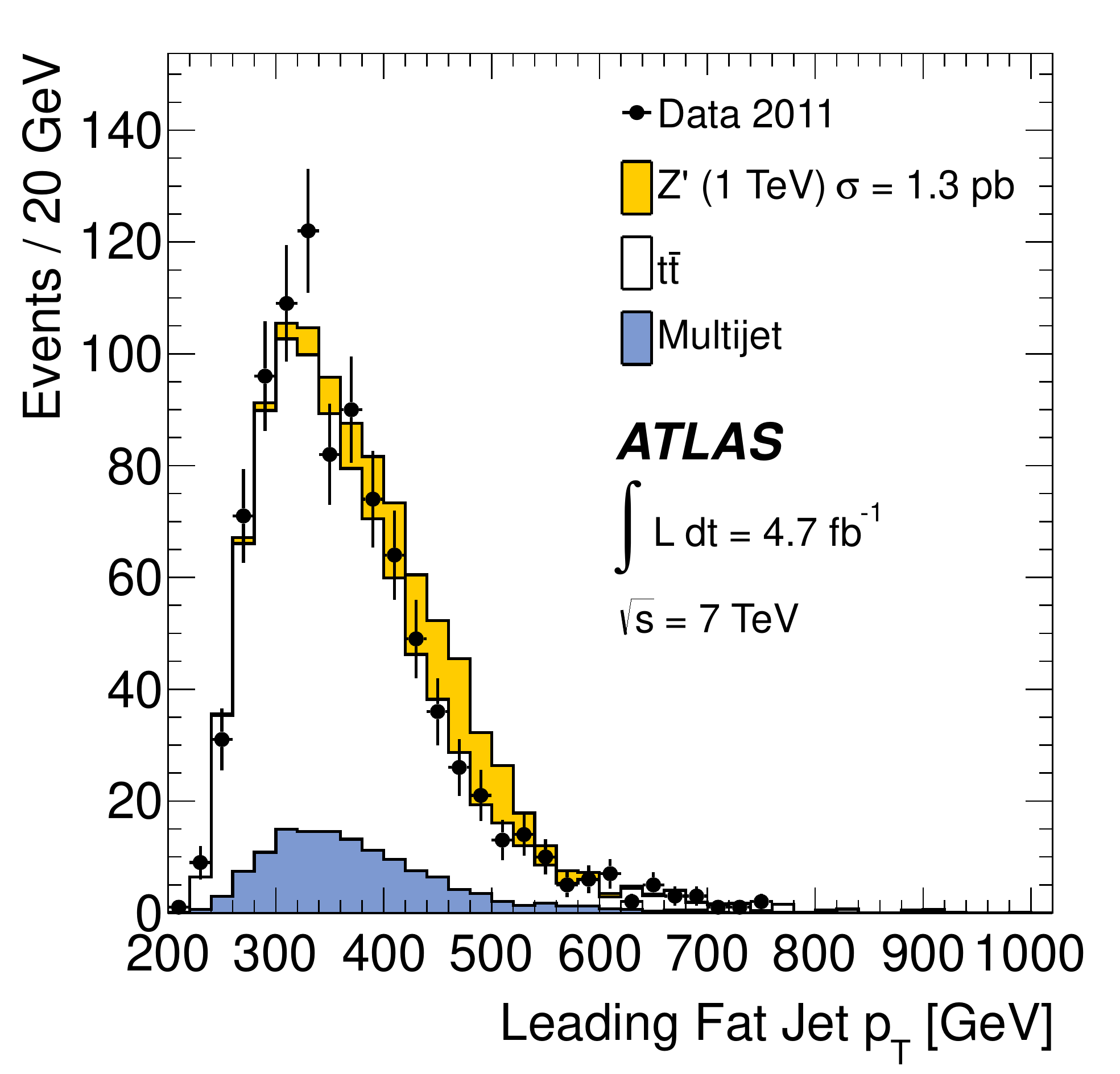}
}
\subfigure[]{
   \includegraphics[width=0.45\textwidth,angle=0]{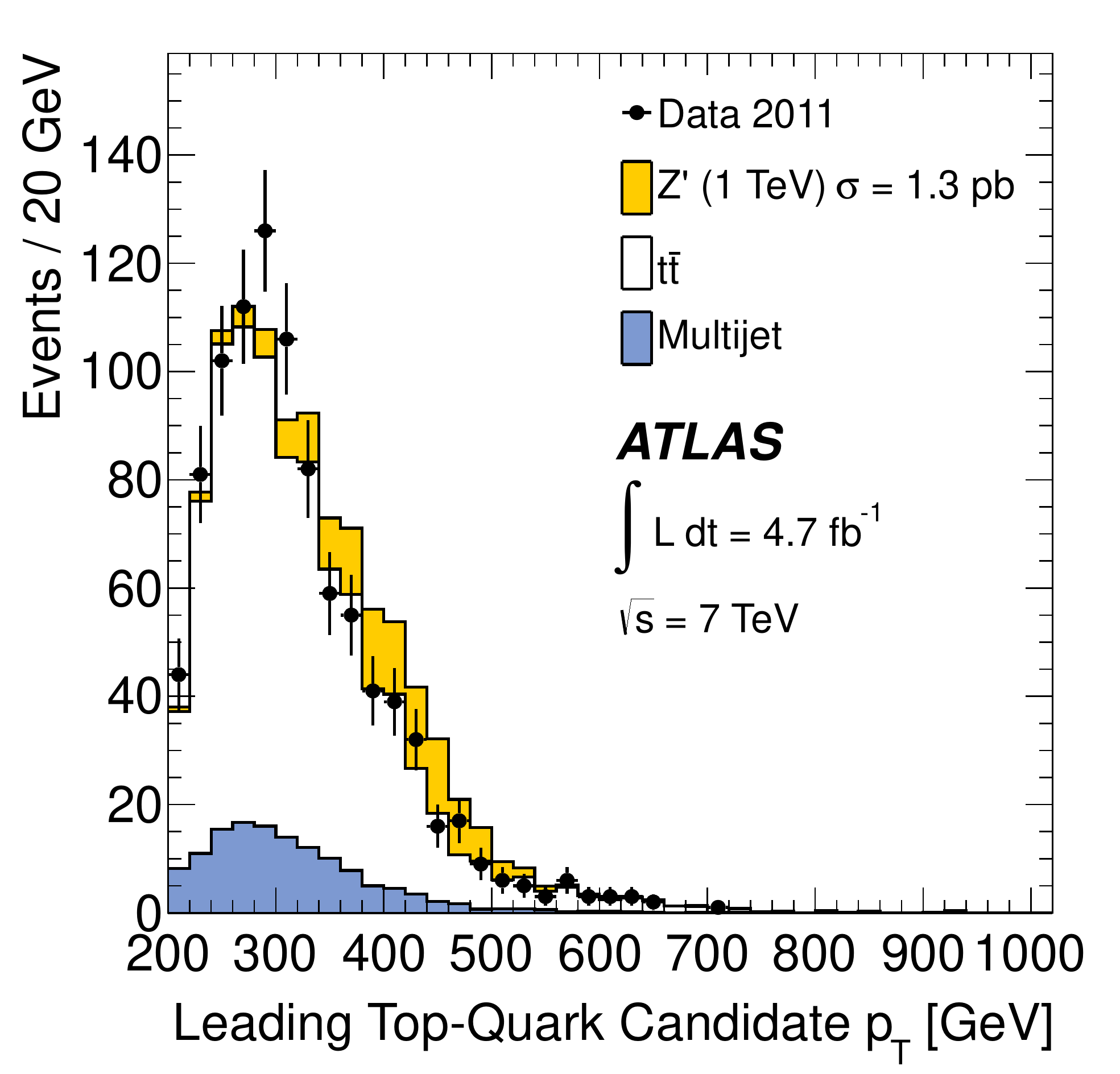}
} \\
\noindent
\subfigure[]{
   \includegraphics[width=0.45\textwidth,angle=0]{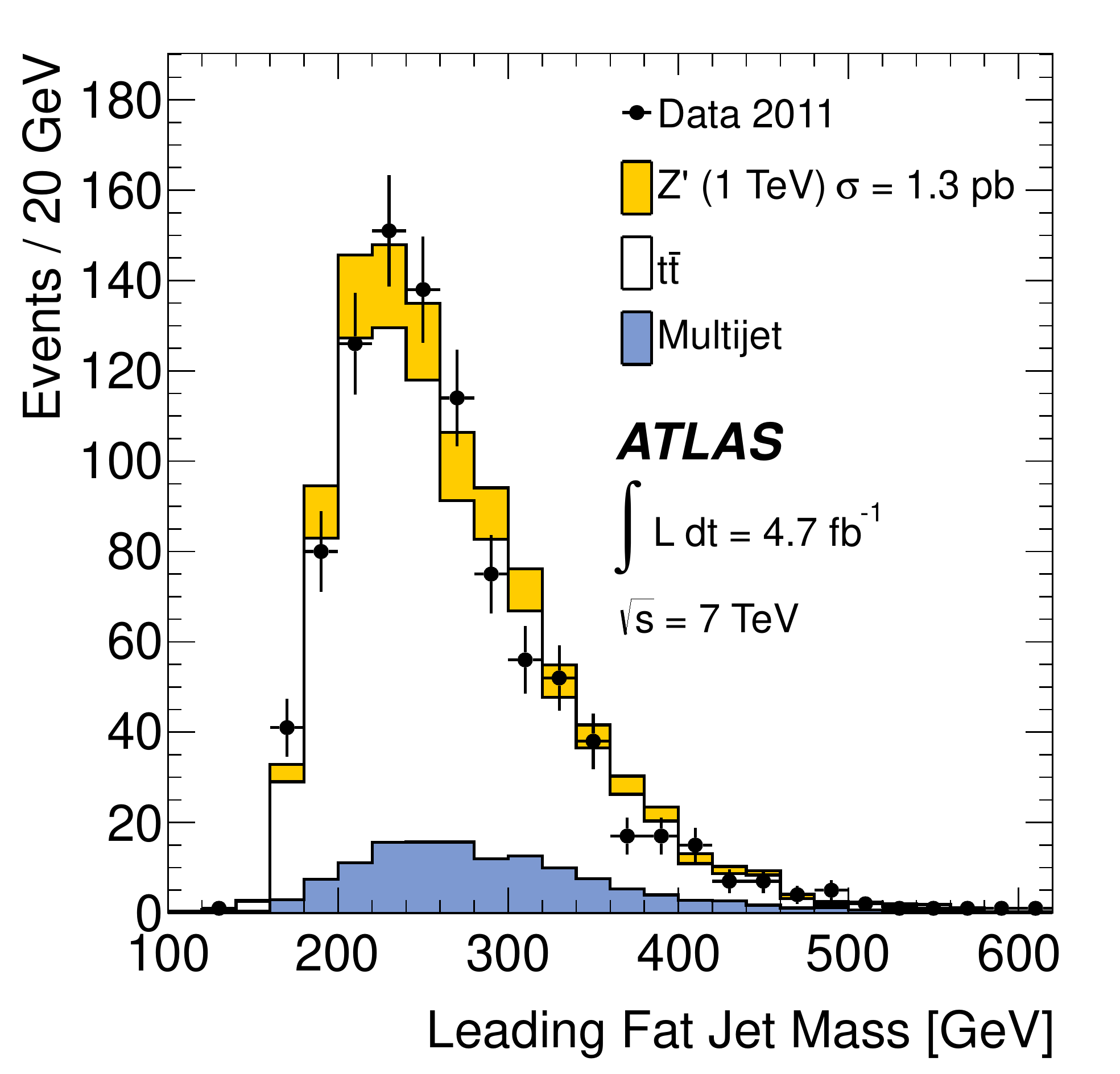}
}
\subfigure[]{
   \includegraphics[width=0.45\textwidth,angle=0]{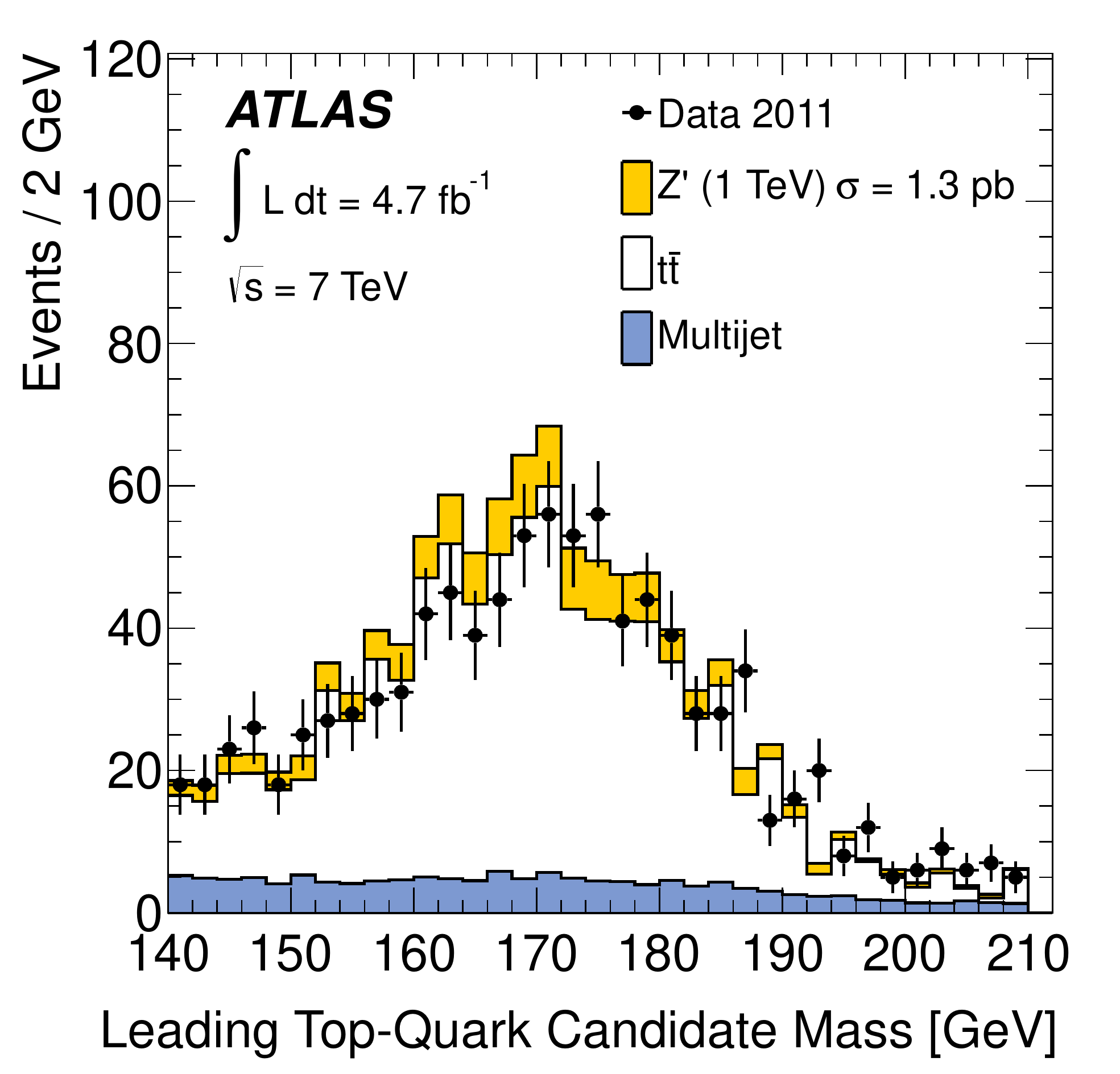}
}
\caption{Control distributions of (a) the leading fat jet \pt, (b) the leading top quark
candidate \pt, (c) the leading \pt fat jet mass, and (d) the leading \pt top quark
candidate mass in the \htt analysis of hadronically decaying \ttbar pairs.
Also shown are contributions from the prediction of SM \ttbar production from simulation (\mcatnlo),
from the data-driven estimate of the remaining SM contributions ({\em Multijet}), and from a hypothetical
\Zprime boson (from \pythia). From~\cite{Aad:2012raa}.}
\label{fig:HTT_controlplots}
\end{figure}

Control plots in the signal region Z of the \pt and mass of the leading \pt fat jet and the top quark candidate
are shown in \figref{HTT_controlplots}. Also shown are the contributions of
SM \ttbar production and the non-top background estimate.
If the correlation between $b$-tags and top tags had a significant impact
on the background estimation, this would show up in the \pt
distributions. Both the \pt and the mass distributions are well described in shape and normalisation by the SM predictions
within the statistical uncertainties of the data.
The impact of the correlation effect therefore has to be smaller than those uncertainties.
Also indicated is the contribution from a \Zprime boson which leads to a clear
overestimation of the data in the leading top quark candidate \pt distribution between $340$ and $540\GeV$
and the leading fat jet \pt between $420$ and $520\GeV$.

The largest systematic uncertainties result from the imperfect knowledge of the
$b$-tagging efficiency ($19\%$ on the \ttbar event yield) and the \ttbar cross section ($12\%$)~\cite{Kasieczka2013}.
The energy scale of the \htt subjets contributes only $8\%$ to the total relative uncertainty.

\begin{figure}[htb]
\centering
\subfigure[]{
   \includegraphics[width=0.45\textwidth,angle=0]{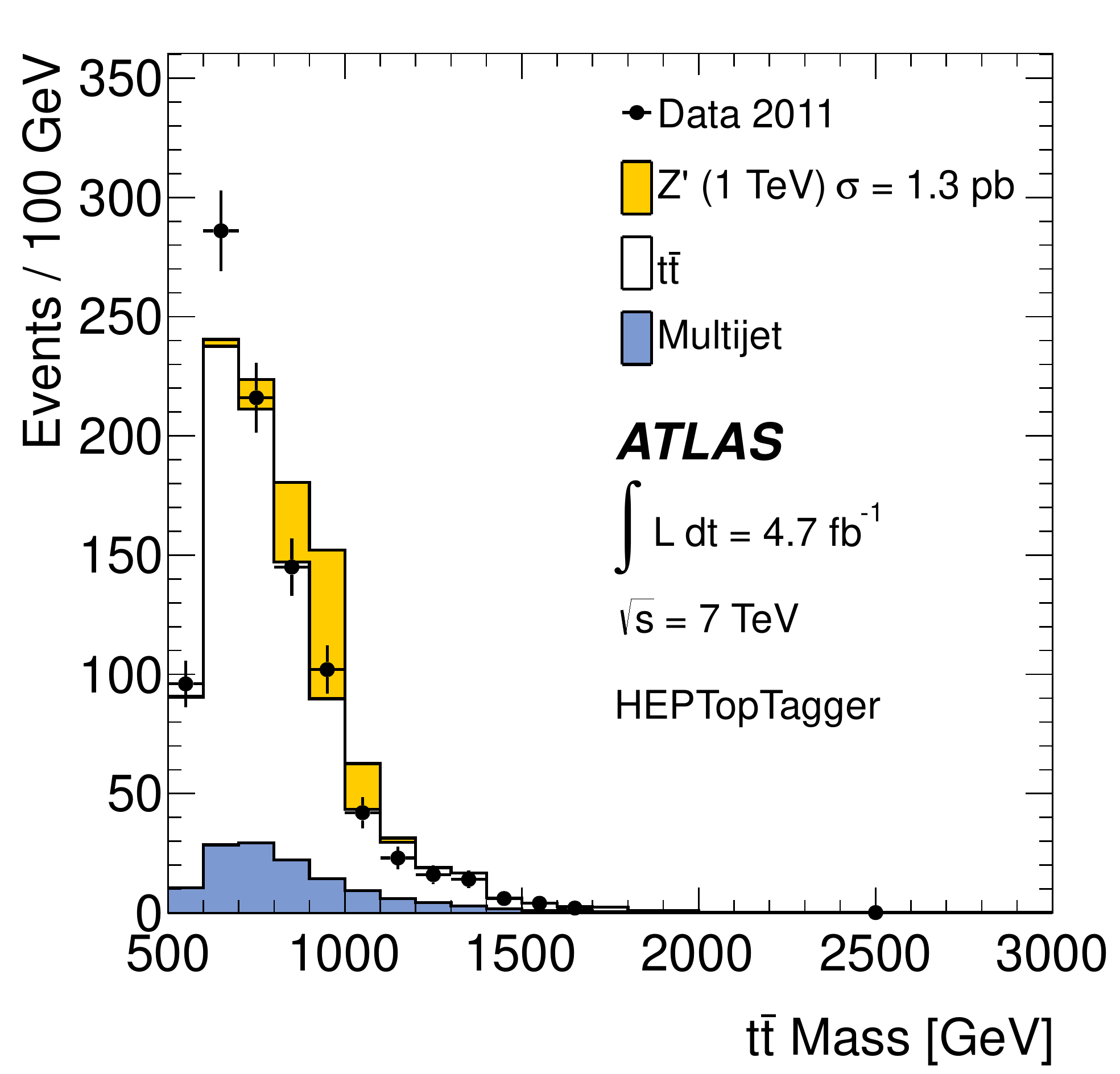}
}
\subfigure[]{
   \includegraphics[width=0.45\textwidth,angle=0]{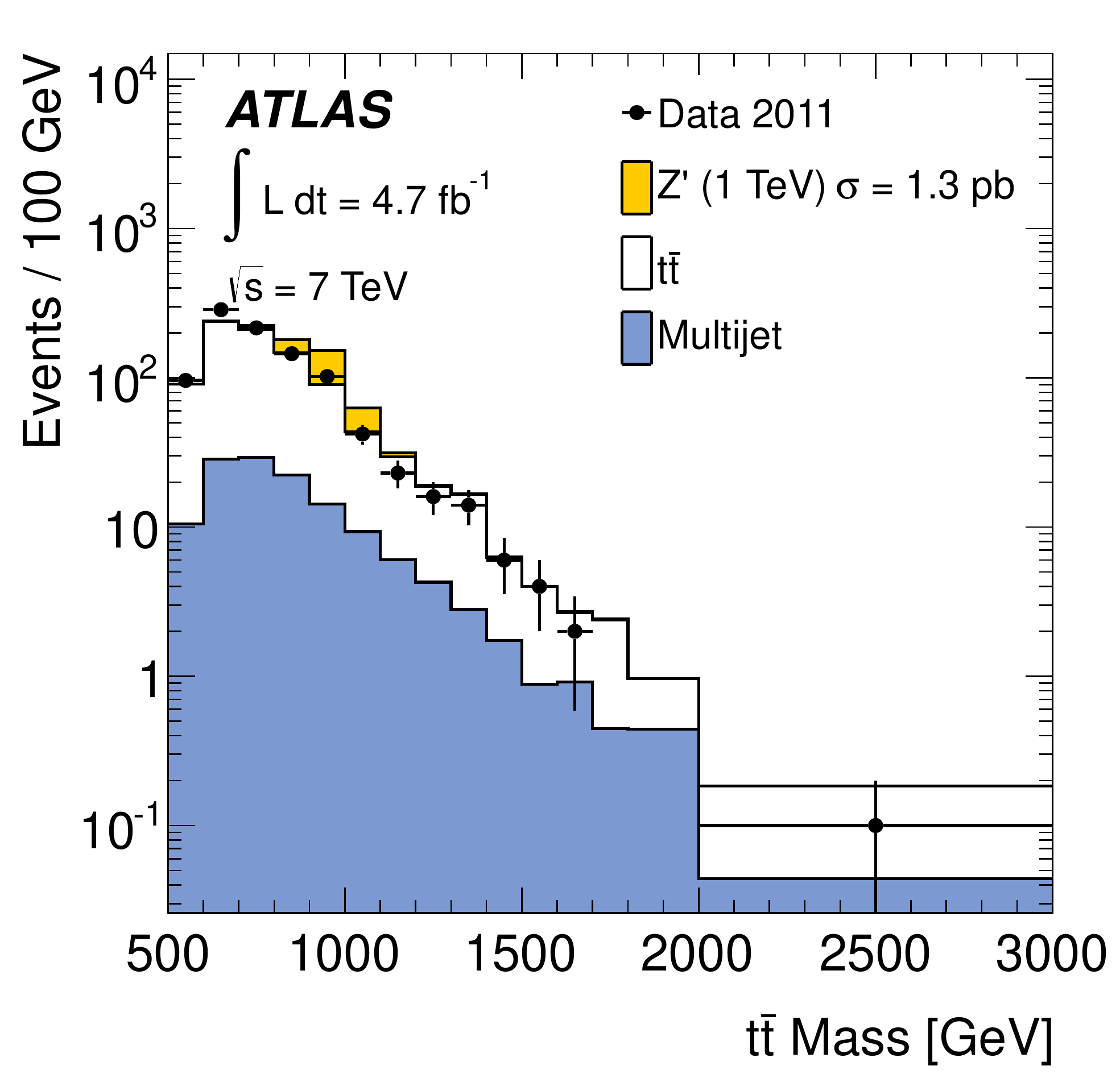}
}
\caption{The \htt reconstructed invariant mass \mtt (a) on a linear vertical
scale and (b) on a logarithmic scale.
Also shown are contributions from the prediction of SM \ttbar production from simulation (\mcatnlo),
from the data-driven estimate of the remaining SM contributions ({\em Multijet}), and from a hypothetical
\Zprime boson (from \pythia). From~\cite{Aad:2012raa}.}
\label{fig:htt_mtt}
\end{figure}

\begin{figure}[hbt]
\centering
\subfigure[]{
   \includegraphics[width=0.45\textwidth,angle=0]{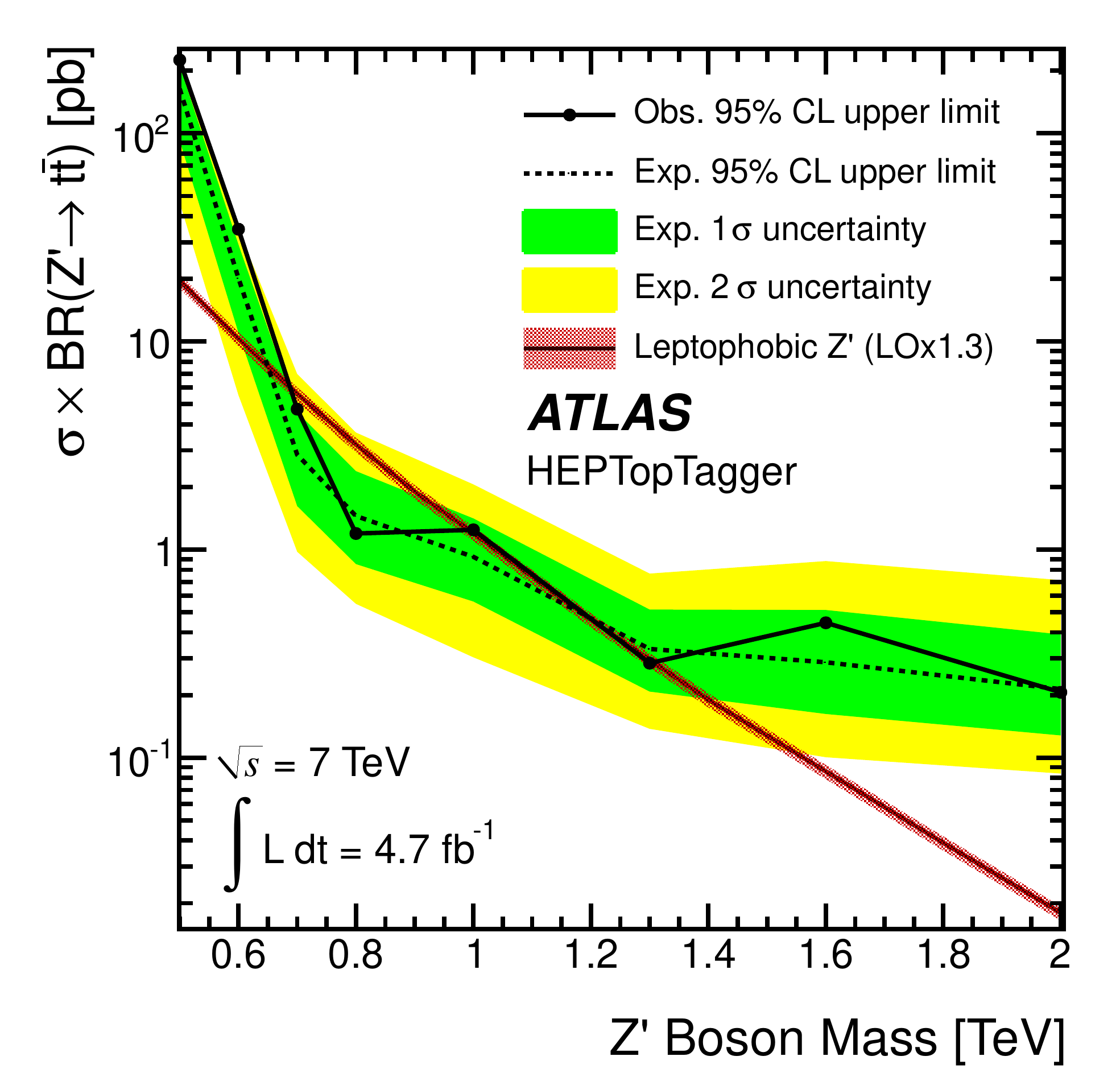}
}
\subfigure[]{
   \includegraphics[width=0.45\textwidth,angle=0]{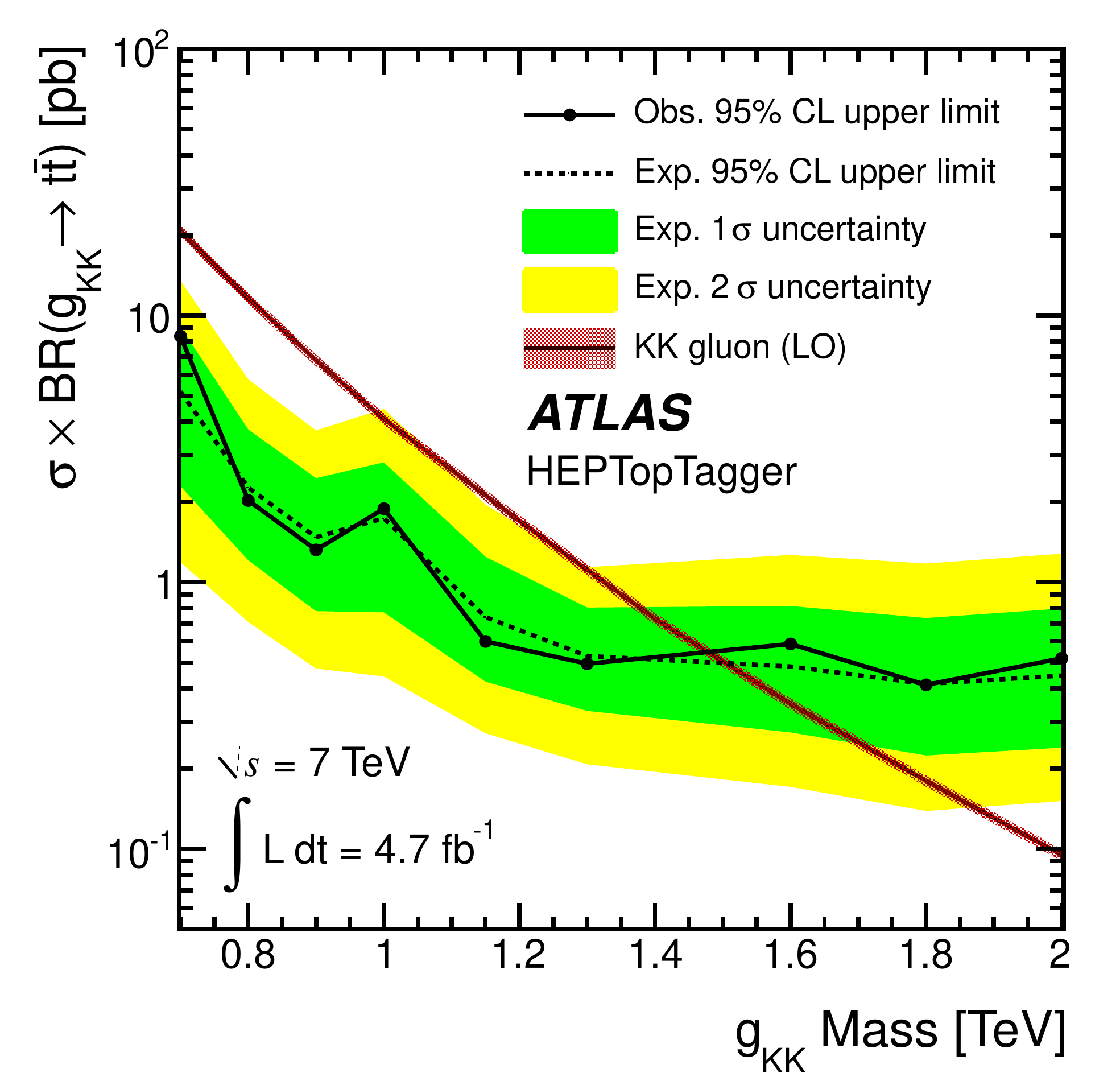}
}
\caption{The $95\%$ credibility level upper limit on the production cross section
times branching ratio into \ttbar for (a) a leptophobic \Zprime boson with
width $\Gamma_{\Zprime}/m_{\Zprime} = 1.2\%$ and (b) a Kaluza-Klein gluon with
width $\Gamma_{\gKK}/m_{\gKK} = 15.3\%$. The limits are obtained from a \htt analysis
of hadronic \ttbar decays. From~\cite{Aad:2012raa}.}
\label{fig:htt_limits}
\end{figure}

The \mtt spectrum is shown in \figref{htt_mtt}. No deviation from the SM
expectation is observed and limits are set on the
production cross section times branching ratio into \ttbar for the \Zprime bosons
and Kaluza-Klein gluon benchmark models described in \secref{ljets}.
The limits are shown in \figref{htt_limits} and exclude the \Zprime boson
with masses $0.70 < m_{\Zprime} < 1.00\TeV$ and $1.28 < m_{\Zprime} < 1.32\TeV$ and
the Kaluza-Klein gluon for $0.70 < m_{\gKK} < 1.48\TeV$ at $95\%$ C.L.
The limits are less stringent than in the lepton+jets case. For low masses, the resolved
selection of the lepton analysis has a higher selection efficiency
than the \htt selection. Compared to the boosted muon selection,
the \htt analysis suffers mainly from an efficiency decrease at high \pt due to the additional
required $b$-tag.

%%%%%%%%%%%%%%%%%%%%%%%%%%%%%%%%%%%%%%%%%%%%%%%%%%%%%%%%%%%%%%%%%%%%%%%%%%%%%%%%

\subsubsection{ATLAS \ttt analysis}
\label{sec:ttt_ana}

ATLAS used the same data set that was used for the lepton+jets (\secref{ljets}) and \htt analyses (\secref{htt})
also for an analysis that employed the \ttt. For every top
quark \pt bin of width $100\GeV$, starting from $450\GeV$, $300,\!000$
templates of parton configurations were calculated and stored.
These numbers were optimised for maximal top quark tagging efficiency at high
background rejection.

The events are triggered by an \akt $R=1.0$ fat jet with $\et>240\GeV$
and are required to have at least two \akt $R=1.0$ fat jets with $|\eta|<2$.
The leading \pt jet is required to have $\pt>500\GeV$ and the subleading jet $\pt>450\GeV$.
These two fat jets must be tagged by the \ttt algorithm.
Each fat jet must have a $b$-tagged \akt $R=0.4$ jet within $\DeltaReta = 1.0$ of the fat jet axis.

The efficiency for selecting events from decays of high mass \Zprime bosons is higher than
for the \htt analysis: $8\%$ for $m_{\Zprime} = 1.6\TeV$,
and $6\%$ for $1.3$ and $2.0\TeV$. The reason for the higher efficiency
is a better top quark finding efficiency when using the \ttt (the same number
of $b$-tags are required in the two analyses).
The \ttt efficiency is $75(5)\%$ for the leading \pt fat jet
and $62(5)\%$ for the subleading fat jet, which are both larger than the plateau efficiency of the \htt ($\approx\!40\%$).
For masses below $1\TeV$, the selection efficiency
is an order of magnitude smaller than in the \htt analysis because of the larger
required fat jet \pt ($500\GeV$ vs. $200\GeV$).

\begin{figure}[hbt]
\centering
\includegraphics[width=0.48\textwidth,angle=0]{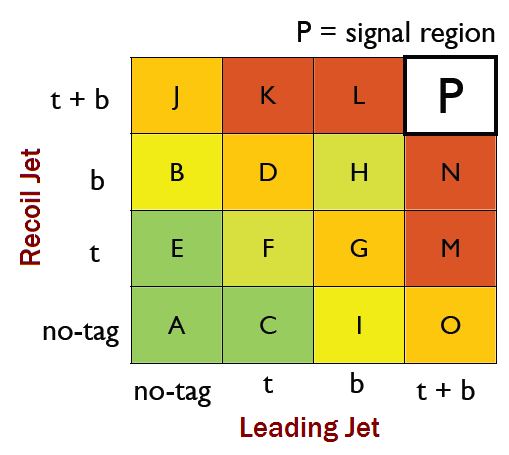}
\caption{Subsamples used to determine the non-\ttbar background in the \ttt
ATLAS analysis of hadronic \ttbar decays. The division is by bottom and top tagging
states of the leading \pt and subleading \pt ({\em Recoil Jet}) fat jets. From~\cite{Aad:2012raa}.}
\label{fig:ttt_BG_samples}
\end{figure}

The SM \ttbar background is determined from simulation. The remaining SM contribution
is determined using a sideband extrapolation method based on the bottom and top
tagging states of the leading and subleading \pt fat jets.
The data are divided into 16 subsamples as shown in \figref{ttt_BG_samples}.
The top tagging fake rates for the leading and subleading \pt
fat jet in multijets events are correlated because the transverse momenta of
the two fat jets are correlated:
the events are dominated by hard $2\to 2$ parton scattering with equal \pt in the
centre-of-mass system of the hard scatter. A simple extrapolation using regions
A, O, and J to estimate the event count in P is therefore not possible.
Instead, the method relies on the assumption that the $b$-tagging and top tagging fake
rates are independent. This assumption is only approximately valid because both rates depend
on \pt as discussed in \secref{htt}. The regions K, L, M, and N can contain sizeable contributions from physics beyond
the standard model and are not used directly.
Instead, the event counts in K and N are estimated from regions with smaller
contamination and these estimates are then used to predict the count in region P:
\begin{eqnarray}
N_{K'} & = & N_J \times N_F/N_E\, , \\
N_{M'} & = & N_F \times N_O/N_C\, , \\
N_{P}  & = & N_{K'} \times N_{M'}/N_F\, .
\end{eqnarray}
A direct comparison of the control regions with those in the \htt analysis is not possible because
the latter made use of all fat jets in the event and not just the two leading \pt ones.
The \htt control region U corresponds to regions C and E, and region W to D and G.
Region V includes region F, region X includes regions K and M, and region Y
includes L and N.
It is interesting that K, L, M, N could not be used directly in the \ttt analysis
because of signal contamination while no such problem was observed in the \htt analysis.
This could be explained if the correlation with $b$-tagging is different
for the two top finding algorithms.

Other combinations of control regions are used in the \ttt analysis to
estimate a systematic uncertainty on the non-\ttbar background.
The dominating systematic uncertainties in the analysis are from $b$-tagging, the fat jet
energy scale, and the SM \ttbar normalisation.

\begin{figure}[hbt]
\centering
\subfigure[]{
   \includegraphics[width=0.45\textwidth,angle=0]{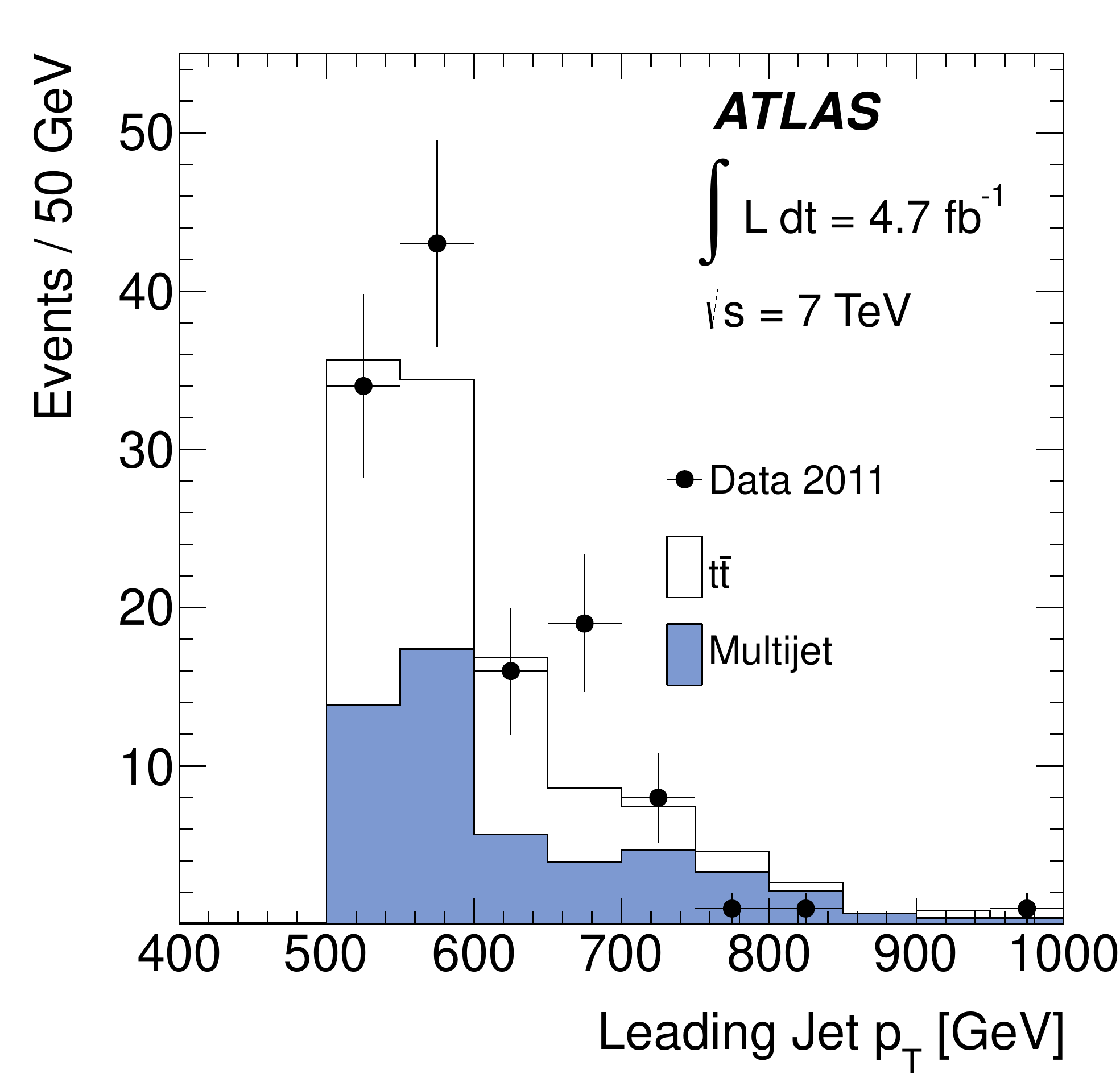}
}
\subfigure[]{
   \includegraphics[width=0.45\textwidth,angle=0]{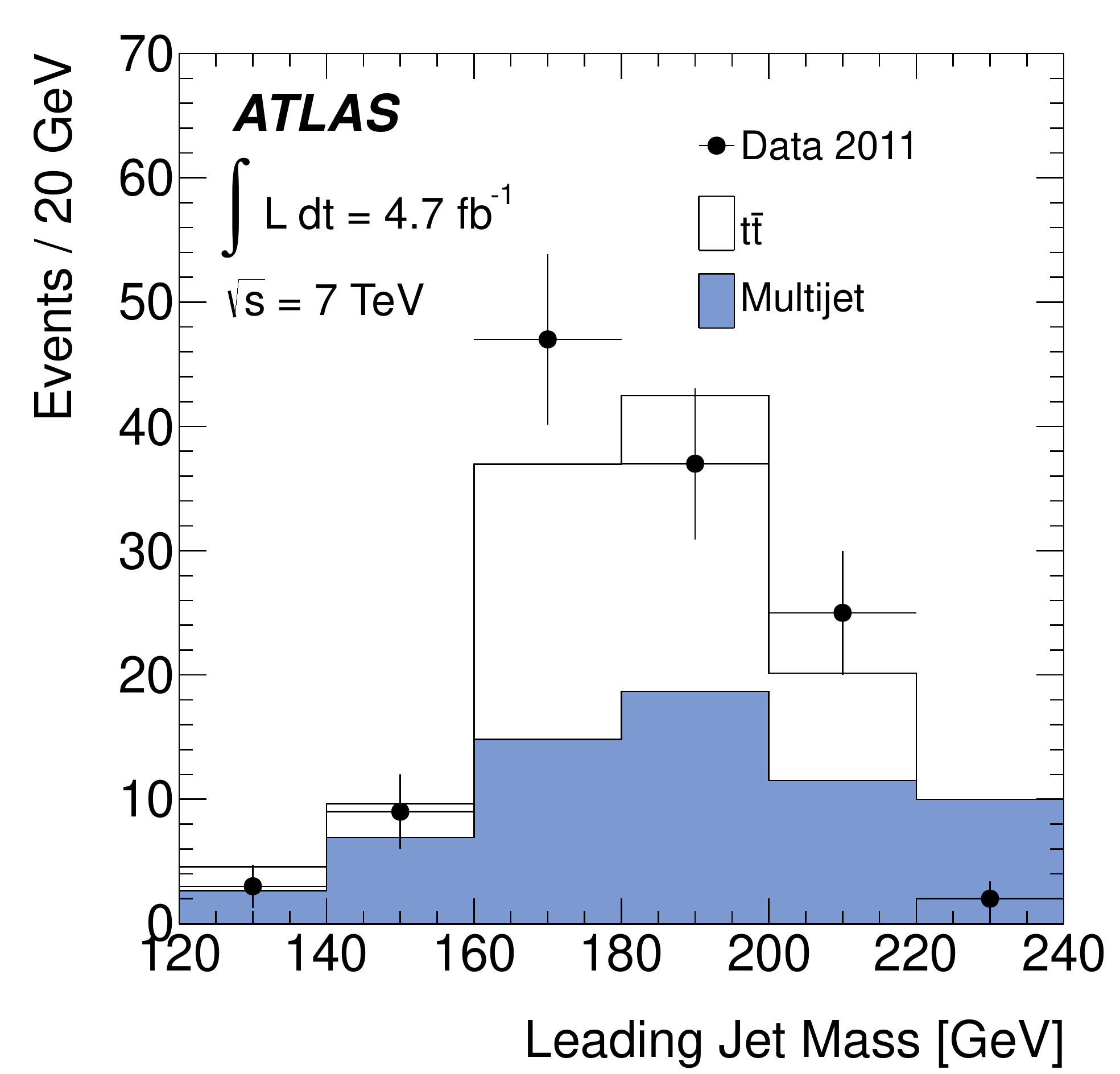}
} \\
\noindent
\subfigure[]{
   \includegraphics[width=0.45\textwidth,angle=0]{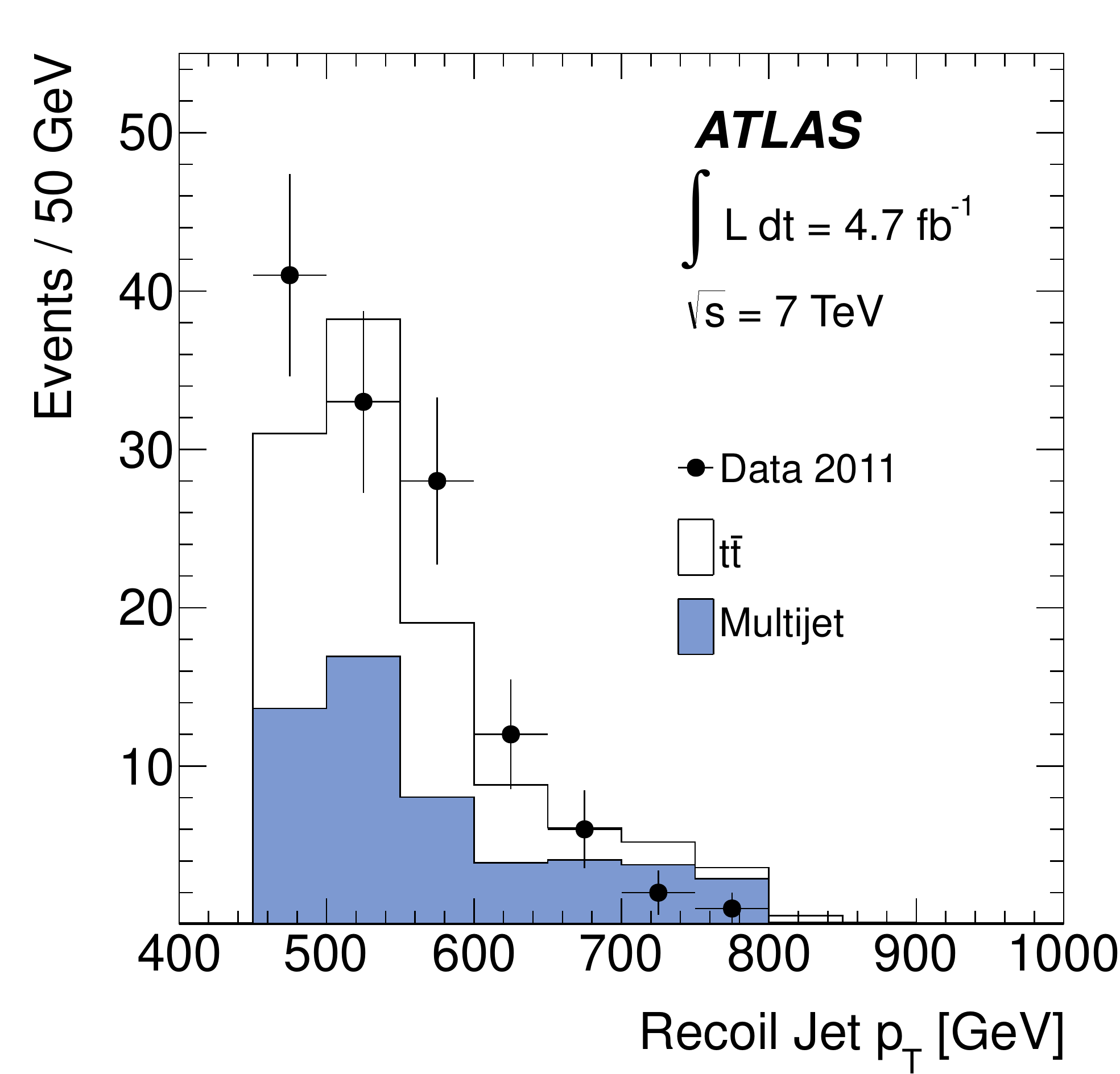}
}
\subfigure[]{
   \includegraphics[width=0.45\textwidth,angle=0]{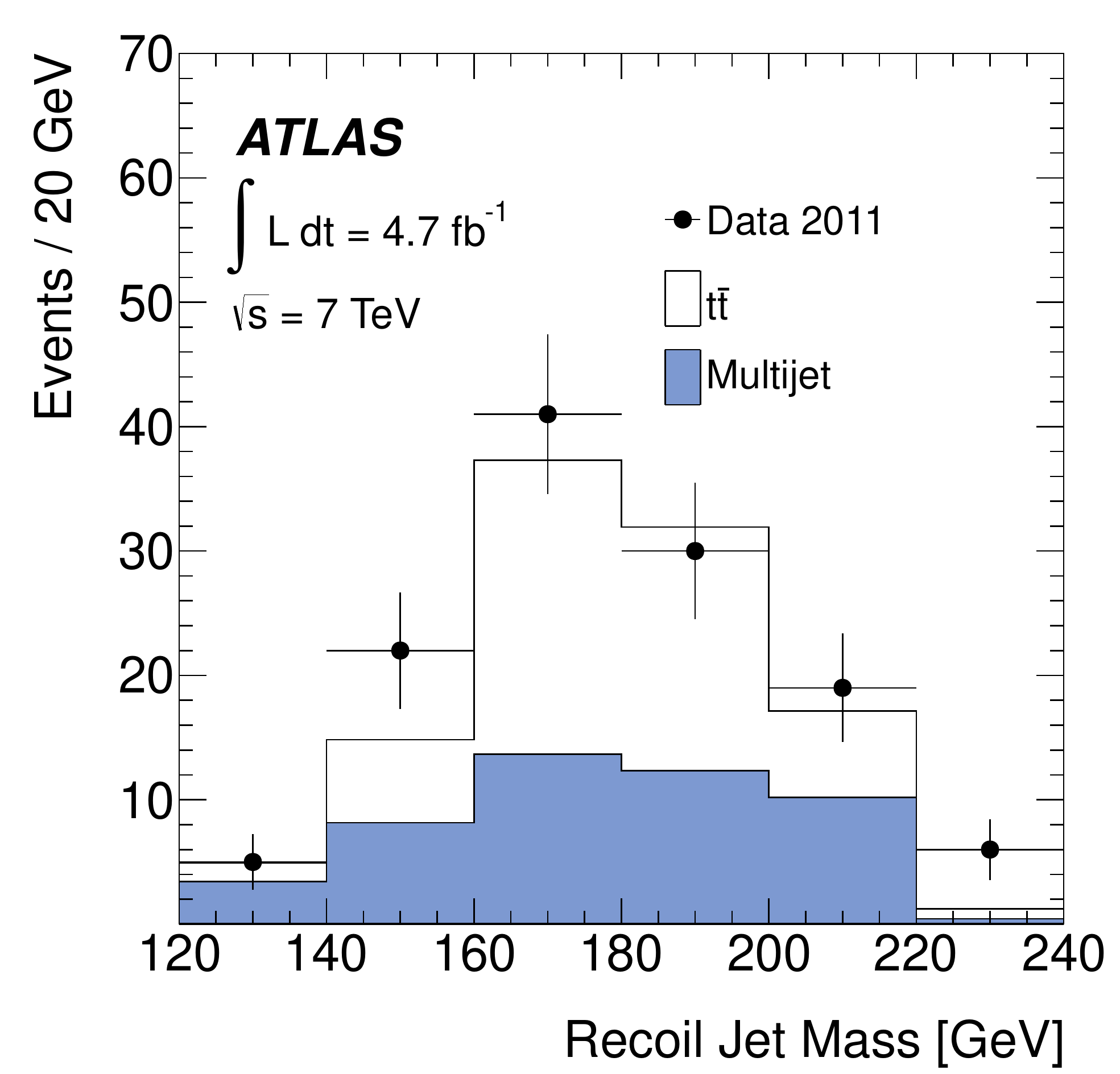}
}
\caption{Distributions of (a) the leading fat jet \pt, (b) the leading \pt fat jet mass,
(c) the subleading fat jet \pt, and (d) the subleading \pt fat jet mass
in the \ttt analysis of hadronically decaying \ttbar pairs.
Also shown are contributions from the prediction of SM \ttbar production from simulation (\mcatnlo),
and from the data-driven estimate of the remaining SM contributions ({\em Multijet}). From~\cite{Aad:2012raa}.}
\label{fig:ttt_controlplots}
\end{figure}

Control plots of the \pt and mass of the two tagged fat jets are shown in \figref{ttt_controlplots}.
The statistical uncertainties of the data are large because of the high fat jet \pt.
For high leading \pt fat jet mass the estimated multijet contribution overestimates the data
significantly whereas in all other bins it is compatible with the data within the
statistical data uncertainty.
The ratio of \ttbar to multijet events in the fat jet \pt spectrum is $\approx\!1$
which has to be compared to a ratio of $\approx\!4$ in the \htt analysis which also
holds for fat jet $\pt > 500\GeV$.
In \cite{Aad:2012raa}, the \ttt mis-tag rate for a light quark or gluon fat jet
is quoted to be $\approx\!10\%$. The corresponding \htt fake rate is $2.5\%$~\cite{Aad:2013gja,ATLAS-CONF-2013-084}.

\begin{figure}[hbt]
\centering
\subfigure[]{
   \includegraphics[width=0.45\textwidth,angle=0]{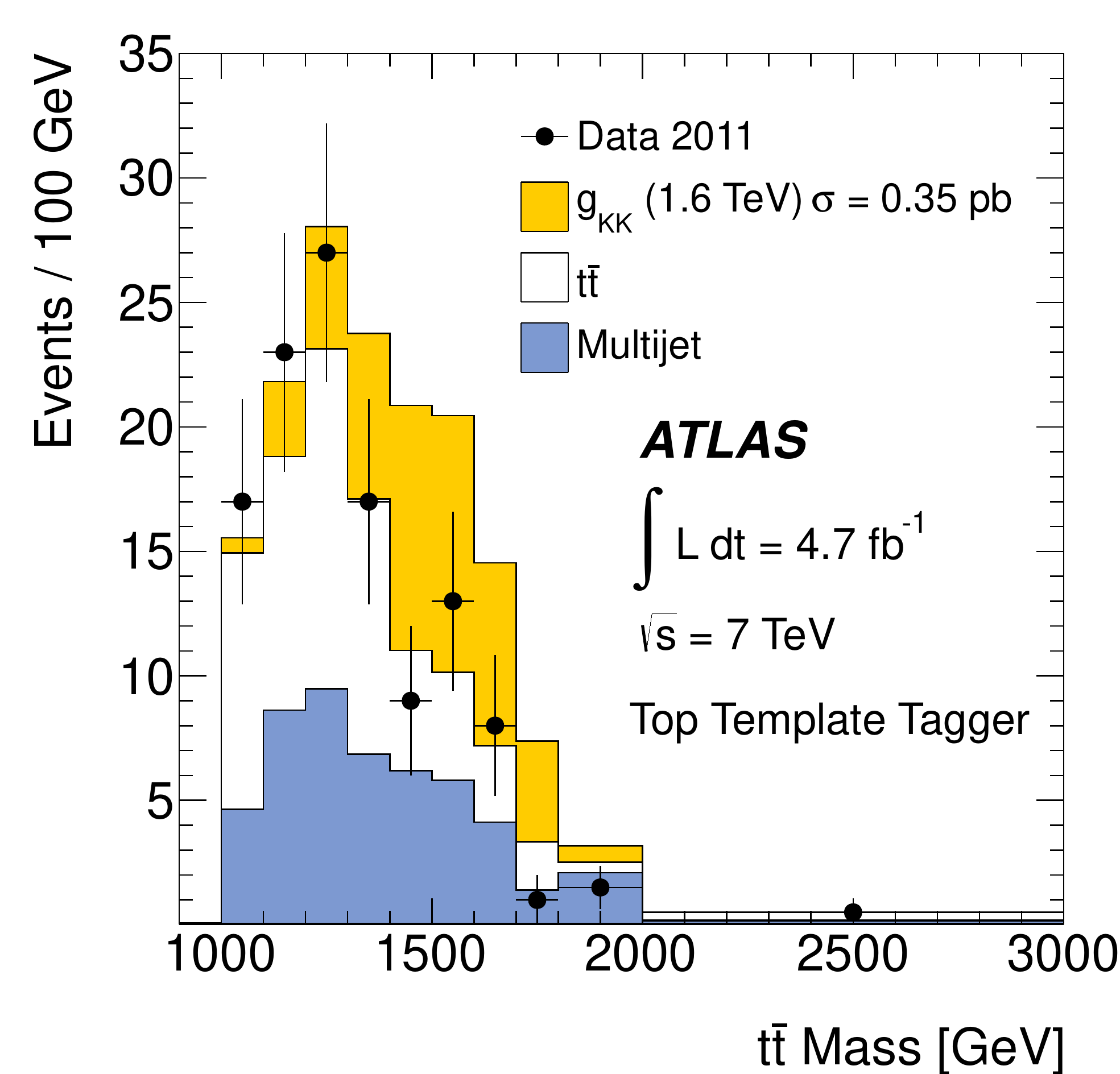}
} \\
\noindent
\subfigure[]{
   \includegraphics[width=0.45\textwidth,angle=0]{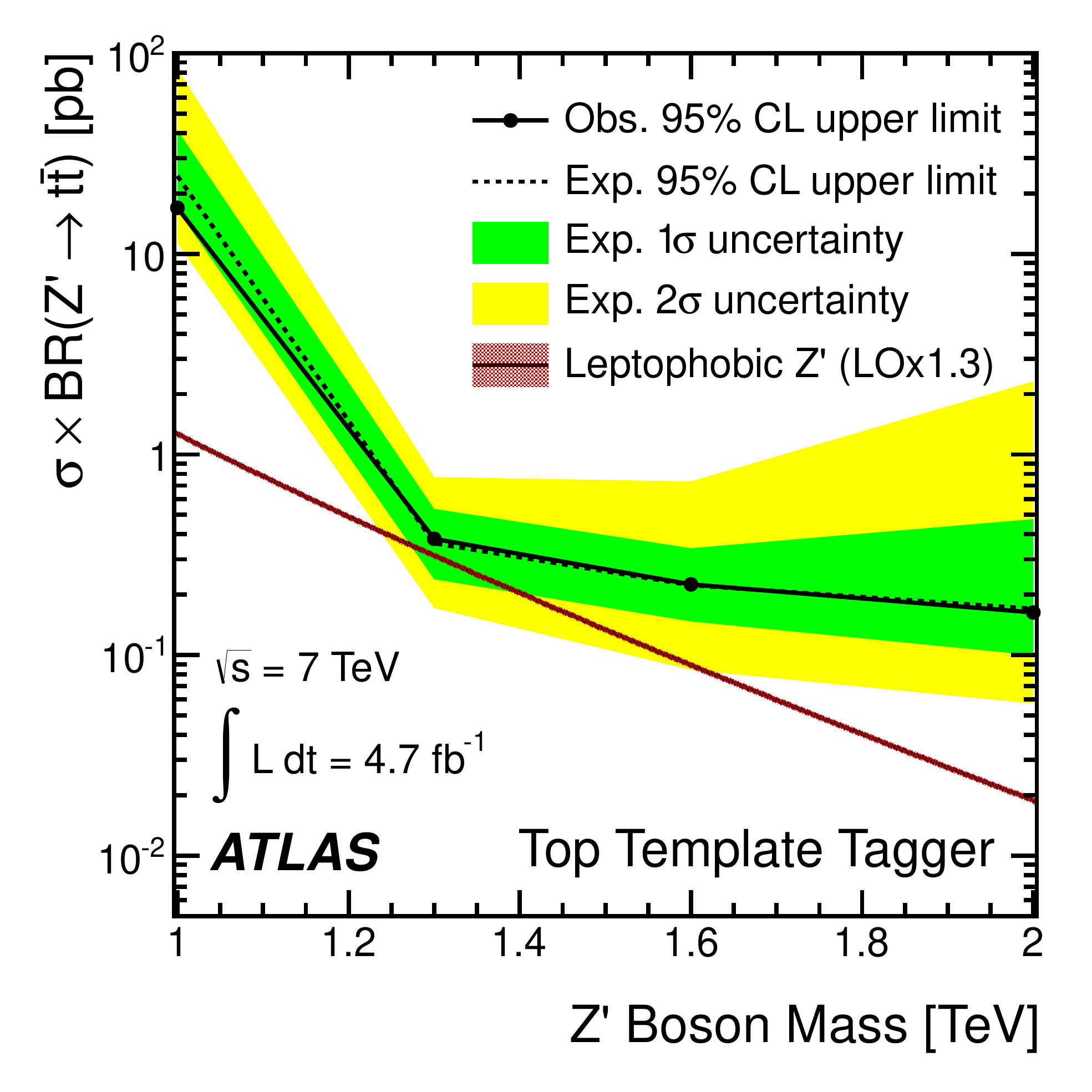}
}
\subfigure[]{
   \includegraphics[width=0.45\textwidth,angle=0]{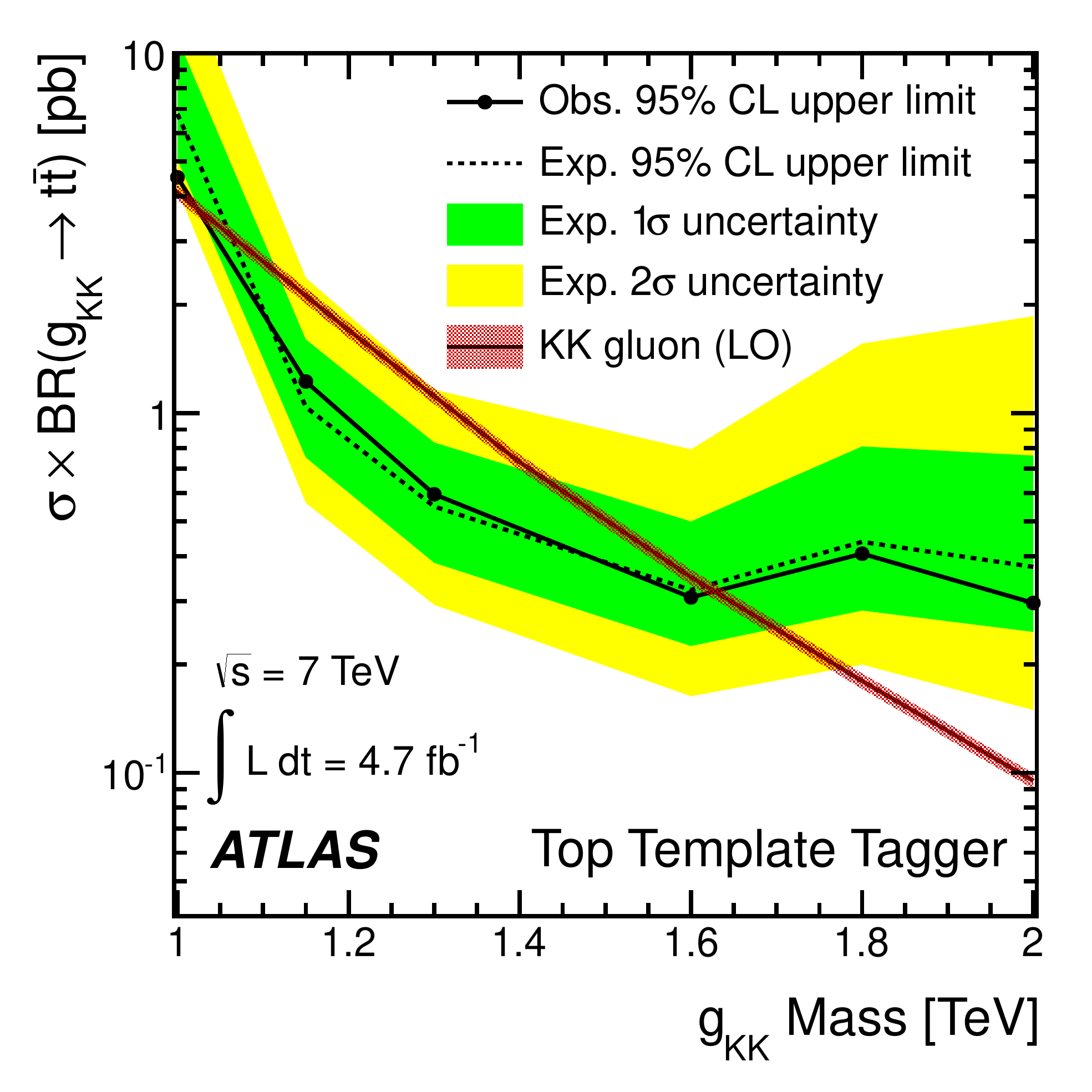}
}
\caption{(a) The invariant mass \mtt reconstructed with the \ttt.
Also shown are contributions from the prediction of SM \ttbar production from simulation (\mcatnlo),
from the data-driven estimate of the remaining SM contributions ({\em Multijet}), and from a hypothetical
Kaluza-Klein gluon (from \pythia).
The $95\%$ credibility level upper limit on the production cross section
times branching ratio into \ttbar for (b) a leptophobic \Zprime boson with
width $\Gamma_{\Zprime}/m_{\Zprime} = 1.2\%$ and (c) a Kaluza-Klein gluon with
width $\Gamma_{\gKK}/m_{\gKK} = 15.3\%$. From~\cite{Aad:2012raa}.}
\label{fig:ttt_results}
\end{figure}

The invariant mass spectrum \mtt is shown in \figref{ttt_results}a. No excess
of events over the SM contributions is observed and limits are set.
No exclusion limits can be set on the masses of the narrow width \Zprime boson benchmark model as shown in \figref{ttt_results}b.
For the Kaluza-Klein gluon model the excluded masses are $1.02 < m_{\gKK} < 1.62$ at $95\%$ C.L. (\figref{ttt_results}c).
The upper limit is better than the one from the \htt due to the higher analysis efficiency
at high mass. The lower mass limit is significantly less stringent than that from
the \htt.

%%%%%%%%%%%%%%%%%%%%%%%%%%%%%%%%%%%%%%%%%%%%%%%%%%%%%%%%%%%%%%%%%%%%%%%%%%%%%%%%

\subsubsection{CMS Top Tagger analysis}
\label{sec:CMS_hadronic}
A search for hadronically decaying \ttbar resonances was published by CMS in~\cite{Chatrchyan:2012ku}
using $5\invfb$ of data collected at $\sqrts = 7\TeV$.\footnote{An analysis of
2012 CMS data~\cite{Chatrchyan:2013lca} was submitted to the preprint archive during the final phase
of the preparation of the manuscript. These results
are not considered in this review.}
Events are triggered by requiring a jet with $\pt > 300\GeV$ ($240\GeV$ in early
2011 running conditions).
Fat jets are selected offline using the \ca algorithm with $R=0.8$.
The detector volume is divided into two hemispheres,
corresponding to positive and negative pseudorapidities, respectively.
The fat jet multiplicities in the two hemispheres are used to classify the events into two categories:
the `1+1' channel with one fat jet in either hemisphere and the `1+2' channel
for which one hemisphere has at least two fat jets.

The fat jets in single-fat-jet hemispheres must be tagged by the CMS Top Tagger
algorithm discussed in \secref{cms_tagger}.
The top quark candidate \pt is required to be larger than $350\GeV$.

One of the fat jets that share a hemisphere in 1+2 events must
be tagged as coming from hadronic \W boson decay.
The \W boson tag requires the pruned jet mass to be between
$60$ and $100\GeV$ (the pruning parameters $\zcut$ and $\Dcut$ are presumably chosen
to correspond to the values in the original publication~\cite{Ellis:2009su}
which are given in \secref{pruning}) and two subjets with a mass drop of at least $60\%$.
The fat jet closest to the \W boson candidate jet in $(\eta,\phi)$ space is taken
to be the $b$-jet (although no $b$-tagging algorithm is used).
The combination of the four-momenta of these two fat jets has to yield
a mass in the range $140$ to $250\GeV$ (\Wpb tag).

\begin{figure}[hbt]
\centering
\subfigure[]{
   \includegraphics[width=0.6\textwidth,angle=0]{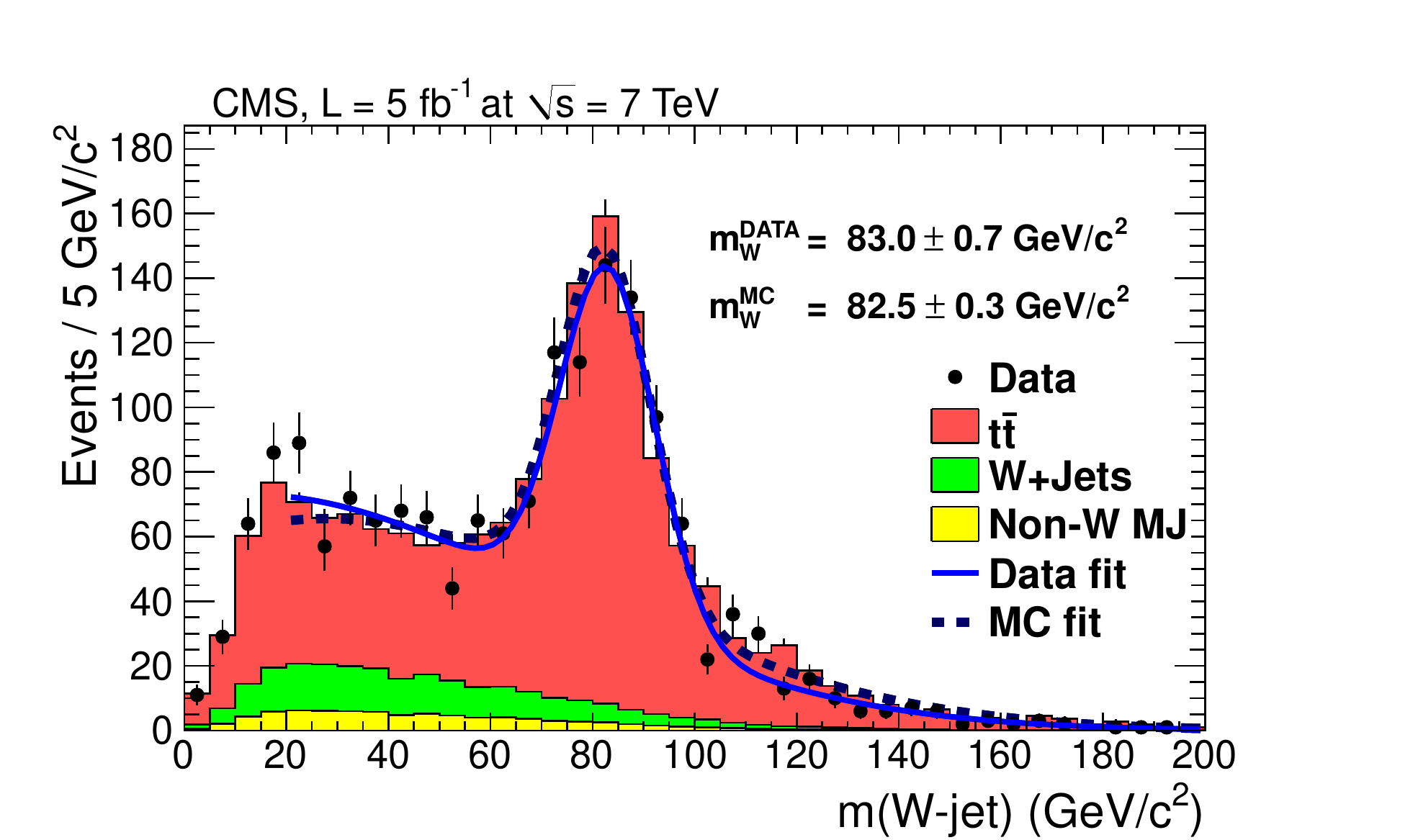}
}

\centering
\subfigure[]{
   \includegraphics[width=0.6\textwidth,angle=0]{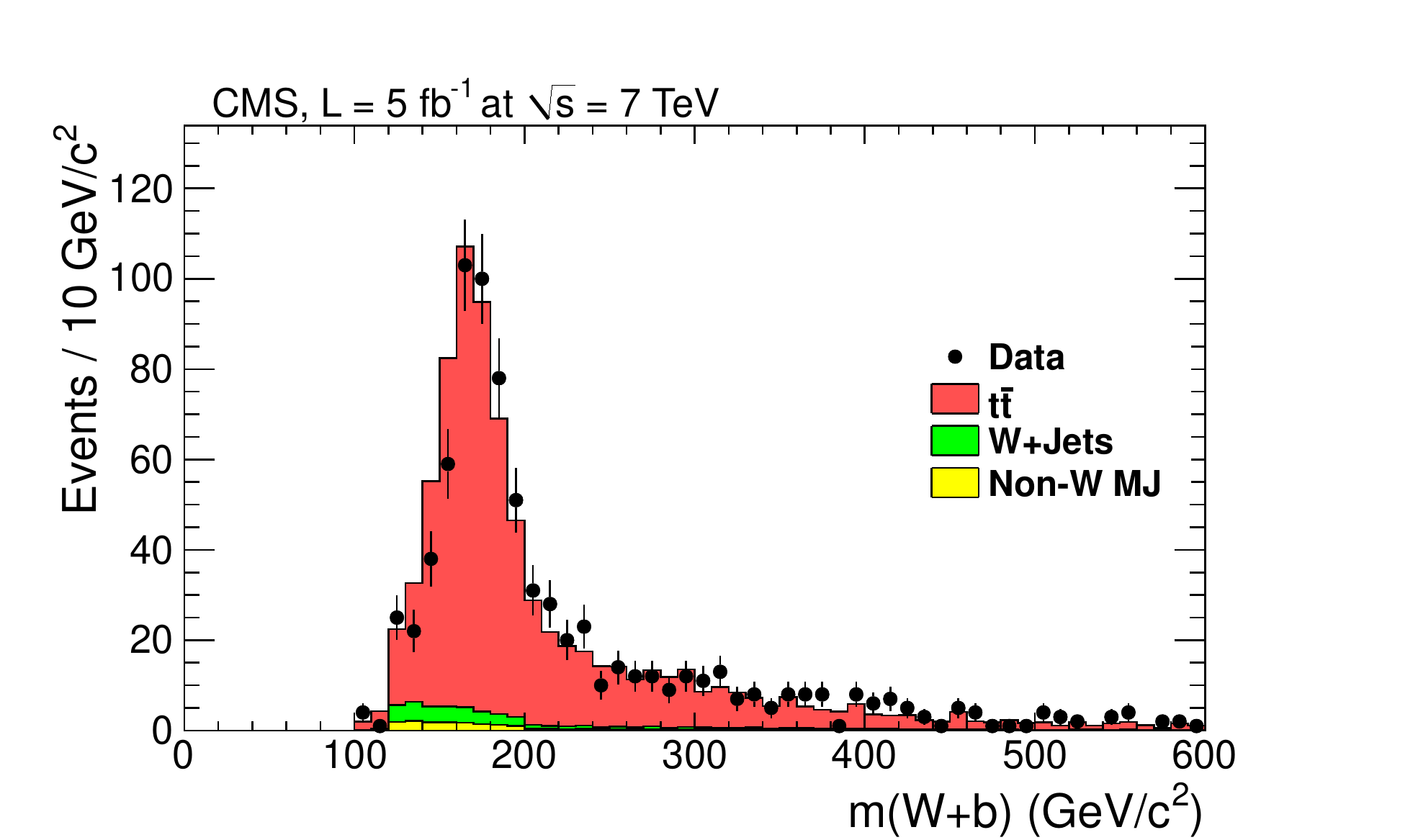}
}
\caption{(a) The mass of the leading mass \W boson candidate pruned fat jet
in the hemisphere opposite to that of the muon in semileptonic \ttbar events.
(b) The invariant mass of the combination of the \W boson candidate fat jet
and the closest fat jet. Also shown are simulated events from SM \ttbar production
(\madgraphpythia). The multijet distribution is obtained from data by
reversing the muon isolation criteria. The \Wpjets shape is assumed to be identical
to the multijet shape and the normalisation is obtained from the
inclusive \W boson production cross section.
From~\cite{Chatrchyan:2012ku}.}
\label{fig:cms_jes_wpeak}
\end{figure}

The extend to which the selection is modelled in simulation is tested using an event sample
of semileptonic \ttbar decays with a single final state muon and at least one $b$-tag.
The limited number of events does not permit the validation of the simulated top tagging efficiency.
This would require significantly large numbers of fat jets with high \pt to capture all top quark decay products.
Instead, the simulated jet energy scale and the selection efficiency was validated
for the \Wpb tag.
\figref{cms_jes_wpeak}a shows the distribution of the leading $\W\!$-tagged fat jet mass in the
hemisphere opposite the one containing the muon.
A clear peak is visible near the \W boson mass.
The jet mass is given by the invariant mass of the two subjets and the peak
position therefore reflects the subjet energy scale.
The peak position is extracted through a fit and the position
in simulation agrees with that in data within $1\%$ which corresponds to the statistical uncertainties of the fits.
The \W tag selection efficiency is determined by comparing the number of events in
the mass window from $60$ to $100\GeV$ in \figref{cms_jes_wpeak}a with the number
of all events in the semileptonic \ttbar sample. The efficiencies for data and simulation
are $49\%$ and $50\%$, respectively, with relative uncertainties of $2\%$.
The mass drop efficiency is determined in a similar way ($64\%$)
and the overall simulated efficiency overestimates the efficiency by $3\%\pm3\%$.

The mass of the combination of the \W boson candidate and the closest fat jet
in semileptonic \ttbar events is shown in \figref{cms_jes_wpeak}b.
A clear peak is visible near the top quark mass with a tail to larger masses.
The position of the peak is approximately described by the simulation (mostly SM \ttbar production),
although the prediction seems shifted to lower values by $\approx\!10\GeV$.
This shift corresponds to $6\%$ at the top quark mass, much more than the
$1\%$ uncertainty that was extracted from the description of the \W boson mass
peak. The larger discrepancy for the top quark mass peak could be related to
the fact that the $b$-jet in $t\to bqq$ decay has on average the
largest \pt of all decay quark jets (see for example \figref{HTTsubst}a and its discussion).
The uncertainty on the simulation of the higher \pt subjets therefore seems to be larger
than that for the lower \pt subjets from \W boson decay.
Using only the \W boson mass peak to determine the subjet energy scale
uncertainty is therefore not recommended because it does not sample
the full phase space relevant to a top quark analysis.

\begin{figure}[hbt]
\centering
\includegraphics[width=0.6\textwidth,angle=0]{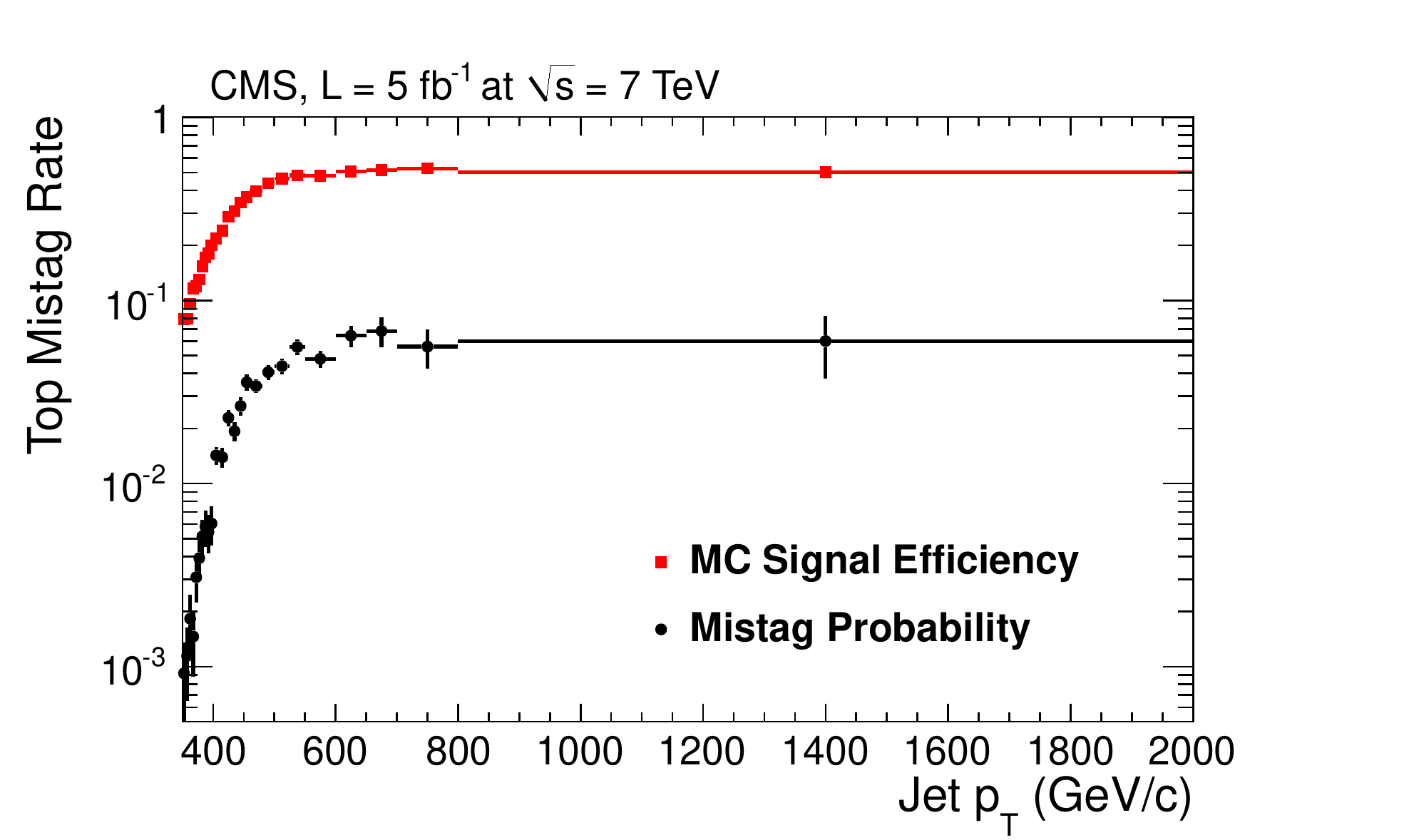}
\caption{The CMS Top Tagger efficiency for \ca $R=0.8$ fat jets when no other
fat jets are in the same hemisphere. Shown is the fake rate of tagging fat jets
from hard light quarks or gluons determined from data and the efficiency to
tag fat jets in simulated $\Zprime \rightarrow \ttbar$ events from \madgraph (the label on the vertical
axis corresponds only to the fake rate). From~\cite{Chatrchyan:2012ku}.}
\label{fig:cms_efficiency}
\end{figure}

A data control region is used to determine the fake top tagging rate for the
multijets background.
The control region is defined by selecting 1+2 events in which the \W boson candidate
and the \Wpb candidate satisfy the mass window requirements to make the selection
kinematically similar to the signal region but the mass drop requirement on the
\W candidate is inverted to remove signal-like events. The events in this control region are then
used to study the tagging of the fat jet in the hemisphere opposite the \Wpb candidate.
The fake top tagging rate is shown in \figref{cms_efficiency}. It is approximately $0.1\%$ at the threshold
and plateaus at $6\%$ for $\pt>600\GeV$.
The rise in the mis-tag rate is due to increased QCD radiation at large \pt
which can fake the top quark decay pattern.
The top quark tagging efficiency is also shown. It is $8\%$ at fat jet $\pt = 350\GeV$,
rises to $20\%$ at $400\GeV$ and reaches a plateau of $50\%$ for $\pt > 500\GeV$.
The efficiency rise is a geometric effect that results from the decay products
being more collimated at large \pt, eventually all being contained inside the
fat jet with $R=0.8$.

To determine the multijet contribution, the selection cuts are relaxed by requiring
only one top tagged fat jet ({\em loose selection}).
The \Wpb tag is required in 1+2 events and for 1+1 events a random one of the
two fat jets is required to be top tagged.
These samples are dominated by multijet events.
The multijet contribution remaining after the second top tag requirement
is determined by weighting these events with the fake rate at
the \pt of the fat jet in the hemisphere opposite to the first tag (`probe jet').
However, the probe jets have a systematically smaller
mass than those in the signal region because the majority of them are not tagged.
To correct for this kinematic bias, the mass of the probe jet is ignored and
replaced by a mass drawn randomly from the distribution of simulated multijet
events in the range $140$ to $250\GeV$.
A systematic uncertainty is assigned by taking half the difference between
the \mtt distributions resulting from the described procedure and when not correcting
the kinematic bias.

\begin{figure}[hbt]
\centering
\subfigure[]{
   \includegraphics[width=0.58\textwidth,angle=0]{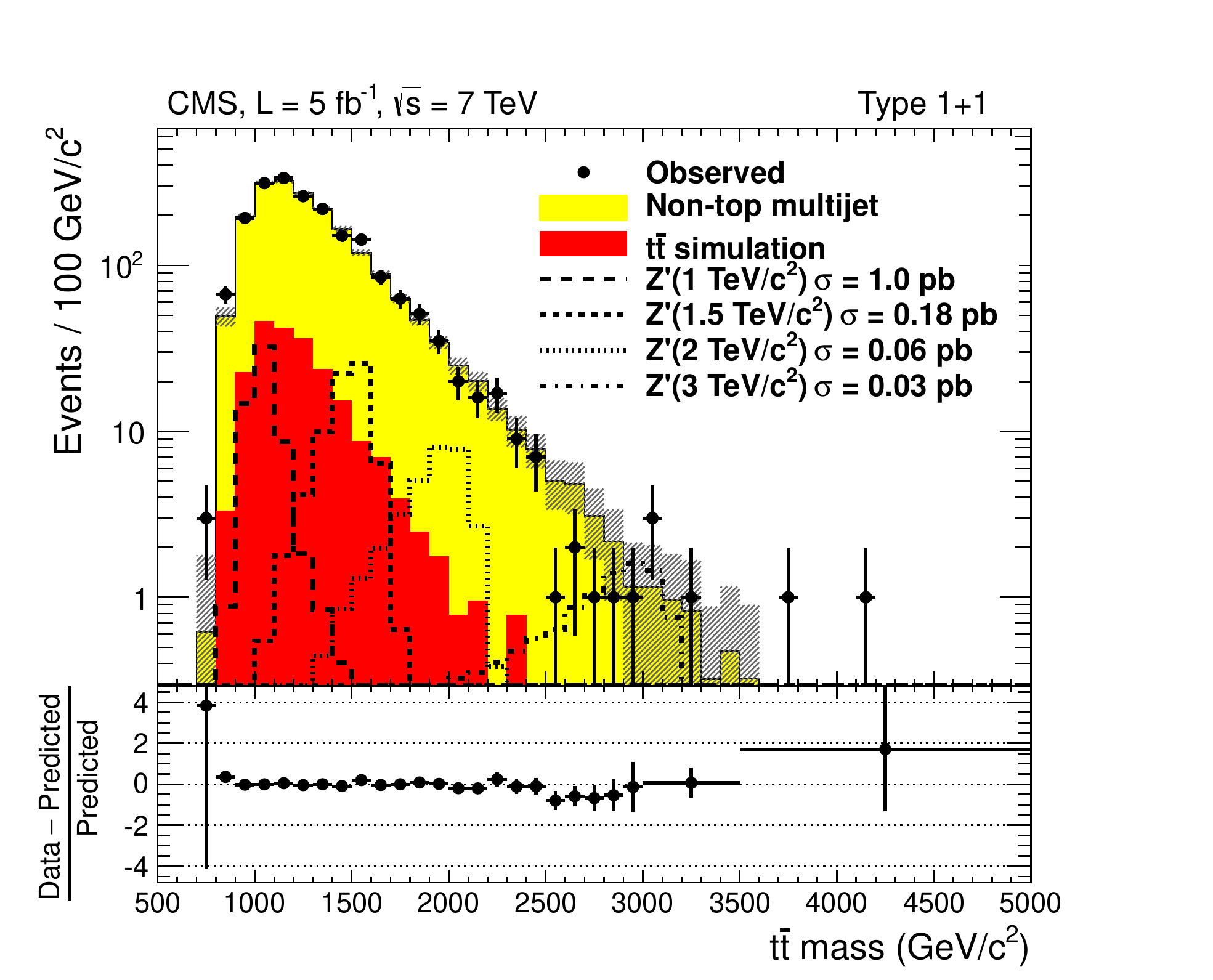}
} \\
\noindent
\subfigure[]{
   \includegraphics[width=0.58\textwidth,angle=0]{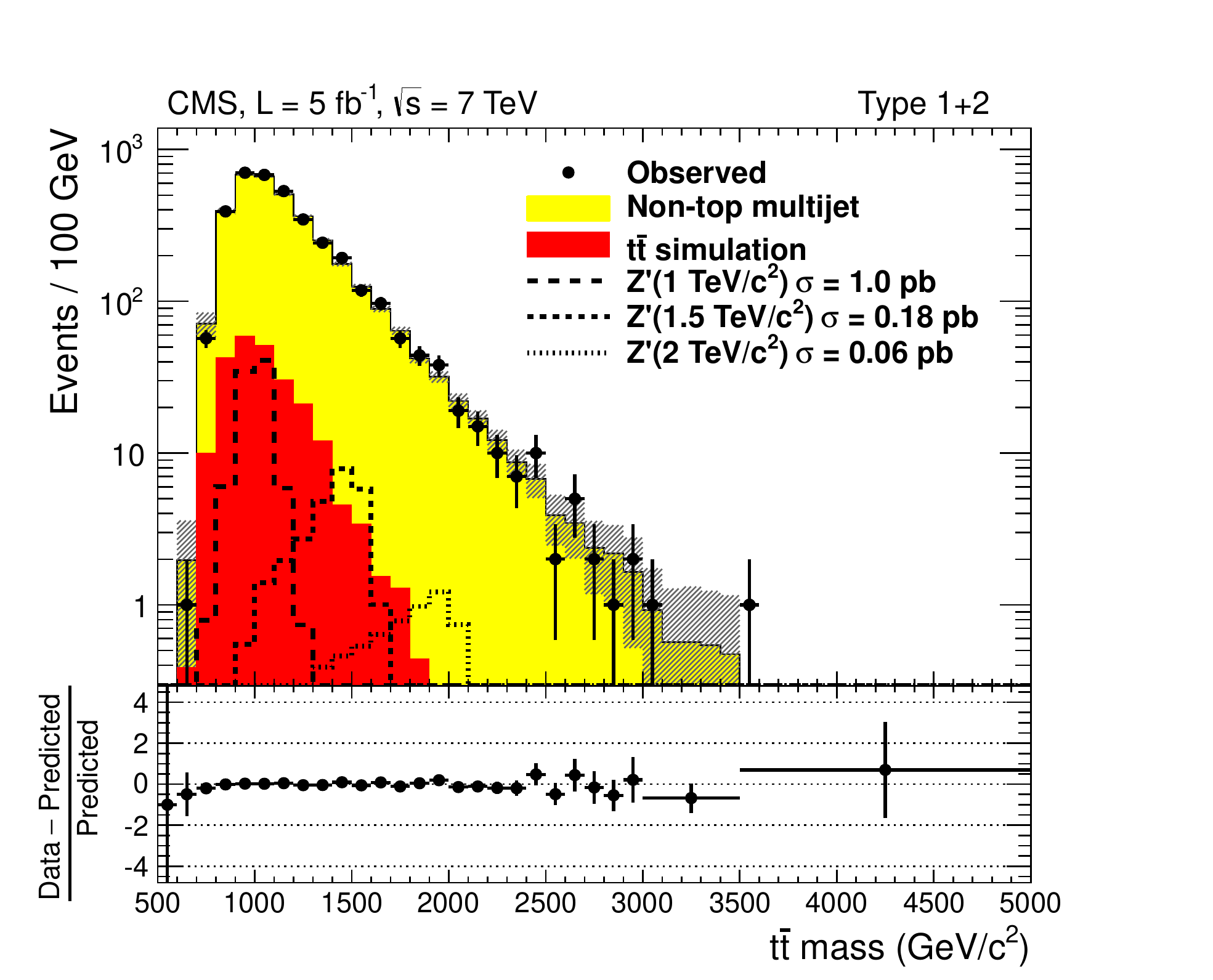}
}
\caption{The reconstructed \mtt spectrum for (a) 1+1 events and (b) 1+2 events
using the CMS Top Tagger.
Also shown are SM contributions from multijet events determined from data and
simulated \ttbar production (from \madgraphpythia). The spectra from $\Zprime \rightarrow \ttbar$
decays for several masses of a hypothetical \Zprime boson with width $\Gamma_{\Zprime}/m_{\Zprime} = 1\%$ are indicated (from \madgraph).
From~\cite{Chatrchyan:2012ku}.}
\label{fig:cms_mtt}
\end{figure}

The measured \mtt distributions in 1+1 and 1+2 events are shown in \figref{cms_mtt}.
The largest SM contribution is from multijets and only $3$--$13\%$ from
\ttbar production, depending on the mass and the event class.
At masses around $1\TeV$, the background uncertainty is $5\%$ and the largest systematic uncertainty
is the trigger uncertainty of $13\%$ for 1+1 events and $20\%$ for 2+2 events.
The trigger uncertainty is taken to be half the trigger inefficiency predicted by simulation.
For masses larger than $2.5\TeV$, the biggest uncertainty results from the finite statistics
of the single-tag multijets sample.
No significant excess over the SM contributions is observed and limits are set.

The uncertainty on the multijet background is remarkably small.
For comparison, in the ATLAS \htt analysis, the multijet background uncertainty is $14\%$ (\secref{htt}).
ATLAS took the approach to reduce the multijet background through $b$-tag requirements.
The $19\%$ relative uncertainty on the $b$-tagging simulation at high \pt is the main contributor
to the total uncertainty in the ATLAS \htt analysis.
Given the small uncertainty found by CMS, it seems that foregoing the use
of $b$-tagging at high \pt is the better strategy until improved $b$-tagging
algorithms will be developed for that region.
But since the relative background uncertainty for CMS is a factor of $\approx\!3$ smaller
than the one for ATLAS, examining the origins of this small uncertainty is warranted.

As described above, the multijet contribution in the signal region is obtained
by weighting fat jet pairs in a loose selection by the fake rate.
In the mass ranges $0.9 < \mtt < 1.1\TeV$ and $1.3 < \mtt < 2.4\TeV$,
the statistical error on the estimated background in the
final selection is quoted to be better than $1\%$. This follows from the
high statistics in the loose selection. The background uncertainty is dominated by
the systematic error which is described as resulting ``from the systematic uncertainty assigned to
the procedure for modifying the probe-jet masses'' and which is $3$--$5\%$,
depending on the mass range and the event topology (1+1, 1+2)~\cite{Chatrchyan:2012ku}.
The statistical uncertainty on the fake rate is quoted to be
``ranging from $<\!\!1\%$ at $1\TeV/c^2$ to $\approx\!\!10\%$ at $3\TeV/c^2$''~\cite{Chatrchyan:2012ku}.
Unfortunately, no mention is made of how the small uncertainties on the fake rate are arrived at;
the error bars in \figref{cms_efficiency} correspond to much larger uncertainties.

\begin{figure}[hbt]
\centering
\subfigure[]{
   \includegraphics[width=0.58\textwidth,angle=0]{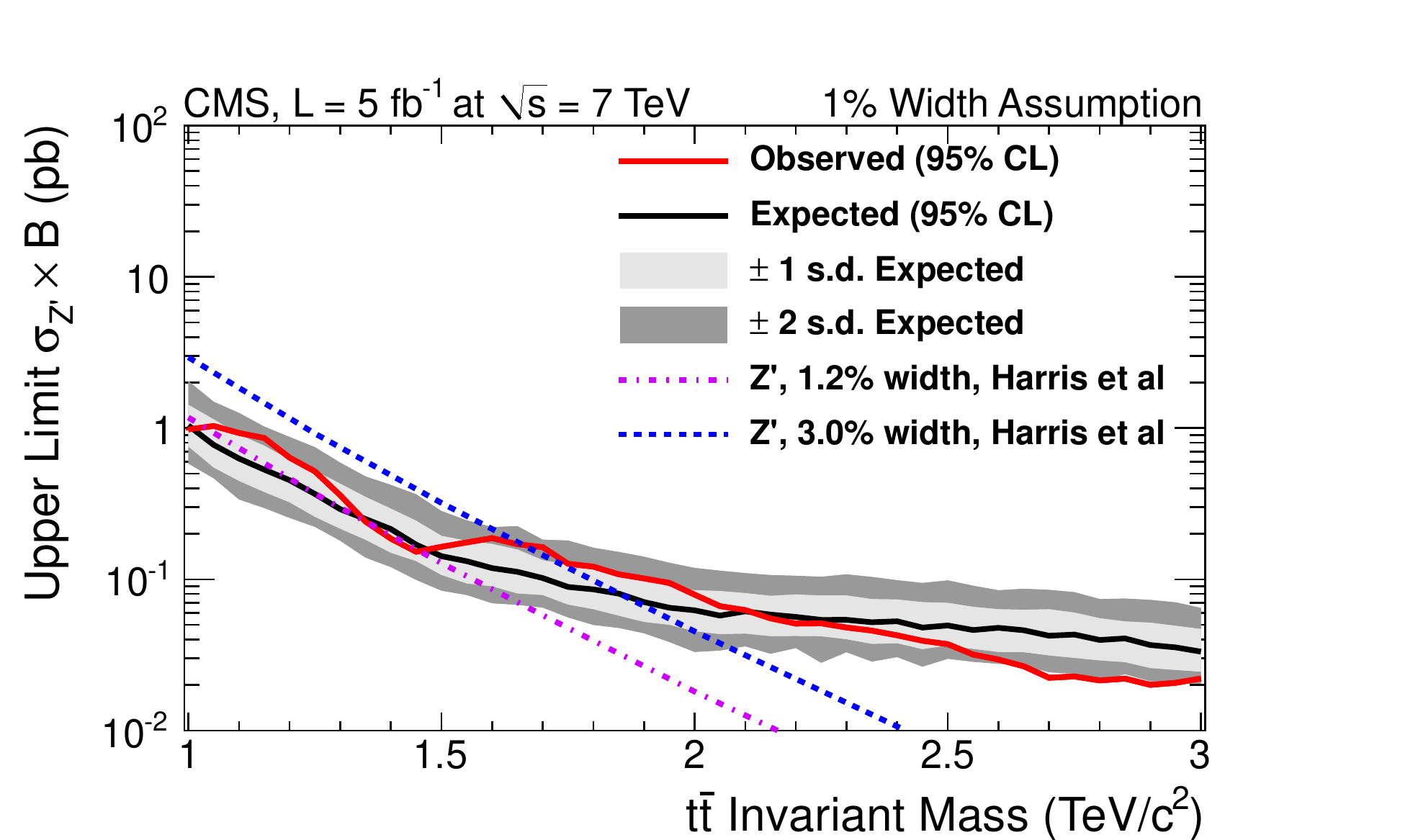}
}
\subfigure[]{
   \includegraphics[width=0.58\textwidth,angle=0]{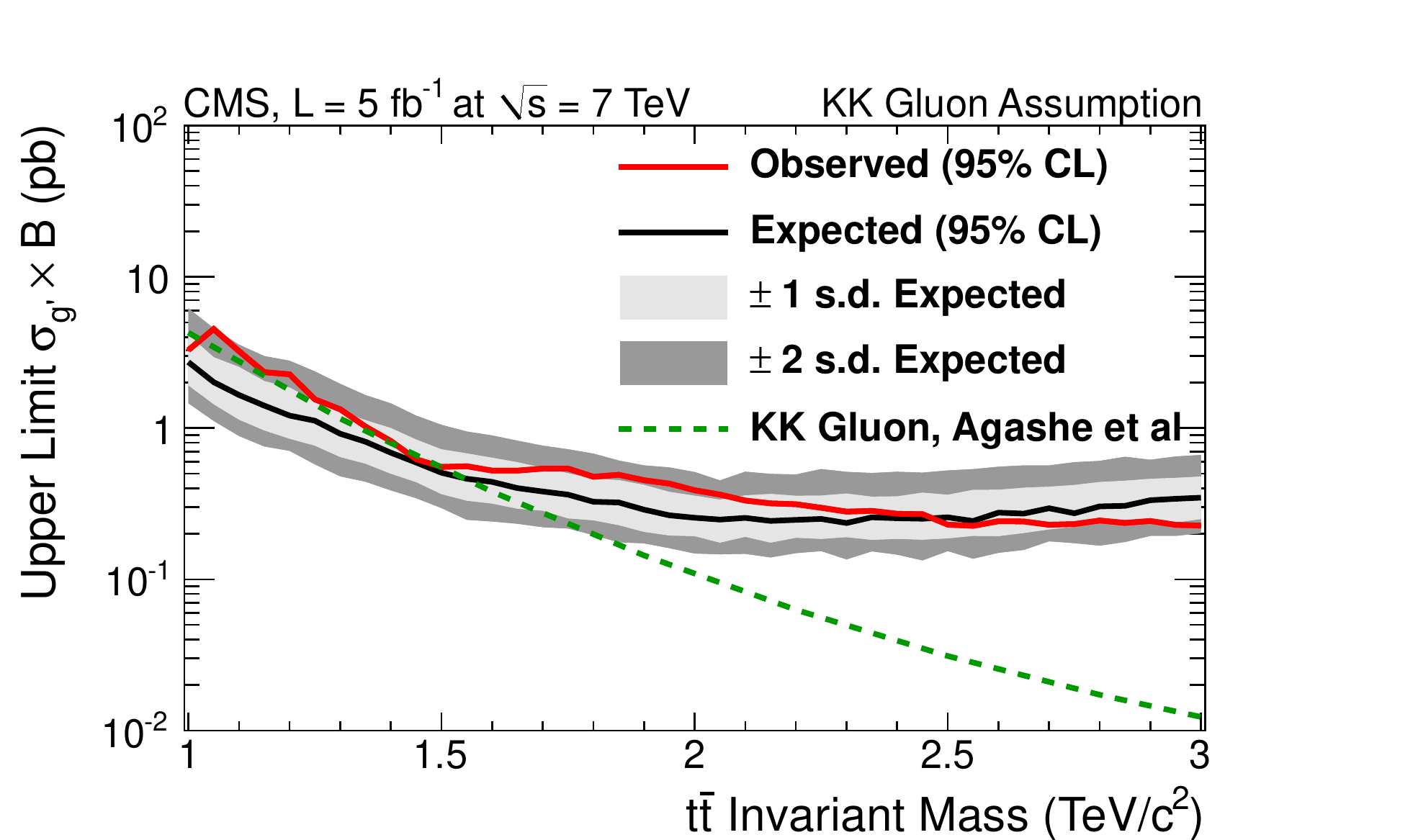}
}
\caption{The $95\%$ confidence level upper limit on the production cross section
times branching ratio into \ttbar for (a) a leptophobic \Zprime boson with
width $\Gamma_{\Zprime}/m_{\Zprime} = 1.2\%$ and $3.0\%$ and (b) a Kaluza-Klein gluon with
width $\Gamma_{\gKK}/m_{\gKK} = 15.3\%$. From~\cite{Chatrchyan:2012ku}.}
\label{fig:cms_limits}
\end{figure}

\figref{cms_limits}a and b show the limits for the narrow-width \Zprime boson
and the Kaluza-Klein gluon, respectively.
Below $1\TeV$ the multijet background is prohibitively large and
the selection is inefficient for the signal models. No limits are therefore
evaluated for this parameter space.
 The limit for the \Zprime model is calculated assuming a width of $1.0\%$
which is much smaller than the \mtt detector resolution of $\approx \! 10\%$.
Any resonance width smaller than $10\%$ will be broadened by detector resolution.
For this reason, it is possible to compare
the \Zprime $1.2\%$ width model with the $1.0\%$ limit.

The $1.2\%$ width \Zprime boson model is the benchmark model also used in the
ATLAS analyses (Figures~\ref{fig:ljets_limits}, \ref{fig:htt_limits}, and \ref{fig:ttt_results})
and the cross section predictions are identical (including a scaling of the leading order prediction by a factor of $1.3$).
This model can be excluded at $95\%$ confidence level at $m_{\Zprime} = 1\TeV$ and
in the range $\approx\!1.33$--$1.46\TeV$. Compared to the \htt analysis, the upper mass limit
is $140\GeV$ higher. The \htt analysis, on the other hand, provides a limit for masses below $1\TeV$.

The Kaluza-Klein gluon model used by CMS has a cross section that is
$\approx\!10\%$ higher at $m_{\gKK} = 1.5\TeV$ than the model used by ATLAS.
The CMS exclusion limit is a small interval near $1\TeV$ and $1.42$--$1.5\TeV$. In this latter
mass range, the ATLAS Kaluza-Klein gluon model cannot be excluded by the CMS analysis.

\subsection{Summary of \ttbar resonance searches}

\begin{figure}[hbt]
\centering
\subfigure[]{
   \includegraphics[width=0.45\textwidth,angle=0]{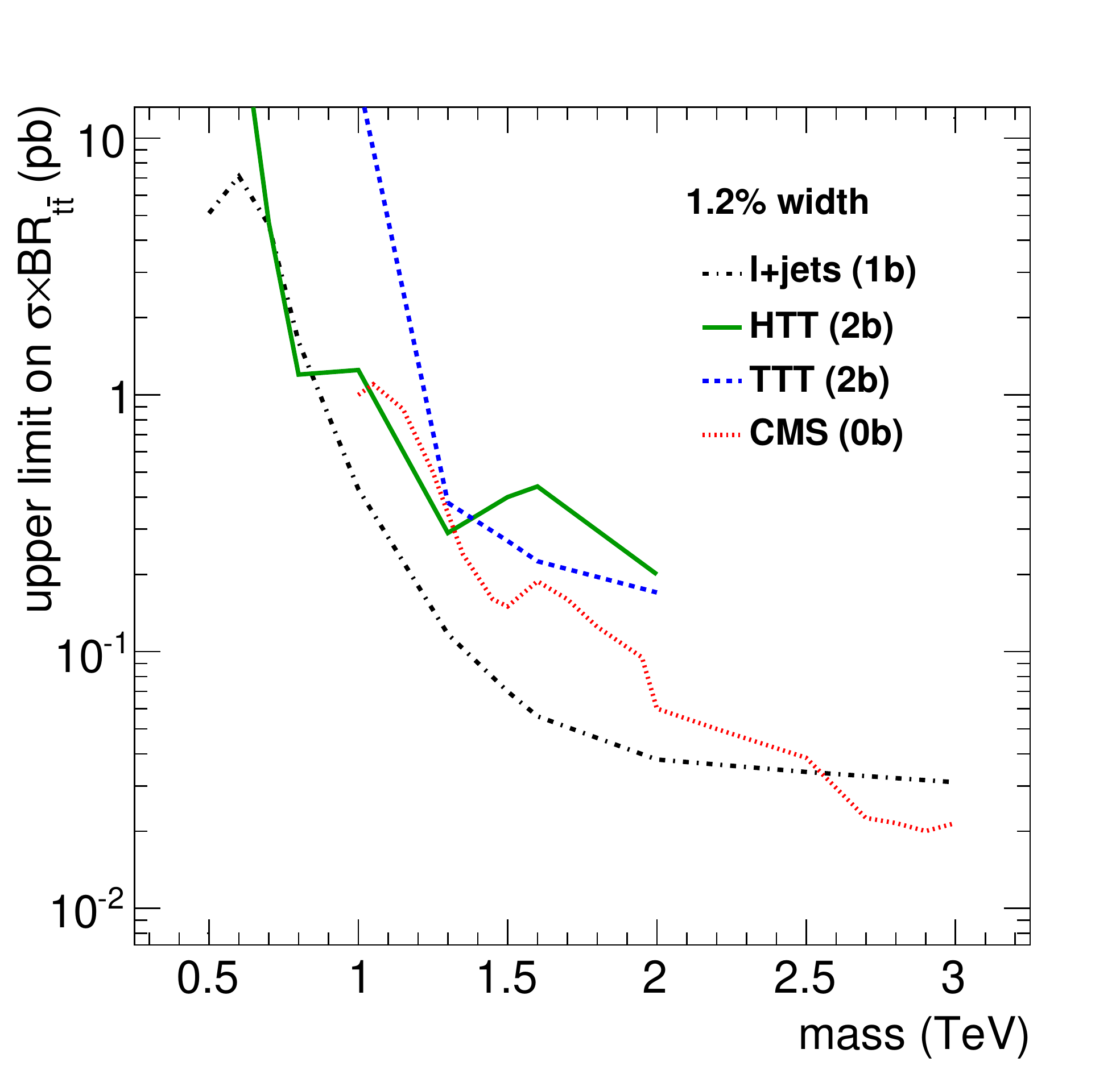}
}
\subfigure[]{
   \includegraphics[width=0.45\textwidth,angle=0]{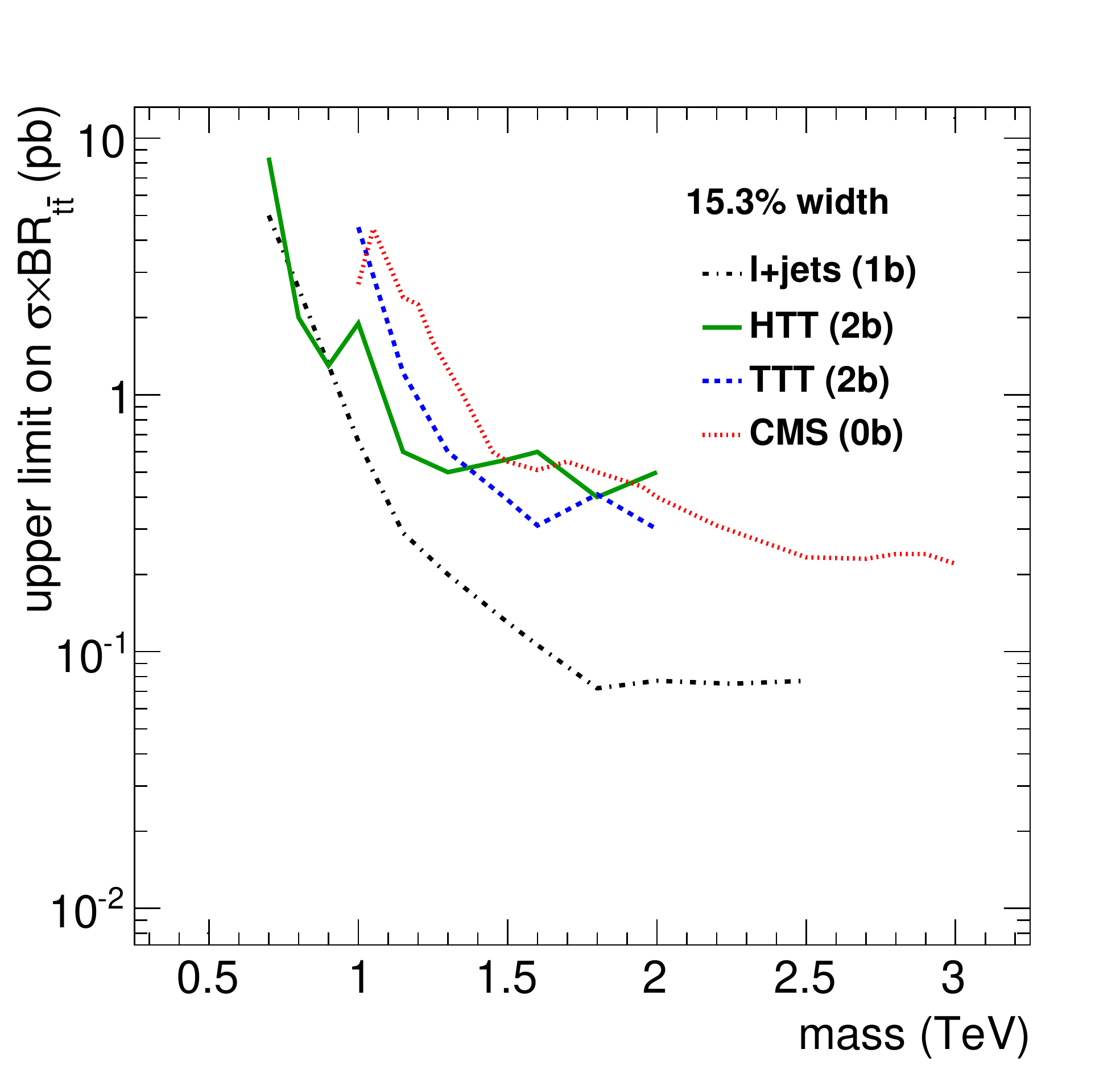}
} \\
\noindent
\subfigure[]{
   \includegraphics[width=0.45\textwidth,angle=0]{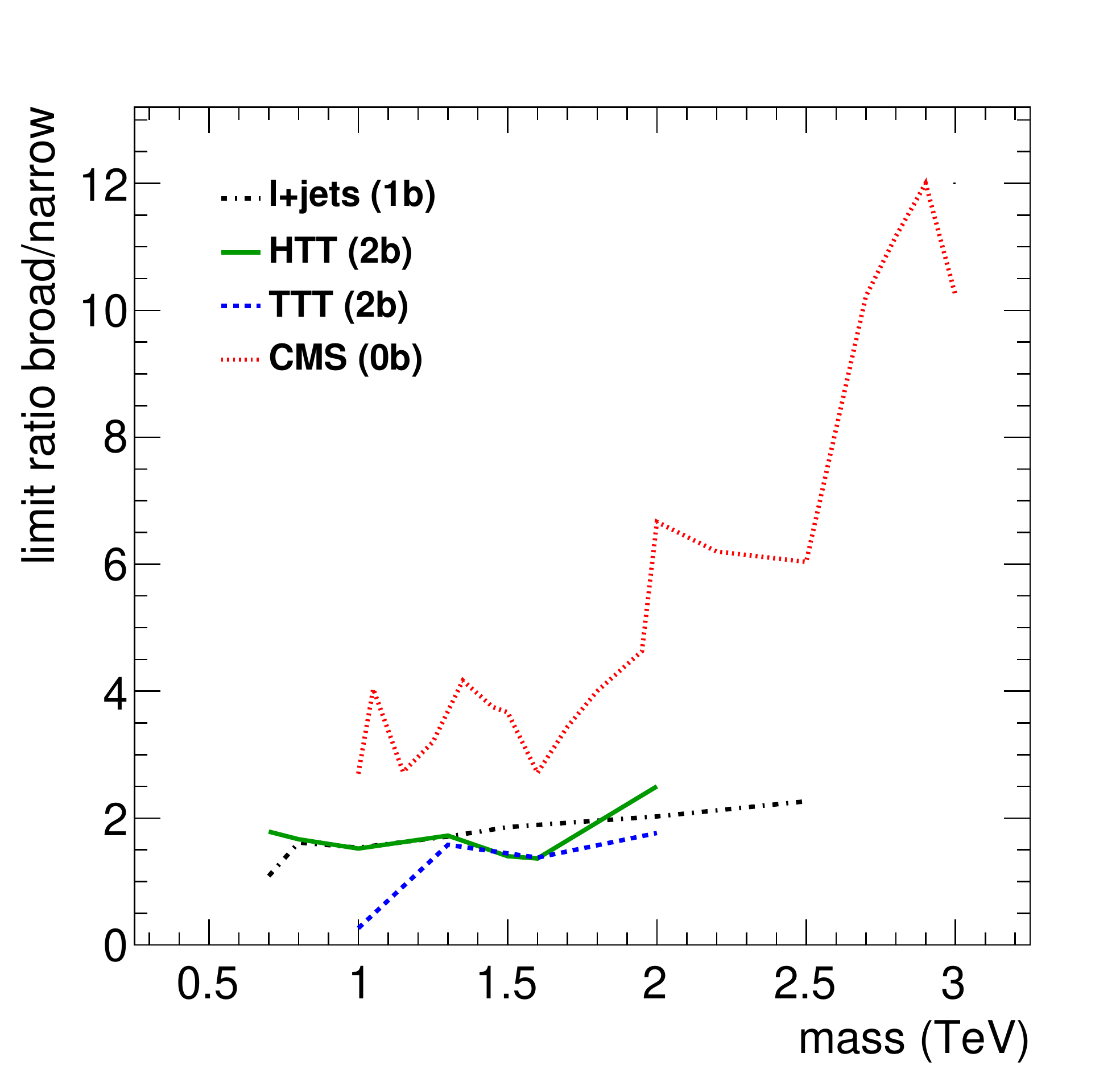}
}
\caption{The observed $95\%$ C.L. upper limits on the production cross section
times branching ratio into \ttbar for (a) a narrow resonance (width $\Gamma/m = 1.2\%$)
for which the reconstructed width is given by the detector resolution of $\approx\!10\%$
and (b) a broader resonance of width $15.3\%$.
The ratio of the limit of the broader to the narrower resonance is shown in (c).
The limits are taken from the ATLAS analyses
using lepton+jets~\cite{Aad:2013nca}, the hadronic ATLAS analyses~\cite{Aad:2012raa} using the
\htt (label `HTT') and the \ttt (`TTT'),
and the hadronic CMS analysis using their top tagger~\cite{Chatrchyan:2012ku}.}
\label{fig:limits_resonances}
\end{figure}

\figref{limits_resonances} shows a comparison of the observed limits in the \ttbar resonance searches
as a function of the mass $m$ of the new particle.
The limits are shown for the case of a narrow resonance ($\Gamma/m = 1.2\%$),
for which the reconstructed width is given by the detector resolution of $\approx\!10\%$
and a broader resonance of width $15.3\%$.

The best limits for masses between $1$ and $2.5\TeV$ are given by the semileptonic analysis.
For masses larger than $1.3\TeV$ the limits are below $0.1$~pb.
For the broad resonance, the semileptonic
limit is approximately a factor of $3$ better than the ones from the fully hadronic
analyses.
\figref{limits_resonances}c shows
the ratios of the limit for the broad resonance to the limit for the narrow resonance.
For the semileptonic analysis, the narrow resonance limit is better by a factor of $1.5$ at $1\TeV$ and $2.2$ at $2.5\TeV$.
The fact that the mass exclusion range is larger for the Kaluza-Klein gluon model (cf.~\secref{ljets})
is due to the higher production cross section of that model.
All analyses set better limits on the narrower
resonance, except for the \ttt at $1\TeV$.

The leptonic analysis uses
a cut on the splitting scale \DOneTwo to tag the hadronically decaying
top quark. This method has a high efficiency and a low rejection.
A larger rejection is not needed because
the multijet background is inherently
lower than in the other analyses due to the lepton requirement.
The analysis also uses only a single $b$-tag.
The tagging of bottom quarks uses settings that have been optimised at low bottom quark \pt and which
may not be optimal at high \pt because of the collimated topology with close-by
jets.
Also the uncertainties on the $b$-tagging efficiency
and fake rate is large at high \pt, thereby reducing the sensitivity of the
searches.

The CMS Top Tagger analysis, which does not use $b$-tagging,
has the next best limit between $1.3$ and $2.5\TeV$ for the narrow resonance.
For masses larger than $2.5\TeV$ is gives the best limit and reaches $0.02$~pb
at $3\TeV$. The sensitivity to the broad resonance is significantly worse.
The CMS tagger shows the largest dependence on the width with the limits being
a factor 3 to 12 worse for the broader resonance.

The \htt analysis limit is as good as the semileptonic limit
for $700<m<800\GeV$. Below this range, the resolved semileptonic analysis provides
a better limit. Up to $1.3\TeV$ the \htt provides the best limit of all
hadronic analyses, then the \ttt provides a comparable or better limit.

It is reassuring that the different analyses give similar limits, despite the
wide range of approaches for tagging top quarks and their different systematic
uncertainties.
The evaluation of systematic uncertainties in boosted top quark analyses
is often limited by the statistics of available high \pt top quarks in data control samples.
Rather than taking the statistical uncertainties as an upper limit for the systematic
errors, the low \pt uncertainties are often taken if no clear trend is visible within
the large statistical errors when extrapolating to high \pt.
High statistics datasets that will become available in the future
will allow to better constrain these uncertainties.

%%%%%%%%%%%%%%%%%%%%%%%%%%%%%%%%%%%%%%%%%%%%%%%%%%%%%%%%%%%%%%%%%%%%%%%%%%%%%%%%

\subsection{Search for SUSY in high jet multiplicity events}
\label{sec:searchsusy}

In addition to the searches for \ttbar resonances, substructure
techniques have also been used to search for SUSY.
The strategy of the analysis in~\cite{Aad:2013wta} is to
look for an excess of events beyond the SM prediction in a final state
with at least seven \akt $R=0.4$ jets (small-$R$ jets), missing
energy, and no leptons.
This final state is especially sensitive to a SUSY model like that used in
\secref{highmultstudy}, in which top squarks decay to top quarks and neutralinos.
The analysis uses a conventional selection ({\em $b$-tag stream}) and
a selection that requires the sum of the masses of all fat jets to be larger
than two times the top quark mass.

ATLAS data collected at $\sqrt{s} = 8\TeV$ are used, corresponding to a luminosity of $20.3(6)\invfb$.
Jets are built from calorimeter clusters and multijet triggers are used to select the events, requiring at least five jets with $\et>55\GeV$ or
at least six jets with $\et>45\GeV$. Events with isolated muons or electrons with $\pt>10\GeV$, that
are separated from the nearest jet by at least $\DeltaReta = 0.4$, are vetoed.
Multiple signal regions are defined, with different jet and $b$-tag multiplicities,
to enhance the sensitivity to different SUSY models.

For the $b$-tag stream, the small-$R$ jets are required to have $|\eta|<2.0$. Seven signal regions are
defined using jet $\pt > 50\GeV$: six regions with exactly eight or exactly nine jets in the event,
subdivided by $b$-tag multiplicity ($0$, $1$, $\ge2$) using the MV1 algorithm.
One additional signal region is formed by events with more than nine jets without a $b$-tag requirement.
Another six signal regions are defined by $\pt > 80\GeV$ and exactly seven or at least eight jets,
subdivided in $b$-tag multiplicity.

For the fat jet stream, the small-$R$ jets are required to have $|\eta|<2.8$.
The fat jets are not constructed from clusters but by running the \akt algorithm
with $R=1.0$ over small-$R$ jets with $\pt > 20\GeV$.
These fat jets are required to have $\pt> 100\GeV$ and $|\eta|<1.5$.
The scalar sum \sumfjm of the masses of all fat jets is required to be larger than
$340$ or $420\GeV$ for events with at least eight, nine, or ten small-$R$ jets
with $\pt>50\GeV$, thereby giving six signal regions.

\begin{figure}[hbt]
\centering
\subfigure[]{
   \includegraphics[width=0.45\textwidth,angle=0]{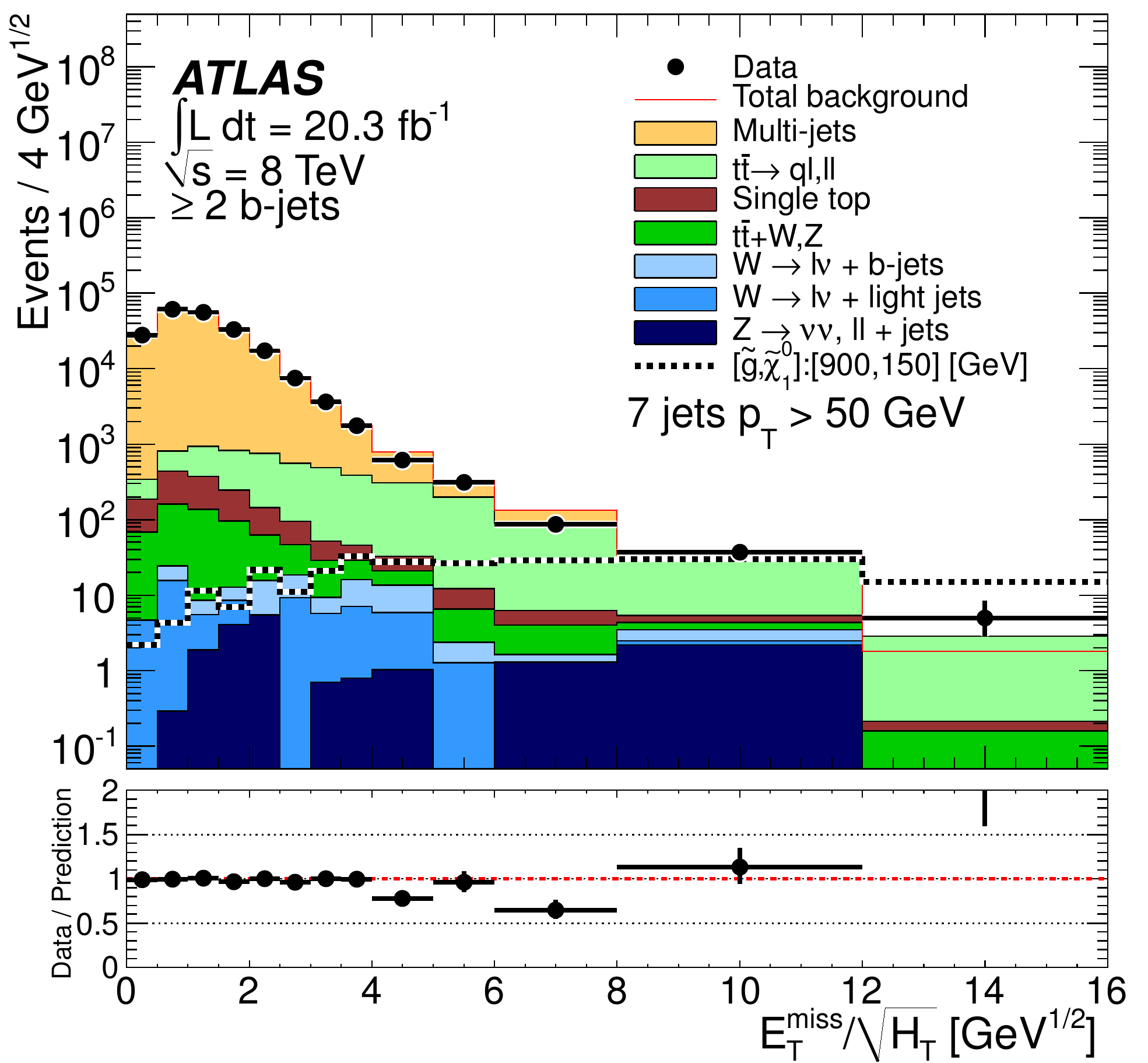}
}
\subfigure[]{
   \includegraphics[width=0.45\textwidth,angle=0]{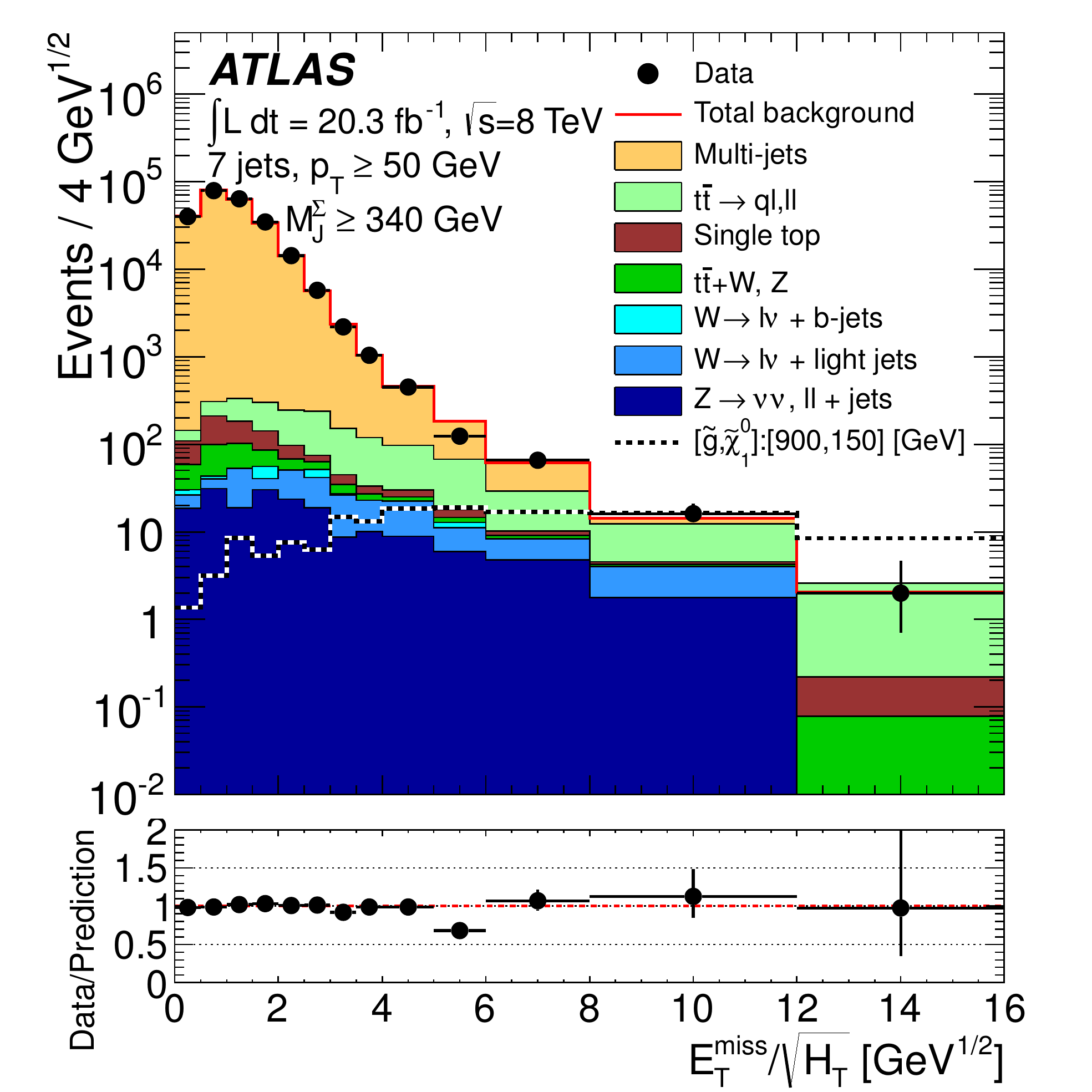}
}
\caption{The \ETmSignif distribution in control regions with exactly seven
\akt $R=0.4$ jets with $\pt>50\GeV$. The multijet prediction is taken
from the measured distribution with exactly six jets and after subtracting the
simulated leptonic backgrounds. The \ttbar and \Wpjets backgrounds are generated
with \sherpa and normalised in control regions. Single-top production
is generated with \mcatnlo in the $s$-channel and with \acermc in the $t$-channel.
Also shown is the distribution of a SUSY model with $m(\tilde{g}) = 900\GeV$ and
$m(\lsp) = 150\GeV$, generated with \madgraphherwigpp. The distribution is shown after requiring
(a) at least two $b$-tags and (b) the sum of the masses of \akt $R=1.0$ jets in the event
to exceed $340\GeV$.
 From~\cite{Aad:2013wta}.}
\label{fig:susy_controlbg}
\end{figure}

In each of the 19 signal regions, $\ETmiss/\sqrtHT > 4\sqrtGeV$ is required,
in which \HT is the scalar sum of the \pt of all small-$R$ jets with $\pt > 40\GeV$ and $|\eta|<2.8$.
For multijet ATLAS events, the shape of the $\ETmiss/\sqrtHT$ distribution
is found to be approximately independent of the small-$R$ jet multiplicity \njet
and of \sumfjm.
The multijet background in the signal regions can therefore be determined
using $\ETmSignif$ templates from multijet-dominated control regions at lower \njet.
The multijet background template is taken from $\njet=6$ events and the other backgrounds
are subtracted from the template.
Separate templates are used for the different $b$-tag multiplicities because
neutrinos from heavy quark decay contribute to \ETmiss but not to \HT.
A correction is applied to account for small changes in the ratio that result from
non-jet contributions to \ETmiss which do not appear in \HT. The correction amounts
to reweighting the distribution of non-jet \pt in the control region to that
in the signal region.
The closure of this method is tested for intervals at smaller $\ETmiss/\sqrtHT$ than in the signal
regions and observed deviations are used as systematic uncertainties.
The closure uncertainties typically are $5$ to $15\%$, reaching up to $\approx\!50\%$
in control regions with small statistics.
\figref{susy_controlbg} shows the \ETmSignif distribution in control regions
with (a) $\njet = 7$ for the
selection with at least two $b$-tags and (b) the requirement $\sumfjm>340\GeV$.
The data are well described for $\ETmSignif < 4\sqrtGeV$,
where multijet events dominate. The closure uncertainties are determined in this region.
The leptonic \ttbar background becomes important at higher values where deviations from
the data are seen in some bins. The \ttbar background carries an uncertainty of $\approx\!50\%$ when
combining experimental and theoretical sources as discussed below. Within this uncertainty,
the description is compatible with the data.
The control regions can potentially be contaminated with New Physics events,
as illustrated in the figure.
A simplified SUSY model is used with pair-produced gluinos, each decaying to $t \bar{t} \lsp$, as
introduced in \secref{highmultstudy}. The masses chosen for the contributions
in the figure are $m(\tilde{g}) = 900\GeV$ and $m(\lsp) = 150\GeV$, which
are excluded by the analysis of the signal regions as discussed below.
The possible contamination
is therefore smaller than that shown in \figref{susy_controlbg} ($20$--$30$ events per $4\sqrtGeV$).

The most important leptonic backgrounds are from \Wpjets events and \ttbar production
and the dominant contributions result from decays to $\tau$ leptons which decay hadronically.
The decays to electrons and muons are suppressed because of the electron and muon vetos.
These backgrounds are taken from \sherpa simulations which are fitted to data in control regions.
The control regions contain exactly one isolated muon or electron and the lepton four-momentum is used to create
an additional jet that is included in the calculation of \HT. In this way, \ETmSignif is made similar
to its value in events with hadronic $\tau$ decay. A simultaneous fit is
performed to the signal and control regions, as discussed below, to determine
the SM contribution and potential contributions from New Physics.

The normalisation of the background simulation for events with $Z\rightarrow \nu \nu$ decay is taken
from data
using events with same-flavour, opposite-charge lepton pairs (electrons or muons)
which combine to an invariant
mass close to the \Z boson mass ($80$ to $100\GeV$).
The two transverse lepton momenta are added
to \ETmiss and the prediction in the region $\ETmSignif > 4\sqrtGeV$ is
normalised to the data in control regions that have relaxed \njet requirements.
The same normalisation factor is then used in the signal region.
Experimental systematic uncertainties are dominated by the modelling of the jet energy scale and
resolution which change the event yield by $20$ to $30\%$. Smaller uncertainties
of $\approx\!10\%$ result from $b$-tagging. For the \Wpjets events, the theoretical
uncertainties are smaller than the experimental ones.
For \ttbar production, two theoretical contributions each give rise to an uncertainty of $25$--$30\%$:
i) variations of the renormalisation and factorisation scales by factors $2$ and $0.5$ and
ii) variations of the gluon fusion production cross section.
To account for higher order terms
not included in \sherpa, the probability for events to be initiated by gluon fusion
is increased by $37\%$ while the other processes are reduced by a common factor
to keep the total cross section unchanged. This procedure improves the description
of the data in the control regions. The difference in the event yield before
and after the $37\%$ correction is used as a systematic uncertainty.

\begin{figure}[hbt]
\centering
\subfigure[]{
   \includegraphics[width=0.45\textwidth,angle=0]{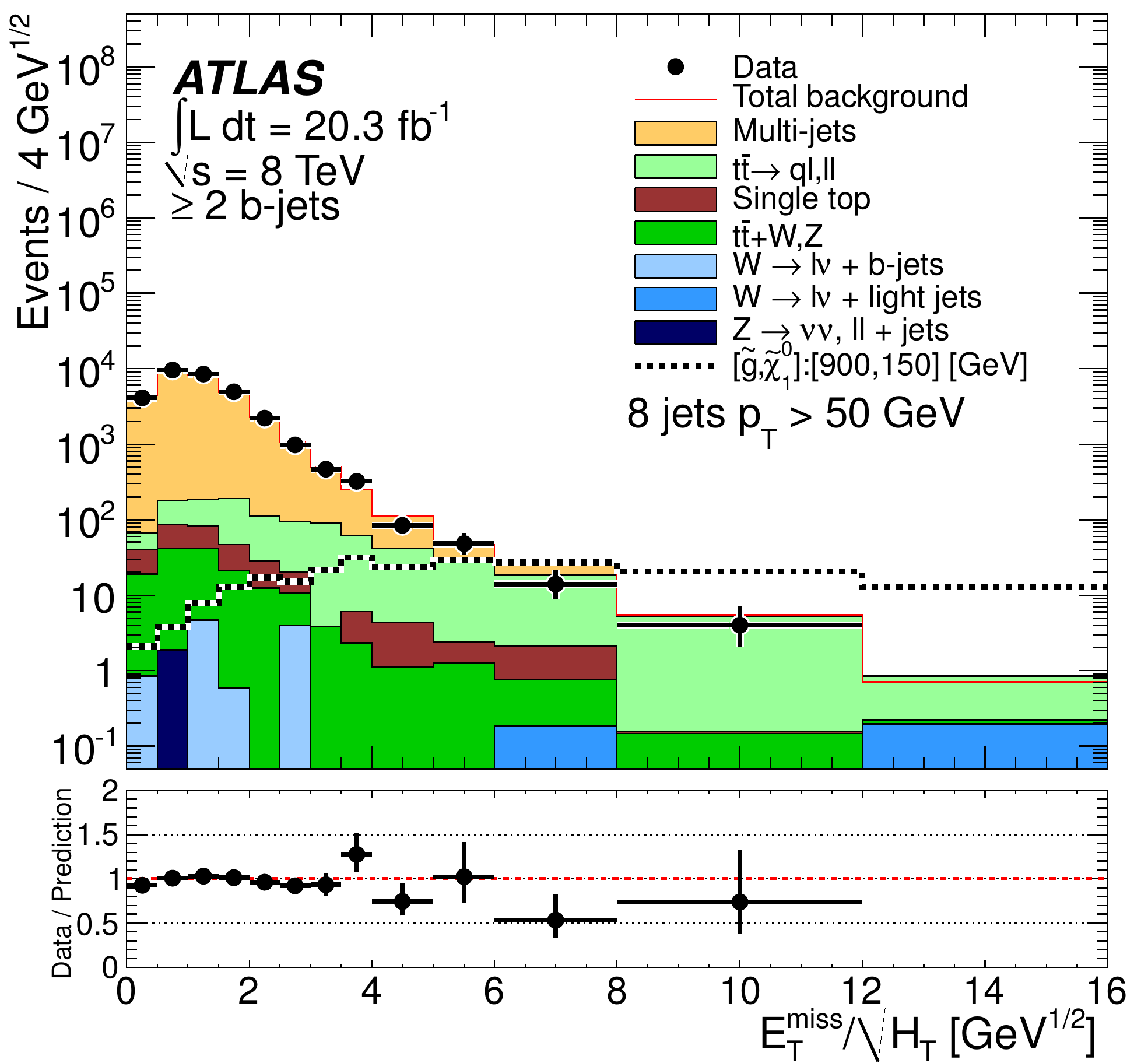}
}
\subfigure[]{
   \includegraphics[width=0.45\textwidth,angle=0]{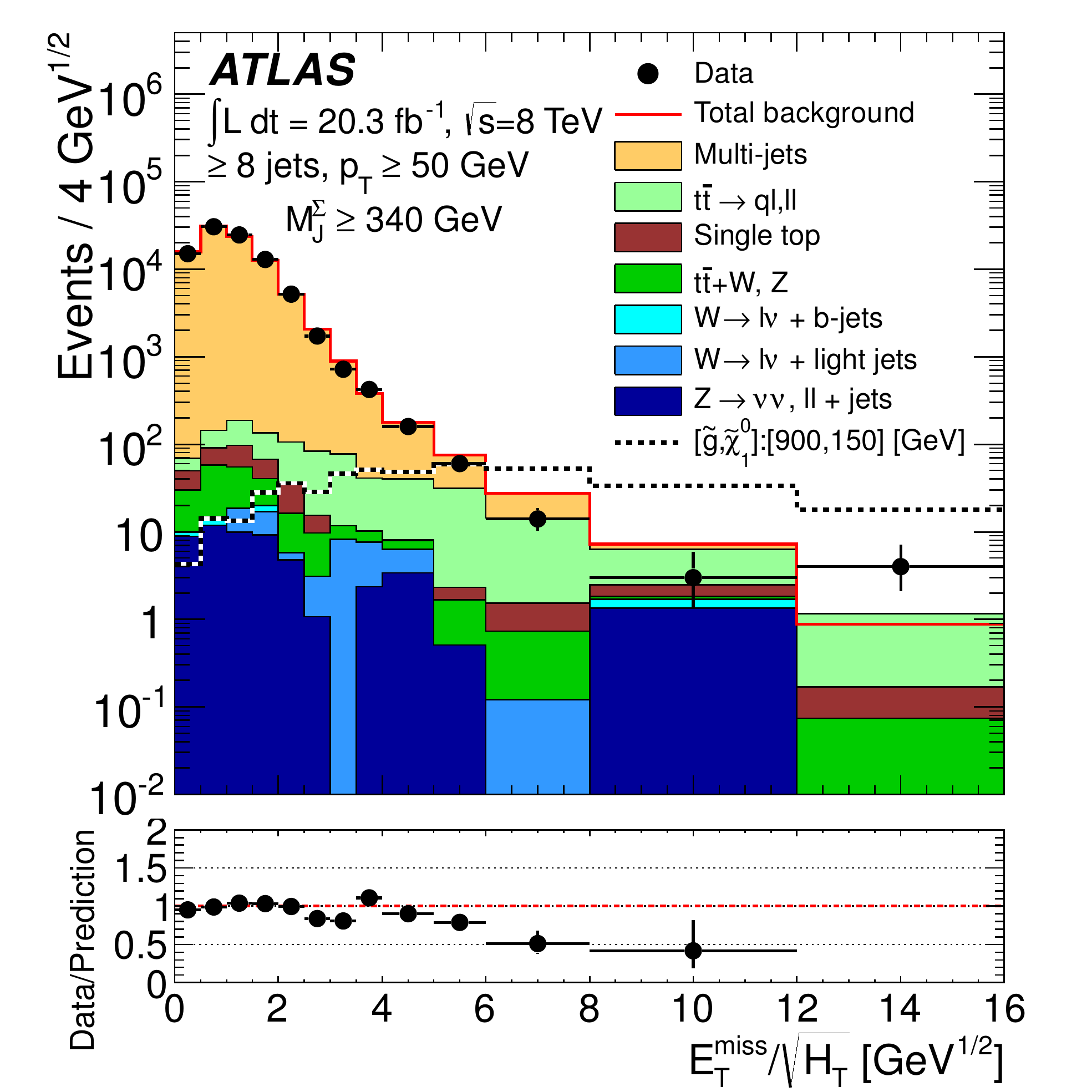}
}
\caption{The \ETmSignif distribution in signal regions with (a) exactly eight
\akt $R=0.4$ jets with $\pt>50\GeV$ and at least two $b$-tags and
(b) at least eight \akt $R=0.4$ jets with $\pt>50\GeV$ and
the sum of the masses of \akt $R=1.0$ jets in the event
exceeding $340\GeV$.
The multijet prediction is taken
from the measured distribution with exactly six jets and after subtracting the
simulated leptonic backgrounds.
The \ttbar and \Wpjets backgrounds are generated
with \sherpa and normalised in control regions. Single-top production
is generated with \mcatnlo in the $s$-channel and with \acermc in the $t$-channel.
Also shown is the distribution of a SUSY model with $m(\tilde{g}) = 900\GeV$ and
$m(\lsp) = 150\GeV$, generated with \madgraphherwigpp.
From~\cite{Aad:2013wta}.}
\label{fig:susy_signalregion}
\end{figure}

Fits are performed to determine the compatibility of the SM with the measured
distributions. These fits are performed separately for each of the 19 signal regions.
The SM simulations are fitted to the data simultaneously in one signal region and
all control regions that have at least two expected events.
The fit results for the normalisations of the SM predictions in the control samples
are consistent with the Monte Carlo normalisations.
In each signal region, the SM prediction is consistent with the measured events.
\figref{susy_signalregion} shows the \ETmSignif distribution for two example
signal regions. There is clearly no room for the indicated SUSY events.

\begin{figure}[hbt]
\centering
\subfigure[]{
   \includegraphics[width=0.45\textwidth,angle=0]{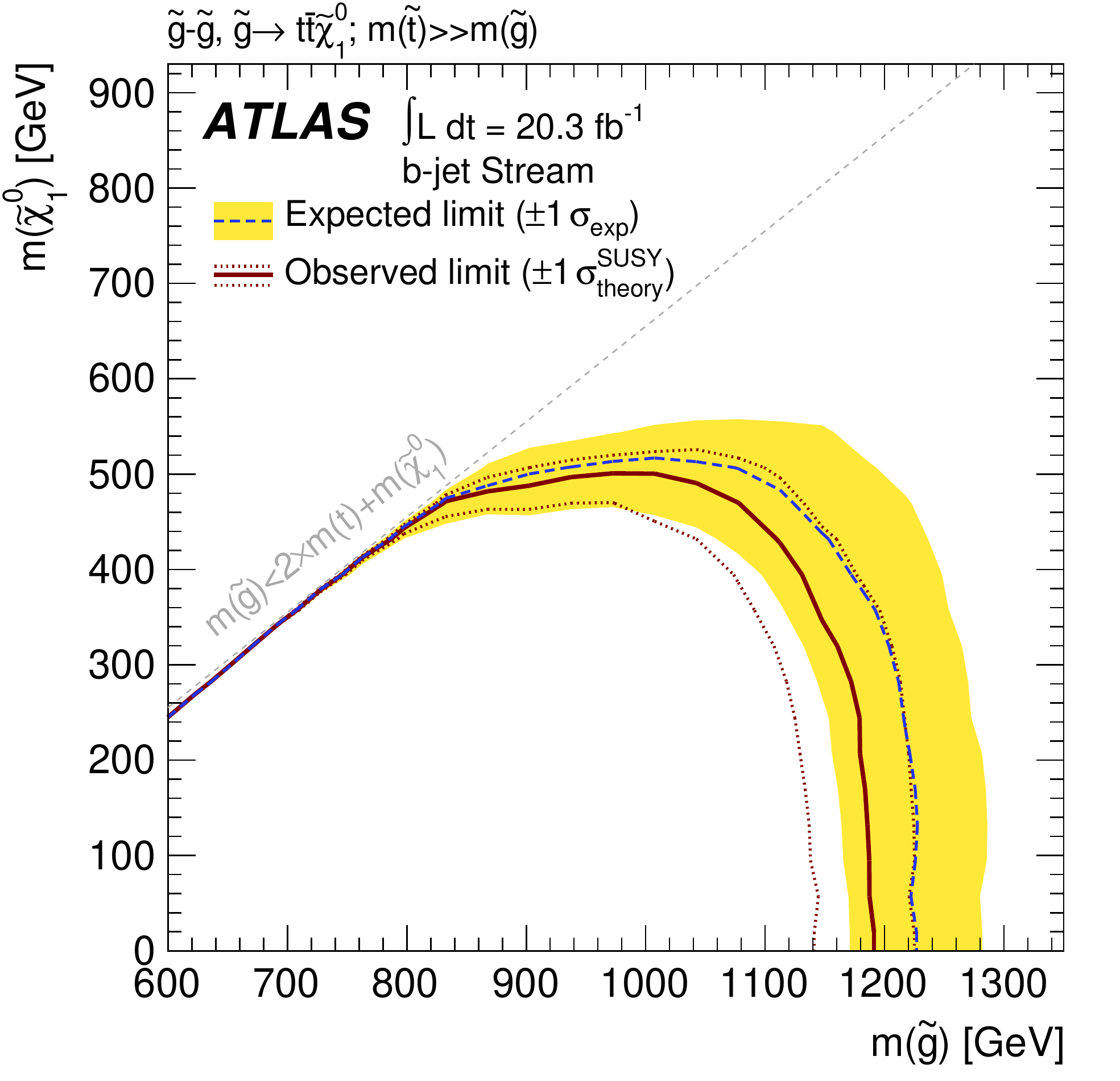}
}
\subfigure[]{
   \includegraphics[width=0.45\textwidth,angle=0]{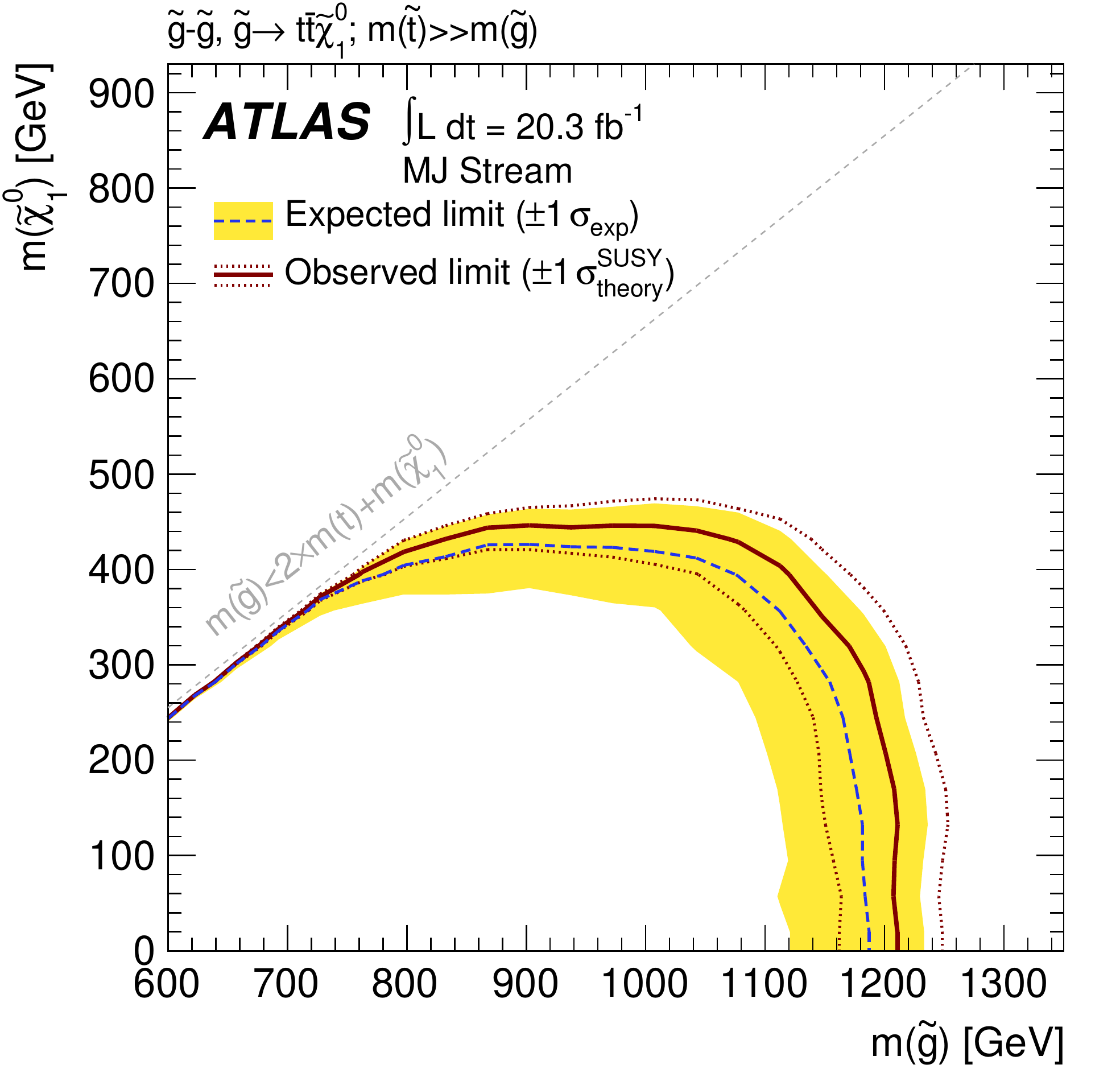}
}
\caption{The $95\%$ confidence level exclusion curve (small masses are excluded)
for the simplified SUSY model $\tilde{g}\rightarrow t \, \bar{t}\, \lsp$
in the gluino-LSP mass plane
for (a) the analysis stream that uses $b$-tags and (b) the stream that
uses the sum of fat jet masses. For each stream, the limit from the signal region
with the best expected limit is used.
The theory error band contains variations of the scales and PDFs in the SUSY prediction.
The experimental band contains all other uncertainties.
 From~\cite{Aad:2013wta}.}
\label{fig:susy_limits_separate}
\end{figure}

The data are interpreted in terms of different SUSY models by
fitting the SM and SUSY predictions to the measured distributions in the signal
and control regions. Upper limits, corresponding to
 $95\%$ confidence level, are set on the masses of new particles.
One limit is determined for every analysis stream by taking the observed limit in the signal region with the best expected
limit.
The limits for the simplified SUSY model are shown separately for the $b$-tag stream and the fat jet stream
in \figref{susy_limits_separate}. Gluino masses below $1210\GeV$ and LSP
masses below $480\GeV$ are excluded.
The observed limits on the gluino mass are similar for the two methods.
The expected limit is better for the $b$-tag stream:
by $\approx\!40\GeV$ for the gluino mass and by $\approx\!80\GeV$ for the LSP mass.
The limit difference is covered by the uncertainty bands but since the uncertainties
are strongly correlated between the two methods, the difference is significant.

Why does the fat jet mass method not improve the significance as suggested by
the study in \secref{highmultstudy}?
The most important reasons, the high jet multiplicity, the
boost of the top quarks, and the size of the fat jets, are discussed in the following.

The study showed that the performance of substructure methods degrades
at high jet multiplicity because a large number of jets can mimic the
top quark decay signature. For prong-based taggers, larger jet combinatorics
increase the chance to hit the top quark and \W boson mass windows.
For the mass cut used in the SUSY analysis, high jet multiplicity
creates a large fat jet mass  because
the fat jet then consists of more small-$R$ jets, all of which themselves have
a mass. The shift in the fat jet mass for background events when moving from low
to high \njet is shown in \figref{highmult_fj}b.
When substructure methods are used, the optimal sensitivity to
New Physics may therefore be reached at a lower jet multiplicity than in conventional analyses.

Substructure methods work better at high boost because then more or all of the
decay products are contained in the fat jet.
The study in \secref{highmultstudy} was carried out for $\sqrt{s} = 14\TeV$ while the analysis uses
the available
$\sqrt{s}=8\TeV$ data. The higher collision energy implies an increased
production cross section for SUSY particles and also a larger \pt for these
particles. The last point implies larger top quark \pt.
In addition, the gluino mass used in the study is $\approx\!100\GeV$ larger
than the exclusion limit, again increasing the boost.
Lastly, the study used
\ca fat jets with $R=1.5$. These fat jets collect more of the decay products
than the \akt $R=1.0$ jets used in the analysis and thereby allow for a better
separation of signal from background.
The \pt spectrum of the top quarks used in the study (\figref{gluinogluino}b)
and the distance between the top quark decay quarks as shown in \figref{topsize} as a function of
\pt together make it clear that only a small fraction of the top decays are
fully captured by the $R=1.0$ jets.

In summary, substructure methods have been applied in a SUSY search and the
obtained limits are close to the ones obtained by conventional methods.
The greatest benefit of top tagging is achieved at low jet multiplicities
and better results may be achieved by re-optimising the analysis cuts.
When the LHC centre-of-mass energy is increased to $13$ and then to $14\TeV$,
the boost of the hypothetical particles and the top quarks to which they decay
will be larger, thereby improving the power of fat jet techniques.

\section{Tagging Highly Boosted Top Quarks}
\label{sec:highpt}

{\em Parts of this section are taken from~\cite{Schaetzel:2013vka}, having been written by the author.}
\vglue0.8em
\noindent
The separate identification of the decay
products of highly boosted top quarks becomes experimentally challenging
when the detector granularity does not allow to resolve the individual
particle jets.
This is particularly an issue if jets are reconstructed using
calorimeter information alone as it is currently done with ATLAS data.
If two top quark decay jets are so close that they
do not leave separate clusters then top taggers based on identifying
the 3-prong decay structure will fail.
This problem was addressed in~\cite{Schaetzel:2013vka}
where a new method was proposed, the \hptt,
which combines tracking and calorimeter information.

\begin{figure}[hbt]
\centering
\includegraphics[width=0.45\textwidth]{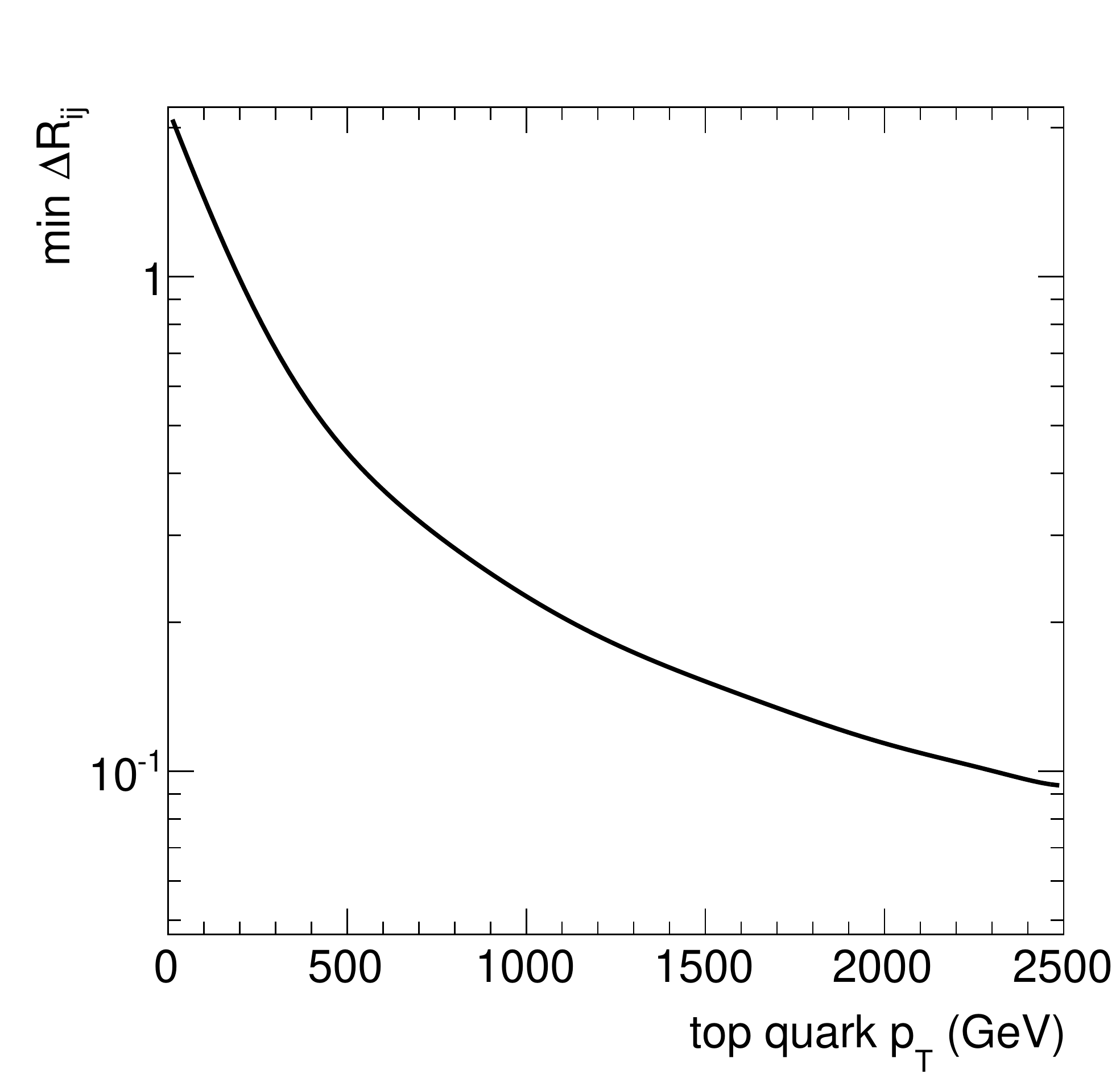}
\caption{Angular separation $\DeltaReta = \sqrt{(\Delta\eta)^2+(\Delta\phi)^2}$
of the two closest quarks in the top quark decay $t\rightarrow b q q'$ as a
function of the top quark \pt in SM \ttbar production generated with \pythia~8.
From~\cite{Schaetzel:2013vka}.}
\label{fig:minSeparation}
\end{figure}

As discussed in \secref{atlas_detector}, the minimal distance between
two clusters in a hadronic calorimeter with $0.1\times 0.1$ cells in ($\eta,\phi)$
is $\DeltaReta = 0.2$.
\figref{minSeparation} shows the angular
separation \DeltaReta of the two
closest final state quarks in hadronic top quark decay $t\rightarrow b q q$ as a function of the
top quark \pt. For $\pt = 1.12\TeV$ the separation is $0.2$ and
calorimeter resolution issues should become apparent around that \pt or even
earlier if the particle level jets that correspond to the quarks are
not collimated enough to hit only a single cell.

The angular resolution of a tracking detector is much better (cf. \secref{atlas_detector}).
The challenge for a prong-based tagger that is based on tracks is that the number
of charged particles fluctuates from jet to jet. Formulating
kinematic constraints to reject background is therefore difficult. The solution
proposed in~\cite{Schaetzel:2013vka} is to use the energy $E_{\rm jet}$
measured for the fat jet in the calorimeter and compare it with the energy $E_{\rm tracks}$ of
the tracks associated with the fat jet. Charged and neutral particles are included in $E_{\rm jet}$.
The ratio $\alpha_j = E_{\rm jet}/E_{\rm tracks}$ can be used to correct for fluctuations in
the charged particle fraction in the jet.

The input to the \hptt algorithm is a \ca $R=0.8$ calorimeter jet. The algorithm
then uses the tracks with $\pt > 500\MeV$ that are associated with the jet and
combines them to a track jet. The structure of the track jet is examined in a way similar to the \htt.
The track jet is decomposed into hard substructure objects using
an iterative mass drop procedure with $\mu = 0.8$ which ends when all substructure objects
have a mass of $20\GeV$ or less.
A notable difference to the \htt is that the \hptt does not try out all triplets
of substructure objects. Instead, all track jet constituents that are part of the
substructure objects are included in the filtering at once.
The filter radius is given by $R_{\mathrm{filt}}=\max (0.05,\min(\Delta R_{ij}/2))$
in which $\min(\Delta R_{ij})$ is the smallest pair-wise distance of all substructure objects.
The constituents of the four hardest filter jets are clustered into three subjets.
The subjet momenta are scaled by $\alpha_j$ before kinematic cuts are applied.

ATLAS subjet calibrations and subjet simulation uncertainties exist for radius
parameters $R$ between $0.2$ and $1.2$~\cite{Aad:2013gja}. Jets with a smaller
radius parameter approach the minimal hadronic cluster size.
At high top quark \pt, the subjets are more collimated and their size is smaller
than $R=0.2$. This problem is illustrated in~\cite{Schaetzel:2013vka} by
modifying the \htt to explicitly require all subjets to have $R\ge 0.2$ and setting the minimal
filter radius and the minimal distance in the mass drop procedure to $0.2$.
This modified algorithm is labelled \http.

\begin{figure}[hbt]
\centering
\subfigure[]{
   \includegraphics[width=0.45\textwidth]{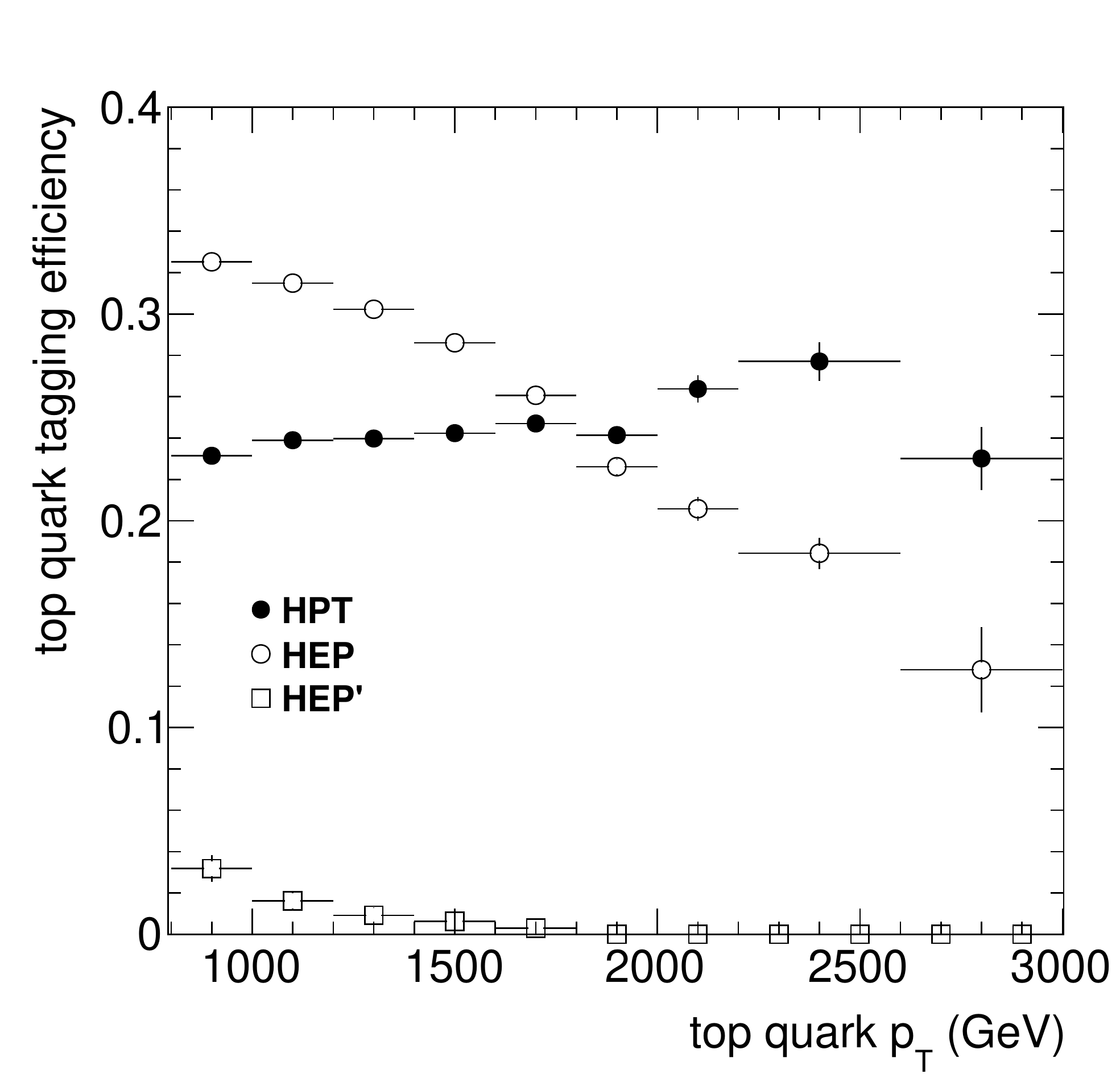}
}
\subfigure[]{
   \includegraphics[width=0.45\textwidth]{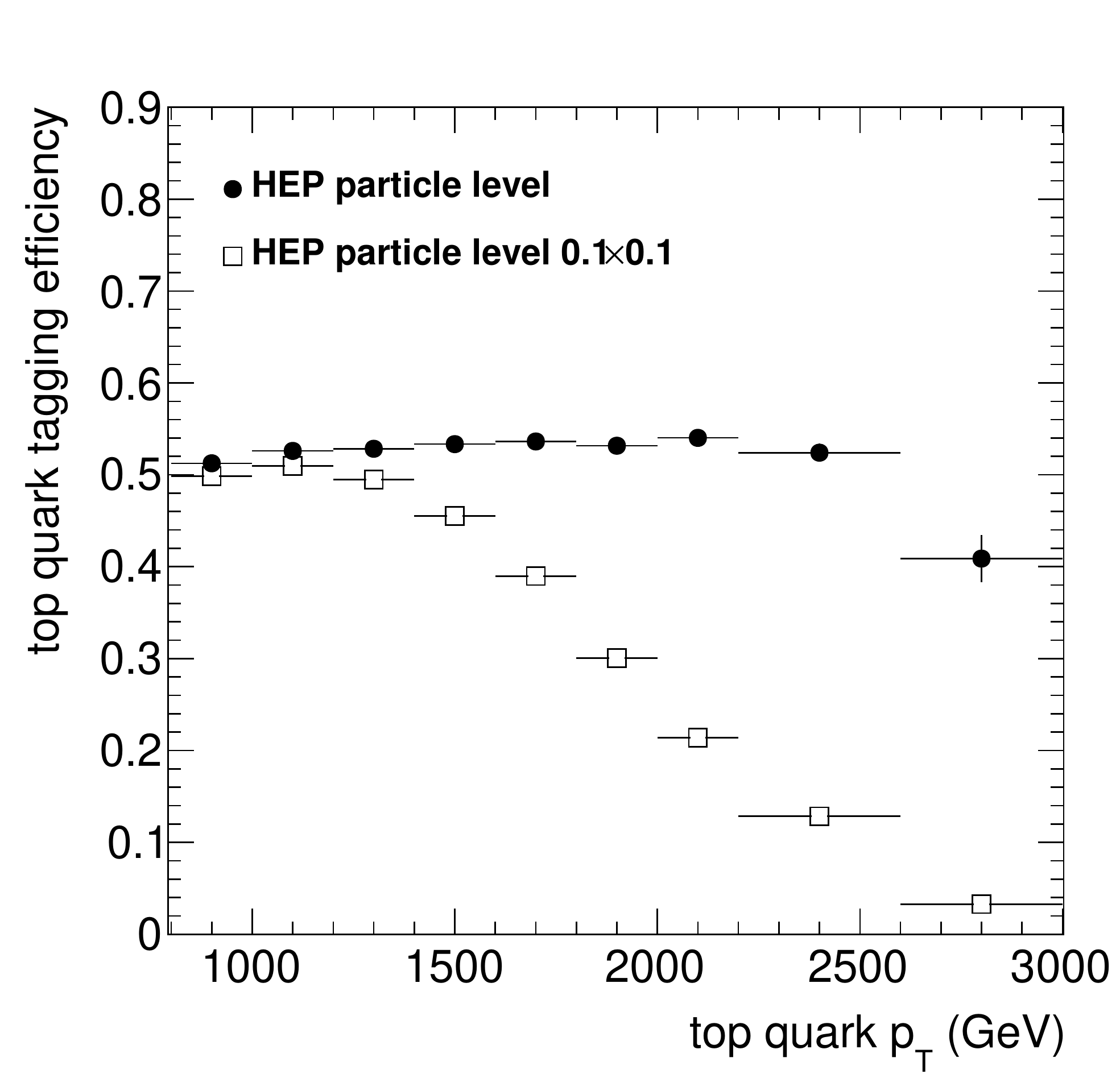}
}
\caption{Efficiencies for tagging top quarks using a) calorimeter cells and b)
stable particles in SM \ttbar events generated with \pythia~8.
For a), the events have been passed through the \delphes simulation of the ATLAS detector.
Shown are results from the \hptt (HPT), the \htt (HEP)
and from a modified \htt (HEP$'$) that requires subjets to have $R\ge 0.2$.
For the open markers in b), the particles are granularised into $0.1\times 0.1$
cells in $(\eta,\phi)$. From~\cite{Schaetzel:2013vka}.}
\label{fig:efftagging}
\end{figure}

The top quark tagging efficiencies
of the \hptt, the \htt, and the \http are shown in \figref{efftagging}a
as a function of the top quark \pt.
A top quark is taken to be tagged if a reconstructed top quark
candidate is found within $\DeltaReta=0.6$ of the top quark.
The generated events are passed through the Delphes simulation of the ATLAS detector.
The efficiency of the \http for finding top quarks using calorimeter
cells is less than $4\%$ for $\pt>800\GeV$. This implies that
with the present available ATLAS jet calibrations and uncertainties
it is not possible to find top quarks at high \pt. To obtain calibrations and uncertainties
also for jets with $R<0.2$ we suggested to use the reconstructed top mass peak
in \ttbar{} events. The position of the peak can be used for calibration and the
difference between simulation and data can serve to estimate the simulation uncertainty.
At higher top quark \pt the fraction of subjets with small $R$
will be higher. This effect can be studied by binning the mass distribution
in the top quark candidate \pt.

The \hptt efficiency is stable at $\approx 24\%$ up to $3\TeV$ in \pt.
For the \htt, the efficiency drops from $\approx 32\%$ for $800<\pt<1000\GeV$ to $\approx 13\%$
for $2600<\pt<3000\GeV$ due to the segmentation of the calorimeter which prevents
all three top quark decay particle jets from being reconstructed.
This is evident from \figref{efftagging}b where the top quark finding
efficiency is shown for the \htt at the particle level. When the constituents of the
fat jets are stable particles, the \htt efficiency is stable
at $53\%$. If the particles are granularised into $(\eta, \phi)$ cells of size
$0.1\times 0.1$, the efficiency starts to drop at a top quark \pt of $1.2\TeV$.

The minimal radius parameter for the filter track jets in the \hptt is taken to be 0.05.
It remains to be shown what the simulation uncertainties are
for such jets in the ATLAS or CMS detector. Smaller radius parameters than for calorimeter jets are possible
for track jets because of the better tracking resolution. Calibrations
for these jets may be obtained from hadronic decays of boosted \W bosons
or boosted top quarks.

Ultimately, at very high top quark \pt, the track reconstruction will struggle
to resolve the tracks left by nearby particles. The limit is reached when
hits are so close that they become part of the same reconstructed track.
The track reconstruction efficiency suffers as a consequence.
The ATLAS tracking efficiency
is $80\%$ for $\pt = 500\MeV$ and rises to $\approx\!86\%$ for
$\pt = 10\GeV$ and higher~\cite{Aad:2010rd,Aad:2010ac}.
The results in~\cite{Schaetzel:2013vka} are obtained with
a reduced tracking efficiency of $78\%$ to take the close-by effect into account.
The impact of the tracking efficiency on the \hptt performance is small,
however, because of the scaling to the calorimeter energy ($\alpha_j$).

\begin{figure}[hbt]
\centering
\subfigure[]{
   \includegraphics[width=0.45\textwidth]{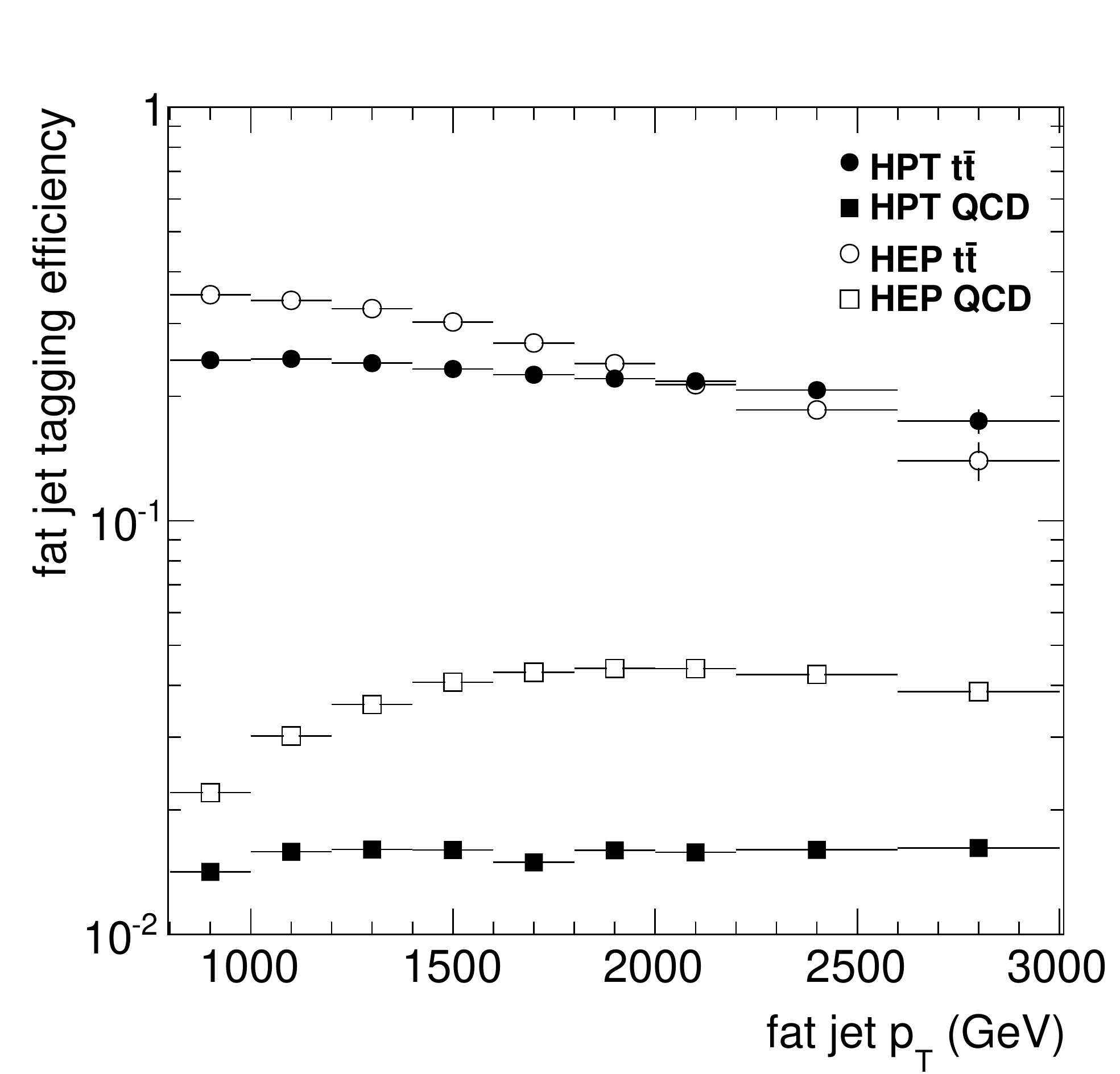}
}
\subfigure[]{
   \includegraphics[width=0.45\textwidth]{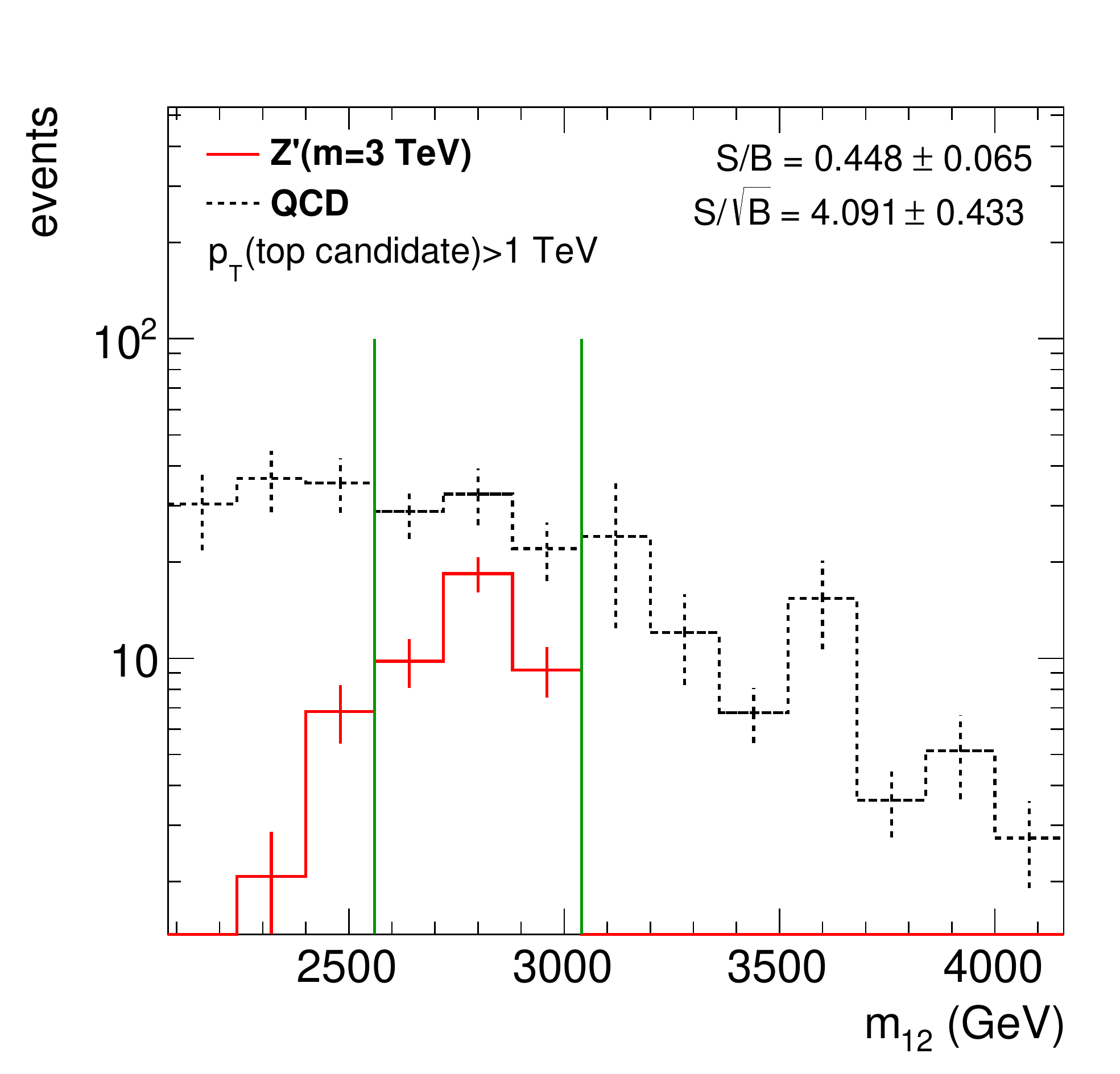}
}
\caption{(a) Efficiencies for tagging \ca $R=0.8$ fat jets
using the \hptt (HPT) and the \htt (HEP). Shown are simulations for \ttbar
events and multijet background (QCD).
(b) Invariant mass of the two leading \pt top quark candidates,
reconstructed with the \hptt from 300~fb$^{-1}$
of decays of \Zprime bosons of mass $3\TeV$, produced in $pp$ collisions
at $\sqrt{s}=14\TeV$. Also shown is the background from QCD dijet production. The signal to noise ratio
$S/B$ and the significance $S/\sqrt{B}$ are given for the indicated mass window.
All events are generated with \pythia~8 and have been passed through the \delphes simulation of the ATLAS detector.
From~\cite{Schaetzel:2013vka}.}
\label{fig:eff_fj}
\end{figure}

The tagging rate is shown as a function of fat jet \pt in \figref{eff_fj}a for
\ttbar signal and multijet background.
For \ttbar events, the values are similar to the top quark tagging efficiencies.
The fake rate, defined as the probability to tag fat jets originating from
light quarks or gluons, is stable at $1.6\%$ for the \hptt while it increases
for the \htt from $\approx 2\%$ for $\pt=800\GeV$ to $4.5\%$ for $\pt=2\TeV$.
The \hptt fake rate (and also the efficiency) is reduced with respect to the \htt
because not all possible triplets of substructure objects are tried.
The signal-to-background ratio obtained with the \hptt is better at high \pt
than that of the \htt.

An example of an application of the \hptt in a search for New Physics is shown in \figref{eff_fj}b.
The signal is a leptophobic topcolor \Zprime boson that decays to two top quarks~\cite{Harris:1999ya}.
The width of the resonance is set to $\Gamma_{\Zprime}/m_{\Zprime}= 3.2\%$
and the mass is $3\TeV$. The production cross section in $pp$ collisions at $\sqrt{s}=14\TeV$
is $3.5$~fb. Shown are distributions of the invariant mass $m_{12}$ of the two leading \pt top quark
candidates for signal and multijet background for $300\invfb$. This is the
dominant background; SM \ttbar production is smaller by a factor of $\approx 0.1$.
The only imposed requirement is that there be at least two tagged \ca $R=0.8$ jets in the
event.
The number of signal ($S$) and background events ($B$) are compared in a
mass window at the expected signal position.
The signal-to-background ratio is $S/B = 0.45(7)$ and the significance
$S/\sqrt{B} = 4.1(4)$ in the window $2560<m_{12}<3040\GeV$.
The discovery of such a \Zprime boson with the \hptt is therefore within
reach. The uncertainties are statistical and dominated by the finite number of simulated background events.
For comparison, with the same generated events, the significance when using the \htt is only $3.3(3)$ and
the difference to the \hptt results directly from the different fat jet tagging
efficiencies shown in \figref{eff_fj}a.
The sensitivity might be improved by applying $b$-quark tagging if the related
systematic uncertainties are small enough at high \pt.

The \hptt is a new algorithm to find boosted top quarks with transverse
momentum $\pt>1\TeV$. It combines track and calorimeter information and can find
top quarks via their 3-prong decay because the finer
spatial resolution of tracking detectors allows the separation of close-by particle jets
that merge in the calorimeter. This tool will prove useful in analyses looking
for massive particles that decay to top quarks.

\section{Conclusions and Outlook}
Jet structure methods that are used to identify boosted top quarks
in LHC data and analyses which employ these techniques have been reviewed.
The techniques are based on large (fat) jets which are used to capture all products
of hadronic top quark decay $t\to b qq$.

Different classes of top quark finders are used.
Energy flow taggers are based on differences in the energy sharing
between particles for signal and background.
The structure of background fat jets is predominantly given by
soft QCD parton splitting in which the original parton retains most of the energy.
The energy is more equally shared between the decay products
in heavy particle decay. Substructure variables such as \kt splitting scales
and jet mass are sensitive to this difference and cuts on these variables
can be used to reject background.
The top quark finding efficiency and the background rejection
can be adjusted continuously by changing the cut value.
These taggers are used in analyses with a relatively low
background level. There the ability to tag top quarks with high
efficiency is essential.

The second class of finders makes explicit use of the 3-prong signature of
hadronic top quark decay, which often manifests itself as three
distinct subjets within the large jet.
Three subjets are reconstructed and then used to test kinematic relations
such as the \W boson mass constraint.
The efficiency of the prong-based tagger is lower than
what is possible with energy flow taggers. This is because
(a) three subjets need to be identified, and (b) the subjets have to pass the
kinematic cuts.
The advantage of the prong-based taggers is the small fake rate:
background fat jets rarely have three hard subjets or the subjets fail the kinematic
cuts. A prominent example is the \htt with an efficiency of $40\%$
and a fake rate of $2.5\%$ for fat jets with $\pt > 400\GeV$.

The fat jets are susceptible to contributions from underlying event
and pile-up. These contributions scale with the area of the jet which is
approximately proportional to $R^2$.
The jet grooming methods that have been devised to overcome this
problem have proven to be remarkably effective, as demonstrated, for example,
by the pile-up stability of the reconstructed top quark mass.

The structure of background (non-top) fat jets is described within
$10$--$20\%$ by Monte Carlo simulations.
Substructure variables such as the jet mass are sensitive to
variations of the parameters in the models for hadronisation and underlying event generation.
The good description is possible only with parameters tuned to minimum bias
LHC data.
The best description is obtained when using \herwigpp.
Significant discrepancies between simulation and data are observed in tails of
distributions such as \kt splitting scales and \Nsjn for \pythia and \powhegpythia.
These discrepancies are improved by jet grooming, indicating that the
problem lies in the simulation of soft energy deposits.
Whether the difference between \herwigpp and \pythia
is due to the hadronisation model, the underlying event model,
the parton shower, or a combination of all three is unclear and more studies are needed.
For example, determining to what extend the jet substructure measurements can be
used to improve the tuning of the models seems to be a topic worth pursuing.

The structure of fat jets from top quark decay is well described
by simulation. Differences to the data are at the level of $5\%$ for the fat jet mass.
Even distribution of variables that involve many analysis steps,
like the subjet invariant mass ratios of the \htt, are well predicted.
This makes it possible to apply the substructure techniques in searches for New
Physics.

Limits on a technicolor \Zprime boson and a Kaluza-Klein gluon have been extended
beyond the Tevatron limits using substructure techniques.
Substructure techniques have a better \ttbar mass resolution
than conventional (resolved) top quark reconstruction methods at high \ttbar masses
where the top quark decay products are contained in the fat jet.
A resolved analysis combines all small-$R$ jets without exploiting
the boosted topology and often picks up a wrong small-$R$ jet as a top quark decay jet.
This then spoils the reconstructed top quark mass. By considering
only the subjets inside a fat jet, the combinatorics is much reduced,
and the mass is reconstructed more precisely.

%%%%%%%%%%%%%%%%%%%%%%%%%%%%%%%%%% Outlook %%%%%%%%%%%%%%%%%%%%%%%%%%%%%%%%%%%%%
The search for new particles was the original impetus in~\cite{Seymour:1993mx}
and~\cite{Butterworth:2002tt} and substructure methods have so far been used
in such searches in events with top quarks at high or intermediate \pt.
These searches will continue with higher statistics and at larger centre-of-mass
energies. The LHC Run 2 is scheduled to start in spring 2015 with $\sqrt{s} = 13\TeV$.
At this high energy, the production cross section for massive particles rises
and the particles will be more boosted.
The angular resolution of the
ATLAS and CMS calorimeters can become a limiting factor as described in \secref{highpt}
and more emphasis will likely have to be placed on the use of tracking information.

The second major driving force behind substructure techniques is the recovery
of hadronic decay channels of the Higgs boson.
Without substructure methods, these channels seem unpromising given the large
multijet background.
In the seminal paper~\cite{Butterworth:2008iy}, the expected significance
for a Higgs boson signal at $m_H = 125\GeV$ is $\approx\!5\sigma$ at $\sqrt{s} = 14\TeV$ and $L=30\invfb$.
The analysis used \ca $R=1.2$ fat jets with $\pt>200\GeV$ to find boosted $H\to b\bar{b}$ decays
in events with $VH$ production ($V=W,Z$). In approximately $5\%$ of the
$VH$ events, the Higgs boson has $\pt>200\GeV$.
In analyses of the LHC Run 1 data ($5\invfb$ at $\sqrt{s}=7\TeV$ and $20\invfb$ at $8\TeV$),
substructure methods do not increase the sensitivity to the Higgs boson.
This is due to reduced statistics at the smaller collision energy.
However, for LHC Run 2 there
exists a very distinct probability to see the Higgs boson in the $b\bar{b}$ channel.

Events with \ttH production can be used to determine the
\ttH Yukawa coupling. This process
has been studied in~\cite{Plehn:2009rk} for $H\to b\bar{b}$ with
one leptonic and one hadronic top quark decay. With $100\invfb$ of data at $14\TeV$, a significance
of $\approx\!4\sigma$ was found for a Higgs of mass $120\GeV$.
Also this analysis becomes feasible in Run 2.

Only substructure methods that have so far been used in physics analyses (as opposed to
studies) were shown in this review. There are many more recently proposed methods. Examples are
shower deconstruction~\cite{Soper:2011cr,Soper:2012pb} and \qjets~\cite{Ellis:2012sn},
which statistically analyse the clustering history in jets.
\qjets have been studied with ATLAS~\cite{ATLAS-CONF-2013-087} and
for an efficiency to tag boosted hadronically decaying \W bosons
of $50\%$, the rejection of multijet background is 15.

The field of jet substructure is new and it evolves through
a close collaboration of theoretical and experimental particle physicists.
In the short time of its existence it has become a proven integral part of
the analysis tool set at the LHC.
Jet structure techniques will become more important
in the next years with the LHC running at full design energy and they will enhance the physics potential
and reach at the energy frontier.

%%%%%%%%%%%%%%%%%%%%%%%%%%%%%%%%%%%%%%%%%%%%%%%%%%%%%%%%%%%%%%%%%%%%%%%%%%%%%%
% Bibliography
%%%%%%%%%%%%%%%%%%%%%%%%%%%%%%%%%%%%%%%%%%%%%%%%%%%%%%%%%%%%%%%%%%%%%%%%%%%%%%
\bibliographystyle{atlasnote}
\bibliography{text}

\providecommand{\href}[2]{#2}\begingroup\raggedright\begin{thebibliography}{100}

\bibitem{Aad:2012tfa}
{ATLAS} Collaboration, {\em {Observation of a new particle in the search for
  the Standard Model Higgs boson with the ATLAS detector at the LHC}\/},
  \href{http://dx.doi.org/10.1016/j.physletb.2012.08.020}{Phys. Lett. {\bf
  B716} (2012)  1--29},
\href{http://arxiv.org/abs/1207.7214}{{\tt arXiv:1207.7214 [hep-ex]}}.
%%CITATION = ARXIV:1207.7214;%%.

\bibitem{Chatrchyan:2012ufa}
{CMS} Collaboration, {\em {Observation of a new boson at a mass of 125 GeV with
  the CMS experiment at the LHC}\/},
  \href{http://dx.doi.org/10.1016/j.physletb.2012.08.021}{Phys. Lett. {\bf
  B716} (2012)  30--61},
\href{http://arxiv.org/abs/1207.7235}{{\tt arXiv:1207.7235 [hep-ex]}}.
%%CITATION = ARXIV:1207.7235;%%.

\bibitem{Seymour:1993mx}
M.~H. Seymour, {\em {Searches for new particles using cone and cluster jet
  algorithms: A Comparative study}\/},
\href{http://dx.doi.org/10.1007/BF01559532}{Z. Phys. {\bf C62} (1994)
  127--138}.
%%CITATION = ZEPYA,C62,127;%%.

\bibitem{Butterworth:2008iy}
J.~M. Butterworth, A.~R. Davison, M.~Rubin, and G.~P. Salam, {\em {Jet
  substructure as a new Higgs search channel at the LHC}\/},
  \href{http://dx.doi.org/10.1103/PhysRevLett.100.242001}{Phys. Rev. Lett. {\bf
  100} (2008)  242001},
\href{http://arxiv.org/abs/0802.2470}{{\tt arXiv:0802.2470 [hep-ph]}}.
%%CITATION = ARXIV:0802.2470;%%.

\bibitem{Plehn:2009rk}
T.~Plehn, G.~P. Salam, and M.~Spannowsky, {\em {Fat Jets for a Light Higgs}\/},
   \href{http://dx.doi.org/10.1103/PhysRevLett.104.111801}{Phys. Rev. Lett.
  {\bf 104} (2010)  111801},
\href{http://arxiv.org/abs/0910.5472}{{\tt arXiv:0910.5472 [hep-ph]}}.
%%CITATION = ARXIV:0910.5472;%%.

\bibitem{Abdesselam:2010pt}
A.~Abdesselam, E.~B. Kuutmann, U.~Bitenc, G.~Brooijmans, J.~Butterworth, et
  al., {\em {Boosted objects: A Probe of beyond the Standard Model physics}\/},
   \href{http://dx.doi.org/10.1140/epjc/s10052-011-1661-y}{Eur. Phys. J. {\bf
  C71} (2011)  1661},
\href{http://arxiv.org/abs/1012.5412}{{\tt arXiv:1012.5412 [hep-ph]}}.
%%CITATION = ARXIV:1012.5412;%%.

\bibitem{Altheimer:2012mn}
A.~Altheimer, S.~Arora, L.~Asquith, G.~Brooijmans, J.~Butterworth, et al., {\em
  {Jet Substructure at the Tevatron and LHC: New results, new tools, new
  benchmarks}\/},  \href{http://dx.doi.org/10.1088/0954-3899/39/6/063001}{J.
  Phys. {\bf G39} (2012)  063001},
\href{http://arxiv.org/abs/1201.0008}{{\tt arXiv:1201.0008 [hep-ph]}}.
%%CITATION = ARXIV:1201.0008;%%.

\bibitem{Plehn:2011tg}
T.~Plehn and M.~Spannowsky, {\em {Top Tagging}\/},
  \href{http://dx.doi.org/10.1088/0954-3899/39/8/083001}{J. Phys. {\bf G39}
  (2012)  083001},
\href{http://arxiv.org/abs/1112.4441}{{\tt arXiv:1112.4441 [hep-ph]}}.
%%CITATION = ARXIV:1112.4441;%%.

\bibitem{Schilling:2012dx}
F.-P. Schilling, {\em {Top Quark Physics at the LHC: A Review of the First Two
  Years}\/},  \href{http://dx.doi.org/10.1142/S0217751X12300165}{Int. J. Mod.
  Phys. {\bf A27} (2012)  1230016},
\href{http://arxiv.org/abs/1206.4484}{{\tt arXiv:1206.4484 [hep-ex]}}.
%%CITATION = ARXIV:1206.4484;%%.

\bibitem{Beringer:1900zz}
{Particle Data Group}, {\em {Review of Particle Physics}\/},
\href{http://dx.doi.org/10.1103/PhysRevD.86.010001}{Phys. Rev. {\bf D86} (2012)
   010001}.
%%CITATION = PHRVA,D86,010001;%%.

\bibitem{PDG2013web}
{Particle Data Group}, {\em 2013 partial update for the 2014 edition\/},
\newblock \url{http://pdg.lbl.gov}.

\bibitem{Aad:2011eu}
{ATLAS} Collaboration, {\em {Measurement of the Inelastic Proton-Proton
  Cross-Section at $\sqrt{s}=7$ TeV with the ATLAS Detector}\/},
  \href{http://dx.doi.org/10.1038/ncomms1472}{Nature Commun. {\bf 2} (2011)
  463},
\href{http://arxiv.org/abs/1104.0326}{{\tt arXiv:1104.0326 [hep-ex]}}.
%%CITATION = ARXIV:1104.0326;%%.

\bibitem{Martin:2009iq}
A.~Martin, W.~Stirling, R.~Thorne, and G.~Watt, {\em {Parton distributions for
  the LHC}\/},  \href{http://dx.doi.org/10.1140/epjc/s10052-009-1072-5}{Eur.
  Phys. J. {\bf C63} (2009)  189--285},
\href{http://arxiv.org/abs/0901.0002}{{\tt arXiv:0901.0002 [hep-ph]}}.
%%CITATION = ARXIV:0901.0002;%%.

\bibitem{Aliev:2010zk}
M.~Aliev, H.~Lacker, U.~Langenfeld, S.~Moch, P.~Uwer, et al., {\em {HATHOR:
  HAdronic Top and Heavy quarks crOss section calculatoR}\/},
  \href{http://dx.doi.org/10.1016/j.cpc.2010.12.040}{Comput. Phys. Commun. {\bf
  182} (2011)  1034--1046},
\href{http://arxiv.org/abs/1007.1327}{{\tt arXiv:1007.1327 [hep-ph]}}.
%%CITATION = ARXIV:1007.1327;%%.

\bibitem{Campbell:2006wx}
J.~M. Campbell, J.~Huston, and W.~Stirling, {\em {Hard Interactions of Quarks
  and Gluons: A Primer for LHC Physics}\/},
  \href{http://dx.doi.org/10.1088/0034-4885/70/1/R02}{Rept. Prog. Phys. {\bf
  70} (2007)  89},
\href{http://arxiv.org/abs/hep-ph/0611148}{{\tt arXiv:hep-ph/0611148
  [hep-ph]}}.
%%CITATION = HEP-PH/0611148;%%.

\bibitem{SM_xs_summary}
{ATLAS} Collaboration, {\em Physics Summary Plots\/},  2013.
\newblock
  \url{https://twiki.cern.ch/twiki/bin/view/AtlasPublic/CombinedSummaryPlots}.

\bibitem{QuadtLiss2010}
T.~M. Liss and A.~Quadt,
  \href{http://dx.doi.org/10.1103/PhysRevD.86.010001}{{\em The Top Quark\/}, }
  in \cite{Beringer:1900zz},
p.~596.
\newblock
%%CITATION = PHRVA,D86,010001;%%.

\bibitem{Salam_boost}
G.~P. Salam, {\em Theory of Fat Jets and Jet Substructure\/},  Invited talk at
  Higgs Hunting 2012, July, 2012.
\newblock \url{http://www.higgshunting.fr/}.

\bibitem{Plehn:2010st}
T.~Plehn, M.~Spannowsky, M.~Takeuchi, and D.~Zerwas, {\em {Stop Reconstruction
  with Tagged Tops}\/},  \href{http://dx.doi.org/10.1007/JHEP10(2010)078}{JHEP
  {\bf 1010} (2010)  078},
\href{http://arxiv.org/abs/1006.2833}{{\tt arXiv:1006.2833 [hep-ph]}}.
%%CITATION = ARXIV:1006.2833;%%.

\bibitem{Hill:2002ap}
C.~T. Hill and E.~H. Simmons, {\em {Strong dynamics and electroweak symmetry
  breaking}\/},  \href{http://dx.doi.org/10.1016/S0370-1573(03)00140-6}{Phys.
  Rept. {\bf 381} (2003)  235--402},
\href{http://arxiv.org/abs/hep-ph/0203079}{{\tt arXiv:hep-ph/0203079
  [hep-ph]}}.
%%CITATION = HEP-PH/0203079;%%.

\bibitem{Fayet:1976et}
P.~Fayet, {\em {Supersymmetry and Weak, Electromagnetic and Strong
  Interactions}\/},
\href{http://dx.doi.org/10.1016/0370-2693(76)90319-1}{Phys. Lett. {\bf B64}
  (1976)  159}.
%%CITATION = PHLTA,B64,159;%%.

\bibitem{Fayet:1977yc}
P.~Fayet, {\em {Spontaneously Broken Supersymmetric Theories of Weak,
  Electromagnetic and Strong Interactions}\/},
\href{http://dx.doi.org/10.1016/0370-2693(77)90852-8}{Phys. Lett. {\bf B69}
  (1977)  489}.
%%CITATION = PHLTA,B69,489;%%.

\bibitem{Farrar:1978xj}
G.~R. Farrar and P.~Fayet, {\em {Phenomenology of the Production, Decay, and
  Detection of New Hadronic States Associated with Supersymmetry}\/},
\href{http://dx.doi.org/10.1016/0370-2693(78)90858-4}{Phys. Lett. {\bf B76}
  (1978)  575--579}.
%%CITATION = PHLTA,B76,575;%%.

\bibitem{Fayet:1979sa}
P.~Fayet, {\em {Relations Between the Masses of the Superpartners of Leptons
  and Quarks, the Goldstino Couplings and the Neutral Currents}\/},
\href{http://dx.doi.org/10.1016/0370-2693(79)91229-2}{Phys. Lett. {\bf B84}
  (1979)  416}.
%%CITATION = PHLTA,B84,416;%%.

\bibitem{Dimopoulos:1981zb}
S.~Dimopoulos and H.~Georgi, {\em {Softly Broken Supersymmetry and SU(5)}\/},
\href{http://dx.doi.org/10.1016/0550-3213(81)90522-8}{Nucl. Phys. {\bf B193}
  (1981)  150}.
%%CITATION = NUPHA,B193,150;%%.

\bibitem{Schmaltz:2005ky}
M.~Schmaltz and D.~Tucker-Smith, {\em {Little Higgs review}\/},
  \href{http://dx.doi.org/10.1146/annurev.nucl.55.090704.151502}{Ann. Rev.
  Nucl. Part. Sci. {\bf 55} (2005)  229--270},
\href{http://arxiv.org/abs/hep-ph/0502182}{{\tt arXiv:hep-ph/0502182
  [hep-ph]}}.
%%CITATION = HEP-PH/0502182;%%.

\bibitem{ArkaniHamed:2001nc}
N.~Arkani-Hamed, A.~G. Cohen, and H.~Georgi, {\em {Electroweak symmetry
  breaking from dimensional deconstruction}\/},
  \href{http://dx.doi.org/10.1016/S0370-2693(01)00741-9}{Phys. Lett. {\bf B513}
  (2001)  232--240},
\href{http://arxiv.org/abs/hep-ph/0105239}{{\tt arXiv:hep-ph/0105239
  [hep-ph]}}.
%%CITATION = HEP-PH/0105239;%%.

\bibitem{Weinberg:1975gm}
S.~Weinberg, {\em {Implications of Dynamical Symmetry Breaking}\/},
\href{http://dx.doi.org/10.1103/PhysRevD.13.974}{Phys. Rev. {\bf D13} (1976)
  974--996}.
%%CITATION = PHRVA,D13,974;%%.

\bibitem{Susskind:1978ms}
L.~Susskind, {\em {Dynamics of Spontaneous Symmetry Breaking in the
  Weinberg-Salam Theory}\/},
\href{http://dx.doi.org/10.1103/PhysRevD.20.2619}{Phys. Rev. {\bf D20} (1979)
  2619--2625}.
%%CITATION = PHRVA,D20,2619;%%.

\bibitem{MandlShaw}
F.~Mandl and G.~Shaw, {\em Quantum Field Theory}.
\newblock 1984.

\bibitem{Hill:1991at}
C.~T. Hill, {\em {Topcolor: Top quark condensation in a gauge extension of the
  standard model}\/},
\href{http://dx.doi.org/10.1016/0370-2693(91)91061-Y}{Phys. Lett. {\bf B266}
  (1991)  419--424}.
%%CITATION = PHLTA,B266,419;%%.

\bibitem{Hill:1994hp}
C.~T. Hill, {\em {Topcolor assisted technicolor}\/},
  \href{http://dx.doi.org/10.1016/0370-2693(94)01660-5}{Phys. Lett. {\bf B345}
  (1995)  483--489},
\href{http://arxiv.org/abs/hep-ph/9411426}{{\tt arXiv:hep-ph/9411426
  [hep-ph]}}.
%%CITATION = HEP-PH/9411426;%%.

\bibitem{Harris:1999ya}
R.~M. Harris, C.~T. Hill, and S.~J. Parke, {\em {Cross-section for topcolor
  $Z_t^\prime$ decaying to \ttbar}\/},
\href{http://arxiv.org/abs/hep-ph/9911288}{{\tt arXiv:hep-ph/9911288
  [hep-ph]}}.
%%CITATION = HEP-PH/9911288;%%.

\bibitem{Randall:1999ee}
L.~Randall and R.~Sundrum, {\em {A Large mass hierarchy from a small extra
  dimension}\/},  \href{http://dx.doi.org/10.1103/PhysRevLett.83.3370}{Phys.
  Rev. Lett. {\bf 83} (1999)  3370--3373},
\href{http://arxiv.org/abs/hep-ph/9905221}{{\tt arXiv:hep-ph/9905221
  [hep-ph]}}.
%%CITATION = HEP-PH/9905221;%%.

\bibitem{Randall:1999vf}
L.~Randall and R.~Sundrum, {\em {An Alternative to compactification}\/},
  \href{http://dx.doi.org/10.1103/PhysRevLett.83.4690}{Phys. Rev. Lett. {\bf
  83} (1999)  4690--4693},
\href{http://arxiv.org/abs/hep-th/9906064}{{\tt arXiv:hep-th/9906064
  [hep-th]}}.
%%CITATION = HEP-TH/9906064;%%.

\bibitem{Morrissey:2009tf}
D.~E. Morrissey, T.~Plehn, and T.~M. Tait, {\em {Physics searches at the
  LHC}\/},  \href{http://dx.doi.org/10.1016/j.physrep.2012.02.007}{Phys. Rept.
  {\bf 515} (2012)  1--113},
\href{http://arxiv.org/abs/0912.3259}{{\tt arXiv:0912.3259 [hep-ph]}}.
%%CITATION = ARXIV:0912.3259;%%.

\bibitem{GiudiceWells2010}
G.~F. Giudice and J.~D. Wells,
  \href{http://dx.doi.org/10.1103/PhysRevD.86.010001}{{\em Extra Dimensions\/},
  } in \cite{Beringer:1900zz},
p.~1354.
\newblock
%%CITATION = PHRVA,D86,010001;%%.

\bibitem{Sellerholm}
A.~Sellerholm, {\em {Kaluza-Klein Theories}\/},  2005.
\newblock \url{http://courses.theophys.kth.se/5A1381/reports/sellerholm.pdf}.

\bibitem{Kaluza:1921tu}
T.~Kaluza, {\em {Zum Unit\"atsproblem der Physik}\/},  Sitzungsber. Preuss.
  Akad. Wiss. Berlin (Math.Phys.) {\bf 1921} (1921)  966--972.
\url{https://archive.org/details/sitzungsberichte1921preussi}.
%%CITATION = SPWPA,1921,966;%%.

\bibitem{Klein:1926tv}
O.~Klein, {\em {Quantentheorie und f\"unfdimensionale
  Relativit\"atstheorie}\/},
\href{http://dx.doi.org/10.1007/BF01397481}{Z. Phys. {\bf 37} (1926)
  895--906}.
%%CITATION = ZEPYA,37,895;%%.

\bibitem{Lillie:2007yh}
B.~Lillie, L.~Randall, and L.-T. Wang, {\em {The Bulk RS KK-gluon at the
  LHC}\/},  \href{http://dx.doi.org/10.1088/1126-6708/2007/09/074}{JHEP {\bf
  0709} (2007)  074},
\href{http://arxiv.org/abs/hep-ph/0701166}{{\tt arXiv:hep-ph/0701166
  [hep-ph]}}.
%%CITATION = HEP-PH/0701166;%%.

\bibitem{Sterman:1977wj}
G.~F. Sterman and S.~Weinberg, {\em {Jets from Quantum Chromodynamics}\/},
\href{http://dx.doi.org/10.1103/PhysRevLett.39.1436}{Phys. Rev. Lett. {\bf 39}
  (1977)  1436}.
%%CITATION = PRLTA,39,1436;%%.

\bibitem{Arnison:1983gw}
{UA1} Collaboration, {\em {Hadronic Jet Production at the CERN Proton -
  anti-Proton Collider}\/},
\href{http://dx.doi.org/10.1016/0370-2693(83)90254-X}{Phys. Lett. {\bf B132}
  (1983)  214}.
%%CITATION = PHLTA,B132,214;%%.

\bibitem{Salam:2009jx}
G.~P. Salam, {\em {Towards Jetography}\/},
  \href{http://dx.doi.org/10.1140/epjc/s10052-010-1314-6}{Eur. Phys. J. {\bf
  C67} (2010)  637--686},
\href{http://arxiv.org/abs/0906.1833}{{\tt arXiv:0906.1833 [hep-ph]}}.
%%CITATION = ARXIV:0906.1833;%%.

\bibitem{Dokshitzer:1997in}
Y.~L. Dokshitzer, G.~Leder, S.~Moretti, and B.~Webber, {\em {Better jet
  clustering algorithms}\/},
  \href{http://dx.doi.org/10.1088/1126-6708/1997/08/001}{JHEP {\bf 08} (1997)
  001},
\href{http://arxiv.org/abs/hep-ph/9707323}{{\tt arXiv:hep-ph/9707323
  [hep-ph]}}.
%%CITATION = HEP-PH/9707323;%%.

\bibitem{Wobisch:1998wt}
M.~Wobisch and T.~Wengler, {\em {Hadronization corrections to jet cross
  sections in deep- inelastic scattering}\/},
\href{http://arxiv.org/abs/hep-ph/9907280}{{\tt arXiv:hep-ph/9907280
  [hep-ph]}}.
%%CITATION = HEP-PH/9907280;%%.

\bibitem{WobischPhD}
M.~Wobisch, {\em Measurement and QCD Analysis of Jet Cross Sections in
  Deep-Inelastic Positron-Proton Collisions at $\sqrt{s}$ = 300 GeV}.
\newblock PhD thesis, {Rheinisch-Westf\"alische Technische Hochschule Aachen,
  Germany}, 2000.
\newblock \url{http://www-h1.desy.de/psfiles/theses/h1th-201.pdf}.

\bibitem{Catani:1991hj}
S.~Catani, Y.~L. Dokshitzer, M.~Olsson, G.~Turnock, and B.~Webber, {\em {New
  clustering algorithm for multi-jet cross-sections in $e^{+}e^{-}$
  annihilation}\/},
\href{http://dx.doi.org/10.1016/0370-2693(91)90196-W}{Phys. Lett. {\bf B 269}
  (1991)  432}.
%%CITATION = PHLTA,B269,432;%%.

\bibitem{Ellis1993}
S.~D. Ellis and D.~E. Soper, {\em {Successive combination jet algorithm for
  hadron collisions}\/},
  \href{http://dx.doi.org/10.1103/PhysRevD.48.3160}{Phys. Rev. {\bf D 48}
  (1993)  3160},
\href{http://arxiv.org/abs/hep-ph/9305266}{{\tt arXiv:hep-ph/9305266
  [hep-ph]}}.
%%CITATION = HEP-PH/9305266;%%.

\bibitem{Catani1993}
S.~Catani, Y.~L. Dokshitzer, M.~Seymour, and B.~Webber, {\em {Longitudinally
  invariant $k_{\perp}$ clustering algorithms for hadron hadron collisions}\/},
\href{http://dx.doi.org/10.1016/0550-3213(93)90166-M}{Nucl. Phys. {\bf B 406}
  (1993)  187}.
%%CITATION = NUPHA,B406,187;%%.

\bibitem{Cacciari:2008gp}
M.~Cacciari, G.~P. Salam, and G.~Soyez, {\em {The anti-k$_t$ jet clustering
  algorithm}\/},  \href{http://dx.doi.org/10.1088/1126-6708/2008/04/063}{JHEP
  {\bf 04} (2008)  063},
\href{http://arxiv.org/abs/0802.1189}{{\tt arXiv:0802.1189 [hep-ph]}}.
%%CITATION = ARXIV:0802.1189;%%.

\bibitem{Marchesini:1991ch}
G.~Marchesini, B.~Webber, G.~Abbiendi, I.~Knowles, M.~Seymour, et al., {\em
  {HERWIG: A Monte Carlo event generator for simulating hadron emission
  reactions with interfering gluons. Version 5.1 - April 1991}\/},
\href{http://dx.doi.org/10.1016/0010-4655(92)90055-4}{Comput. Phys. Commun.
  {\bf 67} (1992)  465--508}.
%%CITATION = CPHCB,67,465;%%.

\bibitem{Altarelli:1977zs}
G.~Altarelli and G.~Parisi, {\em {Asymptotic Freedom in Parton Language}\/},
\href{http://dx.doi.org/10.1016/0550-3213(77)90384-4}{Nucl. Phys. {\bf B126}
  (1977)  298}.
%%CITATION = NUPHA,B126,298;%%.

\bibitem{ATLAS:2012am}
{ATLAS} Collaboration, {\em {Jet mass and substructure of inclusive jets in
  $\sqrt{s}$ = 7 TeV $pp$ collisions with the ATLAS experiment}\/},
  \href{http://dx.doi.org/10.1007/JHEP05(2012)128}{JHEP {\bf 1205} (2012)
  128},
\href{http://arxiv.org/abs/1203.4606}{{\tt arXiv:1203.4606 [hep-ex]}}.
%%CITATION = ARXIV:1203.4606;%%.

\bibitem{Butterworth:2002tt}
J.~Butterworth, B.~Cox, and J.~R. Forshaw, {\em {$W W$ scattering at the CERN
  LHC}\/},  \href{http://dx.doi.org/10.1103/PhysRevD.65.096014}{Phys. Rev. {\bf
  D65} (2002)  096014},
\href{http://arxiv.org/abs/hep-ph/0201098}{{\tt arXiv:hep-ph/0201098
  [hep-ph]}}.
%%CITATION = HEP-PH/0201098;%%.

\bibitem{Ellis:2007ib}
S.~Ellis, J.~Huston, K.~Hatakeyama, P.~Loch, and M.~Tonnesmann, {\em {Jets in
  hadron-hadron collisions}\/},
  \href{http://dx.doi.org/10.1016/j.ppnp.2007.12.002}{Prog. Part. Nucl. Phys.
  {\bf 60} (2008)  484--551},
\href{http://arxiv.org/abs/0712.2447}{{\tt arXiv:0712.2447 [hep-ph]}}.
%%CITATION = ARXIV:0712.2447;%%.

\bibitem{Dasgupta:2007wa}
M.~Dasgupta, L.~Magnea, and G.~P. Salam, {\em {Non-perturbative QCD effects in
  jets at hadron colliders}\/},
  \href{http://dx.doi.org/10.1088/1126-6708/2008/02/055}{JHEP {\bf 0802} (2008)
   055},
\href{http://arxiv.org/abs/0712.3014}{{\tt arXiv:0712.3014 [hep-ph]}}.
%%CITATION = ARXIV:0712.3014;%%.

\bibitem{Krohn:2009th}
D.~Krohn, J.~Thaler, and L.-T. Wang, {\em {Jet Trimming}\/},
  \href{http://dx.doi.org/10.1007/JHEP02(2010)084}{JHEP {\bf 1002} (2010)
  084},
\href{http://arxiv.org/abs/0912.1342}{{\tt arXiv:0912.1342 [hep-ph]}}.
%%CITATION = ARXIV:0912.1342;%%.

\bibitem{Ellis:2009su}
S.~D. Ellis, C.~K. Vermilion, and J.~R. Walsh, {\em {Techniques for improved
  heavy particle searches with jet substructure}\/},
  \href{http://dx.doi.org/10.1103/PhysRevD.80.051501}{Phys. Rev. {\bf D80}
  (2009)  051501},
\href{http://arxiv.org/abs/0903.5081}{{\tt arXiv:0903.5081 [hep-ph]}}.
%%CITATION = ARXIV:0903.5081;%%.

\bibitem{Ellis:2009me}
S.~D. Ellis, C.~K. Vermilion, and J.~R. Walsh, {\em {Recombination Algorithms
  and Jet Substructure: Pruning as a Tool for Heavy Particle Searches}\/},
  \href{http://dx.doi.org/10.1103/PhysRevD.81.094023}{Phys. Rev. {\bf D81}
  (2010)  094023},
\href{http://arxiv.org/abs/0912.0033}{{\tt arXiv:0912.0033 [hep-ph]}}.
%%CITATION = ARXIV:0912.0033;%%.

\bibitem{LHCstatus}
L.~Ponce, {\em LHC Accelerator Performance and Plans\/},  165$^{th}$ CERN
  Council Meeting, Dec 2012.
\newblock \url{https://indico.cern.ch/conferenceDisplay.py?confId=218449}.

\bibitem{mu_lumi}
{ATLAS} Collaboration, {\em Public Luminosity Results\/},  2013.
\newblock
  \url{https://twiki.cern.ch/twiki/bin/view/AtlasPublic/LuminosityPublicResults}.

\bibitem{Aad:2008zzm}
{ATLAS} Collaboration, {\em {The ATLAS Experiment at the CERN Large Hadron
  Collider}\/},
\href{http://dx.doi.org/10.1088/1748-0221/3/08/S08003}{JINST {\bf 3} (2008)
  S08003}.
%%CITATION = JINST,3,S08003;%%.

\bibitem{Collaboration:2010knc}
{ATLAS} Collaboration, {\em {Performance of the ATLAS Detector using First
  Collision Data}\/},  \href{http://dx.doi.org/10.1007/JHEP09(2010)056}{JHEP
  {\bf 1009} (2010)  056},
\href{http://arxiv.org/abs/1005.5254}{{\tt arXiv:1005.5254 [hep-ex]}}.
%%CITATION = ARXIV:1005.5254;%%.

\bibitem{Aad:2010ac}
{ATLAS} Collaboration, {\em {Charged-particle multiplicities in pp interactions
  measured with the ATLAS detector at the LHC}\/},
  \href{http://dx.doi.org/10.1088/1367-2630/13/5/053033}{New J. Phys. {\bf 13}
  (2011)  053033},
\href{http://arxiv.org/abs/1012.5104}{{\tt arXiv:1012.5104 [hep-ex]}}.
%%CITATION = ARXIV:1012.5104;%%.

\bibitem{Aad:2012vm}
{ATLAS} Collaboration, {\em {Single hadron response measurement and calorimeter
  jet energy scale uncertainty with the ATLAS detector at the LHC}\/},
  \href{http://dx.doi.org/10.1140/epjc/s10052-013-2305-1}{Eur. Phys. J. {\bf
  C73} (2013)  2305},
\href{http://arxiv.org/abs/1203.1302}{{\tt arXiv:1203.1302 [hep-ex]}}.
%%CITATION = ARXIV:1203.1302;%%.

\bibitem{Chatrchyan:2008aa}
{CMS} Collaboration, {\em {The CMS experiment at the CERN LHC}\/},
\href{http://dx.doi.org/10.1088/1748-0221/3/08/S08004}{JINST {\bf 3} (2008)
  S08004}.
%%CITATION = JINST,3,S08004;%%.

\bibitem{CMS-PAS-PFT-09-001}
{CMS} Collaboration, {\em Particle-Flow Event Reconstruction in CMS and
  Performance for Jets, Taus, and MET\/},
  \href{https://cds.cern.ch/record/1194487}{CMS-PAS-PFT-09-001}.
  \url{https://cds.cern.ch/record/1194487}.

\bibitem{CMS-PAS-PFT-10-002}
{CMS} Collaboration, {\em Commissioning of the Particle-Flow reconstruction in
  Minimum-Bias and Jet Events from pp Collisions at 7 TeV\/},
  \href{https://cds.cern.ch/record/1279341}{CMS-PAS-PFT-10-002}.
  \url{https://cds.cern.ch/record/1279341}.

\bibitem{pythia}
T.~Sjostrand, S.~Mrenna, and P.~Z. Skands, {\em {PYTHIA 6.4 Physics and
  Manual}\/},  \href{http://dx.doi.org/10.1088/1126-6708/2006/05/026}{JHEP {\bf
  0605} (2006)  026},
\href{http://arxiv.org/abs/hep-ph/0603175}{{\tt arXiv:hep-ph/0603175
  [hep-ph]}}.
%%CITATION = HEP-PH/0603175;%%.

\bibitem{pythia8}
T.~Sjostrand, S.~Mrenna, and P.~Z. Skands, {\em {A Brief Introduction to PYTHIA
  8.1}\/},  \href{http://dx.doi.org/10.1016/j.cpc.2008.01.036}{Comput. Phys.
  Commun. {\bf 178} (2008)  852--867},
\href{http://arxiv.org/abs/0710.3820}{{\tt arXiv:0710.3820 [hep-ph]}}.
%%CITATION = ARXIV:0710.3820;%%.

\bibitem{pythiapartonshower}
R.~Corke and T.~Sjostrand, {\em {Improved Parton Showers at Large Transverse
  Momenta}\/},  \href{http://dx.doi.org/10.1140/epjc/s10052-010-1409-0}{Eur.
  Phys. J. {\bf C 69} (2010)  1--18},
  \href{http://arxiv.org/abs/1003.2384}{{\tt arXiv:1003.2384 [hep-ph]}}.

\bibitem{Sjostrand:2006su}
T.~Sjostrand, {\em {Monte Carlo Generators}\/},
\href{http://arxiv.org/abs/hep-ph/0611247}{{\tt arXiv:hep-ph/0611247
  [hep-ph]}}.
%%CITATION = HEP-PH/0611247;%%.

\bibitem{Andersson:1983ia}
B.~Andersson, G.~Gustafson, G.~Ingelman, and T.~Sjostrand, {\em {Parton
  Fragmentation and String Dynamics}\/},
\href{http://dx.doi.org/10.1016/0370-1573(83)90080-7}{Phys. Rept. {\bf 97}
  (1983)  31--145}.
%%CITATION = PRPLC,97,31;%%.

\bibitem{AUET2B}
{ATLAS} Collaboration, {\em ATLAS tunes of PYTHIA 6 and PYTHIA 8 for MC11\/},
  \href{https://cds.cern.ch/record/1363300}{ATL-PHYS-PUB-2011-009}.
  \url{https://cds.cern.ch/record/1363300}.

\bibitem{AMBT1}
{ATLAS} Collaboration, {\em New ATLAS event generator tunes to 2010 data\/},
  \href{https://cdsweb.cern.ch/record/1345343}{ATL-PHYS-PUB-2011-008}.
  \url{https://cdsweb.cern.ch/record/1345343}.

\bibitem{Marchesini:1987cf}
G.~Marchesini and B.~Webber, {\em {Monte Carlo Simulation of General Hard
  Processes with Coherent QCD Radiation}\/},
\href{http://dx.doi.org/10.1016/0550-3213(88)90089-2}{Nucl. Phys. {\bf B310}
  (1988)  461}.
%%CITATION = NUPHA,B310,461;%%.

\bibitem{Corcella:2002jc}
G.~Corcella, I.~Knowles, G.~Marchesini, S.~Moretti, K.~Odagiri, et al., {\em
  {HERWIG 6.5 release note}\/},
\href{http://arxiv.org/abs/hep-ph/0210213}{{\tt arXiv:hep-ph/0210213
  [hep-ph]}}.
%%CITATION = HEP-PH/0210213;%%.

\bibitem{Webber:1983if}
B.~Webber, {\em {A QCD Model for Jet Fragmentation Including Soft Gluon
  Interference}\/},
\href{http://dx.doi.org/10.1016/0550-3213(84)90333-X}{Nucl. Phys. {\bf B238}
  (1984)  492}.
%%CITATION = NUPHA,B238,492;%%.

\bibitem{jimmy}
J.~M. Butterworth, J.~R. Forshaw, and M.~H. Seymour, {\em {Multiparton
  interactions in photoproduction at HERA}\/},
  \href{http://dx.doi.org/10.1007/s002880050286}{Z. Phys. C {\bf 72} (1996)
  637},
\href{http://arxiv.org/abs/9601371}{{\tt arXiv:9601371 [hep-ph]}}.
%%CITATION = HEP-PH/9601371;%%.

\bibitem{AUET1}
{ATLAS} Collaboration, {\em First tuning of HERWIG/JIMMY to {ATLAS} data\/},
  \href{http://cdsweb.cern.ch/record/1303025}{ATL-PHYS-PUB-2010-014}.
  \url{http://cdsweb.cern.ch/record/1303025}.

\bibitem{Herwigpp}
M.~Bahr et al., {\em Herwig++ physics and manual\/},
  \href{http://dx.doi.org/10.1140/epjc/s10052-008-0798-9}{Eur. Phys. J. {\bf C
  58} (2008)  639}, \href{http://arxiv.org/abs/0803.0883}{{\tt arXiv:0803.0883
  [hep-ph]}}.

\bibitem{Bahr:2008dy}
M.~Bahr, S.~Gieseke, and M.~H. Seymour, {\em {Simulation of multiple partonic
  interactions in Herwig++}\/},
  \href{http://dx.doi.org/10.1088/1126-6708/2008/07/076}{JHEP {\bf 0807} (2008)
   076},
\href{http://arxiv.org/abs/0803.3633}{{\tt arXiv:0803.3633 [hep-ph]}}.
%%CITATION = ARXIV:0803.3633;%%.

\bibitem{mcatnlo}
S.~Frixione and B.~R. Webber, {\em {Matching NLO QCD computations and parton
  shower simulations}\/},
  \href{http://dx.doi.org/10.1088/1126-6708/2002/06/029}{JHEP {\bf 06} (2002)
  029},
\href{http://arxiv.org/abs/hep-ph/0204244}{{\tt arXiv:hep-ph/0204244
  [hep-ph]}}.
%%CITATION = HEP-PH/0204244;%%.

\bibitem{Frixione:2007vw}
S.~Frixione, P.~Nason, and C.~Oleari, {\em {Matching NLO QCD computations with
  Parton Shower simulations: the POWHEG method}\/},
  \href{http://dx.doi.org/10.1088/1126-6708/2007/11/070}{JHEP {\bf 11} (2007)
  070}, \href{http://arxiv.org/abs/0709.2092}{{\tt arXiv:0709.2092 [hep-ph]}}.

\bibitem{Krauss:2001iv}
F.~Krauss, R.~Kuhn, and G.~Soff, {\em {AMEGIC++ 1.0: A Matrix element generator
  in C++}\/},  \href{http://dx.doi.org/10.1088/1126-6708/2002/02/044}{JHEP {\bf
  0202} (2002)  044},
\href{http://arxiv.org/abs/hep-ph/0109036}{{\tt arXiv:hep-ph/0109036
  [hep-ph]}}.
%%CITATION = HEP-PH/0109036;%%.

\bibitem{Gleisberg:2003xi}
T.~Gleisberg, S.~Hoeche, F.~Krauss, A.~Schalicke, S.~Schumann, et al., {\em
  {SHERPA 1. alpha: A Proof of concept version}\/},
  \href{http://dx.doi.org/10.1088/1126-6708/2004/02/056}{JHEP {\bf 0402} (2004)
   056},
\href{http://arxiv.org/abs/hep-ph/0311263}{{\tt arXiv:hep-ph/0311263
  [hep-ph]}}.
%%CITATION = HEP-PH/0311263;%%.

\bibitem{Gleisberg:2008ta}
T.~Gleisberg, S.~Hoeche, F.~Krauss, M.~Schonherr, S.~Schumann, et al., {\em
  {Event generation with SHERPA 1.1}\/},
  \href{http://dx.doi.org/10.1088/1126-6708/2009/02/007}{JHEP {\bf 0902} (2009)
   007},
\href{http://arxiv.org/abs/0811.4622}{{\tt arXiv:0811.4622 [hep-ph]}}.
%%CITATION = ARXIV:0811.4622;%%.

\bibitem{Mangano:2002ea}
M.~L. Mangano, M.~Moretti, F.~Piccinini, R.~Pittau, and A.~D. Polosa, {\em
  {ALPGEN, a generator for hard multiparton processes in hadronic
  collisions}\/},  \href{http://dx.doi.org/10.1088/1126-6708/2003/07/001}{JHEP
  {\bf 0307} (2003)  001},
\href{http://arxiv.org/abs/hep-ph/0206293}{{\tt arXiv:hep-ph/0206293
  [hep-ph]}}.
%%CITATION = HEP-PH/0206293;%%.

\bibitem{Stelzer:1994ta}
T.~Stelzer and W.~Long, {\em {Automatic generation of tree level helicity
  amplitudes}\/},
  \href{http://dx.doi.org/10.1016/0010-4655(94)90084-1}{Comput. Phys. Commun.
  {\bf 81} (1994)  357--371},
\href{http://arxiv.org/abs/hep-ph/9401258}{{\tt arXiv:hep-ph/9401258
  [hep-ph]}}.
%%CITATION = HEP-PH/9401258;%%.

\bibitem{Alwall:2007st}
J.~Alwall, P.~Demin, S.~de~Visscher, R.~Frederix, M.~Herquet, et al., {\em
  {MadGraph/MadEvent v4: The New Web Generation}\/},
  \href{http://dx.doi.org/10.1088/1126-6708/2007/09/028}{JHEP {\bf 0709} (2007)
   028},
\href{http://arxiv.org/abs/0706.2334}{{\tt arXiv:0706.2334 [hep-ph]}}.
%%CITATION = ARXIV:0706.2334;%%.

\bibitem{Kersevan:2004yg}
B.~P. Kersevan and E.~Richter-Was, {\em {The Monte Carlo event generator AcerMC
  version 2.0 with interfaces to PYTHIA 6.2 and HERWIG 6.5}\/},
\href{http://arxiv.org/abs/hep-ph/0405247}{{\tt arXiv:hep-ph/0405247
  [hep-ph]}}.
%%CITATION = HEP-PH/0405247;%%.

\bibitem{Agostinelli:2002hh}
{GEANT4} Collaboration, {\em {GEANT4: A Simulation toolkit}\/},
\href{http://dx.doi.org/10.1016/S0168-9002(03)01368-8}{Nucl. Instrum. Meth.
  {\bf A506} (2003)  250--303}.
%%CITATION = NUIMA,A506,250;%%.

\bibitem{RichterWas:2002ch}
E.~Richter-Was, {\em {AcerDET: A Particle level fast simulation and
  reconstruction package for phenomenological studies on high $p_T$ physics at
  LHC}\/},
\href{http://arxiv.org/abs/hep-ph/0207355}{{\tt arXiv:hep-ph/0207355
  [hep-ph]}}.
%%CITATION = HEP-PH/0207355;%%.

\bibitem{Ovyn:2009tx}
S.~Ovyn, X.~Rouby, and V.~Lemaitre, {\em {DELPHES, a framework for fast
  simulation of a generic collider experiment}\/},
\href{http://arxiv.org/abs/0903.2225}{{\tt arXiv:0903.2225 [hep-ph]}}.
%%CITATION = ARXIV:0903.2225;%%.

\bibitem{deFavereau:2013fsa}
J.~de~Favereau, C.~Delaere, P.~Demin, A.~Giammanco, V.~Lemaître, et al., {\em
  {DELPHES 3, A modular framework for fast simulation of a generic collider
  experiment}\/},
\href{http://arxiv.org/abs/1307.6346}{{\tt arXiv:1307.6346 [hep-ex]}}.
%%CITATION = ARXIV:1307.6346;%%.

\bibitem{PGS}
J.~Conway et al., {\em {PGS 4 -- Pretty Good Simulation of high energy
  collisions}\/},  2006.
\newblock
  \url{http://www.physics.ucdavis.edu/~conway/research/software/pgs/pgs4-general.htm}.

\bibitem{Cacciari:2005hq}
M.~Cacciari and G.~P. Salam, {\em {Dispelling the $N^{3}$ myth for the $k_t$
  jet-finder}\/},
  \href{http://dx.doi.org/10.1016/j.physletb.2006.08.037}{Phys. Lett. {\bf
  B641} (2006)  57--61},
\href{http://arxiv.org/abs/hep-ph/0512210}{{\tt arXiv:hep-ph/0512210
  [hep-ph]}}.
%%CITATION = HEP-PH/0512210;%%.

\bibitem{Cacciari:2011ma}
M.~Cacciari, G.~P. Salam, and G.~Soyez, {\em {FastJet User Manual}\/},
  \href{http://dx.doi.org/10.1140/epjc/s10052-012-1896-2}{Eur. Phys. J. {\bf
  C72} (2012)  1896},
\href{http://arxiv.org/abs/1111.6097}{{\tt arXiv:1111.6097 [hep-ph]}}.
%%CITATION = ARXIV:1111.6097;%%.

\bibitem{Aad:2011he}
{ATLAS} Collaboration, {\em {Jet energy measurement with the ATLAS detector in
  proton-proton collisions at $\sqrt{s}$ = 7 TeV}\/},
  \href{http://dx.doi.org/10.1140/epjc/s10052-013-2304-2}{Eur. Phys. J. {\bf
  C73} (2013)  2304},
\href{http://arxiv.org/abs/1112.6426}{{\tt arXiv:1112.6426 [hep-ex]}}.
%%CITATION = ARXIV:1112.6426;%%.

\bibitem{ATLAS-CONF-2013-004}
{ATLAS} Collaboration, {\em Jet energy scale and its systematic uncertainty in
  proton-proton collisions at $\sqrt{s}$ = 7 TeV with ATLAS 2011 data\/},
  \href{http://cdsweb.cern.ch/record/1509552}{ATLAS-CONF-2013-004}.
  \url{http://cdsweb.cern.ch/record/1509552}.

\bibitem{ATLAS-CONF-2013-084}
{ATLAS} Collaboration, {\em Performance of boosted top quark identification in
  2012 ATLAS data\/},
  \href{https://cds.cern.ch/record/1571040}{ATLAS-CONF-2013-084}.
  \url{https://cds.cern.ch/record/1571040}.

\bibitem{Cacciari:2007fd}
M.~Cacciari and G.~P. Salam, {\em {Pileup subtraction using jet areas}\/},
  \href{http://dx.doi.org/10.1016/j.physletb.2007.09.077}{Phys. Lett. {\bf
  B659} (2008)  119--126},
\href{http://arxiv.org/abs/0707.1378}{{\tt arXiv:0707.1378 [hep-ph]}}.
%%CITATION = ARXIV:0707.1378;%%.

\bibitem{ATLAS-CONF-2013-083}
{ATLAS} Collaboration, {\em Pile-up subtraction and suppression for jets in
  ATLAS\/},  \href{https://cds.cern.ch/record/1570994}{ATLAS-CONF-2013-083}.
  \url{https://cds.cern.ch/record/1570994}.

\bibitem{Cacciari:2008gn}
M.~Cacciari, G.~P. Salam, and G.~Soyez, {\em {The Catchment Area of Jets}\/},
  \href{http://dx.doi.org/10.1088/1126-6708/2008/04/005}{JHEP {\bf 04} (2008)
  005},
\href{http://arxiv.org/abs/0802.1188}{{\tt arXiv:0802.1188 [hep-ph]}}.
%%CITATION = ARXIV:0802.1188;%%.

\bibitem{Aad:2013gja}
{ATLAS} Collaboration, {\em {Performance of jet substructure techniques for
  large-$R$ jets in proton-proton collisions at $\sqrt{s}$ = 7 TeV using the
  ATLAS detector}\/},  \href{http://dx.doi.org/10.1007/JHEP09(2013)076}{JHEP
  {\bf 1309} (2013)  076},
\href{http://arxiv.org/abs/1306.4945}{{\tt arXiv:1306.4945 [hep-ex]}}.
%%CITATION = ARXIV:1306.4945;%%.

\bibitem{htt_jes}
S.~{Sch\"atzel}, {\em Calibration of HEPTopTagger Jets in ATLAS\/},
  \href{https://cds.cern.ch/record/1483209}{ATL-COM-PHYS-2012-1461}.
  \url{https://cds.cern.ch/record/1483209}. Accessible only for ATLAS members.

\bibitem{htt_jer}
S.~{Sch\"atzel}, {\em Jet \pt Resolution for HEPTopTagger Jets in ATLAS\/},
  \href{https://cds.cern.ch/record/1483229}{ATL-COM-PHYS-2012-1463}.
  \url{https://cds.cern.ch/record/1483229}. Accessible only for ATLAS members.

\bibitem{ATLAS-CONF-2012-043}
{ATLAS} Collaboration, {\em Measurement of the b-tag Efficiency in a Sample of
  Jets Containing Muons with 5~fb$^{-1}$ of Data from the ATLAS Detector\/},
  \href{https://cds.cern.ch/record/1435197}{ATLAS-CONF-2012-043}.
  \url{https://cds.cern.ch/record/1435197}.

\bibitem{ATLAS-CONF-2012-097}
{ATLAS} Collaboration, {\em Measuring the b-tag efficiency in a top-pair sample
  with 4.7~fb$^{-1}$ of data from the ATLAS detector\/},
  \href{https://cds.cern.ch/record/1460443}{ATLAS-CONF-2012-097}.
  \url{https://cds.cern.ch/record/1460443}.

\bibitem{ATLAS-CONF-2012-040}
{ATLAS} Collaboration, {\em Measurement of the Mistag Rate with 5~fb$^{-1}$ of
  Data Collected by the ATLAS Detector\/},
  \href{https://cds.cern.ch/record/1435194}{ATLAS-CONF-2012-040}.
  \url{https://cds.cern.ch/record/1435194}.

\bibitem{Chatrchyan:2011ds}
{CMS} Collaboration, {\em {Determination of Jet Energy Calibration and
  Transverse Momentum Resolution in CMS}\/},
  \href{http://dx.doi.org/10.1088/1748-0221/6/11/P11002}{JINST {\bf 6} (2011)
  P11002},
\href{http://arxiv.org/abs/1107.4277}{{\tt arXiv:1107.4277 [physics.ins-det]}}.
%%CITATION = ARXIV:1107.4277;%%.

\bibitem{Abbott:1998xw}
{D0} Collaboration, {\em {Determination of the absolute jet energy scale in the
  D0 calorimeters}\/},
  \href{http://dx.doi.org/10.1016/S0168-9002(98)01368-0}{Nucl. Instrum. Meth.
  {\bf A424} (1999)  352--394},
\href{http://arxiv.org/abs/hep-ex/9805009}{{\tt arXiv:hep-ex/9805009
  [hep-ex]}}.
%%CITATION = HEP-EX/9805009;%%.

\bibitem{CMS-PAS-JME-09-002}
{CMS} Collaboration, {\em The Jet Plus Tracks Algorithm for Calorimeter Jet
  Energy Corrections in CMS\/},
  \href{https://cds.cern.ch/record/1190234}{CMS-PAS-JME-09-002}.
  \url{https://cds.cern.ch/record/1190234}.

\bibitem{Almeida:2010pa}
L.~G. Almeida, S.~J. Lee, G.~Perez, G.~Sterman, and I.~Sung, {\em {Template
  Overlap Method for Massive Jets}\/},
  \href{http://dx.doi.org/10.1103/PhysRevD.82.054034}{Phys. Rev. {\bf D82}
  (2010)  054034},
\href{http://arxiv.org/abs/1006.2035}{{\tt arXiv:1006.2035 [hep-ph]}}.
%%CITATION = ARXIV:1006.2035;%%.

\bibitem{Almeida:2011aa}
L.~G. Almeida, O.~Erdogan, J.~Juknevich, S.~J. Lee, G.~Perez, et al., {\em
  {Three-particle templates for a boosted Higgs boson}\/},
  \href{http://dx.doi.org/10.1103/PhysRevD.85.114046}{Phys. Rev. {\bf D85}
  (2012)  114046},
\href{http://arxiv.org/abs/1112.1957}{{\tt arXiv:1112.1957 [hep-ph]}}.
%%CITATION = ARXIV:1112.1957;%%.

\bibitem{Aad:2012raa}
{ATLAS} Collaboration, {\em {Search for resonances decaying into top-quark
  pairs using fully hadronic decays in $pp$ collisions with ATLAS at $\sqrt{s}$
  = 7 TeV}\/},  \href{http://dx.doi.org/10.1007/JHEP01(2013)116}{JHEP {\bf
  1301} (2013)  116},
\href{http://arxiv.org/abs/1211.2202}{{\tt arXiv:1211.2202 [hep-ex]}}.
%%CITATION = ARXIV:1211.2202;%%.

\bibitem{Alon:2011xb}
R.~Alon, E.~Duchovni, G.~Perez, A.~P. Pranko, and P.~K. Sinervo, {\em {A
  Data-driven method of pile-up correction for the substructure of massive
  jets}\/},  \href{http://dx.doi.org/10.1103/PhysRevD.84.114025}{Phys. Rev.
  {\bf D84} (2011)  114025},
\href{http://arxiv.org/abs/1101.3002}{{\tt arXiv:1101.3002 [hep-ph]}}.
%%CITATION = ARXIV:1101.3002;%%.

\bibitem{Aaltonen:2011pg}
{CDF} Collaboration, {\em {Study of Substructure of High Transverse Momentum
  Jets Produced in Proton-Antiproton Collisions at $\sqrt{s}$ = 1.96 TeV}\/},
  \href{http://dx.doi.org/10.1103/PhysRevD.85.091101}{Phys. Rev. {\bf D85}
  (2012)  091101},
\href{http://arxiv.org/abs/1106.5952}{{\tt arXiv:1106.5952 [hep-ex]}}.
%%CITATION = ARXIV:1106.5952;%%.

\bibitem{Kaplan:2008ie}
D.~E. Kaplan, K.~Rehermann, M.~D. Schwartz, and B.~Tweedie, {\em {Top Tagging:
  A Method for Identifying Boosted Hadronically Decaying Top Quarks}\/},
  \href{http://dx.doi.org/10.1103/PhysRevLett.101.142001}{Phys. Rev. Lett. {\bf
  101} (2008)  142001},
\href{http://arxiv.org/abs/0806.0848}{{\tt arXiv:0806.0848 [hep-ph]}}.
%%CITATION = ARXIV:0806.0848;%%.

\bibitem{CMS-PAS-JME-09-001}
{CMS} Collaboration, {\em A Cambridge-Aachen (C-A) based Jet Algorithm for
  boosted top-jet tagging\/},
  \href{https://cds.cern.ch/record/1194489}{CMS-PAS-JME-09-001}.
  \url{https://cds.cern.ch/record/1194489}.

\bibitem{Chatrchyan:2012ku}
{CMS} Collaboration, {\em {Search for anomalous t t-bar production in the
  highly-boosted all-hadronic final state}\/},
  \href{http://dx.doi.org/10.1007/JHEP09(2012)029}{JHEP {\bf 1209} (2012)
  029},
\href{http://arxiv.org/abs/1204.2488}{{\tt arXiv:1204.2488 [hep-ex]}}.
%%CITATION = ARXIV:1204.2488;%%.

\bibitem{Plehn:2011sj}
T.~Plehn, M.~Spannowsky, and M.~Takeuchi, {\em {How to Improve Top Tagging}\/},
   \href{http://dx.doi.org/10.1103/PhysRevD.85.034029}{Phys. Rev. {\bf D85}
  (2012)  034029},
\href{http://arxiv.org/abs/1111.5034}{{\tt arXiv:1111.5034 [hep-ph]}}.
%%CITATION = ARXIV:1111.5034;%%.

\bibitem{Aad:2012ky}
{ATLAS} Collaboration, {\em {Measurement of the W boson polarization in top
  quark decays with the ATLAS detector}\/},
  \href{http://dx.doi.org/10.1007/JHEP06(2012)088}{JHEP {\bf 1206} (2012)
  088},
\href{http://arxiv.org/abs/1205.2484}{{\tt arXiv:1205.2484 [hep-ex]}}.
%%CITATION = ARXIV:1205.2484;%%.

\bibitem{Gieseke:2012ft}
S.~Gieseke, C.~Rohr, and A.~Siodmok, {\em {Colour reconnections in
  Herwig++}\/},  \href{http://dx.doi.org/10.1140/epjc/s10052-012-2225-5}{Eur.
  Phys. J. {\bf C 72} (2012)  2225},
\href{http://arxiv.org/abs/1206.0041}{{\tt arXiv:1206.0041 [hep-ph]}}.
%%CITATION = ARXIV:1206.0041;%%.

\bibitem{Schaetzel:2013vka}
S.~{Sch\"atzel} and M.~Spannowsky, {\em {Tagging highly boosted top quarks}\/},
   \href{http://dx.doi.org/10.1103/PhysRevD.89.014007}{Phys. Rev. {\bf D89}
  (2014)  014007},
\href{http://arxiv.org/abs/1308.0540}{{\tt arXiv:1308.0540 [hep-ph]}}.
%%CITATION = ARXIV:1308.0540;%%.

\bibitem{Aad:2013nca}
{ATLAS} Collaboration, {\em {A search for \ttbar resonances in the lepton plus
  jets final state with ATLAS using 4.7 fb$^{-1}$ of pp collisions at
  $\sqrt{s}$ = 7 TeV}\/},
  \href{http://dx.doi.org/10.1103/PhysRevD.88.012004}{Phys. Rev. {\bf D88}
  (2013)  012004},
\href{http://arxiv.org/abs/1305.2756}{{\tt arXiv:1305.2756 [hep-ex]}}.
%%CITATION = ARXIV:1305.2756;%%.

\bibitem{ATLAS-CONF-2013-052}
{ATLAS} Collaboration, {\em A search for \ttbar resonances in lepton plus jets
  events with ATLAS using 14~fb$^{-1}$ of proton-proton collisions at
  $\sqrt{s}$ =8 TeV\/},
  \href{http://cdsweb.cern.ch/record/1547568}{ATLAS-CONF-2013-052}.
  \url{http://cdsweb.cern.ch/record/1547568}.

\bibitem{Hook:2012fd}
A.~Hook, E.~Izaguirre, M.~Lisanti, and J.~G. Wacker, {\em {High Multiplicity
  Searches at the LHC Using Jet Masses}\/},
  \href{http://dx.doi.org/10.1103/PhysRevD.85.055029}{Phys. Rev. {\bf D85}
  (2012)  055029},
\href{http://arxiv.org/abs/1202.0558}{{\tt arXiv:1202.0558 [hep-ph]}}.
%%CITATION = ARXIV:1202.0558;%%.

\bibitem{Hedri:2013pvl}
S.~El~Hedri, A.~Hook, M.~Jankowiak, and J.~G. Wacker, {\em {Learning How to
  Count: A High Multiplicity Search for the LHC}\/},
  \href{http://dx.doi.org/10.1007/JHEP08(2013)136}{JHEP {\bf 1308} (2013)
  136},
\href{http://arxiv.org/abs/1302.1870}{{\tt arXiv:1302.1870 [hep-ph]}}.
%%CITATION = ARXIV:1302.1870;%%.

\bibitem{Abazov:2011gv}
{D0} Collaboration, {\em {Search for a Narrow $t\bar{t}$ Resonance in
  $p\bar{p}$ Collisions at $\sqrt{s}$ = 1.96 TeV}\/},
  \href{http://dx.doi.org/10.1103/PhysRevD.85.051101}{Phys. Rev. {\bf D85}
  (2012)  051101},
\href{http://arxiv.org/abs/1111.1271}{{\tt arXiv:1111.1271 [hep-ex]}}.
%%CITATION = ARXIV:1111.1271;%%.

\bibitem{Aaltonen:2011ts}
{CDF} Collaboration, {\em {A Search for resonant production of $t\bar{t}$ pairs
  in $4.8\ \rm{fb}^{-1}$ of integrated luminosity of $p\bar{p}$ collisions at
  $\sqrt{s}$ = 1.96 TeV}\/},
  \href{http://dx.doi.org/10.1103/PhysRevD.84.072004}{Phys. Rev. {\bf D84}
  (2011)  072004},
\href{http://arxiv.org/abs/1107.5063}{{\tt arXiv:1107.5063 [hep-ex]}}.
%%CITATION = ARXIV:1107.5063;%%.

\bibitem{Aaltonen:2011vi}
{CDF} Collaboration, {\em {Search for resonant production of $t\bar{t}$
  decaying to jets in $p\bar p$ collisions at $\sqrt{s}$ = 1.96 TeV}\/},
  \href{http://dx.doi.org/10.1103/PhysRevD.84.072003}{Phys. Rev. {\bf D84}
  (2011)  072003},
\href{http://arxiv.org/abs/1108.4755}{{\tt arXiv:1108.4755 [hep-ex]}}.
%%CITATION = ARXIV:1108.4755;%%.

\bibitem{Aaltonen:2012af}
{CDF} Collaboration, {\em {Search for Resonant $t\bar{t}$ Production in the
  Semi-leptonic Decay Mode Using the Full CDF Data Set}\/},
  \href{http://dx.doi.org/10.1103/PhysRevLett.110.121802}{Phys. Rev. Lett. {\bf
  110} (2013)  121802},
\href{http://arxiv.org/abs/1211.5363}{{\tt arXiv:1211.5363 [hep-ex]}}.
%%CITATION = ARXIV:1211.5363;%%.

\bibitem{Harris:2011ez}
R.~M. Harris and S.~Jain, {\em {Cross Sections for Leptophobic Topcolor Z'
  Decaying to Top-Antitop}\/},
  \href{http://dx.doi.org/10.1140/epjc/s10052-012-2072-4}{Eur. Phys. J. {\bf
  C72} (2012)  2072},
\href{http://arxiv.org/abs/1112.4928}{{\tt arXiv:1112.4928 [hep-ph]}}.
%%CITATION = ARXIV:1112.4928;%%.

\bibitem{Lillie:2007ve}
B.~Lillie, J.~Shu, and T.~M. Tait, {\em {Kaluza-Klein Gluons as a Diagnostic of
  Warped Models}\/},  \href{http://dx.doi.org/10.1103/PhysRevD.76.115016}{Phys.
  Rev. {\bf D76} (2007)  115016},
\href{http://arxiv.org/abs/0706.3960}{{\tt arXiv:0706.3960 [hep-ph]}}.
%%CITATION = ARXIV:0706.3960;%%.

\bibitem{Agashe:2006hk}
K.~Agashe, A.~Belyaev, T.~Krupovnickas, G.~Perez, and J.~Virzi, {\em {LHC
  Signals from Warped Extra Dimensions}\/},
  \href{http://dx.doi.org/10.1103/PhysRevD.77.015003}{Phys. Rev. {\bf D77}
  (2008)  015003},
\href{http://arxiv.org/abs/hep-ph/0612015}{{\tt arXiv:hep-ph/0612015
  [hep-ph]}}.
%%CITATION = HEP-PH/0612015;%%.

\bibitem{Djouadi:2007eg}
A.~Djouadi, G.~Moreau, and R.~K. Singh, {\em {Kaluza-Klein excitations of gauge
  bosons at the LHC}\/},
  \href{http://dx.doi.org/10.1016/j.nuclphysb.2007.12.024}{Nucl. Phys. {\bf
  B797} (2008)  1--26},
\href{http://arxiv.org/abs/0706.4191}{{\tt arXiv:0706.4191 [hep-ph]}}.
%%CITATION = ARXIV:0706.4191;%%.

\bibitem{Agashe:2007zd}
K.~Agashe, H.~Davoudiasl, G.~Perez, and A.~Soni, {\em {Warped Gravitons at the
  LHC and Beyond}\/},
  \href{http://dx.doi.org/10.1103/PhysRevD.76.036006}{Phys. Rev. {\bf D76}
  (2007)  036006},
\href{http://arxiv.org/abs/hep-ph/0701186}{{\tt arXiv:hep-ph/0701186
  [hep-ph]}}.
%%CITATION = HEP-PH/0701186;%%.

\bibitem{Chatrchyan:2012cx}
{CMS} Collaboration, {\em {Search for resonant $t\bar{t}$ production in
  lepton+jets events in $pp$ collisions at $\sqrt{s}$ = 7 TeV}\/},
  \href{http://dx.doi.org/10.1007/JHEP12(2012)015}{JHEP {\bf 1212} (2012)
  015},
\href{http://arxiv.org/abs/1209.4397}{{\tt arXiv:1209.4397 [hep-ex]}}.
%%CITATION = ARXIV:1209.4397;%%.

\bibitem{Chatrchyan:2013lca}
{CMS} Collaboration, {\em {Searches for anomalous ttbar production in pp
  collisions at $\sqrt{s}$=8 TeV}\/},
\href{http://arxiv.org/abs/1309.2030}{{\tt arXiv:1309.2030 [hep-ex]}}.
%%CITATION = ARXIV:1309.2030;%%.

\bibitem{Kasieczka2013}
G.~Kasieczka, {\em Search for Resonances Decaying into Top Quark Pairs Using
  Fully Hadronic Decays in pp Collisions with ATLAS at $\sqrt{s}$ = 7 TeV}.
\newblock PhD thesis, {Universit\"at Heidelberg, Germany}, 2013.
\newblock
  \url{http://www.physi.uni-heidelberg.de/Publications/Kasieczka-Doktor.pdf}.

\bibitem{Aad:2013wta}
{ATLAS} Collaboration, {\em {Search for new phenomena in final states with
  large jet multiplicities and missing transverse momentum at $\sqrt{s}$ = 8
  TeV proton-proton collisions using the ATLAS experiment}\/},
\href{http://arxiv.org/abs/1308.1841}{{\tt arXiv:1308.1841 [hep-ex]}}.
%%CITATION = ARXIV:1308.1841;%%.

\bibitem{Aad:2010rd}
{ATLAS} Collaboration, {\em {Charged-particle multiplicities in $pp$
  interactions at $\sqrt{s}$ = 900 GeV measured with the ATLAS detector at the
  LHC}\/},  \href{http://dx.doi.org/10.1016/j.physletb.2010.03.064}{Phys. Lett.
  {\bf B688} (2010)  21--42},
\href{http://arxiv.org/abs/1003.3124}{{\tt arXiv:1003.3124 [hep-ex]}}.
%%CITATION = ARXIV:1003.3124;%%.

\bibitem{Soper:2011cr}
D.~E. Soper and M.~Spannowsky, {\em {Finding physics signals with shower
  deconstruction}\/},
  \href{http://dx.doi.org/10.1103/PhysRevD.84.074002}{Phys. Rev. {\bf D84}
  (2011)  074002},
\href{http://arxiv.org/abs/1102.3480}{{\tt arXiv:1102.3480 [hep-ph]}}.
%%CITATION = ARXIV:1102.3480;%%.

\bibitem{Soper:2012pb}
D.~E. Soper and M.~Spannowsky, {\em {Finding top quarks with shower
  deconstruction}\/},
  \href{http://dx.doi.org/10.1103/PhysRevD.87.054012}{Phys. Rev. {\bf D87}
  (2013)  054012},
\href{http://arxiv.org/abs/1211.3140}{{\tt arXiv:1211.3140 [hep-ph]}}.
%%CITATION = ARXIV:1211.3140;%%.

\bibitem{Ellis:2012sn}
S.~D. Ellis, A.~Hornig, T.~S. Roy, D.~Krohn, and M.~D. Schwartz, {\em {Q-jets:
  A Non-Deterministic Approach to Tree-Based Jet Substructure}\/},
  \href{http://dx.doi.org/10.1103/PhysRevLett.108.182003}{Phys. Rev. Lett. {\bf
  108} (2012)  182003},
\href{http://arxiv.org/abs/1201.1914}{{\tt arXiv:1201.1914 [hep-ph]}}.
%%CITATION = ARXIV:1201.1914;%%.

\bibitem{ATLAS-CONF-2013-087}
{ATLAS} Collaboration, {\em Performance and Validation of Q-Jets at the ATLAS
  Detector in pp Collisions at $\sqrt{s}$ = 8 TeV in 2012\/},
  \href{http://cds.cern.ch/record/1572981}{ATLAS-CONF-2013-087}.
  \url{http://cds.cern.ch/record/1572981}.

\end{thebibliography}\endgroup

\clearpage

\section*{Danksagung}
\addcontentsline{toc}{section}{Danksagung}
\begin{otherlanguage}{german} 
Viele der Ergebnisse, die in dieser Habilitationsschrift dokumentiert sind,
wurden w\"ahrend der zur\"uckliegenden vier Jahre in der ATLAS-Arbeitsgruppe 
von Prof. Dr. Andr\'e Sch\"oning am Physikalischen Institut der Universit\"at 
Heidelberg erarbeitet.

Ich danke Prof. Sch\"oning f\"ur die Weitsicht, mit der er fr\"uhzeitig
das Potential des Felds {\em Jetstruktur} erkannt hat, und f\"ur das Herstellen
des Kontakts zu Prof. Plehn vom Theoretischen Institut.
Ich danke ihm auch f\"ur das mir entgegengebrachte Vertrauen, die Forschung 
eigenst\"andig zu betreiben, f\"ur Anregungen und Diskussionen sowie f\"ur seine 
uneingeschr\"ankte Unterst\"utzung bei der systematischen Ausweitung des
Arbeitsgebiets und dem Aufbau der Arbeitsgruppe.

Ich danke den Mitgliedern der HEPTopTagger-Arbeitsgruppe am Physikalischen
Institut, Dr. Gregor Kasieczka, Dr. Christoph Anders, Maddalena Giulini
und David Sosa, f\"ur die sehr fruchtbare Zusammenarbeit und den hohen Einsatz,
mit dem sie publikationsreife Ergebnisse erarbeiten.

Ich danke den Theoretischen Physikern Prof. Dr. Tilman Plehn, Dr. Michael Spannowsky,
Dr. Michihisa Takeuchi und Torben Schell f\"ur die exzellente Zusammenarbeit. 
Insbesondere die Diskussionen mit Prof. Plehn und Dr. Spannowsky haben entscheidend
zur Qualit\"at der Analysen beigetragen.

Ich danke auch den Mitgliedern der Arbeitsgruppe {\em Jet Substructure} der
ATLAS-Kolla\-bo\-ra\-ti\-on, insbesondere Dr. David Miller, Dr. Adam Davison und Dr. Emily Thompson
f\"ur Diskussionen, Hilfe und Unterst\"utzung.

Nicht zuletzt danke ich meiner Familie f\"ur ihren R\"uckhalt und ihre Unterst\"utzung.
\end{otherlanguage}

\end{document}